\documentclass{article}
\usepackage{amsmath}
\usepackage{amsfonts}
\usepackage{amssymb}
\usepackage{booktabs}
\usepackage{epubtk}
\usepackage{graphicx,epstopdf}
\usepackage[USenglish]{babel}
\usepackage{ifpdf}
\usepackage{color}
\usepackage[numbers]{natbib}
\usepackage[noblocks]{authblk}
\usepackage[normalem]{ulem}
\usepackage{soul}
\usepackage{multirow}

\renewcommand\citep{\cite}
\renewcommand\citet{\cite}
\renewcommand\Affilfont{\normalsize}

\makeatletter
  \@addtoreset{equation}{section}
  
  \@addtoreset{section}{part}
  \renewcommand{\thesection}{\thepart.\arabic{section}}
\makeatother 

\usepackage{array}

\def\d{\mathrm{d}}
\newcommand{\weff}{w_{\mathrm{eff}}}

\def\ba{\begin{eqnarray}}
\def\ea{\end{eqnarray}}

\def\d{\mathrm{d}}

\def\mn{_{\mu \nu}}

\def\({\left(}
\def\){\right)}
\def\p{\partial}

\def\vk{\vec{k}}
\def\fnl{f_{\mathrm{NL}}}

\newcommand{\data}{d}

\newcommand{\mdl}{\mathcal{M}}
\newcommand{\params}{{\Theta}}

\newcommand{\lnlike}{\ln \mathcal{L}}
\newcommand{\fid}{\params_*}
\newcommand{\dr}{\mathrm{d}}

\newcommand{\Uf}{\mathcal{U}}
\newcommand{\EU}{\mathcal{EU}}
\newcommand{\Df}{{D}_f}
\newcommand{\be}{\begin{equation}}
\newcommand{\ee}{\end{equation}}

\def\d{\mathrm{d}}

\newcommand{\Vsur}{V_{\mathrm{survey}}}
\newcommand{\DA}{D\!_A(z)}
\newcommand{\hz}{H(z)}

\newcommand{\Veff}{V_{\mathrm{eff}}}

\newcommand{\metE}{\ensuremath{\tilde{g}}}
\newcommand{\metM}{\ensuremath{g}}
\newcommand{\RiemE}{\ensuremath{\tilde{R}}}
\newcommand{\volE}{\ensuremath{\sqrt{-\metE}}}
\newcommand{\volM}{\ensuremath{\sqrt{-\metM}}}
\newcommand{\metS}{\ensuremath{\hat{g}}}
\newcommand{\connE}{\ensuremath{\tilde{\nabla}}}

\newcommand{\Vp}{\ensuremath{\frac{\mathrm{d}V}{\mathrm{d}\mu}}}

\newcommand{\kmax}{k_{\max}}

\newcommand{\CD}{\mathcal{D}}
\newcommand{\CQ}{\mathcal{Q}}
\newcommand{\CR}{\mathcal{R}}
\newcommand{\average}[1]{\left\langle #1 \right\rangle_\CD}

\newcommand{\symbolref}[1]{p.\ \pageref{#1}}

\newcommand{\newtext}[1]{{\color{blue}#1}}
\definecolor{darkgreen}{rgb}{0.0, 0.5, 0.13}
\newcommand{\reviewtext}[1]{{\color{darkgreen}#1}}

\begin{document}

\title{Cosmology and Fundamental Physics with the Euclid Satellite}

\def\headerauthor{Luca Amendola et al. (The Euclid Theory Working Group)}

\author{
Luca Amendola,
Stephen Appleby,
Anastasios Avgoustidis, 
David Bacon,
Tessa Baker, 
Marco Baldi, 
Nicola Bartolo,
Alain Blanchard,
Camille Bonvin, 
Stefano Borgani,
Enzo Branchini,
Clare Burrage,
Stefano Camera,
Carmelita Carbone, 
Luciano Casarini, 
Mark Cropper,
Claudia de Rham, 
J\"org P.\ Dietrich,
Cinzia Di Porto,
Ruth Durrer,
Anne Ealet,
Pedro G.\ Ferreira,
Fabio Finelli, 
Juan Garc{\'{\i}}a-Bellido,
Tommaso Giannantonio, 
Luigi Guzzo,
Alan Heavens, 
Lavinia Heisenberg, 
Catherine Heymans, 
Henk Hoekstra,
Lukas Hollenstein, 
Rory Holmes, 
Ole Horst, 
Zhiqi Hwang,
Knud Jahnke,
Thomas D.\ Kitching,
Tomi Koivisto, 
Martin Kunz,
Giuseppe La Vacca, 
Eric Linder,
Marisa March, 
Valerio Marra,
Carlos Martins,
Elisabetta Majerotto, 
Dida Markovic,
David Marsh, 
Federico Marulli, 
Richard Massey, 
Yannick Mellier,
Francesco Montanari,
David F.\ Mota, 
Nelson J.\ Nunes,
Will Percival, 
Valeria Pettorino,
Cristiano Porciani,
Claudia Quercellini,  
Justin Read,  
Massimiliano Rinaldi, 
Domenico Sapone,  
Ignacy Sawicki,
Roberto Scaramella, 
Constantinos Skordis,  
Fergus Simpson, 
Andy Taylor, 
Shaun Thomas, 
Roberto Trotta,
Licia Verde,
Filippo Vernizzi,
Adrian Vollmer, 
Yun Wang,
Jochen Weller,
Tom Zlosnik \\
(The Euclid Theory Working Group)}

\affil[1]{\,Please contact euclidtheoryreview@gmail.com for questions and comments.}

\medskip

\date{}
\maketitle

\begin{abstract}
Euclid is a European Space Agency medium-class mission selected for
launch in 2020 within the Cosmic
Vision 2015\,--\,2025 program. The main goal of Euclid is to
understand the origin of the accelerated expansion of the universe.
Euclid will explore the expansion
history of the universe and the evolution of cosmic structures  by
measuring shapes and red-shifts of
galaxies as well as the distribution
of clusters of galaxies over a large fraction of
the sky.

Although the main driver for Euclid is the nature of dark energy,
Euclid science covers a vast range of topics, from cosmology to
galaxy evolution to planetary research.
In this review we focus on  cosmology and fundamental physics, with a strong
emphasis on science beyond the current standard models. We discuss
five broad topics:
dark energy and modified gravity, dark matter, initial conditions,
basic assumptions and questions of methodology
in the data analysis.

This review has been planned and carried out within Euclid's Theory
Working Group and is meant to provide
a  guide to the scientific themes that will underlie the activity of
the group during the preparation of the Euclid mission.
\end{abstract}

\epubtkKeywords{dark energy, cosmology, galaxy evolution}


\newpage

\tableofcontents

\newpage

\section*{Credits version 2 (2016)}
\subsubsection*{Euclid Theory Working Group Editorial Board (2016):}

Valeria Pettorino (editor in chief)\\
Tessa Baker\\
Stefano Camera\\
Adrian Vollmer\\
Elisabetta Majerotto\\
Martin Kunz (Euclid Theory Working Group Coordinator)\\
Luca Amendola (Euclid Theory Working Group Coordinator)

\subsubsection*{Corresponding authors (2016):}

Luca Amendola\\
Tessa Baker\\
Marco Baldi \\
Stefano Camera\\
Thomas D.\ Kitching\\
Martin Kunz\\
Elisabetta Majerotto \\
Valerio Marra \\
Valeria Pettorino\\
Ignacy Sawicki \\
Licia Verde\\

\vspace{1cm}

\subsubsection*{Contributing Authors (2016):}

Luca Amendola,
Stephen Appleby,
Anastasios Avgoustidis, 
David Bacon,
Tessa Baker, 
Marco Baldi, 
Nicola Bartolo,
Alain Blanchard,
Camille Bonvin, 
Stefano Borgani,
Enzo Branchini,
Clare Burrage,
Stefano Camera,
Carmelita Carbone, 
Luciano Casarini, 
Mark Cropper,
Claudia de Rham, 
J\"org P.\ Dietrich,
Cinzia Di Porto,
Ruth Durrer,
Anne Ealet,
Pedro G.\ Ferreira,
Fabio Finelli, 
Juan Garc{\'{\i}}a-Bellido,
Tommaso Giannantonio, 
Luigi Guzzo,
Alan Heavens, 
Lavinia Heisenberg, 
Catherine Heymans, 
Henk Hoekstra,
Lukas Hollenstein, 
Rory Holmes, 
Ole Horst, 
Zhiqi Hwang,
Knud Jahnke,
Thomas D.\ Kitching,
Tomi Koivisto, 
Martin Kunz,
Giuseppe La Vacca, 
Eric Linder,
Marisa March, 
Valerio Marra,
Carlos Martins,
Elisabetta Majerotto, 
Dida Markovic,
David Marsh, 
Federico Marulli, 
Richard Massey, 
Yannick Mellier,
Francesco Montanari,
David F.\ Mota, 
Nelson J.\ Nunes,
Will Percival, 
Valeria Pettorino,
Cristiano Porciani,
Claudia Quercellini,  
Justin Read,  
Massimiliano Rinaldi, 
Domenico Sapone,  
Ignacy Sawicki,
Roberto Scaramella, 
Constantinos Skordis,  
Fergus Simpson, 
Andy Taylor, 
Shaun Thomas, 
Roberto Trotta,
Licia Verde,
Filippo Vernizzi,
Adrian Vollmer, 
Yun Wang,
Jochen Weller,
Tom Zlosnik.
\vspace{1cm}

\noindent
\emph{DISCLAIMER: This is not an official Euclid document and its
  content reflects solely the views of the contributing authors. Forecasts are not updated in this version, with respect to 2012.}

\newpage

\section*{Credits version 1 (2012)}
\subsubsection*{Euclid Theory Working Group Editorial Board (2012):}

Valeria Pettorino (editor in chief)\\
Tessa Baker\\
Stefano Camera\\
Elisabetta Majerotto\\
Marisa March\\
Cinzia Di Porto\\
Martin Kunz (Euclid Theory Working Group Coordinator)\\
Luca Amendola (Euclid Theory Working Group Coordinator)

\subsubsection*{Corresponding authors (2012):}

Luca Amendola\\
Stefano Camera\\
Cinzia Di Porto\\
Pedro G.\ Ferreira\\
Juan Garc{\'{\i}}a-Bellido\\
Thomas D.\ Kitching\\
Martin Kunz\\
Valeria Pettorino\\
Cristiano Porciani\\
Roberto Trotta\\
Licia Verde\\
Yun Wang

\vspace{1cm}

\subsubsection*{Contributing Authors (2012):}

Luca Amendola, Stephen Appleby, David Bacon, Tessa Baker, Marco Baldi, Nicola Bartolo, Alain Blanchard, Camille Bonvin, Stefano Borgani, Enzo Branchini, Clare Burrage, Stefano Camera, Carmelita Carbone, Luciano Casarini, Mark Cropper, Claudia de Rham, Cinzia Di Porto, Anne Ealet, Pedro G. Ferreira, Fabio Finelli, Juan Gar{\'{\i}}a-Bellido, Tommaso Giannantonio, Luigi Guzzo, Alan Heavens, Lavinia Heisenberg, Catherine Heymans, Henk Hoekstra, Lukas Hollenstein, Rory Holmes, Ole Horst, Knud Jahnke, Thomas D. Kitching, Tomi Koivisto, Martin Kunz, Giuseppe La Vacca, Marisa March, Elisabetta Majerotto, Dida Markovic, David Marsh, Federico Marulli, Richard Massey, Yannick Mellier, David F. Mota, Nelson J. Nunes, Will Percival, Valeria Pettorino, Cristiano Porciani, Claudia Quercellini, Justin Read, Massimiliano Rinaldi, Domenico Sapone, Roberto Scaramella, Constantinos Skordis, Fergus Simpson, Andy Taylor, Shaun Thomas, Roberto Trotta, Licia Verde, Filippo Vernizzi, Adrian Vollmer, Yun Wang, Jochen Weller, Tom Zlosnik.

\vspace{1cm}

\newpage

\section*{List of acronyms}
\ifpdf
\addcontentsline{toc}{subsection}{List of acronyms}\else
\fi

\begin{tabular}{ll}
 AGN & Active Galactic Nucleus\\
 ALP & Axio-Like Particle\\
 BAO & Baryonic Acoustic Oscillations\\
 BBKS & Bardeen--Bond--Kaiser--Szalay	\\ 
 BOSS & Baryon Oscillation Spectroscopic Survey	\\ 
 BPol & B-Polarization Satellite	\\ 
 BigBOSS & Baryon Oscillation Spectroskopic Survey	\\ 
 CAMB & Code for Anisotropies in the Microwave Background \\ 
 CDE & Coupled Dark Energy\\
 CDM & Cold Dark Matter\\
 CDMS & Cryogenic Dark Matter Search	\\ 
 CL & Confidence Level\\
CLASS & Cosmic Linear Anisotropy Solving System\\
 CMB & Cosmic Microwave Background\\
 COMBO-17 & Classifying Objects by Medium-Band Observations \\ 
 COSMOS & Cosmological Evolution Survey	\\ 
 CPL & Chevallier--Polarski--Linder	\\ 
 CQ & Coupled Quintessence\\
 CRESST & Cryogenic Rare Event Search with Superconducting
Thermometers\\
 DE & Dark Energy\\
 DES & Dark Energy Survey \\ 
 DETF & Dark Energy Task Force\\
 DGP & Dvali--Gabadadze--Porrati	\\ 
 DM & Dark Matter\\
 EBI & Eddington--Born--Infeld	\\ 
 EDE & Early Dark Energy\\
 EMT & Energy-Momentum Tensor \\
 EROS & Exp\'{e}rience pour la Recherche d'Objets Sombres	\\ 
 eROSITA & Extended ROentgen Survey with an Imaging Telescope Array \\ 
 FCDM & Fuzzy Cold Dark Matter\\
 FFT & Fast Fourier Transform\\
 FLRW & Friedmann--Lema{\^{\i}}tre--Robertson--Walker	\\ 
 FoM & Figure of Merit\\
 FoG & Fingers of God\\
 GEA & Generalized Einstein-Aether\\
 GR & General Relativity\\
 HETDEX & Hobby-Eberly Telescope Dark Energy Experiment \\ 
 ICM & Intracluster Medium\\
 IH & Inverted Hierarchy\\
 IR & Infrared\\
 ISW & Integrated Sachs--Wolfe 	\\ 
\ifpdf
\end{tabular}

\clearpage
\begin{tabular}{ll}
\else
\fi
 KL & Kullback--Leibler divergence	\\ 
 LCDM & Lambda Cold Dark Matter\\
 LHC & Large Hadron Collider\\
 LRG & Luminous Red Galaxy\\
 LSB & Low Surface Brightness\\
 LSS & Large Scale Structure\\
 LSST & Large Synoptic Survey Telescope	\\ 
 LTB & Lema\^{i}tre--Tolman--Bondi	\\ 
 MACHO & MAssive Compact Halo Object\\
 MCMC & Markov Chain Monte Carlo \\ 
 MCP & Mini-Charged Particles\\
 MF & Mass Function\\
 MG & Modified Gravity\\
 MOND & MOdified Newtonian Dynamics\\
 MaVaNs & Mass Varying Neutrinos \\
 NFW & Navarro--Frenk--White	\\ 
 NH & Normal Hierarchy\\
 PCA & Principal Component Analysis \\ 
 PDF & Probability Distribution Function \\ 
 PGB & Pseudo-Goldstein Boson\\
 PKDGRAV & Parallel K-D tree GRAVity code	\\ 
 PPF & Parameterized Post-Friedmann\\
 PPN & Parameterized Post-Newtonian\\
 PPOD & Predictive Posterior Odds Distribution\\
 PSF & Point Spread Function	\\ 
 QCD & Quantum ChromoDynamics \\ 
 RDS & Redshift Space Distortions \\ 
 RG & Renormalization Group\\
 SD & Savage--Dickey\\
 SDSS & Sloan Digital Sky Survey\\
 SIDM & Self Interacting Dark Matter\\
 SN & Supernova\\
 TeVeS & Tensor Vector Scalar\\
 UDM & Unified Dark Matter\\
 UV & Ultra Violett\\
 WDM & Warm Dark Matter\\
 WFXT & Wide-Field X-Ray Telescope\\
 WIMP & Weakly Interacting Massive Particle\\
 WKB & Wentzel--Kramers--Brillouin	\\ 
 WL & Weak Lensing\\
 WLS & Weak Lensing Survey\\
 WMAP & Wilkinson Microwave Anisotropy Probe\\
 XMM-Newton & X-ray Multi-Mirror Mission	\\ 
 vDVZ & van Dam--Veltman--Zakharov \\ 
\end{tabular}

\newpage
\section*{List of symbols}
\ifpdf
\addcontentsline{toc}{subsection}{List of symbols}\else
\fi

\begin{tabular}{lll}
$c_a$ & Adiabatic sound speed & \symbolref{symbol:ad.sound.speed}  \\
$D_A(z)$ & Angular diameter distance  & \symbolref{eq:ang} \\
$\;\raise1.0pt\hbox{/}\hskip-6pt\partial$ & Angular spin raising operator
& \symbolref{eq:spinraising} \\
$\Pi^i_j$ & Anisotropic stress perturbation tensor  & \symbolref{symbol:stressperturbation} \\
$\sigma$ & Uncertainty  & {} \\
$Bo$ & Bayes factor  & \symbolref{eq:bayesfactor} \\
$b$ & Bias (ratio of galaxy to total matter perturbations)  &
\symbolref{symbol:bias} \\
$B_\Phi(k_1,k_2,k_3)$ & Bispectrum of the Bardeen's  potential  & \symbolref{bispectrum} \\
$g(X)$ & Born--Infeld kinetic term  & \symbolref{UDML} \\
$\mathbf{b}$ & Bulleticity  & \symbolref{symbol:bulleticity} \\
$\zeta$ & Comoving curvature perturbation  & \symbolref{eq:zeta} \\
$r(z)$ & Comoving distance & {} \\
$\mathcal{H}$ & Conformal Hubble parameter, $\mathcal{H} = aH$  & \symbolref{symbol:conformaltime}  \\
$\eta,\tau$ & Conformal time  & \symbolref{symbol:conformaltime} \\
$\kappa$ & Convergence  & \symbolref{eq:kappadef} \\
$t$ & Cosmic time  & \symbolref{CQ_euler} \\
$\Lambda$ & Cosmological constant  & {} \\
$\mathbf{\Theta}$ & Cosmological parameters  & \symbolref{eq:likelihood_i} \\
$r_c$ & Cross over scale  & \symbolref{symbol:rc} \\
$\square$ & d'Alembertian, $\square=\mathbf{\nabla}^2$  & {} \\
$F$ & Derivative of $f(R)$  & \symbolref{symbol:FR} \\
$\theta$ & Divergence of velocity field  & \symbolref{ij_EM} \\
$\mu$ & Direction cosine & \symbolref{eq:Pgini} \\
$\pi$ & Effective anisotropic stress  & \symbolref{eq:real-sigma} \\
$\eta(a,k)$ & Effective anisotropic stress parameterization  & \symbolref{mod_constr} \\
$\rho$ & Energy density  & {} \\
$T_{\mu\nu}$ & Energy momentum tensor  & \symbolref{EMT} \\
$w$ & Equation of state  & \symbolref{phi_bkg} \\
$F_{\alpha\beta}$ & Fisher information matrix  & \symbolref{eq:Falphabeta} \\

$\sigma_8$ & Fluctuation amplitude at 8 km/s/Mpc  & {} \\
$u^\mu$ & Four-velocity  & \symbolref{EMT} \\
$\Omega_m$ & Fractional matter density  & {} \\
$f_{\mathrm{sky}}$ & Fraction of sky observed  & \symbolref{eq:3-fm} \\
$\Delta_M$ & Gauge invariant comoving density contrast  & \symbolref{symbol:DeltaM} \\
$\tau(z)$ & Generic opacity parameter  & \symbolref{symbol:tau} \\
$\varpi$ & Gravitational slip parameter  & \symbolref{symbol:varpi} \\
$G(a)$ & Growth function/Growth factor  & \symbolref{eq:growthdef} \\
\ifpdf
\end{tabular}

\clearpage
\begin{tabular}{lll}
\else
\fi
$\gamma$ & Growth index/Shear  & \symbolref{gamma_Delta}/\symbolref{eq:kappadef} \\
$f_g$ & Growth rate  & \symbolref{def-growth-rate} \\
$b_{\mathrm{eff}}$ & Halo effective linear bias factor  & \symbolref{symbol:halobias} \\
$h$ & Hubble constant in units of 100~km/s/Mpc  & {} \\
$H(z)$ & Hubble parameter  & {} \\
$\xi_i$ & Killing field   & \symbolref{eq:killing} \\
$\delta_{ij}$ & Kronecker delta  & {} \\
$f(R)$ & Lagrangian in modified gravity  & \symbolref{fRaction} \\
$P_l(\mu)$ & Legendre polynomials  & \symbolref{symbol:legendre} \\
$\mathcal{L}(\mathbf{\Theta})$ & Likelihood function  & \symbolref{eq:likelihood_i} \\
$\beta(z)$ & Linear redshift-space distortion parameter  & \symbolref{symbol:beta} \\
$D_L(z)$ & Luminosity distance  & \symbolref{symbol:luminosity} \\
$Q(a,k)$ & Mass screening effect  & \symbolref{mod_constr} \\
$\delta_m$ & Matter density perturbation  & {} \\
$g_{\mu\nu}$ & Metric tensor  & \symbolref{symbol:metric} \\
$\mu$ & Modified gravity function: $\mu=Q/\eta$  & \symbolref{eq:muQeta} \\
$C_\ell$ & Multipole power spectrum  & \symbolref{symbol:Cell} \\
$G$ & Newton's gravitational constant  & {} \\
$N$ & Number of e-folds, $N=\ln a$  & \symbolref{symbol:efolds} \\
$P(k)$ & Matter power spectrum  & {} \\
$p$ & Pressure  & {} \\
$\delta p$ & Pressure perturbation  & {} \\
$\chi(z)$ & Radial, dimensionless comoving distance  & \symbolref{chiz} \\
$z$ & Redshift  & {} \\
$R$ & Ricci scalar  & {} \\
$\phi$ & Scalar field  & \symbolref{symbol:phi} \\
$A$ & Scalar potential  & \symbolref{pert_0_ds} \\
$\Psi,\Phi$ & Scalar potentials  & \symbolref{symbol:scalarpotentials} \\
$n_s$ & Scalar spectral index  & \symbolref{eq:index_ns} \\
$a$ & Scale factor  & {} \\
$f_a$ & Scale of Peccei--Quinn symmetry breaking  & \symbolref{symbol:f_a} \\
$\ell$ & Spherical harmonic multipoles  & \\
$c_s$ & Sound speed  & \symbolref{symbol:c_s} \\
$\Sigma$ & Total neutrino mass/Inverse covariance matrix/PPN parameter  &
\symbolref{symbol:Sigma}/\symbolref{symbol:invcov}/\symbolref{Sigma-wl-isw} \\
$H_T^{ij}$ & Trace-free distortion  & \symbolref{symbol:H_T} \\
$T(k)$ & Transfer function  & \symbolref{symbol:Tk} \\
$B_i$ & Vector shift  & \symbolref{symbol:vectorshift} \\
$\mathbf{k}$ & Wavenumber  & {} \\
\end{tabular}

\newpage
\section*{Introduction}
\ifpdf
\addcontentsline{toc}{section}{Introduction}\else
\fi

Euclid%
\epubtkFootnote{Continuously updated information on Euclid is
  available on \url{http://www.euclid-ec.org}.}
\citep{euclidredbook, 2009ExA....23...17R, 2009ExA....23...39C} is an ESA medium-class mission selected
for the second launch slot (expected for 2020) of the Cosmic
Vision 2015\,--\,2025 program. The main goal of Euclid is to
understand the physical origin of the accelerated expansion of the universe. Euclid
is a satellite equipped with a 1.2~m telescope and
three imaging and spectroscopic instruments working in the visible and
near-infrared wavelength domains. These instruments will explore the expansion
history of the universe and the evolution of cosmic structures
 by measuring shapes and redshifts of galaxies over a large fraction of
the sky. The satellite will be launched by a Soyuz ST-2.1B rocket and
transferred to the L2 Lagrange point for a six-year mission that will
cover at least 15\,000 square degrees of sky. Euclid plans to image
a billion galaxies and measure nearly 100 million galaxy redshifts.

These impressive numbers will allow Euclid to realize a detailed
reconstruction of
 the clustering of galaxies out to a redshift 2 and the pattern of
light distortion from weak lensing to redshift 3. The two main probes, redshift
clustering and weak lensing,
are complemented by a number of additional cosmological probes: cross
correlation between
the cosmic microwave background and the large scale structure;
abundance and properties of galaxy clusters and strong
lensing and possible luminosity distance through
supernovae Ia. To extract
the maximum of information also in the nonlinear regime of
perturbations, these probes will require accurate high-resolution
numerical simulations.
Besides cosmology, Euclid will provide an exceptional dataset for
galaxy evolution, galaxy structure, and
planetary searches. All Euclid data will be publicly released after a
relatively short proprietary period
and will constitute for many years the ultimate survey database for
astrophysics.

A huge enterprise like Euclid requires highly considered planning in terms
not only of technology
but also for the scientific exploitation of future data. Many ideas and
models that today seem to be
abstract exercises for theorists will in fact finally become testable with
the Euclid surveys.
The main science driver of Euclid is clearly the nature of dark
energy, the enigmatic
substance that is driving the accelerated expansion of the universe.
As we discuss
in detail in Part~\ref{dark-energy}, under the label ``dark energy''
we include a wide
variety of hypotheses, from extradimensional physics to higher-order
gravity, from new fields and new
forces to large violations of homogeneity and isotropy. The simplest
explanation, Einstein's
famous cosmological constant, is still currently acceptable from the
observational point of view,
but is not the only one, nor necessarily the most satisfying, as
we will argue.
Therefore, it is important to identify the main observables that will
help distinguish the cosmological
constant from the alternatives and to forecast Euclid's performance in
testing the various models.

Since clustering and weak lensing also depend on the properties of
dark matter, Euclid is a dark matter
probe as well. In Part~\ref{dark-matter} we focus on the models of dark
matter that can be tested with Euclid data,
from massive neutrinos to ultra-light scalar fields. We show that
Euclid can measure the neutrino mass to a very
high precision, making it one of the most sensitive neutrino experiments
of its time, and it can help identify new light fields in the cosmic fluid.

The evolution of perturbations depends not only on the fields and forces
active during the cosmic eras, but also
on the initial conditions. By reconstructing the initial conditions we open
a window on the inflationary physics that created the
perturbations, and allow ourselves the
chance of determining
whether a single inflaton drove the expansion or a mixture of fields.
In Part~\ref{ini-cond} we review the choices of initial conditions and their
impact on Euclid science. In particular we discuss deviations from
simple scale invariance,
mixed isocurvature-adiabatic initial conditions,
non-Gaussianity, and the combined forecasts of Euclid and CMB experiments.

Practically all of cosmology is built on the Copernican Principle, a very fruitful idea
postulating a homogeneous and isotropic background. Although this assumption has been
confirmed time and again since the beginning of modern cosmology,
Euclid's capabilities can push
the test to new levels. In Part~\ref{testing} we challenge some of
the basic cosmological assumptions and
predict how well Euclid can constrain them. We explore the basic
relation between luminosity and angular
diameter distance that holds in any metric theory of gravity if
the universe is transparent to light,
and the existence of
large violations of homogeneity and isotropy, either due to local
voids or to the cumulative stochastic
effects of perturbations, or to intrinsically anisotropic vector
fields or spacetime geometry.

Finally, in Part~\ref{statistical} we review some of the
statistical methods that are used to forecast the performance
of probes like Euclid,
and we discuss some possible future developments.

This review has been planned and carried out within Euclid's Theory
Working Group and is meant to provide
a  guide to the scientific themes that will underlie the activity of
the group during the preparation of the mission.
At the same time, this review will help us and the community at large
to identify the areas
that deserve closer attention, to improve  the development of Euclid
science and to offer new
scientific challenges and opportunities.

\newpage
\part{Dark Energy}
\label{dark-energy}


\section{Introduction}

With the discovery of cosmic acceleration at the end of the 1990s, and its
possible explanation in terms of a cosmological constant, cosmology has returned to its roots in
Einstein's famous 1917 paper that simultaneously inaugurated modern cosmology and the history of the constant $\Lambda$.
Perhaps cosmology is approaching a robust and all-encompassing standard model, like its cousin, 
the very successful standard model of particle physics. 
In this scenario, the cosmological standard model could essentially close the search
for a broad picture of cosmic evolution, leaving to future generations only
the task of filling in a number of important, but not crucial, details.

The cosmological constant is still in remarkably good agreement with
almost all cosmological data more than ten years after the observational discovery of the 
accelerated expansion rate of the universe. However, our
knowledge of the universe's evolution is so incomplete that it would be premature
to claim that we are close to understanding the ingredients of the cosmological standard model. 
If we ask ourselves what we know for certain about
the expansion rate at redshifts larger than unity, or the growth rate of
matter fluctuations, or about the properties of gravity on large scales and at early times, or
about the influence of extra dimensions (or their absence) on our four dimensional world,
the answer would be surprisingly disappointing. 

Our present knowledge can be succinctly summarized as follows: we live in a universe that is consistent with the presence of a cosmological constant in the field equations of general relativity, and as of 2016, the value of this constant corresponds to a fractional energy density today of $\Omega_{\Lambda}\approx 0.7$.
However, far from being disheartening, this current lack of knowledge points to an exciting future. 
A decade of research on dark energy has
taught many cosmologists that this ignorance can be overcome by the same tools
that revealed it, together with many more that have been developed in recent years.

Why then is the cosmological constant not the end of the story as far as cosmic
acceleration is concerned? There are at least three reasons. The first is that we have no simple
way to explain its small but non-zero value. In fact, its value is unexpectedly
small with respect
to any physically meaningful scale, except the \emph{current} horizon scale. The second reason is that this value is not only small, but also
surprisingly close to another unrelated quantity, the \emph{present} matter-energy density. That this
happens just by coincidence is hard to accept, as the matter density is diluted rapidly with the expansion
of space. Why is it that we happen to live at the precise, fleeting epoch when the energy densities of matter and the cosmological constant are of comparable magnitude? Finally, observations of coherent acoustic
oscillations in the cosmic microwave background (CMB)
have turned the notion of accelerated expansion in the very early universe (inflation) into an integral 
part of the cosmological standard model. Yet the simple truth that we exist as observers demonstrates that this early accelerated expansion was of a finite duration, and hence cannot be ascribable to a true, constant
$\Lambda$; this sheds doubt on the nature of the current accelerated expansion.
The very fact that
we know so little about the past dynamics of the universe forces us to enlarge
the theoretical parameter space and to consider phenomenology that a simple
cosmological constant cannot accommodate.

These motivations have led many scientists to challenge one of the most basic tenets
of physics: Einstein's law of gravity. Einstein's theory of general relativity (GR)
 is a supremely successful theory on
scales ranging from the size of our solar system down to micrometers, the shortest distances at which GR has been probed in the laboratory
so far. Although specific predictions about such
diverse phenomena as the gravitational redshift of light, energy loss from
binary pulsars, the rate of precession of the perihelia of bound orbits, and
light deflection by the sun are not unique to GR, it must be
regarded as highly significant that GR is consistent with each
of these tests and more. We can securely state that GR has been tested
to high accuracy \textit{at these distance scales}.

The success of GR on larger scales is less clear. On astrophysical
and cosmological scales, tests of GR are complicated by the existence of 
invisible components like dark matter and by the effects of spacetime
geometry. 
We do not know whether the physics underlying the apparent cosmological constant originates from modifications to GR (i.e., an extended theory of gravity), or from a new fluid or field in our universe that we have not yet detected directly. The latter phenomena are generally referred to as `dark energy' models.

If we only consider observations of the expansion rate of the universe we cannot discriminate between a theory of modified gravity and a dark-energy model. However, it is likely that these two alternatives will cause perturbations around the `background' universe to behave differently. Only by improving our knowledge of the growth of structure in the universe
can we hope to progress towards breaking the degeneracy between dark energy and modified gravity. Part~\ref{dark-energy} of this review
is dedicated to this effort.
We begin with a review of the background and linear perturbation equations in a general
setting, defining quantities that will be employed throughout. We then explore the 
nonlinear effects of dark energy, making use of analytical tools such as the spherical collapse model, perturbation theory and numerical $N$-body simulations.
We discuss a number of competing models proposed in literature and demonstrate what the Euclid survey
will be able to tell us about them. For an updated review of present cosmological constraints on a variety of Dark Energy and Modified Gravity models, we refer to the Planck 2015 analysis \cite{2015arXiv150201590P}.

\section{Background evolution}


Most of the calculations in this review are performed in the Friedmann--Lema\^\i tre--Robertson--Walker (FLRW) metric
\begin{equation}
\mathrm{d}s^2=-\mathrm{d}t^2+{a(t)^2}(\frac{\mathrm{d}r^2}{1-kr^2}+r^2 \, \mathrm{d}\theta^2+r^2\sin^2\theta \, \mathrm{d}\phi^2 ) \,,
\end{equation}
where $a(t)$ is the scale factor (normalized to $a=1$ today) and $k$ the spatial curvature. The usual symbols for
the Hubble function $H=\dot{a}/a$ and the density fractions $\Omega_x$, where $x$ stands
for the component, are employed. We characterize the components with the subscript $M$ or $m$ for matter,
$\gamma$ or $r$ for radiation, $b$ for baryons, \reviewtext{$k$ or} $K$ for curvature and $\Lambda$ for the cosmological constant.
Whenever necessary for clarity, we append a subscript $0$ to denote the present epoch, e.g., $\Omega_{M,0}$.
Sometimes the conformal time $\eta=\int \mathrm{d}t/a$ and the conformal Hubble
function $\mathcal{H}=aH= \mathrm{d}a/(a\mathrm{d}\eta)$ \label{symbol:conformaltime}
 are employed. Unless otherwise stated, we denote with a dot derivatives w.r.t.\ cosmic time $t$ (and sometimes
we employ the dot for derivatives w.r.t.\ conformal time $\eta$) while we use a prime for
derivatives with respect to $\ln a$.

The energy density due to a cosmological constant with $p=-\rho$ is obviously constant over time. This can easily be seen from the covariant
conservation equation $T_{\mu;\nu}^\nu=0$ for the homogeneous and isotropic
FLRW metric,
\begin{equation}
\dot{\rho} + 3 H (\rho+p) = 0 \,.
\end{equation}
However, since we also observe radiation with $p=\rho/3$ and non-relativistic
matter for which $p\approx 0$, it is natural to assume that the dark energy is
not necessarily limited to a constant energy density, but that it could be
dynamical instead.

One of the simplest models that explicitly realizes such a dynamical dark
energy scenario is described by a minimally-coupled canonical scalar field evolving in a
given potential. For this reason, the very concept of dynamical dark energy
is often associated with this scenario, and in this context it is called `quintessence'
\citep{Wetterich_1988, Ratra:1987rm}. In the following, the scalar field will be indicated with $\phi$.
Although in this simplest framework the dark energy does not
interact with other species and influences spacetime only through its energy density
and pressure, this is not the only possibility and we will encounter more general models later on. 
The homogeneous energy density and pressure of the scalar field $\phi$ are
defined as
\begin{equation}
\label{phi_bkg}
 \rho_{\phi} = \frac{{\dot\phi}^2}{2 } + V(\phi) \,, \quad
p_{\phi} = \frac{{\dot\phi}^2}{2 } - V(\phi) \,, \quad w_{\phi} =
\frac{p_{\phi}}{\rho_{\phi}} \,,
\end{equation}
and $w_\phi$ is called the equation-of-state parameter. 
Minimally-coupled dark-energy models can allow for attractor solutions \citep{Copeland:1997et, Liddle:1998xm,
Steinhardt:1999nw}: if an attractor exists, depending on the potential $V(\phi)$ in which
dark energy rolls, the trajectory of the scalar field in
the present regime converges to the path given by the attractor, though starting from a wide set
of different initial conditions for $\phi$ and for its first derivative $\dot\phi$.
Inverse power law and exponential potentials are typical examples of potential
that can lead to attractor solutions.
As constraints on $w_\phi$ become tighter \citep[e.g.,][]{Komatsu:2010fb}, the allowed
range of initial conditions to follow into the attractor solution shrinks, so
that minimally-coupled quintessence is actually constrained to have very flat
potentials. The flatter the potential, the more minimally-coupled quintessence mimics a
cosmological constant, the more it suffers from the same fine-tuning and
coincidence problems that affect a $\Lambda$CDM scenario
\citep{Matarrese:2004xa}. 

However, when GR is modified or when an
interaction with other species is active, dark energy may very well have a non-negligible
contribution at early times. Therefore, it is important, already at the background level, 
to understand the best way to characterize the main features of the evolution of quintessence and dark energy in general, pointing out which parameterizations are more suitable
and which ranges of parameters are of interest to disentangle quintessence or modified gravity from a cosmological constant scenario.

In the following we briefly discuss how to describe the cosmic expansion rate
in terms of a small number of parameters. This will set the stage for the
more detailed cases discussed in the subsequent sections. Even within specific physical models it is often 
convenient to reduce the information to a few phenomenological parameters.

Two important points are left for later: from Eq.~(\ref{phi_bkg}) we can easily see that $w_\phi\geq-1$ as
long as $\rho_\phi>0$, i.e., uncoupled canonical scalar field dark energy never crosses $w_\phi=-1$. However, this is
not necessarily the case for non-canonical scalar fields or for cases where GR is modified. We postpone to Section~\ref{sec:crossing} the discussion of how to parametrize
this `phantom crossing' to avoid singularities, as it also requires the study
of perturbations.

The second deferred part on the background expansion concerns a basic statistical question: 
what is a sensible precision target for a measurement of dark energy, e.g., of its equation 
of state? In other words, how close to $w_\phi=-1$ should we go
before we can be satisfied and declare that dark energy is the cosmological constant? We will address this question in Section~\ref{genericproperties}.

\subsection{Parametrization of the background evolution}
\label{parametrization-of-the-background-evolution}

If one wants to parametrize the equation of state of dark energy, two general 
approaches are possible. 
The first is to start from a set of dark-energy models 
given by the theory and to find parameters describing their $w_\phi$ as accurately as 
possible. Only later one can try and include as many theoretical models as possible 
in a single parametrization.
In the context of scalar-field dark-energy models (to be discussed in Section~\ref{quintessence}),  \cite{Crittenden:2007yy} 
parametrize the case of slow-rolling fields, \cite{Scherrer:2007pu} study
thawing quintessence, \cite{Hrycyna:2007mq} and \cite{Chiba:2010cy} include 
non-minimally coupled fields,  \cite{Setare:2008sf} quintom quintessence,
\cite{Dutta:2008qn} parametrize hilltop quintessence, \cite{Chiba:2009nh} 
extend the quintessence parametrization to a class of $k$-essence models,  
\cite{Huang:2010zr} study a common parametrization for quintessence and phantom fields.
Another convenient way to parametrize the presence of a non-negligible homogeneous dark
energy component at early times (usually labeled as EDE) was presented in
\cite{Wetterich:2004pv}. We recall it here because we will refer to this example in Section~\ref{quintessence_ede}. 
In this case the equation of state is parametrized as:
\begin{equation} {w}_X (z) = \frac{{ w}_0}{1+b \ln{(1+z)}},
\label{w_ede_par} \end{equation}
where $b$ is a constant related to the amount of dark energy at early
times, i.e.,
\begin{equation}
b = - \frac{3 w_0}{\ln{\frac{1-\Omega_{e}}{\Omega_{e}}} +
\ln{\frac{1-\Omega_{m,0}}{\Omega_{m,0}}}} \,.
\end{equation}

Here the subscripts `$0$' and `$e$' refer to quantities calculated today or
early times, respectively. With regard to the latter parametrization, we note that concrete 
theoretical and realistic models involving a non-negligible energy component
at early times are often accompanied by further important
modifications (as in the case of interacting dark energy), not always included
in a  parametrization of the sole equation of state such as (\ref{w_ede_par})
(for further details see Section~\ref{nonlinear-aspects} on nonlinear aspects of dark energy
and modified gravity).

The second approach is to start from a simple expression of $w$ without 
assuming any specific dark-energy model (but still 
checking afterwards whether known theoretical dark-energy models can be represented).
This is what has been done by \cite{Huterer:2000mj, Maor:2000jy, 
Weller:2000pf} (linear and logarithmic parametrization in $z$), 
\cite{chevallier01}, \cite{Linder:2002et} (linear and power law 
parametrization in $a$), \cite{Douspis:2006rs}, \cite{Bassett:2004wz} 
(rapidly varying equation of state).

The most common parametrization, widely employed in this review,
is the linear equation of state \citep{chevallier01,Linder:2002et}
\begin{equation}
\label{CPL}
w_X(a)=w_0+w_a (1-a) \,,
\end{equation}
where the subscript $X$ refers to the generic dark-energy constituent.
While this parametrization is useful as a toy model
in comparing the forecasts for different dark-energy projects,
it should not be taken as all-encompassing. In general a dark-energy model can introduce further significant terms in the effective $w_X(z)$ that cannot be mapped onto the simple form of Eq.~(\ref{CPL}).

An alternative to model-independent constraints
is measuring the dark-energy density $\rho_X(z)$
(or the expansion history $H(z)$) as a free function
of cosmic time \citep{WangGarnavich01,Tegmark02,Daly03}.
Measuring $\rho_X(z)$ has advantages over measuring the dark-energy
equation of state $w_X(z)$ as a free function; $\rho_X(z)$ is more
closely related to observables, hence is more tightly 
constrained for the same number of redshift bins 
used \citep{WangGarnavich01,WangFreese04}.
Note that $\rho_X(z)$ is related to $w_X(z)$ as follows \citep{WangGarnavich01}:
\begin{equation}
\frac{\rho_X(z)}{\rho_X(0)} = \exp\left\{ \int_0^z \, \mathrm{d}z'\, \frac{3
    [1+w_X(z')]}{1+z'} \right\} \,. 
\end{equation} 
Hence, parametrizing dark energy with $w_X(z)$ implicitly assumes that 
$\rho_X(z)$ does not change sign in cosmic time.
This precludes whole classes of dark-energy models in which $\rho_X(z)$ 
becomes negative in the future (``Big Crunch'' models, see 
\cite{WangLinde04} for an example) \citep{WangTegmark04}.

Note that the measurement of $\rho_X(z)$ is straightforward
once $H(z)$ is measured from baryon acoustic oscillations, and $\Omega_m$ is constrained
tightly by the combined data from galaxy clustering, weak lensing, and cosmic microwave background data
-- although strictly speaking this requires a choice of perturbation evolution for the dark energy as well, and in addition one that
is not degenerate with the evolution of dark matter perturbations; see \cite{Kunz:2007rk}.

Another useful possibility is to adopt the principal component approach 
\citep{Huterer:2002hy}, which avoids any assumption about the 
form of $w$ and assumes it to be constant or linear in redshift bins, 
then derives which combination of parameters is best 
constrained by each experiment.

For a cross-check of the results using more complicated parameterizations,
one can use simple polynomial parameterizations of
$w$ and $\rho_{\mathrm{DE}}(z)/\rho_{\mathrm{DE}}(0)$ \citep{Wang:2008zh}.

\section{Perturbations}
\label{sec:dof}

This section is devoted to a discussion of linear perturbation theory
in dark-energy models. Since we will discuss a number of non-standard models
in later sections, we present here the main equations in a general form that
can be adapted to various contexts. This section will identify
which perturbation functions the Euclid survey~\cite{euclidredbook} will try to measure and how they can help
us to characterize the nature of dark energy and the properties of gravity.

\subsection{Cosmological perturbation theory}
\label{sec:cosmo_perts}

Here we provide the perturbation equations in a dark-energy dominated universe 
for a general fluid, focusing  on 
scalar perturbations. 

For simplicity, we consider a flat universe containing only (cold dark) 
matter and dark energy, so that the Hubble parameter is given by
\begin{equation}
H^2 =\left( \frac{1}{a} \frac{\mathrm{d}a}{\mathrm{d}t} \right) ^2 = 
H_{0}^{2}\left[\Omega_{m_0} a^{-3}+\left(
1- \Omega_{m_0} \right)\exp\left(
-3\int_1^a\frac{1+w(a')}{a'} \, \mathrm{d}a \right) \right] .
\end{equation}
We will consider linear perturbations on a spatially-flat background
model, defined by the line of element
\begin{equation}
\mathrm{d}s^{2} = a^{2} \left[ -\left( 1+2A\right) \,
\mathrm{d}\eta^{2}+2B_{i} \, \mathrm{d}\eta \, \mathrm{d}x^{i}+\left( \left(
1+2H_{L}\right) \delta_{ij}+2H_{Tij} \right) \, \mathrm{d}x_{i} \, \mathrm{d}x^{j} \right] ,
\label{pert_0_ds}
\end{equation}
where $A$ is the scalar potential; $B_{i}$\label{symbol:vectorshift} a vector shift; $H_{L}$ is the
scalar perturbation to the spatial curvature; $H_{T}^{ij}$ is the trace-free
distortion to the spatial metric; $\mathrm{d}\eta= \mathrm{d}t/a$ is the
conformal time. \label{symbol:H_T}

We will assume that the universe is filled with perfect fluids only,
so that the energy momentum tensor takes the simple form
\begin{equation}
T^{\mu\nu}=\left( \rho+p\right) u^{\mu}u^{\nu} +p\ g^{\mu\nu}+\Pi^{\mu\nu} \,,
\label{EMT}
\end{equation}
where $\rho$ and $p$ are the density and the pressure of the 
fluid respectively, $u^{\mu}$ is the four-velocity and $\Pi^{\mu\nu}$ 
is the anisotropic-stress perturbation tensor that represents 
the traceless component of the $T_{j}^{i}$.
\label{symbol:stressperturbation}

The components of the perturbed energy momentum tensor can be written as:
\begin{eqnarray}
T_{0}^{0} &=& - \left( \bar\rho + \delta\rho \right) \\
T_{j}^{0} &=& \left( \bar\rho + \bar{p} \right) \left( v_{j} - B_{j} \right) \\
T_{0}^{i} &=& \left( \bar\rho + \bar{p} \right) v^{i} \\
T_{j}^{i} &=& \left( \bar{p} + \delta{p} \right) \delta_{j}^{i}
+ \bar{p}\ \Pi_{j}^{i}.
\end{eqnarray}
Here $\bar\rho$ and $\bar p$ are the energy density and pressure of the
homogeneous and isotropic background universe,
$\delta\rho$ is the density perturbation, $\delta p$ is the pressure perturbation, $v^{i}$ is the velocity vector. 
Here we want to investigate only the scalar modes of the perturbation equations.
So far the treatment of the matter and metric is fully general and applies 
to any form of matter and metric. We now choose the Newtonian gauge (also known as the longitudinal or Poisson gauge), characterized by zero non-diagonal metric terms (the shift vector $B_{i}=0$ and $H_{T}^{ij}=0$) and by two \label{symbol:scalarpotentials}
scalar potentials $\Psi$ and $\Phi$; the metric Eq.~(\ref{pert_0_ds}) then becomes
\begin{equation}
\mathrm{d}s^{2} = a^{2} \left[ -\left( 1+2\Psi \right) \,
\mathrm{d}\eta^{2} + \left( 1-2\Phi\right) \, \mathrm{d}x_{i} \, \mathrm{d}x^{i} \right] .
\label{pert_newton_ds}
\end{equation}
The advantage of using the Newtonian gauge is that the metric tensor $g_{\mu\nu}$ is diagonal
and\label{symbol:metric} 
this simplifies the calculations. This choice not only simplifies the calculations 
but is also the most intuitive one as the observers are attached to the points in 
the unperturbed frame; as a consequence, they will detect a velocity field 
of particles falling into the clumps of matter and will measure their gravitational potential, represented directly by $\Psi$; $\Phi$ corresponds to the perturbation to the spatial curvature. 
Moreover, as we will see later, the Newtonian gauge is the 
best choice for observational tests (i.e., for perturbations smaller than the horizon).

In the conformal Newtonian gauge, and in Fourier space, the first-order perturbed Einstein
equations give \citep[see][for more details]{Ma:1995ey}:
\begin{eqnarray}
k^2\Phi + 3\frac{\dot{a}}{a} \left( \dot{\Phi} + \frac{\dot{a}}{a}\Psi\right) &=& -4\pi G a^2 \sum_{\alpha}\bar{\rho}_{\alpha}\delta_{\alpha} \,,\label{ein-cona}\\
k^2 \left( \dot{\Phi} + \frac{\dot{a}}{a}\Psi \right)&=& 4\pi G a^2 \sum_{\alpha}(\bar{\rho}_{\alpha}+\bar{p}_{\alpha}) \theta_{\alpha}\,,\label{ein-conb}\\
\ddot{\Phi} + \frac{\dot{a}}{a} (\dot{\Psi}+2\dot{\Phi})+\left(2\frac{\ddot{a}}{a} - \frac{\dot{a}^2}{a^2}\right)\Psi+ \frac{k^2}{3} (\Phi-\Psi)
&=& 4\pi G a^2 \sum_{\alpha}\delta p_{\alpha}\,,\label{ein-conc}\\
k^2(\Phi-\Psi) &=& 12\pi G a^2 \sum_{\alpha}\left(\bar{\rho}_{\alpha}+\bar{p}_{\alpha}\right)\pi_{\alpha}\,,\label{ein-cond}
\end{eqnarray}
where a dot denotes $d/d\eta$, $\delta_\alpha=\delta\rho_\alpha/\bar{\rho}_\alpha$, the index $\alpha$ indicates a sum over all matter components in the universe and 
$\pi$ is related to $\Pi_{j}^{i}$ through:
\begin{equation}
\left(\bar{\rho}+\bar{p}\right)\pi = -\left(\hat{k}_i\hat{k}_j-\frac{1}{3}\delta_{ij}\right)\Pi_{j}^{i}.
\end{equation}
The energy-momentum tensor components in the Newtonian gauge become:
\begin{eqnarray}
T_{0}^{0} &=& -\left( \bar\rho + \delta\rho \right) \label{00_EM}\\
ik_i T_{0}^{i} &=& -ik_i T_{i}^{0} = \left(\bar\rho + \bar{p} \right) \theta  \label{0i_EM}\\
T_{j}^{i} &=& \left( \bar p + \delta p \right) \delta_{j}^{i} +\bar{p}\Pi_{j}^{i} \label{ij_EM}
\end{eqnarray}
where we have defined the variable $\theta=ik_j v^j$ that represents the divergence
of the velocity field.

Perturbation equations for a single fluid are obtained taking
the covariant derivative of the perturbed energy momentum tensor, 
i.e., $T_{\nu;\mu}^{\mu}=0$. We have
\begin{eqnarray}
\dot\delta &=& -\left( 1+w \right) \left( \theta - 3\dot\Phi \right)
-3\frac{\dot a}{a} \left( \frac{\delta p}{\bar\rho} - w\delta \right)~~~~~~~~~~~~~~~~~~~{\mathrm{for}}~~~~\nu=0 \label{d_pert}\\
\dot\theta &=& -\frac{\dot a}{a} \left( 1-3w \right) \theta -
\frac{\dot{w}}{1+w}\theta +k^{2}\frac{\delta{p}/\bar\rho}{1+w} + k^{2}\Psi - k^2\pi~~~~~{\mathrm{for}}~~~~\nu=i \label{t_pert}.
\end{eqnarray}
The equations above are valid for any fluid. The evolution of the 
perturbations depends on the characteristics of the fluids considered, i.e., we need to 
specify the equation of state parameter $w$, the pressure perturbation $\delta p$ and the 
anisotropic stress $\pi$. 
For instance, if we want to study how matter perturbations evolve, we simply 
substitute $w=\delta p = \pi = 0$ (matter is pressureless) in the above equations. 
However, Eqs.~(\ref{d_pert})\,--\,(\ref{t_pert}) depend on the 
gravitational potentials $\Psi$ and $\Phi$, which in 
turn depend on the evolution of the perturbations of the other fluids. 
For instance, if we assume that the universe is filled by dark matter and dark energy 
then we need to specify $\delta p$ and $\pi$ for the dark energy. 

The problem here is not only to parameterize the pressure perturbation and the 
anisotropic stress for the dark energy (there is not a unique way to
do it, see below, especially Section~\ref{sec:crossing} for what to do when $w$ crosses $-1$) but rather that we need to run the perturbation equations 
for each model we assume, making predictions and compare the results with observations. 
Clearly, this approach takes too much time. 
In the following Section~\ref{mg_growth_params} we show a general approach to understanding the observed late-time 
accelerated expansion of the universe through the evolution of the 
matter density contrast. 

In the following, whenever there is no risk of confusion, we remove the overbars
from the background quantities.

\subsection{Modified growth parameters}
\label{mg_growth_params}

Even if the expansion history, $H(z)$, of the
FLRW background has been measured (at least up to redshifts $\sim 1$ by supernova data, i.e., via the luminosity distance), it
is not possible yet to identify the physics causing the recent acceleration of
the expansion of the universe. Information on the growth of structure at
different scales and different redshifts is needed to discriminate between
models of dark energy (DE) and modified gravity (MG). 
A definition of what we mean by DE and MG will be postponed to Section~\ref{models-of-modified-gravity}.

An alternative to testing predictions of specific theories is to parameterize
the possible departures from a fiducial model. Two conceptually-different
approaches are widely discussed in the literature:
\begin{itemize}
\item \emph{Model parameters} capture the degrees of freedom of DE/MG and modify
the evolution equations of the energy-momentum content of the fiducial model.
They can be associated with physical meanings and have uniquely-predicted
behavior in specific theories of DE and MG.
\item \emph{Trigger relations} are derived directly from observations and only
hold in the fiducial model. They are constructed to break down if the fiducial
model does not describe the growth of structure correctly.
\end{itemize}
As the current observations favor concordance cosmology, the fiducial model
is typically taken to be spatially flat FLRW in GR with cold dark matter and a
cosmological constant, hereafter referred to as $\Lambda$CDM.

For a large-scale structure and weak lensing survey the crucial quantities are
the matter-density contrast and the gravitational potentials and we therefore focus on scalar perturbations in the Newtonian gauge with the metric (\ref{pert_newton_ds}).

We describe the matter perturbations using the gauge-invariant comoving density
contrast $\Delta_M\equiv\delta_M+3aH \theta_M/k^2$ where $\delta_M$ and
$\theta_M$ are the\label{symbol:DeltaM}
matter density contrast and the divergence of the fluid velocity for matter, respectively. 
The discussion can be generalized to include multiple fluids.

In $\Lambda$CDM, after radiation-matter equality there is no anisotropic stress
present and the Einstein constraint equations become
\begin{equation}
  -k^2 \Phi = 4\pi G a^2 \rho_M \Delta_M \,, \qquad \Phi=\Psi \,.
  \label{lcdm_eeq}
\end{equation}
These can be used to reduce the energy-momentum conservation of matter simply to
the second-order growth equation
\begin{equation}
  \Delta_M''+\left[2+(\ln H)'\right]\Delta_M' = \frac{3}{2}\Omega_M(a)\Delta_M \,.
  \label{lcdm_geq}
\end{equation}
Primes denote derivatives with respect to $\ln a$ and we define the time-dependent fractional matter density as $\Omega_M(a)\equiv8\pi
G\rho_M(a)/(3H^2)$.  Notice that
the evolution of $\Delta_M$ is driven by $\Omega_M(a)$ and is scale-independent
throughout (valid on sub- and super-Hubble scales after radiation-matter
equality). We define the growth factor $G(a)$  as  $\Delta=\Delta_0G(a)$. This is
very well approximated by the expression
\begin{equation} G(a)\approx 
\exp\left\{ \int_1^a
\frac{\mathrm{d}a'}{a'}\left[\Omega_M(a')^\gamma\right] \right\} \label{def_gf}
\end{equation}
and
\begin{equation}
f_g\equiv \frac{\d\log G}{\d\log a}\approx \Omega_M(a)^\gamma\label{def-growth-rate}
\end{equation}
defines the growth rate and the growth index $\gamma$ that is found to be $\gamma_{\Lambda}\simeq 0.545$
for the $\Lambda$CDM solution \citep[see][]{wang98, Linder:2005in,
Huterer:2006mva,Ferreira:2010sz}.

Clearly, if the actual theory of structure growth is not the $\Lambda$CDM
scenario, the constraints (\ref{lcdm_eeq}) will be modified, the growth equation
(\ref{lcdm_geq}) will be different, and finally the growth factor (\ref{def_gf})
is changed, i.e., the growth index is different from $\gamma_\Lambda$ and may
become time and scale dependent. Therefore, the inconsistency of these three
points of view can be used to test the $\Lambda$CDM paradigm.

\subsubsection{Two new degrees of freedom}

Any generic modification of the dynamics of scalar perturbations with respect to
the simple scenario of a smooth dark-energy component that only alters the
background evolution of $\Lambda$CDM can be represented by introducing two new
degrees of freedom in the Einstein constraint equations. We do this by replacing
(\ref{lcdm_eeq}) with
\begin{equation}
  -k^2 \Phi = 4\pi G Q(a,k) a^2 \rho_M \Delta_M \,, \qquad 
\Phi=\eta(a,k)\Psi \,.
  \label{mod_constr}
\end{equation}
Non-trivial behavior of the two functions $Q$ and $\eta$ can be due to a
clustering dark-energy component or some modification to GR. In MG models the
function $Q(a,k)$ represents a mass screening effect due to local modifications
of gravity and effectively modifies Newton's constant. In dynamical DE models $Q$
represents the additional clustering due to the perturbations in the DE. On the
other hand, the function $\eta(a,k)$ parameterizes the effective anisotropic
stress introduced by MG or DE, which is absent in $\Lambda$CDM.

Given an MG or DE theory, the scale- and time-dependence of the functions $Q$ and
$\eta$ can be derived and predictions projected into the $(Q,\eta)$ plane. This
is also true for interacting dark sector models, although in this case the
identification of the total matter density contrast (DM plus baryonic matter)
and the galaxy bias become somewhat contrived \citep[see, e.g.,][for
an overview of predictions for different MG/DE models]{Song:2010rm}.

Using the above-defined modified constraint equations (\ref{mod_constr}), the
conservation equations of matter perturbations can be expressed in the following
form (see~\cite{Pogosian:2010tj})
\begin{eqnarray}
  \Delta_M' &=& -\frac{1/\eta-1+(\ln Q)'}{x_Q^2+\frac{9}{2}\Omega_M}\,
\frac{9}{2}\Omega_M \Delta_M
    - \frac{x_Q^2-3(\ln H)'/Q}{x_Q^2+\frac{9}{2}\Omega_M}\, \frac{\theta_M}{aH}
\nonumber \\
  \theta_M' &=& -\theta_M - \frac{3}{2}aH\Omega_M
\frac{Q}{\eta}\Delta_M,
\end{eqnarray}
where we define $x_Q\equiv k/(aH\sqrt{Q})$. Remember $\Omega_M=\Omega_M(a)$ as defined
above. Notice that it is $Q/\eta$ that modifies the source term of the
$\theta_M$ equation and therefore also the growth of $\Delta_M$. Together with
the modified Einstein constraints (\ref{mod_constr}) these evolution equations
form a closed system for $(\Delta_M,\theta_M,\Phi,\Psi)$ which can be solved for
given $(Q,\eta)$.

The influence of the Hubble scale is modified by $Q$, such that now the size of
$x_Q$ determines the behavior of $\Delta_M$; on ``sub-Hubble'' scales, $x_Q\gg
1$, we find
\begin{equation}
  \Delta_M''+\left[2+(\ln H)'\right]\Delta_M' = \frac{3}{2}\Omega_M(a)
\frac{Q}{\eta} \Delta_M \,
\end{equation}
and $\theta_M=-aH\Delta_M'$. The growth equation is only modified by the factor
$Q/\eta$ on the RHS with respect to $\Lambda$CDM (\ref{lcdm_geq}). On
``super-Hubble'' scales, $x_Q\ll 1$, we have
\begin{eqnarray}
  \Delta_M' &=& -\left[1/\eta-1+(\ln Q)'\right] \Delta_M + \frac{2}{3\Omega_M}\frac{(\ln
H)'}{aH}\frac{1}{Q} \theta_M,
\nonumber \\
  \theta_M' &=& -\theta_M - \frac{3}{2}\Omega_M\,aH
\frac{Q}{\eta}\Delta_M \,.
\end{eqnarray}
$Q$ and $\eta$ now create an additional drag term in the $\Delta_M$ equation,
except if $\eta>1$ when the drag term could flip sign. \cite{Pogosian:2010tj}
also showed that the metric potentials evolve independently and
scale-invariantly on super-Hubble scales as long as $x_Q\to 0$ for $k \to 0$.
This is needed for the comoving curvature perturbation, $\zeta$,  to be constant
on super-Hubble scales.

Many different names and combinations of the above defined functions $(Q,\eta)$
have been used in the literature, some of which are more closely related to
actual observables and are less correlated than others in certain situations
\citep[see, e.g.,][]{Amendola:2007rr, Mota:2007sz,Song:2010rm, Pogosian:2010tj, Daniel:2010ky,
Daniel:2010yt, Ferreira:2010sz}.

For instance, as observed above, the combination $Q/\eta$ modifies the source
term in the growth equation. Moreover, peculiar velocities are following
gradients of the Newtonian potential, $\Psi$, and therefore the comparison of
peculiar velocities with the density field is also sensitive to $Q/\eta$. So we
define
\begin{equation}\label{eq:muQeta}
  \mu \equiv Q\,/\,\eta \qquad \Rightarrow \qquad
  -k^2 \Psi = 4\pi G a^2 \mu(a,k) \rho_M \Delta_M \,.
\end{equation}

Weak lensing and the integrated Sachs--Wolfe (ISW) effect, on the
other hand, are measuring $(\Phi+\Psi)/2$, which is related to the
density field via 
\begin{equation}
  \Sigma \equiv \frac{1}{2}Q(1+1/\eta) = \frac{1}{2}\mu(\eta+1)
   \qquad \Rightarrow \qquad
  -k^2 (\Phi+\Psi) = 8\pi G a^2 \Sigma(a,k) \rho_M \Delta_M \,.
\label{Sigma-wl-isw}
\end{equation}
A summary of different other variables used was given by~\cite{Daniel:2010ky}.
For instance, the gravitational slip parameter \label{symbol:varpi} introduced by~\cite{Caldwell:2007cw} and widely used is related through $\varpi\equiv
1/\eta-1$. Recently~\cite{Daniel:2010yt} used $\{{\cal G}\equiv\Sigma,\
\mu\equiv Q,\ {\cal V}\equiv\mu\}$, while~\cite{Bean:2010zq} defined
$R\equiv1/\eta$. All these variables reflect the same two degrees of freedom
additional to the linear growth of structure in $\Lambda$CDM. 

Any combination of two variables out of $\{Q,\eta,\mu,\Sigma,\ldots\}$ is a
valid alternative to $(Q,\eta)$. It turns out that the pair $(\mu,\Sigma)$ is
particularly well suited when CMB, WL and LSS data are combined as it is less
correlated than others \citep[see][]{Zhao:2010dz, Daniel:2010yt,Axelsson:2011gt}.

\subsubsection{Parameterizations and non-parametric approaches}

So far we have defined two free functions that can encode any departure of the
growth of linear perturbations from $\Lambda$CDM. However, these free
functions are not measurable, but have to be inferred via their impact on the
observables. Therefore, one needs to specify a parameterization of, e.g.,
$(Q,\eta)$ such that departures from $\Lambda$CDM can be quantified.
Alternatively, one can use non-parametric approaches to infer the time and
scale-dependence of the modified growth functions from the observations.

Ideally, such a parameterization should be able to capture all relevant physics
with the least number of parameters. Useful parameterizations can be motivated
by predictions for specific theories of MG/DE \citep[see][]{Song:2010rm} and/or by
pure simplicity and measurability \citep[see][]{Amendola:2007rr}. For instance,
\cite{Zhao:2010dz} and \cite{Daniel:2010ky} use scale-independent
parameterizations that model one or two smooth transitions of the modified
growth parameters as a function of redshift. \cite{Bean:2010zq} also adds a
scale dependence to the parameterization, while keeping the time-dependence a simple power law:
\begin{eqnarray}
  Q(a,k) &\equiv& 1 + \left[ Q_0e^{-k/k_c} +
Q_\infty(1-e^{-k/k_c})-1\right]\,a^s\,,
\nonumber \\
  \eta(a,k)^{-1} &\equiv& 1 + \left[ R_0e^{-k/k_c} +
R_\infty(1-e^{-k/k_c})-1\right]\,a^s\,,
\end{eqnarray}
with constant $Q_0$, $Q_\infty$, $R_0$, $R_\infty$, $s$ and $k_c$. Generally,
the problem with any kind of parameterization is that it is
difficult -- if not impossible -- for it to be flexible enough to
describe all possible modifications.

Daniel et al.~\cite{Daniel:2010ky, Daniel:2010yt} investigate the
modified growth parameters binned in $z$ and $k$. The functions are
taken constant in each bin. This approach is simple and only mildly
dependent on the size and number of the bins. However, the bins can be
correlated and therefore the data might not be used in the most
efficient way with fixed bins. Slightly more sophisticated than simple
binning is a principal component analysis (PCA) of the binned (or
pixelized) modified growth functions. In PCA uncorrelated linear
combinations of the original pixels are constructed. In the limit of a
large number of pixels the model dependence disappears. At the moment
however, computational cost limits the number of pixels to only a
few. Zhao et al.~\cite{Zhao:2009fn, Zhao:2010dz} employ a PCA in the
$(\mu,\eta)$ plane and find that the observables are more strongly
sensitive to the scale-variation of the modified growth parameters
rather than the time-dependence and their average values. This
suggests that simple, monotonically or mildly-varying
parameterizations as well as only time-dependent parameterizations are
poorly suited to detect departures from $\Lambda$CDM.


\subsubsection{Trigger relations}

A useful and widely popular trigger relation is the value of the growth index
$\gamma$ in $\Lambda$CDM. It turns out that the value of $\gamma$ can also be fitted
also for simple DE models and sub-Hubble evolution in some MG models \citep[see,
e.g.,][]{Linder:2005in, Huterer:2006mva, Linder:2007hg, Linder:2009kq,
Nunes:2004wn, Ferreira:2010sz}. For
example, for a non-clustering perfect fluid DE model with equation of state
$w(z)$ the growth factor $G(a)$ given in (\ref{def_gf}) with the fitting formula
\begin{equation}\label{eq:growthdef}
  \gamma = 0.55 + 0.05\left[1+w(z=1)\right]
\end{equation}
is accurate to the $10^{-3}$ level compared with the actual solution of the
growth equation (\ref{lcdm_geq}). Generally, for a given solution of the growth
equation the growth index can simply be computed using
\begin{equation}
  \gamma(a,k) = \frac{\ln(\Delta_M')-\ln\Delta_M}{\ln \Omega_M(a)} \,.
  \label{gamma_Delta}
\end{equation}
The other way round, the modified gravity function $\mu$ can be computed for a
given $\gamma$ \citep{Pogosian:2010tj}
\begin{equation}
  \mu = \frac{2}{3}\Omega_M^{\gamma-1}(a) \left[ 
  \Omega_M^\gamma(a) +2 +(\ln H)' -3\gamma +\gamma'\ln\gamma \right] .
\end{equation}

The fact that the value of $\gamma$ is quite stable in most DE models but
strongly differs in MG scenarios means that a large deviation from
$\gamma_\Lambda$ signifies the breakdown of GR, a substantial DE clustering
or a breakdown of another fundamental hypothesis like near-homogeneity. 
Furthermore, using the
growth factor to describe the evolution of linear structure is a very simple and
computationally cheap way to carry out forecasts and compare theory with data.
However, several drawbacks of this approach can be identified:
\begin{itemize}
\item As only one additional parameter is introduced, a second parameter, such as
$\eta$, is needed to close the system and be general enough to capture all
possible modifications.

\item The growth factor is a solution of the growth equation on sub-Hubble
scales and, therefore, is not general enough to be consistent on all scales.

\item The framework is designed to describe the evolution of the matter density
contrast and is not easily extended to describe all other energy-momentum
components and integrated into a CMB-Boltzmann code.
\end{itemize}

\subsection{Phantom crossing}
\label{sec:crossing}

In this section we pay attention to the evolution of the perturbations of a general 
dark-energy fluid with an evolving equation of state parameter $w$.
Current limits on the equation of state parameter $w=p/\rho$
of the dark energy indicate that $p\approx -\rho$, and so do not
exclude $p<-\rho$, a region of parameter space often called
\emph{phantom energy}. Even though the region for which $w<-1$ may be unphysical at the
quantum level, it is still important to probe
it, not least to test for coupled dark energy and alternative theories of gravity or
higher dimensional models that can give rise to an effective or apparent
phantom energy. 

Although there is no problem in considering $w<-1$ for the background evolution, there
are apparent divergences appearing in the perturbations 
when a model tries to cross the limit $w=-1$. This is a potential headache for
experiments like Euclid that directly probe the perturbations through measurements
of the galaxy clustering and weak lensing. To analyze the Euclid data, we need to
be able to consider models that cross the phantom divide $w=-1$ at the level
of first-order perturbations (since the only dark-energy model that has
no perturbations at all is the cosmological constant).

However, at the level of cosmological first-order perturbation 
theory, there is no fundamental limitation that prevents an effective
fluid from crossing the phantom divide.

As $w \rightarrow -1$ the terms in Eqs.~(\ref{d_pert}) and (\ref{t_pert}) 
containing $1/(1+w)$ will generally
diverge. This can be avoided by replacing $\theta$ with a new variable
$V$ defined via $V=\rho\left( 1+w \right) \theta$. This corresponds to rewriting
the $0$-$i$ component of the energy momentum tensor as 
$ik_j T_{0}^{j}= V$, which avoids problems if $T_{0}^{j}\neq0$ when
$\bar{p}=-\bar\rho$. 
Replacing the time derivatives by a derivative with respect
to the logarithm of the scale factor $\ln a$ (denoted by a prime), we obtain 
\citep{Ma:1995ey,Hu:2004xd, Kunz:2006wc}:
\begin{eqnarray}
\delta' &=& 3(1+w) \Phi' - \frac{V}{H a} 
- 3 \left(\frac{\delta p}{\bar\rho}-w \delta \right) \label{eq:delta} \\
V' &=& -(1-3w) V+ \frac{k^2}{H a} \frac{\delta p}{\bar\rho}
+(1+w) \frac{k^2}{H a}\left( \Psi -\pi\right)\,.  \label{eq:v}
\end{eqnarray}
In order to solve Eqs.~(\ref{eq:delta}) and (\ref{eq:v}) 
we still need to specify the expressions
for $\delta p$ and $\pi$, quantities that characterize the physical, intrinsic nature 
of the dark-energy fluid at first order in perturbation theory. 
While in general the anisotropic stress 
plays an important role as it gives a measure of how 
the gravitational potentials $\Phi$ and $\Psi$ differ, we will set it
in this section to zero, $\pi=0$. Therefore, we will focus on the
form of the pressure perturbation. There are two important special
cases: barotropic fluids\footnote{As pointed out in \cite{Vikman:2004dc} barotropic fluids where the energy conservation
equation defines the evolution can in any case not cross $w=-1$ as this is a fixed point of the evolution.}, which have no internal degrees of freedom
and for which the pressure perturbation is fixed by the evolution of
the average pressure, and non-adiabatic fluids like, e.g., scalar fields
for which internal degrees of freedom can change the pressure
perturbation.

\subsubsection{Parameterizing the pressure perturbation}
\label{sec:pressure-perturbation}

\paragraph*{Barotropic fluids.}
We define a fluid to be barotropic if the pressure $p$ depends strictly
only on the energy density $\rho$: $p=p(\rho)$. These fluids 
have only adiabatic perturbations, so that they are often called adiabatic.
We can write their pressure as
\begin{equation}
p(\rho) = p(\bar{\rho}+\delta\rho) 
= p(\bar{\rho}) + \left.\frac{\mathrm{d}p}{{\mathrm{d}}\rho}\right|_{\bar{\rho}} \delta\rho
+ O\left[(\delta\rho)^2\right].
\label{eq:baro_exp}
\end{equation}
Here $p(\bar\rho) = \bar{p}$ is the pressure of the isotropic and homogeneous
part of the fluid. The second term
in the expansion (\ref{eq:baro_exp}) can be re-written as
\begin{equation}
\label{c_a_def}
\left.\frac{\mathrm{d}p}{\mathrm{d}\rho}\right|_{\bar{\rho}}
= \frac{\dot{\bar{p}}}{\dot{\bar{\rho}}} = w - \frac{\dot{w}}{3 aH(1+w)} 
\equiv c_a^2,
\end{equation}
where we used the equation of state and the conservation
equation for the dark-energy density in the background.
We notice that the adiabatic sound speed $c_a^2$ \label{symbol:ad.sound.speed}will necessarily diverge
for any fluid where $w$ crosses $-1$. 

However, for a perfect barotropic fluid the adiabatic sound speed $c_{a}^2$ turns out to
be the physical propagation speed of perturbations. Therefore, it should never be negative ($c_{a}^{2}<0$) -- otherwise classical, and possible quantum, instabilities appear (superluminal propagation, $c_{a}^{2}>1$, may be acceptable as the fluid is effectively a kind of ether that introduces a preferred frame, see \cite{Babichev:2007dw}). 
Even worse, the pressure perturbation
\begin{equation}
\delta p = c_{a}^{2} \delta\rho = \left( w - \frac{\dot{w}}{3 aH(1+w)} \right) \delta\rho
\end{equation}
will necessarily diverge if $w$ crosses $-1$ and $\delta\rho\neq0$.
Even if we find a way to stabilize the pressure perturbation, 
for instance an equation of state parameter that crosses the $-1$ limit with zero slope ($\dot{w}$), there will always be 
the problem of a negative speed of sound that prevents these
models from being viable dark-energy candidates \newtext{\cite{Vikman:2004dc,Kunz:2006wc}}.

\paragraph*{Non-adiabatic fluids.}\label{sec:non_ad}
To construct a model that can cross the phantom divide, we therefore
need to violate the constraint that $p$ is a unique function of $\rho$. 
At the level of first-order perturbation theory, this amounts
to changing the prescription for $\delta p$, which now becomes an arbitrary
function of $k$ and $t$.
One way out of this problem is to choose an appropriate gauge 
where the equations are simple; one choice is, for instance, the rest 
frame of the fluid where the pressure perturbation reads (in this frame) 
\begin{equation}
\hat{\delta p} = \hat{c}_{s}^{2}\hat{\delta\rho}\label{eq:dp-rest},
\end{equation}
where now the $\hat{c}_{s}^{2}$ is the speed with which fluctuations 
in the fluid propagate, i.e., the sound speed. 
We can write Eq.~(\ref{eq:dp-rest}), with an appropriate gauge transformation, 
in a form suitable for the Newtonian frame, i.e., for Eqs.~(\ref{eq:delta}) and (\ref{eq:v}). 
We find that the pressure perturbation is given by \citep{2002PhRvL..88l1301E, Bean:2003fb, 2003PhRvD..68j3501C}
\begin{equation}
\delta p = \hat{c}_s^2 \delta\rho + 3 a H\left(a\right)\left(\hat{c}_s^2 - c_{a}^{2}\right) \bar{\rho}\frac{V}{k^2} .
\label{eq:dp_rest}
\end{equation}
The problem here is the presence of $c_{a}^2$, which goes to 
infinity at the crossing and it is impossible that this term
stays finite except if $V\rightarrow 0$ fast enough 
or $\dot{w}=0$, but this is not, in general, the case.

This divergence appears because for $w=-1$ the energy momentum tensor Eq.~(\ref{EMT}) 
reads $T^{\mu\nu}=pg^{\mu\nu}$. 
Normally the four-velocity $u^{\mu}$ is the time-like 
eigenvector of the energy-momentum tensor, but now all vectors are 
eigenvectors. So the problem of fixing a unique rest-frame is no longer 
well posed. Then, even though the pressure perturbation looks fine for \emph{the 
observer in the rest-frame}, because it does not
diverge, the badly-defined gauge transformation to the Newtonian frame does, as it also contains 
$c_{a}^{2}$.

\subsubsection{Regularizing the divergences}

We have seen that neither barotropic fluids nor canonical scalar fields, for which the
pressure perturbation is of the type~(\ref{eq:dp_rest}), can cross the phantom divide. 
However, there is a simple model \citep[called the quintom model][]{Feng:2004ad,Hu:2004kh}
consisting of two fluids of the same type as in the previous Section~\ref{sec:pressure-perturbation} but with
a constant $w$ on either side of $w=-1$\footnote{While the scalar field with a constant $w<-1$ does not exhibit instabilities at the level of classical perturbation theory, it is generally a ghost and unstable at the quantum level. It is however possible to construct stable models without ghosts that can cross the phantom barrier, for example based on Kinetic Gravity Braiding \cite{Pujolas:2011he,Deffayet:2010qz}. As mentioned below, also the effective dark energy in coupled dark energy -- dark matter models can cross the phantom barrier.}. The combination of the two fluids then
effectively crosses the phantom divide if we start with $w_{\mathrm{tot}}>-1$, as the energy density in the 
fluid with $w<-1$ will grow faster, so that this fluid will eventually dominate and 
we will end up with $w_{\mathrm{tot}}<-1$.

The perturbations in this scenario were analyzed in detail in \citet{Kunz:2006wc}, 
where it was shown that in addition to the rest-frame contribution, one also has
relative and non-adiabatic perturbations. All these contributions apparently diverge
at the crossing, but their sum stays finite. When parameterizing the perturbations
in the Newtonian gauge as
\begin{equation}
\delta p(k,t) = \gamma(k,t)\, \delta\rho(k,t) \label{d_p_gamma}
\end{equation}
the quantity $\gamma$ will, in general, have a complicated time and scale
dependence. The conclusion of the analysis is that
indeed single canonical scalar fields with pressure perturbations of the type (\ref{eq:dp_rest})
in the Newtonian frame cannot cross $w=-1$, but that this is not the most general
case. More general models have \textit{a priori} no problem crossing the phantom divide,
at least not with the classical stability of the perturbations.

Kunz and Sapone~\cite{Kunz:2006wc} found that a good approximation to the quintom model
behavior can be found by regularizing the adiabatic sound speed in the gauge
transformation with
\begin{equation}
c_a^2 = w - \frac{\dot{w}(1+w)}{3 H a [(1+w)^2 + \lambda]} \label{eq:caeff}
\end{equation}
where $\lambda$ is a tunable parameter which determines how close to $w=-1$ the regularization kicks in. A value of $\lambda \approx 1/1000$ should work reasonably well. However, the final
results are not too sensitive on the detailed regularization prescription.

This result appears also related to the behavior found for coupled dark-energy models (originally 
introduced to solve the coincidence problem) where 
dark matter and dark energy interact not only through gravity \citep{Amendola:1999er}. The
effective dark energy in these models can also cross the phantom divide without divergences \citep{Huey:2004qv,Das:2006cc,Kunz:2007rk}.

The idea is to insert (by hand) a term in the continuity equations 
of the two fluids
\begin{eqnarray} 
&&\dot{\rho}_{M}+3H\rho_{M}=\lambda\\
&&\dot{\rho}_{x}+3H\left(1+w_{x}\right)\rho_{x}=-\lambda,
\end{eqnarray}
where the subscripts $m,x$ refer to dark matter and dark energy, respectively.
In this approximation, the adiabatic sound speed $c_{a}^{2}$ reads
\begin{equation} 
c_{a,x}^{2}= \frac{\dot{p}_{x}}{\dot{\rho}_{x}} = w_{x}-\frac{\dot{w_{x}}}{3aH\left(1+w_{x}\right)+
\lambda/\rho_{x}},
\end{equation}
which stays finite at crossing as long as $\lambda\neq 0$.

However in this class of models there are other instabilities 
arising at the perturbation level regardless of the coupling 
used, \citep[cf.][]{Valiviita:2008iv}.

\subsubsection{A word on perturbations when $w=-1$}

Although a cosmological constant has $w=-1$ and no perturbations, the converse is not
automatically true: $w=-1$ does not necessarily imply that there are no perturbations.
It is only when we set from the beginning (in the calculation): 
\begin{eqnarray}
p &=& -\rho\\
\delta p &=& -\delta\rho\\
\pi &=& 0 \,,
\end{eqnarray}
i.e., $T^{\mu\nu} \propto g^{\mu\nu}$, that we have as a solution $\delta = V =0$. 

For instance, if we set $w=-1$ and $\delta p = \gamma\delta\rho$ 
(where $\gamma$ can be a generic function)
in Eqs.~(\ref{eq:delta}) and (\ref{eq:v}) we have 
$\delta\neq 0$ and $V\neq 0$. 
However, the solutions are decaying modes due to the 
$-\frac{1}{a}\left(1-3w\right)V$ term so they are not important  
at late times; but it is interesting to notice that they are 
in general not zero.

As another example, if we have a non-zero anisotropic stress $\pi$ then the 
Eqs.~(\ref{eq:delta})\,--\,(\ref{eq:v}) will have a source term that will influence the growth 
of $\delta$ and $V$ in the same way as $\Psi$ does (just because they appear in the same way). 
The $\left(1+w\right)$ term in front of $\pi$ should not worry us as we can always define 
the anisotropic stress through
\begin{equation}
\rho\left(1+w\right) \pi = -\left(\hat{k}_{i}\hat{k}_{j}-\frac{1}{3}\delta_{ij}\right)
\Pi^{i}_{\,j} \,,
\label{eq:real-sigma}
\end{equation}
where $\Pi^{i}_{\,j}\neq 0$ when $i\neq j$ is the \emph{real} traceless 
part of the energy momentum tensor, probably the quantity we need to look at:
as in the case of $V=(1+w) \theta$, there is no need for $\Pi \propto (1+w)\pi$ to
vanish when $w=-1$.

It is also interesting to notice that when $w = -1$ the 
perturbation equations tell us 
that dark-energy perturbations are not influenced
through $\Psi$ and $\Phi'$ (see Eq.~(\ref{eq:delta}) and (\ref{eq:v})). Since
$\Phi$ and $\Psi$ are the quantities directly entering the metric, they must
remain finite, and even much smaller than $1$ for perturbation theory to hold.
Since, in the absence of direct couplings, the dark energy only feels the other
constituents through the terms $(1+w)\Psi$ and $(1+w)\Phi'$, it decouples
completely in the limit $w=-1$ and just evolves on its own. But its perturbations
still enter the Poisson equation and so the dark matter perturbation will
feel the effects of the dark-energy perturbations.

Although this situation may seem contrived, it might be that the acceleration of the universe is just an 
observed effect as a consequence of a modified theory of gravity. As was shown in~\cite{Kunz:2006ca}, any 
modified gravity theory can be described as an effective fluid both at 
background and at perturbation level; in such a situation it is imperative to 
describe its perturbations properly as this effective fluid may manifest unexpected behavior.

\section{Generic properties of dark energy and modified gravity models}
\label{genericproperties}

This section explores some generic issues that are not necessarily a feature of any particular model. We will recall the properties of particular classes of models as examples, leaving the details of the model description to section~\ref{models-of-modified-gravity}.

We begin by discussing the general implications of modelling dark energy as an extra degree of freedom, instead of the cosmological constant. We then discuss how the literature tends to categorize models into models of dark energy and models of modified gravity. We focus on the expansion of the cosmological background and ask what precision of measurement is necessary in order to make definite statements about large parts of the interesting model space. Then we address the issue of dark-energy perturbations, their impact on observables and how they can be used to distinguish between different classes of models. Finally, we present some general consistency relations among the perturbation variables that all
models of modified gravity should fulfill.

\subsection{Dark energy as a degree of freedom}
\label{sec:de-as-dof}

De Sitter spacetime, filled with only a cosmological constant, is static, undergoes no evolution. It is also invariant under Lorentz transformations. When other sources of energy-momentum are added into this spacetime, the dynamics occurs on top of this static background, or better to say -- vacuum. This is to say that the cosmological constant is a form of dark energy which has no dynamics of its own and the value of which is fixed for all frames and coordinate choices.

A dynamical model for acceleration implies the existence of some change of the configuration in space or time. It is no longer a gravitational vacuum. In the case of a perfectly homogeneous and isotropic universe, the evolution can only be a function of time. In reality, the universe has neither of these properties and therefore the configuration of any dynamical dark energy \emph{must} also be inhomogeneous. Whether the inhomogeneities are small is a model-dependent statement.

It is important to stress that there exists \emph{no} such thing as a modified gravity theory with no extra degrees of freedom beyond the metric. All models which seemingly involve just the metric degrees of freedom in some modified sense (say $f(R)$ or $f(G)$), in fact can be shown to be equivalent to general relativity plus an extra scalar degree of freedom with some particular couplings to gravity \cite{Chiba:2003ir, Kobayashi:2011nu}. Modifications such as massive gravity increase the number of polarisations.

In the context of $\Lambda$CDM, it has proven fruitful to consider the dynamics of the universe in terms of a perturbation theory: a separation into a background, linear and then higher-order fluctuations, each of increasingly small relevance (see section~\ref{sec:cosmo_perts}). These perturbations are thought to be seeded with a certain amplitude by an inflationary era at early times. Gravitational collapse then leads to a growth of the fluctuations, eventually leading to a breakdown of the perturbation theory; however, for dark matter in $\Lambda$CDM, this growth is only large enough to lead to non-linearity at smaller scales.

When dynamical DE is introduced, it must be described by at least one new (potentially more) degree of freedom. In principle, in order to make any statements about any such theory, one must specify the initial conditions on some space-like hypersurface and then the particular DE model will describe the subsequent evolutionary history. Within the framework of perturbation theory, initial conditions must be specified for \emph{both} the background and the fluctuations. The model then provides a set of \emph{related} evolution equations at each order.

We defer the discussion of the freedom allowed at particular orders to the appropriate sections below (section~\ref{sec:twpswmw} for the background, section~\ref{subsection:anisotropic_stress} for the perturbations). Here, let us just stress that since DE is a full degree of freedom, its initial conditions will contain both adiabatic and isocurvature modes, which may or may not be correlated, depending on their origin and which may or may not survive until today, depending on the particular model. Secondly, the non-linearity in the DE configuration is in principle independent of the non-linearity in the distribution of dark matter and will depend on both the particular model and the initial conditions. For example, the chameleon mechanism present in many non-minimally coupled models of dark energy acts to break down DE perturbation theory in higher-density environments (see section~\ref{sec:Screening}). This breakdown of linear theory is environment-dependent and only indirectly related to non-linearities in the distribution of dark matter.

Let us underline that the absolute and \emph{unique} prediction of $\Lambda$CDM is that $\Lambda$ is constant in space and time and therefore does not contribute to fluctuations at any order. Any violation of this statement at any one order, if it cannot be explained by astrophysics, is sufficient evidence that the acceleration is not caused by vacuum energy.

\subsection{A definition of modified gravity}

In this review we often make reference to DE and MG models. Although in an increasing number of publications a similar dichotomy
is employed, there is currently no consensus on where to draw the line between the
two classes. Here we will introduce an operational definition for
the purpose of this document.

Roughly speaking, what most people have in mind when talking about
standard dark energy are models of minimally-coupled scalar fields with standard kinetic
energy in 4-dimensional Einstein gravity, the only
functional degree of freedom being the scalar potential.
Often, this class of model is referred to simply as ``quintessence''.
However, when we depart from
this picture a simple classification is not easy to draw.
One problem is that, as we have seen in the previous sections, both at
background
and at the perturbation level, different models can have the same observational
signatures \citep{Kunz:2006ca}. This problem is not due to the use of perturbation theory:
any modification to Einstein's equations can be interpreted as
standard Einstein gravity
with a modified ``matter'' source, containing an arbitrary mixture of
scalars, vectors and tensors \citep{Hu:2007pj,Kunz:2008wt}.

Therefore, we could simply abandon any attempt to distinguish between DE and MG, and just analyse different models, comparing their properties and phenomenology.
However, there is a possible classification that helps us set targets
for the observations, which is often useful in concisely communicating 
the results of complex arguments.  In this review, we will use the following
notation: 
\begin{itemize}
\item \emph{Standard dark energy:} These are models in which dark energy lives in 
standard Einstein gravity \textit{and} does not
cluster appreciably on sub-horizon scales and does not carry anisotropic stress. As already noted, the
prime example of a standard dark-energy model is a
minimally-coupled scalar field with standard kinetic energy, for which
the sound speed
equals the speed of light. 
\item \emph{Clustering dark energy:} 
In clustering dark-energy models, there
is an additional contribution to the Poisson equation due to the 
dark-energy perturbation, which induces $Q \neq 1$. However, in this class 
we require $\eta=1$,
i.e., no extra effective anisotropic stress is induced by the extra dark component.
A typical example is a k-essence model with a low
sound speed, $c_s^2\ll 1$. 
\item \emph{Modified gravity models:} 
These are models where from the start the Einstein equations
are modified, for example scalar-tensor and $f(R)$ type theories,
Dvali--Gabadadze--Porrati (DGP) as
well as interacting dark energy, in which effectively a fifth force is
introduced in addition to gravity. Generically they change the clustering and/or induce
a non-zero anisotropic stress. Since our definitions are based
on the phenomenological parameters, we also add dark-energy models that 
live in Einstein's gravity
but that have
non-vanishing anisotropic stress into this class since they cannot be distinguished
by cosmological observations.
\end{itemize}

Notice that both clustering dark energy and explicit modified gravity models
lead to deviations from what is often called `general relativity' (or, like here, standard dark energy) in the literature when
constraining extra perturbation parameters like the growth index $\gamma$.
For this reason we generically call both of these classes MG models. 
In other words, in this review we use the simple and by now extremely
popular (although admittedly somewhat
misleading) expression ``modified gravity'' to denote models in which
gravity is modified and/or dark energy clusters or interacts with other fields.
Whenever we feel useful, we will  remind the reader of the actual meaning of the
expression ``modified gravity'' in this review.

Therefore, on sub-horizon scales and at first order in perturbation theory
our definition of MG is
straightforward: models with $Q=\eta=1$ (see Eq.~\ref{mod_constr}) are standard DE,
otherwise they are MG models. In this sense the definition above
is rather convenient: we can use it to quantify, for instance, how well
Euclid will distinguish between standard dynamical dark energy and
modified gravity
by forecasting the errors on $Q,\eta$ or on related quantities like
the growth index $\gamma$.

On the other hand, it is clear that this definition is only a
practical way to group
different models and should not be taken as a fundamental one. We do not try to
set a precise threshold on, for instance, how much dark energy should
cluster before we call it
modified gravity: the boundary between the classes is therefore left
undetermined but we think
this will not harm the understanding of this document.

\subsection{The background: to what precision should we measure \textit{w}?}
\label{sec:twpswmw}

The effect of dark energy on background expansion is to add a new source of energy density. The chosen model will have dynamics which will cause the energy density to evolve in a particular manner. On the simplest level, this evolution is a result of the existence of intrinsic hydrodynamical pressure of the dark energy fluid which can be described by the instantaneous equation of state. Alternatively, an interaction with other species can result in a non-conservation of the DE EMT and therefore change the manner in which energy density evolves (e.g. coupled dark energy). Taken together, all these effects add up to result in an \emph{effective} equation of state for DE which drives the expansion history of the universe.

It is important to stress that all background observables are geometrical in nature and therefore can only be measurements from \emph{curvatures}. It is not possible to disentangle the dark energy and dark matter in a model independent matter and therefore only the measurement of the Hubble parameter up to a normalization factor, $H(z)/H_0$, and the spatial curvature $\Omega_{k0}$ can be obtained in a DE-model independent manner. In particular, the measurement of the dark-matter density, $\Omega_{m0}$, becomes possible only on choosing some parameterization for $w_\text{eff}$ (e.g.\ a constant) \cite{Amendola:2012ky}. One must therefore always be mindful that extracting DE properties from background measurements is limited to constraining the coefficients of a chosen \emph{parameterization} of the \emph{effective} equation of state for the DE component, rather than being measurements of the actual effective $w$ and definitely not the intrinsic $w$ of the dark energy.

Given the above complications, two crucial questions are often asked in the context of
dark-energy surveys:
\begin{itemize}
\item Since current measurements of the expansion history appear so consistent with $w=-1$, do we not already know that the dark energy is a cosmological constant?
\item To which precision should we measure $w$? Or equivalently,
why is the Euclid target precision of about 0.01 on $w_0$ and 0.1
on $w_a$ interesting?
\end{itemize}

We will now attempt to answer these questions at least
partially. First, we address the question of what the measurement of $w$ can tell us about the viable model space of DE. Then we examine whether
we can draw useful lessons from inflation. Finally, we will look at
what we can learn from arguments based on Bayesian model comparison.

In the first part, we will argue that whereas any detection of a deviation from $\Lambda$CDM expansion history immediately implies that acceleration is not driven by a cosmological constant, the converse is not true, even if $w=-1$ exactly. We will also argue that a detection of a phantom equation of state, $w<-1$, would reveal that gravity is not minimally coupled or that dark energy interacts and immediately eliminate the perfect-fluid models of dark energy, such as quintessence.

Then we will see that for single field slow-roll inflation models we effectively measure $w \sim -1$ with percent-level accuracy (see Figure~\ref{fig:w_k}); however, the deviation from a scale-invariant spectrum means that we nonetheless observe a dynamical evolution and, thus, a deviation from an exact and constant equation of state of $w=-1$. 
Therefore, we know that inflation was not due to a cosmological constant; we also know that we can see no deviation from  a de Sitter expansion for a precision smaller than the one Euclid will reach.

In the final part, we will consider the Bayesian evidence in favor of a true cosmological constant if we keep finding $w=-1$; we will see that for priors on $w_0$ and $w_a$ of order unity, a precision like the one for Euclid is necessary to favor a true cosmological constant decisively. We will also discuss how this conclusion changes depending on the choice of priors.

\subsubsection{What can a measurement of $w$ tell us?}
\label{sec:w-tellsus}

The prediction of $\Lambda$CDM is that $w=-1$ exactly at all times. Any detection of a deviation from this result immediately disqualifies the cosmological constant as a model for dark energy.

The converse is not true, however. Simplest models of dynamical dark energy, such as quintessence (section~\ref{quintessence}) can approach the vacuum equation of state arbitrarily closely, given sufficiently flat potentials and appropriate initial conditions. An equation of state $w=-1$ at all times is inconsistent with these models, but this may never be detectable.

Moreover, there exist classes of models, e.g.\ shift-symmetric k-\emph{essence} with de-Sitter attractors, which have equation of state $w=-1$ exactly, once the attractor is approached. Despite this, the acceleration is not at all driven by a cosmological constant, but by a perturbable fluid which has vanishing sound speed and can cluster. Such models can only be differentiated from a cosmological constant by the measurements of perturbations, if at all, see section~\ref{subsection:anisotropic_stress}.

Beyond eliminating the cosmological constant as a mechanism for acceleration, measuring $w>-1$ is not by itself very informative as to the nature of dark energy. Essentially all classes of models can evolve with such an equation of state given appropriate initial conditions (which is \emph{not} to say that any evolution history can be produced by any class of models). On the other hand, the observation of a phantom equation of state, $w<-1$, at any one moment in time is hugely informative as to the nature of gravitational physics. It is well known that any such background made up of either a perfect fluid or a minimally coupled scalar field suffers from gradient instabilities, ghosts or both \cite{Dubovsky:2005xd}. Therefore such an observation immediately implies that either gravity is non-minimally coupled and therefore there is a fifth force, that dark energy is not a perfect fluid, that dark energy interacts with other species, or that dynamical ghosts are not forbidden by nature, perhaps being stabilized by a mechanism such as ghost condensation \cite{Arkani-Hamed-etal:2004}. Any of these would provide a discovery in itself as significant as excluding a cosmological constant.

In conclusion, we aim to measure $w$ since it is the most direct way of disproving that acceleration is caused by a cosmological constant. However, if it turns out that no significant deviation can be detected this does not imply that the cosmological constant is the mechanism for dark energy. The clustering properties must then be verified and found to not disagree with $\Lambda$CDM predictions. 

\subsubsection{Lessons from inflation}
\label{sec:lessons-inflation}
In all probability the observed late-time acceleration of the universe
is not the first period of accelerated expansion that occurred during its evolution: the current standard
model of cosmology incorporates a much earlier phase with $\ddot{a}>0$,
called inflation. Such a period provides a natural mechanism for generating several
properties of the universe: spatial flatness, gross homogeneity and isotropy on scales beyond naive causal horizons and nearly scale-invariant initial fluctuations.

The first lesson to draw from inflation is that it cannot have been due to a pure cosmological constant. This is immediately clear since inflation actually ended and therefore there had to be some sort of time evolution. We can go even further: since de Sitter spacetime is static, no curvature perturbations are produced in this case (the fluctuations are just unphysical gauge modes) and therefore an exactly scale-invariant power spectrum would have necessitated an alternative mechanism. 

The results obtained by the Planck collaboration from the first year of data imply that the initial spectrum of fluctuations is not scale invariant, but rather has a tilt given by $n_\text{s} = 0.9608\pm 0.0054$ and is consistent with no running and no tensor modes \cite{Ade:2013zuv}. This is consistent with the final results from WMAP \cite{Hinshaw:2012aka}. It is surprisingly difficult to create this observed fluctuation spectrum in alternative scenarios that are strictly causal and only act on sub-horizon scales \citep{Spergel:1997vq, Scodeller:2009iu}.

Let us now translate what the measured properties of the initial power spectrum of fluctuations imply for a observer existing during the inflationary period. We will assume that inflation was driven by one of the simple models (i.e.\ with sound speed $c_\text{s}=1$). Following the analysis in \citet{Ilic:2010zp}, we notice that
\begin{equation} \label{eq:w_lessons}
1+w = - \frac{2}{3} \frac{\dot{H}}{H^2} = \frac{2}{3} \varepsilon_H \,,
\end{equation}
where $\epsilon_H \equiv 2 M^2_{\mathrm{Pl}}(H'/H)^2$ and where the prime denotes a derivative with respect to the inflaton field. 

This equation of state is directly related to the tensor-to-scalar ratio through $r \sim 24(1+w)$. Since no tensor modes have been detected thus far, no deviation from $w=-1$ has been seen either. In fact the Planck 95\% limit is $r\lesssim 0.1$ implying that $1+w\lesssim0.04$. Moreover, the spectral tilt itself is also related to the rate of change of $w$, 
\begin{align}
\frac{d \ln(1+w)}{d N} &= 2 (\eta_H - \varepsilon_H) \label{eq:dw}\\
2\eta_H &= (n_s-1)+4\varepsilon_H \notag
\end{align}
where $\eta_H \equiv 2 M^2_{\mathrm{Pl}} H''/H $. Thus, if $n_s\neq1$ we have
that either $\eta_H\neq0$ or $\varepsilon_H\neq0$, and consequently
either $w\neq-1$ or $w$ is not constant at the pivot scale.
 
We can rewrite Eq.~(\ref{eq:w_lessons}) as
\begin{equation}
(1+w) = - \frac{1}{6} (n_s-1) + \frac{\eta_H}{3}
\approx 0.007 +  \frac{\eta_H}{3} \,. \label{eq:w-tilt}
\end{equation}
Without tuning, it is natural for $\eta_H\sim \mathcal{O}(\varepsilon_H^2)$. However, classes of models exist where $\eta_H\sim \varepsilon_H$. Thus, given the observations of the scale dependence of the initial curvature fluctuations, we can conclude that $1+w$ should lie between 0.005 and 0.04, which is well within the current experimental bounds on the DE equation of state and roughly at the limit of Euclid's sensitivity. We have plotted the allowed values of $w$ as a function of scale in Figure~\ref{fig:w_k}.

We should note that there are classes of models where the cancellation between $\eta_H$ and the tilt in Eq.~\eqref{eq:w-tilt} is indeed natural which is why one cannot give a lower limit for the amplitude of primordial gravitational waves and $w$ lies arbitrarily close to $-1$. On the other hand, the observed period of inflation is probably in the middle of a long slow-roll phase. By Eq.~\eqref{eq:dw}, this cancellation would only happen at one moment in time. We have plotted the typical evolution of $w$ in inflation in Figure~\ref{fig:w_N}.

Despite being the only other physically motivated period of acceleration, inflation does occur at a very different energy scale, between 1~MeV and GUT scale $10^{16}$~GeV, while the energy scale for dark energy is $10^{-3}$~eV. We should therefore be wary about pushing the analogy too far.

\epubtkImage{wk.png}{%
\begin{figure}
\centerline{\scalebox{0.7}{\includegraphics{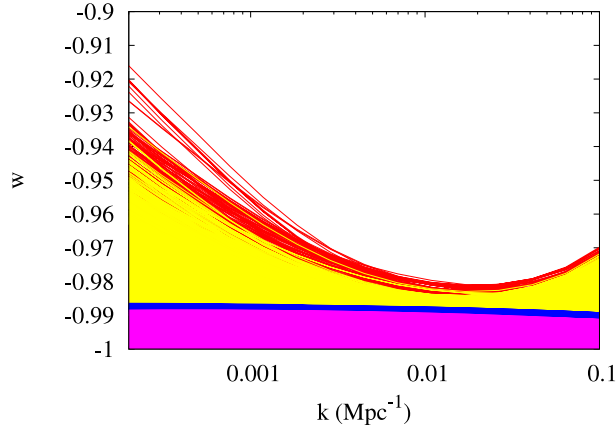}}}
\caption{The evolution of $w$ as a function of the comoving scale $k$,
  using only the 5-year WMAP CMB data. Red and yellow are the 95\% and
  68\% confidence regions for the LV formalism. Blue and purple are
  the same for the flow-equation formalism. From the outside inward,
  the colored regions are red, yellow, blue, and purple. Image
  reproduced by permission from~\cite{Ilic:2010zp}; copyright by APS.}
\label{fig:w_k}
\end{figure}}

\epubtkImage{wN.png}{%
\begin{figure}
\centerline{\scalebox{0.8}{\includegraphics{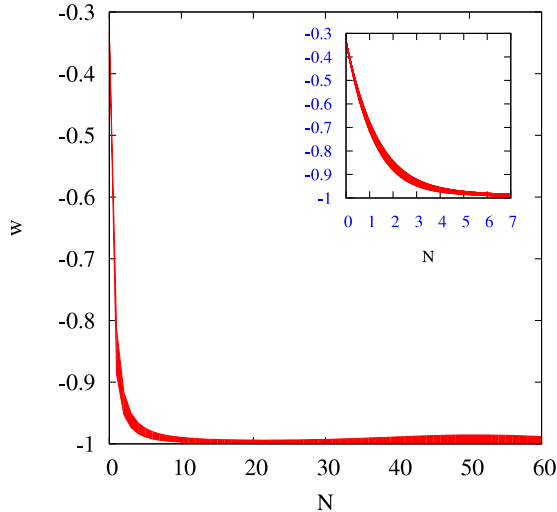}}}
\caption{The complete evolution of $w(N)$, from the flow-equation
  results accepted by the CMB likelihood. Inflation is made to end at
  $N=0$ where $w(N=0)=-1/3$ corresponding to $\epsilon_H(N=0)=1$. For
  our choice of priors on the slow-roll parameters at $N=0$, we find
  that $w$ decreases rapidly towards $-1$ (see inset) and stays close
  to it during the period when the observable scales leave the horizon
  ($N\approx 40\mbox{\,--\,}60$). Image reproduced by permission
  from~\cite{Ilic:2010zp}; copyright by APS.}
\label{fig:w_N}
\end{figure}}

\subsubsection{When should we stop? Bayesian model comparison}
\label{sec:BayesStop}

In section~\ref{sec:w-tellsus}, we explained that the measurement of the equation of state $w$ can exclude some classes of models, including the cosmological constant of $\Lambda$CDM. However, most \emph{classes} of models allow the equation of state to be arbitrarily close to that of vacuum energy, $w=-1$, while still representing completely different physics. Since precision cannot be infinite, we need to propose an algorithm to determine how well this property should be measured.  As we showed in section~\ref{sec:lessons-inflation} above, inflation provides an example of a period that acceleration that, if it occurred at late times would have been judged as consistent with $w=-1$ given today's constraints. We therefore should require a better measurement, but how much better?

We approach the answer to this question from the perspective of Bayesian evidence: at what precision does the non-detection of a deviation of the background expansion history signifies that we should prefer the simpler null hypothesis that $w=-1$.

In our Bayesian framework, the first model, the null hypothesis $M_0$, posits that the background expansion is due to an extra component of energy density that has equation of state $w=-1$ at all times. The other models
assume that the dark energy is dynamical in a way that is well parametrized either by 
an arbitrary constant $w$ (model $M_1$) or by a linear fit $w(a)=w_0+(1-a) w_a$ (model $M_2$).

Here we are using the constant and linear parametrization of $w$ because on the one
hand we can consider the constant $w$ to be an effective quantity, averaged over redshift with the appropriate weighting
factor for the observable, see \citet{Simpson:2006bd},
and on the other hand because the
precision targets for observations are conventionally phrased in terms of the figure
of merit (FoM) given by $1/\sqrt{|{\mathrm{Cov}}(w_0,w_a)|}$. We will, therefore, find a
direct link between the model probability and the FoM. It would be an interesting
exercise to repeat the calculations with a more general model, using e.g. PCA, although
we would expect to reach a similar conclusion.

Bayesian model comparison aims to compute the relative model probability
\begin{equation}
\frac{P(M_0|d)}{P(M_1|d)} = \frac{P(d|M_0)}{P(d|M_1)} \frac{P(M_0)}{P(M_1)} 
\end{equation}
where we used Bayes formula and where $B_{01}\equiv P(d|M_0)/P(d|M_1)$ is called
the Bayes factor. The Bayes factor is the amount by which our relative
belief in the two models is modified by the data, with $\ln B_{01} >
\reviewtext{0\ (<0)}$ indicating a preference for model 0 (model 1). Since model $M_0$ is nested in $M_1$ at the point
$w=-1$ and in model $M_2$ at $(w_0=-1,w_a=0)$, we can use the Savage--Dickey (SD) density ratio \citep[e.g.][]{Trotta:2005ar}.
Based on SD, the Bayes factor between the two models is just the ratio of
posterior to prior at $w=-1$ or at $(w_0=-1,w_a=0)$, marginalized over all other parameters.

Let us start by following~\cite{BMICRTetal} and consider the Bayes factor $B_{01}$ between a cosmological
constant model $w=-1$ and a free but constant effective $w$.
If we assume that the data are
compatible with $\weff=-1$ with an uncertainty $\sigma$, then the
Bayes factor in favor of a cosmological constant is given by
\begin{equation} \label{eq:B}
 B = \sqrt{\frac{2}{\pi}}\frac{\Delta_{+} + \Delta_{-}}{\sigma}
 \left[\text{erfc}\left(-\frac{\Delta_+}{\sqrt{2}\sigma}\right)
- \text{erfc}\left(\frac{\Delta_-}{\sqrt{2}\sigma}\right)
  \right]^{-1},
\end{equation}
where for the evolving dark-energy model we have adopted a flat
prior in the region $-1 - \Delta_{-} \leq \weff \leq -1+\Delta_+$
and we have made use of the Savage--Dickey density ratio formula
\citep[see][]{Trotta:2005ar}. The prior, of total width $\Delta =
\Delta_+ + \Delta_-$, is best interpreted as a factor describing
the predictivity of the dark-energy model under consideration. In what follows we will consider example benchmark three models as alternatives to $w=-1$:
\begin{itemize}
\item \emph{Fluid-like}: we assume that the acceleration is driven by a fluid the background configuration of which satisfies both the strong energy condition and the null energy condition, i.e.\ we have that
$\Delta_+ = 2/3, \Delta_- = 0$.

\item \emph{Phantom}: phantom models violate the null energy condition, i.e.\
are described by $\Delta_+ = 0, \Delta_- > 0$, with the latter
being possibly rather large.

\item \emph{Small departures}: We assume that the equation of state is very close to that of vacuum energy, as seems to have been the case during inflation: $\Delta_+ = \Delta_- = 0.01$.
\end{itemize}
A model with a large $\Delta$ will be more generic and
less predictive, and therefore is disfavored by the Occam's razor
of Bayesian model selection, see Eq.~\eqref{eq:B}. According to the Jeffreys' scale for the strength
of evidence, we have a moderate (strong) preference for the
cosmological constant model for $2.5 < \ln B_{01} < 5.0$ ($\ln B_{01}>5.0$),
corresponding to posterior odds of 12:1 to 150:1 (above
150:1).

\epubtkImage{LnB30_col.png}{%
\begin{figure}[htbp]
\centerline{\includegraphics[width=0.5\textwidth]{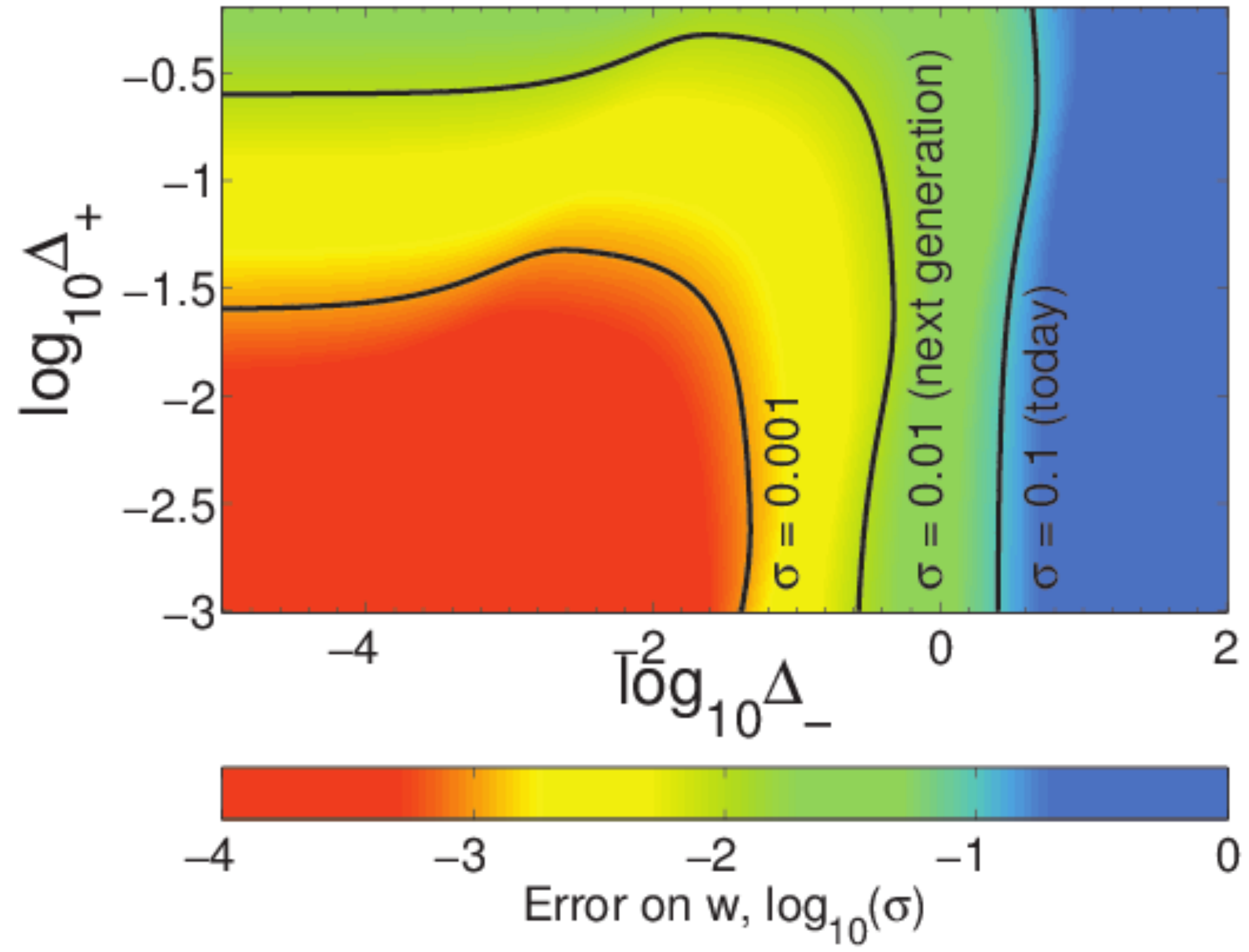}}
\caption{Required accuracy on $\weff = -1$ to obtain strong
evidence against a model where $-1 - \Delta_{-} \leq \weff \leq
-1+\Delta_+$ as compared to a cosmological constant model, $w=-1$.
For a given $\sigma$, models to the right and above the contour
are disfavored with odds of more than 20:1.}
\label{fig:sigma}
\end{figure}}

\begin{table}[htbp]
\caption{Strength of evidence disfavoring the three example benchmark
models against a $\Lambda$CDM expansion history, using an indicative
accuracy on $w=-1$ from present data, $\sigma \sim 0.1$.}
\label{table:evidence}
\centering
\begin{tabular}{l l l }
\toprule
 Model & $(\Delta_+ , \Delta_- )$ & $\ln B$ today ($\sigma = 0.1$)\\
\midrule
 Phantom & $( 0, 10)$      & $4.4$ (strongly disfavored)\\
 Fluid-like & $(2/3,  0)$ & $1.7$ (slightly disfavored) \\
 Small departures & $( 0.01, 0.01)$ & $0.0$ (inconclusive) \\
\bottomrule
\end{tabular}
\end{table}

We plot in Figure~\ref{fig:sigma} contours of constant observational
accuracy $\sigma$ in the model predictivity space
$(\Delta_-,\Delta_+)$ for $\ln B = 3.0$ from Eq.~\eqref{eq:B},
corresponding to odds of 20 to 1 in favor of a cosmological
constant (slightly above the ``moderate'' threshold. The figure can be interpreted as giving the space of
extended models that can be significantly disfavored with respect
to $w=-1$ at a given accuracy. The results for the 3 benchmark models mentioned above
(fluid-like, phantom or small departures from $w=-1$) are
summarized in Table~\ref{table:evidence}.  Instead, we can ask the question 
which precision needs to reached to support $\Lambda$CDM at a given
level. This is shown in Table~\ref{table:sigma} for odds 20:1 and 150:1. We see that
to rule out a fluid-like model, which also covers the parameter space expected for canonical
scalar field dark energy, we need to reach a precision comparable to the one that
the Euclid satellite is expected to attain.

\begin{table}[htbp]
\caption{Required precision $\sigma$ of the value of $w$ for future surveys in order to disfavor
  the three benchmark models against $w=-1$ for two different
  strengths of evidence.}
\label{table:sigma}
\centering
\begin{tabular}{l ll l }
\toprule
 Model & $(\Delta_+ , \Delta_- )$ & \multicolumn{2}{c}{Required $\sigma$ for odds} \\
 ~     & ~                        & $>20:1$  & $>150:1$  \\
\midrule
 Phantom & $( 0, 10)$      & $0.4$           & $5\cdot10^{-2}$\\
 Fluid-like & $(2/3,  0)$ & $3\cdot10^{-2}$ & $3\cdot10^{-3}$\\
 Small departures & $( 0.01, 0.01)$ & $4\cdot10^{-4}$ & $5\cdot10^{-5}$\\
\bottomrule
\end{tabular}
\end{table}

By considering the model $M_2$ we can also provide a direct link with the target DETF FoM:
Let us choose (fairly arbitrarily) a flat probability distribution for the prior, of width
$\Delta w_0$ and $\Delta w_a$ in the dark-energy parameters, so that the value of the prior is $1/(\Delta w_0 \Delta w_a)$
everywhere. Let us assume that the likelihood is Gaussian in $w_0$ and $w_a$ and centered
on $\Lambda$CDM (i.e., the data fully supports $\Lambda$ as the dark energy). 

As above, we need to distinguish different cases depending on the width of the prior.
If you accept the argument of the previous section that we expect only a small deviation from $w=-1$,
and set a prior width of order 0.01 on both $w_0$ and $w_a$, then the posterior is dominated by the
prior, and the ratio will be of order 1 if the future data is compatible with $w=-1$. Since the precision of the experiment is
comparable to the expected deviation, both $\Lambda$CDM and evolving dark energy are equally probable (as
argued above and shown for model $M_1$ in Table~\ref{table:evidence}), 
and we have to wait for a detection of $w\neq-1$ or a significant further 
increase in precision (cf.\ the last row in Table~\ref{table:sigma}).

However, one often considers a much wider range for $w$, for example the fluid-like model with $w_0\in [-1/3,-1]$
and $w_a \in [-1,1]$ with equal probability (and neglecting some subtleties near $w=-1$). If the likelihood
is much narrower than the prior range, then the value of the normalized posterior at $w=-1$ will be
$2/(2\pi\sqrt{|{\mathrm{Cov}}(w_0,w_a)|}={\mathrm{FoM}}/\pi$ (since we excluded $w<-1$, else it would half this value).
The Bayes factor is then given by
\begin{equation}
B_{01} = \frac{\Delta w_0 \Delta w_a {\mathrm{FoM}}}{\pi} \,.
\end{equation}
For the prior given above, we end up with $B_{01} \approx 4 {\mathrm{FoM}}/(3\pi) \approx 0.4 {\mathrm{FoM}}$. In order
to reach a ``decisive'' Bayes factor, usually characterized as $\ln B > 5$ or $B > 150$, we thus need a figure of
merit exceeding 375. Demanding that Euclid achieve a FoM $> 500$ places us, therefore, on the safe side and
allows to reach the same conclusions (the ability to favor the $\Lambda$CDM expansion history decisively \emph{if} the data is in full
agreement with $w=-1$) under small variations of the prior as well.
\\

To summarize, the most direct effect of dynamical dark energy is the modification of the expansion history. We used inflation as a dark-energy prototype to show that the current experimental
bounds of $w \approx -1.0 \pm 0.1$ are not yet sufficient to significantly favor a parameter-free $\Lambda$CDM expansion history: we showed that we need to reach a percent-level accuracy both
to have any chance of observing a deviation of $w$ from $-1$ if the dark energy is similar to inflation, and because it is at this point that a $w=-1$ expansion history beings to be favored decisively for prior widths of order $1$.

We do not expect to be able to improve much our knowledge with a lower-precision measurement of $w$, unless dark energy is significantly different from $w=-1$ either at late times or, for example, owing to a significant early-dark-energy component \cite{pettorino_etal_2013}. A large deviation would be the preferred situation for Euclid, as then we would be able to observe the evolution of dark energy rather than just a fixed state, which would be much more revealing. However, even if the expansion history matches that of $\Lambda$CDM to some arbitrary precision, this does not imply that the cosmological constant is accelerating the universe. Even on such configurations a large amount of freedom exists which can then only be tested by investigating the evolution of large-scale structure, to which we now turn.

\subsection{Dark-energy: linear perturbations and growth rate} 
\label{subsection:anisotropic_stress}

Without a given model for dark energy, the evolution of its perturbations is not determined by the background expansion history. As we have explained in section~\ref{sec:de-as-dof}, the cosmological constant is the only form of dark energy which does not carry fluctuations at all, with all dynamical DE models clustering to a larger or smaller extent. Since both dark matter and dark energy must interact (at least) through gravity, the existence of these fluctuations would alter the geodesics on which pressureless dark matter moves and therefore also change the clustering history of the dark matter. This implies that the appropriate evolution history for perturbations is another consistency check that $\Lambda$CDM must satisfy over and above the matching background expansion history.

In order to meaningfully discuss dark-energy fluctuations, we must specify the following:
\begin{itemize}
	\item the field content of the dark-energy sector
	\item the initial conditions for the fluctuations
	\item either the initial conditions for the DE background configuration and subsequent evolution or a measurement of the background expansion history as discussed in section~\ref{sec:twpswmw}
	\item the rules for evolving the fluctuations (i.e.\ the model of dark energy)
\end{itemize}
A scalar-vector-tensor decomposition of the perturbations on the FLRW background can be performed, where each of the spins evolves independently at linear order. Since general relativity only contains tensors as dynamical degrees of freedom, any dynamics in the scalar and vector modes is determined by the matter content. Therefore fluctuations of dark energy provide a source for metric degrees of freedom. 

Typically, a vector or tensor degree of freedom will contain all the lower helicities and therefore source all the perturbations of lower spin. For example, a vector dark energy (e.g.\ Einstein-Aether or TeVeS), will in general source both vector and scalar perturbations. Higher-spin perturbations would affect polarization predictions for the CMB if the dark energy contributed a significant part of the energy density during recombination \cite{Lim:2004js}, but otherwise are unconstrained and appear largely uninvestigated in the literature, where most attention is paid to scalar modes even in models containing higher-spin matter. If the dark energy itself contains multiple degrees of freedom, the perturbations will also feature internal modes which do not change any of the potential observables, such as the gravitational potentials, and only affect how they evolve in time. 

Each of the new dynamical modes must be given appropriate initial conditions. Typically, they should be set during inflation, where the dark energy plays the role of a spectator field(s). In particular, the dark energy will contribute to the scalar adiabatic mode, which is constant on scales larger than the cosmological horizon. In addition, it will introduce new isocurvature modes with respect to matter and radiation. In general, these only decay if the dark energy interacts with the other matter components and equilibrates, in particular if the dark energy features a tracker in its evolution history \cite{Malquarti:2002iu}. These isocurvature modes affect the CMB and are strongly constrained, but again only if the dark energy is a significant fraction of total energy density during recombination, such as in early dark energy models. Otherwise, the isocurvature modes  don't become relevant until the late universe where they can affect structure formation or at least the magnitude of the ISW effect on the CMB \cite{Gordon:2004ez}. If the dark-energy is not coupled to the inflaton and is not involved in reheating, the isocurvature modes are likely to be statistically uncorrelated.

In practice, for the purpose of the late universe, the assumption is made that the isocurvature modes are not present and only the scalar adiabatic mode is considered. Let us take this point of view for the remainder of this section. Therefore what we discuss now are the possible rules for evolving the linear scalar perturbations. 

As discussed in section~\ref{mg_growth_params}, a ``closure" relation between the dark matter density perturbation $\Delta_M$ and the scalar gravitational potentials $\Phi$ and $\Psi$ is enough to describe the evolution of all these variables \cite{Sawicki:2012re}. We can define two variables $Q$ and $\eta$,
\begin{equation}
  -k^2 \Phi = 4\pi G Q(a,k) a^2 \rho_M \Delta_M \,, \qquad 
\Phi=\eta(a,k)\Psi \,.
  \label{mod_constr2}
\end{equation}
which provide such closure relations. However, these two variables are actually just a recasting of the equations: they effectively parameterize the particular solutions that are realized by the universe, rather than necessarily saying anything in particular about the model: $Q$ describes the energy density perturbations of dark energy in the realized configuration, while $\eta$ is a statement about the anisotropic stress carried by that configuration. It should be clear that at any \emph{one} particular redshift it is always be possible to arrange the dark-energy configuration such that $Q=\eta=1$.

In $\Lambda$CDM, the dark-matter density perturbation $\Delta_M$ and the gravitational potential $\Phi$ are related through a constraint, if we ignore the other components in the universe, such as baryons and radiation. In dynamical dark-energy models with a single additional degree of freedom, the $\Delta_M$ equation is one of two coupled second-order differential equations, with time- and scale-dependent coefficients determined by the model. 

At this point, one often employs the quasi-static approximation, i.e.\ neglecting all time derivatives in the dynamical equation for $\Phi$, which is equivalent to treating it as a constraint and therefore the dark energy purely as an effective modification of the constraint structure of general relativity, rather than as a full degree of freedom.  This sort of approximation seems to be valid inside the sound horizon for the dark energy at least for some models, although it is not proven that it works in general. Under this quasi-static approximation, in any (single) scalar-tensor theory \cite{Sawicki:2012re}
\begin{equation}
	Q = h_1 \left( \frac{ a^2 H^2 + k^2 h_5} {a^2 H^2 + k^2 h_3} \right)\,, \quad \eta = h_2 \left( \frac {a^2 H^2 + k^2 h_4} {a^2 H^2 + k^2 h_5} \right) \,,
\end{equation}
where the $h_i$ are essentially arbitrary functions of time only, determined by the action of the model. In fact, a more general argument given in Ref.~\cite{Silvestri:2013ne}, proves that simply requiring quasi-staticity and locality for the theory of dark energy implies that both the functions $Q$ and $\eta$ are ratios of polynomials of $\left(k/aH\right)^2$ with coefficients purely functions of time. Theories which break these assumptions, can have a different structure, e.g.\ DGP where contributions appear at order $k/aH$ as a result of the existence of a singular brane source, \cite{Amin:2007wi}.

Let us now discuss under what circumstances the two functions $Q$ and $\eta$ deviate from their $\Lambda$CDM values considerably and therefore would presumably significantly change observables. 

\subsubsection{Anisotropic stress: $\eta\neq 1$}
\label{sec:aniso}
A deviation of $\eta$ from 1 results from anisotropic stress at first order in perturbations of the fluid. This occurs in the early universe owing to relativistic neutrinos, but is negligible at late times. Note that at second order in perturbations anisotropic stress is always present (e.g. \cite{Ballesteros:2011cm}).

The existence of anisotropic stress is a frame-dependent question. Models such as $f(R)$ gravity  which exhibit anisotropic stress, can be redefined through a frame transformation to have none. In addition to specifying a model, we must therefore fix a frame in order to discuss this properly. The natural frame to pick is the Jordan frame of the baryons. This is defined as the frame related to the metric on the geodesics of which visible matter propagates when in free fall. In many modified-gravity models, this is also the Jordan frame for dark matter, i.e.\ gravity-like forces couple to just the EMT, irrespective of species.  All observations of perturbations to be performed by Euclid are those of the motion of galaxies and light which directly probes the Jordan-frame metric for these species. 

Given the fixed baryon Jordan frame, the anisotropic stress appears whenever the effective Planck mass is not constant, i.e.\ whenever the normalization of the kinetic term of gravitons is time dependent, or when the speed of tensor modes is different from the speed of light. Anisotropic stress therefore is a probe of the nature of the action for gravitational waves.  This occurs whenever there are dynamical degrees of freedom which are coupled non-minimally to gravity. For example, the $f(R)$ action, seemingly without any additional dynamical degrees of freedom, can be Legendre transformed into the equivalent \cite{Chiba:2003ir}
\begin{equation}
	S=\frac {1}{2\kappa^2} \int \mathrm{d}^4x\sqrt{-g} \left[ \phi R + V(\phi) + \mathcal{L}_m(\Psi_m) \right]\,,
\end{equation}
with $V(\phi)\equiv f - R(\phi)f_R$, where the coupling between gravity and the scalar $\phi=f_R$ is explicit.

On the other hand, many Coupled Dark Energy models  are constructed to be  very similar to $f(R)$, but introduce a split between the dark matter and visible matter frames. When the visible matter is subdominant gravitationally, the growth of dark matter perturbations in these two classes of models should be very similar. However, all the measurements are always performed through the galaxies and weak lensing and therefore observations are different; in particular, there is no anisotropic stress in CDE models \cite{Motta:2013cwa}. 

When dealing with multiple degrees of freedom, it is in principle possible to tune them in such a way that the time-variation of the Planck mass cancels out and therefore there would be no anisotropic stress despite non-minimal coupling to gravity in the baryon Jordan frame. However, if the action for the two degrees of freedom is of a different model class, it is not clear whether it is possible to perform this cancellation during more than one era of the evolution of the universe, e.g.\ matter domination (see \cite{Saltas:2010tt} for the case of $f(R,G)$ gravity \footnote{All $f(G)$ models with $G$ the Gauss-Bonnet term as in fact a particular choice of non-minimally coupled scalar-tensor theories, a subclass of the Horndeski lagrangian, see Ref.~\cite{Kobayashi:2011nu}.}).

\subsubsection{Clustering: $Q\neq 1$}
\label{sec:clustering}

The implication of DE clustering, $Q\neq1$ is that the dark matter perturbations are dressed with dark energy perturbations. This means that the effect of some particular density of matter is to curve space differently than in $\Lambda$CDM. On scales where $Q$ is a constant, the dark-energy distribution follows that of the dark matter precisely.  $QG_\text{N}$ is an effective Newton's constant for non-relativistic matter.

If the curvature of space sourced by the DM changes, then so does the gravitational force acting on the dark matter. This implies that given a fixed background expansion history, the growth rate of the perturbations is different, see section \eqref{mg_growth_params}.
\\

As discussed in section~\ref{sec:de-as-dof}, only a cosmological constant is not perturbed at all. Therefore only in this case do we have $Q=1$ exactly up to relativistic corrections near the cosmological horizon, $k/aH\sim 1$. 

However, when the dark energy comprises a single dynamical degree of freedom and has an EMT of perfect-fluid form (e.g.\ k-essence is the most general model of this type, with quintessence \ref{quintessence} as a subclass) and the background expansion history is very close to $\Lambda$CDM, the exact equations for the DE/DM system coupled through gravity can be written as \cite{Motta:2013cwa}
\begin{align}
\Phi'' &+ \left( 4 + \frac{H'}{H} + 3c_\text{a}^2 \right) \Phi' + \left( 3+ 2\frac{H'}{H}+3c_\text{a}^2 \right)\Phi  + \left( \frac{c_\text{s}k}{aH} \right)^2 \Phi = \label{Phievol}\\
& - \frac{3}{2}\Omega_\text{m} \left(c_\text{s}^2 \delta_\text{m} + 3(c_\text{a}^2-c_\text{s}^2) \frac{a^2 H}{k^2} \theta_\text{m}\right) \notag\\
\delta_\text{m}' &+ H^{-1}\theta_\text{m} = 3\Phi' \qquad \theta_\text{m}' + 2\theta_\text{m} = \frac{k^2}{a^2H}\Phi
\end{align}
with $c_\text{a}^2\equiv p'_\text{DE}/\rho_\text{DE}'$ the adiabatic sound speed and the prime denoting differentiation w.r.t.~$\ln a$. Deep inside the sound horizon, $c_\text{s}k/aH\gg 1$, the standard Poisson constraint can be recovered from Eq.~\eqref{Phievol}, i.e.\ $Q=1$. The growth rate of the dark matter perturbation is then fully determined by the background expansion and is only different from the $\Lambda$CDM one when the expansion history deviates significantly. This is the standard case of quintessence, for which the sound speed $c_\text{s}=1$.

In the opposite limit of clustering dark energy, $c_\text{s}=0$, which is typical of k-essence models such as the ghost-condensate \cite{Arkani-Hamed-etal:2004} or dusty dark energy \cite{Lim:2010yk}, there is no sound horizon, just as for dark-matter dust, and in principle the dark energy clusters. If the background expansion is now very close to $\Lambda$CDM (i.e.\ $w\approx-1$ and $c_\text{a}^2\approx 0$), Eq.~\eqref{Phievol} reduces to the standard equation in $\Lambda$CDM. If the initial conditions are adiabatic, the evolution of both the potential and of the dark matter density is the same as in $\Lambda$CDM, i.e.\ $Q=1$ again. Any deviations are purely a result of a different background expansion history. The above implies that given a very similar expansion history to $\Lambda$CDM, dark-energy models comprising a single degree of freedom that is minimally coupled do not significantly cluster \cite{Sapone:2009} if the initial conditions are adiabatic and there is no anisotropic stress. 
\\

For significant clustering, a coupling of dark energy to gravity or dark matter is required, with a strength similar to that of gravity (or possibly to other species, with appropriately stronger couplings to compensate for the relatively smaller energy density). All models which exhibit significant anisotropic stress also cluster significantly, since the anisotropic stress is a sign of non-minimal coupling to gravity, see section~\ref{sec:aniso}. This implies that models such as coupled dark energy also cluster, since they are effectively scalar-tensor models with non-universal couplings to matter.

If the couplings are universal, the most general class of models where $Q\neq1$ while there is no anisotropic stress are kinetic gravity braiding models of dark energy \cite{Deffayet:2010qz}, which are a class of imperfect fluids \cite{Pujolas:2011he}. The effective Planck mass is constant in these models, however there is still non-minimal coupling to gravity on the level of the equations of motion. This implies that they cluster significantly \cite{Kimura:2010di}.
\\

In summary, given a fixed background expansion history close to $\Lambda$CDM, the appearance of anisotropic stress is a sign of a modification of the action for gravitational waves in the Jordan frame of baryons: either a time-varying effective Planck mass, i.e.\ a normalization scale for graviton kinetic terms or a deviation of the speed of  gravitational waves from that of light.  On the other hand, a detection of significant clustering, resulting in a growth rate significantly deviating from the $\Lambda$CDM one, is a sign of coupling of the dark-energy to gravity or some of the species with a strength similar to that of gravity.

\subsection{Parameterized frameworks for theories of modified gravity}

As explained in earlier sections of this report, modified-gravity
models cannot be distinguished from dark-energy models by using solely
the FLRW background equations. But by comparing the background
expansion rate of the universe with observables that depend on linear
perturbations of an FRW spacetime we can hope to distinguish between
these two categories of explanations. An efficient way to do this is
via a parameterized, model-independent framework that describes
cosmological perturbation theory in modified gravity. We present here
one such framework, the parameterized post-Friedmann formalism
\cite{Baker2013:vx}%
\epubtkFootnote{Not to be confused with a different formalism of the
  same name by other authors \cite{Hu:2007pj}.}
that implements possible extensions to the linearized gravitational field equations.

The parameterized post-Friedmann approach (PPF) is inspired by the parameterized post-Newtonian (PPN) formalism \cite{WillNordtvedt1972,Will1971}, which uses a set of parameters to summarize leading-order deviations from the metric of GR. PPN was developed in the 1970s for the purpose of testing of alternative gravity theories in the solar system or binary systems, and is valid in weak-field, low-velocity scenarios. 
PPN itself cannot be applied to cosmology, because we do not know the exact form of the linearized metric for our Hubble volume. Furthermore, PPN can only test for constant deviations from GR, whereas the cosmological data we collect contain inherent redshift dependence.

For these reasons the PPF framework is a parameterization of the gravitational field equations (instead of the metric) in terms of a set of functions of redshift. A theory of modified gravity can be analytically mapped onto these PPF functions, which in turn can be constrained by data.

We begin by writing the perturbed Einstein field equations for spin-0 (scalar) perturbations in the form:
\begin{equation}
\label{abc}
\delta G_{\mu\nu} \;=\; 8\pi G\,\delta T_{\mu\nu}+\delta
U_{\mu\nu}^{\mathrm{metric}}+\delta
U_{\mu\nu}^{\mathrm{d.o.f}}+\mathrm{\ gauge\ invariance\ fixing\ terms}\,,
\end{equation}
where $\delta T_{\mu\nu}$ is the usual perturbed stress-energy tensor of all cosmologically-relevant fluids. The tensor $\delta U_{\mu\nu}^{\mathrm{metric}}$ holds new terms that may appear in a modified theory, containing perturbations of the metric (in GR such perturbations are entirely accounted for by $\delta G_{\mu\nu}$). $\delta U_{\mu\nu}^{\mathrm{d.o.f.}}$ holds perturbations of any new degrees of freedom that are introduced by modifications to gravity. A simple example of the latter is a new scalar field, such as introduced by scalar-tensor or Galileon theories. However, new degrees of freedom could also come from spin-0 perturbations of new tensor or vector fields, St$\ddot{\mathrm{u}}$ckelberg fields, effective fluids and actions based on curvature invariants (such as $f\left(R\right)$ gravity).

In principle there could also be new terms containing matter perturbations on the RHS of Eq.~(\ref{abc}). However, for theories that maintain the weak equivalence principle -- i.e., those with a Jordan frame where matter is uncoupled to any new fields -- these matter terms can be eliminated in favor of additional contributions to $\delta U_{\mu\nu}^{\mathrm{metric}}$ and $\delta U_{\mu\nu}^{\mathrm{d.o.f.}}$.

The tensor $\delta U_{\mu\nu}^{\mathrm{metric}}$ is then expanded in terms of two gauge-invariant perturbation variables $\hat\Phi$ and $\hat\Gamma$. $\hat\Phi$ is one of the standard gauge-invariant Bardeen potentials, while $\hat\Gamma$ is the following combination of the Bardeen potentials: $\hat\Gamma=1/k (\dot{\hat\Phi}+{\cal H}\hat\Psi)$. We use $\hat\Gamma$ instead of the usual Bardeen potential $\hat\Psi$ because $\hat\Gamma$ has the same derivative order as $\hat\Phi$ (whereas $\hat\Psi$ does not). We then deduce that the only possible structure of $\delta U_{\mu\nu}^{\mathrm{metric}}$ that maintains the gauge-invariance of the field equations is a linear combination of $\hat\Phi$, $\hat\Gamma$ and their derivatives, multiplied by functions of the cosmological background (see Eqs.~(\ref{FE1})\,--\,(\ref{FE4}) below). 

$\delta U_{\mu\nu}^{\mathrm{d.o.f.}}$ is similarly expanded in a set of gauge-invariant potentials $\{\hat\chi_i\}$ that contain the new degrees of freedom. \cite{Baker2013:vx} presented an algorithm for constructing the relevant gauge-invariant quantities in any theory.

For concreteness we will consider here a theory that contains only one new degree of freedom and is second-order in its equations of motion (a generic but not watertight requirement for stability, see \cite{Woodard07}). Then the four components of Eq.~(\ref{abc}) are:
\begin{align}
-a^2\delta G^0_0&=8\pi a^2 G\,\rho_M\delta_M+A_0 k^2\hat\Phi+F_0k^2\hat\Gamma+\alpha_0k^2\hat\chi+\alpha_1k\dot{\hat\chi}+k^3 M_{\Delta}(\dot\nu+2\epsilon)\label{FE1}\\ 
-a^2\delta G^0_i&=\nabla_i\left[8\pi a^2 G\,\rho_M (1+\omega_M)\theta_M+B_0 k\hat\Phi+I_0k\hat\Gamma+\beta_0 k\hat\chi+\beta_1\dot{\hat\chi}+k^2M_{\Theta}(\dot\nu+2\epsilon)\right]\label{FE2}\\
a^2\delta G^i_i&=3\,8\pi a^2 G\,\rho_M\Pi_M+C_0 k^2\hat\Phi+C_1 k\dot{\hat\Phi}+J_0k^2\hat\Gamma+J_1 k\dot{\hat\Gamma}+\gamma_0 k^2\hat\chi+\gamma_1 k \dot{\hat\chi}+\gamma_2 \ddot{\hat\chi}\\
&\quad +k^3M_P (\dot\nu+2\epsilon)\label{FE3}\\
a^2\delta \hat{G}^i_j&=8\pi a^2 G\,\rho_M (1+\omega_M)\Sigma_M+ D_0\hat\Phi+\frac{D_1}{k} \dot{\hat\Phi}+K_0\hat\Gamma+\frac{K_1}{k}\dot{\hat\Gamma}+\epsilon_0\hat\chi+\frac{\epsilon_1}{k}\dot{\hat\chi}+\frac{\epsilon_2}{k^2} \ddot{\hat\chi}  \label{FE4}
\end{align}
where $\delta \hat{G}^i_j=\delta G^i_j-\frac{\delta^i_j}{3}\delta G^k_k$. Each of the lettered coefficients in Eqs.~(\ref{FE1})\,--\,(\ref{FE4}) is a function of cosmological background quantities, i.e., functions of time or redshift; this dependence has been suppressed above for clarity. Potentially the coefficients could also depend on scale, but this dependence is not arbitrary \cite{Silvestri2013:rt}). These PPF coefficients are the analogy of the PPN parameters; they are the objects that a particular theory of gravity `maps onto', and the quantities to be constrained by data. Numerous examples of the PPF coefficients corresponding to well-known theories are given in \cite{Baker2013:vx}.

The final terms in Eqs.~(\ref{FE1})\,--\,(\ref{FE3}) are present to ensure the gauge invariance of the modified field equations, as is required for any theory governed by a covariant action. The quantities $M_\Delta$, $M_\Theta$ and $M_P$ are all pre-determined functions of the background. $\epsilon$ and $\nu$ are off-diagonal metric perturbations, so these terms vanish in the conformal Newtonian gauge. The gauge-fixing terms should be regarded as a piece of mathematical book-keeping; there is no constrainable freedom associated with them.

One can then calculate observable quantities -- such as the weak lensing kernel or the growth rate of structure $f(z)$ -- using the parameterized field equations (\ref{FE1})\,--\,(\ref{FE4}). Similarly, they can be implemented in an Einstein--Boltzmann solver code such as \textsc{camb} \cite{CAMB} to utilize constraints from the CMB. If we take the divergence of the gravitational field equations (i.e., the unperturbed equivalent of Eq.~(\ref{abc})), the left-hand side vanishes due to the Bianchi identity, while the stress-energy tensor of matter obeys its standard conservation equations (since we are working in the Jordan frame). Hence the $U$-tensor must be separately conserved, and this provides the necessary evolution equation for the variable $\hat\chi$:
\begin{align}
\label{Uconsv}
\delta\left(\nabla^\mu\left[U_{\mu\nu}^{\mathrm{metric}}+U_{\mu\nu}^{\mathrm{d.o.f.}}\right]\right)&=0.
\end{align}  
Eq.~(\ref{Uconsv}) has two components. If one wishes to treat theories with more than two new degrees of freedom, further information is needed to supplement the PPF framework.

The full form of the parameterized equations~(\ref{FE1})\,--\,(\ref{FE4}) can be simplified in the `quasistatic regime', that is, significantly sub-horizon scales on which the time derivatives of perturbations can be neglected in comparison to their spatial derivatives~\cite{Hu:2007pj}. Quasistatic lengthscales are the relevant stage for weak lensing surveys and galaxy redshift surveys such as those of Euclid. A common parameterization used on these scales has the form:
\begin{align}
2\nabla^2\Phi&=8\pi a^2G\,\mu(a,k)\,{\bar \rho}_M\Delta_M, \label{Poisson}\\
\frac{\Phi}{\Psi}&=\gamma(a,k), \label{slip}
\end{align}
where $\{\mu,\gamma\}$ are two functions of time and scale to be
constrained. This parameterization has been widely employed
\cite{Bertschinger:2008zb, Daniel:2010yt, Linder:2007hg, Bean:2010zq,
  Pogosian:2010tj, Zhao:2010dz, Dossett:2011vu,Hojjati:2011df,
  Hojjati:2011vp}. It has the advantages of simplicity and somewhat
greater physical transparency: $\mu (a,k)$ can be regarded as
describing evolution of the effective gravitational constant, while
$\gamma (a,k)$ can, to a certain extent, be thought of as acting like
a source of anisotropic stress (see
Section~\ref{subsection:anisotropic_stress}).

Let us make a comment about the number of coefficient functions employed in the PPF formalism. One may justifiably question whether the number of unknown functions in Eqs.~(\ref{FE1})\,--\,(\ref{FE4}) could ever be constrained. In reality, the PPF coefficients are not all independent. The form shown above represents a fully agnostic description of the extended field equations. However, as one begins to impose restrictions in theory space (even the simple requirement that the modified field equations must originate from a covariant action), constraint relations between the PPF coefficients begin to emerge. These constraints remove freedom from the parameterization.

Even so, degeneracies will exist between the PPF coefficients. It is likely that a subset of them can be well-constrained, while another subset have relatively little impact on current observables and so cannot be tested. In this case it is justifiable to drop the untestable terms. Note that this realization, in itself, would be an interesting statement -- that there are parts of the gravitational field equations that are essentially unknowable.

Finally, we note that there is also a completely different, complementary approach to parameterizing modifications to gravity. Instead of parameterizing the linearized field equations, one could choose to parameterize the perturbed gravitational action. This approach has been used recently to apply the standard techniques of effective field theory to modified gravity; see~\cite{Battye2012:kd,Bloomfield2012:as,Gubitosi2013:pv} and references therein.

\section{Models of dark energy and modified gravity}
\label{models-of-modified-gravity}
In this section we review a number of popular models of dynamical DE and MG. This section is  more
technical than the rest and it is meant to provide a quick but self-contained
review of the current research in the theoretical foundations of DE models. The selection of models is of course somewhat arbitrary but we have tried to cover the most well-studied cases and those that introduce new and interesting observable phenomena.

\subsection{Quintessence}
\label{quintessence}

In this review we refer to scalar field models with canonical kinetic energy 
in Einstein's gravity as ``quintessence models''. 
Scalar fields are obvious candidates for dark energy, as they are for the inflaton, for many reasons: they are the simplest fields since they lack internal degrees of freedom,  do not introduce
preferred directions, are typically weakly clustered (as discussed later on), and can easily drive an accelerated expansion.
If the kinetic energy has a canonical form, the only degree
of freedom is then provided by the field potential (and of course by the initial conditions).
The typical requirement is that the potentials are flat
enough to lead to the slow-roll inflation today with an energy scale
$\rho_{\mathrm{DE}}\simeq10^{-123}m_{\mathrm{pl}}^{4}$ and a mass scale
$m_{\phi}\lesssim10^{-33}\mathrm{\ eV}$. 

Quintessence models are the prototypical DE models
\citep{1998PhRvL..80.1582C} and as such are the most studied
ones. Since they have been explored in many reviews of DE, we limit
ourselves here to a few remarks.%
\epubtkFootnote{This subsection is based on~\cite{Amendola2010}.}

The quintessence model is  described by the action
\begin{eqnarray}
S=\int{\mathrm{d}}^{4}x\sqrt{-g}\,\left[\frac{1}{2\kappa^{2}}R+{\cal
    L}_{\phi}\right]+S_{M}\,,\qquad{\cal
  L}_{\phi}=-\frac{1}{2}g^{\mu\nu}\partial_{\mu}\phi\partial_{\nu}\phi-V(\phi)\,,\label{action}
\end{eqnarray}
 where $\kappa^{2}=8\pi G$ and $R$ is the Ricci scalar and  $S_{M}$ is the matter action.
The fluid satisfies the continuity equation
\begin{eqnarray}
\dot{\rho}_{M}+3H(\rho_{M}+p_{M})=0\,.\label{rhomeq}
\end{eqnarray}
The energy-momentum tensor of quintessence is
\begin{eqnarray}
T_{\mu\nu}^{(\phi)} & = & -\frac{2}{\sqrt{-g}}\frac{\delta(\sqrt{-g}{\cal L}_{\phi})}{\delta g^{\mu\nu}}\\
 & = & \partial_{\mu}\phi\partial_{\nu}\phi-g_{\mu\nu}\left[\frac{1}{2}g^{\alpha\beta}\partial_{\alpha}\phi\partial_{\beta}\phi+V(\phi)\right]\,.\end{eqnarray}
As we have already seen, in a FLRW background, the energy density $\rho_{\phi}$
and the pressure $p_{\phi}$ of the field are \begin{eqnarray}
\rho_{\phi}=-{T_{0}^{0}}^{(\phi)}=\frac{1}{2}\dot{\phi}^{2}+V(\phi)\,,\quad p_{\phi}=\frac{1}{3}{T_{i}^{i}}^{(\phi)}=\frac{1}{2}\dot{\phi}^{2}-V(\phi)\,,\end{eqnarray}
 which give the equation of state \begin{eqnarray}
w_{\phi}\equiv\frac{p_{\phi}}{\rho_{\phi}}=\frac{\dot{\phi}^{2}-2V(\phi)}{\dot{\phi}^{2}+2V(\phi)}\,.\label{wphiqui}
\end{eqnarray}
In the flat universe, Einstein's equations give
 the following equations of motion:
\begin{eqnarray}
 &  & H^{2}=\frac{\kappa^{2}}{3}\left[\frac{1}{2}\dot{\phi}^{2}+V(\phi)+\rho_{M}\right]\,,\label{Heq1}\\
 &  &
   \dot{H}=-\frac{\kappa^{2}}{2}\left(\dot{\phi}^{2}+\rho_{M}+p_{M}\right)\,,\label{Heq2}
\end{eqnarray}
where $\kappa^{2}=8\pi G$. The variation of the action~(\ref{action})
with respect to $\phi$ gives
\begin{eqnarray}
\ddot{\phi}+3H\dot{\phi}+V_{,\phi}=0\,,\label{Klein}
\end{eqnarray}
where $V_{,\phi}\equiv{\mathrm{d}}V/{\mathrm{d}}\phi$. 

During radiation or matter dominated epochs, the energy density $\rho_{M}$
of the fluid dominates over that of quintessence, i.e., $\rho_{M}\gg\rho_{\phi}$.
If the potential is steep so that the condition $\dot{\phi}^{2}/2\gg V(\phi)$
is always satisfied, the field equation of state is given by $w_{\phi}\simeq1$
from Eq.~(\ref{wphiqui}). In this case the energy density of the
field evolves as $\rho_{\phi}\propto a^{-6}$, which decreases much
faster than the background fluid density.

The condition $w_{\phi}<-1/3$ is required to realize the late-time
cosmic acceleration, which translates into the condition $\dot{\phi}^{2}<V(\phi)$.
Hence the scalar potential needs to be shallow enough for the field
to evolve slowly along the potential. This situation is similar to
that in inflationary cosmology and it is convenient to introduce the
following slow-roll parameters \citep{2006RvMP...78..537B}
\begin{equation}
\epsilon_{s}\equiv\frac{1}{2\kappa^{2}}\left(\frac{V_{,\phi}}{V}\right)^{2} \,,
\qquad \eta_{s}\equiv\frac{V_{,\phi\phi}}{\kappa^{2}V} \,.
\label{slowroll}
\end{equation}
If the conditions $\epsilon_{s}\ll1$ and $|\eta_{s}|\ll1$ are satisfied,
the evolution of the field is sufficiently slow so that $\dot{\phi}^{2}\ll V(\phi)$
and $|\ddot{\phi}|\ll|3H\dot{\phi}|$ in Eqs.~(\ref{Heq1}) and (\ref{Klein}).

From Eq.~(\ref{Klein}) the deviation of $w_{\phi}$ from $-1$
is given by 
\begin{equation}
\label{eq:pot_w}
1+w_{\phi}=\frac{V_{,\phi}^{2}}{9H^{2}(\xi_{s}+1)^{2}\rho_{\phi}} \,,
\end{equation}
where $\xi_{s}\equiv\ddot{\phi}/(3H\dot{\phi})$. This shows that
$w_{\phi}$ is always larger than $-1$ for a positive potential and energy density.
In the slow-roll limit, $|\xi_{s}|\ll1$ and $\dot{\phi}^{2}/2\ll V(\phi)$,
we obtain $1+w_{\phi}\simeq2\epsilon_{s}/3$ by neglecting the matter
fluid in Eq.~(\ref{Heq1}), i.e., $3H^{2}\simeq\kappa^{2}V(\phi)$.
The deviation of $w_{\phi}$ from $-1$ is characterized by the slow-roll
parameter $\epsilon_{s}$. It is also possible to consider Eq.~(\ref{eq:pot_w})
as a prescription for the evolution of the potential given $w_\phi(z)$ and to
reconstruct a potential that gives a desired evolution of the equation of
state (subject to $w\in[-1,1]$). This was used, for example, in~\cite{Bassett:2002qu}.

However, in order to study the evolution of the perturbations of a quintessence field
it is not even necessary to compute the field evolution explicitly. Rewriting the perturbation
equations of the field in terms of the perturbations of the density contrast $\delta_\phi$ and
the velocity $\theta_\phi$ in the conformal Newtonian gauge, one finds \cite[see, e.g.,][Appendix A]{Kunz:2006wc} that they
correspond precisely to those of a fluid, (\ref{d_pert}) and (\ref{t_pert}), with $\pi=0$ and 
$\delta p = c_s^2 \delta\rho + 3 a H (c_s^2-c_a^2) (1+w) \rho \theta/k^2$ with $c_s^2=1$. The adiabatic sound speed, $c_a$, is defined in Eq.~(\ref{c_a_def}).
The large value of the sound speed $c_s^2$, equal to the speed of light, means that
quintessence models do not cluster significantly inside the horizon \citep[see][and Section~\ref{soundspeed} for a detailed analytical discussion of quintessence clustering
and its detectability with future probes, for arbitrary $c_s^2$]{Sapone:2009,Sapone:2010}.

Many quintessence potentials have been proposed in the literature. A simple
crude classification divides them into two classes, (i) ``freezing" models
and (ii) ``thawing'' models
\citep{2005PhRvL..95n1301C}. In class (i) the field was rolling along the
potential in the past, but the movement gradually slows down after
the system enters the phase of cosmic acceleration. The representative
potentials that belong to this class are

\vspace{0.2cm}
 \underline{\textbf{(i) Freezing models}} 
\begin{itemize}
\item $V(\phi)=M^{4+n}\phi^{-n}\quad(n>0)$\,, 
\item $V(\phi)=M^{4+n}\phi^{-n}\exp(\alpha\phi^{2}/m_{\mathrm{pl}}^{2})$\,. 
\end{itemize}
The former potential does not possess a minimum and hence the field
rolls down the potential toward infinity.
This appears, for example, in the fermion condensate model as a dynamical
supersymmetry breaking \citep{1999PhRvD..60f3502B}. The latter potential has
a minimum at which the field is eventually trapped (corresponding
to $w_{\phi}=-1$). This potential can be constructed in the framework
of supergravity \citep{1999PhLB..468...40B}.

\vspace{0.3cm}
In thawing models (ii) the field (with mass $m_{\phi}$) has been frozen
by Hubble friction (i.e., the term $H\dot{\phi}$ in Eq.~(\ref{Klein})) until recently and
then it begins to evolve once $H$ drops below $m_{\phi}$. The equation
of state of DE is $w_{\phi}\simeq-1$ at early times, which
is followed by the growth of $w_{\phi}$. The representative potentials
that belong to this class are

\vspace{0.2cm}
 \underline{\textbf{(ii) Thawing models}} 
\begin{itemize}
\item $V(\phi)=V_{0}+M^{4-n}\phi^{n}\quad(n>0)$\,, 
\item $V(\phi)=M^{4}\cos^{2}(\phi/f)$\,. 
\end{itemize}
The former potential is similar to that of chaotic inflation ($n=2,4$)
used in the early universe (with $V_{0}=0)$ \citep{Linde:1983},
while the mass scale $M$ is very different. The model with
$n=1$ was  proposed 
by~\cite{2003JCAP...10..015K} in connection with the possibility to allow
for negative values of $V(\phi)$. The universe will collapse in the
future if the system enters the region with $V(\phi)<0$. The latter
potential appears as a potential for the Pseudo-Nambu--Goldstone Boson
(PNGB)\index{Pseudo-Nambu-Goldstone Boson (PNGB)}. This was introduced
by~\cite{1995PhRvL..75.2077F} in response to the first
tentative suggestions that the universe may be dominated by the cosmological
constant. In this model the field is nearly frozen at the potential
maximum during the period in which the field mass $m_{\phi}$ is smaller
than $H$, but it begins to roll down around the present ($m_{\phi}\simeq H_{0}$).

Potentials can also be classified in several other ways, e.g., on the basis
of the existence of special solutions. For instance,
tracker solutions have approximately constant $w_{\phi}$ and $\Omega_{\phi}$ along
 special attractors. A wide range of initial conditions converge
to a common, cosmic evolutionary tracker. Early DE models
contain instead solutions in which DE was not negligible even during the last scattering.
While in the specific Euclid forecasts section (\ref{euclid-forecasts}) we will not explicitly consider these models, it is worthwhile to note that 
the combination of  observations of the CMB and of large scale structure (such as Euclid) can dramatically constrain these models drastically improving the inverse area figure of merit compared to current constraints, as discussed in~\cite{Huterer:2006mv}.

\subsection{K-essence}
\label{sec:k-essence}

In a quintessence model it is the potential energy of a scalar field that leads to the late-time acceleration of the expansion of the universe; the alternative, in which the kinetic energy of the scalar field which dominates, is known as k-essence.  Models of k-essence are characterized by an action for the scalar field of the following form
\begin{equation}
S=\int \mathrm{d}^4 x\;\sqrt{-g}p(\phi,X) \,,
\end{equation}
where $X=(1/2)g^{\mu\nu}\nabla_{\mu}\phi\nabla_{\nu}\phi$.  The energy density of the scalar field is given by 
\begin{equation}
\rho_{\phi}=2X\frac{\mathrm{d}p}{\mathrm{d}X}-p \,,
\end{equation}
and the pressure is simply $p_{\phi}=p(\phi,X)$.  Treating the k-essence scalar as a perfect fluid, this means that k-essence has the equation of state
\begin{equation}
w_{\phi}=\frac{p_{\phi}}{\rho_{\phi}}=-\frac{p}{p-2X p,_{X}} \,,
\end{equation}
where the subscript $,_{X}$ indicates a derivative with respect to
$X$.  Clearly, with a suitably chosen $p$ the scalar can have an appropriate equation of state to allow it to act as dark energy.

The dynamics of the k-essence field are given by a continuity equation
\begin{equation}
\dot{\rho}_{\phi}=-3H(\rho_{\phi}+p_{\phi}) \,,
\end{equation}
or equivalently by the scalar equation of motion
\begin{equation}
G^{\mu\nu}\nabla_{\mu}\nabla_{\nu}\phi+2X\frac{\partial^2p}{\partial X \partial \phi}-\frac{\partial p}{\partial \phi}=0 \,,
\end{equation}
where
\begin{equation}
G^{\mu\nu}=\frac{\partial p}{\partial X}g^{\mu\nu}+\frac{\partial^2 p}{\partial X^2}\nabla^{\mu}\phi\nabla^{\nu}\phi \,.
\end{equation}
For this second order equation of motion to be hyperbolic, and hence physically meaningful, we must impose
\begin{equation}
1+2X\frac{p,_{XX}}{p,_X}>0 \,.
\end{equation}
K-essence was first proposed by~\cite{Armendariz-PiconMukhanovSteinhardt2000, 2001PhRvD..63j3510A}, where it was also shown that tracking solutions to this equation of motion, which are attractors in the space of solutions, exist during the radiation and matter-dominated eras for k-essence in a similar manner to quintessence.

The speed of sound for k-essence fluctuation is 
\begin{equation}
c_s^2 =\frac{p,_{X}}{p,_X+2Xp,_{XX}} \,.
\end{equation}
So that whenever the kinetic terms for the scalar field are not linear in $X$, the speed of sound of fluctuations differs from unity.  It might appear concerning that superluminal fluctuations are allowed in k-essence models (and even necessarily arise in models where k-essence dark energy solves the coincidence problem \cite{Bonvin:2006vc}). However it was shown in~\cite{Babichev:2007dw} that this does not lead to any causal paradoxes.

\subsection{Coupled dark-energy models}
\label{mg:cde}

A first class of models in which dark energy shows dynamics, in
connection with the presence of a fifth force different from gravity, is the
case of `interacting dark energy': we consider the
possibility that dark energy, seen as a dynamical scalar field, may
interact with other components in the universe. This class of models
effectively enters in the ``explicit modified gravity models'' in the
classification above, because the gravitational attraction between dark matter particles is modified by the presence of a fifth force. However, we note that the anisotropic stress for DE is still zero in the Einstein frame, while it is, in general, non-zero in the Jordan frame. In some cases (when a universal coupling is present) such an interaction can be explicitly recast in a non-minimal coupling to gravity, after a redefinition of the metric and matter fields (Weyl scaling).
We would like to identify
whether interactions (couplings) of dark energy with matter fields, neutrinos or
gravity itself can affect the universe in an observable way.

In this subsection we give a general description of the following main interacting
scenarios:
\begin{enumerate}
 \item  couplings between dark energy and baryons;
 \item \label{de_dm}  couplings between dark energy and dark matter (coupled
quintessence);
 \item \label{de_nu} couplings between dark energy and neutrinos (growing
neutrinos, MaVaNs);
 \item  universal couplings with all species (scalar-tensor theories and $f(R)$). 
\end{enumerate}
In all these cosmologies the coupling introduces a fifth force, in addition to standard
gravitational attraction. 
The presence of a new force, mediated by the DE scalar field (sometimes called
the `cosmon' \cite{Wetterich_1988}, seen as the mediator of a cosmological interaction) has several
implications and can significantly modify the process of structure formation.
We will discuss cases~(\ref{de_dm}) and (\ref{de_nu}) in 
Section~\ref{dark-matter}.

In these scenarios the presence of the additional interaction couples the evolution of components 
that in the standard $\Lambda$-FLRW would evolve independently. 
The stress-energy tensor
${T{^{\mu}}}_{\nu}$ of each species is, in general, not conserved -- only
the total stress-energy tensor is. 
Usually, at the level of the Lagrangian, the coupling is introduced by allowing
the mass $m$ of matter fields to depend on a scalar field $\phi$ via a function
$m(\phi)$ whose choice specifies the interaction.\label{symbol:phi}
This wide class of cosmological models can be described by the following action:
\begin{equation}
  \label{mg:cde:action} {\cal S} = \int{\mathrm{d}^4x \sqrt{-g} \left[-\frac{1}{2}\partial^\mu 
\phi \partial_\mu \phi - U(\phi) - m(\phi)\bar{\psi}\psi + {\cal L}_{\mathrm{kin}}[\psi]\right] } ,
\end{equation}
where $U(\phi)$ is the potential in which the scalar field $\phi$ rolls, $\psi$
describes matter fields, and $g$ is defined in the usual way as the determinant
of the metric tensor, whose background expression is $g_{\mu \nu} = \mathrm{diag}[-a^2,
a^2, a^2, a^2]$.

For a general treatment of background and perturbation equations we refer to
\cite{Kodama:1985bj, Amendola:1999er, Amendola:2003wa, Pettorino:2008ez}. Here
the coupling of the dark-energy scalar field to a generic matter component
(denoted by index $\alpha$) is treated as an external source $Q_{(\alpha)\mu}$ in the Bianchi
identities:
\begin{equation}
\nabla_{\nu}T_{(\alpha)\mu}^{\nu}=Q_{(\alpha)\mu}\,,\label{tensor_conserv_alpha}
\end{equation}
with the constraint \begin{equation}
\sum_{\alpha}Q_{(\alpha)\mu}=0\label{Q_conserv_total} \,.\end{equation}

The zero component of (\ref{tensor_conserv_alpha}) gives the background conservation equations:
\begin{eqnarray}
 \label{cons_phi} \frac{\d\rho_{\phi}}{\d\eta} = -3 {\cal H} (1 + w_\phi) \rho_{\phi} +
\beta(\phi) \frac {\d\phi}{\d\eta} (1-3 w_{\alpha}) \rho_{\alpha} \,, \\
\label{cons_gr} \frac{\d\rho_{\alpha}}{\d\eta} = -3 {\cal H} (1 + w_{\alpha}) \rho_{\alpha} -
\beta(\phi) \frac{\d\phi}{\d\eta} (1-3 w_{\alpha}) \rho_{\alpha} \,,
\end{eqnarray}
for a scalar field $\phi$ coupled to one single fluid $\alpha$ with a function $\beta(\phi)$, which in general may not be constant. 
The choice of the mass function $m(\phi)$ corresponds to a choice of
$\beta(\phi)$ and equivalently to a choice of the source $Q_{(\alpha)\mu}$ and
specifies the strength of the coupling according to the following relations:
\begin{equation} \label{mass_def}
Q_{(\phi)\mu}=\frac{\partial\ln{m(\phi)}}{\partial\phi}
T_{\alpha}\,\partial_{\mu}\phi \,, \,
m_{\alpha}=\bar{m}_\alpha ~ e^{-{\beta(\phi)}{\phi}} \,,
\end{equation}
where $\bar{m}_\alpha$ is the constant Jordan-frame bare mass. 
The evolution of dark energy is related to the trace
$T_{\alpha}$ and, as a consequence, 
to density and pressure of the species $\alpha$. We note that a description of
the coupling via an action such as~(\ref{mg:cde:action}) 
is originally motivated by the wish to modify GR with an extension
such as scalar-tensor theories.
In general, one of more couplings can be active \citep{Brookfield:2007au}.

As for perturbation equations, it is possible to include the
coupling in a modified Euler equation: 
\begin{eqnarray}
& &\frac {\d\mathbf{v}_{\alpha}}{\d\eta} + \left({\cal H} - {\beta(\phi)}
  \frac{\d\phi}{\d\eta} \right) \mathbf{v}_{\alpha} - \mathbf{\nabla} \left[\Phi_\alpha + \beta \phi \right] = 0 \,.
\end{eqnarray}
The Euler equation in cosmic time ($\mathrm{d}t = a\, \mathrm{d}\tau$) can also be
rewritten in the form of an acceleration equation for
particles at position $\mathbf{r}$:
\begin{equation}
\label{CQ_euler}
\dot{\mathbf{v}}_{\alpha} = -\tilde{H}\mathbf{v}_{\alpha} - \mathbf{\nabla}\frac{\tilde{G}_{\alpha}{m}_{\alpha}}{r} \,.
\end{equation}
The latter expression explicitly contains all the main ingredients that affect
dark-energy interactions:
\begin{enumerate}
 \item a fifth force $\mathbf{\nabla} \left[\Phi_\alpha + \beta \phi
\right]$ with an effective $\tilde{G}_{\alpha} = G_{N}[1+2\beta^2(\phi)]$ ;
\item a velocity dependent term $\tilde{H}\mathbf{v}_{\alpha} \equiv H \left(1 -
{\beta(\phi)} \frac{\dot{\phi}}{H}\right) \mathbf{v}_{\alpha}$
\item a time-dependent mass for each particle $\alpha$, evolving according to
(\ref{mass_def}).
\end{enumerate} 

The relative significance of these key ingredients can lead to a variety of
potentially observable effects, especially on structure formation. We will recall some of
them in the following subsections as well as, in more detail, for two specific
couplings in the dark matter section (\ref{dms:de_dm}, \ref{dms:de_nu}) of this
report. 

\subsubsection{Dark energy and baryons}
\label{de_baryons}

A coupling between dark energy and baryons is active when the baryon mass is a
function of the dark-energy scalar field: $m_b = m_b(\phi)$. 
Such a coupling is
constrained to be very small: main bounds come from tests
of the equivalence principle and solar system constraints
\citep{Bertotti:2003rm}. 
More in general, depending on the coupling, bounds on the variation of fundamental constants over cosmological time-scales
may have to be considered (\cite{Marra:2005ma, Dent:2008gx, Dent:2008vd, Martins:2010gu} and references therein). 
It is presumably very difficult to have significant
cosmological effects due to a coupling to baryons only. However, uncoupled baryons can
still play a role in the presence of a coupling to dark matter (see Section~\ref{nonlinear-aspects} on nonlinear aspects). 

\subsubsection{Dark energy and dark matter}

An interaction between dark energy and dark matter (CDM) is active when CDM mass
is a function of the dark-energy scalar field: $m_c = m_c(\phi)$. In this case
the coupling is not affected by tests on the equivalence principle and solar-system constraints and can therefore be stronger than the one with baryons.
One may argue that dark-matter particles are themselves coupled to baryons, which leads, through quantum
corrections, to direct coupling between dark energy and baryons. The strength of such couplings can still be small and was discussed in \cite{Dent:2008vd} for the case of neutrino--dark-energy couplings. Also, quantum corrections are often recalled to spoil the flatness of a quintessence potential. 
However, it may be misleading to calculate quantum corrections up to a cutoff scale, as contributions above the cutoff can possibly compensate terms below the cutoff, as discussed in \cite{2008PhRvD..77j3505W}. 

Typical values of $\beta$ presently allowed by observations (within current CMB 
data) are within the range $0< \beta < 0.06$ (at 95\% CL for a constant coupling and an exponential potential) 
\citep{Bean:2008ac, amendola_etal_2003, Amendola:2003wa, amendola_quercellini_2003}, or possibly more 
\citep{LaVacca:2009yp, Kristiansen:2009yx} if neutrinos are taken into account or for
 more realistic time-dependent choices of the coupling. This 
framework is generally referred to as `coupled quintessence' (CQ).
Various choices of couplings have been investigated in literature, including
constant and varying $\beta(\phi)$ \citep{Amendola:1999er, Mangano:2002gg,
Amendola:2003wa, Koivisto:2005nr, Guo:2007zk, Quartin:2008px, quercellini_etal_2008, Pettorino:2008ez,Gannouji:2010fc}.

The presence of a coupling (and therefore, of a fifth force acting among dark-matter particles) modifies the background expansion and linear perturbations \citep{amendola1999, Amendola:1999er, Amendola:2003wa}, therefore affecting CMB and cross-correlation of CMB and LSS \citep{amendola_quercellini_2003, Amendola:2003wa, amendola_etal_2003, amendola_quercellini_2004, Bean:2008ac, LaVacca:2009yp, Kristiansen:2009yx, Xia:2009zzb, Mainini:2010ng, Amendola:2011ie}.

Furthermore, structure formation itself is modified \citep{maccio_etal_2004, Manera:2005ct, Koivisto:2005nr, Mainini:2006zj, Sutter:2007ky, Abdalla:2007rd,Mota:2008ne, Bertolami:2007tq, Wintergerst:2010ui, Baldi_etal_2010, Baldi:2010vv, Baldi_Pettorino_2010, Baldi:2010ks, Li:2010zw, Li:2010eu, Baldi:2010pq, Zhao:2010dz, Marulli:2011jk}.

An alternative approach, also investigated in the literature \citep{Mangano:2002gg,Valiviita:2008iv,Valiviita:2009nu,Majerotto:2009np,
Gavela:2009cy,Gavela:2010tm,CalderaCabral:2008bx,Schaefer:2008ku,CalderaCabral:2009ja}, where the authors consider as a starting point 
Eq.~(\ref{tensor_conserv_alpha}): the coupling is then introduced by choosing directly a covariant stress-energy tensor on the RHS of the equation, treating dark energy as a fluid and 
in the absence of a starting action. The advantage of this
approach is that a good parameterization allows us to investigate several models of dark energy 
at the same time. Problems connected to instabilities of some parameterizations or to the definition
of a physically-motivated speed of sound for the density fluctuations can be found in \cite{Valiviita:2008iv}.
It is also possible to both take a covariant form for the coupling and a
quintessence dark-energy scalar field, starting again directly from Eq.~(\ref{tensor_conserv_alpha}). This has been done, e.g., in \cite{Boehmer:2008av},
\cite{Boehmer:2009tk}.
At the background level only, \cite{Chimento:2003iea}, \cite{Chimento:2005xa}, \cite{delCampo:2006vv} and \cite{Olivares:2006jr} have also considered
which background constraints can be obtained when starting from a fixed present ratio of dark energy and dark matter. The disadvantage of this approach is that it is not clear how to perturb a coupling that has been defined as a background quantity. 

A Yukawa-like interaction was investigated \citep{Farrar:2003uw, Das:2006cc}, pointing out that coupled dark energy behaves as a fluid with an effective equation of state $w \lesssim -1$, though staying well defined and without the presence of ghosts \citep{Das:2006cc}. 

For an illustration of observable effects related to dark-energy--dark-matter interaction see also Section~(\ref{dms:de_dm}) of this report.

\subsubsection{Dark energy and neutrinos}

A coupling between dark energy and neutrinos can be even stronger than the one
with dark matter and as compared to gravitational strength. Typical values of
$\beta$ are order 50\,--\,100 or even more, such that even the small fraction of
cosmic energy density in neutrinos can have a substantial influence on the time
evolution of the quintessence field. In this scenario neutrino masses change in
time, depending on the value of the dark-energy scalar field $\phi$. 
Such a coupling has been investigated within MaVaNs \citep{Fardon:2003eh, Peccei:2004sz, Bi:2004ns, Afshordi:2005ym,
Weiner:2005ac, Das:2006ht, Takahashi:2006jt, Spitzer:2006hm, Bjaelde:2007ki, Brookfield:2005td, Brookfield:2005bz} and more recently within growing neutrino 
cosmologies \citep{Amendola2008b, Wetterich:2007kr, Mota:2008nj, Wintergerst:2009fh, Wintergerst:2010ui,
Pettorino:2010bv, Brouzakis:2010md, Baldi:2011es}.
In this latter case, DE properties are related to the neutrino mass and to a cosmological event, i.e., neutrinos becoming non-relativistic.
This leads to the formation of stable neutrino lumps \citep{Mota:2008nj, Wintergerst:2009fh,Baldi:2011es} at very large scales only ($\sim$~100~Mpc and beyond) as well as to signatures in the CMB spectra \citep{Pettorino:2010bv}. For an illustration of observable effects related to this case see Section~(\ref{dms:de_nu}) of this report.

\subsubsection{Scalar-tensor theories}

Scalar-tensor theories \citep{Wetterich_1988, Hwang:1990re, Hwang:1990jh, Damour_Gibbons_Gundlach_1990, casas_etal_1991, casas_etal_1992, Wetterich:1994bg,
Uzan:1999ch, Perrotta:1999am, Faraoni:2000wk, Boisseau:2000pr, Riazuelo:2001mg,
Perrotta:2002sw, Schimd:2004nq, Matarrese:2004xa, Pettorino:2004zt,
Pettorino:2005pv, Capozziello:2007iu, Appleby:2010dx} extend GR by introducing a non-minimal
coupling between a scalar field (acting also as dark energy) and the metric
tensor (gravity); they are also sometimes referred to as `extended
quintessence'. 
We include scalar-tensor theories among `interacting cosmologies' because, via a
Weyl transformation, they are equivalent to a GR framework
(minimal coupling to gravity) in which the dark-energy scalar field $\phi$ is
coupled (universally) to all species \citep{Wetterich_1988, Maeda:1988ab, Wands:1993uu, EspositoFarese:2000ij, Pettorino:2008ez, Catena:2006bd}. In other words, these theories correspond
to the case where, in action~(\ref{mg:cde:action}), the mass of all species
(baryons, dark matter, \dots) is a function $m=m(\phi)$ with the same coupling for
every species $\alpha$. Indeed, a description of the coupling via an action such
as (\ref{mg:cde:action}) is originally motivated by extensions of GR
such as scalar-tensor theories.
Typically the strength of the scalar-mediated interaction is required to be
orders of magnitude weaker than gravity (\cite{Lee:2010zy}, \cite{Pettorino:2004zt} and references therein for recent constraints).  
It is possible to tune this coupling
to be as small as is required -- for example by choosing a suitably flat potential $V(\phi)$ for the scalar field. However, this leads back to naturalness and fine-tuning problems.

In Sections~\ref{fr-general} and \ref{massive-gr-higher-dm} we will discuss in more detail a number of ways in which new scalar degrees of freedom can
naturally couple to standard model fields, while still being in agreement with
observations. We mention here only that the presence of
  chameleon mechanisms \citep{Brax:2004qh, Mota:2010uy, Mota:2006fz,Brax:2008hh, Hui:2009kc,
Brax:2010kv, Davis:2011qf} can, for example, modify the coupling depending on the environment. In
this way, a small (screened) coupling in high-density regions, in agreement with
observations, is still compatible with a bigger coupling ($\beta \sim 1$) active
in low density regions. In other words, a dynamical mechanism ensures that the effects of the coupling are
screened in laboratory and solar system tests of gravity.

Typical effects of scalar-tensor  theories on CMB and structure
formation include:
\begin{itemize}
 \item enhanced ISW \citep{Pettorino:2004zt, Giannantonio:2009zz, Zhao:2010dz};
 \item violation of the equivalence principle: extended objects such as galaxies
do not all fall at the same rate \citep{amendola_quercellini_2004, Hui:2009kc}.
\end{itemize}

However, it is important to remark that screening mechanisms are meant to protect the scalar field in high-density regions (and therefore allow for bigger couplings in low density environments) but they do not address problems related to self-acceleration of the DE scalar field, which still usually require some fine-tuning to match present observations on $w$.
$f(R)$ theories, which can be mapped into a subclass of scalar-tensor theories, 
will be discussed in more detail in Section~\ref{fr-general}.

\label{cde_eq}

\subsection{\textit{f(R)} gravity}
\label{fr-general}

In parallel to models with extra degrees of freedom in the matter sector, such 
as interacting quintessence (and k-essence, not treated here), another promising approach to the
late-time acceleration enigma is to modify the left-hand side of the Einstein
equations and invoke new degrees of freedom, belonging this time to the
gravitational sector itself.
One of the simplest and most popular extensions of GR and a known example of 
modified gravity models is the $f(R)$
gravity\index{f(R) gravity@$f(R)$ gravity} in which the 4-dimensional
action is given by some generic function $f(R)$ of the Ricci scalar
$R$\index{Ricci scalar} (for an introduction see, e.g., \cite{Amendola2010}): 
\begin{equation}
S=\frac{1}{2\kappa^{2}}\int{\mathrm{d}}^{4}x\sqrt{-g}f(R)+S_{m}(g_{\mu\nu},\Psi_{m})
\,,\label{fRaction}
\end{equation}
where as usual $\kappa^{2}=8\pi G$, and $S_{m}$ is a matter action
with matter fields $\Psi_{m}$. Here $G$ is a \emph{bare} gravitational
constant\index{gravitational constant!bare}: we will see that the 
observed value will in general be different. As mentioned in the
previously, it is possible to show that 
 $f(R)$ theories can be mapped into a subset of scalar-tensor theories and, therefore, to 
a class of interacting scalar field dark-energy models universally coupled to all species. When seen 
in the Einstein frame \citep{Wetterich_1988, Maeda:1988ab, Wands:1993uu, EspositoFarese:2000ij, Pettorino:2008ez, Catena:2006bd},
action~(\ref{fRaction}) can, therefore, be related to the action~(\ref{mg:cde:action}) shown previously. 
Here we describe $f(R)$ in the Jordan frame: the matter fields in
$S_{m}$ obey standard conservation equations and, therefore, the metric
$g_{\mu\nu}$ corresponds to the physical frame (which here is the
Jordan frame).  
 
There are two approaches to deriving field equations
from the action (\ref{fRaction}). 
\begin{itemize}
\item \textbf{(I) The metric formalism}

The first approach is the \emph{metric formalism}\index{metric formalism}
in which the connections $\Gamma_{\beta\gamma}^{\alpha}$ are the
usual connections defined in terms of the metric $g_{\mu\nu}$. The
field equations can be obtained by varying the action (\ref{fRaction})
with respect to $g_{\mu\nu}$:
\begin{equation}
F(R)R_{\mu\nu}(g)-\frac{1}{2}f(R)g_{\mu\nu}-\nabla_{\mu}\nabla_{\nu}F(R)+g_{
\mu\nu}\square F(R)=\kappa^{2}T_{\mu\nu}\,,\label{fREin}
\end{equation}
where $F(R)\equiv\partial f/\partial R$ (we also use the notation
$f_{,R}\equiv\partial f/\partial R,\, f_{,RR}\equiv\partial^{2}f/\partial
R^{2}$), \label{symbol:FR}
and $T_{\mu\nu}$ is the matter energy-momentum tensor.
The trace\index{trace} of Eq.~(\ref{fREin}) is given by 
\begin{equation}
3\,\square F(R)+F(R)R-2f(R)=\kappa^{2}T\,,\label{trace}
\end{equation}
where $T=g^{\mu\nu}T_{\mu\nu}=-\rho+3P$. Here $\rho$ and $P$ are
the energy density and the pressure of the matter, respectively.

\item \textbf{(II) The Palatini formalism}

The second approach is the \emph{Palatini
 formalism}\index{Palatini formalism}, where $\Gamma_{\beta\gamma}^{\alpha}$
and $g_{\mu\nu}$ are treated as independent variables. Varying the
action (\ref{fRaction}) with respect to $g_{\mu\nu}$ gives
\begin{equation}
F(R)R_{\mu\nu}(\Gamma)-\frac{1}{2}f(R)g_{\mu\nu}=\kappa^{2}T_{\mu\nu} \,,
\label{fRpala}
\end{equation}
 where $R_{\mu\nu}(\Gamma)$ is the Ricci tensor corresponding to
the connections $\Gamma_{\beta\gamma}^{\alpha}$. In general this
is different from the Ricci tensor $R_{\mu\nu}(g)$ corresponding
to the metric connections. Taking the trace of Eq.~(\ref{fRpala}),
we obtain
\begin{equation}
F(R)R-2f(R)=\kappa^{2}T \,,
\label{trace2}
\end{equation}
where $R(T)=g^{\mu\nu}R_{\mu\nu}(\Gamma)$ is directly related to
$T$. Taking the variation of the action (\ref{fRaction}) with respect
to the connection, and using Eq.~(\ref{fRpala}), we find 
\begin{eqnarray}
R_{\mu\nu}(g)-\frac{1}{2}g_{\mu\nu}R(g) & = & 
\frac{\kappa^{2}T_{\mu\nu}}{F}-\frac{FR(T)-f}{2F}g_{\mu\nu}
+\frac{1}{F}(\nabla_{\mu}\nabla_{\nu}F-g_{\mu\nu}\square F)\nonumber \\
 &  & -\frac{3}{2F^{2}}\left[\partial_{\mu}F\partial_{\nu}F
 -\frac{1}{2}g_{\mu\nu}(\nabla F)^{2}\right]\,.
\label{equationpala}
\end{eqnarray}

\end{itemize}
In GR\index{general relativity} we have $f(R)=R-2\Lambda$ 
and $F(R)=1$, so that the term $\square F(R)$
in Eq.~(\ref{trace}) vanishes. 
In this case both the metric and the Palatini formalisms give the relation
$R=-\kappa^{2}T=\kappa^{2}(\rho-3P)$, which means that the Ricci
scalar $R$ is directly determined by the matter (the trace $T$).

In modified gravity models where $F(R)$ is a function of $R$, the
term $\square F(R)$ does not vanish in Eq.~(\ref{trace}).
This means that, in the metric formalism, there is a propagating scalar
degree of freedom, $\psi\equiv F(R)$. The trace equation (\ref{trace})
governs the dynamics of the scalar field $\psi$ -- dubbed
``scalaron''\index{scalaron}
\cite{Sta80}. In the Palatini formalism the kinetic term $\square F(R)$
is not present in Eq.~(\ref{trace2}), which means that the scalar-field
degree of freedom does not propagate freely \cite{Amarzguioui:2005zq,Li:2007xw,Li:2008bma,Li:2008fa}.

The de~Sitter point\index{de Sitter!point} corresponds to a vacuum
solution at which the Ricci scalar is constant. Since $\square F(R)=0$
at this point, we get
\begin{equation}
F(R)R-2f(R)=0 \,,
\label{fRdeSitter}
\end{equation}
which holds for both the metric and the Palatini formalisms. Since
the model $f(R)=\alpha R^{2}$ satisfies this condition, it possesses
an exact de~Sitter solution \citep{Sta80}.

It is important to realize that the dynamics of $f(R)$ dark-energy models is
different depending
on the two formalisms. Here we confine ourselves to  the metric case only; details of a viable model in unifying the metric and Palatini formalism can be found in \cite{Harko_Koivisto_2012}.

Already in the early 1980s it was known that the model $f(R)=R+\alpha R^{2}$
can be responsible for inflation\index{inflation} in the early universe
\citep{Sta80}. This comes from the fact that the presence of the quadratic
term $\alpha R^{2}$ gives rise to an asymptotically exact de~Sitter
solution. Inflation ends when the term $\alpha R^{2}$ becomes smaller
than the linear term $R$. Since the term $\alpha R^{2}$ is negligibly
small relative to $R$ at the present epoch, this model is not suitable
to realizing the present cosmic acceleration.

Since a late-time acceleration requires modification for small $R$,
models of the type $f(R)=R-\alpha/R^{n}$ ($\alpha>0,n>0$) were proposed
as a candidate for dark energy \citep{Capo1,CDTT,Nojiri03}.
While the late-time cosmic acceleration is possible in these models,
it has become clear that they do not satisfy local gravity constraints
because of the instability associated with negative values of $f_{,RR}$
\citep{Chiba:2003ir,Dolgov,Woodard,Olmo,Faraoni}. Moreover a standard matter
epoch is not present because of a large coupling between the Ricci
scalar and the non-relativistic matter \citep{APT07}.

Then, we can ask what are the conditions for the viability of $f(R)$
dark-energy  models in the metric formalism. In the following we first
present such conditions and then explain step by step why they are
required.

\begin{itemize}
\item (i) $f_{,R}>0$ for $R\ge R_{0}~(>0)$, where $R_{0}$ is the Ricci
scalar at the present epoch. Strictly speaking, if the final attractor
is a de~Sitter point with the Ricci scalar $R_{1}~(>0)$, then the
condition $f_{,R}>0$ needs to hold for $R\ge R_{1}$.

This is required to avoid a negative effective gravitational constant.

\item (ii) $f_{,RR}>0$ for $R\ge R_{0}$.

This is required for consistency with local gravity tests\index{local gravity
constraints}
\citep{Dolgov,Olmo,Faraoni,Navarro}, for the presence of the matter-dominated
epoch\index{matter-dominated epoch} \citep{APT07,AGPT}, and for the
stability of cosmological perturbations \citep{Carroll06,Song07,Bean07,Faulkner}.

\item (iii) $f(R)\to R-2\Lambda$ for $R\gg R_{0}$.

This is required for consistency with local gravity tests
\citep{2008PhLB..660..125A,Hu07,Star07,Appleby,Tsuji08}
and for the presence of the matter-dominated epoch \citep{AGPT}.

\item (iv) $0<\frac{Rf_{,RR}}{f_{,R}}(r=-2)<1$ at $r=-\frac{Rf_{,R}}{f}=-2$.

This is required for the stability of the late-time de~Sitter point
\citep{Muller,AGPT}.

\end{itemize}
For example, the model $f(R)=R-\alpha/R^{n}$ ($\alpha>0$, $n>0$)
does not satisfy the condition (ii).

Below we list some viable $f(R)$ models that satisfy the above conditions.
\begin{eqnarray}
 &  & {\mathrm{(A)}}~f(R)=R-\mu R_{c}(R/R_{c})^{p}\qquad{\mathrm
{with}}~~0<p<1,~~\mu,R_{c}>0\,,\label{Amodel}\\
 &  & {\mathrm{(B)}}~f(R)=R-\mu
R_{c}\frac{(R/R_{c})^{2n}}{(R/R_{c})^{2n}+1}\qquad{\mathrm
{with}}~~n,\mu,R_{c}>0\,,\label{Bmodel}\\
 &  & {\mathrm{(C)}}~f(R)=R-\mu
R_{c}\left[1-\left(1+R^{2}/R_{c}^{2}\right)^{-n}\right]\qquad{\mathrm
{with}}~~n,\mu,R_{c}>0\,,\label{Cmodel}\\
 &  & {\mathrm{(D)}}~f(R)=R-\mu R_{c}{\mathrm{tanh}}\,(R/R_{c})\qquad{\mathrm
{with}}~~\mu,R_{c}>0\,.\label{Dmodel}
\end{eqnarray}
The models (A), (B), (C), and (D) have been proposed in \citet{AGPT},
\citet{Hu07}, \citet{Star07}, and \citet{Tsuji08}, respectively. A
model similar to (D) has been also proposed in \citet{Appleby},
while a generalized model encompassing (B) and (C) has been studied
in~\cite{mir09}. In model (A), the power $p$ needs to
be close to 0 to satisfy the condition (iii). In models (B) and
(C) the function $f(R)$ asymptotically behaves as $f(R)\to R-\mu
R_{c}[1-(R^{2}/R_{c}^{2})^{-n}]$
for $R\gg R_{c}$ and hence the condition (iii) can be satisfied even
for $n={\cal O}(1)$. In model (D) the function $f(R)$ rapidly
approaches $f(R)\to R-\mu R_{c}$ in the region $R\gg R_{c}$. These
models satisfy $f(R=0)=0$, so the cosmological constant vanishes in
the flat spacetime.

Let us consider the cosmological dynamics of $f(R)$ gravity in the metric
formalism. It is possible to carry out a general analysis without
specifying the form of $f(R)$. In the flat FLRW spacetime the Ricci
scalar is given by
\begin{equation}
R=6(2H^{2}+\dot{H}) \,,
\end{equation}
 where $H$ is the Hubble parameter. As a matter action $S_{m}$
we take into account non-relativistic matter and radiation, which
satisfy the usual conservation equations $\dot{\rho}_{m}+3H\rho_{m}=0$
and $\dot{\rho}_{r}+4H\rho_{r}=0$ respectively. From Eqs.~(\ref{fREin})
and (\ref{trace}) we obtain the following equations
\begin{eqnarray}
3FH^{2} & = &
\kappa^{2}\,(\rho_{m}+\rho_{r})+(FR-f)/2-3H\dot{F}\,,\label{FRWfR1}\\
-2F\dot{H} & = &
\kappa^{2}\left[\rho_{m}+(4/3)\rho_{r}\right]+\ddot{F}-H\dot{F}\,.\label{FRWfR2}
\end{eqnarray}
We introduce the dimensionless variables:
\begin{equation}
x_{1}\equiv-\frac{\dot{F}}{HF}\,,\quad x_{2}\equiv-\frac{f}{6FH^{2}}\,,\quad
x_{3}\equiv\frac{R}{6H^{2}}\,,\quad
x_{4}\equiv\frac{\kappa^{2}\rho_{r}}{3FH^{2}}\,,
\end{equation}
together with the following quantities
\begin{equation}
\Omega_{m}\equiv\frac{\kappa^{2}\rho_{m}}{3FH^{2}}=1-x_{1}-x_{2}-x_{3}-x_{4}
\,, \qquad 
\Omega_{r}\equiv x_{4}\,,\qquad\Omega_{\mathrm{DE}}\equiv
x_{1}+x_{2}+x_{3}\,.
\label{fromedef}
\end{equation}
It is straightforward to derive the following differential equations
\citep{AGPT}:
\begin{eqnarray}
 x_{1}' & = &
-1-x_{3}-3x_{2}+x_{1}^{2}-x_{1}x_{3}+x_{4} \,,\label{x1fR}\\
 x_{2}' & = &
\frac{x_{1}x_{3}}{m}-x_{2}(2x_{3}-4-x_{1}) \,,\label{x2fR}\\
 x_{3}' & = &
-\frac{x_{1}x_{3}}{m}-2x_{3}(x_{3}-2) \,,\label{x3fR}\\
x_{4}' & = &
-2x_{3}x_{4}+x_{1}x_{4} \,,\label{x4fR}
\end{eqnarray}
where the prime denotes $\d/\d\ln a$ and
\begin{eqnarray}
m & \equiv & \frac{\d\ln F}{\d\ln R}=\frac{Rf_{,RR}}{f_{,R}}\,,\label{eq:charact}\\
r & \equiv & -\frac{\d\ln f}{\d\ln R}=-\frac{Rf_{,R}}{f}=\frac{x_{3}}{x_{2}}\,.
\label{mdef}
\end{eqnarray}
From Eq.~(\ref{mdef}) one can express $R$ as a function of
$x_{3}/x_{2}$. Since $m$ is a function of $R$, it follows that
$m$ is a function of $r$, i.e., $m=m(r)$. The $\Lambda$CDM model,
$f(R)=R-2\Lambda$, corresponds to $m=0$. Hence the quantity $m$
characterizes the deviation\index{deviation parameter} from the $\Lambda$CDM
model. Note also that the model, $f(R)=\alpha R^{1+m}-2\Lambda$,
gives a constant value of $m$. The analysis using Eqs.~(\ref{x1fR})\,--\,(\ref{x4fR})
is sufficiently general in the sense that the form of $f(R)$ does
not need to be specified.

The effective equation of state\index{effective equation of state!for $f(R)$
models}
of the system (i.e., $p_{\mathrm{tot}}/\rho_{\mathrm{tot}}$) is
\begin{equation}
w_{\mathrm{eff}}=-\frac{1}{3}(2x_{3}-1)\,. \label{ldef}
\end{equation}

The dynamics of the full system can be investigated by analyzing the stability
properties of the critical phase-space points as in, e.g., \cite{AGPT}. 
The general conclusions is that only models with a characteristic function $m(r)$ positive and
close to $\Lambda$CDM, i.e., $m\ge 0$, are cosmologically viable. That is, only for these
models one finds a sequence of a long decelerated matter epoch followed by a stable
accelerated attractor.

The perturbation equations have been derived in, e.g., \cite{Hwang:2001qk,Tsujikawa:2007tg}.
Neglecting the contribution of radiation one has
\begin{align}
\delta_{m}'' & +\left(x_{3}-\frac{1}{2}x_{1}\right)\delta_{m}'-\frac{3}{2}(1-x_{1}-x_{2}-x_{3})\delta_{m}\nonumber \\
 & =\frac{1}{2}\biggl[\left\{ \frac{k^{2}}{x_{5}^{2}}-6+3x_{1}^{2}-3x_{1}'-3x_{1}(x_{3}-1)\right\} \delta\tilde{F}\nonumber \\
 & ~~~~+3(-2x_{1}+x_{3}-1)\delta\tilde{F}'+3\delta\tilde{F}''\biggr]\,,\\
\delta\tilde{F}'' & +(1-2x_{1}+x_{3})\delta\tilde{F}'\nonumber \\
 & +\left[\frac{k^{2}}{x_{5}^{2}}-2x_{3}+\frac{2x_{3}}{m}-x_{1}(x_{3}+1)-x_{1}'+x_{1}^{2}\right]\delta\tilde{F}\nonumber \\
 & ~~~~=(1-x_{1}-x_{2}-x_{3})\delta_{m}-x_{1}\delta_{m}'\,,
\end{align}
where $\delta\tilde{F}\equiv\delta F/F$, and the new variable $x_{5}\equiv aH$
satisfies
\begin{equation}
x_{5}'=(x_{3}-1)\, x_{5} \,.
\label{x5eq}
\end{equation}
The perturbation $\delta F$ can be written as $\delta F=f_{,RR}\delta R$ and, therefore, $\delta\tilde{F}=m\delta R /R$.
These equations can be integrated numerically to derive the behavior of $\delta_m$ at all scales.
However, at sub-Hubble scales they can be simplified and the following expression for the two MG functions $Q,\eta$ of Eq.~(\ref{mod_constr})
can be obtained:
\begin{eqnarray}
Q &=& 1-{ \frac{k^{2}}{ 3(a^2M^2 + k^{2})}
}\nonumber\\
\eta &=&   1- {\frac{2k^{2} }{3a^2M^2 +
4k^{2}} }
\end{eqnarray}
where
\begin{equation}
\label{frr-mass}
 M^{2} = {\frac{1}{3 f_{,RR}} }.
\end{equation}
Note that in the $\Lambda$CDM limit $f_{,RR}\to 0$ and $Q,\eta\to 1$.

These relations can be straightforwardly generalized. In \cite{2010PhRvD..82b3524D}
the perturbation equations for the $f(R)$ Lagrangian  have been extended to include coupled scalar fields 
and their kinetic energy $X\equiv -\phi_{,\mu}\phi^{\mu}/2$, 
resulting in a $f(R,\phi,X)$-theory. In the slightly simplified case in which
$f(R,\phi,X)=f_1(R,\phi)+f_2(\phi,X)$, with arbitrary functions $f_1,2$, one obtains
\begin{eqnarray}
Q &=&
-\frac{1}{F} \frac{(1+2r_1)(f_{,X}+2r_2)+2F_{,\phi}^2/F}
{(1+3r_1)(f_{,X}+2r_2)+3F_{,\phi}^2/F}\,,\nonumber\\
\eta &=&  \frac{(1+2r_1)(f_{,X}+2r_2)+2F_{,\phi}^2/F}
{(1+4r_1)(f_{,X}+2r_2)+4F_{,\phi}^2/F}\,,
\end{eqnarray}
where the notation $f_{,X}$ or $F_{,\phi}$ denote differentiation wrt $X$ or $\phi$, respectively,
and where
$
r_1 \equiv \frac{k^2}{a^{2}} \frac{m}{R}
$ and
$
r_2 \equiv \frac{a^2}{k^2}M_\phi^2\,
$, $M_{\phi}=-f_{,\phi\phi}/2$ being the scalar field effective mass. In the same paper
\cite{2010PhRvD..82b3524D} an extra term proportional to $X\Box\phi$ in the Lagrangian is also taken into account.

Euclid forecasts for the $f(R)$ models will be presented in Section~\ref{fRforecastconstraints}.

\subsection{Massive gravity and higher-dimensional models}
\label{massive-gr-higher-dm}

Instead of introducing new scalar degrees of freedom such as in $f(R)$ theories, another philosophy in modifying gravity is to modify the graviton itself.
In this case the new degrees of freedom belong to the
gravitational sector itself; examples include massive gravity and higher-dimensional
frameworks, such as the Dvali--Gabadadze--Porrati (DGP) model \citep{dvali00} and its extensions. The new degrees of freedom can be responsible for a late-time
acceleration of the universe, as is summarized below for a choice of selected models.
We note here that while such self-accelerating solutions are interesting in their own right, they
do not tackle the old cosmological constant problem:
why the observed cosmological constant is so
much smaller than expected in the first place.
Instead of answering this question directly, an alternative approach is the
idea of
degravitation \citep[see][]{Degravitation1,Degravitation2,Degravitation3,Degravitation4},
where the cosmological
constant could be as large as expected from standard field theory,
but would simply gravitate very little (see the paragraph in
Section~\ref{sec:Self-Acceleration} below).

\subsubsection{Modified Cosmology}
\label{sec:Self-Acceleration}

Infrared Modifications of Gravity are of great interest for Cosmology as they can affect the evolution of the Universe in two different ways.

\paragraph*{Self-Acceleration.}
The first interest in modifications of gravity is the possibility of self-acceleration where the late-time acceleration of the Universe is not sourced by a cosmological constant or dark energy but rather by the graviton itself. This interesting phenomenology was first encountered in the DGP model as is explained below and was later shown to be also present in the Galileon, massive gravity and bi-gravity. Technically speaking if the Galileon is considered as a scalar field in its own right then the acceleration of the Universe is due to a new scalar degree of freedom and lies in the category of dark energy. However massive gravity
and higher-dimensional models of gravity often behave as a Galileon model in some limit, where the Galileon plays the role of one of the graviton's own degree of freedom, in this sense Galileon models are often also thought of models of self-acceleration.

\paragraph*{Degravitation.}
\label{sec:Degravitation}

The idea behind degravitation is to modify gravity in the IR, such that the
vacuum energy could have a weaker effect on the geometry, and therefore
reconcile a natural value for the vacuum energy as expected from particle
physics with the observed late-time acceleration. Such modifications of gravity
typically arise in models of massive gravity \citep{Degravitation1,Degravitation2,Degravitation3,Degravitation4}, i.e., where gravity is mediated by a massive spin-2 field.
The extra-dimensional DGP scenario presented below, represents a specific model of soft mass gravity, where gravity weakens at large
distance, with a force law going as $1/r$. Nevertheless, this weakening is too
weak to achieve degravitation and tackle the cosmological constant problem.
However, an obvious way out is to extend the DGP model to higher
dimensions, thereby diluting gravity more efficiently at large distances. This
is achieved in models of cascading gravity, as is presented below. An
alternative to cascading gravity is to work directly with theories of constant
mass gravity (hard mass graviton).

\subsubsection{Models of Infrared Modifications of Gravity}
Infrared modifications of gravity usually weaken the effect of gravity on cosmological scales, i.e. the propagation of gravitational waves is affected at distances and time-scales that are of the order of the size and age of the current Universe. These infrared modifications of general relativity are united by the common feature of invoking new degrees of freedom which could be used to either explain the recent acceleration of the Hubble expansion or tackle the cosmological constant problem. Below we will discuss different models which share these features.

\paragraph*{DGP.}
\label{sec:DGP}

The DGP model is one of the important infrared (IR) modified theories of gravity.
From a four-dimensional point of view this corresponds effectively to a theory
in which the graviton acquires a soft mass $m$. In this braneworld model our
visible universe is confined to a brane of four dimensions embedded into a five-dimensional bulk. At small distances, the four-dimensional gravity is recovered due
to an intrinsic Einstein--Hilbert term sourced by the brane curvature causing a
gravitational force law that scales as $r^{-2}$. At large scales the
gravitational force law asymptotes to an $r^{-3}$ behavior.
 The cross over scale $r_c=m^{-1}$ is given by the ratio of the Planck masses in
four ($M_4$) and five ($M_5$) dimensions.\label{symbol:rc}
 One can study perturbations around flat spacetime and compute the
gravitational exchange amplitude between two conserved sources, which does not
reduce to the GR result even in the limit m$\to0$. However, the successful
implementation of the Vainshtein mechanism for decoupling the additional modes
from gravitational dynamics at sub-cosmological scales makes these theories
still very attractive \citep{Vainshtein:1972sx}. Hereby, the Vainshtein effect
is realized through the nonlinear interactions of the helicity-0 mode $\pi$,
as will be explained in further detail below.
 Thus, this vDVZ discontinuity does not appear close to an astrophysical source
where the $\pi$ field becomes nonlinear and these nonlinear effects of $\pi$
restore predictions to those of GR. This is most easily understood in the limit
where $M_4, M_5\to\infty$ and $m\to 0$ while keeping the strong coupling scale
$\Lambda=(M_4m^2)^{1/3}$ fixed. This allows us to treat the usual helicity-2
mode of gravity linearly while treating the helicity-0 mode $\pi$
nonlinearly. The resulting effective action is then
 \ba
 \label{eq.DGP_decoupling}
 \mathcal{L}_{\pi}=3 \pi \Box \pi -\frac{1}{\Lambda^3}(\partial \pi)^2 \Box \pi\,,
 \ea
where interactions already become important at the scale $\Lambda \ll M_{\mathrm
{Pl}}$ \citep{Luty:2003vm}.

Furthermore, in this model, one can recover an interesting range of cosmologies,
in particular a modified Friedmann equation with a self-accelerating solution.
The Einstein equations thus obtained reduce to the following modified Friedmann
equation in a homogeneous and isotropic metric \citep{Deffayet:2001pu}
\ba
\label{eq:Friedmann_DGP}
H^2\pm m H=\frac{8\pi G}{3}\rho\,,
\ea
such that at higher energies one recovers the usual four-dimensional behavior,
$H^2\sim \rho$, while at later time corrections from the extra dimensions kick
in. As is clear in this Friedmann equation, this braneworld scenario holds two
branches of cosmological solutions with distinct properties. The
self-accelerating branch (minus sign) allows for a de Sitter behavior $H={\mathrm
{const}}=m$ even in the absence of any cosmological constant $\rho_{\Lambda}=0$
and as such it has attracted a lot of attention. Unfortunately, this branch suffers
from a ghost-like instability. The normal branch (the plus sign) instead slows
the expansion rate but is stable. In this case a cosmological constant is
still required for late-time acceleration, but it provides significant
intuition for the study of degravitation.
%
%
\paragraph*{The Galileon.}
Even though the DGP model is interesting for several reasons like giving the
Vainshtein effect a chance to work, the self-acceleration solution unfortunately introduces
extra ghost states as outlined above. However, it has been generalized to a ``Galileon'' model,
which can be considered as an effective field theory for the helicity-0 field $\pi$. Galileon models are invariant under shifts of the field $\pi$ and shifts of the gradients of $\pi$ (known as the Galileon symmetry), meaning that a Galileon model is invariant under the transformation
\begin{equation}
\pi \rightarrow \pi + c +v_{\mu}x^{\mu} \,,
\label{eq:Galileonsymmetry}
\end{equation}
for arbitrary constant $c$ and $v_{\mu}$.
 In induced gravity braneworld models, this symmetry is naturally inherited from
the five-dimensional Poincar\'{e} invariance \cite{deRham:2010eu}. The Galileon theory relies strongly on this symmetry to
constrain the possible structure of the effective $\pi$ Lagrangian, and insisting that the effective field theory for $\pi$ bears no ghost-like instabilities further restricts the possibilities \cite{Nicolis:2008in}. It can be shown that there
exist only five derivative interactions which preserve the Galilean symmetry in flat spacetime without introducing
ghosts. In curved spacetimes the situation is more subtle, see \cite{Deffayet_etal_2009} for details. In flat spacetime, the interactions are symbolically of the form $\mathcal{L}_{\pi}^{(1)}=\pi$ and $\mathcal{L}_{\pi}^{(n)}=(\partial \pi)^2(\partial\partial \pi)^{n-2}$, for $n = 2,\ldots 5$. A general Galileon Lagrangian can be constructed as a linear combination of these Lagrangian operators. The effective action for the DGP scalar (\ref{eq.DGP_decoupling}) can be seen to be a combination of $\mathcal{L}_{\pi}^{(2)}$ and $\mathcal{L}_{\pi}^{(3)}$. Such interactions have been shown to naturally arise from Lovelock invariants in
the bulk of generalized braneworld models \cite{deRham:2010eu}. However, the Galileon does not necessarily require a higher-dimensional origin and can be consistently treated as a four-dimensional effective field theory.

As shown in~\cite{Nicolis:2008in}, such theories can allow for self-accelerating de~Sitter solutions without any ghosts, unlike
in the DGP model. In the presence of compact sources, these solutions can support spherically-symmetric, Vainshtein-like nonlinear perturbations that are also stable against small fluctuations.
However, this is constrained to the subset of the third-order Galileon, which contains only $\mathcal{L}_{\pi}^{(1)}$, $\mathcal{L}_{\pi}^{(2)}$ and $\mathcal{L}_{\pi}^{(3)}$ \cite{Mota:2010bs}.

The fact that they give rise to second order equations of motion, have a symmetry and allow for healthy self-accelerating solutions, have initiated a wealth of investigations in cosmology. Moreover the non-renormalization theorem makes them theoretically very interesting since once the parameters in the theory are tuned by observational constraints they are radiatively stable. This means that the coefficients governing the Galileon interactions are technically natural.

%
%
\paragraph*{``Generalized galileons" and Horndeski interactions.}
The Galileon terms described above form a subset of the
``generalized Galileons''. A generalized Galileon model allows
nonlinear derivative interactions of the scalar field $\pi$ in the
Lagrangian while insisting that the equations of motion remain at most
second order in derivatives, thus removing any ghost-like
instabilities. However, unlike the pure Galileon models, generalized
Galileons do not impose the symmetry of
Eq.~(\ref{eq:Galileonsymmetry}). These theories were first written
down by Horndeski~\cite{Horndeski:1974wa}.
They are a linear combination of
Lagrangians constructed by multiplying the Galileon Lagrangians
$\mathcal{L}_{\pi}^{(n)}$ by an arbitrary scalar function of the
scalar $\pi$ and its first derivatives. Just like the Galileon,
generalized Galileons can give rise to cosmological acceleration and
to Vainshtein screening. However, as they lack the Galileon symmetry
these theories are not protected from quantum corrections.
The non-renormalization theorem is lost and hence the technical naturalness. Even if the naive covariantization of the Galileon interactions on non-flat backgrounds break the Galileon symmetry explicitly, one can successfully generalize the Galileon interactions to  maximally symmetric backgrounds \cite{oai:arXiv.org:1104.0155,Trodden:2011xh}.
It is also worth mentioning that a given subclass of these Horndeski interactions can also be constructed within the context of massive gravity from covariantizing its decoupling limit \cite{oai:arXiv.org:1106.3312}.
Many other theories can also be found within the spectrum of generalized Galileon
models, including k-essence.
Recently a new way to maintain a generalized Galileon symmetry on curved spacetimes was proposed in~\cite{Gabadadze:2012tr,Trodden:2012qe} by coupling massive gravity to a higher-dimensional DBI Galileon as in~\cite{deRham:2010eu}. In~\cite{Hinterbichler:2013dv} it was shown that such a generalized covariant Galileon model can lead to stable self-accelerating solutions.\\
Even if the scalar fields are by far the most extensively explored fields in cosmology, there are also motivations for the exploration of the role of vector fields or higher p-forms in general.   Inspired by the Horndeski interactions of the scalar field, one can construct the most general vector-tensor interactions with non-minimal coupling giving rise to second order equations of motion\cite{Jimenez:2013qsa}.

%
%

\paragraph*{Cascading gravity.}
Cascading gravity is an explicit realization of the idea of degravitation,
where gravity behaves as a high-pass filter, allowing sources with
characteristic wavelength (in space and in time) shorter than a characteristic
scale $r_c$ to behave as expected from GR, but weakening the
effect of sources with longer wavelengths.
This could explain why a large cosmological constant does not backreact as much
as
anticipated from standard GR. Since the DGP model does not
modify gravity enough in the IR, ``cascading gravity'' relies on the
presence of at least two infinite extra dimensions, while our world
is confined on a four-dimensional brane \citep{deRham:2007rw}. Similarly as in
DGP, four-dimensional gravity is recovered at short distances thanks to an
induced Einstein--Hilbert term on the brane with associated Planck scale $M_4$.
The brane we live in is then embedded in a five-dimensional brane, which bears a
five-dimensional Planck scale $M_5$, itself embedded in six dimensions (with
Planck scale $M_6$). From a four-dimensional perspective, the relevant scales
are the 5d and 6d masses $m_4=M_5^3/M_4^2$ and $m_5=M_6^4/M_5^3$, which
characterize the transition from the 4d to 5d and 5d to 6d behavior
respectively.

Such theories embedded in more-than-one extra dimensions involve at least one
additional scalar field that typically enters as a ghost. This ghost is
independent of the ghost present in the self-accelerating branch of DGP but is
completely generic to any
codimension-two and higher framework with brane localized kinetic terms. However, there
are two ways to cure the ghost, both of which are natural when
considering a
realistic higher codimensional scenario, namely smoothing out the brane, or
including a brane tension \citep{deRham:2007rw,deRham:2007xp,deRham:2010rw}.

When properly taking into account the issue associated with the
ghost, such models give rise to a theory of massive gravity (soft mass graviton)
composed of one
helicity-2 mode, helicity-1 modes that decouple and 2 helicity-0
modes. In order for this theory to be consistent with standard
GR in four dimensions, both helicity-0 modes should
decouple from the theory. As in DGP, this decoupling does not happen in a
trivial way, and relies on a phenomenon of strong coupling. Close
enough to any source, both scalar modes are strongly coupled and therefore
freeze.

The resulting theory appears as a theory of a massless spin-2 field in
four-dimensions,
in other words as GR.
If $r\ll m_5$ and for $m_6\le m_5$, the respective Vainshtein scale or strong
coupling scale, i.e., the distance from the source $M$
within which each mode is strongly coupled is $r_{i}^3=M/m_i^2
M_4^2$, where $i=5,6$. Around a source $M$, one recovers
four-dimensional gravity for $r\ll r_{5}$, five-dimensional
gravity for $r_{5}\ll r \ll r_{6}$ and finally six-dimensional
gravity at larger distances $r\gg r_{6}$.

The extension of Cascading gravity to higher dimensions also show the presence of solutions which allow for arbitrarily large cosmological constant without leading to any cosmic acceleration of the 3+1 brane \cite{deRham:2009wb}, hence providing a first ingredient towards tackling the cosmological constant problem.

\paragraph*{Massive gravity.}

While laboratory experiments, solar systems tests and cosmological
observations have all been in complete agreement with GR
for almost a century now, these bounds do not eliminate the
possibility for the graviton to bear a small hard mass $m\lesssim
6.10^{-32}\mathrm{\ eV}$ \citep{Goldhaber:2008xy}. The question of whether or not gravity
could be mediated by a hard-mass graviton
is not only a purely fundamental but could potentially have interesting observational implications and help with the late-time acceleration of the Universe and the original Cosmological Constant problem. Since the degravitation
mechanism is also expected to be present if
the graviton bears a hard mass, such models can play an important role for late-time cosmology, and more precisely when the age of the universe becomes on the
order of the graviton Compton wavelength. See \cite{deRham:2014zqa} for a recent review on massive gravity and related models.

Lorentz invariant  theories of hard massive gravity can be free of
any ghost-like pathologies in the decoupling limit where $M_{\mathrm{Pl}}\to \infty$
and $m\to 0$ keeping the scale $\Lambda_{3}^3=M_{\mathrm{Pl}} m^2$ fixed
\citep{deRham:2010ik,deRham:2010kj}. The decoupling limit
provides a good framework to understand the implications of a small
graviton mass. Unlike a massless spin-2 field, which only bears two
polarizations, a massive one bears five of them, namely two helicity-2 modes, two
helicity-1 modes which decouple, and one helicity-0 mode (denoted as $\pi$). As
in the braneworld models presented previously, this helicity-0 mode behaves as a
scalar field with specific derivative interactions of the form
\begin{equation}
\mathcal{L}_{\pi}=h^{\mu\nu}\left(X^{(1)}\mn+\frac{1}{\Lambda_{3}^{3}}
X^{(2)}\mn+\frac{1}{\Lambda_{3}^{6}} X^{(3)}\mn\right) \,.
\label{eq:massiveGravity}
\end{equation}
Here, $h_{\mu\nu}$ denotes the canonically-normalized (rescaled by $M_{\mathrm{pl}}$)
tensor field perturbation (helicity-2 mode), while $X^{(1)}\mn,X^{(2)}\mn,$ and
 $X^{(3)}\mn$ are
respectively, linear, quadratic and cubic in the helicity-0 mode $\pi$.
Importantly, they are all transverse (for instance, $X^{(1)}\mn \propto \eta_
{\mu\nu}\square\pi - \partial_{\mu} \partial_{\nu} \pi$). Not only do these
interactions automatically satisfy the Bianchi identity, as they should to
preserve diffeomorphism invariance, but they are also at most second order in
time derivatives. Hence, the interactions (\ref{eq:massiveGravity}) are linear
in the helicity-2 mode, and are free of any ghost-like pathologies. Therefore, such
interactions are very similar in spirit to the Galileon ones, and bear
the same internal symmetry (\ref{eq:Galileonsymmetry}), and present very similar physical properties. The stability of spherically symmetric configurations forces the $X^{(3)}\mn$ term to be absent \cite{Berezhiani:2013dca}. This represents a tuning of the parameters of the original theory but since these parameters are radiatively stable, this is not a self-tuning \cite{deRham:2012ew,deRham:2013qqa}. In that case one recovers an Einstein frame picture for which the
interactions are specific Galileon ones
\ba
\label{galgen}
\mathcal{L} &=& \frac{M_{\mathrm{Pl}}^2}{2}\sqrt{-g}R +\frac 32 \pi \Box \pi
+\frac{3\beta}{2\Lambda_{3}^{3}}(\p \pi)^{2} \Box \pi
+\frac{\beta^{2}}{2\Lambda_{3}^{6}}(\p \pi)^{2}\left(
(\partial_{\alpha}\partial_{\beta} \pi)^{2}-(\Box \pi)^{2} \right) \nonumber\\
&& +\mathcal{L}_{\mathrm{mat}}[\psi, \tilde g\mn]\,,
\ea
where $\beta$ is an arbitrary constant and matter fields $\psi$ do not couple to
the metric $g\mn$ but to $\tilde g\mn=g\mn+\pi\eta\mn+\frac{\beta}{\Lambda_{3}^3}
\partial_{\mu} \pi \partial_{\nu} \pi$.
Here again, the recovery of GR in the UV is possible via a strong coupling
phenomena, where the interactions for $\pi$ are already important at the scale
$\Lambda_{3}\ll M_{\mathrm{Pl}}$, well before the interactions for the usual helicity-2
mode. This strong coupling, as well as the peculiar coupling to matter sources,
have distinguishable features in cosmology as is explained below
\citep{Afshordi:2008rd,Jain:2010ka}.

Spherically Symmetric solutions in the decoupling limit were considered in \cite{Berezhiani:2013dw}. Stability of this solutions requires the parameter $\beta$ to be positive definite which sets another constraint of the parameters of the original theory. Furthermore it was also shown that the solutions \reviewtext{are} asymptotic to a non-trivial FRW solution which is independent of the source at infinity. Notice however that these solutions are valid within the decoupling limit of massive gravity. At very large distances from the source, the decoupling limit is no longer valid, at the graviton mass takes over. At distances comparable to the graviton's Compton wavelength one expects any solutions to be reach a Yukawa-like type of behaviour and so the space-time to be asymptotically flat, although this has not been shown explicitly in any cosmological solution.

\paragraph*{Massive Gravity Cosmology.}
As in the studies of the spherically symmetric solutions mentioned above, a considerable amount of insight into the cosmological solutions can be gained from the decoupling limit analysis. Considering the de Sitter geometry as being a small perturbation about Minkowski space-time, one can construct self-accelerating solutions which are at leading order indistinguishable from a standard $\Lambda$CDM model. The helicity-0 degree of freedom of massive gravity forms a condensate whose energy density sources self-acceleration \cite{oai:arXiv.org:1010.1780}. However, as mentioned above, the solutions found in the decoupling limit could be considered just as a transient state of the full solution. In addition, that specific cosmological solution found in the decoupling limit suffers from pathologies since the vector fields lose their kinetic terms.

Beyond the decoupling limit, it has been shown that there is a no-go theorem against the existence of flat and closed FRW solutions, i.e. if the reference metric is chosen to be Minkowski then there is no flat/closed FRW solutions in the full theory beyond the decoupling limit \cite{PhysRevD.84.124046}. The constraint needed for the absence of the Boulware-Deser ghost actually  forbids the existence of homogeneous and isotropic cosmological solutions. Despite this no-go, there still exists non-FRW solutions that are approximately homogeneous and isotropic locally within domains of the size of inverse graviton mass. These solutions can be used to put constraints on the magnitude of the graviton mass coming from the consistency with known constraints on homogeneity and isotropy. This kind of solutions demands the successful implementation of the Vainshtein mechanism in the cosmological evolution which so far has not been investigated in detail in the literature.

The no-go theorem for the existence of flat/closed FRW solutions does not apply to the case of open FRW solutions \cite{oai:arXiv.org:1109.3845}. Unfortunately, non-linear perturbations around this open FRW background are unstable making these solutions phenomenologically unviable.

A possible way out of these problems is to consider a more general reference metric. Indeed, if one takes the reference metric to be de Sitter, then one can construct FRW solutions. Nonetheless, these solutions bring other problems along due to the Higuchi bound, which imposes the mass of the graviton to be $m^2 > H^2$ which is in conflict with observational constraints in our cosmological past.
Promoting the reference metric to a FRW metric leads to a generalized Higuchi bound and one encounters similar problems \cite{Fasiello:2013woa}.

Finally another more natural possibility is the presence of inhomogeneous and even possibility anisotropies at large distance scales.
Recently there has been a considerable amount of work devoted to this studies and it is beyond the scope of this review to detail them all. We simply refer to~\cite{Volkov:2013roa} for a recent review and some of the most general solutions.  
 
 Such inhomogeneities/anisotropies are indeed to be expected on distance scales larger than the observable Universe. After all one of the main motivations of inflation is to ensure that such inhomogeneities/anisotropies are diluted in our observable Universe, but if inflation lasted a minimum number of e-folds such inhomogeneities/anisotropies would also be expected in General Relativity.

The first type  of inhomogeneous solutions corresponds to the case where only the St\"uckelberg fields (or new degrees of freedom) carry order unity inhomogeneities while the metric remains isotropic and homogeneous. The inhomogeneities are then effectively unobservable since  matter only couples to the metric and not directly to the St\"uckelberg fields.

Solutions where the metric itself carries explicit inhomogeneities while remaining isotropic have also been explored. These solutions can be constructed in such a way that the effective impact of the metric remains homogeneous and isotropic on short distance scales. In some of these cases, the mass term effectively plays the role of a cosmological constant leading to self-accelerating solutions.

Anisotropic solutions have been explored in \cite{Gumrukcuoglu:2012aa} and subsequent litterature, for which the observed anisotropy remains small at short distance scales. The presence of the anisotropy also allow for stable self-accelerating solutions.

These represents special cases of exact solutions found in massive gravity although it is understood that the most general solution is likely to differ from these exact cases by carrying order one inhomogeneity or anisotropy or both at large distances which would requires  numerical methods to be solved. This is still very much work in progress.

\paragraph*{Extensions of Massive Gravity.}

Different extensions of massive gravity have been introduced which could lead to an enriched phenomenology. First the mass can be promoted to a function of a new scalar field \cite{Huang:2012pe}. This allows for more interesting cosmology and some stable self-accelerating solutions. In this model the graviton mass could be effectively larger at earlier cosmological time, which implies that it can have an interesting phenomenology both at early and late times.

Another extension of massive gravity which also includes a new scalar field is the quasi-dilaton  \cite{D'Amico:2012zv} and its extension \cite{DeFelice:2013dua}, where the extra scalar field satisfies a specific symmetry and its interactions are thus radiatively stable. 
In the original quasi-dilaton model the self-accelerating solution has a ghost and is unstable, however this issue is avoided in the extended quasi-dilaton proposed in \cite{DeFelice:2013dua}. Moreover new types of stable self-accelerating solutions were recently found in \cite{Gabadadze:2014kaa}. Similarly as in massive gravity, the decoupling limit solution must have a completion in the full theory although it might require some level of inhomogeneity at large distance scales, which are screened at small distance scales via the Vainshtein mechanism.

A different version of massive gravity has been proposed in \cite{2013arXiv1307.3898M}, based on a non-local modification of Einstein's gravity that avoids the introduction of a second metric. The resulting cosmological model has a phantom behavior and it has been shown to be consistent with the expansion observational data.

\paragraph*{Bi- and Multi-Gravity.}

Unlike DGP or cascading gravity, models of massive gravity require the presence of a reference metric. The dynamics of this reference metric can be included and leads to a model of bi-gravity where two metrics, say $g\mn$ and $f\mn$ with their own Einstein-Hilbert kinetic terms respectively $M_g^2\sqrt{-g}R[g\mn]$ and $M_f^2\sqrt{-f}R[f\mn]$ in addition to interactions between the two-metrics which takes precisely the same form as the potential term in massive gravity \cite{Hassan:2011zd}. In this form bi-gravity was shown to be ghost free  so long as different species of matter couple to either one of both metrics. The absence of ghost when some species couple to both metrics $f$ and $g$ at the same time has not been proven but is feasible.

Bi-gravity has two metrics and yet only one copy of diffeomorphism invariance. The second copy of diffeomorphism can be restored by introducing three St\"uckelberg fields similarly as in massive gravity and can be thought of as the three additional degrees of freedom in addition to the two degrees of freedom present in metric. This leads to a total of seven degrees of freedom: two in an effectively massless spin-2 field and five in an effectively massive spin-2 field. Notice that both the massive and the massless modes are a combination of $g\mn$ and $f\mn$.

Among these three additional degrees of freedom, one counts a helicity-0 mode which satisfies the same properties as in massive gravity. In particular this helicity-0 mode behaves as a Galileon in a similar decoupling limit and is screened via a Vainshtein mechanism.

The cosmology of bi-gravity was investigated for instance in \cite{Volkov:2011an,vonStrauss:2011mq,Comelli:2011zm} and subsequent literature (see \cite{deRham:2014zqa} and \cite{Volkov:2013roa} for a review.)
Unlike in massive gravity, both metrics can take a FLRW form and lead to an interesting new cosmology. For instance in \cite{Akrami:2012vf,Akrami:2013pna} explicit self-accelerating solutions were provided in the absence of a cosmological constant. These solutions were tested against $\Lambda$CDM solutions using data from supernovae, CMB and
large scale structure. For some parameters of the theory the best-fit chi-square is competitive to that of $\Lambda$CDM. The explicit Friedman equation for these parameters was derived in
\cite{Fasiello:2013woa}
\ba
H^2=\frac{1}{6M_g^2}\(\rho+\sqrt{\rho^2+\frac{12m^4 M_g^6}{M_f^2}}\)\,,
\ea
assuming that matter only couples to the metric $g\mn$, and has an effective energy density $\rho$. In this case the scale $M_g$ is essentially the Planck scale. The scale $m$ governs the interactions between both metrics $g\mn$ and $f\mn$. In this case the self-accelerating solution can be shown to be stable.

Recently it has also been shown that  a simple form  of bi-gravity that depends on a single parameter (the minimal model) allows for stable self-accelerating solutions with distinguishable features from $\Lambda$CDM and an effective equation of state for small redshift $\omega(z)\approx -1.22 ^{+0.02}_{0.02} -0.64^{+0.05}_{-0.04}z/(1+z)$ \cite{Konnig:2014dna}.

Bi-gravity was also shown to be extendable to an arbitrary number of interacting metrics in \cite{Hinterbichler:2012cn}, which would lead to multiple Galileon in its decoupling limit.

\subsection{Effective Field Theory of Dark Energy}

One of the most productive recent ideas in dark-energy cosmology has
been the employing of effective field-theory methods originally developed
for inflation \citep{Creminelli:2006xe,Cheung:2007st} to limit the
space of possible parameterisations of gravity to that obtainable
from local actions with a fixed number of degrees of freedom and also
to describe the perturbation evolution in different models of modified
gravity using a common approach \citep{Gubitosi:2012hu,Bloomfield:2012ff,Gleyzes:2013ooa}.

We refer the reader to e.g. the review \citep{Gleyzes:2014rba} for
details, here mentioning only the rough principles. The EFT of DE
approach depends on choosing an FRW cosmological background as well
as being able to pick one of the degrees of freedom of the model to
be used as a clock on this background. This means that the approach
is most directly applicable to models with at least one scalar degree
of freedom, where the background configuration of the scalar field
evolves monotonically (i.e.~does not oscillate during the evolution).
When this is possible, the scalar will play a role of a goldstone
boson of the broken time symmetry in cosmology, its field value will
define a time slicing (a unitary gauge). The symmetries of the FRW
background must then also be the symmetries of the action which describes
the evolution of fluctuations on the cosmological background: the
action for perturbations must obey time-reparameterisation invariance
and the remaining unbroken diffeomorphism invariance of the spatial
slice. 

The Arnowitt-Deser-Misner (ADM) 3+1 split is the natural choice to
employ in this approach. One forms from the full space-time metric
$g_{\mu\nu}$ a three-dimensional spatial metric 
\[
h_{\mu\nu}=g_{\mu\nu}+u_{\mu}u_{\nu}
\]
by projecting out a time direction defined by the timelike scalar
field gradient, $u_{\mu}\equiv-\partial_{\mu}\phi/\sqrt{-\partial_{\alpha}\phi\partial^{\alpha}\phi}$.
With this choice of slicing, one can then use the ADM coordinates,
describing the metric through
\[
\mathrm{d}s^{2}=-N^{2}\mathrm{d}t^{2}+h_{ij}(\mathrm{d}x^{i}+N^{i}\mathrm{d}t)(\mathrm{d}x^{j}+N^{j}\mathrm{d}t)\,,
\]
with $N$ called the lapse and the vector $N^{i}$ — the shift. In
order to preserve the appropriate symmetries, the action for gravity
and the goldstone boson must then be a scalar formed out of only these
geometrical quantities and the spatial covariant derivative compatible
with $h_{\mu\nu}$, $D_{i}$,
\[
S_{g}=\int\mathrm{d}^{4}x\sqrt{-g}L(N,K_{ij},R_{ij},h_{ij},D_{i};t)\,,
\]
where $K_{ij}$ is the extrinsic curvature of the spatial slice and
$R_{ij}$ its intrinsic curvature. If other degrees of freedom are
present in the problem, then terms mixing these ADM geometrical quantities
and the additional fields can also appear.

One then writes down various operators for the FRW background, the
quadratic action for fluctuations and, in principle, higher-order
actions, if non-linearities are of interest. The symmetry of the cosmological
background is such that the coefficients of all these operators are
allowed to be arbitrary functions of time $t$, but the resulting
scale dependence is given by the particular operators in a fixed manner.
Note that since the ADM curvatures do not contain time derivatives,
but only spatial derivatives $D_{i}$, by using them, one does not
introduce new degrees of freedom through higher time-derivatives.
Thus one avoids this complication of the usual covariant approach
for actions. However, higher-order spatial derivatives are generated
through this approach.

The simplest application is to universally coupled scalar-tensor theories,
modifications of gravity which contain no more than one extra scalar
degree of freedom. For the background, the end result is that an arbitrary
expansion history $H(t)$ can be generated, and thus the background
should be thought of as an input of the EFT approach (e.g. $w=-1$
or any other such choice). The question of determining the theory
of gravity is then one of constraining the behaviour of perturbations
and growth of structure on this chosen background. Linear structure
formation is then determined by the quadratic action for fluctuations.

At quadratic order, one can write down multiple operators, thus a
choice of basis for them must be made. A particularly useful one is
the one introduced in ref. \citep{Bellini:2014fua}, so-called $\alpha$-functions,
since it uses operators most closely related to physical properties
of the dark energy. In this basis, the most general action for perturbations
in a scalar-tensor theory which does not contain derivatives higher
than second in the equation of motion for the propagating degrees
of freedom can be written \citep{Gleyzes:2014rba}
\begin{align*}
S_{2} & =\int\mathrm{d}t\mathrm{d}^{3}xa^{3}\frac{M_{*}^{2}}{2}\left[\delta K_{ij}\delta K^{ij}-\delta K^{2}+(1+\alpha_{\text{T}})\delta_{2}R+\right.\\
 & \left.+\alpha_{\text{K}}H^{2}\delta N^{2}+4\alpha_{\text{B}}H\delta K\delta N+(1+\alpha_{\text{H}})\delta R\delta N\right]\,,
\end{align*}
where $M_{*}$ is the effective Planck mass and the $\alpha_{i}$
are all dimensionless functions of time that are in principle arbitrary
and the $\delta$'s signify fluctuations of quantities away from their
background value. The GR limit can be recovered by taking all the
$\alpha_{i}\rightarrow0$ and $M_{*}=\text{const}$. Other operators
can be added to the the quadratic action, but they will invariably
result in higher derivatives in the equations of motion.

The $\alpha$ functions play a role in modifying the properties of
the perturbations and growth of structure. In particular they can
be divided into two classes: 
\begin{enumerate}
\item Non-minimal coupling of gravity. These functions modify both the scalar
and the tensor propagation \citep{2014PhRvL.113s1101S}: 

\begin{enumerate}
\item $M_{*}^{2}(t)$\emph{, the effective Planck mass. }$M_{*}^{2}$ is
the normalisation of the kinetic term for gravitons. It encodes the
strength of  the gravitational force/space-time curvature produced
by a fixed amount of energy . Large-scale structure is sensitive only
to the time variation of the Planck mass,
\begin{equation}
\alpha_{\textrm{M}}\equiv\frac{\mathrm{d}\ln M_{*}^{2}}{\mathrm{d}\ln a}\,,\label{eq:running def}
\end{equation}
or the \textit{Planck-mass run rate}.
\item $\alpha_{\textrm{T}}(t)$, \textit{tensor speed excess}. This parameter
denotes the difference in the speed propagation of gravitational waves
compared to the speed of light, i.e.~$\alpha_{\textrm{T}}=c_{\textrm{T}}^{2}-1$.
It applies to modes propagating on cosmological scales and is quite
weakly constrained despite the recent detection from LIGO \citep{Abbott:2016blz,Blas:2016qmn}.
\end{enumerate}
\item Kinetic terms. The scalar mode is in addition affected by the following
three functions:

\begin{enumerate}
\item $\alpha_{\textrm{K}}(t)$, \textit{kineticity}. Coefficient of the
kinetic term for the scalar d.o.f.~before demixing (see ref.~\citep{Bellini:2014fua}).
Increasing this function leads to a relative increase of the kinetic
terms compared to the gradient terms and thus a lower sound speed
for the scalar field. This creates a sound horizon smaller than the
cosmological horizon: super-sound-horizon the scalar does not have
pressure support and clusters similarly to dust. Inside, it is arrested
and eventually can enter a quasi-static configuration \citep{Sawicki:2015zya}.
When looking only at the quasi-static scales, inside the sound horizon,
this function cannot be constrained \citep{Gleyzes:2015rua}. This
is the only term present in the simplest DE models, e.g.~quintessence
and in perfect-fluid dark energy.
\item $\alpha_{\textrm{B}}(t)$, \textit{braiding}. This operator gives
rise to a new mixing of the scalar field and the extrinsic curvature
of the spatial metric, $K$. This leads to a modification of the coupling
of matter to the curvature, independent and additional to any change
in the Planck mass. This is typically interpreted as an additional
fifth force between massive particles and can be approximated as a
modification of the effective Newton's constant for perturbations.
It is present in archetypal modified gravity models such as $f(R)$
gravity (see \citep{Bellini:2014fua} for details). A purely conformal
coupling of the scalar to gravity leads to the universal property
$\alpha_{\text{M}}+\alpha_{\text{B}}=0$. 
\item $\alpha_{\text{H}}(t)$, \emph{beyond Horndeski}. This term is generated
by a kinetic mixing of the scalar with the intrinsic curvature $R$.
It results in third-order derivatives in the equations of motion,
but which cancel once all the constraints are solved for. It produces
a coupling of the gravitational field to the velocity of the matter.
\end{enumerate}
Note that either $\alpha_{\text{M}}$, $\alpha_{\text{T}}$ or $\alpha_{\text{H}}$
must not vanish in order for gravitational slip to be generated by
perfect-fluid matter sources. such a case, the equation of motion
for the propagation of gravitational waves is also modified.
\end{enumerate}
In order for the chosen background $H(t)$ to be stable in the model
of gravity given by the choice of $\alpha$ functions, certain algebraic
conditions on the $\alpha$ functions must be satisfied (the perturbations
must not be ghosty and the sound speed squared needs to be positive,
for both scalars and tensors; see e.g. \citep{Bellini:2014fua,Gleyzes:2014rba}
for details).

In addition to providing a method for exploring new models of gravity,
the EFT of DE approach allows one to describe linear perturbations
in a very efficient and unified fashion. It is enough to obtain the
appropriate functions $H(t)$ and $\alpha_{i}$ for any particular
model and structure formation in that model can be fully solved for.
In particular, Horndeski models of gravity (or the generalised galileon
models) determined by the four free functions of the scalar field
$G_{2,3,4,5}$ can be encoded in the EFT language through \citep{Bellini:2014fua}
\begin{align}
M_{*}^{2}\equiv & 2\left(G_{4}-2XG_{4X}+XG_{5\phi}-\dot{\phi}HXG_{5X}\right)\label{eq:planckmass}\\
H^{2}M_{*}^{2}\alpha_{\textrm{K}}\equiv & 2X\left(K_{X}+2XK_{XX}-2G_{3\phi}-2XG_{3\phi X}\right)+\label{eq:omega2}\\
 & +12\dot{\phi}XH\left(G_{3X}+XG_{3XX}-3G_{4\phi X}-2XG_{4\phi XX}\right)+\nonumber \\
 & +12XH^{2}\left(G_{4X}+8XG_{4XX}+4X^{2}G_{4XXX}\right)-\nonumber \\
 & -12XH^{2}\left(G_{5\phi}+5XG_{5\phi X}+2X^{2}G_{5\phi XX}\right)+\nonumber \\
 & +4\dot{\phi}XH^{3}\left(3G_{5X}+7XG_{5XX}+2X^{2}G_{5XXX}\right)\nonumber \\
HM_{*}^{2}\alpha_{\textrm{B}}\equiv & 2\dot{\phi}\left(XG_{3X}-G_{4\phi}-2XG_{4\phi X}\right)+\label{eq:omega3}\\
 & +8XH\left(G_{4X}+2XG_{4XX}-G_{5\phi}-XG_{5\phi X}\right)+\nonumber \\
 & +2\dot{\phi}XH^{2}\left(3G_{5X}+2XG_{5XX}\right)\nonumber \\
M_{*}^{2}\alpha_{\textrm{T}}\equiv & 2X\left(2G_{4X}-2G_{5\phi}-\left(\ddot{\phi}-\dot{\phi}H\right)G_{5X}\right)\label{eq:omega4}\\
\alpha_{\text{H}}= & 0
\end{align}
The operator $\alpha_{\text{H}}$ only appears in the so-called beyond
Horndeski models introduced in ref. \citep{Gleyzes:2014dya}.

Since Horndeski theories include as subclasses the majority of the
popular models of modified gravity, including perfect-fluid dark energy,
linear structure formation in all these models can be solved for in
a unified manner by obtaining these $\alpha$ functions. This method
is now employed in the publicly available Boltzmann codes EFTCAMB
\citep{Hu:2013twa}, used also in the Planck analysis on Dark Energy and Modified Gravity \cite{2015arXiv150201590P}, and \texttt{hi\_class} \citep{Zumalacarregui:2016pph}.

\subsection{Observations and Screening mechanisms}
\label{sec:Screening}

All models of modified gravity presented in this section have in
common the presence of at least one additional helicity-0 degree of freedom
that is not an arbitrary scalar, but descends from a full-fledged spin-two
field. As such it has no potential and enters the Lagrangian
via very specific derivative terms fixed by symmetries. However, tests of gravity
severely
constrain the presence of additional scalar degrees of freedom.
Interestingly this degree of freedom would severly affect the behavior of voids and could potentially help reducing the tension between Planck and supernovae. Euclid could detect such an effect at the $5\sigma$ confidence level \cite{Spolyar:2013maa}. Outside voids,
as it is well
known, in theories of massive gravity
the helicity-0 mode can evade fifth-force constraints in the vicinity of matter
if the helicity-0 mode interactions
are important enough to freeze out the field fluctuations
\citep{Vainshtein:1972sx}. This Vainshtein mechanism is similar in spirit but
different in practice to the chameleon and symmetron mechanisms presented in
detail below.
One key difference relies on the presence of
derivative interactions rather than a specific potential. So, rather
than becoming massive in dense regions, in the Vainshtein mechanism
the helicity-0 mode becomes weakly coupled to matter (and light, i.e.,
sources in general) at high energy. This screening of scalar mode can
yet have distinct signatures in cosmology and in particular for
structure formation.

\paragraph*{Different classes of screening.}
While quintessence introduces a new degree of freedom to explain the late-time acceleration of the universe, the idea behind modified gravity is instead to tackle the core of the cosmological constant problem and its tuning issues as well as screening any fifth forces that would come from the introduction of extra degrees of freedom.
As mentioned in Section~\ref{de_baryons},
the strength with which these new degrees of freedom can
couple to the fields of the standard model is very tightly constrained by
searches for fifth forces and violations of the weak equivalence principle.
Typically the strength of the scalar mediated interaction is required to be
orders of magnitude weaker than gravity. It is possible to tune this coupling
to be as small as is required, leading however to additional naturalness
problems. Here we discuss in more detail a number of ways in which new scalar degrees of freedom can
naturally couple to standard model fields, whilst still being in agreement with
observations, because a dynamical mechanism ensures that their effects are
screened in laboratory and solar system tests of gravity. This is done by
making some property of the field dependent on the background environment under
consideration. These models typically fall into two classes; either the field
becomes massive in a dense environment so that the scalar force is suppressed
because the Compton wavelength of the interaction is small, or the coupling to
matter becomes weaker in dense environments to ensure that the effects of the
scalar are suppressed. Both types of behavior require the presence of
nonlinearities.

\paragraph*{Density dependent masses: The chameleon.}

The chameleon \citep{Khoury:2003rn} is the archetypal model of a scalar field
with a mass that depends on its environment, becoming heavy in dense
environments and light in diffuse ones. The ingredients for construction of a
chameleon model are a conformal coupling between the scalar field and the matter
fields of the standard model, and a potential for the scalar field, which
includes relevant self-interaction terms.

In the presence of non-relativistic matter these two pieces conspire to give
rise to an effective potential for the scalar field
\begin{equation}
V_{\mathrm{eff}}(\phi) = V(\phi)+\rho A(\phi),
\end{equation}
where $V(\phi)$ is the bare potential, $\rho$ the local energy density and
$A(\phi)$ the conformal coupling function. For suitable choices of $A(\phi)$
and $V(\phi)$ the effective potential has a minimum and the position of the
minimum depends on $\rho$. Self-interaction terms in $V(\phi)$ ensure that the
mass of the field in this minimum also depends on $\rho$ so that the field
becomes more massive in denser environments.

The environmental dependence of the mass of the field allows the chameleon to
avoid the constraints of fifth-force experiments through what is known as the
thin-shell effect. If a dense object is embedded in a diffuse background the
chameleon is massive inside the object. There, its Compton wavelength is small.
If the Compton wavelength is smaller than the size of the object, then the scalar
mediated force felt by an observer at infinity is sourced, not by the entire
object, but instead only by a thin shell of matter (of depth the Compton
wavelength) at the surface. This leads to a natural suppression of the force
without the need to fine tune the coupling constant.

\paragraph*{The Vainshtein Mechanism.}
In models such as DGP\newtext{,}  the Galileon, Cascading gravity, massive gravity and bi- or multi-gravity, the effects of the scalar field(s) are
screened by the Vainshtein mechanism \citep{Vainshtein:1972sx,Deffayet:2001uk}, see also \cite{Babichev:2013usa} for a recent review on the Vainshtein mechanism.
This occurs when nonlinear, higher-derivative operators are present in the
Lagrangian for a scalar field, arranged in such a way that the equations of
motion for the field are still second order, such as the interactions presented
in Eq.~(\ref{eq.DGP_decoupling}).

In the presence of a massive source the nonlinear terms force the suppression
of the scalar force in the vicinity of a massive object. The radius within
which the scalar force is suppressed is known as the Vainshtein radius. As an
example in the DGP model the Vainshtein radius around a massive object of mass
$M$ is
\begin{equation}
r_{\star}\sim \left( \frac{M}{4\pi M_{\mathrm{Pl}}}\right)^{1/3}\frac{1}{\Lambda} \,,
\end{equation}
where $\Lambda$ is the strong coupling scale introduced in section
\ref{sec:DGP}. For the Sun, if $m\sim 10^{-33}\mathrm{\ eV}$, or in other words,
$\Lambda^{-1}=1000\mathrm{\ km}$, then the Vainshtein radius is
$r_{\star} \sim 10^2\mathrm{\ pc}$.

Inside the Vainshtein radius, when the nonlinear, higher-derivative terms
become important they cause the kinetic terms for scalar fluctuations to become
large. This can be interpreted as a relative weakening of the coupling between
the scalar field and matter. In this way the strength of the interaction is
suppressed in the vicinity of massive objects.

Related to the Vainshtein mechanism but slight more general is the screening via a disformal coupling between the scalar field and the stress-energy tensor $\p_\mu \pi \p_\nu \pi T^{\mu\nu}$ \cite{ARXIV:1205.3167} as is present in DBI-braneworld types of models \cite{deRham:2010eu} and massive gravity \cite{deRham:2010ik}.

\paragraph*{The Symmetron.}
The symmetron model \citep{Hinterbichler:2010es} is in many ways similar to the
chameleon model discussed above. It requires a conformal coupling between the
scalar field and the standard model and a potential of a certain form. In the
presence of non-relativistic matter this leads to an effective potential for the
scalar field
\begin{equation}
V_{\mathrm{eff}}(\phi)
=-\frac{1}{2}\left(\frac{\rho}{M^2}-\mu^2\right)\phi^2+\frac{1}{4}\lambda
\phi^4 \,,
\end{equation}
where $M$, $\mu$ and $\lambda$ are parameters of the model, and $\rho$ is the
local energy density.

In sufficiently dense environments, $\rho>\mu^2M^2$, the field sits in a minimum
at the origin. As the local density drops the symmetry of the field is
spontaneously broken and the field falls into one of the two new minima with a
non-zero vacuum expectation value. In high-density symmetry-restoring environments, the scalar field
vacuum expectation value should be near zero and fluctuations of the field should not couple to
matter. Thus, the symmetron force in the exterior of a massive object is
suppressed because the field does not couple to the core of the object.

\paragraph*{The Olive--Pospelov model.}
The Olive--Pospelov model \citep{Olive:2007aj} again uses a scalar conformally
coupled to matter. In this construction both the coupling function and the
scalar field potential are chosen to have quadratic minima. If the background
field takes the value that minimizes the coupling function, then fluctuations of
the scalar field decouple from matter. In non-relativistic environments the
scalar field feels an effective potential, which is a combinations of these two
functions. In high-density environments the field is very close to the value
that minimizes the form of the coupling function. In low-density
environments the field relaxes to the minimum of the bare potential. Thus, the
interactions of the scalar field are suppressed in dense environments.


\subsection{Einstein Aether and its generalizations}
\label{sec:EinAet}

In 1983 it was suggested by Milgrom \citep{Milgrom1983a} that the
emerging evidence for the presence of dark matter in galaxies
could follow from a modification either to how `baryonic' matter
responded to the Newtonian
gravitational field it created
or to how the gravitational field was related to the baryonic matter density.
Collectively these ideas are referred to as MOdified Newtonian Dynamics
(MOND).  By way of illustration, MOND may be considered as a
modification to the non-relativistic Poisson equation:
\begin{equation}
\label{eq:mond}
\nabla\cdot\left(\mu\left(\frac{|\nabla\Psi|}{a_{0}}\right)\nabla\Psi\right)=
4\pi G\rho \,,
\end{equation} 
where $\Psi$ is the gravitational potential, $a_{0}$ is a number with dimensions
Length$^{-1}$ and $\rho$ is the baryonic matter density. The number $a_{0}$ is
determined by looking at the dynamics
of visible matter in galaxies \cite{SandersMcGaugh2002}. The function $\mu(x)$
would simply be equal to unity in Newtonian gravity. In MOND, the functional
form is only fixed at its limits: $\mu \rightarrow 1$ as $x \rightarrow \infty$
and $\mu \rightarrow x$ as $x \rightarrow 0$.

We are naturally interested in a relativistic version of such a proposal. The building
block is the perturbed spacetime metric already introduced in Eq.~\ref{pert_newton_ds}
\begin{eqnarray} 
\label{eq:metric}
ds^{2}= -(1+2\Psi)\,
\mathrm{d}t^{2}+(1-2\Phi)a^{2}(t)(\mathrm{d}R^{2}+R^{2}\, \mathrm{d}\Omega^{2}).
\end{eqnarray} 
A simple approach is to introduce a dynamical clock field, which we
will call $A^{\mu}$. If it has solutions aligned with the time-like coordinate $t^{\mu}$ then
it will be sensitive to $\Psi$.  The dynamical 
nature of the field implies that it should have an action that will contain gradients
of the field  and thus potentially scalars formed from gradients of $\Psi$, as
we seek.
A family of covariant actions for the clock field is as follows
\cite{Zlosnik:2006zu}: 
\begin{eqnarray}
\label{geaact}
\nonumber I [g^{ab},A^{a},\lambda] &=&  \frac{1}{16\pi G} \int \mathrm{d}^4x \sqrt{-g}
\left[\frac{1}{\ell^2}  F(K) + \lambda \left(A^a A_a + 1 \right) \right] ,
\end{eqnarray}
where 
\begin{equation}
  K = \ell^2 K^{\mu\nu\gamma\delta} \nabla_\mu A_\nu \nabla_\gamma A_\delta
\end{equation}
with
\begin{equation}
 K^{\mu\nu\gamma\delta} = c_1 g^{\mu\gamma} g^{\nu\delta} + c_2 g^{\mu\nu}
g^{\gamma\delta} + c_3 g^{\mu\delta} g^{\nu\delta} \,.
\end{equation}
The quantity $\ell$ is a number with dimensions of length, the $c_{A}$ are
dimensionless constants, the Lagrange multiplier field $\lambda$ enforces the
unit-timelike constraint on $A^{a}$, 
and $F$ is a function. These models
have been termed Generalized Einstein-Aether (GEA) theories, emphasizing the
coexistence of general covariance and a `preferred' state of rest
in the model, i.e., keeping time with $A^{\mu}$.

Indeed, when the geometry is of the form (\ref{eq:metric}), anisotropic
stresses are negligible and $A^{\mu}$ is aligned with the flow of time
$t^{\mu}$, then one can find appropriate values of the $c_{A}$ and $\ell$ such
that $K$ is dominated by 
a term equal to
$|\nabla\Psi|^{2}/a_{0}^{2}$. This influence then leads to a modification to the
time-time component of Einstein's equations: instead of reducing to Poisson's
equation,
one recovers an equation of the form (\ref{eq:mond}). Therefore the models are
successful covariant realizations of MOND.

Interestingly, in the FLRW limit $\Phi,\Psi\rightarrow 0$, the time-time
component of Einstein's equations in the GEA model becomes a modified Friedmann
equation:
\begin{eqnarray}
\label{modfr}
\beta\left(\frac{H^{2}}{a_{0}^{2}}\right)H^{2} = \frac{8\pi G\rho}{3} \,,
\end{eqnarray} 
where the function $\beta$ is related to $F$ and its derivatives with respect to
$K$.
The dynamics in galaxies prefer a value $a_{0}$ on the order the Hubble
parameter today $H_{0}$ \cite{SandersMcGaugh2002} and so one typically 
gets  a modification to the background expansion with a characteristic scale
$H_{0}$, i.e., the scale associated with modified gravity
models that produce dark-energy effects. Ultimately the GEA model is a
phenomenological one and as such there currently lack deeper reasons to favor
any 
particular form of $F$. However, one may gain insight into the possible
solutions of (\ref{modfr}) by looking at simple forms for
$F$. In~\cite{Zuntz:2010jp} the monomial case $F\propto K^{n_{ae}}$
was considered where the kinetic index $n_{ae}$ was allowed to
vary. Solutions with accelerated expansion were found that could mimic dark energy.

Returning to the original motivation behind the theory, the next step is to look
at the theory on cosmological scales and see whether the GEA
models are realistic alternatives to dark matter. As emphasized, the additional
structure in spacetime is dynamical and so possesses
independent degrees of freedom. As the model is assumed to be uncoupled to other
matter, the gravitational field equations would regard
the influence of these degrees of freedom as a type of dark matter (possibly
coupled non-minimally to gravity, and not necessarily `cold').

The possibility that the model may then be a viable alternative to the dark
sector in background cosmology and linear cosmological 
perturbations has been explored in depth in
\cite{ZlosnikFerreiraStarkman2008, Li:2007vz}
and \cite{Zuntz:2010jp}. As an alternative to dark matter,
it was found that the GEA models could replicate some but not all of the
following features of cold dark matter: 
influence on background dynamics of the universe;
negligible sound speed of perturbations;
growth rate of dark matter `overdensity';
absence of anisotropic stress and contribution to the cosmological Poisson
equation;
effective minimal coupling to the gravitational field.
When compared to the data from large scale structure and the CMB, the model
fared significantly less well than the Concordance Model
and so is excluded. If one relaxes the requirement that the vector field be 
responsible for the effects of cosmological dark matter, one can 
look at the model as one responsible only for the effects of dark energy. It was
found \cite{Zuntz:2010jp} that the current most stringent constraints
on the model's success as dark energy were from constraints on the size of large
scale CMB anisotropy. Specifically, possible variation in $w(z)$ of the
`dark energy' along with new degrees of freedom sourcing anisotropic stress in
the perturbations was found to lead to new, 
non-standard time variation of the potentials $\Phi$ and $\Psi$. These time
variations source large scale anisotropies via the integrated
Sachs--Wolfe effect, and the parameter space of the model is constrained in
avoiding the effect becoming too pronounced. 

In spite of this, given the status of current experimental bounds it is 
conceivable that a more successful alternative to the dark sector may share some
of
these points of departure from the Concordance Model and yet fare significantly
better at the level of the background and linear perturbations.

\subsection{The Tensor-Vector-Scalar theory of gravity}
\label{sec:TeVeS}
Another proposal for a theory of modified gravity arising from Milgrom's observation is the Tensor-Vector-Scalar theory of gravity, or TeVeS.
TeVeS theory is \textit{bimetric} with two frames: the 
``geometric frame'' for the gravitational fields, and the ``physical frame'', for the matter fields. The three gravitational fields are the metric $\metE_{ab}$ (with connection $\tilde{\nabla}_a$) 
that we refer to as the geometric metric, 
the vector field $A_a$ and the scalar field $\phi$.
The action for all matter fields, uses a single physical metric 
$g_{ab}$ (with connection $\nabla_a$). The two metrics are related via an algebraic, disformal relation~\citep{Bekenstein1993} as
\begin{equation}
  \metM_{ab} = e^{-2\phi}\metE_{ab} - 2\sinh(2\phi)A_a A_b \,.
  \label{eq:metric_relation}
\end{equation}
Just like in the Generalized Einstein-Aether theories, the vector field is further enforced to be unit-timelike with respect to the geometric metric, i.e.,
\begin{equation}
 \metE^{ab} A_a A_b = A^a A_a = -1.
\label{eq:A_unit}
\end{equation}
The theory is based on an action $S$, which is split as $S = S_{\metE} + S_A + S_{\phi}+S_m$
where
\begin{equation}
  S_{\metE} = \frac{1}{16\pi G}\int \mathrm{d}^4x \; \volE \; \RiemE \,,
\label{eq:S_EH}
\end{equation}
where $\metE$ and $\RiemE$ are \label{def_detE}\label{def_RE} the determinant and scalar curvature of $\metE_{\mu\nu}$ respectively and
$G$ \label{def_Gbare} is the bare gravitational constant, 
\begin{equation}
  S_A = -\frac{1}{32\pi G} \int \mathrm{d}^4x \; \volE \; \left[ K F^{ab}F_{ab} - 2\lambda (A_a A^a + 1)\right],
\end{equation}
where $F_{ab} = \nabla_a A_b - \nabla_b A_a$ leads to a Maxwellian kinetic term
and $\lambda$ \label{def_lambda} is a Lagrange multiplier ensuring the unit-timelike constraint on $A_a$ and $K$
is a dimensionless constant (note that indices on $F_{ab}$ are raised using the geometric metric, i.e., $F^a_{\;\;b} = \metE^{ac} F_{cb}$) and
\begin{equation}
  S_{\phi} = -\frac{1}{16\pi G} \int \mathrm{d}^4x \volE \left[ 
  \mu \; \metS^{ab}\connE_a\phi \connE_b\phi + V(\mu) \right],
\end{equation}
where $\mu$ \label{def_mu} is a non-dynamical dimensionless scalar field, $\metS^{ab} = \metE^{ab} - A^a A^b$
 and $V(\mu)$ is an arbitrary function that typically depends on a scale $\ell_B$. 
 The matter is coupled only to the physical metric $\metM_{ab}$ and defines the matter stress-energy tensor $T_{ab}$ through
 $\delta S_m = -\frac{1}{2} \int \mathrm{d}^4x \volM \; T_{ab} \; \delta\metM^{ab}$. The TeVeS action can be written entirely in the physical 
frame~\citep{ZlosnikFerreiraStarkman2006,Skordis2009a} or in a diagonal frame~\citep{Skordis2009a} where the scalar and vector fields decouple.

In a Friedmann universe, the cosmological evolution is governed by the Friedmann equation
\begin{equation}
3\tilde{H}^2 = 8\pi G e^{-2\phi} \left( \rho_\phi + \rho\right),
\label{eq_beke_friedmann}
\end{equation}
where $\tilde{H}$ is the Hubble rate in terms of the geometric scale factor, $\rho$ is the physical matter density
 that obeys the energy conservation equation with respect to the physical metric 
and where the scalar field energy density is
\begin{equation}
\rho_\phi = \frac{e^{2\phi}}{16\pi G}\left( \mu \Vp + V \right) 
\end{equation}
Exact analytical and numerical solutions with the Bekenstein free function have been found in~\cite{SkordisEtAl2006}
and in~\cite{DodelsonLiguori2006}. It turns out that 
energy density tracks the matter fluid energy density. 
The ratio of the energy density of the scalar field to that of ordinary matter is approximately constant, so that the scalar field exactly tracks the matter dynamics.
In realistic situations, the radiation era tracker is almost never realized, as has been noted by Dodelson and Liguori, but rather $\rho_\phi$ is subdominant
and slowly-rolling and $\phi \propto a^{4/5}$.
\cite{BourliotEtAl2006} studied more general free functions which have the Bekenstein function as a special case and found a whole range of behavior, from tracking and accelerated expansion to finite time singularities. \cite{Diaz-RiveraSamushiaRatra2006} have 
studied cases where the cosmological TeVeS equations lead to inflationary/accelerated expansion solutions. 

Although no further studies of accelerated expansion in TeVeS have been performed, it is very plausible that certain choices of function will
inevitably lead to acceleration. It is easy to see that the scalar field action has the same form 
as a k-essence/k-inflation~\citep{Armendariz-PiconMukhanovSteinhardt2000} action which has been considered as
a candidate theory for acceleration.
It is unknown in general whether this has similar features as the uncoupled k-essence, although Zhao's study
indicates that this a promising research direction~\citep{Zhao2006a}. 

In TeVeS, cold dark matter is absent. Therefore, in order to get acceptable values for the physical Hubble constant today (i.e., around $H_0 \sim 70\mathrm{\ km/s/Mpc}$), 
we have to supplement the absence of CDM with something else. Possibilities include the scalar field itself, massive neutrinos~\citep{SkordisEtAl2006,FerreiraSkordisZunkel2008}
and a cosmological constant. At the same time, one has to get the right angular diameter distance to recombination~\citep{FerreiraSkordisZunkel2008}.
These two requirements can place severe constraints on the allowed free functions.

\epubtkImage{cl_teves-pk_teves.png}{%
\begin{figure}
\centerline{
\includegraphics[width=6.5cm]{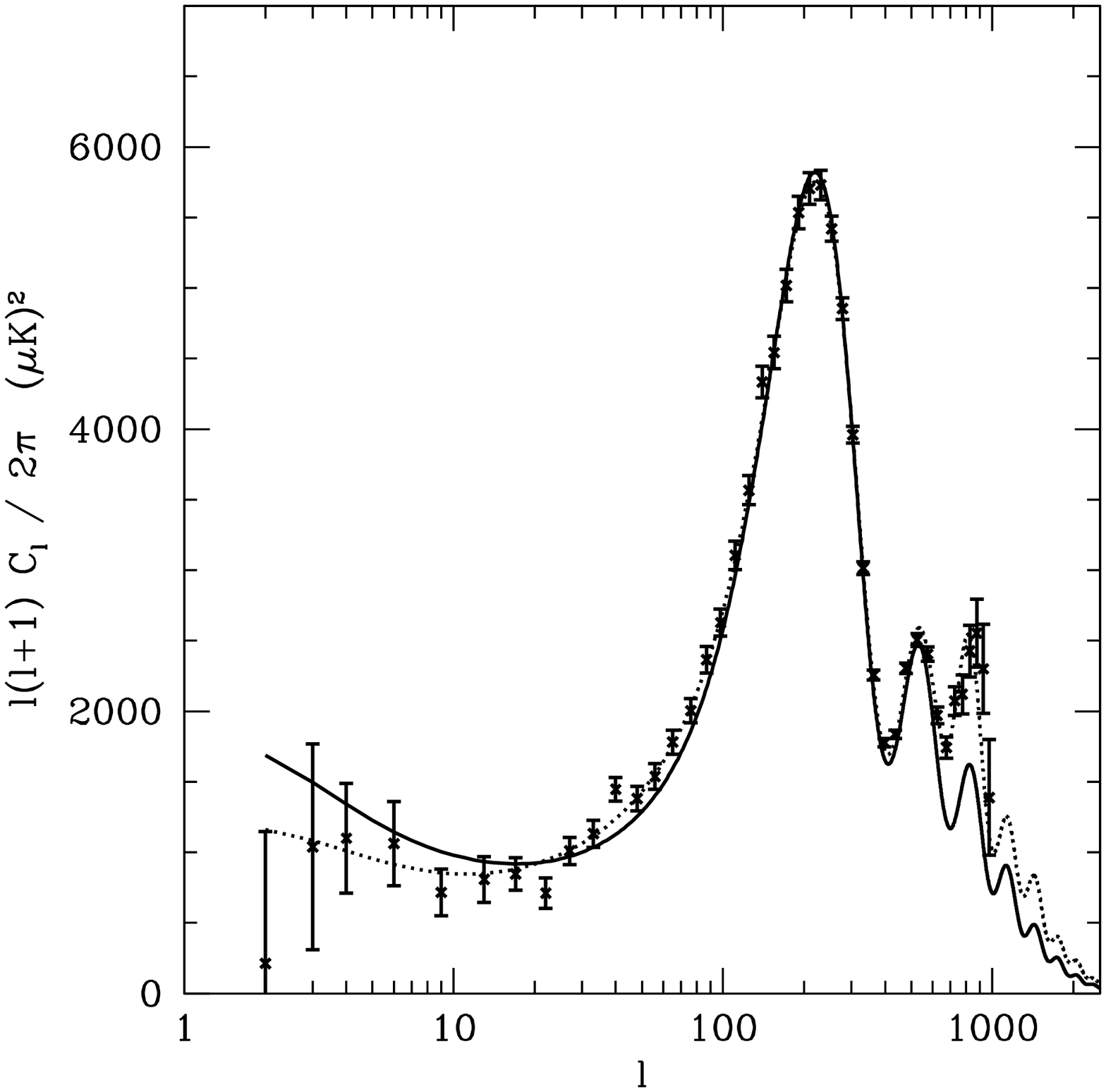}
\includegraphics[width=6.5cm]{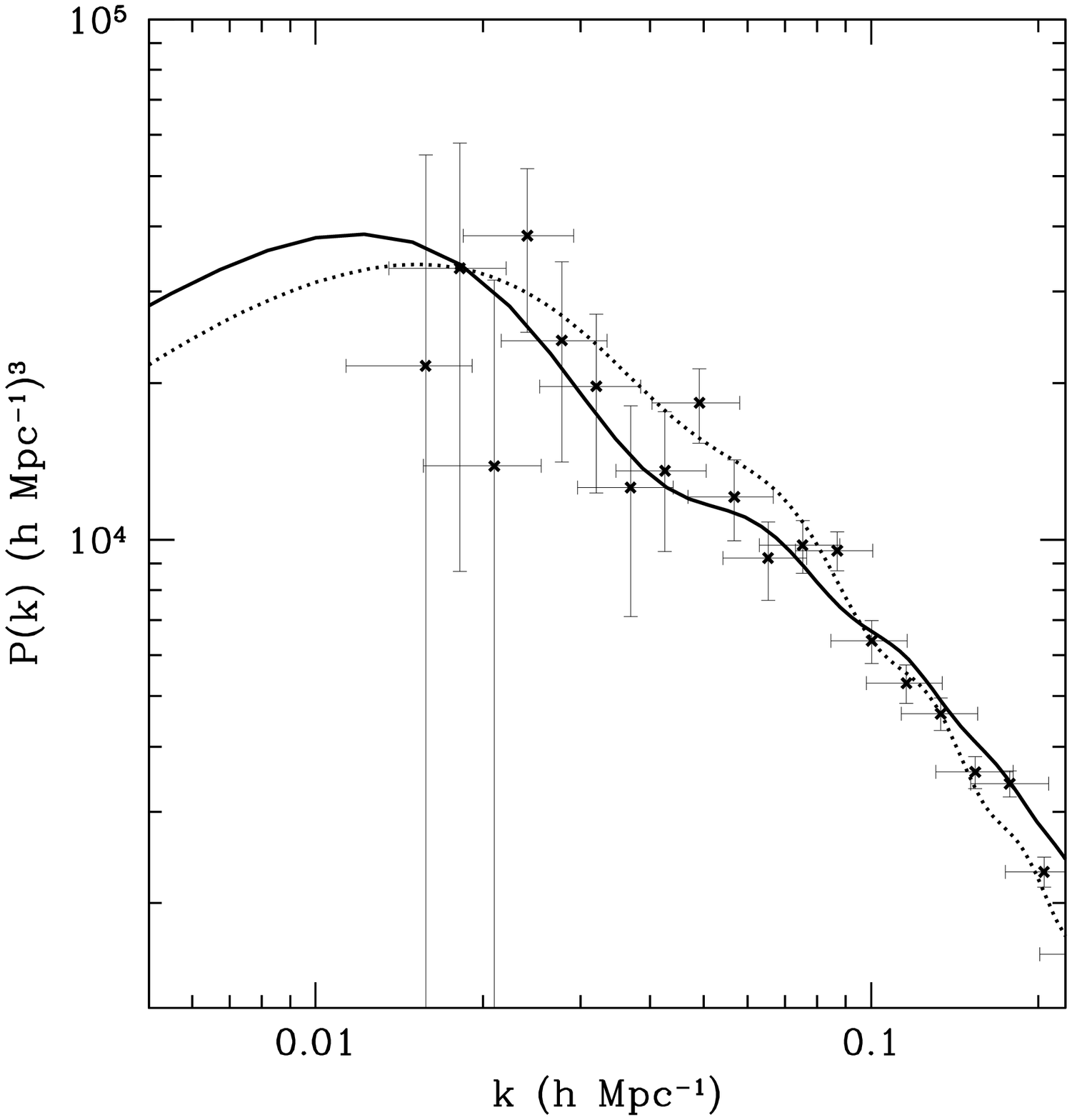}
}
\caption{\emph{Left:} the cosmic microwave background angular power
  spectrum $l(l+1)C_l/(2\pi)$ for TeVeS (solid) and $\Lambda$CDM
  (dotted) with WMAP 5-year data~\citep{NoltaEtAl2008}. \emph{Right:}
  the matter power spectrum $P(k)$ for TeVeS (solid) and $\Lambda$CDM
  (dotted) plotted with SDSS data.}
\label{fig_Pk}
\end{figure}}

Until TeVeS was proposed and studied in detail, MOND-type theories were assumed to be fatally flawed: their lack of a dark matter component would necessarily prevent the formation of large-scale structure
compatible with current observational data. In the case of an Einstein universe,
it is well known that, since baryons are coupled to photons before recombination they do not have enough time to grow into structures on their own. 
In particular, on scales smaller than the diffusion damping scale perturbations in such a universe are exponentially
damped due to the Silk-damping effect. CDM solves all of these problems because it does not couple to photons and therefore can start creating potential
wells early on, into which the baryons fall. 

TeVeS contains two additional fields, which change the structure of the equations significantly. 
The first study of TeVeS predictions for large-scale structure observations was conducted in~\cite{SkordisEtAl2006}.
They found that TeVeS can indeed form large-scale structure compatible with observations depending
on the choice of TeVeS parameters in the free function. In fact the form of the matter power spectrum $P(k)$ in TeVeS looks quite similar to that in
$\Lambda$CDM. Thus TeVeS can produce matter power spectra that cannot be distinguished from $\Lambda$CDM by current observations. 
One would have to turn to other observables to distinguish the two models. The power spectra for TeVeS and $\Lambda$CDM are plotted on the right panel of Figure~\ref{fig_Pk}. \cite{DodelsonLiguori2006} provided an analytical explanation of the growth of structure seen numerically by~\cite{SkordisEtAl2006} and found that the growth in TeVeS is due to the
vector field perturbation.

It is premature to claim (as
in~\cite{SlosarMelchiorriSilk2005,SpergelEtAl2006}) 
that only a theory with CDM can fit CMB observations; a prime example to the
contrary is the EBI theory~\cite{Banadosetal2009}.
Nevertheless, in the case of TeVeS~\cite{SkordisEtAl2006}
numerically solved the linear Boltzmann equation in the case of TeVeS and
calculated the CMB angular power
spectrum for TeVeS. By using initial conditions close to adiabatic the spectrum
thus found provides very poor fit as compared to
the $\Lambda$CDM model (see the left panel of Figure~\ref{fig_Pk}). The CMB
seems to put TeVeS into trouble, at least for the Bekenstein free function.
The result of~\cite{DodelsonLiguori2006} has a further direct consequence. The
difference $\Phi -\Psi$, sometimes named the gravitational slip (see Section~\ref{mg_growth_params}),
 has additional contributions coming from the 
perturbed vector field $\alpha$. 
Since the vector field is required to grow in order to drive structure
formation, it will inevitably lead to a growing $\Phi - \Psi$. 
If the difference $\Phi - \Psi$
 can be measured observationally, it can provide a substantial test of TeVeS
that can distinguish TeVeS from $\Lambda$CDM.

\subsection{Other Models of Interest}

\subsubsection{Models of Varying Alpha}
Whenever a dynamical scalar field is added to a theory, the field will naturally couple to all other fields present, unless a (still unknown) symmetry is postulated to suppress such couplings \cite{carroll}. A coupling to the electromagnetic sector leads to spacetime variations of Nature'€™s fundamental constants, which are constrained both by local atomic clock experiments and by astrophysical observations \cite{constants}. Joint constraints on dynamical dark energy model parametrizations and on the coupling with electromagnetism were obtained in \cite{calabrese}, combining weak lensing and supernova measurements from Euclid with high-resolution spectroscopy measurements from the European Extremely Large Telescope \cite{2015GReGr..47.1843M}. Inter alia it has been shown that in the CPL parametrization of these models, the addition of spectroscopic data (which spans the range $0<z \lesssim4$) inproves constraints from Euclid observables by a factor of 2 for $w_0$ and by one order of magnitude for $w_a$.

\subsubsection{Three-form Dark Energy}
It is possible to construct dark energy models with exotic field content such as spinors \cite{Boehmer:2009aw} 
and three-forms \cite{Koivisto:2009ew,Koivisto:2009fb,Boehmer:2011tp}. Three-form dark energy can reproduce a viable late-time accelerating expansion, and presents particularly interesting signatures when coupled to dark matter \cite{Ngampitipan:2011se,Koivisto:2012xm}. 

\subsubsection{$f(R,P,Q)$ theories}
A generalization of the $f(R)$ theories discussed in section (1.4.6) involves adding  to the usual Einstein-Hilbert term negative powers of the polynomial $aR^{2}+bP+cQ$, where $P\equiv R_{\mu\nu}R^{\mu\nu}$ and $Q=R_{\mu\nu\alpha\beta}R^{\mu\nu\alpha\beta}$ \cite{Carroll:2004de}. Negative powers guarantee that flat space is not a solution to the equations of motion and that there are no conformal transformations that map these models onto quintessence-like models. At late times the equations of motion generically admit attractor solutions of the type $a(t)\sim t^{p}$, with $p>1$ so that $\ddot a(t)>0$. The classical and quantum stability of these solutions has been investigated (see respectively \cite{Cognola:2007vq} and \cite{Cognola:2006sp}) while their compatibility with Solar System tests was studied, for example, in \cite{Nojiri:2006ri}.

\subsubsection{Yang-Mills-Higgs dark energy}
It is known that  the standard model Higgs field, non-minimally coupled to gravity, can act as the inflaton field  \cite{Bezrukov2007}. What is less well-known is that it can also account for dark energy thanks to its SU(2) gauge symmetry. When implemented on a FLRW spacetime, such a symmetry forces the Goldstone components of the Higgs to participate in the background dynamics in a fashion similar to the ``spintessence'' model of dark energy \cite{Boyle:2001du}. In the simplest Abelian U(1) case, it has been demonstrated that accelerated expansion occurs at late times \cite{Rinaldi:2013lsa}. However, the most natural model is obtained by considering a SU(2) standard model-like action of  the Higgs field together with the Yang-Mills fields, minimally coupled to gravity \cite{2015JCAP...10..023R}. Here, it is  that the full dynamical system of equations of Einstein gravity together with the ones from the Yang-Mills-Higgs action has only one stable fixed point in the future that corresponds to an accelerated Universe. The other fixed points are all unstable and correspond to matter- or radiation-dominated Universe. Therefore, in this model, dark energy is simply the result of the (classical) interaction of the non-abelian Yang-Mills-Higgs fields with gravity.

\subsubsection{$f(T)$ Gravity}
$f(T)$ gravity is a generalization of teleparallel gravity, where the torsion scalar $T$, instead of curvature, is responsible for gravitational interactions. In this theory spacetime is endowed with a curvature-free Weitzenbock connection. Thus, torsion acts as a force, allowing for the interpretation of gravity as a gauge theory of the translation group \citep{Arcos:2005ec}. Teleparallel gravity and GR yield completely equivalent dynamics for the choice $f(T)=T$, but differ for any other choice\citep{Ferraro:2008ey, Fiorini:2009ux}. Unlike analogous $f(R)$ theories, $f(T)$ gravity yields equations that remain at second order in field derivatives; however, local Lorentz invariance is lost.

 In $f(T)$ cosmology, structure formation is modified because of a time dependent effective gravitational constant. \citet{Cardone:2012xq} analysed two viable $f(T)$ gravity models and showed that both are in very good agreement with a wide set of data, including SNIa and GRB Hubble diagrams, BAOs at different redshifts, Hubble expansion rate measurements and the WMAP7 distance priors. Yet, that wide dataset is unable to severely constrain the model parameters and hence discriminate amongst the considered $f(T)$ models and the $\Lambda$CDM scenario. Therefore, \citet{Camera:2013bwa} investigated the imprints of $f(T)$ gravity on galaxy clustering and weak gravitational lensing in the context of Euclid, finding that with such a combination of probes it will indeed be possible to tightly constrain $f(T)$ model parameters.

\section{Nonlinear aspects}
\label{nonlinear-aspects}

In this section we discuss how the nonlinear evolution of cosmic structures in
the context of
different non-standard cosmological models can be studied by means of numerical
simulations
based on $N$-body algorithms and of analytical approaches based on the spherical collapse model.

\subsection{\textit{N}-body simulations of dark energy and modified gravity}
\label{Nbody_sims}

Here we discuss the numerical methods presently available
for this type of 
analyses, and
we review the main results obtained so far for different classes of
alternative cosmologies. These can be grouped into models where structure formation is
affected only through
a modified expansion history (such as quintessence and early dark-energy models, Section~\ref{quintessence})
and models where particles experience modified gravitational forces, either for individual particle
species (interacting dark-energy models
and growing neutrino models, 
Section~(\ref{mg:cde}) or for all types of particles in
the universe (modified gravity models).
A general overview on the recent developments in the field of Dark Energy and Modified Gravity N-body simulations can be found in \citep{Baldi_2012b}.

\subsubsection{Quintessence and early dark-energy models}
\label{quintessence_ede}

In general, in the context of flat FLRW cosmologies, any dynamical evolution of the 
dark-energy density ($\rho_{\mathrm{DE}}\ne {\mathrm{const.}} = \rho_{\Lambda }$) determines a modification of the cosmic expansion history with respect to
the standard $\Lambda $CDM cosmology. 
In other words, if the dark energy is a dynamical quantity, i.e., if its equation of state parameter $w\ne -1$ exactly, 
for any given set of cosmological parameters ($H_{0}$, $\Omega_{\mathrm{CDM}}$, $\Omega_{\mathrm{b}}$, $\Omega_{\mathrm{DE}}$, $\Omega_{\mathrm{rad}}$),
the redshift evolution of the Hubble function $H(z)$ will differ from the standard $\Lambda$CDM case $H_{\Lambda}(z)$.

Quintessence models of dark energy \citep{Wetterich_1988,Ratra:1987rm} 
based on a classical scalar field $\phi $ subject to a self-interaction potential $V(\phi)$ have an energy density $\rho_{\phi } \equiv \dot{\phi}^{2}/2 + V(\phi)$ that evolves in time according to the 
dynamical evolution of the scalar field, which is governed by the homogeneous Klein--Gordon equation:
\begin{equation}
\label{kg}
\ddot{\phi } + 3H\dot{\phi } + \frac{\mathrm{d}V}{\mathrm{d}\phi } = 0 \,.
\end{equation}
Here the dot denotes derivation w.r.t.\ ordinary time $t$.

For a canonical scalar field, the equation of state parameter $w_{\phi }\equiv
\rho_{\phi}/p_{\phi}$,
where $p_{\phi}\equiv \dot{\phi}^{2}/2 - V(\phi)$,
will in general be larger than $-1$, and the density of dark energy $\rho_{\phi}$ will consequently be larger than $\rho_{\Lambda }$
at any redshift $z > 0$.
Furthermore, for some simple choices of the
potential function such as those discussed in Section~\ref{quintessence} (e.g., an exponential
potential $V\propto \exp(-\alpha \phi /M_{\mathrm{Pl}})$ or an inverse-power
potential $V\propto (\phi /M_{\mathrm{Pl}})^{-\alpha }$),
scaling solutions for the evolution of the system can be analytically derived.
In particular, for an exponential potential, a scaling solution exists
where the dark energy scales as the dominant cosmic component, with a fractional
energy density
\begin{equation}
\label{EDE_1}
\Omega_{\phi }\equiv \frac{8\pi G \rho_{\phi }}{3H^{2}} = \frac{n}{\alpha^{2}}\,,
\end{equation}
with $n=3$ for matter domination and $n=4$ for radiation domination.
This corresponds to a relative fraction of dark energy at high redshifts, which
is in general not negligible, whereas during matter and radiation domination
$\Omega_{\Lambda }\sim 0$ and, therefore, represents a phenomenon of an early
emergence of dark energy as compared to $\Lambda $CDM where
dark energy is for all purposes negligible until $z\sim 1$.

Early dark energy (EDE) is, therefore, a common prediction of scalar
field models of dark energy, and observational constraints put firm bounds on
the
allowed range of $\Omega_{\mathrm{DE}}$ at early times, and consequently on the
potential slope $\alpha$.

As we have seen in Section~\ref{parametrization-of-the-background-evolution}, a completely phenomenological parametrization of EDE, independent from any
specific model of dynamical dark energy has been proposed by
\citet{Wetterich:2004pv}
as a function of the present dark-energy density $\Omega_{\mathrm{DE}}$, its value
at early times $\Omega_{\mathrm{e}}$, and the present value of the equation of
state parameter
$w_{0}$. 
From Eq.~\ref{w_ede_par}, the full expansion history of the corresponding EDE model
can be derived.

A modification of the expansion history indirectly influences also the growth of density perturbations and ultimately the formation of cosmic structures.
While this effect can be investigated analytically for the linear regime, N-body simulations are required to extend the analysis to the nonlinear stages
of structure formation. For standard Quintessence and EDE models, the only modification that is necessary to implement into standard $N$-body algorithms is the 
computation of the correct Hubble function $H(z)$ for the specific model under investigation, since this is the only way in which these non standard cosmological models can alter
structure formation processes. 

This has been done by the independent studies of \citet{Grossi_2008} and \citet{Francis_Lewis_Linder_2008},
 where a modified expansion history consistent with
EDE models described by the parametrization of Eq.~\ref{w_ede_par} has been implemented in the widely used $N$-body code \textsc{Gadget-2} \citep{gadget-2} and 
the properties of nonlinear structures forming in these EDE cosmologies have been analyzed.
Both studies have shown that the standard formalism for the computation of the halo mass function still holds for EDE models at $z=0$. In other words, both the
standard fitting formulae for the number density of collapsed objects as a function of mass, and their key parameter $\delta_{c} = 1.686$ representing the linear
overdensity at collapse for a spherical density perturbation, remain unchanged also for EDE cosmologies.

The work of \citet{Grossi_2008}, however, 
investigated also the internal properties of collapsed halos in EDE models, finding a slight increase of halo concentrations due to the earlier onset of structure formation
and most importantly a significant increment of the line-of-sight velocity dispersion of massive halos. The latter effect could mimic a higher $\sigma_{8}$ normalization
for cluster mass estimates based on galaxy velocity dispersion measurements and, therefore, represents a potentially detectable signature of EDE models.

Besides determining a different expansion history with respect to the standard $\Lambda $CDM cosmology due to the presence of an EDE component, scalar-field DE cosmologies also predict the existence of spatial  perturbations of the DE density, resulting in a modification of the shape of the matter power spectrum. Even though such density perturbations are suppressed by free-streaming at sub-horizon scales (thereby allowing to discard the effect of DE fluctuations on the dynamical evolution of cosmic structures), they remain frozen to a constant value at super-horizon scales. Therefore, as new large scales continuously enter the causal horizon, they will be affected by the presence of DE perturbations before these are eventually damped by free-streaming. Consequently, DE perturbations are expected to slightly change the large-scale shape of the linear power spectrum, thereby affecting the initial conditions of structure formation \citep{Ma_etal_1999,Alimi_etal_2010}. This has motivated the development of DE N-body simulations with extremely large volumes, comparable or larger to the comoving size of the cosmic horizon, in order to investigate the nonlinear signatures of the large-scale DE perturbations \cite{Alimi_etal_2010,Rasera_etal_2010,Alimi_etal_2012}. Such studies have highlighted that the nonlinear regime of structure formation carries information on the initial conditions of the Universe and keeps memory  of the growth history of density perturbations even for the case of perfectly degenerate linear matter power spectra and $\sigma _{8}$ values. Therefore, nonlinear structure formation processes represent a precious source of information for the highly demanding requirements of precision cosmology.

\subsubsection{Interacting dark-energy models}
\label{cDE}

Another interesting class of non standard dark-energy models, as introduced in Section~\ref{mg:cde}, 
is given by coupled dark energy where a direct interaction
is present between a Quintessence scalar field $\phi$ and other cosmic components, in the 
form of a source term in the background continuity equations:
\begin{eqnarray}
\label{coupled_phi} \frac{\d\rho_{\phi}}{\d\eta} &=& -3 {\cal H} (1 + w_\phi) \rho_{\phi} +
\beta(\phi) \frac {\d\phi}{\d\eta} (1-3 w_{\alpha}) \rho_{\alpha} \,, \\
\label{cons_species} \frac{\d\rho_{\alpha}}{\d\eta} &=& -3 {\cal H} (1 + w_{\alpha}) \rho_{\alpha} -
\beta(\phi) \frac{\d\phi}{\d\eta} (1-3 w_{\alpha}) \rho_{\alpha} \,,
\end{eqnarray}
where $\alpha$ represents a single cosmic fluid coupled to $\phi$.

While such direct interaction with baryonic particles ($\alpha=b$) is tightly constrained
by observational bounds, and while it is suppressed for relativistic particles ($\alpha=r$) by symmetry reasons ($1-3w_{r}=0$), 
a selective interaction with cold dark matter (CDM hereafter) 
or with massive neutrinos is still observationally viable (see Section~\ref{mg:cde}).

Since the details of interacting dark-energy models have been discussed in Section~\ref{mg:cde}, here we simply recall the main features of these models
that have a direct relevance for nonlinear structure formation studies.
For the case of interacting dark energy, in fact, the situation is much more complicated than for the simple EDE scenario discussed above.
The mass of a coupled particle changes in time due to the energy exchange with the dark-energy scalar field $\phi $ according to the equation:
\begin{equation}
\label{mass_var}
m(\phi ) = m_{0}e^{-\int \beta (\phi ') \, \mathrm{d}\phi '}
\end{equation}
where $m_{0}$ is the mass at $z=0$.
Furthermore, the Newtonian acceleration of a coupled particle (subscript $c$) gets modified as:
\begin{equation}
\label{mod_accel}
\dot{\vec{v}}_{c} = -\tilde{H}\vec{v}_{c} - \vec{\nabla }\tilde{\Phi }_{c} - \vec{\nabla }\Phi_{nc} \,.
\end{equation}
where $\tilde{H}$ contains a new velocity-dependent acceleration:
\begin{equation}
\tilde{H}\vec{v}_{c} = H\left( 1-\beta_{\phi }\frac{\dot{\phi }}{H}\right) \vec{v}_{c}\,,
\end{equation}
and where a fifth-force acts only between coupled particles as
\begin{equation}
\label{phitilde}
\tilde{\Phi }_{c} = (1 + 2\beta ^{2})\Phi_{c}\,,
\end{equation}
while $\Phi_{nc}$ represents the gravitational potential due to all massive particles with no coupling to the dark energy that
exert a standard gravitational pull.

As a consequence of these new terms in the Newtonian acceleration equation the growth of density perturbations will
be affected, in interacting dark-energy models, not only by the different Hubble expansion due to the
dynamical nature of dark energy, but also by a direct modification of the effective gravitational interactions at subhorizon scales.
Therefore, linear perturbations of coupled species will grow with a higher rate in these cosmologies
In particular, for the case of a coupling to CDM, a different amplitude of the matter power spectrum
will be reached at $z=0$ with respect to $\Lambda $CDM if a normalization in accordance with CMB measurements at high redshifts is assumed.

Clearly, the new acceleration equation (\ref{mod_accel}) will have an influence also on the formation and evolution of nonlinear structures, and
a consistent implementation of all the above mentioned effects into an $N$-body algorithm is required in order to investigate this regime.

For the case of a coupling to CDM (a coupling with neutrinos will be discussed in the next section) 
this has been done, e.g., by \citet{maccio_etal_2004,Sutter:2007ky} with 1D or 3D grid-based field solvers, and more recently
by means of suitable modifications \citep[by][]{Baldi_etal_2010,Carlesi_etal_2014a} of the TreePM hydrodynamic $N$-body code \textsc{Gadget-2} \citep{gadget-2}, and similarly through a modified version \citep[by][]{Li_Barrow_2010} of the Adaptive Mesh Refinements code \textsc{Ramses} \citep{Teyssier_2001}.

Nonlinear evolution within coupled
quintessence cosmologies has been addressed using various methods of
investigation, such as spherical collapse
\citep{Mainini:2006zj,Wintergerst:2010ui,Manera:2005ct,Koivisto:2005nr,
Sutter:2007ky,Abdalla:2007rd,Bertolami:2007tq} and alternative semi-analytic
methods \citep{Saracco_etal_2010,amendola_quercellini_2004}. $N$-body and
hydro-simulations have also been done
\citep{maccio_etal_2004,Baldi_etal_2010,Baldi:2010vv,Baldi_Pettorino_2010,
Baldi:2010ks,Li:2010zw,Li:2010eu,Baldi:2010pq,Zhao:2010dz}.
We list here briefly the main observable features typical of this class of
models:

\begin{itemize}
\item The suppression of power at small scales in the power spectrum of interacting dark-energy models as compared to $\Lambda$CDM \citep[see e.g.][]{Baldi:2010pq};
\item An enhanced lensing power spectrum as compared to $\Lambda $CDM \citep[see e.g.][]{Beynon_etal_2012};
\item The development of a gravitational bias in the amplitude of density perturbations of uncoupled baryons and coupled CDM particles
defined as $P_{b}(k)/P_{c}(k)<1$, which determines
a significant decrease of the baryonic content of massive halos at low redshifts in accordance with a large number of observations \citep{Baldi_etal_2010,Baldi:2010pq};
\item The increase of the number density of high-mass objects at any redshift as compared to $\Lambda$CDM \citep[see e.g.][]{Baldi_Pettorino_2010,Baldi_2012,Cui_Baldi_Borgani_2012};
\item An enhanced ISW effect
\citep{Amendola:1999er,Amendola:2003wa,Mainini:2010ng}; such effects may be
partially reduced when taking into account nonlinearities, as described in
\citet{Pettorino:2010bv};
\item A modification in the shape of $z$-space distortions \citep[][]{Marulli:2011jk} and an enhanced pairwise infall velocity of colliding massive clusters \citep[][]{Lee_Baldi_2012};
\item A less steep inner core halo profiles (depending on the interplay between
fifth force and velocity-dependent terms)
\citep{Baldi_etal_2010,Baldi:2010vv,Li:2010zw,Li:2010eu,Baldi:2010pq};
\item A lower concentration of the halos
\citep{Baldi_etal_2010,Baldi:2010vv,Li:2010eu};
\item CDM voids are larger and more underdense when a coupling is active \citep{Baldi:2010ks,Sutter_etal_2014}.
\item A modified amplitude and time evolution of the halo bias \citep[see e.g.][]{Marulli:2011jk,Moresco_etal_2014}, which might determine a counterintuitive behaviour in the connection between CDM and halo populations statistics in the context of interacting Dark Energy cosmologies.
\end{itemize}

The analysis has been extended to the case of non-constant coupling functions $\beta (\phi )$ by \citet{Baldi:2010vv}.
As discussed in Section~\ref{Nbody_sims}, when a variable coupling $\beta (\phi )$ 
is active the relative balance of the fifth-force and other dynamical effects depends on
 the specific time evolution of the coupling strength. Under such conditions, some of the above mentioned
results no longer hold. In particular, the \textsc{CoDECS} simulations series \citep[][]{CoDECS} has provided evidence that for various combinations of coupling and self-interaction potential functions -- $V(\phi )$ and $\beta {\phi }$ -- some of the following effects might arise: 

\begin{itemize}
\item Small scale power can be both suppressed and enhanced when a growing coupling function
is considered, depending on the magnitude of the coupling time derivative $\mathrm{d}\beta (\phi )/\mathrm{d}\phi $
\item The inner overdensity of CDM halos, and consequently the halo concentrations, 
can both decrease (as always happens for the case of constant couplings) or increase, again depending on the rate of change of the 
coupling strength \citep[][]{Cui_Baldi_Borgani_2012};
\item The abundance of halo substructures \citep[see][]{Giocoli_etal_2013} as well as the CMB lensing power spectrum \citep[see][]{Carbone_etal_2013} might show both an enhancement or a suppression with respect to $\Lambda $CDM depending on the specific model under exam.
\end{itemize}

More recently, \citet{Carlesi_etal_2014a,Carlesi_etal_2014b} have employed a specific suite of high-resolution N-body simulations to investigate how some of the above mentioned effects depend on the cosmic environment in which CDM halos reside, showing that most of the characteristic observational features of interacting Dark Energy are significantly enhanced for halo populations residing in underdense regions of the Universe. Furthermore, \citet{Carlesi_etal_2014a} also highlighted a positive correlation between the average spin parameter of CDM halos and the coupling constant.

All these effects represent characteristic features of interacting dark-energy models and could provide a direct way to observationally test these scenarios.

A slightly more complex realisation of the interacting Dark Energy scenario has been recently proposed by \citet{Baldi_2012c}, and termed the ``Multi-coupled Dark Energy" model. The latter is characterised by two distinct CDM particle species featuring an opposite constant coupling to a single classical Dark Energy scalar field, and represents the simplest possible realisation of the general multiple-interaction scenario proposed by \citet{Gubser2004,Gubser_Peebles_2004,Brookfield:2007au}. The most noticeable feature of such model is the dynamical screening that effectively suppresses the coupling at the level of the background and linear perturbations evolution, although leaving room for a possible interesting phenomenology at nonlinear scales \citep[see e.g.][]{Piloyan_etal_2013,Piloyan_etal_2014}. Some first N-body simulations of the Multi-coupled Dark Energy scenario have been performed by \citet{Baldi_2013,Baldi_2014}, showing for the first time the halo fragmentation process occurring in these cosmologies as a consequence of the repulsive long-range fifth-force between CDM particles of different types. Higher resolution simulations will be required in order to investigate possible observable effects of this new phenomenon on the shape and abundance of CDM halos at very small scales.

 Alternatively, 
the coupling can be introduced by choosing directly a covariant stress-energy tensor, treating dark 
energy as a fluid in the absence of a starting action \citep{Mangano:2002gg,Valiviita:2008iv,CalderaCabral:2008bx,Schaefer:2008ku,Valiviita:2009nu,
Majerotto:2009np,Gavela:2009cy,CalderaCabral:2009ja,Gavela:2010tm}.

\subsubsection{Growing neutrinos}
\label{cnu}

In case of a coupling between the dark-energy scalar field $\phi $ and the relic fraction of massive neutrinos, all the above basic 
equations (\ref{mass_var})\,--\,(\ref{phitilde}) still hold. However, such models are found to be cosmologically viable only for large negative values of the 
coupling $\beta$ \citep[as shown by][]{Amendola2008b}, that according to Eq.~\ref{mass_var} determines 
a neutrino mass that grows in time (from which these models have been dubbed ``growing neutrinos'').
An exponential growth of the neutrino mass implies that cosmological bounds on the neutrino mass are 
no longer applicable
and that neutrinos remain relativistic much longer than in the standard scenario,
 which keeps them effectively uncoupled until
recent epochs, according to Eqs.~(\ref{coupled_phi} and \ref{cons_species}). However, as soon as neutrinos become non-relativistic at redshift $z_{\mathrm{nr}}$
due to the exponential
growth of their mass, the pressure terms $1-3w_{\nu }$ in
Eqs.~(\ref{coupled_phi} and \ref{cons_species}) no longer vanish
and the coupling with the DE scalar field $\phi $ becomes active.

Therefore, while before $z_{\mathrm{nr}}$ the model behaves as a standard $\Lambda$CDM scenario, after $z_{\mathrm{nr}}$ the non-relativistic
massive neutrinos obey the modified Newtonian equation~(\ref{mod_accel}) and a fast growth of neutrino density perturbation
takes place due to the strong fifth force described by Eq.~(\ref{phitilde}).

The growth of neutrino overdensities in the context of growing neutrinos models has been studied in the linear regime by
\citet{Mota:2008nj}, predicting the formation of very large neutrino lumps at the scale of superclusters and above (10\,--\,100~Mpc/h)
at redshift $z\approx 1$.

The analysis has been extended to the nonlinear regime in \citet{Wintergerst:2009fh} by following the spherical collapse of a neutrino
lump in the context of growing neutrino cosmologies. This study has witnessed the onset of virialization processes in the nonlinear evolution of
the neutrino halo at $z\approx 1.3$, and provided a first estimate of the associated gravitational potential at virialization being of the order of
$\Phi_{\nu }\approx 10^{-6}$ for a neutrino lump with radius $R \approx 15\mathrm{\ Mpc}$.

An estimate of the potential impact of such very large nonlinear structures onto the CMB angular power spectrum through 
the Integrated Sachs--Wolfe effect has been attempted by \citet{Pettorino:2010bv}. This study has shown that the linear
approximation fails in predicting the global impact of the model on CMB anisotropies at low multipoles,
and that the effects under consideration are very sensitive to the details of the transition between the linear and nonlinear regimes 
and of the virialization processes of nonlinear neutrino lumps, 
and that also significantly depend on possible backreaction effects of the evolved neutrino density field onto the local scalar filed evolution.

A full nonlinear treatment by means of specifically designed $N$-body simulations is, therefore, required 
in order to follow in further detail the evolution of a cosmological sample of neutrino lumps beyond virialization,
and to assess the impact of growing neutrinos models onto potentially observable quantities as the low-multipoles
CMB power spectrum or the statistical properties of CDM large scale structures. 
Simulations of the growing neutrino scenario have been performed for the first time by \citet{Baldi:2011es} by means of a modified version of the Gadget-2 code which assumed the linearity of the scalar field spatial perturbations (and consequently of the neutrino mass) and no backreaction of the growth of neutrino lumps on the overall background cosmic expansions. Although such approximations are quite restrictive, the simulations performed by \citet{Baldi:2011es} allowed to follow the evolution of the formation of a few large neutrino structures down to $z\sim 1$, after which neutrino particles start to become relativistic thereby breaking the Newtonian implementation of gravitational dynamics implemented in standard N-body algorithms. Such restrictions and approximations have been subsequently removed by the specific N-body algorithm developed by \citet{Ayaita_Weber_Wetterich_2012} which self-consistently implements both the relativistic evolution of neutrino particles and the backreaction effect on the background cosmology, and which employs a Newton-Gauss-Seidl relaxation scheme to solve for nonlinear spatial fluctuations of the scalar field. Nonetheless, even this more accurate numerical treatment has been so far successfully employed only down to $z\sim 1$ due to its high computational cost.

\subsubsection{Modified gravity}
\label{modgrav}

Modified gravity models, presented in Section~\ref{models-of-modified-gravity}, represent a different perspective to account for the nature of
the dark components of the universe. Although most of the viable modifications of GR are
constructed in order to provide an identical cosmic expansion history to the standard $\Lambda$CDM model,
their effects on the growth of density perturbations could lead to observationally testable predictions capable of
distinguishing modified gravity models from standard GR plus a cosmological constant.

Since a modification of the theory of gravity would affect all test masses in the universe, i.e., including the standard baryonic matter,
an asymptotic recovery of GR for solar system environments, where deviations from GR are tightly 
constrained, is required for all viable modified gravity models. Such ``screening mechanism''
represents the main difference between modified gravity models and the interacting dark-energy scenarios discussed above, by 
determining a local dependence of the modified gravitational laws in the Newtonian limit.
Different modifications of the GR Action integral  might feature different types of screening mechanisms (see Section~\ref{models-of-modified-gravity} for an introduction to modified gravity theories and screening mechanisms) -- as e.g. the ``Chameleon" \citep[][]{Khoury:2003rn}, the ``Symmetron" \citep[][]{Hinterbichler:2010es}, the ``Dilaton" \citep[][]{Damour_Polyakov_1994} or the ``Vainshtein" \citep[][]{Vainshtein:1972sx} mechanisms -- which in turn might require different numerical implementations in order to solve for the fully nonlinear evolution of the additional degrees of freedom associated to the modifications of gravity.

While the linear growth of density perturbations in the context of modified gravity theories can be studied 
\citep[see, e.g.,][]{Hu07,Motohashi_etal_2010,Amarzguioui:2005zq,Appleby:2010dx} by parametrizing
the scale dependence of the modified Poisson and Euler equations in Fourier space (see the discussion in Section~\ref{sec:dof}),
 the nonlinear 
evolution of the additional degrees of freedom of any viable modified gravity scenario makes the implementation of these theories into nonlinear $N$-body algorithms
much more challenging. 
Nonetheless, enormous progress has been made over the past few years in the development of specific N-body codes for various classes of modified gravity cosmologies, such that the investigation of nonlinear structure formation for (at least some) alternative gravitational theories by means of dedicated N-body simulations is becoming a mature field of investigation in computational cosmology.
The first simulations of modified gravity cosmologies, limited to the ``Chameleon" screening mechanism featured by $f(R)$ theories, have been performed by means of mesh-based iterative relaxation schemes 
\citet{Oyaizu_2008,Oyaizu_etal_2008,Schmidt_etal_2009,Khoury_Wyman_2009,Zhao_Li_Koyama_2010,Davis:2011pj,Winther:2011qb}
and showed an enhancement of the power spectrum amplitude at intermediate and small scales. 
These studies also showed that this nonlinear enhancement of small scale power cannot be accurately reproduced
by applying the linear perturbed equations of each specific modified gravity theory to the standard nonlinear
fitting formulae \citep[as, e.g.,][]{Smith2003}.

After these first pioneering studies, a very significant amount of work has been done in both extending the numerical implementation of modified gravity models within N-body algorithms and in using high-resolution N-body simulations to investigate the impact of various modifications of gravity on possible observable quantities. 

Concerning the former aspect, the main advancements have been obtained by extending the range of possible screening mechanisms implemented in the modified gravity nonlinear solvers, in order to include ``Symmetron" \citep[see e.g.][]{Davis:2011pj}, ``Dilaton" \citep[see e.g.][]{Brax_etal_2012}, and Vainshtein-like \citep[see e.g.][for the cases of general Vainshtein as well as for cubic and quartic Galileon models]{Li_Zhao_Koyama_2013,Barreira_etal_2013,Li_etal_2013} mechanisms, and by optimising the nonlinear Poisson solvers. Presently, three main parallel codes for modified gravity models have been developed by independent groups, based on different underlying N-body numerical schemes and procedures, namely the \textsc{ECOSMOG} \citep[][]{Ecosmog}, the \textsc{MG-GADGET} \citep[][]{Puchwein_Baldi_Springel_2013}, and the \textsc{ISIS} \citep[][]{Llinares_Mota_Winther_2013} codes. The latter has also recently provided the first implementation of Chameleon and Symmetron modified gravity theories beyond the quasi-static approximation \citep[][]{Llinares_Mota_2013,Llinares_Mota_2014}.

Concerning the latter aspect, a wide range of results about the impact of various modified gravity theories on several observable quantities have been obtained with the above-mentioned N-body codes, generally finding a good agreement between the different algorithms even though a properly controlled code-comparison study has yet to be performed. Among the most relevant results it is worth mentioning the identification of modified gravity signatures in the large-scale structure statistics \citep[see e.g.][]{Li_Zhao_Koyama_2011,Lombriser_etal_2013,Lombriser_Koyama_Li_2014,Arnold_Puchwein_Springel_2014}, in the environmental dependence of CDM halo properties \citep[][]{Winther:2011qb}, on the large-scale velocity field \citep[see e.g.][]{Li_etal_2012,Jennings_etal_2012,Hellwing_etal_2014}, and on the ISW effect \citep[see e.g.][]{Cai_etal_2014}. Furthermore, a recent study \citep[][]{Baldi_etal_2014} performed with the \textsc{MG-GADGET} code has highlighted the issue of a severe observational degeneracy between the effects of an $f(R)$ modification of gravity and a cosmological background of massive neutrinos 

Despite the huge advancements in the field of nonlinear simulations of modified gravity models achieved in recent years,

higher resolution simulations and new numerical approaches will be necessary in order to extend these results
to smaller scales and to accurately evaluate the deviations of specific models of modified gravity
from the standard GR predictions to a potentially detectable precision level.

\subsection{The spherical collapse model}
\label{the-spherical-collapse-model}

A popular analytical approach to study nonlinear clustering of dark matter
without recurring to $N$-body simulations is the spherical collapse model, first
studied by~\cite{Gunn-Gott:1972}. In this approach, one studies the collapse of
a spherical overdensity and determines its critical overdensity for collapse as
a function of redshift. Combining this information with the extended
Press--Schechter theory (\cite{Press1974a,Bond-etal:1991}; see
\cite{Zentner:2007} for a review) one can provide a statistical model for the
formation of structures which allows to predict the abundance of virialized
objects as a function of their mass. Although it fails to match the details of
$N$-body simulations, this simple model works surprisingly well and can give
useful insigths into the physics of structure formation. Improved models
accounting for the complexity of the collapse exist in the literature and offer
a better fit to numerical simulations. For instance, \cite{Sheth-Tormen:1999}
showed that a significant improvement can be obtained by considering an
ellipsoidal collapse model. Furthermore, recent theoretical developments and new
improvements in the excursion set theory have been undertaken by
\cite{Maggiore-Riotto:2010} and other authors (see e.g., \cite{Shaw:2007tr}).

The spherical collapse model has been generalized to include a cosmological
constant by~\cite{Peebles:1984,Weinberg:1987}. \cite{Lahav-etal:1991} have used
it to study the observational consequences of a cosmological constant on the
growth of perturbations. The case of standard quintessence, with speed of sound
$c_s = 1$, have been studied by~\cite{wang98}.  In this case,
scalar fluctuations propagate at the speed of light and sound waves maintain
quintessence homogeneous on scales smaller than the horizon scale. In the
spherical collapse pressure gradients maintain the same energy density of
quintessence between the inner and outer part of the spherical overdensity, so
that the evolution of the overdensity radius is described by
\begin{equation}
\label{uc_sp}
\frac{\ddot R}{R} = -\frac{4 \pi G}{3} (\rho_m + \bar \rho_Q + 3 \bar p_Q)\,,
\end{equation}
where $\rho_m$ denotes the energy density of dark matter while $\bar \rho_Q$ and
$\bar p_Q$ denote the homogeneous energy density and pressure of the
quintessence field.
Note that,
although this equation looks like one of the Friedmann equations, the dynamics
of $R$ is not the same
as for a FLRW universe. Indeed, $\rho_m$ evolves following the scale factor $R$,
while the quintessence follows the external scale factor $a$, according to the
continuity equation $\dot{\bar \rho}_Q + 3 (\dot{a}/a) (\bar \rho_Q + \bar p_Q)
=0$. 

In the following we will discuss the spherical collapse model in the contest of
other dark energy and modified gravity models.

\subsubsection{Clustering dark energy}

In its standard version, quintessence is described by a minimally-coupled
canonical field, with speed of sound $c_s=1$. As mentioned above, in this case
clustering can only take place on scales larger than the horizon, where sound
waves have no time to propagate. However, observations on such large scales are
strongly limited by cosmic variance and this effect is difficult to observe. A
minimally-coupled scalar field with fluctuations characterized by a practically
zero speed of sound can cluster on all observable scales. There are several
theoretical motivations to consider this case. In the limit of zero sound speed
one recovers the Ghost Condensate theory proposed
by~\cite{Arkani-Hamed-etal:2004} in the context of modification of
gravity, which is invariant under shift symmetry of the field $\phi
\to \phi + {\mathrm{constant}}$. Thus, there is no fine tuning in assuming
that the speed of sound is very small: quintessence models with
vanishing speed of sound should be thought of as deformations of this
particular limit where shift symmetry is recovered. Moreover, it has
been shown that minimally-coupled quintessence with an equation of
state $ w<  -1$ can be free from ghosts and gradient instabilities
only if the speed of sound is very tiny, $| c_s |  \lesssim
10^{-15}$. Stability can be guaranteed by the presence of higher
derivative operators, although their effect is absent on
cosmologically relevant scales \cite{Creminelli-etal:2006,Cheung-etal:2008,Creminelli-etal:2009}.

The fact that the speed of sound of quintessence may vanish opens up new
observational consequences. Indeed, the absence of quintessence pressure
gradients allows instabilities to develop on all scales, also on scales where
dark matter perturbations become nonlinear. Thus, we expect quintessence to
modify the growth history of dark matter not only through its different
background evolution but also by actively participating to the structure
formation mechanism, in the linear and nonlinear regime, and by contributing to
the total mass of virialized halos.

Following~\cite{Creminelli-etal:2010}, in the limit of zero sound speed pressure
gradients are negligible and, as long as the fluid approximation is valid,
quintessence follows geodesics remaining comoving with the dark matter (see also
\cite{Lim-Sawicki-Vikman:2010} for a more recent model with identical
phenomenology). In particular, one can study the effect of quintessence with
vanishing sound speed on the structure formation in the nonlinear regime, in the
context of the spherical collapse model. The zero speed of sound limit
represents the natural counterpart of the opposite case $c_s = 1$. Indeed, in
both cases there are no characteristic length scales associated with the
quintessence clustering and the spherical collapse remains independent of the
size of the object  (see \cite{EggersBjaelde-Wong:2010,Mota:2004pa, Nunes:2004wn} for a study of the
spherical collapse when $c_s$ of quintessence is small but finite).

Due to the absence of pressure gradients quintessence follows dark matter in the collapse and the evolution of the overdensity radius is described by
\begin{equation}
\frac{\ddot R}{R} = -\frac{4 \pi G}{3} (\rho_m + \rho_Q + \bar p_Q) \,,
\end{equation}
where the energy density of quintessence $\rho_Q$ has now a different value
inside and outside the overdensity, while the pressure remains unperturbed. In
this case 
the quintessence inside the overdensity evolves following the internal scale
factor $R$, $\dot{\rho}_Q + 3 (\dot{R}/R) (\rho_Q + \bar p_Q) =0$ and
the comoving regions behave as closed FLRW universes. $R$ satisfies the
Friedmann equation and the spherical collapse can be solved exactly
\cite{Creminelli-etal:2010}. 

\epubtkImage{qratio0-qratio1.png}{%
\begin{figure}[htbp]
\centerline{\includegraphics[scale=1]{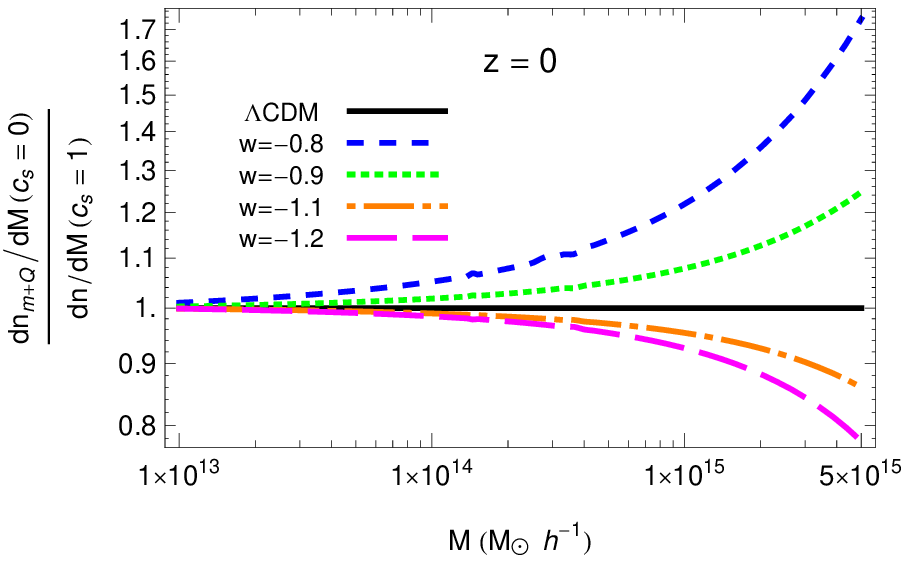}}
\vspace{0.5cm}
\centerline{\includegraphics[scale=1]{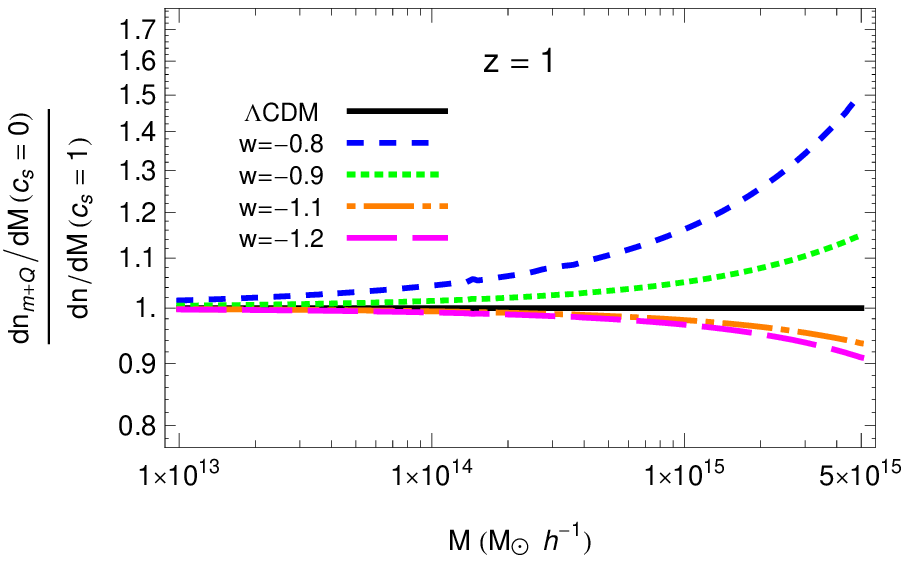}}
\caption{Ratio of the total mass functions, which include the
  quintessence contribution, for $c_s=0$ and $c_s=1$ at $z=0$ (above)
  and $z=1$ (below). Image reproduced by permission
  from~\cite{Creminelli-etal:2010}; copyright by IOP and
  SISSA.}
\label{fig:qmassfunctionratio}
\end{figure}}

Quintessence with zero speed of sound modifies dark matter clustering with
respect to the smooth quintessence case through the linear growth function and
the linear threshold for collapse. Indeed, for $w >-1$ ($w < -1$), it enhances
(diminishes) the clustering of dark matter, the effect being proportional to
$1+w$.
The modifications to the critical threshold of collapse are small and the
effects on the dark matter mass function are dominated by the modification on
the linear dark matter growth function. Besides these conventional effects there
is a more important and qualitatively new phenomenon: quintessence mass adds to
the one of dark matter, contributing to the halo mass by a fraction of order
$\sim (1 + w) \Omega_Q/ \Omega_m$. Importantly, it is possible to show that the
mass associated with quintessence stays constant inside the virialized object,
independently of the details of virialization. Moreover ,the ratio between the
virialization and the turn-around radii is approximately the same as the one for
$\Lambda$CDM computed by~\cite{Lahav-etal:1991}.
In Figure~\ref{fig:qmassfunctionratio} we plot the ratio of the mass function
including the quintessence mass contribution, for the $c_s=0$ case to the smooth
$c_s=1$ case. 
The sum of the two effects is rather large: for values of $w$ still compatible
with the present data and for large masses the difference between the
predictions of the $c_s = 0$ and the $c_s = 1$ cases is of order one.

\subsubsection{Coupled dark energy}

We now consider spherical collapse within coupled dark-energy cosmologies. 
The presence of an interaction that couples the cosmon dynamics to another species
introduces a new force acting between particles (CDM or neutrinos in the examples
mentioned in Section~\ref{mg:cde}) and mediated by dark-energy fluctuations. Whenever such a coupling is active, spherical collapse,
whose concept is intrinsically based on gravitational attraction via the Friedmann equations, has to be suitably modified in order to account for other external forces. As shown in \cite{Wintergerst:2010ui} the inclusion of the fifth force within the spherical collapse picture deserves particular caution. Here we summarize the main results on this topic and
we refer to \cite{Wintergerst:2010ui} for a detailed illustration of spherical collapse in presence of a fifth force. 

If CDM is coupled to a quintessence scalar field as described in Sections~\ref{mg:cde} and \ref{dms:de_dm} of the present document, the full nonlinear evolution equations within the Newtonian limit read:
\begin{eqnarray}
\label{eq:sph_com_ns1}
\dot{\delta}_m &=& -\mathbf{v}_m\,\nabla\delta_m - (1 + \delta_m)\,\nabla\cdot\mathbf{v}_m \\
\dot{\mathbf{v}}_m &=& -(2{\bar H} - \beta\,\dot{\bar\phi})\,\mathbf{v}_m - (\mathbf{v}_m\,\nabla)\mathbf{v}_m \nonumber \\
\label{eq:sph_com_ns2} && - a^{-2}\,\nabla(\Phi - \beta\,\delta\phi) \\
\label{eq:sph_com_poisson} \Delta\delta\phi &=& -\beta\,a^2\,\delta\rho_m \\
\label{eq:sph_com_grav_pot} \Delta\Phi &=& -\frac{a^2}{2}\,\sum_{\alpha} \delta\rho_{\alpha}
\end{eqnarray}
These equations can be derived from the non-relativistic Navier--Stokes equations and from the Bianchi identities written in presence of an external source of the type:
\begin{equation}
\label{eq:ps_cons}\nabla_{\gamma}T_{\mu}^{\gamma} = Q_{\mu} = -\beta T_{\gamma}^{\gamma}  \partial_{\mu}\phi \,,
\end{equation}
where $T^{\gamma}_{\mu}$ is the stress energy tensor of the dark matter fluid and we are using comoving spatial coordinates $\mathbf{x}$ and cosmic time $t$. Note that $\mathbf{v}_m$ is the comoving velocity, related to the peculiar velocities by $\mathbf{v}_m = \mathbf{v}_{pec}/a$.  They are valid for arbitrary quintessence potentials as long as the scalar field is sufficiently light, i.e., $m_\phi^2 \delta\phi = V''(\phi)\delta\phi \ll \Delta\delta\phi$ for the scales under consideration. For a more detailed discussion see \cite{Wintergerst:2010ui}.
Combining the above equations yields to the following expression for the evolution of the matter perturbation $\delta_m$:
\begin{equation}
\label{eq:sph_gen_del}
\ddot\delta_m = -(2{\bar H}-\beta\,\dot{\bar\phi})\,\dot{\delta}_m 
  + \frac{4}{3}\frac{\dot{\delta}_m^2}{1 + \delta_m} + \frac{1 +
    \delta_m}{a^2} \,\Delta\Phi_{\text{eff}} \,,
\end{equation} 
Linearization leads to: 
\begin{equation}
\label{eq:sph_gen_del_lin} \ddot\delta_{m,L} = -(2{\bar H}-\beta\,\dot{\bar\phi})\,\dot{\delta}_{m,L} + a^{-2}\,\Delta\Phi_{\text{eff}} \,.
\end{equation}
where the effective gravitational potential follows the modified Poisson equation:
\begin{equation}
\Delta {\Phi_{\text{eff}}} = -\frac{a^2}{2} {\bar\rho}_m \delta_m \left(1+2 \beta^2\right) \,.
\end{equation}
Eqs.~(\ref{eq:sph_gen_del}) and (\ref{eq:sph_gen_del_lin}) are the two main equations which correctly describe the nonlinear and linear evolution for a coupled dark-energy model. 
They can be used, among other things,
 for estimating the extrapolated linear density contrast at collapse $\delta_c$ in the presence of
a fifth force.
It is possible to reformulate Eqs.~(\ref{eq:sph_gen_del}) and (\ref{eq:sph_gen_del_lin}) into 
an effective spherical collapse: 
\begin{equation}
\label{eq:imp_sc_f2b} \frac{\ddot{R}}{R} = -\beta\,\dot{\phi}\left( H - \frac{\dot{R}}{R}\right) - \frac{1}{6} \sum_{\alpha} \left[{\rho}_{\alpha}(1 + 3 { w}_{\alpha})\right] 
 - \frac{1}{3}\,\beta^2\,\delta\rho_m \,.
\end{equation}
Eq.~(\ref{eq:imp_sc_f2b}) \cite{Mainini:2006zj, Wintergerst:2010ui}, describes the general evolution of
the radius of a spherical overdense region within coupled quintessence. 
Comparing with the standard case (\ref{uc_sp}) we notice the presence of two additional terms: a `friction' term 
and the coupling term $\beta^2\,\delta\rho_m$, the latter being responsible for the additional attractive fifth force.
Note that the 'friction' term is actually velocity dependent and its effects on collapse depend, more realistically, on the
direction of the velocity, information which is not contained within a \emph{spherical} collapse picture and can be treated within simulations \cite{Baldi_Pettorino_2010,Li:2010zw, Baldi:2010vv, Li:2010eu, Baldi:2010pq}.
We stress that it is crucial to include these additional terms in the equations, as derived from the nonlinear equations, in order to correctly account for the presence of a fifth force.
The outlined procedure can easily be generalized to include uncoupled components, for example baryons. 
In this case, the corresponding evolution equation for $\delta_b$, will be fed by $\Phi_{\text{eff}} = \Phi$. 
This yields an evolution equation for the uncoupled scale factor $R_{uc}$ that is equivalent to the standard Friedmann equation.
In Figure~\ref{fig:delta_c_ccdm} we show the linear density contrast at collapse $\delta_c(z_c)$ for three coupled quintessence
 models with $\alpha = 0.1$ and $\beta = 0.05$, $0.1$, $0.15$.

\epubtkImage{delta_c_ccdm.png}{%
\begin{figure}[htbp]
\centerline{\includegraphics[width=85mm,angle=0.]{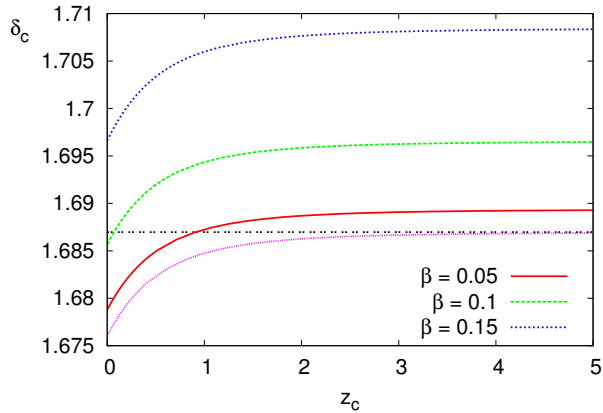}}
\caption{Extrapolated linear density contrast at collapse for coupled
  quintessence models with different coupling strength $\beta$. For
  all plots we use a constant $\alpha = 0.1$. We also depict
  $\delta_c$ for reference $\Lambda$CDM (dotted, pink) and EdS
  (double-dashed, black) models. Image reproduced by permission
  from~\cite{Wintergerst:2010ui}; copyright by APS.}
\label{fig:delta_c_ccdm}
\end{figure}}

An increase of $\beta$ results in an increase of $\delta_c$.
As shown in \cite{Wintergerst:2010ui}, $\delta_c(\beta)$ is well described by a simple quadratic fitting formula,
\begin{equation}
\delta_c(\beta) = 1.686(1 + a\beta^2)\,,a = 0.556 \,,
\end{equation}
valid for small $\beta \lesssim 0.4$ and $z_c \geq 5$. We recall that a nonlinear analysis beyond the spherical collapse method can be addressed by means of the 
time-renormalization-group method, extended to the case of couple quintessence in \cite{Saracco_etal_2010}.

If a coupling between dark energy and neutrinos is present, as described in Sections~\ref{mg:cde} and \ref{dms:de_nu}, bound neutrino structures may form within these models \cite{Brouzakis:2007aq}. It was shown in \cite{Mota:2008nj} that their formation will only start after neutrinos become non-relativistic. A nonlinear treatment of the evolution of neutrino densities is thus only required for very late times, and one may safely neglect neutrino pressure as compared to their density. The evolution equations (\ref{eq:sph_gen_del}) and (\ref{eq:sph_gen_del_lin}) can then also be applied for the nonlinear and linear neutrino density contrast.
The extrapolated linear density at collapse $\delta_c$ for growing neutrino quintessence reflects in all respects 
the characteristic features of this model and results in a $\delta_c$ which looks quite different from standard dark-energy cosmologies.
We have plotted the dependence of $\delta_c$ on the collapse redshift $z_c$ in Figure~\ref{fig:delta_c_cnu} for three values of the coupling. 
The oscillations seen are the result of the oscillations of the neutrino mass caused by the coupling to the scalar field: the latter 
has characteristic oscillations as it approaches the minimum of the effective potential in which it rolls, given by a combination of
the self-interaction potential $U(\phi)$ and the coupling contribution $\beta(1-3{ w}_\nu){\rho}_\nu$. 
Furthermore, due to the strong coupling $\beta$, the average value of $\delta_c$ is found to be substantially higher than 1.686, corresponding to the Einstein de Sitter value, shown in black (double-dashed) in Figure~\ref{fig:delta_c_cnu}.
Such an effect can have a strong impact on structure formation and on CMB \cite{Pettorino:2010bv}.
For the strongly coupled models, corresponding to a low present day neutrino mass $m_\nu(t_0)$, the critical density at collapse is only available for $z_c \lesssim 0.2$, $1$ for $\beta = -560$, $-112$, respectively. This is again a reflection of the late transition to the non-relativistic regime. 
Nonlinear investigations of single lumps beyond the spherical collapse picture was performed in \cite{Wintergerst:2009fh, Brouzakis:2010md}, the latter showing the influence of the gravitational potentials induced by the neutrino inhomogeneities on the acoustic oscillations in the baryonic and dark-matter spectra.

\epubtkImage{delta_c_cnu.png}{%
\begin{figure}[htb]
\centerline{\includegraphics[width=85mm,angle=0.]{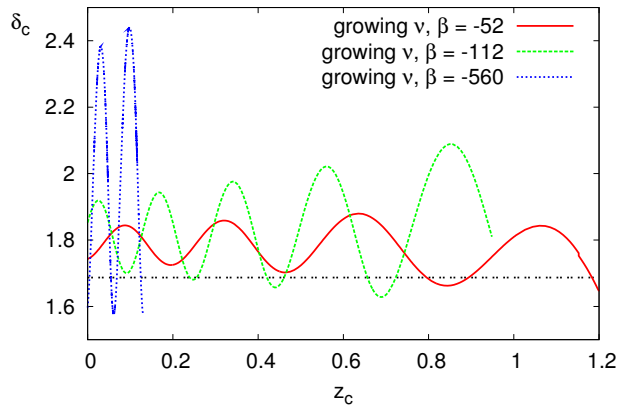}}
\caption{Extrapolated linear density contrast at collapse $\delta_c$
  vs.\ collapse redshift $z_c$ for growing neutrinos with $\beta =
  -52$ (solid, red), $\beta = -112$ (long-dashed, green) and $\beta =
  -560$ (short-dashed, blue). A reference EdS model
  (double-dashed. black) is also shown. Image reproduced by
  permission from~\cite{Wintergerst:2010ui}; copyright by APS.}
\label{fig:delta_c_cnu}
\end{figure}}

\subsubsection{Early dark energy}
\label{sec:early_results}

A convenient way to parametrize the presence of a nonnegligible homogeneous dark-energy component at early times was presented in~\cite{Wetterich:2004pv} and has been illustrated in Section~\ref{parametrization-of-the-background-evolution} of the present review. 
If we specify the spherical collapse equations for this case, the nonlinear evolution of the density contrast follows the evolution equations~(\ref{eq:sph_gen_del}) and (\ref{eq:sph_gen_del_lin}) without the terms related to the coupling. As before, we assume relativistic components to remain homogeneous.
In Figure~\ref{fig:delta_c_ede} we show $\delta_c$ for two models of
early dark energy, namely model I and II, corresponding to the choices
($\Omega_{m,0} = 0.332\,, \quad w_0 = -0.93\,, \quad
\Omega_{\text{DE},e}  = 2\cdot10^{-4}$) and ($\Omega_{m,0} = 0.314\,,
\quad w_0 = -0.99\,, \quad \Omega_{\text{DE},e} = 8\cdot10^{-4}$) respectively.
Results show $\delta_c(z_c = 5) \sim 1.685$ ($\sim 5\cdot10^{-2}\%$)
\cite{Francis:2008ka, Wintergerst:2010ui}.

\epubtkImage{delta_c_ede.png}{%
\begin{figure}[htbp]
\centerline{\includegraphics[width=85mm,angle=0.]{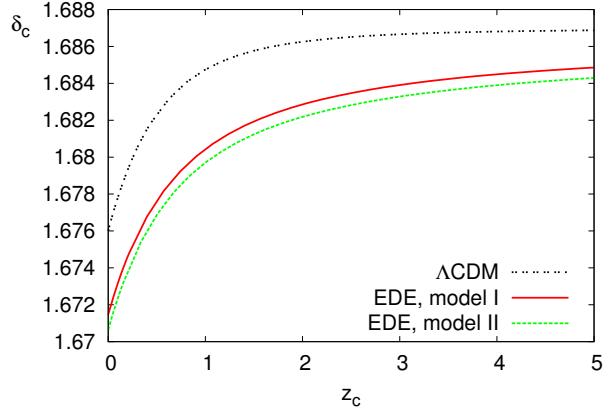}}
\caption{Extrapolated linear density contrast at collapse $\delta_c$
  vs.\ collapse redshift $z_c$ for EDE models I (solid, red) and II
  (long-dashed, green), as well as $\Lambda$CDM (double-dashed,
  black). Image reproduced by permission
  from~\cite{Wintergerst:2010ui}; copyright by APS.}
\label{fig:delta_c_ede}
\end{figure}}

\subsubsection{Universal couplings}
\label{sec:early_results}
In \cite{Kopp13} the authors compute the critical density of collapse for spherically symmetric overdensities in a class of $f(R)$ modified gravity models. They evolve the  Einstein, scalar field and non-linear fluid equations,under the assumptions that system remains quasi-static throughout the collapse. The result of this analysis is a fitting function for the spherical collapse $\delta_{c}$ as a function of collapse redshift, mass of the overdensity and $f_{\rm R0}$.

\begin{align} \label{deltacfit}
\delta_c(z,M,f_{\rm R0})&= \nonumber
\delta^{\Lambda}_c(z) \Bigg\{ 1+ b_2 (1+z)^{-a_3} \left( m_{b} -\sqrt{m_{b}^2+1}\right)+\\
& \qquad \qquad \ + b_3(\tanh m_{b}-1) \Bigg\}\\
m_{b}(z,M,f_{\rm R0})&=(1+z)^{a_3}\left(\log_{10} [M/(M_\odot h^{-1})]-m_1(1+z)^{-a_4}\right) \nonumber\\
m_1(f_{\rm R0}) &=  1.99 \log_{10}f_{\rm R0}+26.21 \nonumber\\
b_2 &= 0.0166\nonumber\\
b_3 (f_{\rm R0}) &=0.0027 \cdot (2.41-\log_{10}f_{\rm R0} ) \nonumber\\
a_3(f_{\rm R0}) &= 1 + 0.99 \exp\left[-2.08 (\log_{10}f_{\rm R0} + 5.57)^2\right]\nonumber\\
a_4(f_{\rm R0}) &= \left(\tanh\left[0.69\cdot (\log_{10}f_{\rm R0} + 6.65) \right] + 1\right) 0.11\nonumber
\end{align}

\begin{figure*}[p!] 
\centering
\includegraphics[width=0.99 \textwidth]{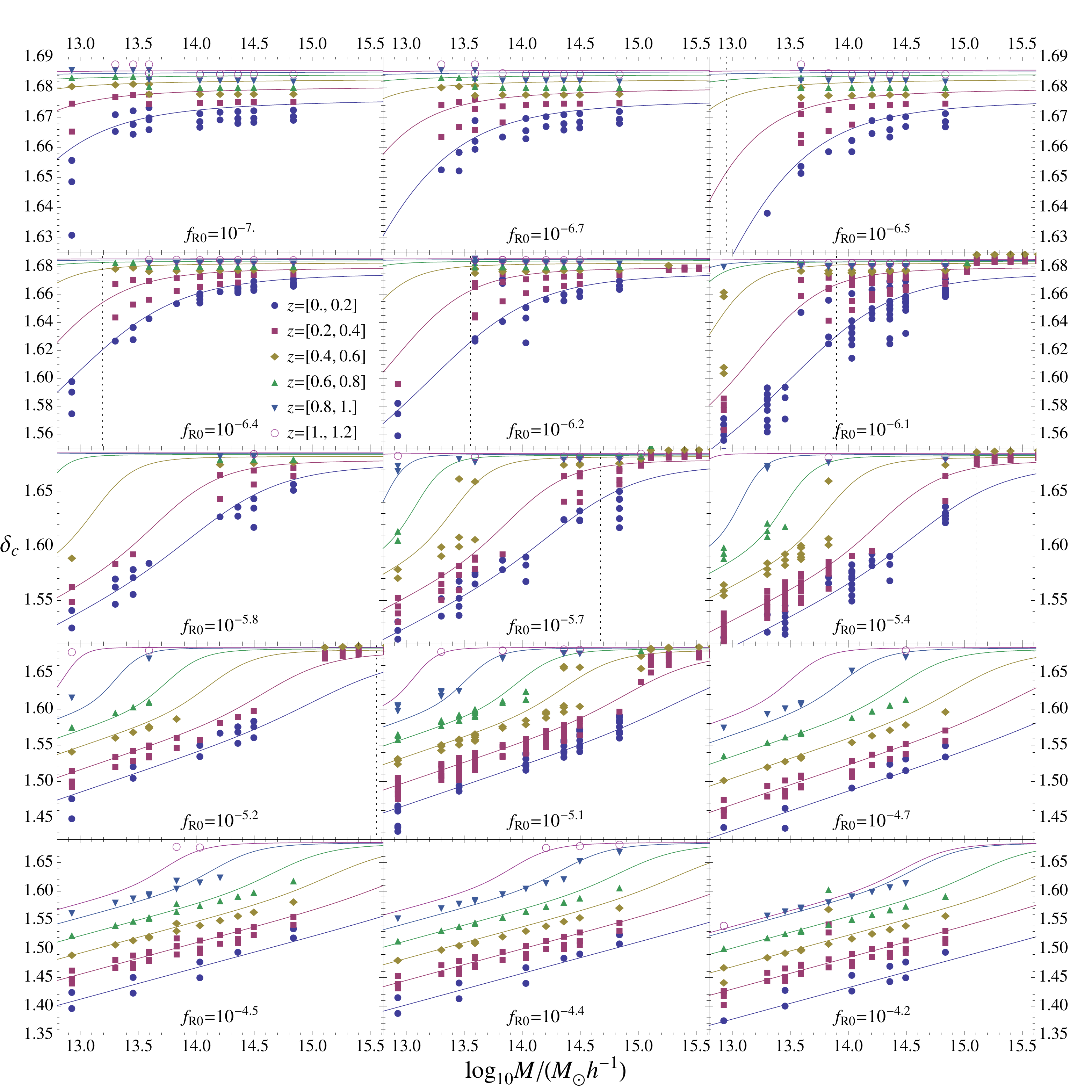}
\caption{$\delta_{c}$ as a function of mass. In each panel we show the results from the numerical analysis (points) and from the fitting function (lines). Figure taken from \cite{Kopp13} \vspace{5cm}}
\label{Fig624-1}
\end{figure*}

The results of the numerical collapse simulation and the fitting function are shown in Figure \ref{Fig624-1}.

In \cite{Kopp13, Achitouv15} the authors extend $\delta_{c}$  into drifting and diffusing barrier within the context of excursion set theory. With this procedure they obtain an `analytical' mass function for $f(R)$ models. The analytic formula for the halo mass function is tested against Monte Carlo random walks for a wide class of moving barriers and can therefore be applied to other modified gravity theories. In addition the results are compared to the results from N-body simulations obtained by the method described in \cite{Puchwein13}

\begin{figure}[ht]
\centering
\begin{tabular}{cc}
\includegraphics[scale=0.35]{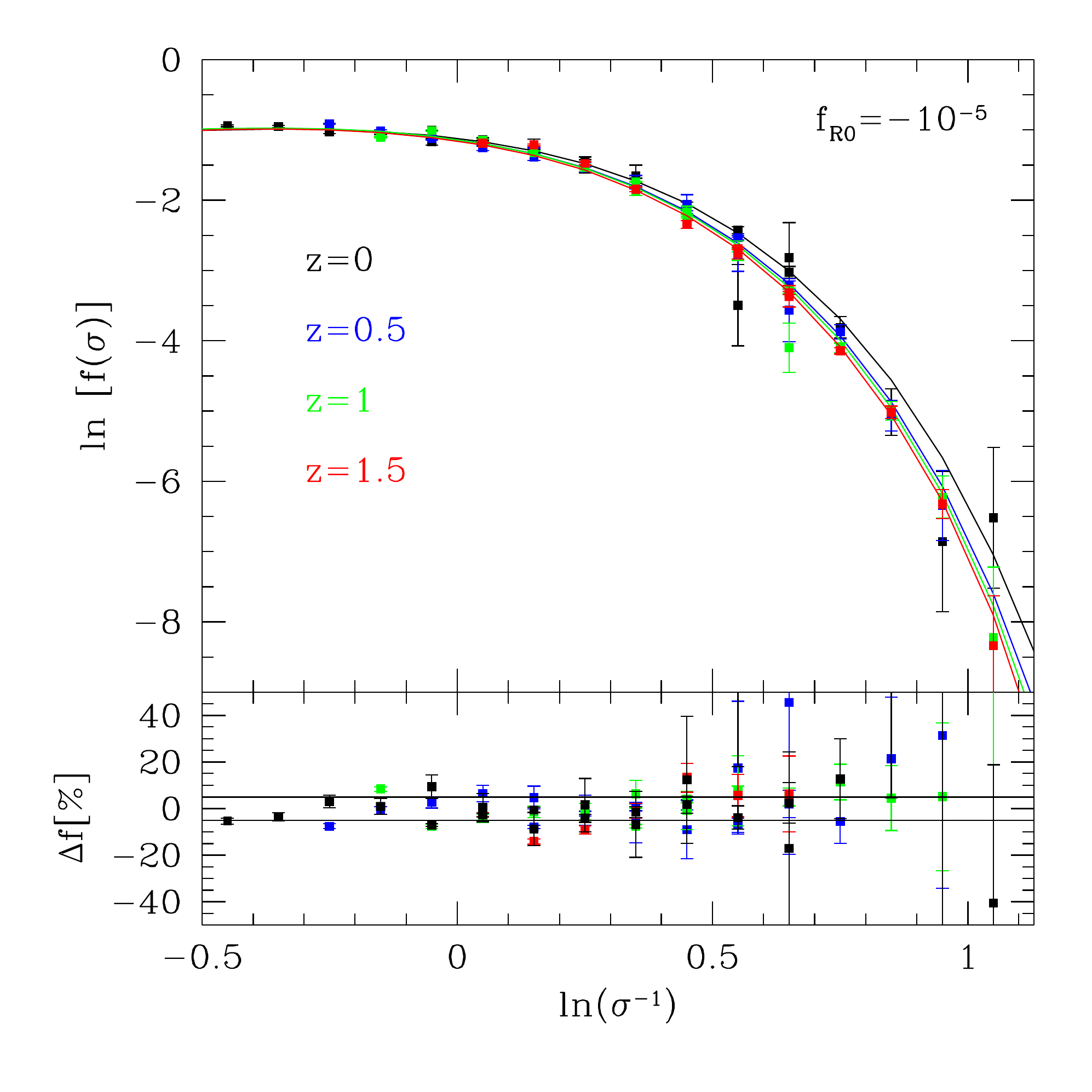}\\
\end{tabular}
\caption{Mass function for $f_{\rm R0}=-10^{-5}$  at redshifts $z=0-1.5$. The solid lines are theoretical predictions, squares with errorbars are from simulations. The lower panel shows the relative difference. The black solid lines show $\pm5\%$ differences. Figure taken from \cite{Achitouv15}}\label{Fig624-2}
\end{figure}

In Figure \ref{Fig624-2} we show an example of the derived mass function computed for $f(R_0)=10^{-5}$ for different redshits (solid lines). The comparison to the simulations (points), shows good agreement, but not high precision agreement as would be required for a detailed cosmological data analysis using the mass function.

\section{Observational properties of dark energy and modified gravity}
\label{observational-properties-of-modified-gravity}

Both scalar field dark-energy models and modifications of gravity can in
principle lead to any desired expansion history $H(z)$, or equivalently any
evolution of the effective dark-energy equation of state parameter $w(z)$. For
canonical scalar fields, this can be achieved by selecting the appropriate
potential $V(\varphi)$ along the evolution of the scalar field $\varphi(t)$, as
was done, e.g., in~\cite{Bassett:2002qu}. For modified gravity models, the same
procedure can be followed for example for $f(R)$ type models
\citep[e.g.][]{Pogosian:2007sw}. The evolution history on its own can thus not
tell us very much about the physical nature of the mechanism behind the
accelerated expansion (although of course a clear measurement showing that $w
\neq -1$ would be a sensational discovery). A smoking gun for modifications of
gravity can thus only appear at perturbation level.

In the next subsections we explore how dark energy or modified gravity
effects can be detected through weak lensing and redshift surveys.

\subsection{General remarks}

Quite generally, cosmological observations fall into two categories: geometrical
probes and structure formation probes. While the former provide a measurement of
the Hubble function, the latter are a test of the gravitational theory in an
almost Newtonian limit on subhorizon scales. Furthermore, possible effects on
the geodesics of test particles need to be derived: naturally, photons follow
null-geodesics while massive particles, which constitute the cosmic large-scale
structure, move along geodesics for non-relativistic particles.

In some special cases, modified gravity models predict a strong deviation from the standard
Friedmann equation as in, e.g., DGP, (\ref{eq:Friedmann_DGP}).
While the Friedmann equation is not known explicitly in more general models of
massive gravity (cascading gravity or hard mass gravity), similar modifications
are expected to arise and provide characteristic features, \citep[see, e.g.,][]{Afshordi:2008rd,Jain:2010ka}) that could distinguish these models
from other scenarios of modified gravity or with additional dynamical degrees of
freedom. 

In general however the most interesting signatures of modified gravity
models are to be found in the perturbation sector. For instance, in DGP, growth functions 
differ from those in dark-energy models by a few percent for identical Hubble
functions, and for that reason, an observation of both the Hubble and the growth
function gives a handle on constraining the gravitational theory,
\citep{Lue:2004rj}. The growth function can be estimated both through weak lensing
and through galaxy clustering and redshift distortions.

Concerning the interactions of light with the cosmic large-scale structure, one
sees a modified coupling in general models and a difference between the metric potentials.
These effects are present in the anisotropy pattern of the CMB, as shown in~\cite{Sawicki:2005cc}, where smaller fluctuations were found on large angular
scales, which can possibly alleviate the tension between the CMB and the
$\Lambda$CDM model on small multipoles where the CMB spectrum acquires smaller
amplitudes due to the ISW-effect on the last-scattering surface, but provides a
worse fit to supernova data. An interesting effect inexplicable in GR is the
anticorrelation between the CMB temperature and the density of galaxies at high
redshift due to a sign change in the integrated Sachs--Wolfe effect.
Interestingly, this behavior is very common in modified gravity theories.

A very powerful probe of structure growth is of course weak lensing, but to evaluate the
lensing effect it is important to understand the nonlinear structure formation
dynamics as a good part of the total signal is generated by small structures.
Only recently has it been possible to perform structure formation simulations
in modified gravity models, although still without a mechanism in which GR is recovered on
very small scales, necessary to be in accordance with local tests of gravity.

In contrast, the number density of collapsed objects relies only little on
nonlinear physics and can be used to investigate modified gravity cosmologies.
One needs to solve the dynamical equations for a spherically symmetric matter
distribution. Modified gravity theories show the feature of lowering the
collapse threshold for density fluctuations in the large-scale structure,
leading to a higher comoving number density of galaxies and clusters of
galaxies.
 This probe is degenerate with respect to dark-energy cosmologies,
which generically give the same trends.

Finally, supernova observations -- able of accurately mapping the expansion history of the universe -- are themselves lensed by foreground matter structures. This extra spread in the Hubble diagram caused by lensing contains precious clustering information, which is encoded in the one-point lensing PDF and can be used to constrain parameters such as the power spectrum normalization $\sigma_8$ or the growth index $\gamma$.
Therefore, forthcoming supernova catalogs can be seen as both geometrical and structure formation probes. It is important to point out that the one-point statistics is independent of and complementary to the methods based on cosmic shear and cluster abundance observables. See \cite{Amendola:2013twa,Quartin:2013moa} for more details and references therein.

\subsection{Observing modified gravity with weak lensing}

The magnification matrix is a $2\times2$ matrix that relates the true
shape of a galaxy to its image. It contains two distinct parts: the
convergence, defined as the trace of the matrix, modifies the size of
the image, whereas the shear, defined as the symmetric traceless part,
distorts the shape of the image. At small scales the shear and the
convergence are not independent. They satisfy a consistency relation,
and they contain therefore the same information on matter density
perturbations. More precisely, the shear and the convergence are both
related to the sum of the two Bardeen potentials, $\Phi+\Psi$,
integrated along the photon trajectory. At large scales however, this
consistency relation does not hold anymore. Various relativistic
effects contribute to the convergence, see \cite{Bonvin:2008ni}. Some
of these effects are generated along the photon trajectory, whereas
others are due to the perturbations of the galaxies redshift. These
relativistic effects provide independent information on the two
Bardeen potentials, breaking their degeneracy. The convergence is
therefore a useful quantity that can increase the discriminatory power
of weak lensing.

The convergence can be measured through its effect on the galaxy number
density, see e.g.~\cite{Broadhurst:1994qu}. The standard method extracts the
magnification from correlations of distant quasars with foreground clusters,
see~\cite{Scranton:2005ci, Menard:2009yb}. Recently, \cite{Zhang:2005pu,
Zhang:2005eb} designed a new method that permits to accurately measure
auto-correlations of the magnification, as a function of the galaxies redshift.
This method potentially allows measurements of the relativistic effects in the
convergence.

\subsubsection{Magnification matrix}

We are interested in computing the magnification matrix $\mathcal{D}_{ab}$ in a perturbed Friedmann
universe. The magnification matrix relates the true shape of a galaxy to its image, and describes
therefore the deformations encountered by a light bundle along its trajectory.
$\mathcal{D}_{ab}$ can be computed by solving Sachs equation, 
see \cite{Sachs:1961zz}, that governs propagation of light in a generic geometry.
The convergence $\kappa$ and the shear $\gamma\equiv \gamma_1+i \gamma_2$
are then defined respectively as the trace and the symmetric traceless part
of $\mathcal{D}_{ab}$
\begin{equation}
\mathcal{D}_{ab}=\frac{\chi_S}{1+z_S}\left(\begin{array}{cc}
1-\kappa-\gamma_1&-\gamma_2\\
-\gamma_2&1-\kappa+\gamma_1 \end{array}\right) . \label{eq:kappadef}
\end{equation}
Here $z_S$ is the redshift of the source and $\chi_S$ is a time coordinate related to 
conformal time $\eta_S$ through $\chi_S=\eta_O-\eta_S$. 

\reviewtext{In this section we} consider a spatially flat ($K=0$) Friedmann universe with scalar perturbations.
We start from the usual longitudinal (or Newtonian) gauge where the metric is given by
\begin{equation}
\label{eq:metricmag}
g_{\mu\nu} \, \mathrm{d}x^\mu \, \mathrm{d}x^\nu = a^2\left[ -(1+2\Psi)d\eta^2 +
(1-2\Phi)\delta_{ij} \, \mathrm{d}x^i \, \mathrm{d}x^j\right] .
\end{equation}
We compute $\mathcal{D}_{ab}$ at linear order in $\Phi$ and $\Psi$ and then
we extract the shear and the convergence. We find, see
\cite{Bonvin:2008ni, Bernardeau:2009bm}
\begin{eqnarray}
\label{eq:gamma}\gamma&=&\frac{1}{2}\int_0^{\chi_S}\mathrm{d}\chi
\frac{\chi_S-\chi}{\chi\chi_S}\;\raise1.0pt\hbox{/}
\hskip-6pt\partial^2(\Phi+\Psi)\label{gamma} \,, \\
\label{eq:kappa}\kappa&=& \frac{1}{2}\int_0^{\chi_S}\mathrm{d}\chi
\frac{\chi_S-\chi}{\chi\chi_S}\;\overline{\raise1.0pt\hbox{/}
\hskip-6pt\partial}\;\raise1.0pt\hbox{/}\hskip-6pt\partial(\Phi+\Psi)
+ \Phi_S-\int_0^{\chi_S}\frac{\mathrm{d}\chi}{\chi_S}(\Phi+\Psi)\label{kappa}\\
&+& \left(\frac{1}{\mathcal{H}_S \chi_S}-1\right)\left(\Psi_S+\mathbf{n}\cdot
\mathbf{v}_S-\int_0^{\chi_S}\mathrm{d}\chi(\dot\Phi+\dot\Psi) \right) ,\nonumber 
\end{eqnarray}
where $\mathbf{n}$ is the direction of observation and $\mathbf{v}_S$ is the peculiar velocity of the source.
Here we are making use of
the angular spin raising $\;\raise1.0pt\hbox{/}\hskip-6pt\partial$ and
lowering $\;\overline{\raise1.0pt\hbox{/}\hskip-6pt\partial}$ operators (see
e.g., \cite{Lewis:2001hp} for a review of the properties of these operators)
defined as
\begin{equation}
\;\raise1.0pt\hbox{/}\hskip-6pt\partial \; {}_s X  \equiv - \sin^s \theta
(\partial_\theta + i \csc \theta \partial_\varphi) (\sin^{-s} \theta) \; {}_s X
\,, 
\qquad \;\overline{\raise1.0pt\hbox{/}\hskip-6pt\partial} \; {}_s X  \equiv -
\sin^{-s} \theta (\partial_\theta - i \csc \theta \partial_\varphi) (\sin^{s}
\theta) \; {}_s X \,,
\label{eq:spinraising}
\end{equation} 
where ${}_s X$ is an arbitrary field of spin $s$ and $\theta$ and $\varphi$ are
spherical coordinates.

Eq.~(\ref{gamma}) and the first term in Eq.~(\ref{kappa}) are the standard contributions of the shear and the
convergence, but expressed here with the full-sky transverse operators
\begin{eqnarray}
\frac{1}{\chi^2}\;\raise1.0pt\hbox{/}\hskip-6pt\partial^2&=&
\frac{1}{\chi^2}\left(\partial_\theta^2-\cot\theta\partial_\theta -\frac{1}{\sin^2\theta}\partial_\varphi\right)
+\frac{2 \mathrm{i}}{\chi^2\sin\theta}\Big(
\partial_\theta\partial_\varphi-\cot\theta\partial_\theta \Big) ,\\
\frac{1}{\chi^2}\;\raise1.0pt\hbox{/}\hskip-6pt\partial\;\overline{\raise1.0pt\hbox{/}\hskip-6pt\partial}&=&
\frac{1}{\chi^2}\left(\partial_\theta^2+\cot\theta\partial_\theta
+\frac{1}{\sin^2\theta}\partial_\varphi\right) .
\end{eqnarray}
In the flat-sky approximation, where $\theta$ is very small, 
$\frac{1}{\chi^2}\;\raise1.0pt\hbox{/}\hskip-6pt\partial\;\overline{\raise1.0pt\hbox{/}\hskip-6pt\partial}$
reduces to the 2D Laplacian $\partial_x^2+\partial_y^2$ and one recovers the standard expression for the convergence.
Similarly, the real part of $\frac{1}{\chi^2}\;\raise1.0pt\hbox{/}\hskip-6pt\partial^2$ that corresponds to $\gamma_1$
reduces to $\partial_y^2-\partial_x^2$ and the imaginary part that corresponds to $\gamma_2$ becomes
$\partial_x\partial_y$.

The other terms in Eq.~(\ref{kappa}) are relativistic corrections to the convergence,
that are negligible at small scales but may become relevant at large scales.
The terms in the first line are intrinsic corrections, generated respectively by the curvature
perturbation at the source position and the Shapiro time-delay. The terms in the
second line are due to the fact that we measure the convergence at a fixed redshift of the source $z_S$
rather that at a fixed conformal time $\eta_S$. Since in a perturbed universe, the observable redshift
is itself a perturbed quantity, this transformation generates additional contribution to the convergence. 
Those are respectively the Sachs--Wolfe contribution, the Doppler contribution and the integrated Sachs--Wolfe
contribution. 
Note that we have neglected the contributions at the observer position since they only give rise
to a monopole or dipole term. The dominant correction to the convergence is due
to the Doppler term. Therefore in the following we are interested in comparing
its amplitude with the amplitude of the standard contribution. To that end we
define $\kappa_{\mathrm{st}}$ and $\kappa_{\mathrm{vel}}$ as 
\begin{eqnarray}
\label{eq:kp} \kappa_{\mathrm{st}}&=&\int_{0}^{\chi_S}\mathrm{d}\chi
\frac{\chi_S-\chi}{2\chi\chi_S}\;\raise1.0pt\hbox{/}
\hskip-6pt\partial\;\overline{\raise1.0pt\hbox{/}\hskip-6pt\partial}
(\Phi+\Psi)\; ,\\ 
\label{eq:kv} \kappa_{\mathrm{vel}}&=&\left(\frac{1}{\mathcal{H}_S
\chi_S}-1\right)\mathbf{n}\cdot \mathbf{v}_S \,.
\end{eqnarray}

\subsubsection{Observable quantities}

The convergence is not directly observable. However it can be measured through
the modifications that it induces on the galaxy number density. Let us introduce
the magnification 
\begin{equation}
\mu=\frac{1}{\det \mathcal{D}} \simeq 1+2 \kappa \,,
\quad \mbox{when } \quad |\kappa|, |\gamma| \ll 1 \,.
\end{equation} 
The magnification modifies the size of a source: $d\Omega_O=\mu d\Omega_S $,
where $d\Omega_S$ is the true angular size of the source and $d\Omega_O$ is the
solid angle measured by the observer, i.e. the size of the image. The
magnification has therefore an impact on the observed galaxy number density. Let
us call $\bar{n}(f)df$ the number of unlensed galaxies per unit solid angle, at
a redshift $z_S$, and with a flux in the range $[f,f+df]$. The magnification
$\mu$ modifies the flux measured by the observer, since it modifies the observed
galaxy surface. It affects also the solid angle of observation and hence the
number of galaxies per unit of solid angle. These two effects combine to give a
galaxy number overdensity, see~\cite{Broadhurst:1994qu, Scranton:2005ci}
\begin{equation}
\label{eq:overdens}
\delta^\mu_g=\frac{n(f)-\bar{n}(f)}{\bar{n}(f)}
\simeq1+2\big(\alpha-1\big)(\kappa_{\mathrm{st}}+\kappa_{\mathrm{vel}}) \,.
\end{equation}
Here $\alpha\equiv -N'(>f_c)f_c/N(f_c)$, where $N(>f_c)$ is the number of
galaxies brighter than $f_c$ and $f_c$ is the flux limit adopted. Hence $\alpha$
is an observable quantity, see e.g.~\cite{Zhang:2005pu, Scranton:2005ci}. Recent
measurements of the galaxy number overdensity $\delta^\mu_g$ are reported
in~\cite{Scranton:2005ci, Menard:2009yb}. The challenge in those measurements is
to eliminate intrinsic clustering of galaxies, which induces an overdensity
$\delta_g^{cl}$ much larger than $\delta_g^\mu$. One possibility to separate
these two effects is to correlate galaxy number overdensities at widely
separated redshifts. One can then measure $\langle
\delta_g^\mu(z_S)\delta_g^{cl}(z_{S'})\rangle$, where $z_S$ is the redshift of
the sources and $z_{S'}<z_S$ is the redshift of the lenses. Another possibility,
proposed by \cite{Zhang:2005pu, Zhang:2005eb}, is to use the unique dependence
of $\delta^\mu_g$ on galaxy flux (i.e., on $\alpha$) to disentangle
$\delta^\mu_g$ from $\delta_g^{cl}$. This method, combined with precise
measurements of the galaxies redshift, allows to measure auto-correlations of
$\delta^\mu_g$, i.e., $\langle \delta_g^\mu(z_S)\delta_g^{\mu}(z_{S'})\rangle$,
either for $z_S\neq z_{S'}$ or for $z_S=z_{S'}$. The velocity contribution,
$\kappa_{\mathrm{vel}}$, has only an effect on $\langle
\delta_g^\mu(z_S)\delta_g^{\mu}(z_{S'})\rangle$. The correlations between
$\delta_g^{cl}(z_{S'})$ and $\mathbf{v}_S$ are indeed completely negligible and
hence the source peculiar velocity does not affect $\langle
\delta_g^\mu(z_S)\delta_g^{cl}(z_{S'})\rangle$. In the following we study in
detail the contribution of peculiar motion to
$\langle\delta_g^\mu(z_S)\delta_g^{\mu}(z_S)\rangle$.

The two components of the convergence $\kappa_{\mathrm{st}}$ and $\kappa_{\mathrm{vel}}$
(and consequently the galaxy number overdensity)
are functions of redshift $z_S$ and direction of observation $\mathbf{n}$. We
can therefore determine the angular power spectrum
\begin{equation}\label{eRF:demu}
\langle \delta^\mu_g(z_S,\mathbf{n})\delta^\mu_g(z_S,\mathbf{n}')\rangle=
\sum_{\ell}\frac{2\ell+1}{4\pi}C_\ell(z_S)P_\ell(\mathbf{n}\cdot
\mathbf{n}') \,.
\end{equation}
The angular power spectrum $C_\ell(z_S)$ contains two contributions, generated
respectively by $\langle \kappa_{\mathrm{st}} \kappa_{\mathrm{st}}\rangle$ and $\langle
\kappa_{\mathrm{vel}} \kappa_{\mathrm{vel}}\rangle$. The cross-term $\langle \kappa_{\mathrm{vel}} \kappa_{\mathrm{st}}\rangle$ is negligible since $\kappa_{\mathrm{st}}$ contains
only Fourier modes with a wave vector $\mathbf{k}_\perp$ perpendicular to the
line of sight (see Eq.~(\ref{eq:kp})), whereas $\kappa_{\mathrm{vel}}$ selects modes
with wave vector along the line of sight (Eq.~(\ref{eq:kv})). 

So far the derivation has been completely generic. Eqs.~(\ref{eq:gamma}) and (\ref{eq:kappa}) are valid in any theory of gravity whose metric can be written as in Eq.~(\ref{eq:metricmag}). To evaluate the angular power spectrum we now have to be more specific. In the following we assume GR, with no anisotropic stress such that $\Phi=\Psi$. We use the Fourier transform convention 
\begin{equation}
\mathbf{v}(\mathbf{x},\chi)=\frac{1}{(2\pi)^3}\int \mathrm{d}^3k \,
\mathbf{v}(\mathbf{k},\chi)e^{i\mathbf{k}\mathbf{x}} \,.
\end{equation}
The continuity equation, see e.g., \cite{Dodelson:2003ft}, allows us to express the peculiar velocity as
\begin{equation}
\label{eq:velocity}
\mathbf{v}(\mathbf{k},\chi)=-i\frac{\dot{G}(a)}{G(a)}\frac{\mathbf{k}}{k^2}
\delta(\mathbf{k},a) \,,
\end{equation}
where $\delta(\mathbf{k},a)$ is the density contrast, $G(a)$ is the growth
function, and $\dot{G}(a)$ its derivative with respect to $\chi$. With this we can
express the angular power spectrum as
\begin{equation}
\label{cv}
C_\ell^{\mathrm{vel}}(z_S)=\frac{16\pi\delta_H^2(\alpha_S-1)^2\dot{G}(a_S)^2}{H_0^4G^2(a=1)}
\left(\frac{1}{\mathcal{H}_S \chi_S}-1\right)^2 \int \mathrm{d}k \, k T^2(k)j_\ell'(k
\chi_S)^2 \,.
\end{equation}
Here $\delta_H$ is the density contrast at horizon and $T(k)$ is the transfer
function defined through, see e.g., \cite{Dodelson:2003ft}
\begin{equation}
\Psi(\mathbf{k},a)=\frac{9}{10}\Psi_p(\mathbf{k})T(k)\frac{G(a)}{a} \,.
\end{equation}
We assume a flat power spectrum, $n_s=1$, for the primordial potential $\Psi_p(\mathbf{k})$. We want to compare
this contribution with the standard contribution 
\begin{equation}
\label{eq:clstand}
C_\ell^{\mathrm{st}}(z_S)=\frac{36\pi\delta_H^2(\alpha_S-1)^2\Omega_m^2\ell^2(\ell+1)^2}{G^2(a=1)}\int\frac{\mathrm{d}k}{k}T^2(k)
\left[\int_0^{\chi_S}\mathrm{d}\chi \frac{\chi_S-\chi}{\chi\chi_S}\frac{G(a)}{a}j_\ell(k\chi)\right]^2 \,.
\end{equation}

\epubtkImage{cl_vel-cl_st.png}{%
\begin{figure}[htbp]
  \centerline{
    \includegraphics[height=5.8cm]{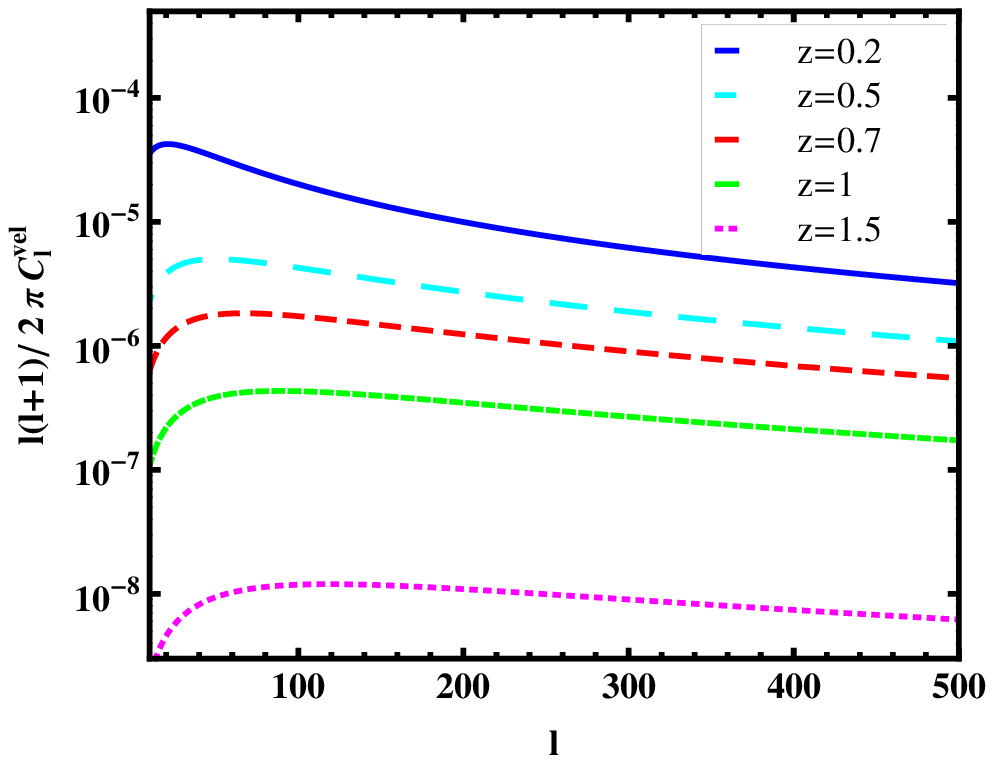}
    \includegraphics[height=5.8cm]{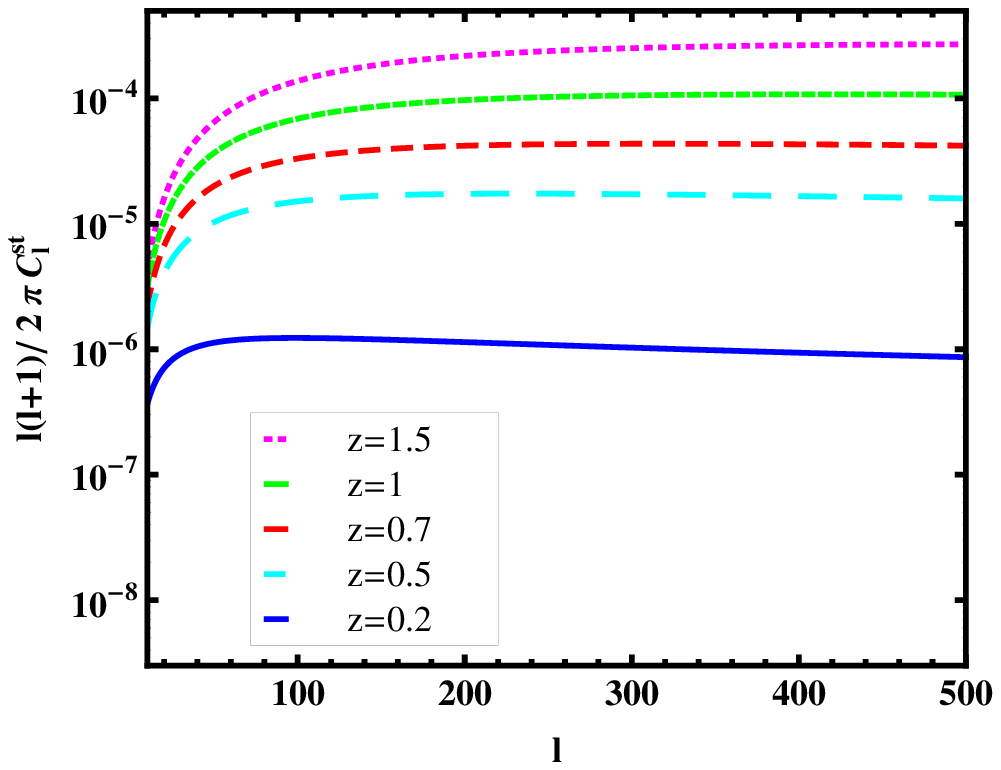}
  }
  \caption{\emph{Left:} The velocity contribution $C_\ell^{\mathrm{vel}}$ as a function
    of $\ell$ for various redshifts. \emph{Right:} The standard contribution $C_\ell^{\mathrm{st}}$ as
a function of $\ell$ for various redshifts.}
 \label{fig:clv-lens}\label{fig:lens}
\end{figure}}

We evaluate $C_\ell^{\mathrm{vel}}$ and $C_\ell^{\mathrm{st}}$ in a $\Lambda$CDM universe
with $\Omega_m = 0.25$, $\Omega_\Lambda = 0.75$ and
$\delta_H=5.7\cdot 10^{-5}$. We approximate the transfer function with the BBKS formula,
see~\cite{Bardeen:1985tr}. In Figure~\ref{fig:clv-lens}, we plot
$C_\ell^{\mathrm{vel}}$ and $C_\ell^{\mathrm{st}}$ for various source redshifts. The
amplitude of $C_\ell^{\mathrm{vel}}$ and $C_\ell^{\mathrm{st}}$ depends on $(\alpha-1)^2$,
which varies with the redshift of the source, the flux threshold adopted, and
the sky coverage of the experiment. Since $(\alpha-1)^2$ influences $C_\ell^{\mathrm{vel}}$ and $C_\ell^{\mathrm{st}}$ in the same way we do not include it in our plot.
Generally, at small redshifts, $(\alpha-1)$ is smaller than 1 and consequently
the amplitude of both $C_\ell^{\mathrm{vel}}$ and $C_\ell^{\mathrm{st}}$ is slightly
reduced, whereas at large redshifts $(\alpha-1)$ tends to be larger than 1 and
to amplify $C_\ell^{\mathrm{vel}}$ and $C_\ell^{\mathrm{st}}$, see
e.g., \cite{Zhang:2005eb}. However, the general features of the curves and more
importantly the ratio between $C_\ell^{\mathrm{vel}}$ and $C_\ell^{\mathrm{st}}$ are not
affected by $(\alpha-1)$. 

Figure~\ref{fig:clv-lens} shows that $C_\ell^{\mathrm{vel}}$ peaks at rather small $\ell$,
between 30 and 120 depending on the redshift. This corresponds to rather
large angle $\theta\sim 90\mbox{\,--\,}360\mathrm{\ arcmin}$. This behavior differs from the standard
term (Figure~\ref{fig:clv-lens}) that peaks at large
$\ell$. Therefore, it is important
to have large sky surveys to detect the velocity contribution. The relative
importance of $C_\ell^{\mathrm{vel}}$ and $C_\ell^{\mathrm{st}}$ de\-pends stron\-gly on
the redshift of the source. At small redshift, $z_S=0.2$, the velocity
contribution is about $4\cdot 10^{-5}$ and is hence larger than the standard
contribution which reaches $10^{-6}$. At redshift $z_S=0.5$, $C_\ell^{\mathrm{vel}}$
is about 20\% of $C_\ell^{\mathrm{st}}$, whereas at redshift $z_S=1$, it is about
$1\,\%$ of $C_\ell^{\mathrm{st}}$. Then at redshift $z_S=1.5$ and above,
$C_\ell^{\mathrm{vel}}$ becomes very small with respect to $C_\ell^{\mathrm{st}}$:
$C_\ell^{\mathrm{vel}} \leq 10^{-4}\,C_\ell^{\mathrm{st}}$. The enhancement of
$C_\ell^{\mathrm{vel}}$ at small redshift together with its fast decrease at large
redshift are due to the prefactor $\left(\frac{1}{\mathcal{H}_S
\chi_S}-1\right)^2$ in Eq.~(\ref{cv}). Thanks to this enhancement we
see that if the magnification can be measured with an accuracy of
10\%, then the velocity contribution is observable up to redshifts
$z\leq 0.6$. If the accuracy reaches 1\% then the velocity
contribution becomes interesting up to redshifts of order 1.

The shear and the standard contribution in the convergence are not independent.
One can easily show that their angular power spectra satisfy the consistency
relation, see \cite{Hu:2000ee}
\begin{equation}
C_\ell^{\kappa\, {\mathrm{st}}}=\frac{\ell(\ell+1)}{(\ell+2)(\ell-1)}C_\ell^\gamma \,.
\end{equation}
This relation is clearly modified by the velocity contribution. Using that the
cross-correlation between the standard term and the velocity term is negligible,
we can write a new consistency relation that relates the observed convergence
$C_\ell^{\kappa\, {\mathrm{tot}}}$ to the shear
\begin{equation}
\label{eq:cgamma}
\frac{\ell(\ell+1)}{(\ell+2)(\ell-1)}C_\ell^\gamma=C_\ell^{\kappa\,
  {\mathrm{tot}}}-C_\ell^{\kappa\, {\mathrm{vel}}} \,.
\end{equation}
Consequently, if one measures both the shear $C_\ell^\gamma$ and the
magnification $C_\ell^{\kappa\, {\mathrm{tot}}}$ as functions of the
redshift, Eq.~(\ref{eq:cgamma}) allows to extract the peculiar
velocity contribution $C_\ell^{\kappa\, {\mathrm{vel}}}$. 
This provides a
new way to measure peculiar velocities of galaxies.

Note that in practice, in weak lensing tomography, the angular power
spectrum is computed in redshift bins and therefore the square bracket
in Eq.~(\ref{eq:clstand}) has to be integrated over the bin 
\begin{equation}
\int_0^\infty \mathrm{d}\chi n_i(\chi)\int_0^{\chi}\mathrm{d}\chi'
\frac{\chi-\chi'}{\chi\chi'}\frac{G(\chi')}{a(\chi')}j_\ell(k\chi') \,,
\end{equation}
where $n_i$ is the galaxy density for the $i$-th bin, convolved with a
Gaussian around the mean redshift of the bin. The integral over
$\chi'$ is then simplified using Limber approximation, i.e.,
\begin{equation}
\int_0^\chi \mathrm{d}\chi' F(\chi')J_\ell(k\chi')\simeq
\frac{1}{k}F\left(\frac{\ell}{k} \right)\theta(k\chi-\ell) \,,
\end{equation}
where $J_\ell$ is the Bessel function of order $\ell$. 
The accuracy of Limber approximation increases with $\ell$. 
Performing a change of coordinate such that $k=\ell/\chi$, Eq.~(\ref{eq:clstand}) 
can be recast in the usual form used in weak lensing tomography, see
e.g., Eq.~(\ref{weak-lensing-non-parametric}).

\subsection{Observing modified gravity with redshift surveys}
\label{dark-energy-and-redshift-surveys}

Wide-deep galaxy redshift surveys have the power to yield information on both
$H(z)$ and $f_{g}(z)$ through measurements of Baryon Acoustic Oscillations 
(BAO) and redshift-space distortions.
In particular, if gravity is not modified and matter is not interacting other
than gravitationally, then a detection of the expansion rate is directly linked
to a unique prediction of the growth rate. Otherwise galaxy redshift surveys
provide a unique and crucial way to make a combined analysis of $H(z)$ and
$f_{g}(z)$ to test gravity. 
As a wide-deep survey, Euclid allows us to measure $H(z)$ directly from BAO, but
also indirectly through the angular diameter distance $D_A(z)$ (and possibly
distance ratios from weak lensing). Most importantly,
Euclid survey enables us to measure the cosmic growth history using two
independent methods: $f_g(z)$ from galaxy clustering, and $G(z)$ from weak
lensing. In the following we discuss the estimation of
$[H(z), D_A(z)$ and $f_g(z)]$ from galaxy clustering.

\label{sec:rs-BAO}
From the measure of BAO in the matter power spectrum or in the 2-point
correlation function one can infer information on the expansion rate of the
universe. In fact, the sound waves imprinted in the CMB can be also detected in
the clustering of galaxies, thereby completing an important test of our theory
of gravitational structure formation. 

\epubtkImage{pk.png}{%
\begin{figure}[htbp]
\centerline{\includegraphics[width=9cm]{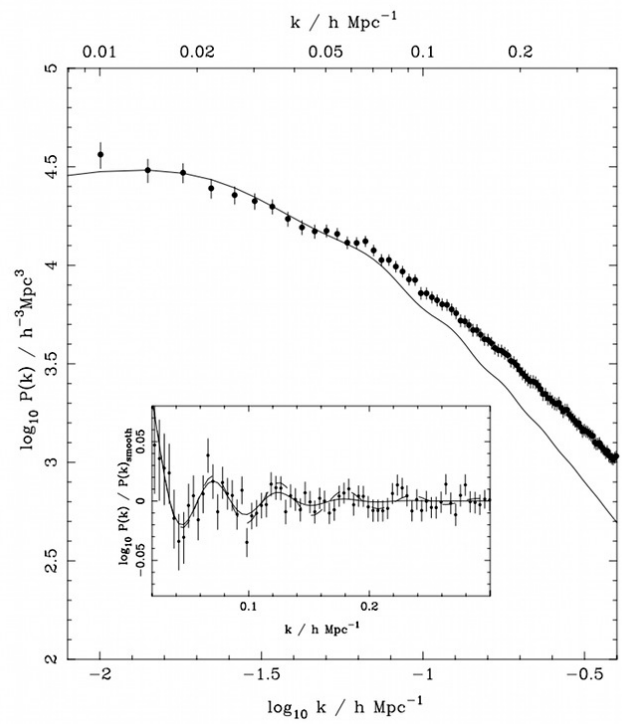}}
\caption{Matter power spectrum form measured from SDSS \citep{2007ApJ...657...51P}}
\label{fig:Pk} 
\end{figure}}

The BAO in the radial and tangential directions offer a way to measure the
Hubble parameter and angular diameter distance, respectively. 
In the simplest FLRW universe the basis to define
distances is the dimensionless, radial, comoving distance: 
\begin{equation} 
\chi(z) \equiv\int_0^z \frac{\mathrm{d}z'}{E(z')}\,. 
\label{chiz}
\end{equation}
The dimensionless version of the Hubble parameter is:
\begin{equation} 
E^2(z) = \Omega_m^{(0)}(1+z)^3 +\Omega_k(1+z)^2 + (\Omega_k-\Omega_m^{(0)})
\exp\left[ \int_0^z
\frac{3(1+w(\tilde{z}))}{1+\tilde{z}}\,\mathrm{d}\tilde{z} \right].
\end{equation}
The standard cosmological distances are related to $\chi(z)$ via
\begin{equation} 
D_A(z) = \frac{c}{H_0 (1+z)
\sqrt{-\Omega_k}}\sin\left(\sqrt{-\Omega_k}\chi(z)\right) 
\label{eq:ang}
\end{equation} 
where the luminosity distance, $D_L(z)$, is given by the distance duality: 
\begin{equation} 
D_L(z) = (1+z)^2 D_A(z). 
\label{eq:dual} 
\end{equation}
The coupling between $D_A(z)$ and $D_L(z)$ persists in any metric theory of
gravity as long as photon number is conserved (see Section~\ref{transparency-and-Etherington-relation} for
cases in which the duality relation is violated). 
BAO yield both $D_A(z)$ and $H(z)$ making use of an almost completely linear
physics (unlike for example SN~Ia, demanding complex and poorly understood
mechanisms of explosions). Furthermore, they provide the chance of constraining
the growth rate through the change in the amplitude of the power spectrum.

The characteristic scale of the BAO is set by the sound horizon at decoupling.
Consequently, one can attain the angular diameter distance and Hubble parameter
separately. This scale along the line of sight ($s_{||}(z)$) measures $H(z)$
through 
$H(z) = c\Delta z/s_{||}(z)$, while the tangential mode measures the angular
diameter distance $D_A(z) = s_{\perp}/\Delta \theta (1+z)$.

One can then use the power spectrum to derive predictions on the parameter
constraining power of the survey (see e.g.,
\cite{2005MNRAS.357..429A,2008Natur.451..541G,Wang08,Wang2010,diporto10}).

In order to explore the cosmological parameter
constraints from a given redshift survey, one needs to specify
the measurement uncertainties of the galaxy power spectrum.
In general, the statistical error on the measurement of the galaxy
power spectrum $P_{\mathrm{g}}(k)$
at a given wave-number bin is \citep{FKP}
\begin{equation}
\left[\frac{\Delta P_{\mathrm{g}}}{P_{\mathrm{g}}}\right]^2=
\frac{2(2\pi)^2 }{\Vsur k^2\Delta k\Delta \mu}
\left[1+\frac1{n_{\mathrm{g}}P_{\mathrm{g}}}\right]^2,
\label{eqn:pkerror}
\end{equation}
where $n_{\mathrm{g}}$ is the mean number density of galaxies, 
$\Vsur$ is the comoving survey volume of the galaxy survey, and $\mu$
is the cosine of the angle between $\mathbf{k}$ and the line-of-sight
direction $\mu = \vec{k}\cdot \hat{r}/k$.

In general, the \emph{observed} galaxy power spectrum is different
from the \emph{true} spectrum, and it can be reconstructed approximately
assuming a reference cosmology (which we consider to be our fiducial
cosmology) as (e.g., \cite{seo03})
\begin{equation}
P_{\mathrm{obs}}(k_{{\mathrm{ref}}\perp},k_{{\mathrm{ref}}\parallel},z)
=\frac {\DA _{\mathrm{ref}} ^2 \hz}{\DA ^2 \hz _{\mathrm{ref}}} P_{\mathrm{g}}(k_{{\mathrm{ref}}\perp},k_{{\mathrm{ref}}\parallel},z)
+P_{\mathrm{shot}}\,,
\label{eq:Pobs}
\end{equation}
where
\begin{equation}
P_{\mathrm{g}}(k_{{\mathrm{ref}}\perp},k_{{\mathrm{ref}}\parallel},z)=b(z)^2\left[1+\beta(z) 
\frac{k_{{\mathrm{ref}}\parallel}^2}{k_{{\mathrm{ref}}\perp}^2+k_{{\mathrm{ref}}\parallel}^2}\right]^2\times
P_{\mathrm{matter}}(k,z)\,.
\label{eq:Pg}
\end{equation}
In Eq.~(\ref{eq:Pobs}), $H(z)$ and $D_A(z)$ are the Hubble parameter and the angular
diameter distance, respectively, and the prefactor 
$(\DA _{\mathrm{ref}} ^2 \hz)/(\DA ^2 \hz _{\mathrm{ref}})$ encapsulates the
geometrical distortions due to the Alcock--Paczynski
effect \citep{seo03,9605017}. Their values in the reference cosmology are
distinguished by the subscript `ref', while those in the true cosmology have no
subscript. $k_\perp$ and $k_\parallel$ are the wave-numbers across and along
the line of sight in the true cosmology, and they are related to the 
wave-numbers calculated assuming the reference
cosmology by
$k_{{\mathrm{ref}}\perp} = k_\perp D_A(z)/D_A(z)_{\mathrm{ref}}$ and
$k_{{\mathrm{ref}}\parallel} = k_\parallel H(z)_{\mathrm{ref}}/H(z)$. 
$P_{\mathrm{shot}}$ is the unknown white shot noise that 
remains even after the conventional shot noise of inverse number density has been 
subtracted \citep{seo03}.
In Eq.~(\ref{eq:Pg}), $b(z)$ is the \emph{linear bias} factor between galaxy and
matter density distributions, $f_g(z)$ is the linear growth\label{symbol:bias}
rate,%
\epubtkFootnote{In presence of massive neutrinos $f_g$ depends also on
  the scale $k$ \citep{0709.0253}.} 
and $\beta(z)=f_g(z)/b(z)$ is the linear \label{symbol:beta}
redshift-space distortion parameter \citep{Kaiser:1987}. 
The linear matter power spectrum $P_{\mathrm{matter}}(k,z)$ in Eq.~(\ref{eq:Pobs}) takes the form
\begin{eqnarray}
P_{\mathrm{matter}}(k,z)=\frac{8\pi^2c^4k_0\Delta^2_{\cal R}(k_0)}{25
  H_0^4\Omega_{m}^2} T^2(k) \left [\frac{G(z)}{G(z=0)}\right]^2
\left(\frac{k}{k_0}\right)^{n_s}e^{-k^2\mu^2\sigma_r^2},
\label{eq:Pm}
\end{eqnarray}
where $G(z)$ is the usual \emph{scale independent} linear
growth-factor in the absence of massive neutrino free-streaming (see Eq.~(25)
in~\cite{Eisenstein_Hu_1997}), whose fiducial value in each redshift bin is computed 
through numerical integration of the differential equations governing the growth
of linear perturbations in presence of dark energy
\citep{astro-ph/0305286} or employing the approximation of
Eq. (\ref{def-growth-rate}). $T(k)$ depends on matter and baryon
densities%
\epubtkFootnote{If we assume that neutrinos have a non-vanishing mass,
  then the transfer function is also redshift-dependent.}
(neglecting dark energy at early times), and is computed in each
redshift bin using a Boltzmann code like \textsc{camb}%
\epubtkFootnote{\url{http://camb.info/}}
\citep{CAMB} or \textsc{cmbfast}.

In Eq.~(\ref{eq:Pm}) a damping factor
$e^{-k^2\mu^2\sigma_r^2}$ has been added,
due to redshift uncertainties, where $\sigma_r=(\partial r/\partial
z)\sigma_z$, $r(z)$ being the comoving
distance \citep{0904.2218,seo03},
and assumed that the power spectrum of primordial curvature
perturbations, $P_{\cal R}(k)$, is
\begin{equation}
\Delta^2_{\cal R}(k) \equiv \frac{k^3P_{\cal R}(k)}{2\pi^2}
= \Delta^2_{\cal R}(k_0)\left(\frac{k}{k_0}\right)^{n_s} \,,
\label{eq:pR}
\end{equation}
where $k_0=0.002/\mathrm{Mpc}$, $\Delta^2_{\cal R}(k_0)|_{\mathrm{fid}}=2.45\times
10^{-9}$ is the dimensionless amplitude of the primordial curvature perturbations
evaluated at a pivot scale $k_0$, and $n_s$ is the scalar spectral
index \citep{arXiv:1001.4635}.

In the limit where the survey volume is much larger than the scale of 
any features in $P_{\mathrm{obs}}(k)$, it has been shown that
the redshift survey Fisher matrix for a given redshift bin can be approximated as \citep{Tegmark:1997rp}
\begin{eqnarray}
F_{ij}^{\mathrm{LSS}}
&=&\int_{-1}^{1} \int_{k_{\min}}^{\kmax}\frac{\partial \ln
  P_{\mathrm{obs}}(k,\mu)}{\partial p_i} \frac{\partial \ln P_{\mathrm{obs}}(k,\mu)}{\partial p_j} 
\Veff(k,\mu) \frac{2\pi k^2 \, \mathrm{d}k \, \mathrm{d}\mu}{2(2\pi)^3} \,,
\label{Fisher}
\end{eqnarray}
where the derivatives are evaluated at the parameter values $p_i$
of the fiducial model,
and $\Veff$ is the effective volume of the survey:
\begin{eqnarray}
\Veff(k,\mu) =
\left [ \frac{{n_{\mathrm{g}}}P_{\mathrm{g}}(k,\mu)}{{n_{\mathrm{g}}}P_{\mathrm{g}}(k,\mu)+1} \right ]^2 \Vsur \,,
\label{V_eff}
\end{eqnarray}
where the comoving number density
$n_{\mathrm{g}}(z)$ is assumed to be spatially constant.
Due to azimuthal symmetry around the line of sight,
the three-dimensional galaxy redshift power spectrum
$P_{\mathrm{obs}}(\vec{k})$ depends only on $k$ and $\mu$, i.e., is reduced
to two dimensions by symmetry \citep{seo03}. The total Fisher matrix
can be obtained by summing over the redshift bins.

To minimize nonlinear effects, one should restrict wave-numbers to the 
quasi-linear regime, e.g., imposing that $\kmax$ is given
by requiring that the variance of matter fluctuations in a sphere of
radius $R$ is, for instance, $\sigma^2(R)=0.25$ for $R=\pi/(2\kmax)$. 
Or one could model the nonlinear distortions as in \cite{eisenstein07}.
On scales larger than ($\sim 100\, h^{-1}\mathrm{\ Mpc}$) where we focus our analysis,
nonlinear effects can be represented in fact as a displacement field in Lagrangian space
modeled by an elliptical Gaussian function. Therefore, following
\cite{eisenstein07,seo07}, to model nonlinear effect
we multiply $P(k)$ by the factor
\begin{equation}
\exp\left\{ 
-k^{2}\left[\frac{(1-\mu^{2})\Sigma_{\perp}^{\,2}}{2}+\frac{\mu^{2}\Sigma_{
\parallel}^{\,2}}{2}\right]\right\} \,,
\label{eq:damping}
\end{equation}
where $\Sigma_{\perp}$ and $\Sigma_{\parallel}$ represent the
displacement across and along the line of sight, respectively. They are related
to the growth factor $G$ and to the growth rate
$f_g$ through $\Sigma_{\perp}=\Sigma_{0}G$ and
$\Sigma_{\parallel}=\Sigma_{0}G(1+f_g)$.
The value of $\Sigma_{0}$ is proportional to $\sigma_{8}$. For a
reference cosmology where $\sigma_{8}=0.8$ \citep{Komatsu:2010fb}, we have
$\Sigma_{0}=11\, h^{-1}\mathrm{\ Mpc}$.

Finally, we note that when actual data are available, the usual
way to measure $\beta=f_g/b$ is by fitting the measured
galaxy redshift-space correlation function $\xi(\sigma,\pi)$ to a model 
\citep{Peebles80}:
\begin{equation}
\xi(\sigma,\pi)= \int_{-\infty}^{\infty}{\mathrm{d}}v\, f(v)\,
\tilde{\xi}(\sigma,\pi-v/H_0) \,,
\end{equation}
where $f(v)$ describes the small-scale random motion (usually modeled by 
a Gaussian that depends on the galaxy pairwise peculiar velocity dispersion),
and $\tilde{\xi}(\sigma,\pi)$ is the model accounting for coherent 
infall velocities:%
\epubtkFootnote{See~\cite{Hamilton92}. $\tilde{\xi}(\sigma,\pi)$ is
  the Fourier transform of $P_s(k)=(1+\beta \mu^2)^2 P_r(k)$
  \citep{Kaiser:1987}.} 
\begin{equation}
\tilde{\xi}(\sigma,\pi)=\xi_0(s) P_0(\mu)
+\xi_2(s) P_2(\mu)+ \xi_4(s) P_4(\mu).
\end{equation}
$P_l(\mu)$ are Legendre polynomials; $\mu=\cos\theta$,\label{symbol:legendre}
where $\theta$ denotes the angle between $\mathbf{r}$ and $\pi$; 
$\xi_0(s)$, $\xi_2(s)$, and $\xi_4(s)$ depend on
$\beta$ and the real-space correlation function $\xi(r)$.

The bias between galaxy and matter distributions can be estimated from
either galaxy clustering, or weak lensing. 
To determine bias, we can assume that the galaxy density perturbation 
$\delta_g$ is related to the matter density perturbation $\delta_m(\mathbf{x})$
as \citep{Fry_Gaztanaga93}:
\begin{equation}
\delta_g= b \delta_m(\mathbf{x})+ b_2 \delta_m^2(\mathbf{x})/2.
\end{equation}

Bias can be derived from galaxy clustering by measuring the galaxy bispectrum:
\begin{eqnarray}
\langle \delta_{g\mathbf{k}_1} \delta_{g\mathbf{k}_2} \delta_{g\mathbf{k}_1}\rangle &=& (2\pi)^3 \left\{P(\mathbf{k}_1) P(\mathbf{k}_2)\left[J(\mathbf{k}_1,\mathbf{k}_2)/b
+b_2/b^2\right] \right. \nonumber\\
& & \left. \hskip 1cm +\mathrm{cyc.}\right\} \delta^D(\mathbf{k}_1+\mathbf{k}_2+\mathbf{k}_3),
\end{eqnarray}
where $J$ is a function that depends on the shape of the
triangle formed by ($\mathbf{k}_1$, $\mathbf{k}_2$, $\mathbf{k}_3$)
in
$\mathbf{k}$ space, but only depends very weakly on cosmology 
\citep{Matarrese_Verde_Heavens97,Verde02}.

In general, bias can be measured from weak lensing through the comparison of the
shear-shear and shear-galaxy correlations functions.
A combined constraint on bias and the growth factor $G(z)$
can be derived from weak lensing by comparing the cross-correlations
of multiple redshift slices.

Of course, if bias is assumed to be linear ($b_2=0$) and scale independent, or is parametrized in
some simple way, e.g.,
with a power law scale dependence, then it is possible to estimate it even from
linear galaxy clustering alone,
as we will see in Section~\ref{gamma-bias-forecasts}.

\subsection{Constraining modified gravity with galaxy--CMB correlations}

Two of the above-mentioned observable signatures of dark energy and modified gravity are especially suitable to study the time evolution of dark energy at the perturbative level: the ISW effect and CMB lensing. Both effects produce sub-dominant secondary anisotropies imprinted on the CMB at late times, and can be measured as a function of redshift by cross-correlating CMB temperature and lensing maps with galaxy surveys, thus allowing a tomographic analysis of the dark energy properties.

\subsubsection{The ISW effect}
The CMB photons freely streaming in the late universe encounter over- and under-densities; their energy will thus change as a function of time as the photons climb in and out of potential wells, but the average net energy gain will be null as long as the potentials are globally constant in time. Since the potentials decay in the presence of cosmic acceleration, a non-zero ISW effect will be produced in this case, corresponding to a temperature anisotropy in the direction $ \hat {\mathbf n}$
\be
\frac{\Delta T}{T} (\hat {\mathbf n}) = \int d \eta \, e^{-\tau(z)} \, \left( \dot \Phi + \dot \Psi \right)  \left[ \eta,  \hat {\mathbf n} (\eta_0 - \eta) \right] \, ,
\ee
where dots indicate derivatives with respect to the conformal time $\eta$, $\tau$ is the optical depth, and $\Phi, \Psi$ are the Newtonian gauge potentials describing the time and space metric perturbations, respectively.
This effect is subdominant with respect to the primary CMB temperature anisotropies produced at primordial times, from which it can however be extracted by cross-correlating the full CMB maps with galaxy catalogues, which are correlated with the ISW signal since the galaxy overdensities also trace the same gravitational potentials \cite{1996PhRvL..76..575C}. 

The ISW has been detected, in agreement with the $\Lambda$CDM predictions, at the $\sim 4\, \sigma$ significance level by cross-correlating \textit{WMAP} and \textit{Planck} CMB data with numerous galaxy catalogues: see \cite{2008PhRvD..78d3519H,2008PhRvD..77l3520G,2012MNRAS.426.2581G} and references therein.

Future galaxy surveys including the \textit{Euclid} satellite are expected to improve current ISW measurements by increasing redshift depth and survey volume, thus allowing a consistent tomographic study from one galaxy survey, as well as by improving the control of systematics; the total signal-to-noise is however not expected to exceed the $\sim 8 \, \sigma$ level \cite{1996PhRvL..76..575C} in the $\Lambda$CDM scenario, since the ISW signal peaks on the largest scales, which are dominated by cosmic variance. The measurement of ISW at high redshift has however a significant discovery potential, as in case exotic dark energy models are correct, the actual level of ISW may be significantly higher.

\subsubsection{CMB lensing}

An additional, complementary observable is provided by CMB lensing \cite{2006PhR...429....1L}. This is a special case of the weak gravitational lensing described above, where the sources are set to the CMB last-scattering surface at redshift $z_{\star} \simeq 1100$. In this case, the primary CMB lensing map is deflected by the intervening large-scale structure by small angles of the order $\sim 2.5$ arcmin, by the effect of a lensing potential $\varphi$ in a direction $ \hat {\mathbf n}$ given by
\be
\varphi (\hat {\mathbf n}) = - \int d \chi \, \frac{\chi_{\ast} - \chi}{\chi_{\ast} \chi} \left(\Phi + \Psi  \right)  \left[ \chi  \hat {\mathbf n}, \eta_0 - \chi \right] \, ,
\ee
where $\chi$ is the conformal distance. This potential is simply related to the convergence $\kappa$ used above in multipole space by $\kappa_{lm} = l(l+1) \varphi_{lm} / 2$.
The effect of lensing on the CMB temperature anisotropies is a smoothing of the peaks an troughs in the angular power spectrum.

Maps of the CMB lensing potential have been reconstructed from higher-order statistics of the CMB temperature maps by the \textit{Planck} \cite{planck2013-p12}, South Pole Telescope \cite{2012ApJ...756..142V} and Atacama Cosmology Telescope \cite{2011PhRvL.107b1301D} surveys; cross-correlations between these lensing maps and galaxy surveys have also been confirmed with these three data sets (see e.g. \cite{2014MNRAS.441L..16G}): such cross-correlations allow once again to study the redshift evolution of the gravitational potentials, and thus the physical properties of the dark sector.

Upcoming and future galaxy surveys leading up to the \textit{Euclid} satellite mission, combined with rapidly improving CMB data, will increase the signal-to-noise of the CMB lensing cross-correlations well beyond the current levels, since the CMB lensing signal is maximum on smaller scales, which are currently dominated by statistical and systematic errors, but not by cosmic variance.

\subsection{Cosmological bulk flows}

As we have seen, the additional redshift induced by the galaxy peculiar velocity field generates the redshift 
distortion in the power spectrum. In this section we discuss
a related effect on the luminosity of the galaxies and on its use
to measure the peculiar velocity in large volumes, the so-called bulk flow.

In the gravitational instability framework, inhomogeneities in the matter distribution 
induce gravitational accelerations $\mathbf{g}$, which result in galaxies having peculiar velocities 
$\mathbf{v}$ that add to the Hubble flow.
In linear theory the peculiar velocity field is proportional to the peculiar acceleration
\begin{equation}
\label{eq:vg}
\mathbf{v}(\mathbf{r})=\frac{2f_g}{3H_0\Omega_m}\mathbf{g}(\mathbf{r})=
\frac{H_0 f_g}{4\pi} \int \delta_m(\mathbf{r^\prime})\frac{(\mathbf{r^\prime} - \mathbf{r})}{|\mathbf{r^\prime}-\mathbf{r}|^3}\,\mathrm{d}^3\mathbf{r^\prime} \,,
\end{equation}
and the bulk flow of a spherical region is solely determined by the gravitational
pull of the dipole of the external mass distribution. For this reason, bulk flows
are reliable indicators to deviations from homogeneity and isotropy on 
large scale, should they exist.

Constraints on the power spectrum and growth rate can be obtained by comparing
the bulk flow estimated from the volume-averaged motion of the sphere
of radius $R$:
\begin{equation}
\label{eq:vbulk}
\mathbf{B}_{R}\equiv \frac{\int \mathbf{v}(\mathbf{x})
  W(\mathbf{x}/R)\,\mathrm{d}^3\mathbf{x}}{\int W(\mathbf{x}/R)\,\mathrm{d}^3\mathbf{x}} \,,
\end{equation}
with expected variance:
\begin{equation}
\label{eq:vbexp}
\sigma^2_{\mathbf{B},{R}}=
\frac{H_0^2 f_g^2}{6\pi^2}\int P(k){\cal W}(kR)^2(k)\,\mathrm{d}k \,,
\end{equation}
where the window function $W(\mathbf{x}/R)$ and its Fourier transform
${\cal W}(kR)$ describe the spatial distribution of the dataset.

Over the years the bulk flows has been estimated from 
the measured peculiar velocities of a large variety of objects ranging from galaxies
\citep{1998ApJ...505L..91G, 1998AJ....116.2632G, 1999ApJ...522....1D,
  2000ApJ...544..636C, 2000ApJ...537L..81D, 2007MNRAS.375..691S}
clusters of galaxies \citep{1994ApJ...425..418L, 1996ApJ...461L..17B, 2004MNRAS.352...61H}
and SN~Ia  \citep{1995ApJ...445L..91R}. Conflicting results triggered by the use of error-prone distance indicators 
have fueled a long lasting controversy on the amplitude and  convergence of the bulk flow, that is still ongoing.
For example, the recent claim of a bulk flow of $407\pm81\mathrm{\ km\ s}^{-1}$ within $R=50$ $h^{-1}\mathrm{\ Mpc}$
\citep{Watkins:2008hf}, inconsistent with expectation from the $\Lambda$CDM model, has been seriously
challenged by the re-analysis of the same data by \cite{2011arXiv1101.1650N} who found a 
bulk flow amplitude consistent with $\Lambda$CDM expectations and from which they 
were able to set the strongest constraints on modified gravity models so far. 
On larger scales, \cite{2010ApJ...712L..81K} claimed the detection of a dipole anisotropy
attributed to the kinetic SZ decrement in the WMAP temperature map at the 
position of X-ray galaxy clusters. When interpreted  as a coherent motion, this signal would indicate 
a  gigantic bulk flow of $1028\pm 265\mathrm{\ km\ s}^{-1}$ within
$R=528$ $h^{-1}\mathrm{\ Mpc}$. This highly debated result has been
seriously questioned by independent analyses of WMAP data \citep[see, e.g.,][]{2010arXiv1011.2781O})

The large, homogeneous dataset expected from Euclid has the potential to settle these issues.
The idea is to measure bulk flows in 
large redshift surveys, based on the apparent, dimming or brightening of galaxies
due to their peculiar motion. The method, originally proposed by \cite{TYS}, has been recently
extended by \cite{2011arXiv1102.4189N}
 who propose to estimate the bulk flow 
by minimizing systematic variations in galaxy luminosities
with respect to a reference luminosity function measured 
from the whole survey.  It turns out that, if applied to the
photo-$z$ catalog expected from Euclid, this method would be able to 
detect at $\> 5 \sigma$ significance
 a bulk flow like the one of  \citep{Watkins:2008hf} 
over  $\sim 50$ independent spherical volumes at $z \ge 0.2$, provided that the
systematic magnitude offset over the corresponding areas in the sky does not exceed 
the expected random magnitude errors of 0.02\,--\,0.04~mag.
Additionally, photo-$z$ or spectral-$z$ could be used to validate or disproof 
with very large ($> 7 \sigma$) significance
the claimed bulk flow detection of \cite{2010ApJ...712L..81K} at $z=0.5$.

Closely related to the bulk flow 
is the Local Group peculiar velocity inferred from the observed CMB dipole
\citep{1990ApJ...349..408J}
\begin{equation}
\label{eq:vlg}
\mathbf{v}_{\mathrm{CMB}}=\mathbf{v}_{\mathrm{LG},R}-\frac{H_0f_g}{3}\mathbf{x}_{c.m.}+\mathbf{B}_{R} \,,
\end{equation}
where $\mathbf{v}_{\mathrm{LG},R}$ is the Local Group velocity resulting from the gravitational pull
of all objects in the sample within the radius $R$, $\mathbf{x}_{\mathrm{c.m.}}$ is the position of
the center of mass of the sample and $\mathbf{v}_{\mathrm{CMB}}$ is the LG velocity  inferred from the 
CMB dipole 
\citep{2003ApJS..148....1B}.
The convergence of $\mathbf{v}_{\mathrm{LG},R}$ with the radius and its alignment with the CMB dipole direction
indicates a crossover to homogeneity
\citep{1991ApJ...376L...1S}
 and allows to constrain the growth rate by comparing 
$\mathbf{v}_{\mathrm{CMB}}$ with $\mathbf{v}_{\mathrm{LG},R}$. The latter can be estimated from the dipole in the distribution 
of objects either using a number-weighting scheme if redshifts are available for all objects of the sample,
or using a flux-weighting scheme. In this second case the fact that both gravitational acceleration and
flux are inversely proportional to the distance allows to compute the dipole from photometric catalogs with no
need to measure redshifts. The drawback is that the information on the convergence scale is lost.

As for the bulk flow case, despite the many measurements of cosmological dipoles using galaxies
\citep{1980ApJ...242..448Y, 1982ApJ...254..437D, 1986AJ.....91..191M, 1992ApJ...397..395S,
1999MNRAS.304..893S, 2006ApJ...645.1043K}
there is still no general consensus on the scale of convergence and even on
the convergence itself. Even the recent analyses of measuring the acceleration of the Local Group from
the 2MASS redshift catalogs provided conflicting results. 
\cite{2006MNRAS.368.1515E} found that the galaxy dipole seems to converge beyond  $R=60\,h^{-1}\mathrm{\ Mpc}$,
whereas \cite{2010ApJ...709..483L} find no convergence within $R=120\,h^{-1}\mathrm{\ Mpc}$.

Once again, Euclid will be in the position to solve this controversy by measuring the galaxy 
and cluster dipoles 
not only at the LG position and out to very large radii, but also in several independent ad truly
all-sky spherical samples carved out from the the observed areas with $|b|>20^{\circ}$.
In particular, coupling photometry with photo-$z$ one expects to be able to estimate
the convergence scale of the flux-weighted dipole over about 100 independent spheres of radius
$200\,h^{-1}\mathrm{\ Mpc}$ out to $z=0.5$ and, beyond that, to compare number-weighted and flux-weighted
dipoles over a larger number of similar volumes using spectroscopic redshifts.

Similarly, the growth rate can be constrained by studying the possibility of a Hubble bubble, a local region of space with a (slightly) different Hubble rate. This study was triggered by the fact that global observables such as Planck and BAO~\citep[][Table 5]{Ade:2013zuv} yield a present-day Hubble constant 9\% lower than local measurements performed by considering recession velocities of objects around us~\cite{Riess:2011yx}. This 2.4$\sigma$ tension could be relieved if the effect of a local Hubble bubble is taken into account, see~\cite{Marra:2013rba} and references therein. With Euclid one will of course use the data the other way around, using observations to constrain the Hubble bubbles (velocity monopoles) at different radii, and so the growth rate of matter structures, similarly to what discussed regarding the bulk flow.

\subsection{Model independent observations}\label{s:model-indep}
As discussed, one of the most powerful statistical tools that can be used to
describe the distribution of matter in the Universe is the power spectrum
$P(k)$, or its Fourier transform $\xi(r)$, the two-point correlation function.
However, comoving distances $r$ and the corresponding wavenumbers $k$ are not
observationally accessible.

When observing galaxies we measure their redshift and angular position.
To convert this into a three-dimensional galaxy
catalog we must make model assumptions in order to relate the observed redshift
to a distance.
For small redshift, $z\ll1$, the simple relation $r(z)=z/H_0$ can be used.
When expressing distances in units of $h^{-1}$Mpc the uncertainty in the measurement of $H_0$ is then absorbed in $h=H_0/(100$Mpc/(km/s)). However, when $z\simeq 1$ the distance $r(z)=\chi(z)/H_0$ depends on the full cosmic expansion history, i.e. on the parameters $\Omega_m$, $\Omega_k$, $\Omega_{DE}=1-\Omega_m-\Omega_k$ and $w_{DE}$, see Eq.~(\ref{chiz}), and wrong assumptions about the distance redshift relation will bias the entire catalog in a non-trivial way.

\begin{figure}[hT]
\includegraphics[width=.5\columnwidth]{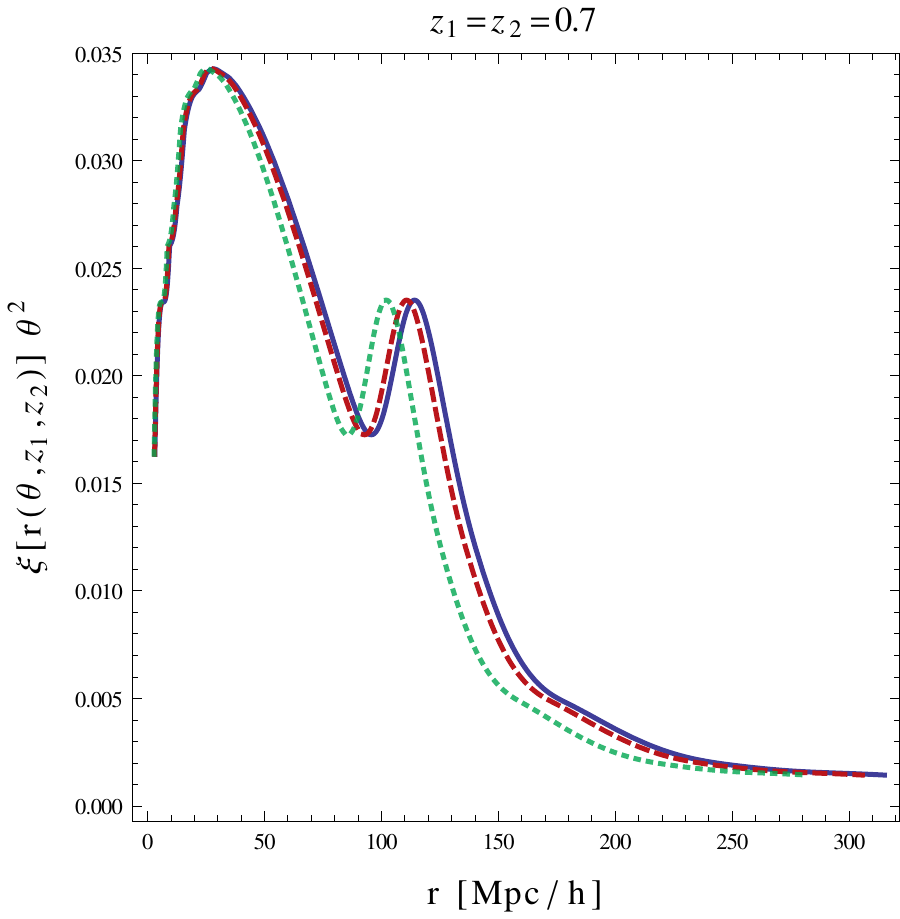}
\includegraphics[width=.5\columnwidth]{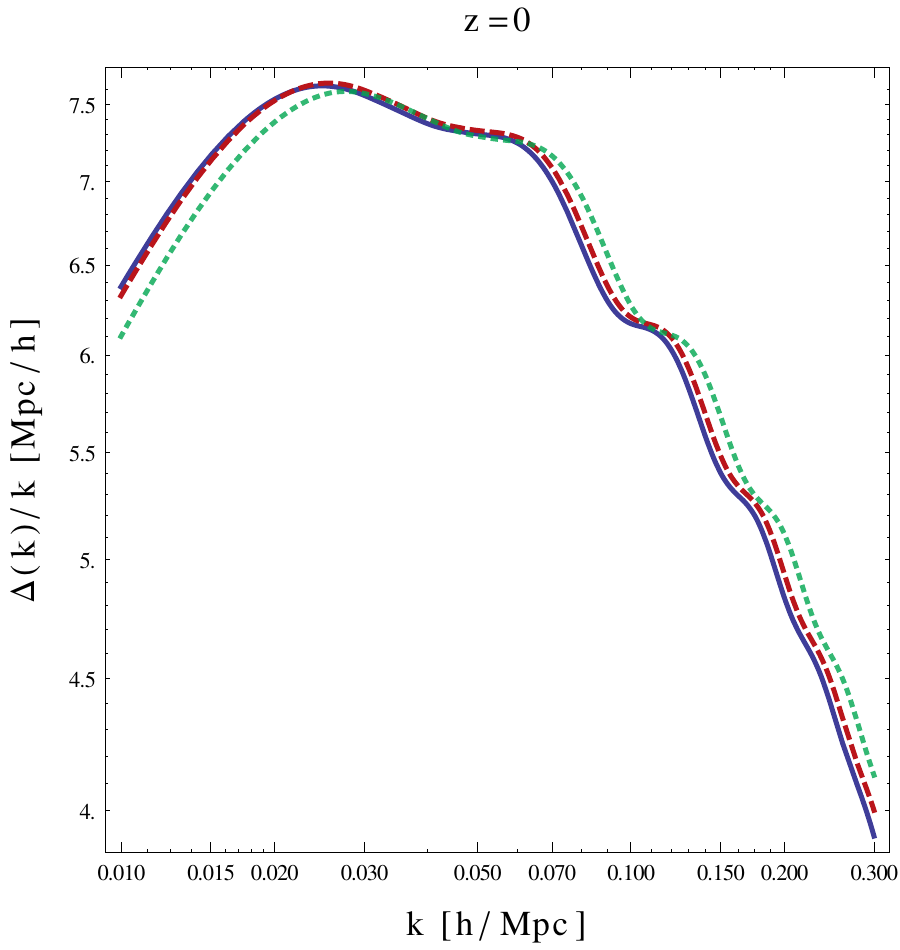}
\caption{Effect of wrong cosmological parameters on the power spectrum. The true one (solid line)
 assumes the cosmological parameters of~\cite{Komatsu:2010fb}
(in particular $\Omega_{\rm m}=0.27$) and takes into account
redshift space distortions. The wrong assumptions $\Omega_{\rm m}=0.3, 0.5$
(dashed and dotted line, respectively) rescale the
correlation function (on the left, multiplied by $\theta^2$ to enhance the BAOs)
and the dimensionless power spectrum (on the right, divided by $k$
to enhance the BAOs).
}
\label{fig:WrongCosmo}
\end{figure}

Assuming an incorrect cosmology causes geometric redshift-distortions, in addition to the dynamical redshift distortions due to the peculiar velocities of galaxies.
In Fig.~\ref{fig:WrongCosmo} we show the effect of assuming a  wrong cosmology.
To illustrate this, we consider a $\Lambda$CDM universe with the cosmological parameters given in Ref.~\cite{Komatsu:2010fb}.
To reconstruct the comoving separation between two galaxies with redshift $z_1$
and $z_2$, respectively, and separated by the angle $\theta$, an observer must
assume a model to reconstruct the relation $r(\theta,z_1,z_2)$ which for vanishing 
curvature, is given by
$r(z_1,z_2,\theta) = \sqrt{r(z_1)^2 + r(z_2)^2 - 2r(z_1)r(z_2)\cos\theta}$,
where the comoving distance $r(z) =\int_0^z dz'/H(z')$ depends on the cosmological model.
Iterative methods are usually applied to converge to the correct cosmology.
The inferred galaxy clustering in a different cosmological model can also be approximately obtained from the fiducial one by a rescaling of the transverse and parallel separations~\cite{Percival:2010,Reid:2012sw}, so that an Alcock-Paczy\'nski test~\cite{Alcock:1979mp} can be performed to select the best fit cosmological model.
Nevertheless, these procedures rely on the assumption of a fiducial cosmology and are not very well suited to {\em measure} cosmological parameters, especially error estimates are not straight forward.

Together with the standard analysis, it is therefore important to determine the truly observed two-point statistics from observations, either in terms of the redshift dependent angular power spectra, $C_\ell(z_1,z_2)$, or in terms of the redshift dependent angular correlations functions  $\xi(\theta,z_1,z_2)$, which are related to the power spectra by Eq.~(\ref{eRF:demu}), and to compare them directly with their theoretically obtained counterparts.
The full expression for $C_\ell(z_1,z_2)$ are found in~\cite{Bonvin:2011bg,Asorey:2012rd,DiDio:2013bqa}, see also~\cite{Challinor:2011bk,Yoo:2009au,Yoo:2010ni}.
The angular correlation function has been studied in \cite{Montanari:2012me,Bertacca:2012tp}, while the role of the lensing magnification, $\kappa$, which dominates the cross-correlations of different redshifts has been investigated in \cite{Montanari:2015rga}.

\begin{figure}[hT]
\centering
\begin{tabular}{cc}
 \includegraphics[width=.5\columnwidth]{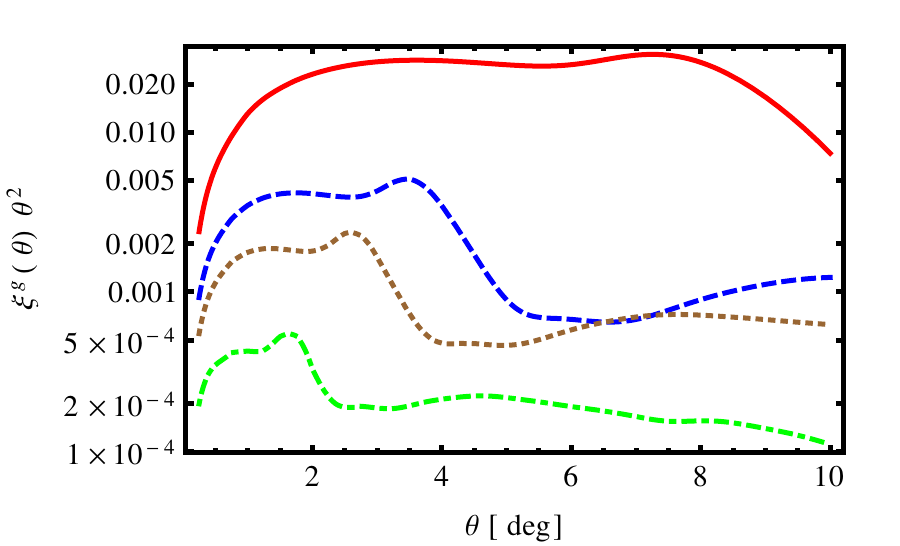} &
 \includegraphics[width=.5\columnwidth]{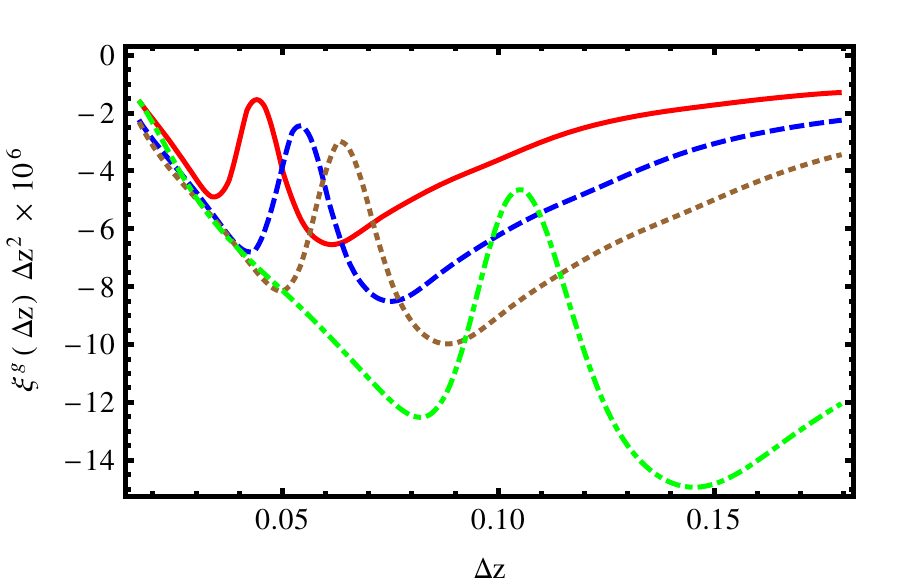} \\
 \includegraphics[width=.46\columnwidth]{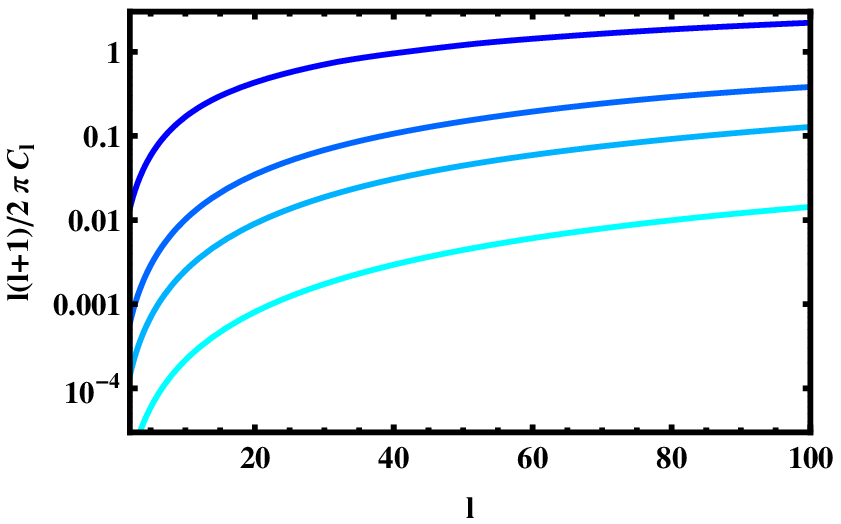} & \includegraphics[width=.46\columnwidth]{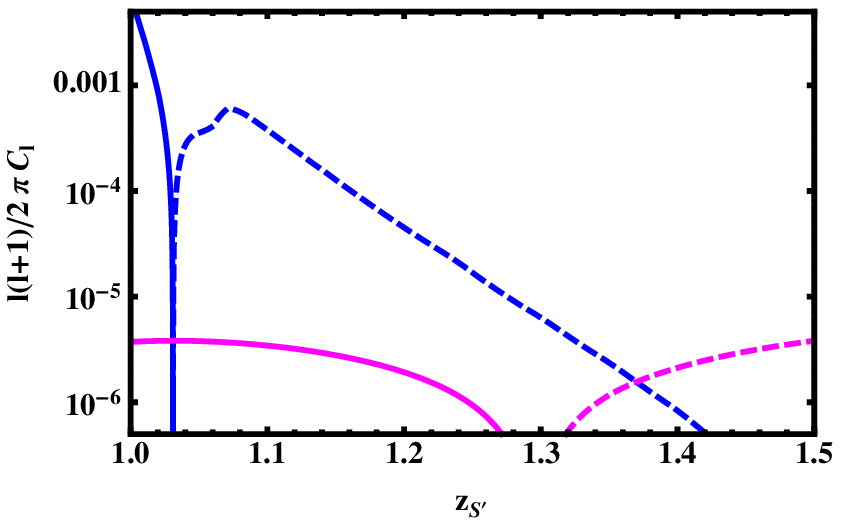} \\
\end{tabular}
\caption{\emph{Top panel:} transverse (on the left) and radial (on the right)
correlation function at $z=0.3, 0.7, 1, 3$ from top to bottom, respectively.
From \cite{Montanari:2012me}.
\emph{Bottom panel:} the transverse power spectra at $z=0.1, 0.5, 1, 3$ from top to bottom, respectively  (on the left), and the radial one for $\ell=20$ and $z_1=1$ as function of $z_2$ (on the right). The standard, non-relativistic terms in blue, the relativistic corrections from lensing in magenta. 
From \cite{Bonvin:2011bg}.
}
\label{fig:xi_Cl}
\end{figure}

The power spectra for $z_1=z_2 = 0.1,~0.5,~1$ and $3$ are shown on Fig.~\ref{fig:xi_Cl}, lower, left panel, and $C_{20}(1,z_2)$ is plotted as a function of $z_2$ in  Fig.~\ref{fig:xi_Cl}, lower, right panel.

These directly observed quantities are functions of three variables, $\theta, z_1$ and $z_2$. 
Therefore they are harder to infer from observations than a function of only one variable. Especially, the shot noise problem is much more severe. However, they also contain more information. They combine in a non-trivial way clustering information given by $\xi(r)$ and geometrical information about the evolution of distances in the Universe via $r(\theta,z_1,z_2)$. The Euclid galaxy survey will be sufficiently big to beat down the significant shot noise problem and profit maximally from this cleaner cut between observations and modeling, see Section~\ref{s:Clfor} for forecasts.

As an illustration of what can be done with this correlation function, we briefly consider the baryon acoustic oscillations (BAOs). The transverse BAOs  at fixed redshift ${z}$ in $\xi(\theta,z,z)$ are shown in Fig.~\ref{fig:xi_Cl}, top left panel.

The radial BAO, the correlation along the line of sight, $\xi(\theta=0,z-\Delta z/2,z+\Delta z/2)$ as a function of the redshift separation $\Delta z$ of galaxy pairs around the reference value ${z}$ are shown in Fig.~\ref{fig:xi_Cl}, top right panel.
To measure the radial correlation function, a spectroscopic determination of the
redshift is required to resolve the BAO feature.
The transverse and radial correlation functions can be used to determine the angular and redshift extension ($\theta_{\rm BAO}(z)$ and $\Delta z_{\rm BAO}(z)$, respectively) of the BAOs as function of redshift, which determine
\begin{equation}
F(z)\equiv (1+z)H(z)D_A(z) = \frac{\Delta z_{\rm BAO}(z)}{\theta_{\rm BAO}(z)} \equiv F^{AP}(z) \,.
\end{equation}
where $H(z)$ is the Hubble parameter and $D_A(z)$ is the angular diameter
distance.
Combining this with a measurement of the luminosity distance
$D_L(z)=(1+z)^2D_A(z)$, e.g. from supernova type 1a data~\cite{Suzuki:2011hu}, we can break the
degeneracy between $H(z)$ and $D_A(z)$.
This allows us to test the  relation
$$  D_A(z) = \frac{1}{z+1}\int_0^z\frac{dz'}{H(z')}
\quad \mbox{or}\quad
F(z) = \int_0^zdz'\frac{H(z)}{H(z')}\,$$
which must be valid, if the geometry of our Universe is close to a flat Friedmann-Lema\^\i tre metric.




\section{Forecasts for Euclid}
\label{euclid-forecasts}


Here\footnote{Forecasts are not updated in this version with respect to 2012.} we describe forecasts for the constraints on modified gravity
parameters which Euclid observations should be able to achieve. We begin
with reviewing the relevant works in literature. Then, after we define
 our ``Euclid model'', i.e., the main specifics of the redshift
and weak lensing survey, we illustrate a number of Euclid forecasts
obtained through a Fisher matrix approach.

\subsection{A review of forecasts for parametrized modified gravity with Euclid}

Heavens et al.~\cite{Heavens-etal:2007} have used Bayesian evidence to distinguish between
models, using the Fisher matrices for the parameters of interest. This study
calculates the ratio of evidences $B$ for a 3D weak lensing analysis of the full
Euclid survey, for a dark-energy model with varying equation of state, and
modified gravity with additionally varying growth parameter $\gamma$. They find
that Euclid can decisively distinguish between, e.g., DGP and dark energy, with
$|\ln B|\simeq 50$. In addition, they find that it will be possible to
distinguish any departure from GR which has a difference in $\gamma$ greater
than $\simeq 0.03$. A phenomenological extension of the DGP
model~\citep{Dvali:2003rk, Afshordi:2008rd} has also been tested with
Euclid. Specifically, \citet{Camera:2011mg} found that it will be
possible to discriminate between this modification to gravity from
$\Lambda$CDM at the $3\sigma$ level in a wide range of angular scale,
approximately $1000\lesssim\ell\lesssim4000$.

\epubtkImage{RB_WL-g0-S0.png}{%
\begin{figure}[htb]
\centerline{\includegraphics[width=9cm]{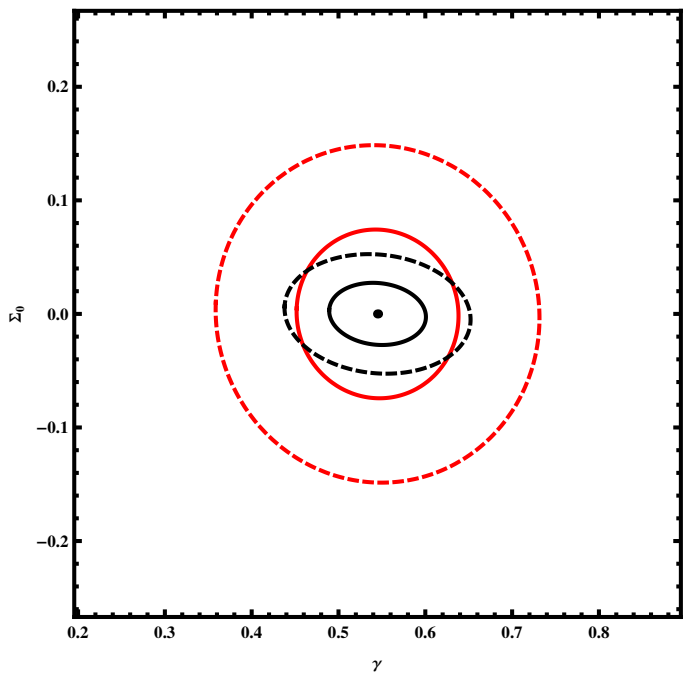}}
\caption{Marginalized $\gamma-\Sigma_0$ forecast for weak lensing only
analysis with Euclid. Here $\Sigma_0$ is defined from $\Sigma =
1+\Sigma_0 a$ and $\Sigma$, defined via Eq.~\ref{Sigma-wl-isw}, is
related to the WL potential. Black contours correspond to
$\ell_{\max}=5000$, demonstrating an error of 0.089$(1\sigma)$ on
$\Sigma_0$, whereas the red contours correspond to $\ell_{\max}=500$
giving an error of 0.034. In both cases, the inner and outer contours
are $1\sigma$ and $2\sigma$ respectively. GR resides
at [0.55, 0], while DGP resides at [0.68, 0].}
 \label{fig:thomas}
\end{figure}}

Thomas et al.~\cite{2009MNRAS.395..197T} construct Fisher matrix forecasts for the Euclid
weak lensing survey, shown in Figure~\ref{fig:thomas}. The constraints obtained
depend on the maximum wavenumber which we are confident in using;
$\ell_{\max}=500$ is relatively conservative as it probes the linear regime where we can
hope to analytically track the growth of structure; $\ell_{\max}=10000$ is more
ambitious as it includes nonlinear power, using the~\cite{Smith2003}
fitting function. This will not be strictly correct, as the fitting function was
determined in a GR context. Note that $\gamma$ is not very sensitive
to $\ell_{\max}$, while $\Sigma_0$, defined in~\cite{Amendola:2007rr}
as $\Sigma = 1 + \Sigma_0 a$ (and where $\Sigma$ is defined in
Eq.~\ref{Sigma-wl-isw}) is measured much more accurately in the nonlinear regime.

Amendola et al.~\cite{Amendola:2007rr} find Euclid weak lensing constraints for a more
general parameterization that includes evolution. In particular, $\Sigma(z)$ is
investigated by dividing the Euclid weak lensing survey into three redshift bins
with equal numbers of galaxies in each bin, and approximating that $\Sigma$ is
constant within that bin. Since $\Sigma_1$, i.e., the value of $\Sigma$ in the $a=1$ bin (present-day) is degenerate with the amplitude of matter
fluctuations, it is set to unity. The study finds that a
deviation from unit $\Sigma$ (i.e., GR) of 3\% can be
detected in the second redshift bin, and a deviation of 10\% is still
detected in the furthest redshift bin.

Beynon et al.~\cite{2009arXiv0910.1480B} make forecasts for modified gravity with Euclid weak
lensing including~\cite{Hu:2007pj} in interpolating between the linear
spectrum predicted by modified gravity, and GR on small scales as required by
Solar System tests. This requires parameters $A$ (a measure of the abruptness of
transitioning between these two regimes), $\alpha_1$ (controlling the
$k$-dependence of the transition) and $\alpha_2$ (controlling the $z$-dependence
of the transition). 

\epubtkImage{RB_EuclidContours.png}{%
\begin{figure}[htbp]
\centerline{\includegraphics[width=0.8\textwidth]{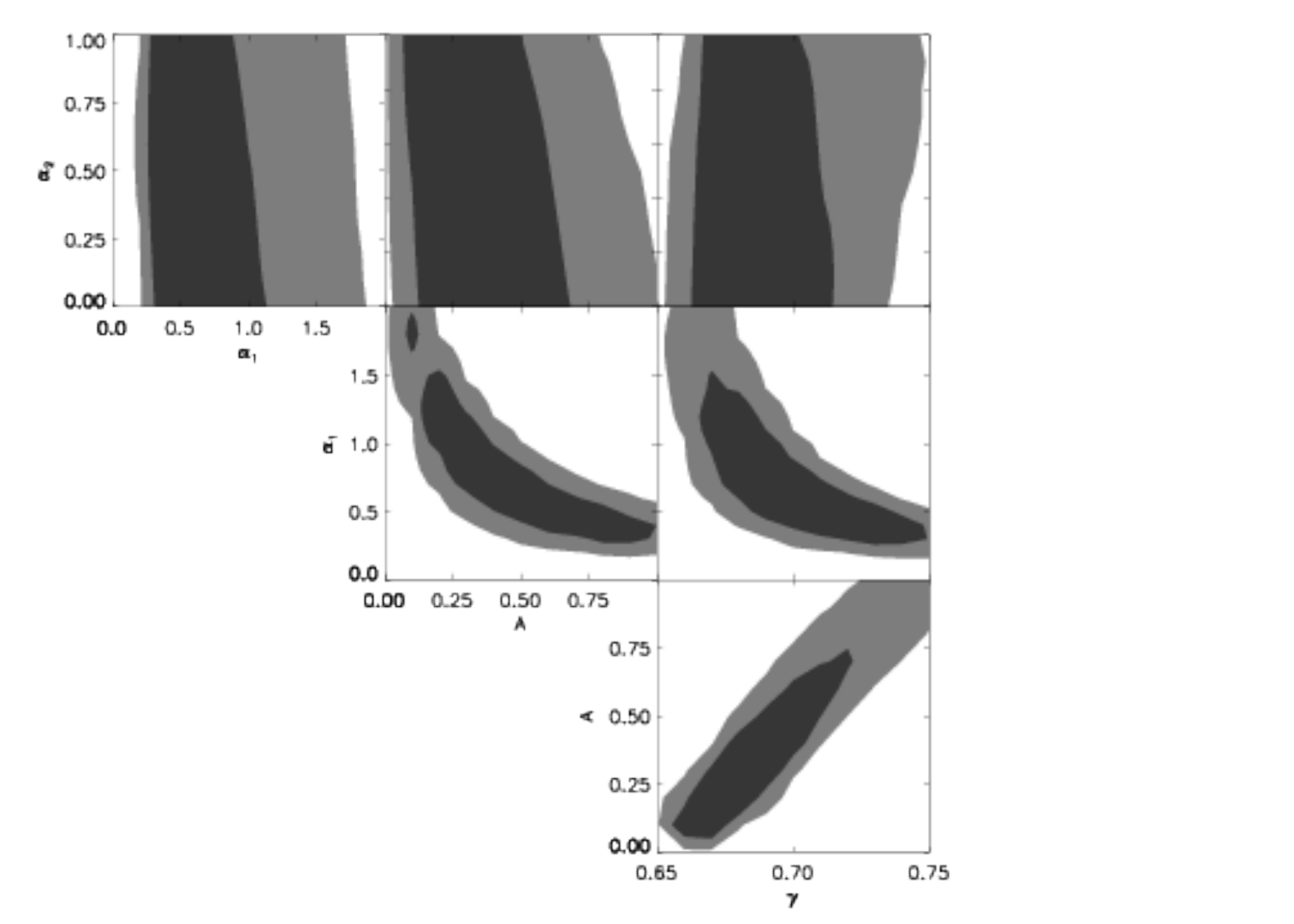}}
\caption{Constraints on $\gamma$, $\alpha_1$, $\alpha_2$ and $A$ from Euclid,
using a DGP fiducial model and 0.4 redshift bins between 0.3 and 1.5 for the
central cosmological parameter values fitting WMAP+BAO+SNe.}
 \label{fig:beynon}
\end{figure}}

The forecasts for modified gravity parameters are shown in
Figure~\ref{fig:beynon} for the Euclid lensing data. Even with this
larger range of parameters to fit, Euclid provides a measurement of
the growth factor $\gamma$ to within 10\%, and also allows some
constraint on the $\alpha_1$ parameter, probing the physics of
nonlinear collapse in the modified gravity model. 

Finally, Song et al.~\cite{Song:2010rm} have shown forecasts for measuring $\Sigma$ and $\mu$ using
both imaging and spectroscopic surveys. They combine 15,000
square-degree lensing data (corresponding to~\cite{euclidyellowbook} rather than to the updated~\cite{euclidredbook}) with the peculiar velocity dispersion measured from redshift
space distortions in the spectroscopic survey, together with stringent
background expansion measurements from the CMB and supernovae. They find that
for simple models for the redshift evolution of $\Sigma$ and $\mu$, both
quantities can be measured to 20\% accuracy.

\subsection{Euclid surveys}
\label{sec:baofm_survey}

The Euclid mission will produce a catalog of up to 30 million galaxy redshifts with $f_{H_\alpha} > 3 \times 10^{-16}$ and 50 million with $f_{H_\alpha} > 2 \times 10^{-16}$
and an imaging survey that should allow to estimate the galaxy ellipticity
of up to 1.5 billion galaxy images with photometric redshifts. Here we discuss these surveys and fix their 
main properties into a ``Euclid model'', i.e., an approximation to the real
Euclid survey that will be used as reference mission in the following.
\\

\noindent\textbf{Modeling the Redshift Survey.}\\
\noindent
The main goals of next generation redshift surveys will be
to constrain the dark-energy parameters and  to explore models
alternative to standard Einstein gravity. For these purposes they will
need to consider very large volumes that encompass
$z\sim 1$, i.e., the epoch at which dark energy  started dominating
the energy budget, spanning a range of epochs large enough to
provide a sufficient leverage to discriminate among
competing models at different redshifts.

Here we consider a survey covering a large fraction of the
extragalactic corresponding to $\sim 15000\mathrm{\ deg}^2$ capable to
measure a large number of galaxy redshifts out to $z\sim 2$. A
promising observational strategy is to target H$\alpha$
emitters at near-infrared  wavelengths (which implies $z>0.5$) since
they guarantee both relatively dense sampling (the space density of
this population is expected to increase out to $z\sim 2$) and an 
efficient method to measure the redshift of the object.
The limiting flux of the survey should be the tradeoff between
the requirement of minimizing the shot noise, the contamination by
other lines (chiefly among them the [O\,{\sc ii}] line), and that of 
maximizing the so-called efficiency $\varepsilon$, i.e., the fraction of
successfully measured redshifts.
To minimize shot noise one should obviously strive for a low flux. Indeed, 
\cite{geach10} found that a limiting flux  $f_{\mathrm{H}\alpha} \ge
1\times10^{-16}\mathrm{\ erg\ cm^{-2}\ s^{-1}}$ would be
able to balance shot noise and cosmic variance
out to $z=1.5$. However, simulated observations of mock  H$\alpha$ galaxy
spectra have shown that $\varepsilon$ ranges between 30\% and 60\% (depending
on the redshift) for a limiting flux $f_{\mathrm{H}\alpha}\ge
3\times10^{-16}\mathrm{\ erg\ cm^{-2}\ s^{-1}}$~\citep{euclidredbook}. Moreover,
contamination from  [O\,{\sc ii}] line drops from 12\% to 1\% when the limiting
flux increases from $1\times10^{-16}$ to
$5\times10^{-16}\mathrm{\ erg\ cm^{-2}\ s^{-1}}$~\citep{geach10}. 

Taking all this into account, in order to reach the top-level science
requirement on the number density of H$\alpha$ galaxies, the average
effective H$\alpha$ line flux limit from a 1-arcsec diameter source
shall be lower than or equal to
$3\times10^{-16}\mathrm{\ erg\ cm^{-2}\ s^{-1}}$. However, a slitless
spectroscopic survey has a success rate in measuring redshifts that is
a function of the emission line flux. As such, the Euclid survey
cannot be characterized by a single flux limit, as in conventional
slit spectroscopy.

We use the number density of H$\alpha$ galaxies at a given redshift, $n(z)$,
estimated using the latest empirical data (see Figure~3.2
of~\cite{euclidredbook}), where the values account for redshift -- and
flux -- success rate, to which we refer as our reference efficiency
$\varepsilon_r$.

However, in an attempt to bracket current uncertainties in modeling
galaxy surveys, we consider two further scenarios, one where the
efficiency is only the half of $\varepsilon_r$ and one where it is
increased by a factor of 40\%. Then we define the following cases:

\begin{itemize}
\item \textit{Reference case (ref.)}. Galaxy number density $n(z)$ which
  include efficiency $\varepsilon_r$ (column $n_2(z)$ in
  Table~\ref{tab:n_z}).

\item \textit{Pessimistic case (pess.)}. Galaxy number density $n(z)\cdot
  0.5$, i.e., efficiency is $\varepsilon_{r}\cdot 0.5$ (column
  $n_3(z)$ in Table~\ref{tab:n_z}).

\item \textit{Optimistic case (opt.)}. Galaxy number density $n(z)\cdot
  1.4$, i.e., efficiency is $\varepsilon_{r}\cdot 1.4$ (column
  $n_1(z)$ in Table~\ref{tab:n_z}).
\end{itemize}

The total number of observed galaxies ranges from $2\cdot10^7$ (pess.) to
$5\cdot10^7$ (opt.) with a central value at $3\cdot10^7$ . For all cases we assume that the  error on the
measured redshift is $\Delta z=0.001(1+z)$, independent of the
limiting flux of the survey. \\

\begin{table}[htbp]
\caption[Expected galaxy number densities in units of
  $(h/\mathrm{Mpc})^{3}$ for Euclid survey.]{Expected galaxy number
  densities in units of $(h/\mathrm{Mpc})^{3}$ for Euclid survey. Let
  us notice that the galaxy number densities $n(z)$ depend on the
  fiducial cosmology adopted in the computation of the survey volume,
  needed for the conversion from the galaxy numbers $dN/dz$ to
  $n(z)$.}
\label{tab:n_z}
\centering
{\small
\begin{tabular}{cccc}
\toprule
$z$  & $n_{1}(z)$ $\times10^{-3}$  & $n_{2}(z)$ $\times10^{-3}$  &
$n_{3}(z)$ $\times10^{-3}$\\
\midrule
0.65\,--\,0.75  & 1.75  & 1.25  & 0.63\\
0.75\,--\,0.85  & 2.68  & 1.92  & 0.96\\
0.85\,--\,0.95  & 2.56  & 1.83  & 0.91\\
0.95\,--\,1.05  & 2.35  & 1.68  & 0.84\\
1.05\,--\,1.15  & 2.12  & 1.51  & 0.76\\
1.15\,--\,1.25  & 1.88  & 1.35  & 0.67\\
1.25\,--\,1.35  & 1.68  & 1.20  & 0.60\\
1.35\,--\,1.45  & 1.40  & 1.00  & 0.50\\
1.45\,--\,1.55  & 1.12  & 0.80  & 0.40\\
1.55\,--\,1.65  & 0.81  & 0.58  & 0.29\\
1.65\,--\,1.75  & 0.53  & 0.38  & 0.19\\
1.75\,--\,1.85  & 0.49  & 0.35  & 0.18\\
1.85\,--\,1.95  & 0.29  & 0.21  & 0.10\\
1.95\,--\,2.05  & 0.16  & 0.11  & 0.06\\
\bottomrule
\end{tabular}}
\end{table}

\noindent\textbf{Modeling the weak lensing survey.}
\noindent
For the weak lensing survey, we assume again a sky coverage of 15,000
square degrees.  For the number density we use the common
parameterization
\begin{equation}
n(z) = z^2 \exp(-(z/z_0)^{3/2}) \,,
\end{equation}
where $z_0 =z_{\mathrm{mean}}/1.412$ is the peak of $n(z)$ and
$z_{\mathrm{mean}}$ the median and typically we assume $z_{\mathrm{mean}}=0.9$
and a surface density of valid images of $n_g=30$ per
arcmin$^2$~\citep{euclidredbook}). We also assume that the photometric
redshifts give an error of $\Delta z=0.05(1+z)$. Other specifications
will be presented in the relevant sections.

\subsection{Forecasts for the growth rate from the redshift survey}
\label{gamma-bias-forecasts}

In this section we forecast the constraints that future observations
can put on the growth rate and on a scale-independent bias, employing
the Fisher matrix method presented in
Section~\ref{dark-energy-and-redshift-surveys}. We use the
representative Euclid survey presented in
Section~\ref{sec:baofm_survey}. We assess how well one can constrain
the bias function from the analysis of the power spectrum itself and
evaluate the impact that treating bias as a free parameter has on the
estimates of the growth factor. We estimate how errors depend on the
parametrization of the growth factor and on the number and type of
degrees of freedom in the analysis. Finally, we explicitly explore the
case of coupling between dark energy and dark matter and assess the
ability of measuring the coupling constant. Our parametrization is
defined as follows. More details can be found in~\cite{diporto10}.

\paragraph*{Equation of state.}
In order to represent the evolution of the
equation of state parameter $w$, we use the popular CPL
parameterization~\citep{chevallier01,Linder:2002et} 
\begin{equation}
w(z)=w_{0}+w_{1}\frac{z}{1+z} \,.
\label{eq:w_CPL}
\end{equation}
As a special case we will also consider the case of a constant $w$.
We refer to this as the $w$-parametrization.

\paragraph*{Growth rate.}
Here we assume that the growth rate, $f_g$, is a function of time but not
of scale. As usual, we use the simple
prescription~\citep{peebles76,Lahav-etal:1991,polarski08,Linder:2005in,wang98}
\begin{equation}
f_g=\Omega_{m}^{\gamma} \,,
\label{eq:standard}
\end{equation}
where $\Omega_{m}(z)$ is the matter density in units of the critical
density as a function of redshift. A value $\gamma\approx0.545$ reproduces
well the $\Lambda$CDM behavior while departures from this value characterize
different models. Here we explore three different parameterizations of $f_g$:

\begin{itemize}

\item \textit{$f$-parameterization}. This is in fact a non-parametric 
model in which the growth rate itself is modeled as a step-wise function
$f_g(z)=f_{i}$, specified in different redshift bins. The errors are derived on
$f_{i}$ in each $i$-th redshift bin of the survey. 

\item \textit{$\gamma$-parameterization}.
As a second case we assume
\begin{equation}
f_g\equiv\Omega_{m}(z)^{\gamma(z)} \,.
\label{eq:s_parametriz}
\end{equation}
where the $\gamma(z)$ function is parametrized as 
\begin{equation}
\gamma(z)=\gamma_{0}+\gamma_{1}\frac{z}{1+z} \,.
\label{eq:gam_CPL}
\end{equation}
As shown by~\cite{wu09,fu09}, this parameterization is more accurate than that
of Eq.~(\ref{eq:standard}) for both $\Lambda$CDM and DGP models.
Furthermore, this parameterization is especially effective to distinguish
between a $w$CDM model (i.e., a dark-energy model with a constant equation of
state) that has a negative $\gamma_{1}$ 
($-0.020\lesssim\gamma_{1}\lesssim-0.016$) and a DGP model that instead, has 
a positive $\gamma_{1}$ ($0.035<\gamma_{1}<0.042$).
In addition, modified gravity models show a strongly evolving
$\gamma(z)$~\citep{gannouji09,motohashi10,fu09}, in contrast with conventional
dark-energy models.
As a special case we also consider $\gamma=$ constant (only when $w$ also is
assumed constant), to compare our results with those of previous works.

\item \textit{$\eta$-parameterization}.
To explore models in which perturbations grow faster than in the 
$\Lambda$CDM case, like in the case of a coupling between dark energy and dark
matter~\citep{diporto08}, we consider a model in which $\gamma$ is constant and
the growth rate varies as
\begin{equation}
f_g\equiv\Omega_{m}(z)^{\gamma}(1+\eta) \,,
\label{eq:eta_paramet}
\end{equation}
where $\eta$ quantifies the strength of the coupling.
The example of the coupled quintessence model worked out by~\cite{diporto08}
illustrates this point. In that model, the numerical solution for the growth
rate can be fitted by the formula~(\ref{eq:eta_paramet}), with 
$\eta=c\beta_{c}^{2}$,
where $\beta_{c}$ is the dark energy-dark matter coupling constant
and best fit values $\gamma=0.56$ and $c=2.1$.
In this simple case, observational constraints over
$\eta$ can be readily transformed into constraints over $\beta_{c}$.

\end{itemize}

\paragraph*{Reference Cosmological Models.}
We assume as reference model a ``pseudo'' $\Lambda$CDM, where the
growth rate values are obtained from Eq.~(\ref{eq:standard}) with
$\gamma=0.545$ and $\Omega_m(z)$ is given by the standard
evolution. Then $\Omega_m(z)$ is completely specified by setting
$\Omega_{m,0}=0.271$, $\Omega_k=0$, $w_0=-0.95$, $w_1=0$. When the
corresponding parameterizations are employed, we choose as fiducial
values $\gamma_{1}=0$ and $\eta=0$, We also assume a primordial slope
$n_s=0.966$ and a present normalization $\sigma_8=0.809$.

One of the goals of this work is to assess whether the
analysis of the power spectrum in redshift-space
can distinguish the fiducial model from alternative
cosmologies, characterized by their own set of parameters
(apart from $\Omega_{m,0}$ which is set equal to 0.27 for all of
them). The alternative models that we consider in this work are:

\begin{itemize}

\item \textit{DGP model}. We consider the flat space case studied
in~\cite{maartens06}.
When we adopt this model then we set
$\gamma_{0}=0.663$, $\gamma_{1}=0.041$~\citep{fu09} or
$\gamma=0.68$~\citep{Linder:2007hg} and $w=-0.8$ when $\gamma$ and $w$ are assumed
constant.

\item \textit{$f(R)$ model}. Here we consider different classes of $f(R)$ models: i) the one proposed in 
\cite{Hu07}, depending on two parameters, $n$ and $\mu$ in Eq.~\ref{Bmodel}, which we fix to
$n=0.5,1,2$ and $\mu=3$. For the model with $n=2$ we assume $\gamma_{0}=0.43$,
$\gamma_{1}=-0.2$, values that apply quite generally in the limit of small
scales (provided they are still linear, see \cite{gannouji09}) or $\gamma=0.4$
and $w=-0.99$. Unless differently specified, we will always refer to this specific model when we mention comparisons to a single $f(R)$ model. ii) The model of Eq.~\ref{Cmodel}, proposed in \cite{Star07}, where we fix $\mu=3$ and $n=2$, which shows a very similar behavior to the previous one. iii) The one proposed in \cite{Tsuji08}, Eq.~\ref{Dmodel}, fixing $\mu=1$.

\item \textit{Coupled dark-energy (CDE) model}. This is the coupled model 
proposed by \cite{Amendola:1999er,Wetterich:1994bg}. In this case we assume
$\gamma_{0}=0.56$, $\eta=0.056$ (this value comes from putting $\beta_{c}=0.16$
as coupling, which is of the order of the maximal value allowed by CMB
constraints)~\citep{amendola_quercellini_2003}. As already explained, this model cannot be
reproduced by a constant $\gamma$. Forecasts on coupled quintessence
based on \cite{Amendola:2011ie, Amendola:1999er, Pettorino:2008ez} are
discussed in more detail in Section~\ref{cdeforecast}.

\end{itemize}

For the fiducial values of the bias parameters in every bin, we assume 
$b(z)=\sqrt{1+z}$ (already used in~\cite{rassat08}) since this function provides
a good fit to H${\alpha}$ line galaxies with luminosity
$L_{\mathrm{H}\alpha}=10^{42}\mathrm{\ erg^{-1}\ s^{-1}\ h^{-2}}$ modeled by~\cite{orsi10}
using the semi-analytic \textit{GALFORM} models of~\cite{baugh05}.
For the sake of comparison, we will also consider the case of constant
$b=1$ corresponding to the rather unphysical case of a redshift-independent
population of unbiased mass tracers.

The fiducial values for $\beta$ are computed through
\begin{equation}
\beta^F(z) = \frac{\Omega_m^F(z)^{\gamma^F}}{b^F(z)}=\frac{f_g^F}{b^F} \,.
\label{eq:beta_sb}
\end{equation}

Now we express the growth function $G(z)$ and the redshift distortion parameter
$\beta(z)$ in terms of the growth rate $f_g$ (see Eqs.~(\ref{eq:gengz}),
(\ref{eq:beta_sb})). When we compute the derivatives of the spectrum
in the Fisher matrix $b(z)$ and $f_g(z)$ are considered as independent
parameters in each redshift bin. In this way we can compute the errors
on $b$ (and $f_g$) self consistently by marginalizing over
all other parameters.

Now we are ready to present the main result of the Fisher matrix analysis .
We note that in all tables below we always quote errors at 68\% probability
level and draw in the plots the probability regions at 68\% and/or 95\% (denoted
for shortness as 1 and 2$\sigma$ values). Moreover, in all figures, all the
parameters that are not shown have been marginalized over or fixed to a fiducial
value when so indicated.

\paragraph*{Results for the $f$-parameterization.}

The total number of parameters that enter in the Fisher matrix analysis is 45:
5 parameters that describe the background cosmology
($\Omega_{m,0}h^{2},\Omega_{b,0}h^{2},$ $h$, $n$, $\Omega_{k}$)
plus 5 $z$-dependent parameters specified in 8 redshift bins evenly spaced in
the range $z=[0.5,2.1]$. They are $P_{\textrm{s}}(z)$, $D(z)$, $H(z)$, $f_g(z)$,
$b(z)$. However, since we are not interested in constraining $D(z)$ and $H(z)$, we always project them to the set of parameters they depend on (as explained in \cite{seo03}) instead of marginalizing over, so extracting more information on the background parameters.

The fiducial growth function $G(z)$ in the $(i+1)$-th redshift bin is evaluated
from a step-wise, constant growth rate $f_g(z)$ as
\begin{equation}
G(z)=\exp\left\{-\int_{0}^{z}f_g(z)\frac{dz}{1+z}\right\} =
\prod_{i}\left(\frac{1+z_{i}}{1+z_{i-1}}\right)^{-f_{i}}\left(\frac{1+z}{1+z_{i}}\right)^{-f_{i+1}}
\,.
\label{eq:gengz}
\end{equation}
To obtain the errors on $s_{i}$ and $b_{i}$ we compute the elements of the
Fisher matrix and marginalize over all other parameters. In this case one is
able to obtain, self-consistently, the error on the bias and on the growth
factor at different redshifts, as detailed in Table~\ref{tab:sigma_bias_s_bint}.
In Figure~\ref{fig:s_b_ref_cp} we show the contour plots at 68\% and
95\% of probability for all the pairs $s(z_i)-b(z_i)$ in several
redshift bins (with $b=\sqrt{1+z}$), where $z_{i}$'s are the central
values of the bins. We do not show the ellipses for all the 14 bins to
avoid overcrowding. 

Table~\ref{tab:sigma_bias_s_bint} illustrates one important result: through the analysis of
the redshift-space galaxy power spectrum in a next-generation
Euclid-like survey, it will be possible to measure galaxy biasing
in $\Delta z=0.1$ redshift bins with less than 1.6\% error, provided
that the bias function is independent of scale. 
We also tested a different choice for the fiducial form of the bias: $b(z)=1$
finding that the precision in
measuring the bias as well as the other parameters has a very little dependence
on the $b(z)$ form.
Given the robustness of the results on the choice of $b(z)$ 
in the following we only consider the $b(z)=\sqrt{1+z}$ case.

In Figures~\ref{fig:s_err_bint_revised1} and \ref{fig:s_err_bint_revised2} we show the errors on
the growth rate $f_g$ as a function of redshift,
overplotted to our fiducial $\Lambda$CDM (green solid
curve). The three sets of error bars are plotted in correspondence of
the 14 redshift bins and refer (from left to right) to the \emph{Optimistic, 
Reference} and \emph{Pessimistic}
cases, respectively. The other curves show the expected growth rate in three alternative cosmological
models: flat DGP (red, longdashed curve), CDE (purple, dot-dashed curve) and different $f(R)$ models (see description in the figure captions).
This plot clearly illustrates the ability of next generation surveys to 
distinguish between alternative models, even in the less favorable choice of
survey parameters.

The main results can be summarized as follows.
\begin{enumerate}

\item The ability of measuring the biasing function is not too sensitive
to the characteristic of the survey ($b(z)$ can be constrained to within 
1\% in the \emph{Optimistic} scenario and up to 1.6\% in the \emph{Pessimistic}
one) provided that the bias function is independent of
scale. Moreover, we checked that the precision in measuring the bias
has a very little dependence on the $b(z)$ form.

\item The growth rate $f_g$ can be estimated to within 1\,--\,2.5\%
in each bin for the \emph{Reference case} survey with
no need of estimating the bias function $b(z)$ from some
dedicated, independent analysis using higher order statistics~\citep{Verde02}
or full-PDF analysis~\citep{sigad00}.

\item The estimated errors on $f_g$ depend weakly on the fiducial
model of $b(z)$.
\end{enumerate}

\epubtkImage{RB_s_b_ref_cp.png}{%
\begin{figure}[htbp]
\centerline{\includegraphics[width=9cm]{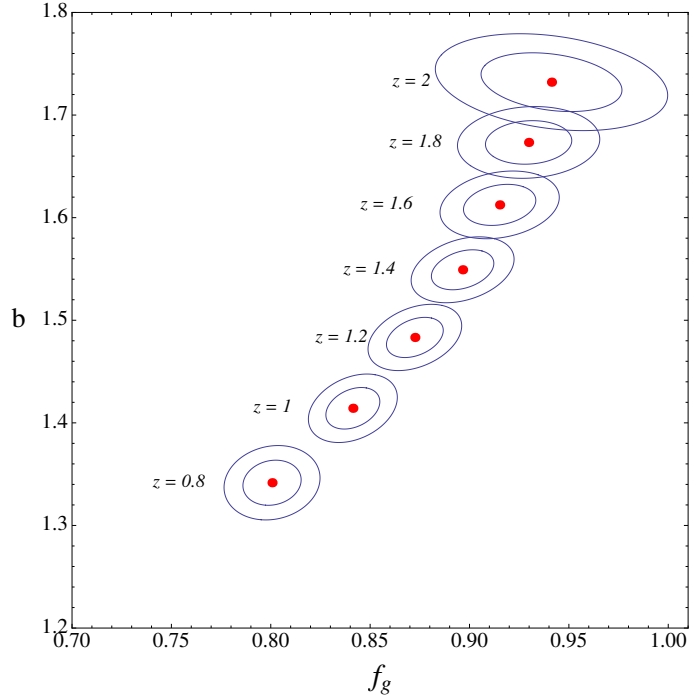}}
\caption{Contour plots at 68\% and 98\% of probability for the pairs
  $s(z_i)-b(z_i)$ in 7 redshift bins (with $b=\sqrt{1+z}$). The
  ellipses are centered on the fiducial values of the growth rate and
  bias parameters, computed in the central values of the bins,
  $z_{i}$.}
\label{fig:s_b_ref_cp}
\end{figure}}

\epubtkImage{fr3n.png}{%
\begin{figure}[htbp]
\centerline{\includegraphics[width=\textwidth]{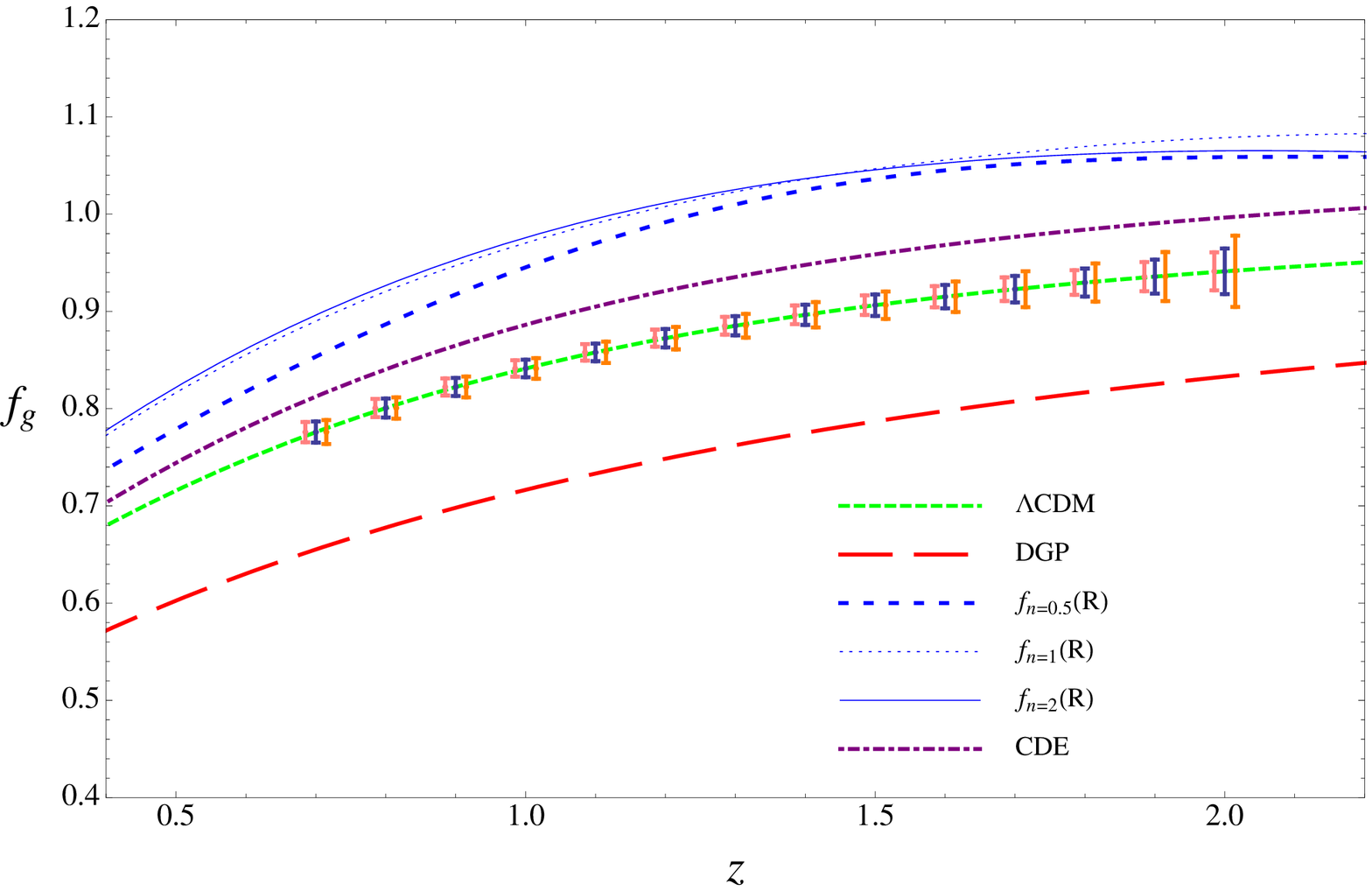}}
\caption{Expected constraints on the growth rates in each redshift
  bin. For each $z$ the central error bars refer to the \emph{Reference
    case} while those referring to the \emph{Optimistic} and \emph{
    Pessimistic} case have been shifted by $-0.015$ and $+0.015$
  respectively. The growth rates for different models are
    also plotted: $\Lambda$CDM (green tight shortdashed curve), flat
    DGP (red longdashed curve) and a model with coupling between dark
    energy and dark matter (purple, dot-dashed curve). The blue
    curves (shortdashed, dotted and solid) represent the $f(R)$ model
    by \cite{Hu07}, Eq.~\ref{Bmodel} with $n=0.5,1,2$ respectively and $\mu=3$. The plot shows that
    it will be possible to distinguish these models with next
    generation data.}
\label{fig:s_err_bint_revised1} 
\end{figure}}

\epubtkImage{fr3models.png}{%
\begin{figure}[htbp]
\centerline{\includegraphics[width=\textwidth]{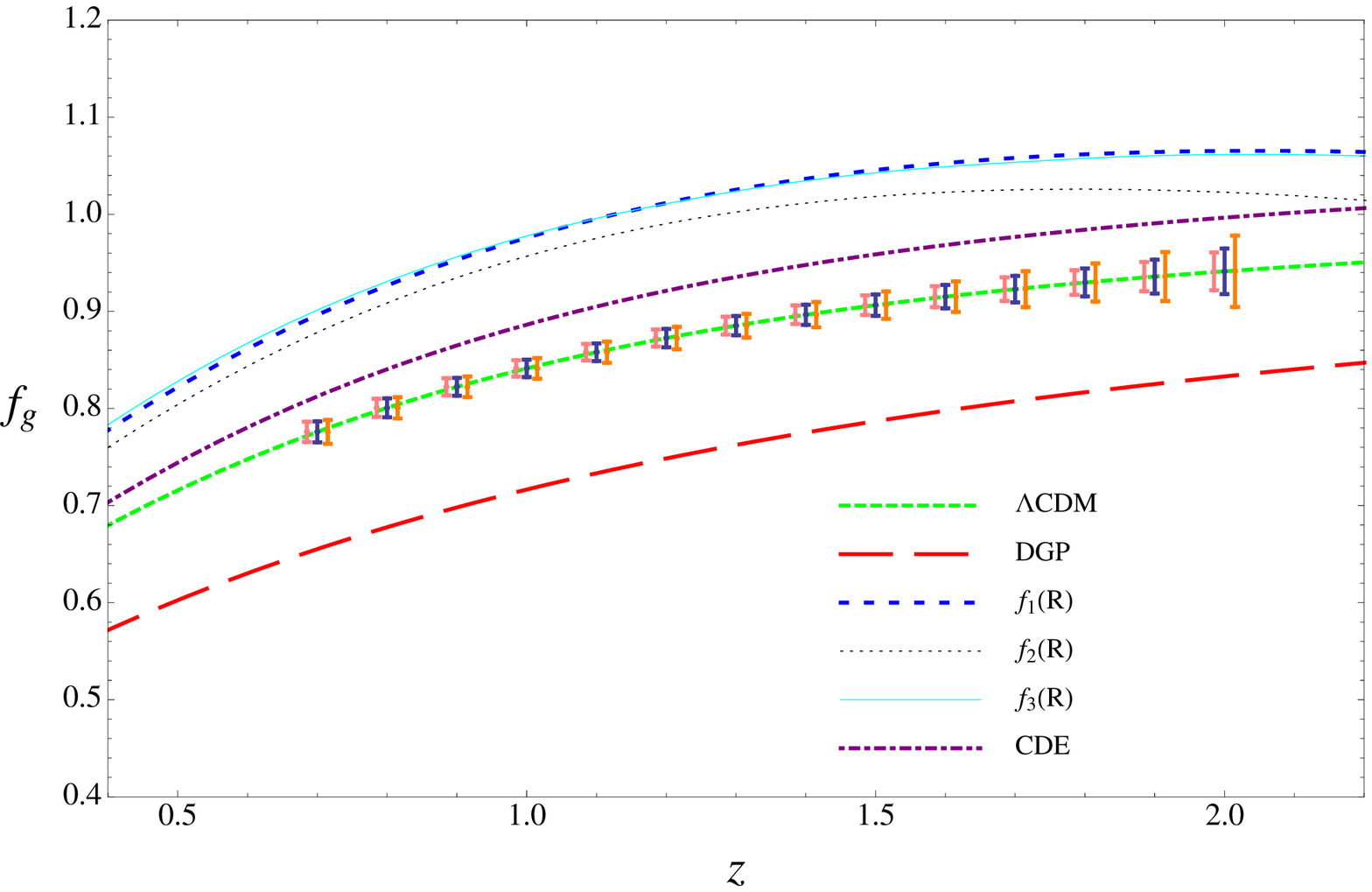}}
\caption{Expected constraints on the growth rates in each redshift
 bin. For each $z$ the central error bars refer to the
  \emph{Reference case} while those referring to the \emph{Optimistic}
  and \emph{Pessimistic} case have been shifted by $-0.015$ and $+0.015$
  respectively. The growth rates for different models are
  also plotted: $\Lambda$CDM (green tight shortdashed curve), flat
    DGP (red longdashed curve) and a model with coupling between dark
    energy and dark matter (purple, dot-dashed curve). Here we plot
    again the $f(R)$ model by \cite{Hu07}, Eq.~\ref{Bmodel}, with $n=2$ and $\mu=3$ (blue shortdashed
   curve) together with the model by \cite{Star07}, Eq.~\ref{Cmodel}, with $n=2$ and $\mu=3$ (cyan solid curve)
    and the one by \cite{Tsuji08}, Eq.~\ref{Dmodel}, with $\mu=1$ (black dotted curve). Also in this
    case it will be possible to distinguish these models with next
    generation data.
}
\label{fig:s_err_bint_revised2}
\end{figure}}

\begin{table}[htbp]
\caption{$1\sigma$ marginalized errors for the bias and the growth
  rates in each redshift bin.}
\label{tab:sigma_bias_s_bint}
\centering
{\small
\begin{tabular}{c ccc c c c ccc}
\toprule 
z & \multicolumn{3}{c}{$\sigma_{b}$} & $b^F$ & $z$ & $f_g^F$ & \multicolumn{3}{c}{$\sigma_{f_g}$}\\
\midrule
~ & ref.  & opt.  & pess.  & ~ & ~ & ~ & ref.  & opt.  & pess.\\
\midrule
0.7 & 0.016 & 0.015 & 0.019 & 1.30 & 0.7 & 0.76  & 0.011  & 0.010 & 0.012\\
0.8 & 0.014 & 0.014 & 0.017 & 1.34 & 0.8 & 0.80  & 0.010  & 0.009 & 0.011\\
0.9 & 0.014 & 0.013 & 0.017 & 1.38 & 0.9 & 0.82  & 0.009  & 0.009 & 0.011\\
1.0 & 0.013 & 0.012 & 0.016 & 1.41 & 1.0 & 0.84  & 0.009  & 0.008 & 0.011\\
1.1 & 0.013 & 0.012 & 0.016 & 1.45 & 1.1 & 0.86  & 0.009  & 0.008 & 0.011\\
1.2 & 0.013 & 0.012 & 0.016 & 1.48 & 1.2 & 0.87  & 0.009  & 0.009 & 0.011\\
1.3 & 0.013 & 0.012 & 0.016 & 1.52 & 1.3 & 0.88  & 0.010  & 0.009 & 0.012\\
1.4 & 0.013 & 0.012 & 0.016 & 1.55 & 1.4 & 0.89  & 0.010  & 0.009 & 0.013\\
1.5 & 0.013 & 0.012 & 0.016 & 1.58 & 1.5 & 0.91  & 0.011  & 0.010 & 0.014\\
1.6 & 0.013 & 0.012 & 0.016 & 1.61 & 1.6 & 0.91  & 0.012  & 0.011 & 0.016\\
1.7 & 0.014 & 0.013 & 0.017 & 1.64 & 1.7 & 0.92  & 0.014  & 0.012 & 0.018\\
1.8 & 0.014 & 0.013 & 0.018 & 1.67 & 1.8 & 0.93  & 0.014  & 0.013 & 0.019\\
1.9 & 0.016 & 0.014 & 0.021 & 1.70 & 1.9 & 0.93  & 0.017  & 0.015 & 0.025\\
2.0 & 0.019 & 0.016 & 0.028 & 1.73 & 2.0 & 0.94  & 0.023  & 0.019 & 0.037\\
\bottomrule
\end{tabular}}
\end{table}

Next, we focus on the ability of determining $\gamma_0$
and $\gamma_1$, in the context of the \emph{$\gamma$-parameterization}
and $\gamma$, $\eta$ in the \emph{$\eta$-parameterization}.
In both cases the Fisher matrix elements have been estimated
by expressing the growth factor as
\begin{eqnarray}
G(z) & = & 
\delta_{0}\exp\left[(1+\eta)\int_{0}^{z}\Omega_{m}(z^{\prime})^{\gamma(z)}\frac{
dz^{\prime}}{1+z^{\prime}}\right] ,
\label{eq:growth_fact_def_gam}
\end{eqnarray}
where  for the \emph{$\gamma$-parameterization} we fix $\eta=0$.

\begin{itemize}

\item \textit{$\gamma$-parameterization}.
We start by considering the  case
of constant $\gamma$ and $w$ in which we set $\gamma=\gamma^F=0.545$ and
$w=w^F=-0.95$. As we will discuss in the next Section, this simple case will
allow us to cross-check our results with those in the literature.
In Figure~\ref{fig:gam_w_b1_dgp} we show the marginalized probability 
regions, at $1$ and $2\sigma$ levels, for  $\gamma$ and $w$.
The regions with different shades of green illustrate the
\emph{Reference case} for the survey whereas the
blue long-dashed and the black short-dashed ellipses
refer to the \emph{Optimistic} and \emph{Pessimistic} cases, respectively.
Errors on $\gamma$ and $w$ are listed in Table~\ref{tab:sigma_gam_w}
together with the corresponding figures of merit [FoM]
defined to be the squared inverse of the Fisher matrix determinant and therefore
equal to the inverse of the
product of the errors in the pivot point, see~\cite{Albrecht2006}.
Contours are centered on the fiducial model. The blue triangle and
the blue square represent the  flat DGP and the $f(R)$ models' predictions,
respectively.
It is clear that, in the case of constant $\gamma$ and $w$, the
measurement of the growth rate in a Euclid-like survey will
allow us to discriminate among these models. These results have been 
obtained by fixing the curvature to its fiducial value $\Omega_k=0$. If instead,
we consider 
curvature as a free parameter and marginalize over, the errors on $\gamma$ and
$w$ increase significantly, as shown in Table~\ref{tab:sigma_gam_w_omk_marg},
and yet the precision is good enough to distinguish the different models.
For completeness, we also computed the fully marginalized errors over the other
cosmological parameters for the reference survey, given in
Table~\ref{tab:cosm_par_errors}.

As a second step we considered the case in which
$\gamma$ and $w$ evolve with redshift according to
Eqs.~(\ref{eq:gam_CPL}) and~(\ref{eq:w_CPL})
and then we marginalized over the parameters $\gamma_{1}$, $w_{1}$ and  
$\Omega_k$.
The marginalized probability contours
are shown  in Figure~\ref{fig:gam_w_margover_gam1w1} in which we have shown
the three survey setups in three different panels to avoid overcrowding.
Dashed contours refer to the $z$-dependent parameterizations while red,
continuous
contours refer to the case of  constant $\gamma$ and $w$ obtained after
marginalizing over $\Omega_k$.
Allowing for time dependency increases the size of the confidence ellipses since
the Fisher matrix analysis now accounts for the additional uncertainties in the
extra-parameters $\gamma_{1}$ and $w_{1}$; marginalized error values are in
columns $\sigma_{{\gamma}_{\textrm{marg},1}}$,
$\sigma_{{w}_{\textrm{marg},1}}$ of Table~\ref{tab:sigma_gam_w_marg}.  
The uncertainty ellipses are now larger and show that DGP and
fiducial models could be distinguished at $>2\sigma$ level only if the redshift survey
parameter will be more favorable than in the \emph{Reference case}.

We have also projected the marginalized ellipses for the parameters $\gamma_{0}$
and $\gamma_{1}$  and calculated their marginalized
errors and figures of merit, which are reported in 
Table~\ref{tab:sigma_gam0_gam1}.
The corresponding uncertainties contours are shown in the right panel of
Figure~\ref{fig:gam_w_b1_dgp}. Once again we overplot the expected values
in the $f(R)$ and DGP scenarios to stress the fact that
one is expected to be able to distinguish among competing models, 
irrespective on the survey's precise characteristics.

\item \textit{$\eta$-parameterization}.

We have repeated the same analysis as for
the \emph{$\gamma$-parameterization} taking into account the 
possibility of coupling between DE and DM, i.e., we have
modeled the growth factor according to Eq.~(\ref{eq:eta_paramet}) and 
the dark-energy equation of state as in Eq.~(\ref{eq:w_CPL}) and 
marginalized over all parameters, including $\Omega_k$. The marginalized errors
are shown in columns $\sigma_{{\gamma}_{\textrm{marg},2}}$,
$\sigma_{{w}_{\textrm{marg},2}}$ of Table~\ref{tab:sigma_gam_w_marg} and the
significance contours are shown in the three panels of
Figure~\ref{fig:gam_w_margover_etaw1} which is analogous to
Figure~\ref{fig:gam_w_margover_gam1w1}. Even if the ellipses are now
larger we note that errors are still small enough to distinguish the
fiducial model from the $f(R)$ and DGP scenarios at $>1\sigma$ and
$>2\sigma$ level respectively.

Marginalizing over all other parameters we can compute the uncertainties in
the $\gamma$ and $\eta$ parameters, as listed in Table~\ref{tab:sigma_gam_eta}.
The relative confidence ellipses are shown in the left panel of
Figure~\ref{fig:gamma_eta_new_past}.
This plot shows that next generation Euclid-like surveys will be able to 
distinguish the reference model with no coupling (central, red dot) to the CDE
model proposed by~\cite{amendola_quercellini_2003} (white square) only
at the $1\mbox{\,--\,}1.5\sigma$ level.

\end{itemize}

\epubtkImage{RB_gam_w__bsqrt_dgp_fr_green_okfix-RB_gam0gam1_bsqrt_3mod_all.png}{%
\begin{figure}[htbp]
\centerline{
\includegraphics[width=0.5\textwidth]{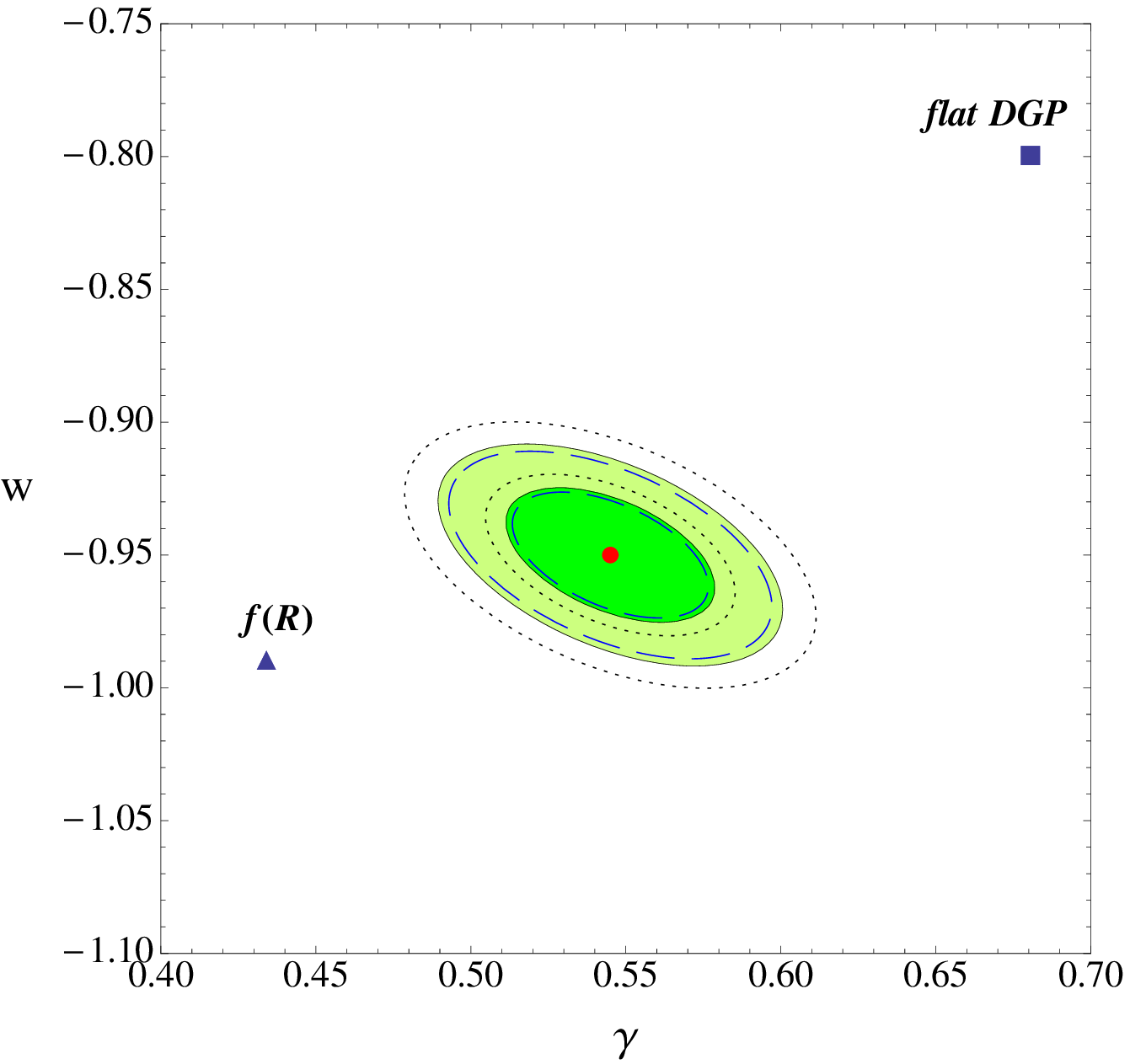} 
\includegraphics[width=0.5\textwidth]{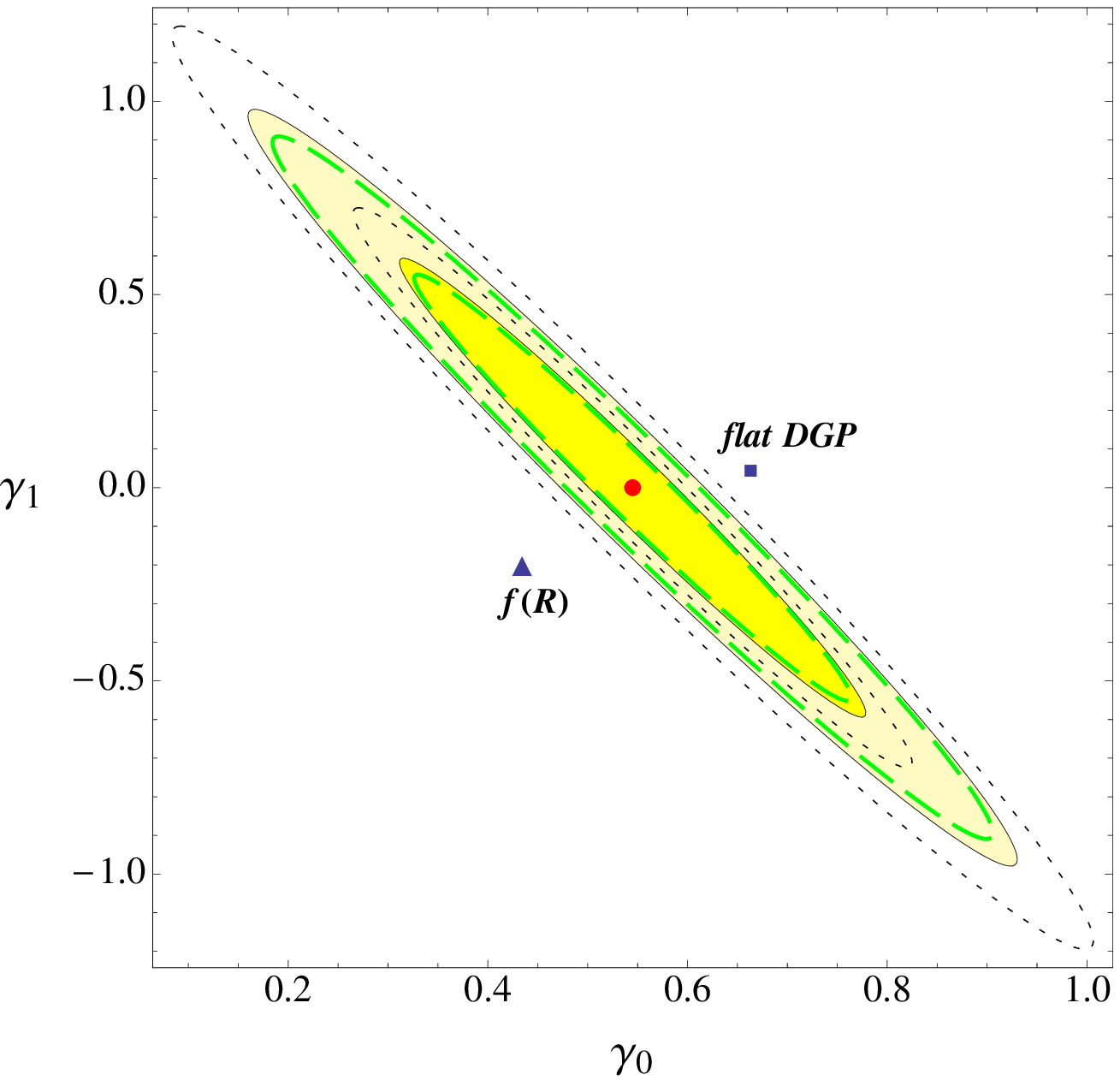} 
}
\caption{$\gamma$-parameterization. \emph{Left panel:} 1 and 2$\sigma$
  marginalized probability regions for constant $\gamma$ and $w$: the
  green (shaded) regions are relative to the \emph{Reference case},
  the blue long-dashed ellipses to the \emph{Optimistic case}, while
  the black short-dashed ellipses are the probability regions for the
  \emph{Pessimistic case}. The red dot marks the fiducial model; two
  alternative models are also indicated for comparison. \emph{Right
    panel:} 1 and 2$\sigma$ marginalized probability regions for the
  parameters $\gamma_{0}$ and $\gamma_{1}$, relative to the
  \emph{Reference case} (shaded yellow regions), to the
  \emph{Optimistic case} (green long-dashed ellipses), and to the
  \emph{Pessimistic case} (black dotted ellipses). Red dots represent
  the fiducial model, blue squares mark the DGP while triangles stand
  for the $f(R)$ model. Then, in the case of
  $\gamma$-parameterization, one could distinguish these three models
  (at 95\% probability).}
\label{fig:gam_w_b1_dgp}
\end{figure}}

\begin{table}[htbp]
\caption{Numerical values for $1\sigma$ constraints on parameters in
  Figure~\ref{fig:gam_w_b1_dgp} and figures of merit. Here we have
  fixed $\Omega_{k}$ to its fiducial value, $\Omega_k = 0$.}
\label{tab:sigma_gam_w}
\centering
\begin{tabular}{ccccc}
\toprule
~ & case  & $\sigma_{\gamma}$  & $\sigma_{w}$ & FoM\\
\midrule
$b=\sqrt{1+z}$    & ref.  & 0.02  & 0.017  & 3052\\
 with             & opt.  & 0.02  & 0.016  & 3509\\
$\Omega_{k}$ fixed & pess.  & 0.026  & 0.02  & 2106\\
\bottomrule
\end{tabular}
\end{table}

\epubtkImage{RB_gam0_w0__bsqrt_marg_over_gam1w1.png}{%
\begin{figure}[htbp]
\centerline{\includegraphics[width=\textwidth]{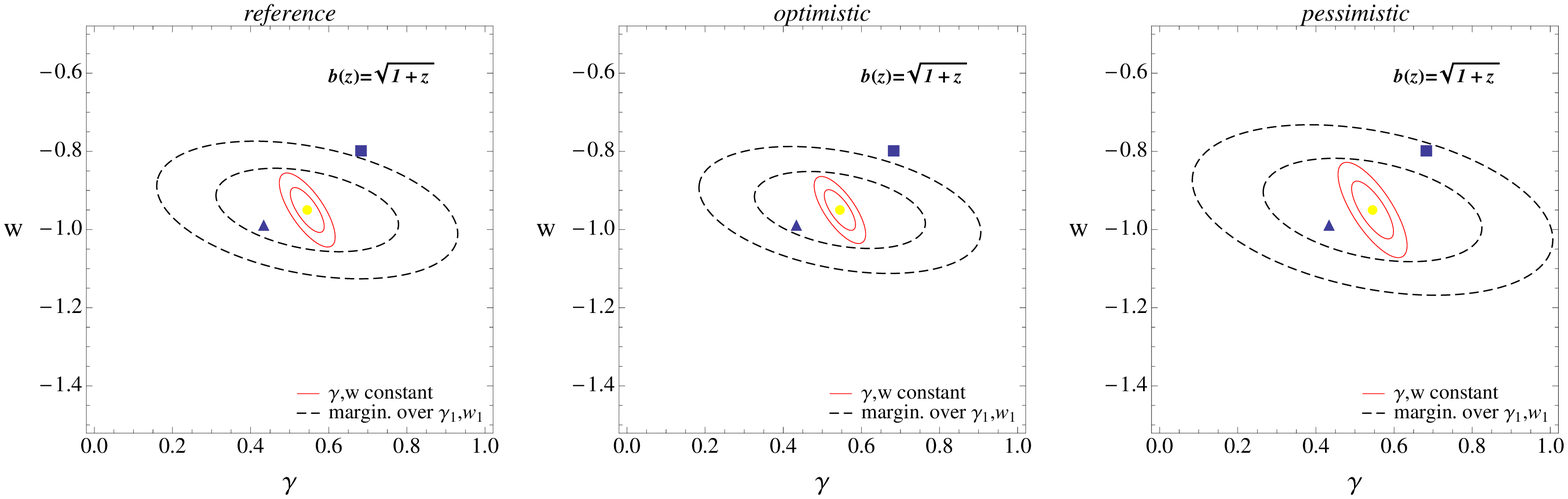}}
\caption{$\gamma$-parameterization. $1$ and $2\sigma$ marginalized
  probability regions obtained assuming constant $\gamma$ and $w$ (red
  solid curves) or assuming the parameterizations (\ref{eq:gam_CPL})
  and (\ref{eq:w_CPL}) and marginalizing over $\gamma_{1}$ and $w_{1}$
  (black dashed curves); marginalized error values are in columns
  $\sigma_{{\gamma}_{\textrm{marg},1}}$,
  $\sigma_{{w}_{\textrm{marg},1}}$ of
  Table~\ref{tab:sigma_gam_w_marg}. Yellow dots represent the fiducial
  model, the triangles a $f(R)$ model and the squares mark the flat
  DGP.}
\label{fig:gam_w_margover_gam1w1}
\end{figure}}

\begin{table}[htbp]
\caption{Numerical values for $1\sigma$ constraints on parameters
  $\gamma$ and $w$ (assumed constant), relative to the red ellipses in
  Figures~\ref{fig:gam_w_margover_gam1w1}, \ref{fig:gam_w_margover_etaw1}
  and figures of merit. Here we have marginalized over $\Omega_{k}$.}
\label{tab:sigma_gam_w_omk_marg}
\centering
\begin{tabular}{ccccc}
\toprule
bias           & case  & $\sigma_{\gamma}$ & FoM\\
\midrule
~              & ref.  & 0.03 & 0.04 & 1342\\
$b=\sqrt{1+z}$ & opt.  & 0.03 & 0.03 & 1589\\
~              & pess. & 0.04 & 0.05 &  864\\
\hline
\end{tabular}
\end{table}

\begin{table}[htbp]
\caption{Numerical values for marginalized $1\sigma$ constraints on
  cosmological parameters using constant $\gamma$ and $w$.}
\label{tab:cosm_par_errors}
\centering
\begin{tabular}{cccccccc}
\toprule
~ & case & $\sigma_{h}$ & $\sigma_{\Omega_m h^2}$  & $\sigma_{\Omega_b h^2}$ &
$\sigma_{\Omega_k}$ & $\sigma_{n_s}$ & $\sigma_{\sigma_8}$ \\
\midrule
$b=\sqrt{1+z}$ & ref.  & 0.007  & 0.002  & 0.0004 & 0.008 & 0.03 & 0.006\\
\bottomrule
\end{tabular}
\end{table}

\begin{table}[htbp]
\caption[$1\sigma$ marginalized errors for parameters $\gamma$ and $w$
  expressed through $\gamma$ and $\eta$ parameterizations.]{$1\sigma$
  marginalized errors for parameters $\gamma$ and $w$ expressed
  through $\gamma$ and $\eta$ parameterizations. Columns
  $\gamma_{0,\mathrm{marg}1},w_{0,\mathrm{marg}1}$ refer to marginalization over
  $\gamma_{1},w_{1}$ (Figure~\ref{fig:gam_w_margover_gam1w1}) while
  columns $\gamma_{0,\mathrm{marg}2},w_{0,\mathrm{marg}2}$ refer to marginalization over
  $\eta,w_{1}$ (Figure~\ref{fig:gam_w_margover_etaw1}).}
\label{tab:sigma_gam_w_marg}
\centering
\begin{tabular}{ccccc|ccc}
\toprule 
bias  & case  & $\sigma_{\gamma_{\mathrm{marg},1}}$  & $\sigma_{w_{\mathrm{marg},1}}$ & FoM  &
$\sigma_{\gamma_{\mathrm{marg},2}}$  & $\sigma_{w_{\mathrm{marg},2}}$  & FoM\\
\midrule 
 & ref.  & 0.15  & 0.07  & 97  & 0.07  & 0.07  & 216\\
$b=\sqrt{1+z}$  & opt.  & 0.14  & 0.06  & 112  & 0.07  & 0.06  & 249\\
 & pess.  & 0.18  & 0.09  & 66  & 0.09  & 0.09  & 147\\
\bottomrule
\end{tabular}
\end{table}

\begin{table}[htbp]
\caption{Numerical values for $1\sigma$ constraints on parameters in
  right panel of Figure~\ref{fig:gam_w_b1_dgp} and figures of merit.}
\label{tab:sigma_gam0_gam1}
\centering
\begin{tabular}{ccccc}
\toprule 
bias  & case  & $\sigma_{\gamma_{0}}$  & $\sigma_{\gamma_{1}}$ & FoM\\
\midrule
~ & ref.  & 0.15  & 0.4  & 87\\
$b=\sqrt{1+z}$  & opt.  & 0.14  & 0.36  & 102\\
~ & pess.  & 0.18  & 0.48  & 58\\
\hline
\end{tabular}
\end{table}

\epubtkImage{RB_gam0_w0__bsqrt_marg_over_etaw1.png}{%
\begin{figure}[htbp]
\centerline{\includegraphics[width=\textwidth]{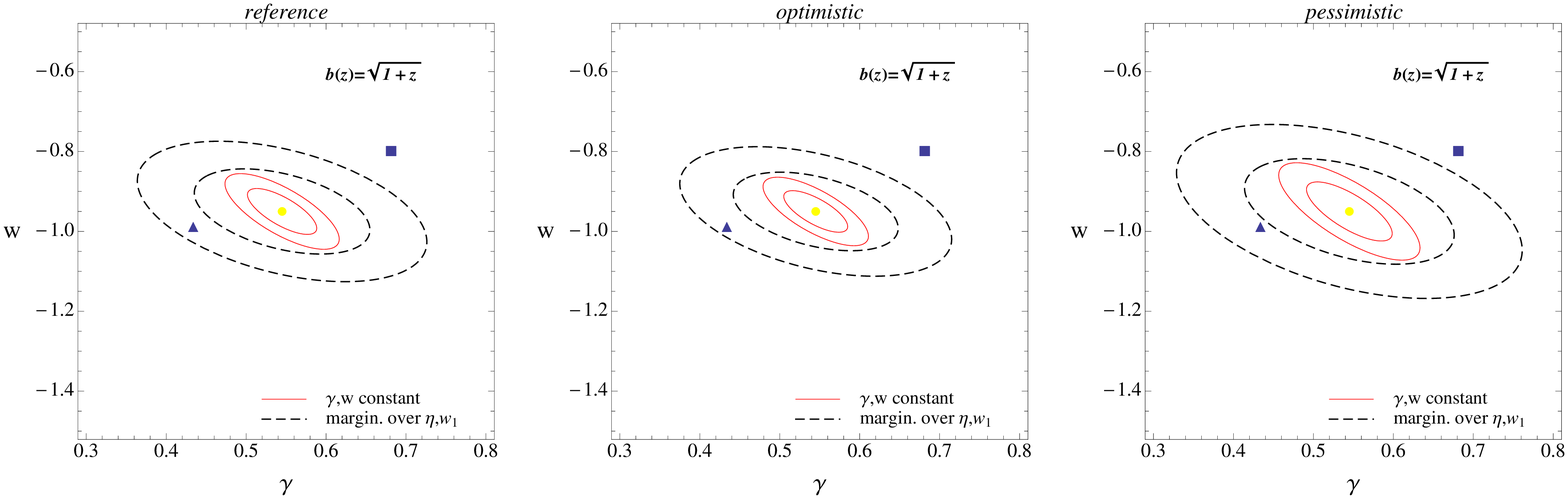}}
\caption{$\eta$-parameterization. 1 and 2$\sigma$ marginalized
  probability regions obtained assuming constant $\gamma$ and $w$ (red
  solid curves) or assuming the parameterizations
  (\ref{eq:eta_paramet}) and (\ref{eq:w_CPL}) and marginalizing over
  $\eta$ and $w_{1}$ (black dashed curves); marginalized error values
  are in columns $\sigma_{{\gamma}_{\textrm{marg},2}}$,
  $\sigma_{{w}_{\textrm{marg},2}}$ of
  Table~\ref{tab:sigma_gam0_gam1}. Yellow dots represent the fiducial
  model, the triangles stand for a $f(R)$ model and the squares mark
  the flat DGP.}
\label{fig:gam_w_margover_etaw1}
\end{figure}}

\epubtkImage{RB_gamma_eta_3cases_bsqrt_blue-gam_eta_present_constraints_v2.png}{%
\begin{figure}[htb]
\centerline{
\includegraphics[height=7cm]{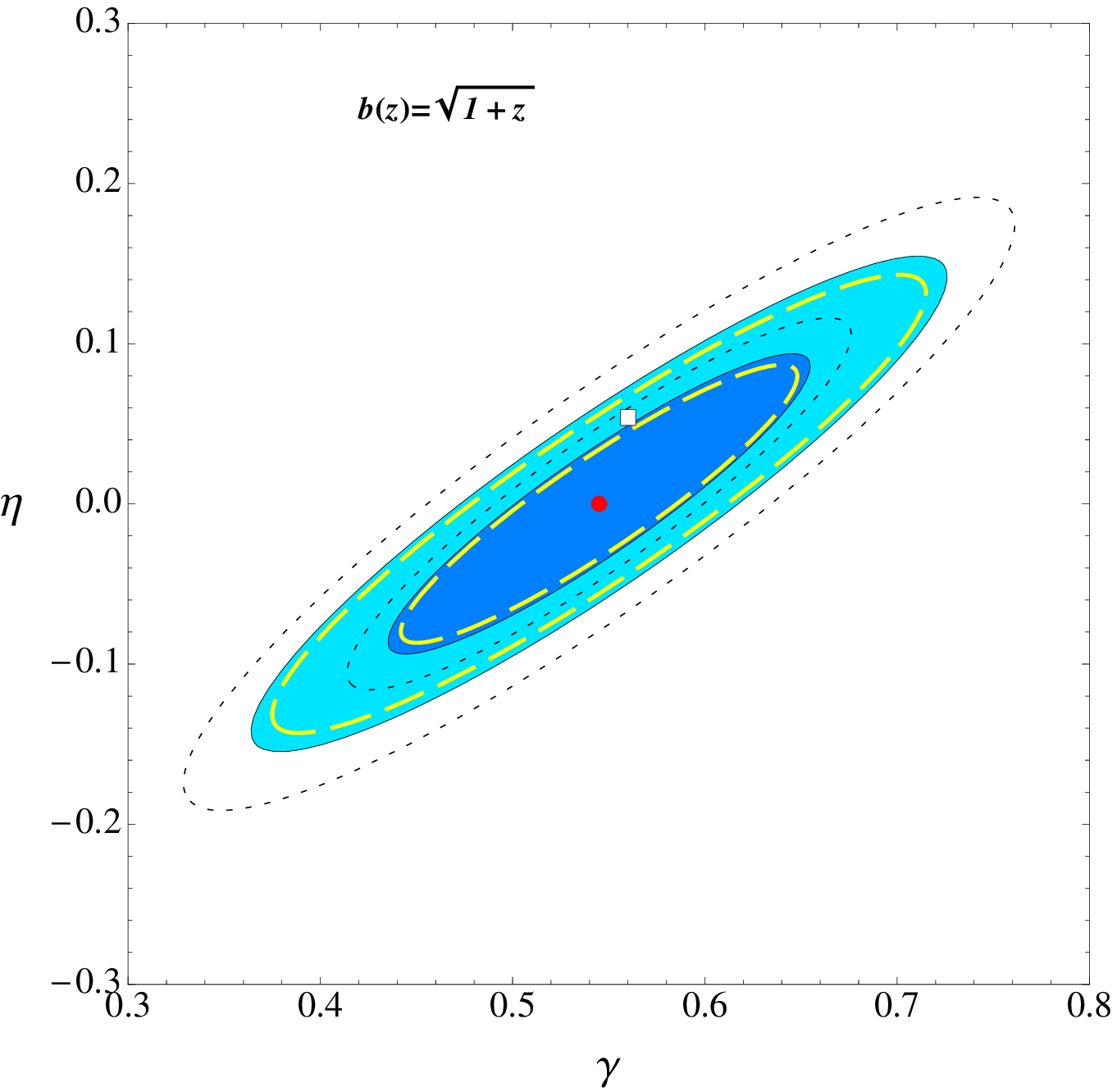}
\includegraphics[height=7cm]{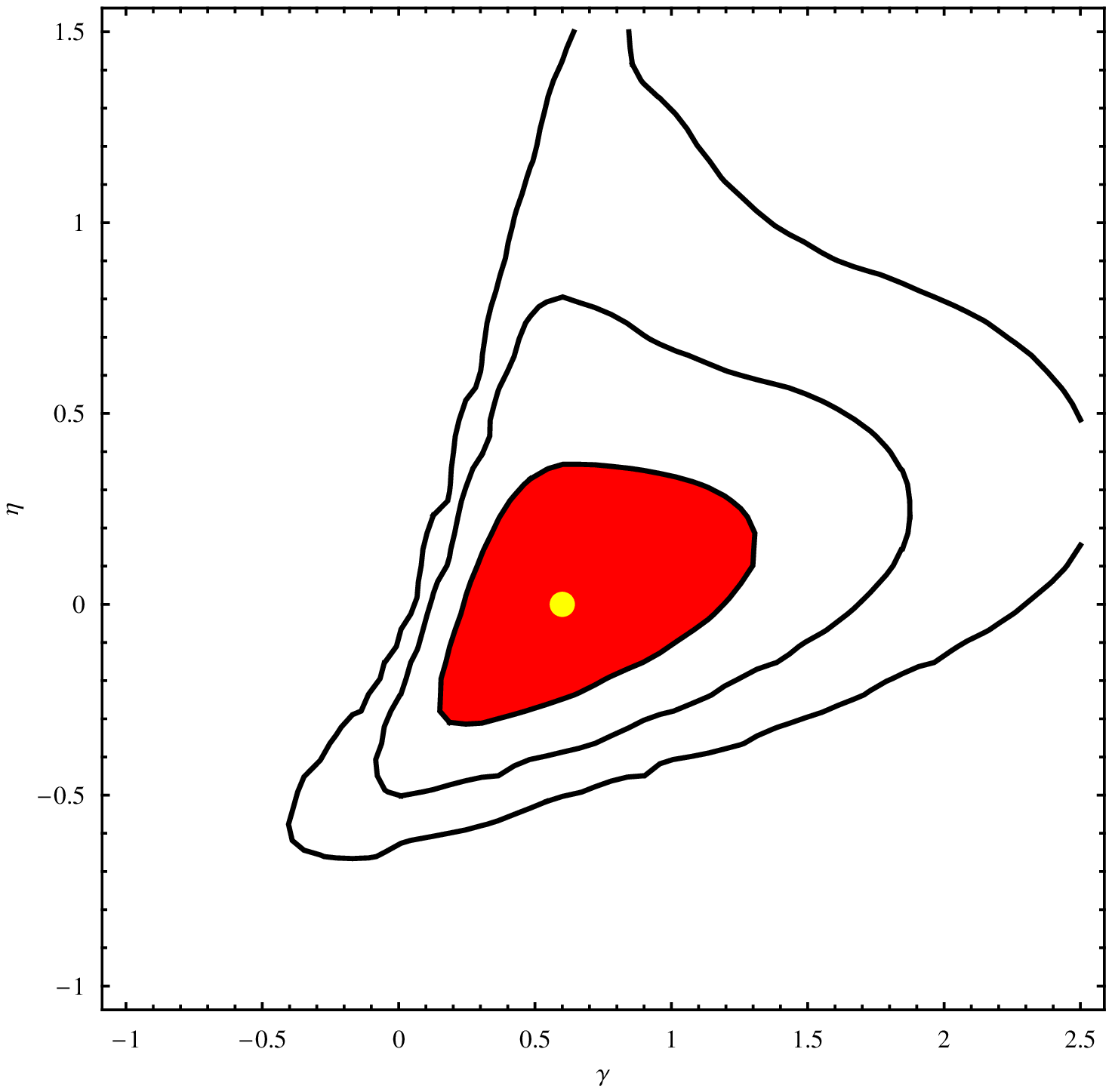}
}
\caption{$\eta$-parameterization. \emph{Left panel:} 1 and 2$\sigma$
  marginalized probability regions for the parameters $\gamma$ and
  $\eta$ in Eq.~(\ref{eq:eta_paramet}) relative to the reference case
  (shaded blue regions), to the optimistic case (yellow long-dashed
  ellipses) and to the pessimistic case (black short-dashed
  ellipses). The red dot marks the fiducial model while the square
  represents the coupling model. \emph{Right panel:} present constraints on
  $\gamma$ and $\eta$ computed through a full likelihood method (here
  the red dot marks the likelihood peak) \citep{diporto08}.}
\label{fig:gamma_eta_new_past}
\end{figure}}

\begin{table}[htbp]
\caption{Numerical values for $1\sigma$ constraints on parameters in
  Figure~\ref{fig:gamma_eta_new_past} and figures of merit.}
\label{tab:sigma_gam_eta}
\centering
\begin{tabular}{ccccc}
\toprule
bias  & case  & $\sigma_{\gamma}$  & $\sigma_{\eta}$ & FoM\\
\midrule 
~ & ref.  & 0.07  & 0.06  & 554\\
$b=\sqrt{1+z}$  & opt.  & 0.07  & 0.06  & 650\\
~ & pess.  & 0.09  & 0.08  & 362\\
\bottomrule
\end{tabular}
\end{table}

\epubtkImage{RB_w0w1_bsqrt_ref_3methods.png}{%
\begin{figure}[htbp]
 \centerline{\includegraphics[width=8cm]{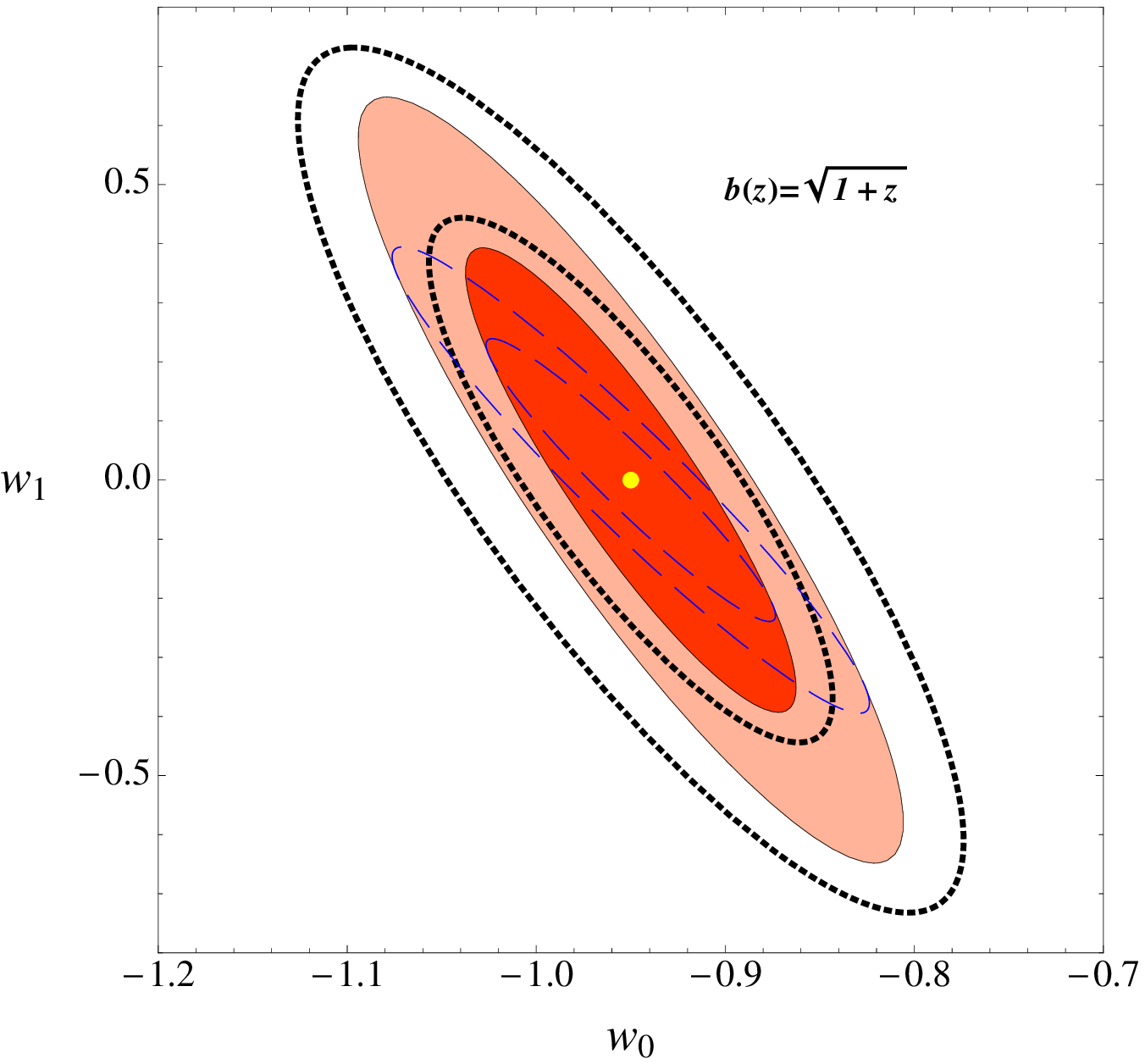}}
\caption{Errors on the equation of state. $1$ and $2\sigma$ marginalized
  probability regions for the parameters $w_{0}$ and $w_{1}$, relative
  to the reference case and bias $b=\sqrt(1+z)$. The blue dashed
  ellipses are obtained fixing $\gamma_{0}, \gamma_{1}$ and
  $\Omega_{k}=0$ to their fiducial values and marginalizing over all
  the other parameters; for the red shaded ellipses instead, we also
  marginalize over  $\Omega_{k}=0$ but we fix $\gamma_{0},
  \gamma_{1}$. Finally, the black dotted ellipses are obtained
  marginalizing over all parameters but $w_{0}$ and $w_{1}$. The
  progressive increase in the number of parameters reflects in a
  widening of the ellipses with a consequent decrease in the figures
  of merit (see Table~\ref{tab:w0_w1}).}
\label{fig:w0_w1}
\end{figure}}

Finally, in order to explore the dependence on the number of parameters and to
compare our results to previous works, we also draw the confidence ellipses for
$w_0$, $w_1$ with three different methods: i) fixing $\gamma_{0}, \gamma_{1}$ and
$\Omega_k$ to their fiducial values and marginalizing over all the
other parameters; ii) fixing only $\gamma_{0}$ and
  $\gamma_{1}$; iii) marginalizing over all parameters but $w_0$,
$w_1$. As one can see in Figure~\ref{fig:w0_w1} and
Table~\ref{tab:w0_w1} this progressive increase in the number of
marginalized parameters reflects in a widening of the ellipses with a
consequent decrease in the figures of merit. These results are in
agreement with those of other authors (e.g., \cite{Wang2010}).

The results obtained in this section can be summarized as follows.

\begin{enumerate}

\item If both $\gamma$ and $w$ are assumed to be constant
and setting  $\Omega_k=0$,
then a redshift survey described by our \emph{Reference case}
will be able to constrain these parameters to within 4\% and 2\%,
respectively.

\item Marginalizing over $\Omega_{k}$ degrades these constraints to 5.3\% and
4\% respectively.

\item If $w$ and $\gamma$ are considered redshift-dependent and
parametrized according to Eqs.~(\ref{eq:gam_CPL}) and~(\ref{eq:w_CPL})
then the errors on $\gamma_{0}$ and $w_{0}$ obtained after
marginalizing over  $\gamma_{1}$ and  $w_{1}$ increase by a
factor $\sim$~7, 5. However, with this
precision we will
be able to distinguish the fiducial model from the  DGP and $f(R)$
scenarios with more than $2\sigma$ and $1\sigma$ significance, respectively.

\item The ability to discriminate these models with a significance above
$2\sigma$ is confirmed by the confidence contours drawn in the
$\gamma_{0}$-$\gamma_{1}$ plane, obtained after marginalizing over all other
parameters.

\item If we allow for a coupling between dark matter and dark  energy,
and  we marginalize over $\eta$ rather than over $\gamma_{1}$,
then the errors on $w_{0}$ are almost identical
to those obtained in the case of the \emph{$\gamma$-parameterization},
while the errors on $\gamma_{0}$ decrease significantly.

However, our ability in separating the fiducial model from the
CDE model is significantly hampered: the confidence contours
plotted in the $\gamma$-$\eta$ plane show that discrimination
can only be performed wit $1\mbox{\,--\,}1.5\sigma$ significance. Yet,
this is still a remarkable improvement over the present situation, as
can be appreciated from Figure~\ref{fig:gamma_eta_new_past} where we
compare the constraints expected by next generation data to the
present ones. Moreover, the \emph{Reference} survey will be able to
constrain the parameter $\eta$ to within 0.06. Reminding that we can
write $\eta=2.1 \beta_c^2$~\citep{diporto08}, this means that the
coupling parameter $\beta_c$ between dark energy and dark matter can
be constrained to within 0.14, solely employing the growth rate
information. This is comparable to existing constraints from the CMB
but is complementary since obviously it is obtained at much smaller
redshifts. A variable coupling could therefore be detected by
comparing the redshift survey results with the CMB ones.

\end{enumerate}

\begin{table}[htbp]
\caption{$1\sigma$ marginalized errors for the parameters $w_{0}$ and
  $w_{1}$, obtained with three different methods (reference case, see
  Figure~\ref{fig:w0_w1}).}
\label{tab:w0_w1}
 \label{tab:sigma_w0_w1}
\centering
\begin{tabular}{cccc}
\toprule 
~ & $\sigma_{w_{0}}$ & $\sigma_{w_{1}}$ & FoM\\
\midrule
$\gamma_{0}, \gamma_{1}$, $\Omega_{k}$ fixed & 0.05 & 0.16 & 430\\
$\gamma_{0},\gamma_{1}$ fixed & 0.06 & 0.26 & 148\\
marginalization over all other parameters & 0.07 & 0.3 & 87\\
\bottomrule
\end{tabular}
\end{table}

It is worth pointing out that, whenever we have performed statistical tests
similar to those already discussed by other authors in the context of a
Euclid-like survey, we did find consistent results.
Examples of this are the values of FoM and errors for $w_0$, $w_1$, similar to
those in~\cite{Wang2010,majerotto11} and the errors on constant $\gamma$ and
$w$~\citep{majerotto11}. However, let us notice that all these values strictly
depend on the parametrizations adopted and on the numbers of parameters fixed
or marginalized over (see, e.g., \cite{rassat08}).

\subsection{Weak lensing non-parametric measurement of expansion and
  growth rate}
\label{weak-lensing-non-parametric}

In this section we apply power spectrum tomography \citep{Hu1999} to the Euclid weak lensing
survey without using any parameterization of the Hubble parameter $H(z)$ as
well as the growth function $G(z)$. Instead, we add the 
fiducial values of those functions at the center of some redshift bins of
our choice to the list of cosmological parameters. Using the Fisher
 matrix formalism, we can forecast the constraints that future
surveys can put on $H(z)$ and $G(z)$. Although such a non-parametric approach is quite common
for as concerns  the equation-of-state ratio $w(z)$ in supernovae
surveys \citep[see, e.g.,][]{Albrecht:2009ct} and also in redshift
surveys \citep{seo03}, it has not been investigated for weak lensing surveys.

The Fisher matrix is given by \citep{1999ApJ...514L..65H}
\begin{equation}
F_{\alpha\beta} = f_\mathrm{sky} \sum_\ell 
\frac{(2\ell+1)\Delta\ell}{2}\frac{\partial P_{ij}(\ell)}{\partial
p_\alpha}C^{-1}_{jk}\frac{\partial P_{km}(\ell)}{\partial
p_\beta}C^{-1}_{mi},
\label{eq:3-fm}
\end{equation}
where $f_\mathrm{sky}$ is the observed fraction of the sky, $C$ is the covariance matrix, $P(\ell)$ is the convergence
power spectrum and $\mathbf p$ is the vector of the parameters defining our
cosmological model. Repeated indices are being summed over from $1$ to $N$,
the number of redshift bins. The covariance matrix is defined as (no
summation over $j$)
\begin{equation}
C_{jk}=P_{jk} + \delta_{jk}\gamma_\mathrm{int}^2 n^{-1}_j,
\end{equation}
where $\gamma_\mathrm{int}$ is the intrinsic galaxy shear and $n_j$ is the
fraction of galaxies per steradian belonging to the $j$-th redshift bin:
\begin{equation}
n_j = 3600 \left( \frac{180}{\pi} \right)^2 
n_\theta\int_{0}^\infty n_j(z){\mathrm{d}} z
\end{equation}
where $n_\theta$ is the galaxy density per arc minute and $n_j(z)$ the galaxy
density for the $j$-th bin, convolved with a gaussian around $\hat z_j$, the
center of that bin, with a width of $\sigma_z(1+\hat z_j)$ in order to
account for errors in the redshift measurement.

For the matter power spectrum we use the
fitting formulae from \cite{Eisenstein_Hu_1997} and for its nonlinear
corrections the results from \cite{Smith2003}. Note that this is where the
growth function enters. The convergence power
spectrum for the $i$-th and $j$-th bin can then be written as 
\begin{equation}
P_{ij}(\ell) = \frac{9H_0^3}{4}\int_0^\infty 
\frac{W_i(z)W_j(z)E^3(z)\Omega_m^2(z)}{(1+z)^4}
P_{\delta_m}\left(\frac{\ell}{\pi r(z)}\right){\mathrm{d}} z.
\label{eq:3-convspec}
\end{equation}
Here we make use of the window function
\begin{equation}
W_i(z) = \int_z^\infty \frac{\mathrm{d}\tilde z}{H(\tilde z)}\left[
1-\frac{r(z)}{r(\tilde z)}
\right] n_i[r(\tilde z)]
\label{eq:3-wind}
\end{equation}
(with $r(z)$ being the comoving distance)
and the dimensionless Hubble parameter
\begin{equation} 
E^2(z) = \Omega_m^{(0)}(1+z)^3 + (1-\Omega_m^{(0)})
\exp\left[ \int_0^z
\frac{3(1+w(\tilde z))}{1+\tilde z}{\mathrm{d}}\tilde z \right].
\label{eq:3-e(z)}
\end{equation}
For the equation-of-state ratio, finally, we use the usual CPL   parameterization.

We determine $N$ intervals in redshift space such that each interval
contains the same amount of galaxies. For this we use the common
parameterization
\begin{equation}
n(z) = z^2 \exp(-(z/z_0)^{3/2}),
\end{equation}
where $z_0 =z_\mathrm{mean}/1.412$ is the peak of $n(z)$ and
$z_\mathrm{mean}$ the
median.
Now we can define $\hat z_i$ as the center of the $i$-th redshift bin and
add $h_i\equiv \log\left({H(\hat z_i)/H_0}\right)$ as well as $g_i\equiv\log
G(\hat z_i)$ to the list of cosmological parameters. The Hubble parameter
and the growth function now become functions of the $h_i$ and $g_i$
respectively:
\begin{align}
H(z;\Omega_m^{(0)},w_0,w_1) &\rightarrow H(z;h_1,\ldots,h_N)\\
G(z;\Omega_m^{(0)},\gamma) &\rightarrow G(z;g_1,\ldots,g_N)
\end{align}
This is being done by linearly interpolating the functions through their
supporting points, e.g., $(\hat z_i,\exp( h_i))$ for $H(z)$. Any function
that depends on either $H(z)$ or $G(z)$ hence becomes a function of the 
$h_i$ and $g_i$ as well.

\begin{table}[htb]
\caption{Values used in our computation. The values of the fiducial
  model (WMAP7, on the left) and the survey parameters (on the right).}
\label{tab1}
\centering
 \begin{tabular}{l|l}
   $\omega_m$ & 0.1341 \\ 
   $\omega_b$ & 0.02258  \\
   $\tau$     & 0.088 \\
   $n_s$      & 0.963 \\
   $\Omega_m$ & 0.266 \\
   $w_0$      & --1  \\
   $w_1$      & 0   \\
   $\gamma$   & 0.547 \\
   $\gamma_{\mathrm{ppn}}$     & 0   \\
   $\sigma_8$  & 0.801 \\
 \end{tabular}
 \hspace{1cm}
\begin{tabular}{l|l}
$f_\mathrm{sky}$ & 0.375 \\
$z_\mathrm{mean}$ & 0.9 \\
$\sigma_z$ & 0.05 \\
$n_\theta$& 30 \\
$\gamma_\mathrm{int}$ & 0.22 \\
\hline
$\ell_{\max}$ & $5\cdot 10^3$\\
$\Delta \log_{10}\ell$ & 0.02
\end{tabular}
\end{table}

The values for our fiducial model (taken from WMAP 7-year data
\citep{Komatsu:2010fb}) and the survey parameters that we chose
for our computation can be found in Table~\ref{tab1}.

As for the sum in Eq.~(\ref{eq:3-fm}), we generally found that with a
realistic upper
limit of $\ell_\mathrm{max} = 5\cdot 10^3$ and a step size of $\Delta \lg \ell =
0.2$ we get the best result in terms of a figure of merit (FoM),
that we defined as 
\begin{equation}
\mathrm{FoM} = \sum \sigma_i^{-2}.
\end{equation}

Note that this is a fundamentally different FoM than the one defined by the
Dark Energy Task Force. Our definition allows for a single large error
without influencing the FoM significantly and should stay almost constant
after dividing a bin arbitrarily in two bins, assuming the error scales
roughly as the inverse of the root of the number of galaxies in a given bin.

\epubtkImage{RB_h-5bins.png}{%
\begin{figure}[htb] 
\centerline{\includegraphics[width=10cm]{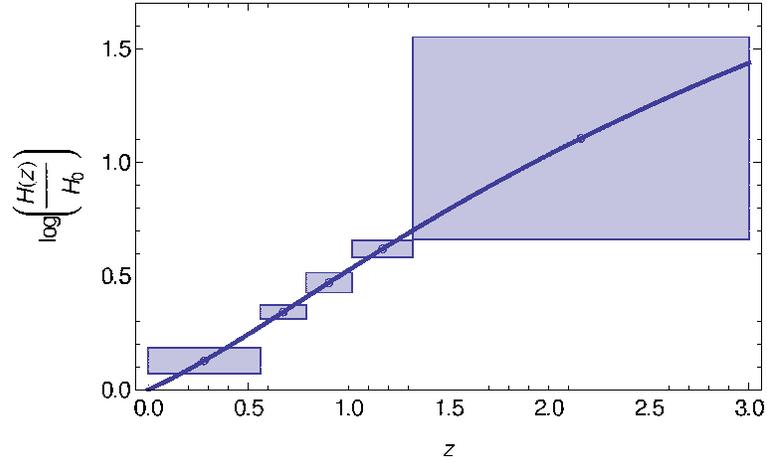}}
\caption{Error bars on the Hubble parameter $H(z)$ with five redshift bins.
The exact height of the error bars respectively are $(0.23, 0.072,
0.089, 0.064, 0.76)$.}
\label{fig:hplot}
\end{figure}}

\epubtkImage{RB_g-3bins.png}{%
\begin{figure}[htb]
\centerline{\includegraphics[width=10cm]{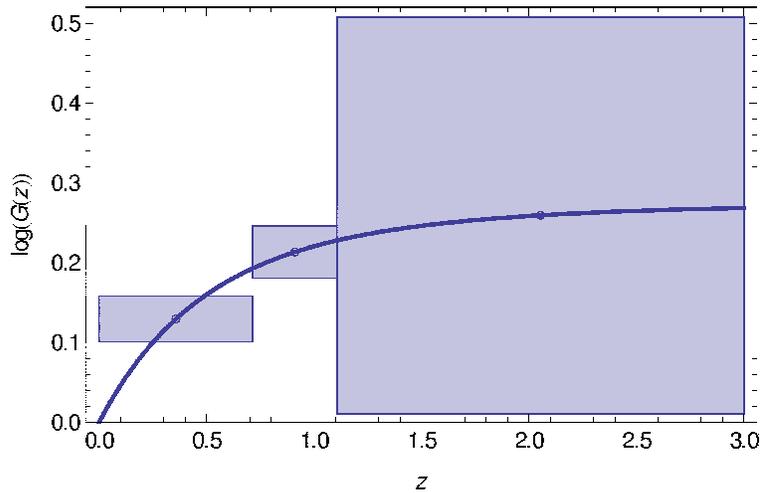}}
\caption{Error bars on the growth function $G(z)$ with three redshift bins
while marginalizing over the $h_i$s. The exact height of the error
bars respectively are $(0.029, 0.033, 0.25)$.}
\label{fig:gplot}
\end{figure}}

We first did the computation with just binning $H(z)$ and using the common
fit for the growth function slope \citep{wang98}
\begin{equation}
\frac{d \log G(z)}{d\log a} = \Omega_m(z)^\gamma,
\end{equation}
yielding the result in Figure~\ref{fig:hplot}. Binning both $H(z)$ and
$G(z)$ and marginalizing over the $h_i$s yields the plot for $G(z)$ seen in
Figure~\ref{fig:gplot}. 

Notice that here we assumed no prior information. Of course one could improve the FoM by taking into account
some external constraints due to other experiments.

\subsection{Testing the non-linear corrections for weak lensing forecasts}
\label{weak-lensing-non-parametric_test}

In order to fully exploit the scientific potential  of the next generation of weak lensing surveys, accurate
predictions of the matter power spectrum are required. The signal-to-noise ratio of the cosmic shear signal
is highest on angular scales of $5-10$ arcminutes, which correspond to physical scales of $\sim 1$ Mpc.
Restricting the analysis to larger scales does not necessarily solve the problem, because the observed
two-point ellipticity correlation functions are still sensitive to small scale structures projected along the
line-of-sight. This may be avoided using a full 3D shear analysis \citep[see][for details]{Castro05, Kitching11b},
but using only the larger scales increases the statistical uncertainties due to cosmic variance.

Currently only N-body simulations allow us to capture the non-linear structure formation, but for a survey
such as {\it Euclid} an accuracy of $\sim1\%$ is needed~\citep{Huterer:2001yu,Huterer:2004tr}. 
This accuracy goes beyond the claimed $\pm 3\%$ uncertainty of the popular \textsc{halofit} code~\citep{Smith2003}.
However, the accuracy can be improved provided the simulations are started with adequate initial conditions,
with a large volume, sufficient time stepping and high mass resolution. For instance \cite{Heitmann10} obtained an
accuracy of $\sim 1\%$ out to $k\sim 1 h$Mpc$^{-1}$ for a gravity-only simulation.

It is important to distinguish between gravity-only simulations, which are used to make the forecasts, and
hydrodynamical simulations that attempt to capture the modifications to the matter power spectrum due to baryon
physics. Although most of the matter in the Universe is indeed believed to be in the form of collissionless cold dark
matter, baryons represent a non-negligible fraction of the total matter content. The distribution of baryons traces
that of the underlying dark matter density field and thus gravity-only simulations should capture most of the structure
formation. Nonetheless, differences in the spatial distribution of baryons with respect to the dark matter is expected to lead
to changes that exceed the required accuracy of 1 per cent.

Various processes, which include radiative cooling, star formation and energy injection from supernovae and
active galactic nuclei,  affect the distribution of baryons. Implementing these processes correctly is difficult, and 
as a consequence the accuracy of hydrodynamic simulations is under discussion. That baryon physics cannot 
be ignored was perhaps most clearly shown in \cite{vanDaalen11} who looked at the changes in the matter
power spectra when different processes are included. This was used by \cite{Semboloni11} to examine the impact
on cosmic shear studies. The results suggest that AGN feedback may lead to a suppression of the power by as much as $10\%$
at $k\sim 1h$Mpc$^{-1}$. 

\cite{Semboloni11} showed that ignoring the baryonic physics leads to biases in the  cosmological parameter estimates
that are much larger than the precision of {\it Euclid}.  In the case of the AGN model,  the bias in $w$ is as much as $40\%$. 
Unfortunately our knowledge of the various feedback processes is still incomplete and we  cannot use the simulations to interpret 
cosmic shear signal. Furthermore, hydrodynamic simulations are too expensive to simulate  large volumes for a range of cosmological 
parameters. To circumvent this problem several approaches have been suggested. For instance, \cite{Bernstein09} proposed to describe 
the changes in the power spectrum by Legendre polynomials, and to marginalise over the nuisance parameters 
\citep[also see][for a similar approach]{Kitching11a}. Although this leads to unbiased estimates for cosmological parameters, the 
precision decreases significantly, by as much as 30\% \citep{Zentner08}. 

Instead \cite{Semboloni11} and \cite{Semboloni13} examined whether it is possible to model the effects of baryon
physics using a halo model approach, in which the baryons and stars are treated separately from the dark matter
distribution. The model parameters, rather than being mere nuisance parameters, correspond to physical quantities
that can be constrained observationally. These works showed that even with this still rather simple approach it is 
possible to reduce the biases in the cosmological parameters to acceptable levels, without a large loss in precision.

The forecasts do not include the uncertainty due to baryon physics, hence the results implicitly assume that 
this can be understood sufficiently well that no loss in precision occurs. This may be somewhat optimistic, as more work is 
needed in the coming years to accurately quantify the impact of baryon physics on the modelling of the matter power spectrum, 
but we note that the initial results are very encouraging. In particular, \cite{Semboloni13} found that requiring consistency
between the two- and three-point statistics can be used to self-calibrate feedback models.

Another complication for the forecasts is the performance of the prescriptions for the non-linear power spectrum
for non-$\Lambda$CDM models. For instance, \cite{McDonald:2005gz} showed that, using \textsc{halofit} for
non-$\Lambda$CDM models, requires suitable corrections. In spite of that, \textsc{halofit} has been often used to calculate the spectra of
models with non-constant DE state parameter $w(z)$.  This procedure was dictated by the lack of appropriate extensions of \textsc{halofit} to
non-$\Lambda$CDM cosmologies.

In this paragraph we quantify the effects of using the \textsc{halofit} code instead of $N$-body outputs for nonlinear corrections for DE
spectra, when the nature of DE is investigated through weak lensing surveys. Using a Fisher-matrix approach, we evaluate the discrepancies
in error forecasts for $w_{0}$, $w_{a}$ and $\Omega_m$ and compare the related confidence ellipses. See~\cite{casarini:2010} for further details.

The weak lensing survey is as specified in Section~\ref{sec:baofm_survey}. Tests are performed assuming three different fiducial cosmologies:
$\Lambda$CDM model ($w_0 = -1$, $w_a = 0$) and two dynamical DE models, still consistent with the WMAP+BAO+SN combination \citep{Komatsu:2010fb} at 95\% C.L. They will be dubbed M1 ($w_0 = -0.67$, $w_a = 2.28$) and M3 ($w_0 = -1.18$, $w_a = 0.89$). In this way we explore the
dependence of our results on the assumed fiducial model. For the other parameters we adopt the fiducial cosmology of Secton~\ref{sec:baofm_survey}.

The derivatives needed to calculate the Fisher matrix are evaluated by extracting the power spectra from the $N$-body simulations of models
close to the fiducial ones, obtained by considering parameter increments $\pm 5\%$. For the $\Lambda$CDM case, two different initial
seeds were also considered, to test the dependence on initial conditions, finding that Fisher matrix results are almost insensitive
to it. For the other fiducial models, only one seed is used. 

$N$-body simulations are performed by using a modified version of
\textsc{pkdgrav} \citep{2001PhDT........21S} able to handle any DE state
equation $w(a)$, with $N^3 = 256^{3}$ particles in a box with side $L
= 256\, h^{-1}\mathrm{\ Mpc}$. Transfer functions generated using the
\textsc{camb} package are employed to create initial conditions, with
a modified version of the PM software by \cite{Klypin:1997sk}, also
able to handle suitable parameterizations of DE.

Matter power spectra are obtained by performing a FFT (Fast Fourier
Transform) of the matter density fields, computed from the particles
distribution through a Cloud-in-Cell algorithm, by using a regular
grid with $N_{g}=2048$. This allows us to obtain nonlinear spectra
in a large $k$-interval. In particular, our resolution allows to work
out spectra up to $k \simeq 10\, h\mathrm{\ Mpc}^{-1}$. However, for $k >
2\mbox{\,--\,}3\, h\mathrm{\ Mpc}^{-1}$ neglecting baryon physics is
no longer accurate
\citep{Jing:2005gm,Rudd:2007zx,Bonometto:2010kz,Zentner:2007,Hearin:2009hz}. For
this reason, we consider WL spectra only up to $\ell_{\max} = 2000$.

Particular attention has to be paid to matter power spectra
normalizations. In fact, we found that, normalizing all models to the
same linear $\sigma_8 (z=0)$, the shear derivatives with respect to
$w_0$, $w_a$ or $\Omega_m$ were largely dominated by the normalization
shift at $z=0$, $\sigma_{8}$ and $\sigma_{8,nl}$ values being quite
different and the shift itself depending on $w_0$, $w_a$ and
$\Omega_m$. This would confuse the $z$ dependence of the growth
factor, through the observational $z$-range. This normalization
problem was not previously met in analogous tests with the Fisher
matrix, as \textsc{halofit} does not directly depend on the DE state
equation.

As a matter of fact, one should keep in mind that, observing the
galaxy distribution with future surveys, one can effectively measure
$\sigma_{8,nl}$, and not its linear counterpart. For these reasons, we
choose to normalize matter power spectra to $\sigma_{8,nl}$, assuming
to know it with high precision.

\epubtkImage{RB_welli3-2000-RB_welli5-2000.png}{%
\begin{figure}
\centerline{
\includegraphics[scale=0.4]{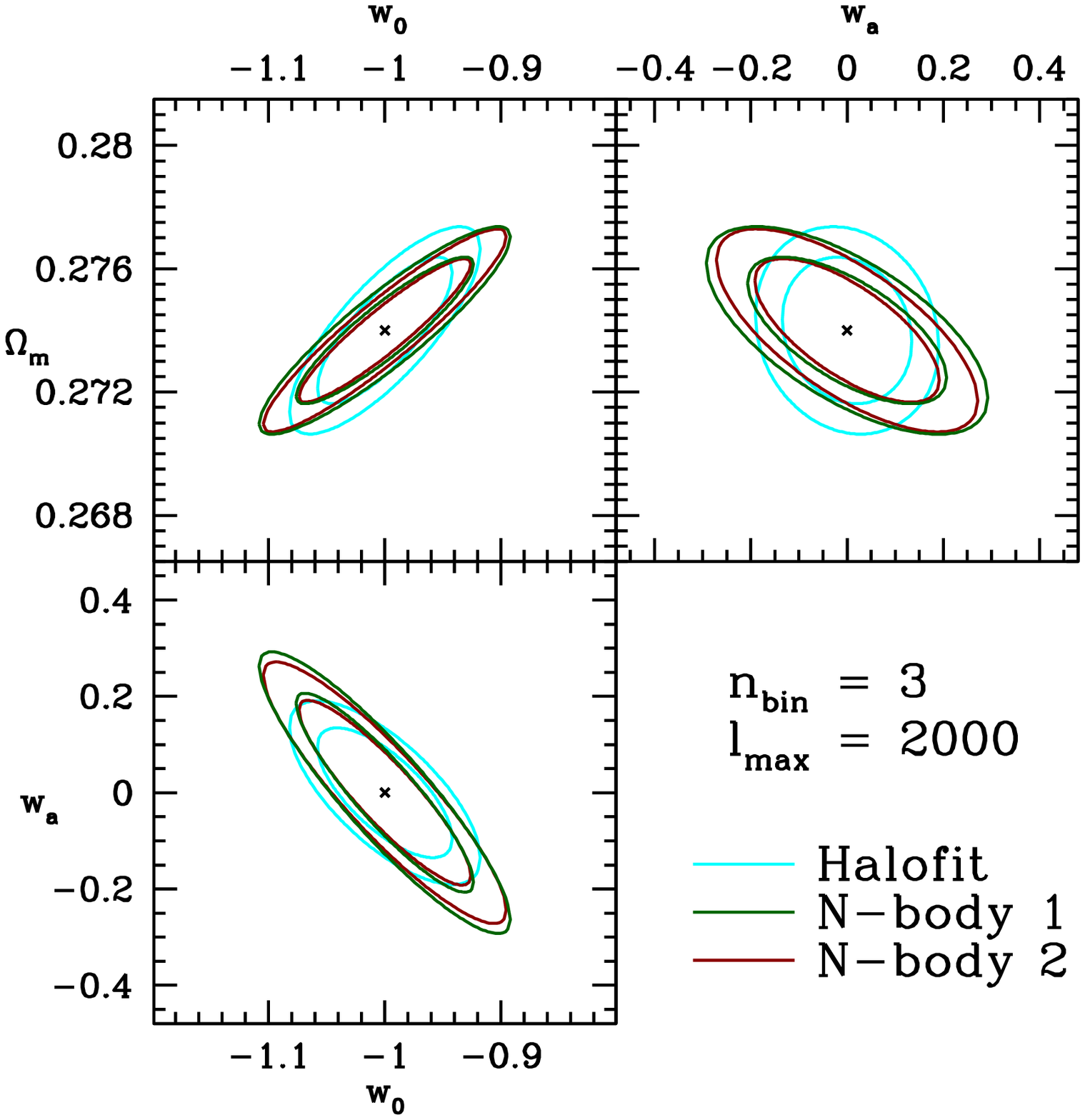}
\hskip1.truecm
\includegraphics[scale=0.4]{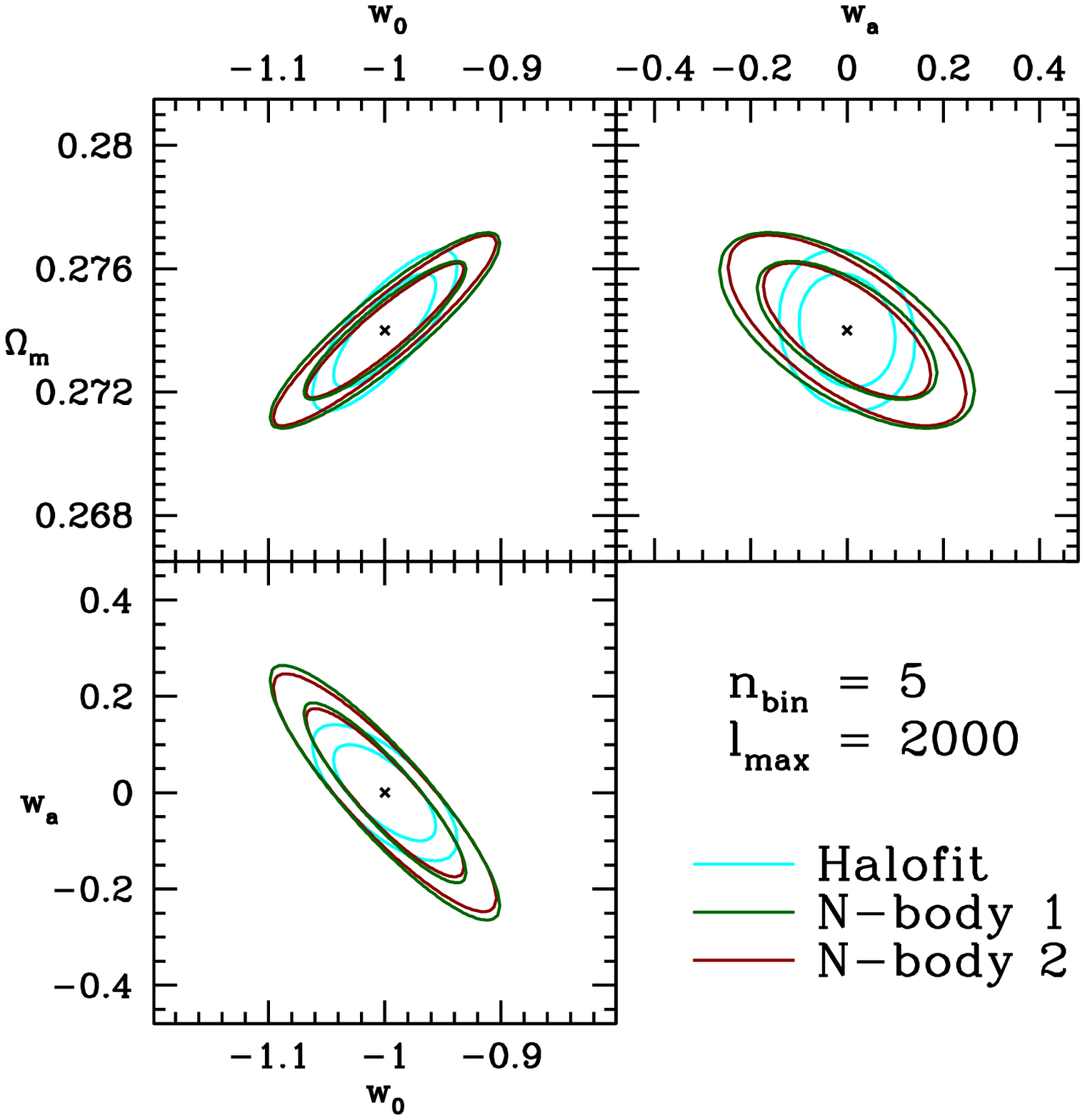}
}
\caption{Likelihood contours, for 65\% and 95\% C.L.,
  calculated including signals up to $\ell \simeq 2000$ for the
  $\Lambda$CDM fiducial. Here simulations and \textsc{halofit} yield
  significantly different outputs.}
\label{lcdm2000}
\end{figure}}

In Figures~\ref{lcdm2000} we show the confidence ellipses, when the
fiducial model is $\Lambda$CDM, in the cases of 3 or 5 bins and with
$\ell_{\max} = 2000$. Since the discrepancy between different seeds are
small, discrepancies between \textsc{halofit} and simulations are truly
indicating an underestimate of errors in the \textsc{halofit} case.

As expected, the error on $\Omega_m$ estimate is not affected by the
passage from simulations to \textsc{halofit}, since we are dealing with
$\Lambda$CDM models only. On the contrary, using \textsc{halofit} leads
to underestimates of the errors on $w_0$ and $w_a$, by a substantial
30\,--\,40\% (see \cite{casarini:2010} for further details).

This confirms that, when considering models different from
$\Lambda$CDM, nonlinear correction obtained through \textsc{halofit} may
be misleading. This is true even when the fiducial model is
$\Lambda$CDM itself and we just consider mild deviations of $w$ from
$-1$.

\epubtkImage{RB_welli2000_w0-067wa-228-RB_welli2000_w0-118wa089.png}{%
\begin{figure}
\centerline{
\includegraphics[scale=0.4]{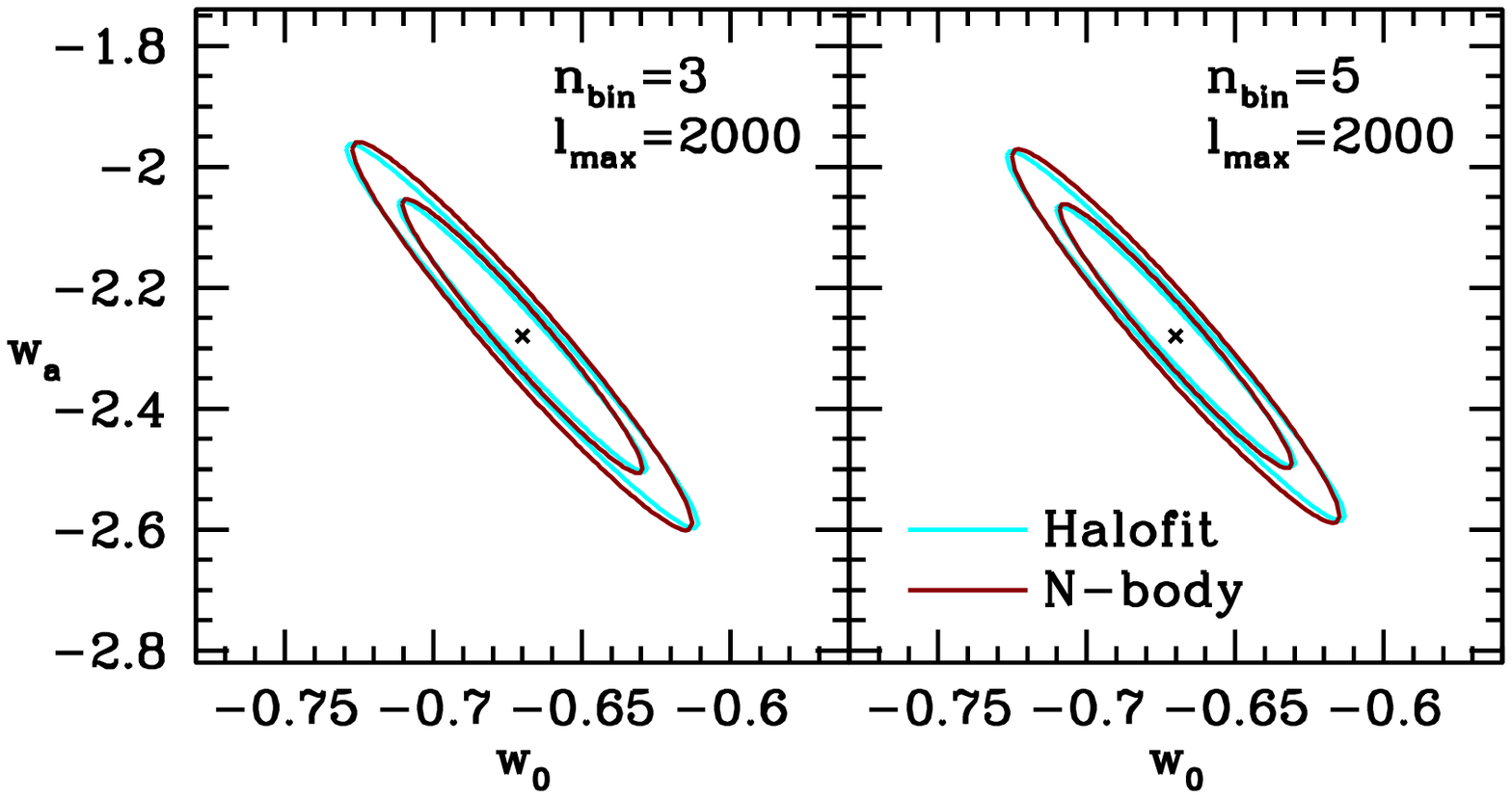}
\hskip1.truecm
\includegraphics[scale=0.4]{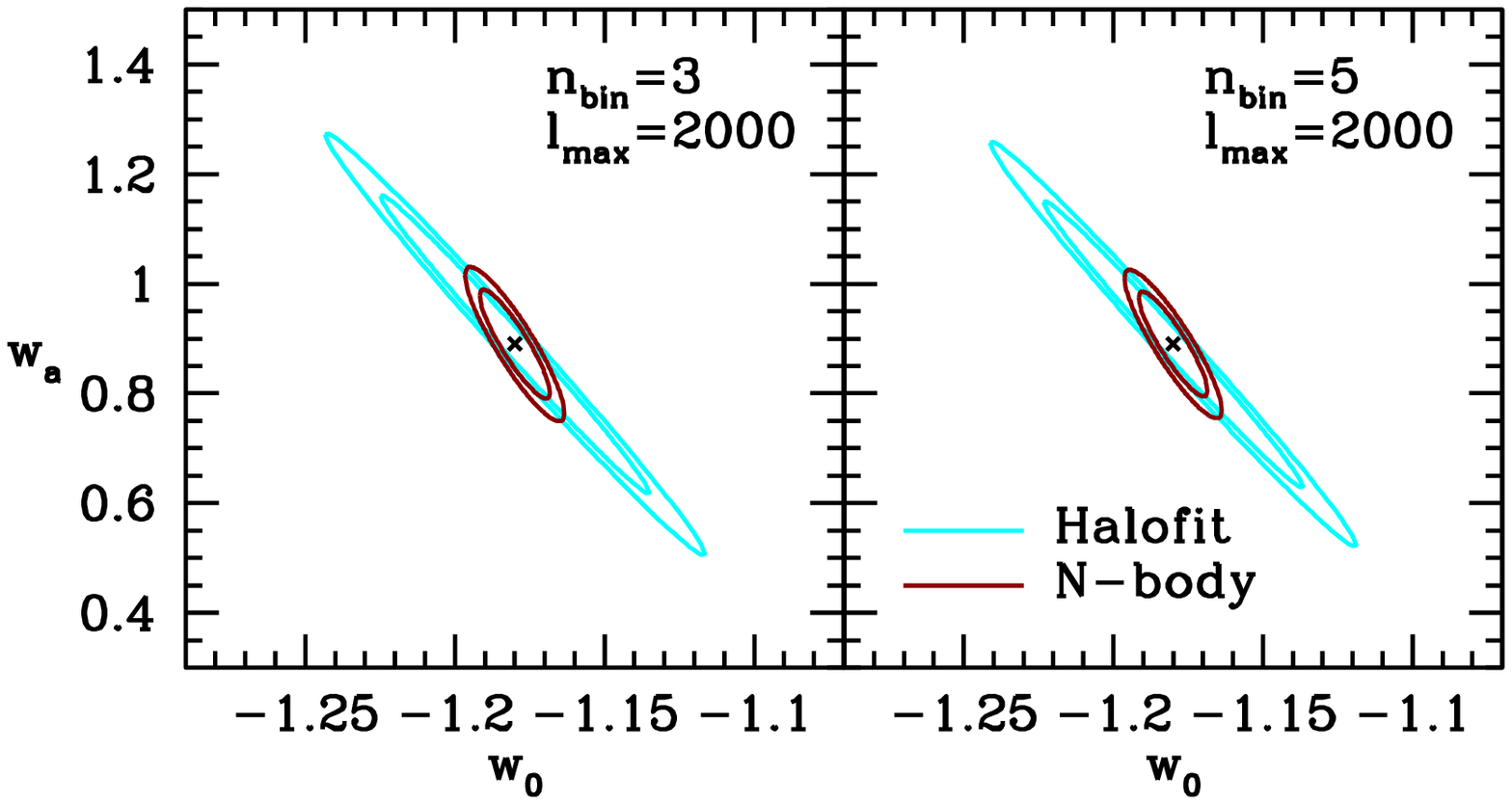}
}
\caption{On the left (right) panel, 1- and 2-$\sigma$ contours for the
  M1 (M3) model. The two fiducial models exhibit quite
  different behaviors.}
\label{mfig}
\end{figure}}

Figure~\ref{mfig} then show the results in the $w_0$-$w_a$ plane,
when the fiducial models are M1 or M3. It is evident that the two
cases are quite different. In the M1 case, we see just quite a mild
shift, even if they are $\cal O$ (10\%) on error predictions. In
the M3 case, errors estimated through \textsc{halofit} exceed simulation
errors by a substantial factor. Altogether, this is a case when
estimates based on \textsc{halofit} are  not trustworthy.

The effect of baryon physics is another nonlinear correction to be
considered. We note that the details of a study on  the impact of
baryon physics on the power spectrum and the parameter estimation can
be found in~\cite{2011MNRAS.417.2020S}

\subsection{Forecasts for the dark-energy sound speed}
\label{soundspeed}

As we have seen  in Section~\ref{sec:cosmo_perts},  when dark
energy clusters, the standard sub-horizon Poisson equation that links matter
fluctuations to the gravitational potential is modified and $Q\ne 1$. The deviation
from unity will depend on the degree of DE clustering and therefore on the sound
speed $c_s$. \label{symbol:c_s}
In this subsection we try to forecast the constraints that Euclid can put
on a constant $c_s$ by measuring $Q$ both via weak lensing and via redshift clustering. Here
 we assume standard Einstein gravity and zero anisotropic stress (and therefore we have $\Psi=\Phi$) and
 we allow $c_{s}$ to 
assume different values in the range 0\,--\,1. 

Generically, while dealing with a non-zero sound speed, 
we have to worry about the sound horizon $k_{sh}=aH/c_{s}$, which 
characterizes the growth of the perturbations; 
then we have at least three regimes with different behavior of the perturbations: 
\begin{enumerate}
\item perturbations larger than the causal horizon (where perturbations 
are not causally connected and their growth is suppressed),
\item perturbations smaller than the causal horizon but larger than 
the sound horizon, $k\ll aH/c_{s}$ (this is the only regime where perturbations are free to 
grow because the velocity dispersion, or equivalently the pressure perturbation, 
is smaller than the gravitational attraction), 
\item perturbations smaller than the sound horizon, $k\gg aH/c_{s}$ (here perturbations stop growing 
because the pressure perturbation is larger than the gravitational attraction).
\end{enumerate}

As we have set the anisotropic stress to zero, the perturbations are
fully described by $Q$. 
The main problem is therefore to find an explicit expression that shows
how $Q$ depends on $c_s$. 
\cite{Sapone:2009} have provided 
the following explicit approximate expression for  $Q\left(k,a\right)$ 
which captures the behavior for both super- and sub-horizon scales: 
\begin{equation}
Q(k,a)=1+\frac{1-\Omega_{M,0}}{\Omega_{M,0}}\frac{(1+w)a^{-3w}}{1-3w+\frac{2}{3}\nu(a)^{2}}.
\label{eq:qtot}
\end{equation}
Here $\nu(a)^{2}=k^{2}c_{s}^2 a/\left(\Omega_{M,0}H_{0}^{2}\right)$
which it is defined through $c_{s}k\equiv\nu aH$ so that $\nu$ counts
how deep a mode is inside the sound horizon.

Eq.~(\ref{eq:qtot}) depends substantially on the value of the 
sound speed or, to put it differently, on the scale considered. 
For scales larger than the sound horizon ($\nu\approx0$), Eq.~(\ref{eq:qtot}) 
scales as $a^{-3w}$ and for $\Omega_{m,0}=0.25$ and $w=-0.8$ we have that 
\begin{equation}
Q-1\approx\frac{3}{17}a^{2.4}\simeq0.18a^{2.4}\,.
\end{equation}
This is not a negligible deviation today, but it decreases rapidly
as we move into the past, as the dark energy becomes less important.%
\epubtkFootnote{For this reason, early dark-energy models can have a
  much stronger impact.}
As a scale enters the sound horizon, $Q-1$ grows with one power of the
scale factor slower (since $\delta_{\mathrm{DE}}$ stops growing),
suppressing the final deviation roughly by the ratio of horizon size
to the scale of interest (as now $\nu^2\gg 1$). In the observable
range, $(k/H_{0})^{2}\approx10^{2}\mbox{\,--\,}10^{4}$. Therefore, if
$c_{s}\approx1$, $Q\to1$ and the dependence on $c_{s}$ is lost. This
shows that $Q$ is sensitive to $c_{s}$ only for small values,
$c_{s}^{2}\lesssim10^{-2}$.

We can characterize the dependence of $Q$ on the main perturbation
parameter $c_{s}^2$ by looking at its derivative, a key quantity for
Fisher matrix forecasts: 
\begin{equation}
\frac{\partial\log Q}{\partial\log
  c_s^2}=-\frac{x}{\left(1+x\right)}\frac{Q-1}{Q} \,,
\label{eq:Qdercs}
\end{equation}
where $x=\frac{2}{3}\nu(a)^{2}/(1-3w)\simeq0.2\nu(a)^{2}$ (with
the last expression being for $w=-0.8$). For the values we are interested in
here, this derivative has a peak at the present epoch at the sound
horizon, i.e., for $c_{s}\approx H_{0}/k$, which in the observable
range of $k$ is $c_{s}\approx.01-.001$, and declines rapidly for
larger $c_{s}$.
This means that the sensitivity of $Q$ to the sound speed can be
boosted by several orders of magnitude as the sound speed is decreased.

There are several observables that depend on $Q$:
\begin{itemize} 

\item The growth of matter perturbations

There are two
ways to influence the growth factor: firstly at background level,
with a different Hubble expansion. Secondly at perturbation level:
if dark energy clusters then the gravitational potential changes because
of the Poisson equation, and this will also affect the growth rate
of dark matter. All these effects can be included in the growth index
$\gamma$ and we therefore expect that $\gamma$ is a function of
$w$ and $c_s^2$ (or equivalently of $w$ and $Q$).

The growth index depends on dark-energy
perturbations (through $Q$) as \citep{Sapone:2009}
\begin{equation}
\gamma=\frac{3\left(1-w-A\left(Q\right)\right)}{5-6w}
\label{eq:gamma-Q}
\end{equation} 
where 
\begin{equation}
A\left(Q\right)=\frac{Q-1}{1-\Omega_{M}\left(a\right)}.
\label{eq:A-Q}
\end{equation}
Clearly here, the key quantity is the derivative of the growth 
factor with respect to the sound speed: 
\begin{equation}
\frac{\partial\log G}{\partial \ln c_s^2}\propto \int_{a_0}^{a_1}{\frac{\partial\gamma}{\partial c_{s}^2}{\mathrm{d}}a}\propto \int_{a_0}^{a_1}{\frac{\partial Q}{\partial c_{s}^2}{\mathrm{d}}a} \propto \int_{a_0}^{a_1}{\left(Q-1\right){\mathrm{d}}a}\,.
\label{eq:Gdercs}
\end{equation}
From the above equation we also notice that the derivative of the growth factor  
does not depend on $Q-1$ like the derivative 
$Q$, but on $Q-Q_{0}$ as it is an integral (being $Q_0$ the 
value of $Q$ today). The growth factor is thus not directly probing
the deviation of $Q$ from unity, but rather how $Q$ evolves over
time, see \cite{Sapone:2010} for more details.

\item Redshift space distortions

The distortion induced by redshift can be expressed in linear theory
by the $\beta$ factor, related to the bias factor and the growth
rate via: 
\begin{equation}
\beta(z,k)=\frac{\Omega_{m}\left(z\right)^{\gamma(k,z)}}{b(z)}\,.
\end{equation}
The derivative of the redshift distortion parameter with respect to
the sound speed is: 
\begin{equation}
\frac{\partial\log\left(1+\beta\mu^{2}\right)}{\partial\log c_s^2}= 
-\frac{3}{5-6w}\frac{\beta\mu^{2}}{1+\beta\mu^{2}}\frac{x}{1+x}\left(Q-1\right)\,.
\label{eq:derbetadcs}
\end{equation}
We see that the behavior versus $c_{s}^{2}$ is similar to the one
for the $Q$ derivative, so the same discussion applies. Once again,
the effect is maximized for small $c_{s}$. 
The $\beta$ derivative is comparable to that of $G$ at $z=0$ but becomes more important 
at low redshifts.

\item Shape of the dark matter power spectrum

Quantifying the impact of the sound speed on the matter power
spectrum is quite hard as we need to run 
Boltzmann codes (such as \textsc{camb}, \cite{CAMB}) in order to get the full impact of 
dark-energy perturbations into the matter power spectrum. 
\cite{Sapone:2010} proceeded in two ways: 
first using the \textsc{camb} output and then considering the analytic expression 
from  \cite{Eisenstein_Hu_1997} (which does not include dark 
energy perturbations, i.e., does not include $c_{s}$).

They find that the impact of the derivative of
the matter power spectrum with respect the sound speed on the final
errors is only relevant if high values of $c_s^2$ are considered; 
by decreasing the sound speed, the results are less and less affected. 
The reason is that for low values of the sound
speed other parameters, like the growth factor, start to be the dominant
source of information on $c_{s}^{2}$.

\end{itemize}

\paragraph*{Impact on weak lensing.}

Now it is possible to investigate the response of weak lensing (WL) to the
dark-energy parameters. Proceeding with a Fisher matrix as in \cite{Amendola:2007rr}, 
the main difference here being that the parameter $Q$ has an
explicit form. Since $Q$ depends on $w$ and $c_s^2$, we can forecast
the precision with which those parameters can be extracted. We can
also try to trace where the constraints come from.
For a vanishing anisotropic stress the WL potential becomes: 
\begin{equation}
k^{2}\left(\Phi+\Psi\right) = -2Q\frac{3H_{0}^{2}\Omega_{M,0}}{2a}\Delta_{M}
\label{eq:WL-potential-Q}
\end{equation}
which can be written, in linear perturbation theory as: 
\begin{equation}
k^{2}\left(\Phi+\Psi\right)=-3H\left(a\right)^{2}a^{3}Q\left(a,k\right)\Omega_{M}\left(a\right)G\left(a,k\right)\Delta_{M}\left(k\right)\,.
\label{eq:phiwl}
\end{equation}

Hence, the lensing potential contains three conceptually different
contributions from the dark-energy perturbations: 
\begin{itemize}
\item The direct contribution of the perturbations to the gravitational
potential through the factor $Q$.
\item The impact of the dark-energy perturbations on the growth rate of
the dark matter perturbations, affecting the time dependence of $\Delta_{M}$,
through $G\left(a,k\right)$.
\item A change in the shape of the matter power spectrum $P(k)$, corresponding
to the dark energy induced $k$ dependence of $\Delta_{M}$.
\end{itemize}
We use the representative Euclid survey presented in Section~\ref{sec:baofm_survey} 
and we extend our survey up to three different redshifts: $z_{\max}=2,3,4$. 
We choose different values of $c_s^2$ and  $w_0 = -0.8$ in order to 
maximize the impact on $Q$: values closer to $-1$
reduce the effect and therefore increase the errors on $c_{s}$.

In Figure~\ref{fig:ellipsesw0cs} we report the $1-\sigma$ confidence 
region for $w_{0},c_s^2$ for two different
values of the sound speed and $z_{\max}$. For high value of the sound
speed ($c_s^2=1$) we find $\sigma(w_{0})=0.0195$ and the relative
error for the sound speed is $\sigma(c_s^2)/c_s^2=2615$. As expected,
WL is totally insensitive to the clustering properties of quintessence
dark-energy models when the sound speed is equal to $1$. The presence
of dark-energy perturbations leaves a $w$ and $c_s^2$ dependent signature
in the evolution of the gravitational potentials through
$\Delta_{\mathrm{DE}}/\Delta_{m}$ and, as already mentioned, the
increase of the $c_s^2$ enhances the suppression of dark-energy
perturbations which brings $Q \rightarrow 1$.

Once we decrease the sound speed then dark-energy perturbations are
free to grow at smaller scales. In Figure~\ref{fig:ellipsesw0cs} the confidence region for $w_{0},c_s^2$ for $c_s^2=10^{-6}$ is 
shown; 
we find $\sigma(w_{0})=0.0286$, $\sigma(c_s^2)/c_s^2=0.132$;
in the last case the error on the measurement of the sound speed is reduced
to the 70\% of the total signal.

\epubtkImage{w0-cs1-wl_RB-w0-cs0000001-wl_RB.png}{%
\begin{figure}
\centerline{
\includegraphics[width=0.48\textwidth]{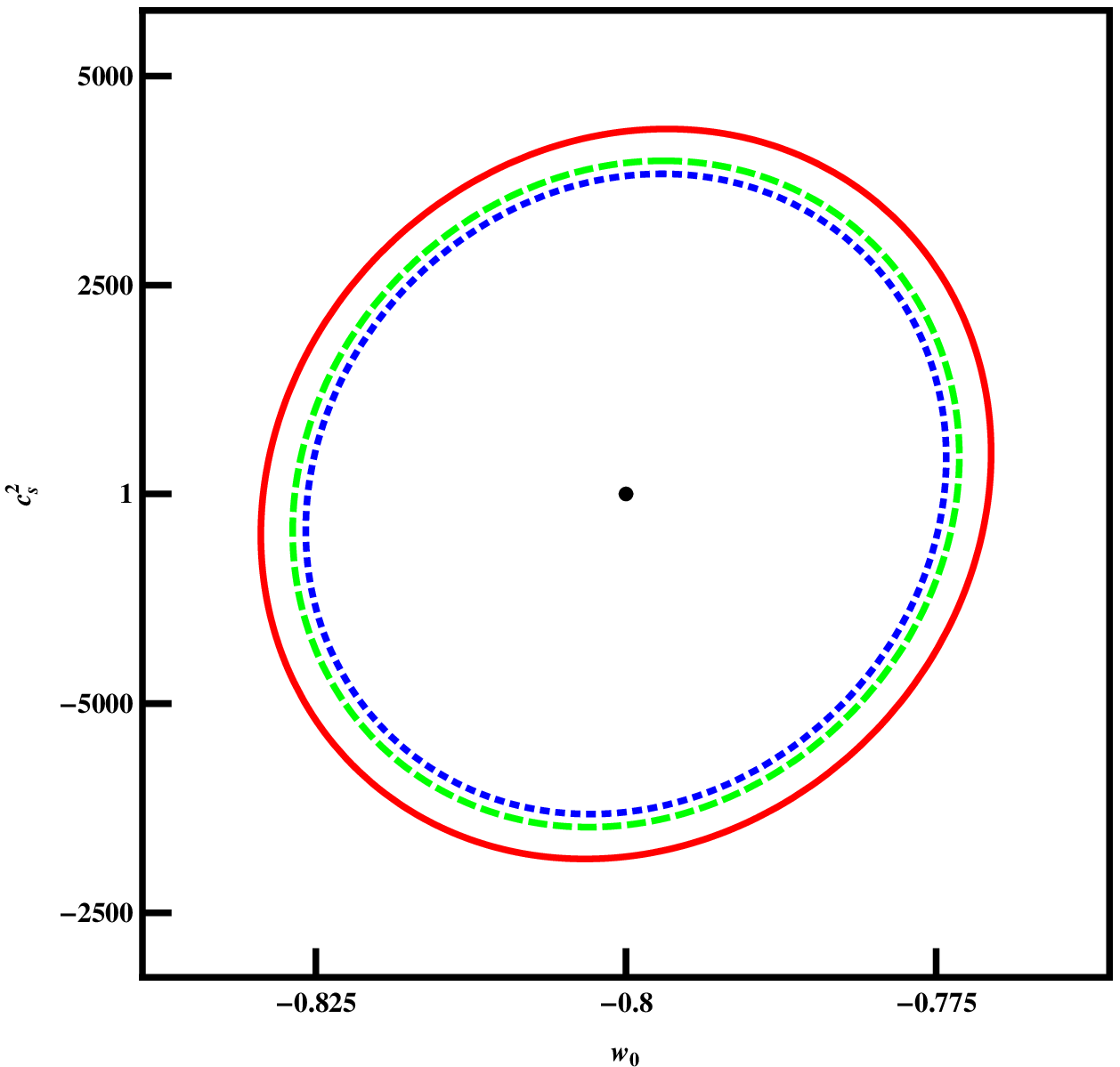}
\hspace{0.1in}
\includegraphics[width=0.48\textwidth]{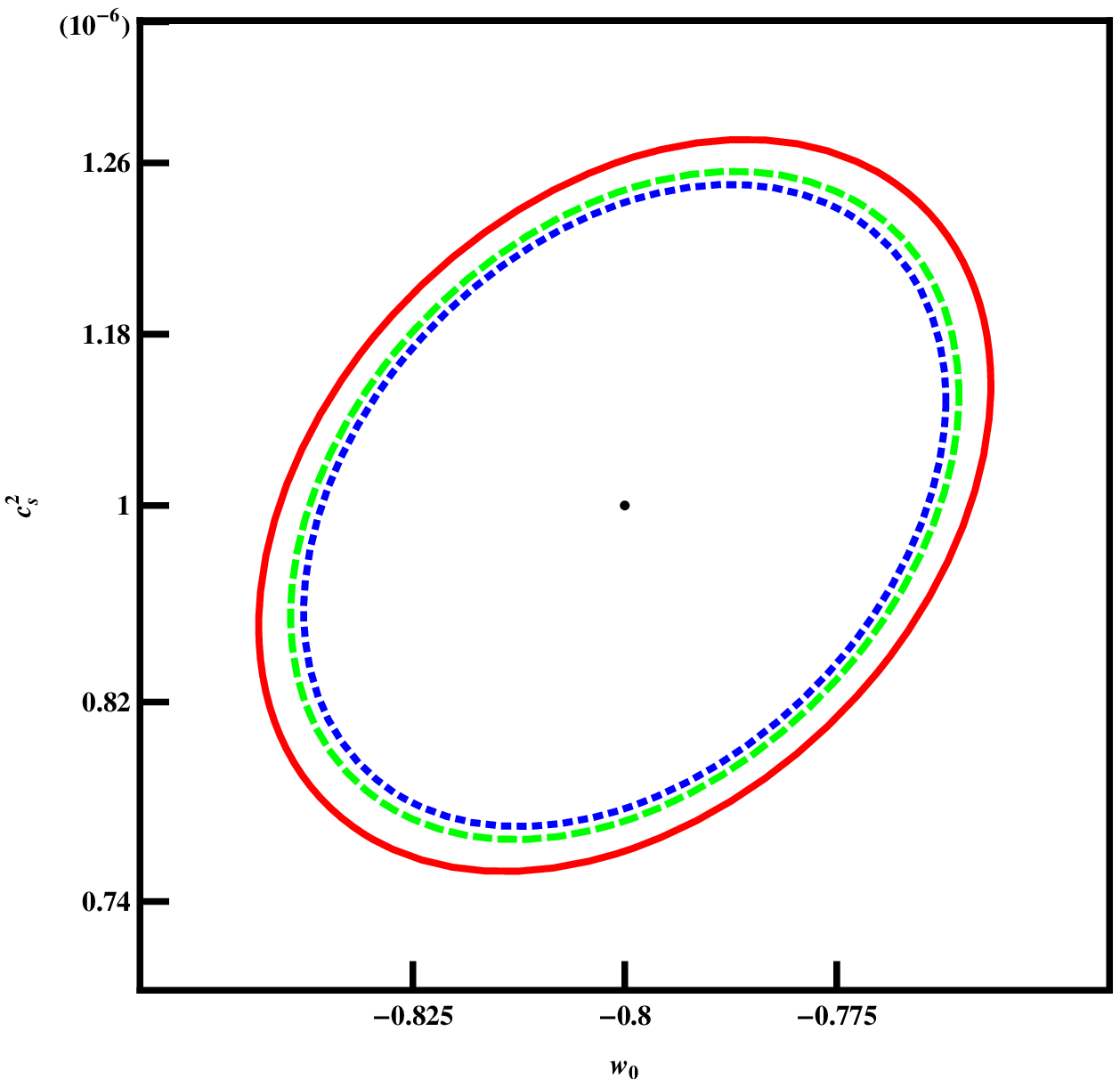}
}
\caption{Confidence region at 68\% for three different values of $z_{\max}=2.5,3.5,4$,
red solid,  green long-dashed and blue dashed contour, respectively. The
left panel shows the confidence region when the sound speed is $c_s^2=1$;
the right panel with the sound speed $c_s^2=10^{-6}$. The 
equation of state parameter is for both cases $w_{0}=-0.8$.}
\label{fig:ellipsesw0cs} 
\end{figure}}

\paragraph*{Impact on galaxy power spectrum.}

We now explore a second probe of clustering, the galaxy power spectrum. The procedure is the same outlined in Section~\ref{dark-energy-and-redshift-surveys}.
We use the representative Euclid survey presented in Section~\ref{sec:baofm_survey}. 
Here too we also consider in addition possible extended surveys to $z_{\max}=2.5$ and $z_{\max}=4$. 

In Figure~\ref{fig:ellipsesw0cs-pk} we report the confidence region
for $w_{0},c_s^2$ for two different values of the sound speed and $z_{\max}$.
For high values of the sound speed ($c_s^2=1$) we find, for our benchmark
survey: $\sigma(w_{0})=0.0133$, and $\sigma(c_s^2)/c_s^2=50.05$. Here
again we find that galaxy power spectrum is not sensitive to the clustering
properties of dark energy when the sound speed is of order unity.
If we decrease the sound speed down to $c_s^2=10^{-6}$ then the errors
are $\sigma(w_{0})=0.0125$, $\sigma(c_s^2)/c_s^2=0.118$.

\epubtkImage{w0-cs1-pk_RB-w0-cs0000001-pk_RB.png}{%
\begin{figure}
\centerline{
\includegraphics[width=0.48\textwidth]{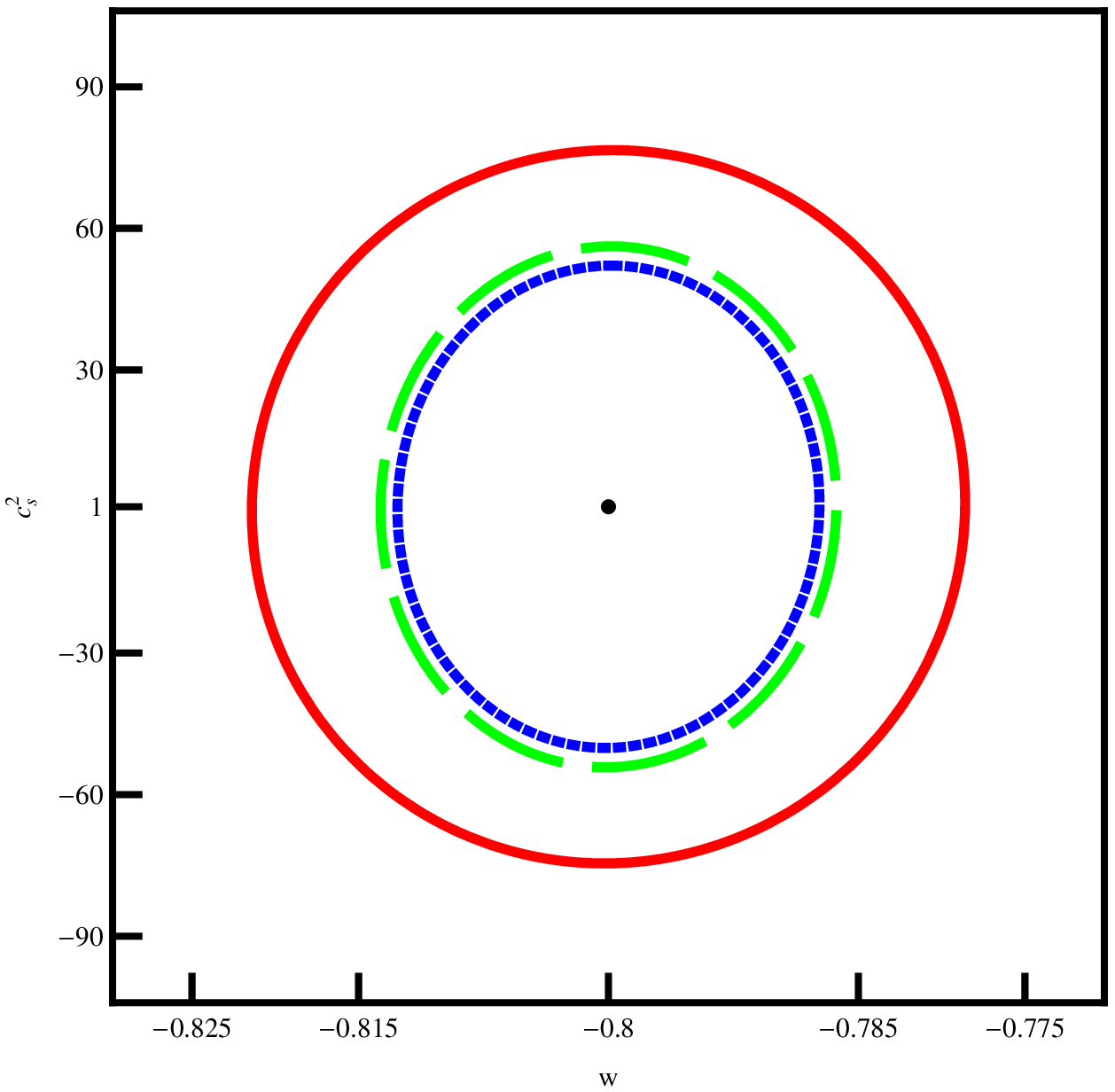}
\hspace{0.1in}
\includegraphics[width=0.48\textwidth]{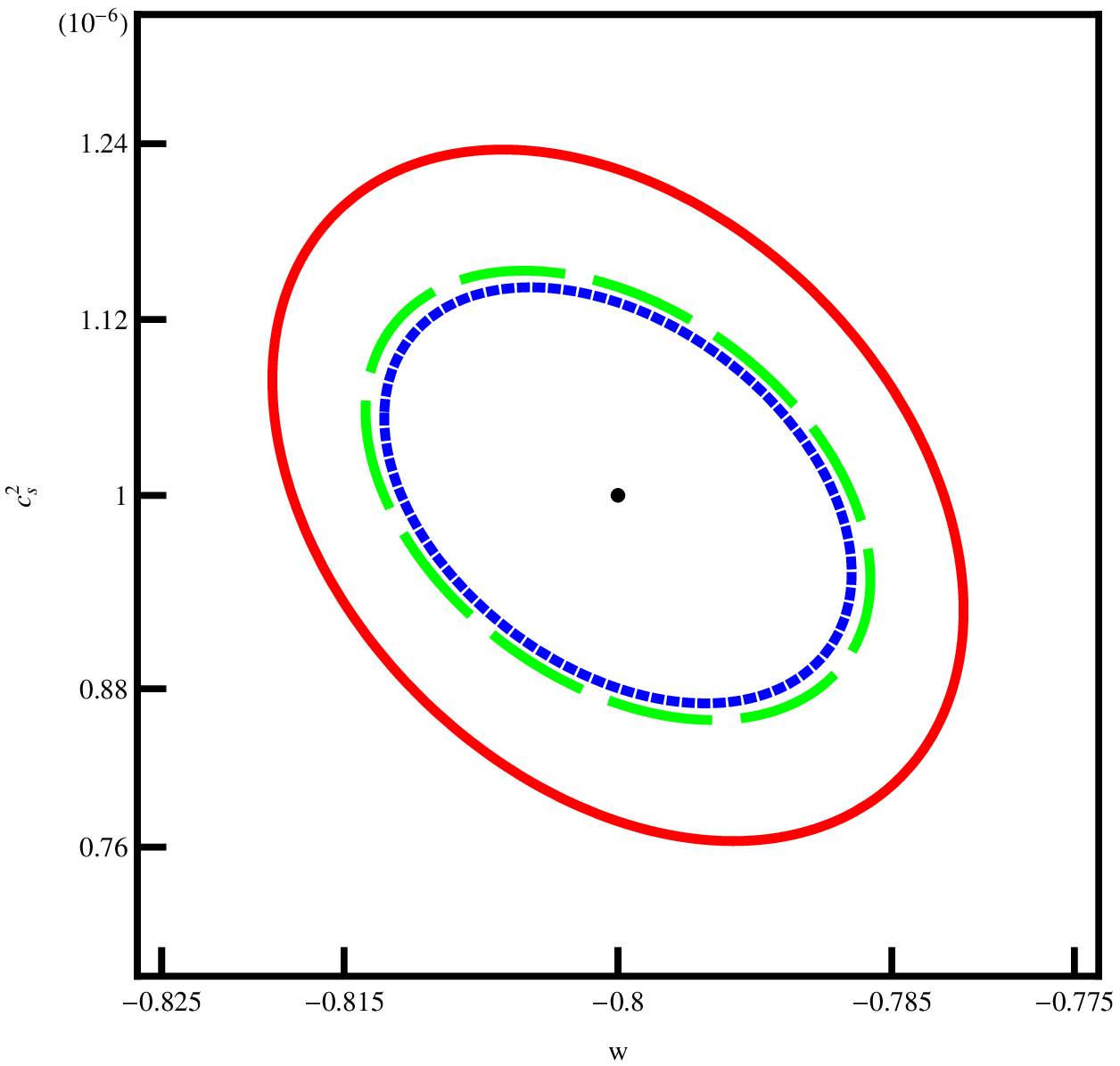}
}
\caption{Confidence region at 68\% for three different values of $z_{\max}=2.5,3.5,4$,
red solid, green long-dashed and blue dashed contour, respectively. The
left panel shows the confidence region when the sound speed is $c_s^2=1$;
the right panel with the sound speed $c_s^2=10^{-6}$. The 
equation of state parameter is for both cases $w_{0}=-0.8$.}
\label{fig:ellipsesw0cs-pk} 
\end{figure}}

In conclusion, as perhaps expected, we find that
dark-energy perturbations have a very small effect on dark matter
clustering unless the sound speed is extremely small, $c_{s}\le 0.01$. 
Let us remind that in order to boost the observable
effect, we always assumed $w=-0.8$; for values closer to $-1$ the
sensitivity to $c_s^2$ is further reduced. As a test,  \cite{Sapone:2010} 
performed the calculation for $w=-0.9$ and $c_s^2=10^{-5}$ and found $\sigma_{c_s^2}/c_s^2=2.6$
and $\sigma_{c_s^2}/c_s^2=1.09$ for WL and galaxy power spectrum experiments,
respectively.

Such small sound speeds are not in contrast with the fundamental expectation
of dark energy being much smoother that dark matter: even with $c_{s}\approx0.01$,
dark-energy perturbations are more than one order of magnitude weaker
than dark matter ones (at least for the class of models investigated
here) and safely below nonlinearity at the present time at all scales.
Models of ``cold'' dark energy are interesting because they can
cross the phantom divide \citep{Kunz:2006wc} and contribute to the cluster
masses \citep{Creminelli-etal:2010} (see also Section~\ref{the-spherical-collapse-model} of this review ). 
Small $c_{s}$ could be constructed for instance
with scalar fields with non-standard kinetic energy terms.

\subsection{Weak lensing constraints on \textit{f(R)} gravity}
\label{fRforecastconstraints}

In this section, we present the Euclid weak lensing forecasts of a
specific, but very popular, class of models, the so-called $f(R)$
models of gravity. As we have already seen in Section~\ref{fr-general}
these models are described by the action
\begin{equation}
 S_{\mathrm{grav}} = \int \sqrt{-g} \, \mathrm{d}^{4}x \left[ \frac{f(R)}{16\pi
G} - {\cal L}_{\mathrm{m}} \right] ,
\end{equation}
where $f(R)$ is an arbitrary function of the Ricci scalar and ${\cal
L}_{\mathrm{m}}$ is the Lagrange density of standard matter and radiation.

In principle one has complete freedom to specify the function $f(R)$, and so any expansion history  can be
reproduced.
However, as discussed in Section~\ref{fr-general}, those that remain viable are the subset
that very closely mimic the standard $\Lambda$CDM background expansion, as this
restricted subclass of models can evade solar system constraints
\citep{Chiba:2003ir,Tsujikawa:2008uc,Lin:2010hk}, have a standard matter era in
which the scale factor evolves according to $a(t) \propto t^{2/3}$
\citep{APT07} and can also be free of ghost and
tachyon instabilities \citep{Nariai:1973eg,Gurovich:1979xg}.

To this subclass belongs  the popular $f(R)$ model proposed by
\citet{Hu07} (\ref{Bmodel}). \citet{Camera:2011ms} demonstrated that Euclid will
have the power of distinguishing between it and $\Lambda$CDM with a
good accuracy. They performed a tomographic analysis using several values of the
maximum allowed wavenumber of the Fisher matrices; specifically, a conservative
value of 1000, an optimistic value of 5000 and a bin-dependent setting,
which increases the maximum angular wavenumber for distant shells and reduces it
for nearby shells. Moreover, they computed the Bayesian expected evidence for
the model of Eq.~(\ref{Bmodel}) over the $\Lambda$CDM model as a function of the
extra parameter $n$. This can be done because the $\Lambda$CDM model is formally
nested in this $f(R)$ model, and the latter is equivalent to the former when
$n=0$. Their results are shown in Figure~\ref{fig:lnB-HS}. For another  Bayesian  evidence analysis of $f(R)$ models  and the added value of probing the growth of structure with galaxy surveys  see also \cite{Song:2007da}.

\epubtkImage{lnB-SDDR-HS-RedBook.png}{%
\begin{figure}[htbp]
\centerline{\includegraphics[width=0.5\textwidth]{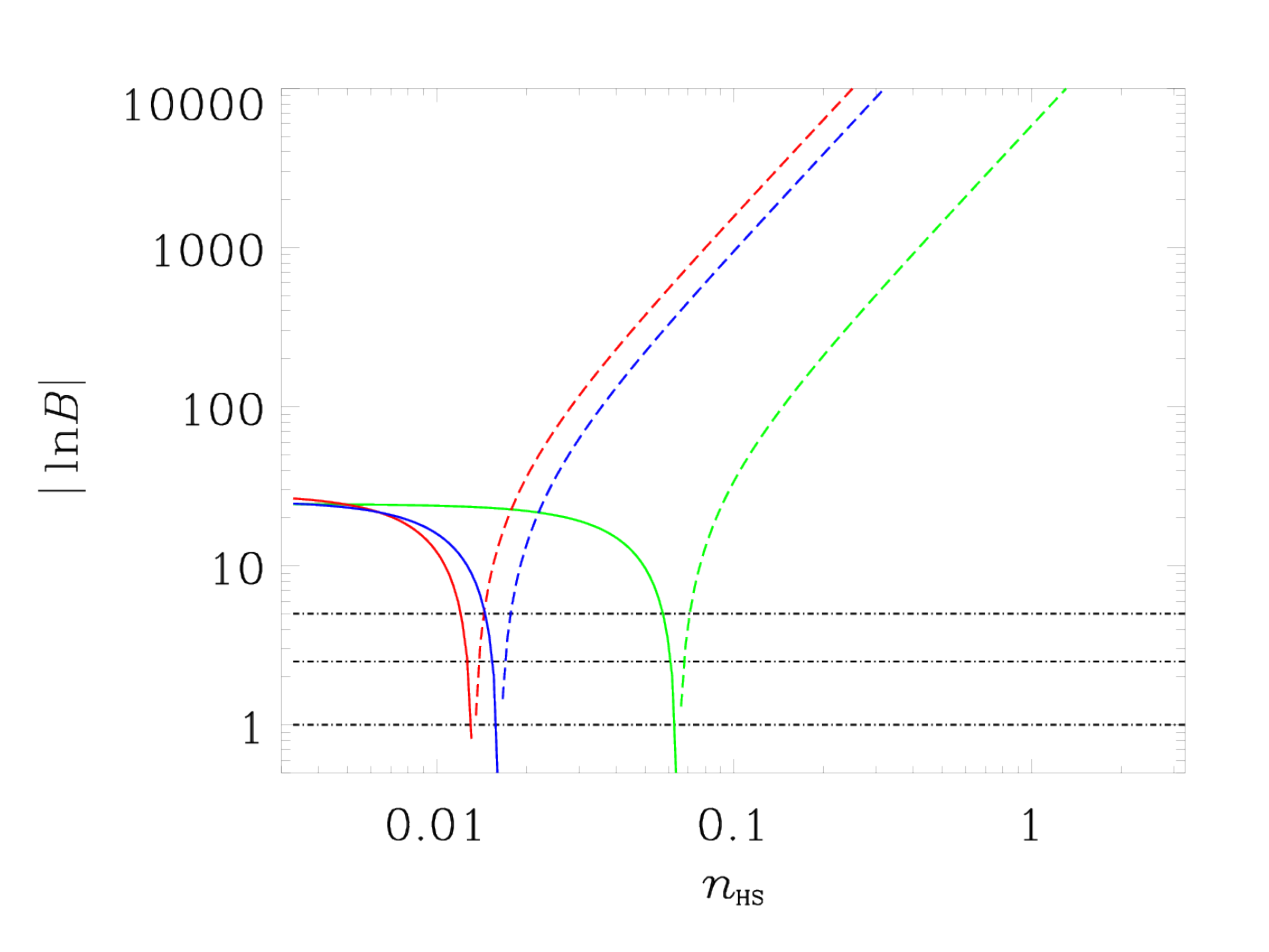}}
\caption{The Bayes factor $\ln B$ for the $f(R)$ model of Eq.~(\ref{Bmodel})
over standard $\Lambda$CDM as a function of the extra parameter $n$. The green,
red and blue curves refer to the conservative, bin-dependent and optimistic
$\ell_{\max}$, respectively. The horizontal lines denote the Jeffreys'
scale levels of significance.
\reviewtext{[Is there a higher resolution version?]}}
\label{fig:lnB-HS}
\end{figure}}

This subclass of $f(R)$ models can be parameterized solely in terms of the mass
of the scalar field, which as we have seen in Eq.~(\ref{frr-mass}) is related to the $f(R)$ functional form via the
relation 
\begin{equation}
 M^{2}(a) = {\frac{1} {3 f_{,RR}[R_{\mathrm{back}}(a)]}}
\end{equation}
where $R$ subscripts denote differentiation with respect to $R$. The function $f_{,RR}$ can be approximated by its standard $\Lambda$CDM form,
\begin{equation}
 {\frac{R_{\mathrm{back}}} {H_{0}^{2}}} \simeq {\frac{3\Omega_{\rm m0}} 
{a^{3}}}  + 12\Omega_{\Lambda} \,,
\end{equation}
valid for $z \lesssim 1000$. The mass $M(a)$ is typically a function
of redshift which decays from a large value in the early universe to its present
day value $M_{0}$.

Whilst these models are practically indistinguishable from $\Lambda$CDM at the
level of background expansion, there is a significant difference in the
evolution of perturbations relative to the standard GR behavior.

The evolution of linear density perturbations in the context of $f(R)$ gravity
is markedly different than in the standard $\Lambda$CDM scenario; $\delta_{\mathrm{m}} \equiv \delta \rho_{\mathrm{m}} /\rho_{\mathrm{m}}$ acquires a nontrivial scale
dependence at late times. This is due to the presence of an additional scale
$M(a)$ in the equations; as any given mode crosses the modified gravity
`horizon' $k = aM(a)$, said mode will feel an enhanced gravitational force due
to the scalar field. This will have the effect of increasing the power of small
scale modes.

Perturbations on sub-horizon scales in the Newtonian gauge evolve approximately
according to
\begin{eqnarray}
 \label{eq:16} & &  \Psi = \left(1 + {\frac{2\bar{K}^{2}} {3 +
2\bar{K}^{2}}}\right) \Phi , \\ \label{eq:17} & & k^{2}\Phi = -4\pi G \left({\frac{3 +
2\bar{K}^{2}} {3 + 3\bar{K}^{2}}}\right)a^{2}\rho_{\mathrm{m}} \delta_{\mathrm{m}} , \\
\label{eq:18}  & & \label{eq:p1} \ddot{\delta}_{\mathrm{m}} + 2H \dot{\delta}_{\mathrm{m}}
- 4\pi G \left({\frac{3 + 4\bar{K}^{2}}{ 3 + 3\bar{K}^{2}}}\right)\rho_{\mathrm{m}}\delta_{\mathrm{m}} = 0 \,,
\end{eqnarray}
where $\bar{K} = k/(aM(a))$. These equations represent a particular
example of a general parameterization introduced in
\cite{Martinelli:2010wn,Bertschinger:2008zb,Zhao:2008bn}. To solve them one
should first parameterize the scalaron mass $M(a)$, choosing a form that broadly
describes the behavior of viable $f(R)$ models. A suitable functional form,
which takes into account the evolution of $M(a)$ in both the matter era and the
late-time accelerating epoch, is given by \cite{Thomas:2011pj}
\begin{equation}
 \label{eq:107}
 M^{2} = M_{0}^{2} \left( {\frac{ a^{-3} + 4 a_{*}^{-3}}
{ 1 + 4 a_{*}^{-3}}}\right)^{2\nu} \,,
 \end{equation}
where $a_{*}$ is the scale factor at matter-$\Lambda$ equality; $a_{*}
= (\Omega_{\rm m0}/\Omega_{\Lambda})^{1/3}$.  There are two modified gravity
parameters; $M_{0}$ is the mass of the scalaron at the present time and $\nu$ is
the rate of increase of $M(a)$ to the past.

\epubtkImage{scalaron_matter_power_proper-scalaron_chameleon_PT_2_proper.png}{%
\begin{figure}[htbp]
  \centerline{
    \includegraphics[width=7.6cm]{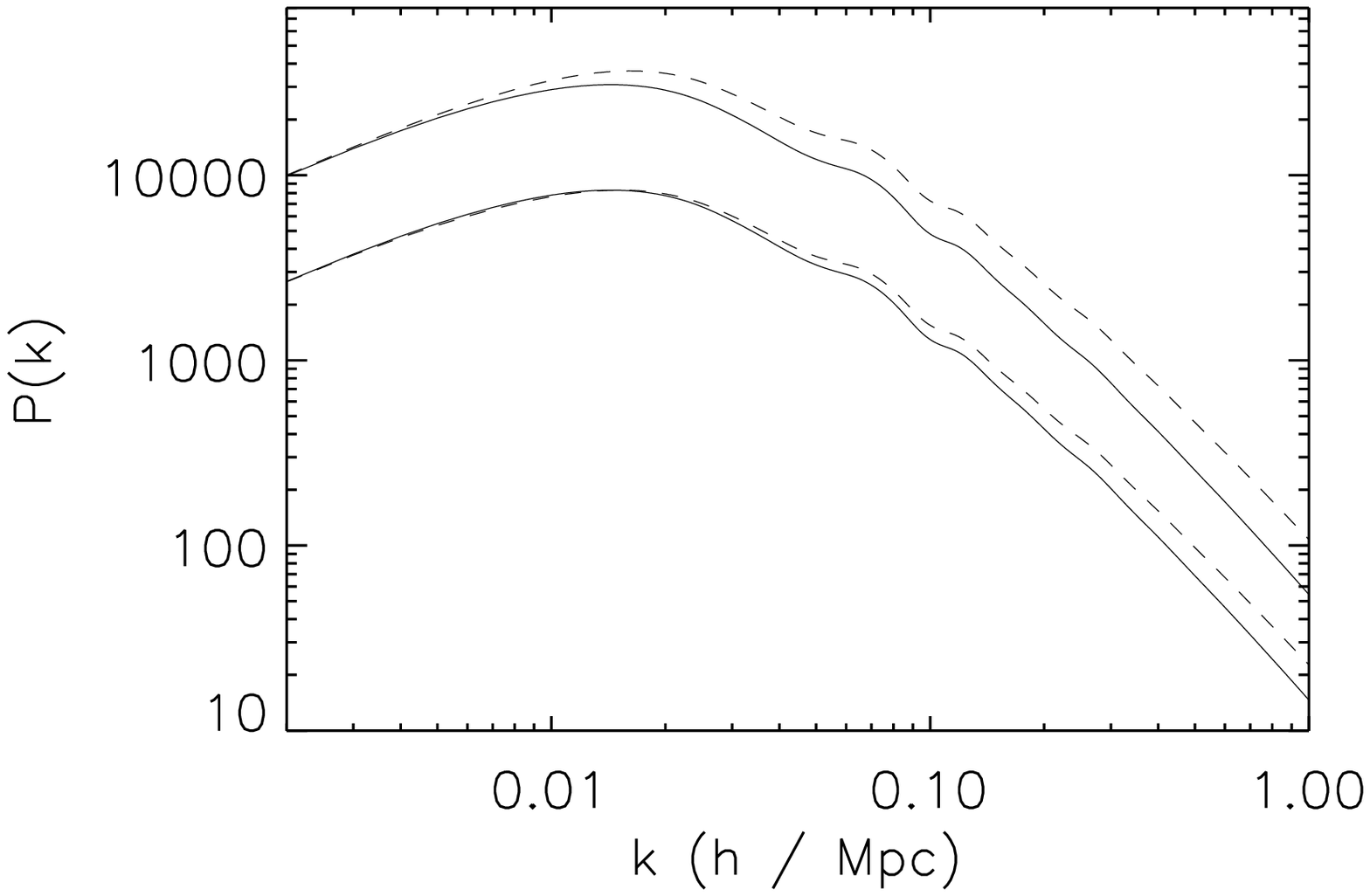}
    \includegraphics[width=7.6cm]{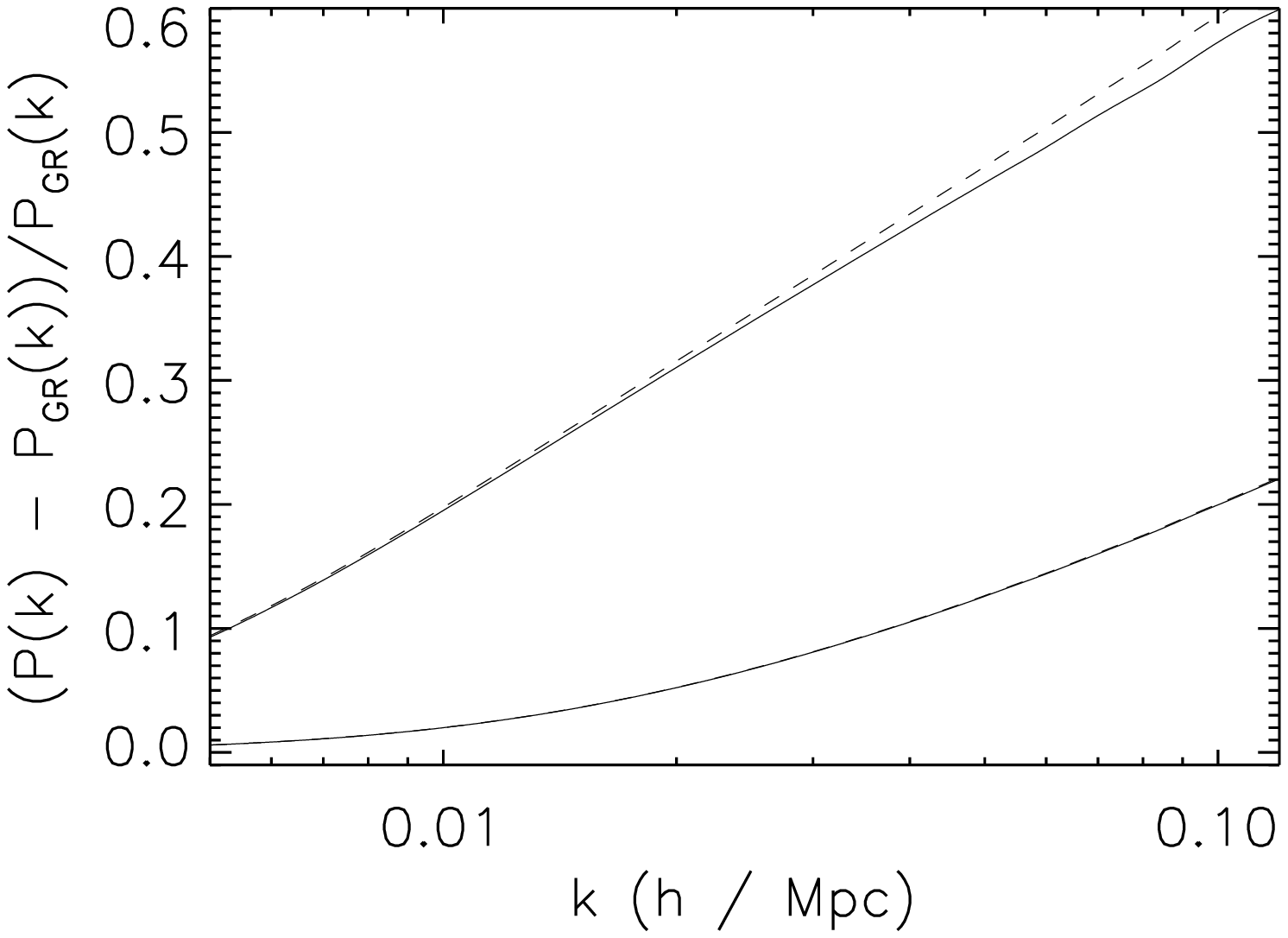}
  }
  \caption{\emph{Left panel:} Linear matter power spectra for
    $\Lambda$CDM (solid line; $M_{0}^{-1}=0$, $\nu=1.5$) and scalaron
    (dashed line; $M^{-1}_{0}=375
    [10^{28}\mathrm{\ h^{-1}\ eV^{-1}}]$, $\nu=1.5$) cosmologies. The
    modification to gravity causes a sizeable scale dependent effect
    in the growth of perturbations. The redshift dependence of the
    scalaron can be seen by comparing the top and bottom pairs of
    power spectra evaluated at redshifts $z=0.0$ and $z=1.5$,
    respectively. \emph{Right panel:} The environmental dependent
    chameleon mechanism can be seen in the mildly nonlinear
    regime. We exhibit the fractional difference $(P(k) -
    P_{\mathrm{GR}}(k))/P_{\mathrm{GR}}(k)$ between the $f(R)$ and GR
    power spectra for the model (\ref{eq:107}) with parameters
    $M^{-1}_{0}=375 [10^{28}\mathrm{\ h^{-1}\ eV^{-1}}]$ and
    $\nu=1.5$. The dashed lines represent linear power spectra ($P(k)$
    and $P_{\mathrm{GR}}(k)$ calculated with no higher order effects)
    and the solid lines are the power spectra calculated to second
    order. We see that the nonlinearities decrease the modified
    gravity signal. This is a result of the chameleon mechanism. The
    top set of lines correspond to $z=0$ and the bottom to $z=0.9$;
    demonstrating that the modified gravity signal dramatically
    decreases for larger $z$. This is due to the scalaron mass being
    much larger at higher redshifts. Furthermore, nonlinear effects
    are less significant for increasing $z$.}
    \label{fig:2b}
\end{figure}}

In Figure~\ref{fig:2b} the linear matter power spectrum is exhibited for this
parameterization (dashed line), along with the standard $\Lambda$CDM power
spectrum (solid line). The observed, redshift dependent tilt is due to the
scalaron's influence on small scale modes, and represents a clear modified
gravity signal. Since weak lensing is sensitive to the underlying matter power
spectrum, we expect Euclid to provide  direct constraints on the mass of the
scalar field.

By performing a Fisher analysis, using the standard Euclid
specifications, \cite{Thomas:2011pj} calculates the expected $f(R)$
parameter sensitivity of the weak lensing survey. By combining Euclid
weak lensing and Planck Fisher matrices, both modified gravity
parameters $M_{0}$ and $\nu$ are shown to be strongly constrained by
the growth data in Figure~\ref{fig:fr_saa}. The expected $1\sigma$
bounds on $M_{0}$ and $\nu$ are quoted as $M_{0} = 1.34 \pm 0.62
\times 10^{-30} [\mathrm{h\ eV}]$, $\nu = 1.5 \pm 0.18$ when using linear data
$l < 400$ only and $M_{0} = 1.34 \pm 0.25 \times 10^{-30} [\mathrm{h\ eV}]$,
$\nu = 1.5 \pm 0.04$ when utilizing the full set of nonlinear modes
$l < 10000$. 

\epubtkImage{power_law_combined_400_to_10000.png}{%
\begin{figure}[htbp]
  \centerline{\includegraphics[width=7.6cm]{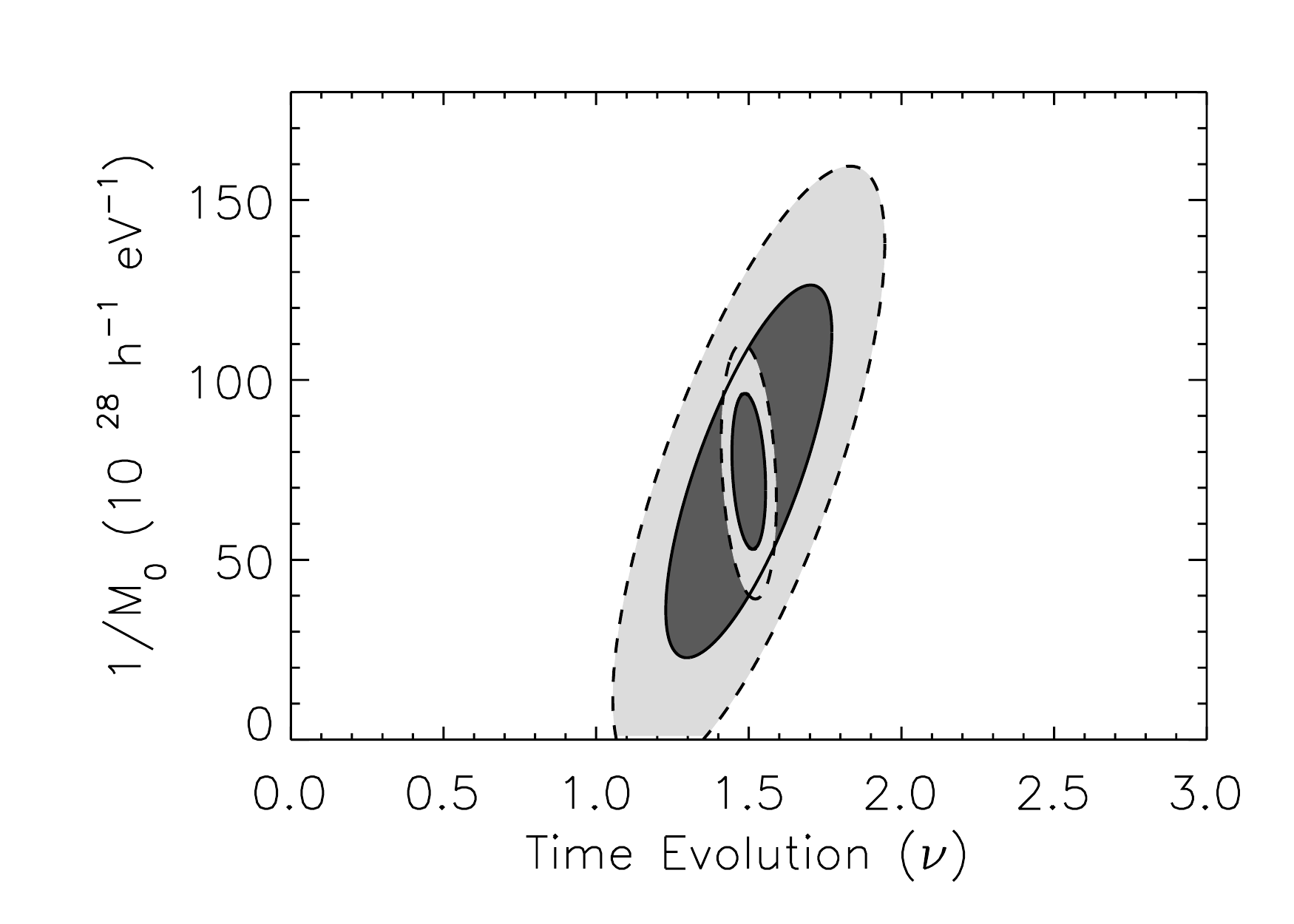}}
  \caption{68\% (dark grey) and 95\% (light grey) projected bounds on
    the modified gravity parameters $M_{0}^{-1}$ and $\nu$ for the
    combined Euclid weak lensing and Planck CMB surveys. The \reviewtext{\st{smaller
    (larger)}} \reviewtext{larger (smaller)} contours correspond to including modes $l = 400 (10000)$
    in the weak lensing analysis.}
  \label{fig:fr_saa}
\end{figure}}

\subsection{Forecast constraints on coupled quintessence cosmologies}
\label{cdeforecast}

In this section we present forecasts for coupled quintessence cosmologies \cite{Amendola:1999er,Wetterich:1994bg,Pettorino:2008ez}, obtained when combining Euclid weak lensing, Euclid redshift survey (baryon acoustic oscillations, redshift distortions and full $P(k)$ shape) and CMB as obtained in Planck (see also the next section for CMB priors). Results reported here were obtained in \cite{Amendola:2011ie} and we refer to it for details on the analysis and Planck specifications 
(for weak lensing and CMB constraints on coupled quintessence with a different coupling see also \cite{martinelli,DeBernardis:2011iw}). 
In \cite{Amendola:2011ie} the coupling is the one described in Section~\ref{cde_eq}, as induced by a scalar-tensor
model. The slope $\alpha$ of the Ratra--Peebles potential is included as an additional parameter and Euclid specifications refer to the Euclid Definition phase \cite{euclidredbook}.

The combined Fisher confidence regions are plotted in
Figure~\ref{fig:cmbbaowl-cont} and the results are in
Table~\ref{tab:combined}. The main result is that future surveys can
constrain the coupling of dark energy to dark matter $\beta^{2}$ to
less than $3\cdot10^{-4}$. Interestingly, some combinations of
parameters (e.g., $\Omega_b$ vs $\alpha$) seem to profit the most from
the combination of the three probes. 

\begin{table}[htbp]
\caption[1-$\sigma$ errors for the set $\Theta\equiv\{\beta^{2},\alpha,\Omega_{c},h,\Omega_{b},n_{s}\,\sigma_{8},\log(A)\}$ of cosmological parameters, combining CMB~+$P(k)$ and CMB~+~$P(k)$~+~WL.]{1-$\sigma$ errors for the set $\Theta\equiv\{\beta^{2},\alpha,\Omega_{c},h,\Omega_{b},n_{s}\,\sigma_{8},\log(A)\}$ of cosmological parameters, combining CMB~+~$P(k)$ (left column) and CMB~+~$P(k)$~+~WL (right column).}
\label{tab:combined}
\centering
{\small
\begin{tabular}{lll}
\toprule
Parameter & $\sigma_{i}$~CMB~+~$P(k)$ & $\sigma_{i}$~CMB~+~$P(k)$~+~WL \\
\midrule
$\beta^{2}$  & 0.00051  & 0.00032 \\
$\alpha$    & 0.055   & 0.032 \\
$\Omega_{c}$ & 0.0037  & 0.0010 \\
$h$         & 0.0080  & 0.0048 \\
$\Omega_{b}$ & 0.00047 & 0.00041 \\
$n_{s}$      & 0.0057  & 0.0049 \\
$\sigma_{8}$ & 0.0049  & 0.0036 \\
$\log(A)$   & 0.0051  & 0.0027 \\
\bottomrule
\end{tabular}}
\end{table}

\epubtkImage{cmbbaowl_beta2_plot_8par_colorv2_amendolaetal2011.png}{%
\begin{figure}
\centerline{\includegraphics[width=\textwidth]{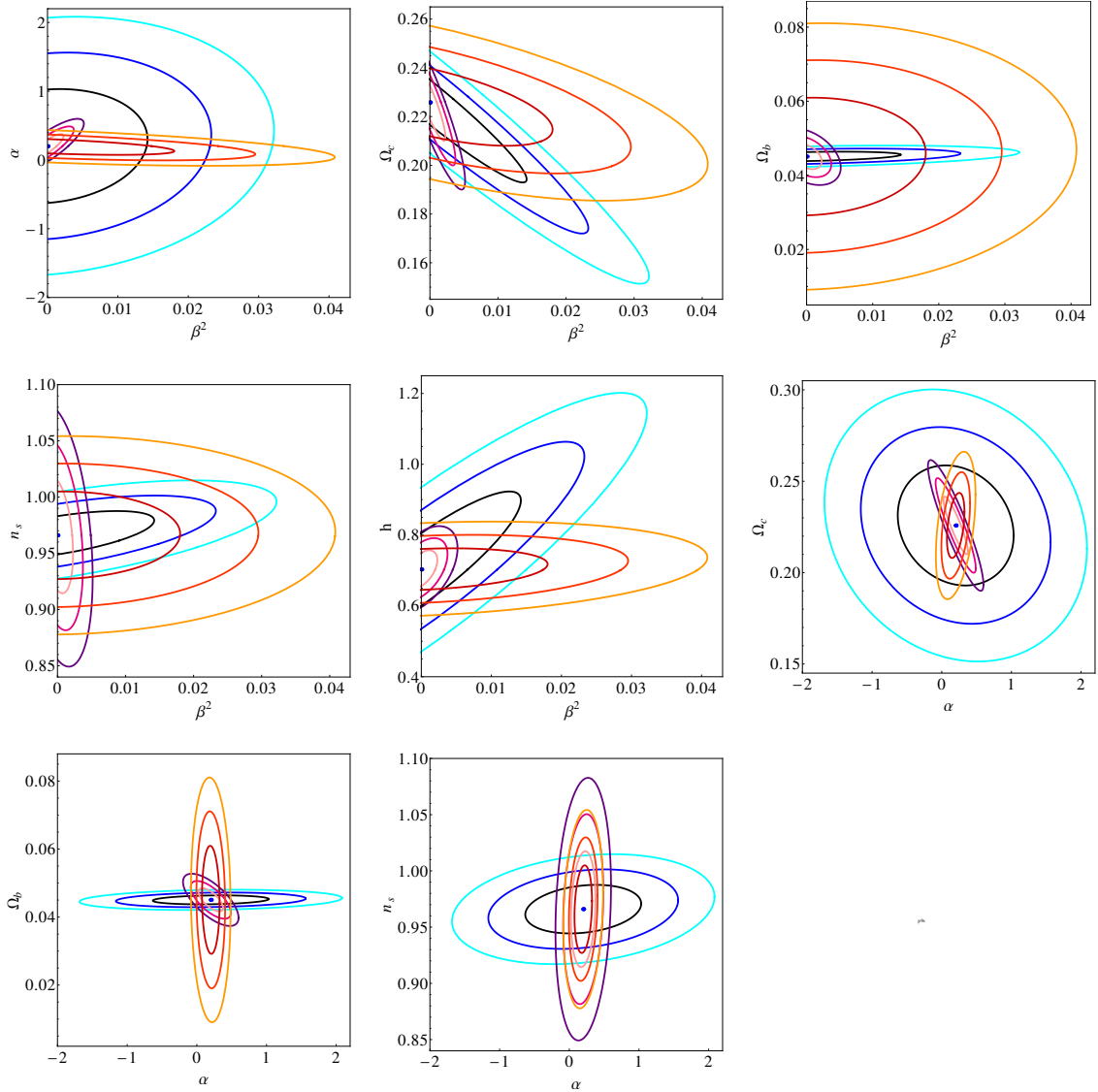}}
\caption{Comparison among predicted confidence contours for the
  cosmological parameter set
  $\Theta\equiv\{\beta^{2},\alpha,\Omega_{c},h,\Omega_{b},n_{s},\sigma_{8},\log(A)\}$
  using CMB (Planck, blue contours), $P(k)$ (pink-violet contours)
  and weak lensing (orange-red contours) with Euclid-like
  specifications. Image reproduced by permission
  from~\cite{Amendola:2011ie}, copyright by APS.}
\label{fig:cmbbaowl-cont}
\end{figure}}

We can also ask whether a better knowledge of the parameters $\{\alpha,\Omega_{c},h,\Omega_{b},n_{s},\sigma_{8},\log(A)\}$,
obtained by independent future observations, can give us better constraints
on the coupling $\beta^{2}$. In Table~\ref{tab:fixedpar}  we show the errors
on $\beta^{2}$ when we have a better knowledge of only one other
parameter, which is here fixed to the reference value. All remaining
parameters are marginalized over. 

It is remarkable to notice that the combination of CMB, power spectrum
and weak lensing is already a powerful tool and a better knowledge
of one parameter does not improve much the constraints on $\beta^{2}$.
CMB alone, instead, improves by a factor 3 when $\Omega_{c}$ is known
and by a factor 2 when $h$ is known. The power spectrum is mostly
influenced by $\Omega_{c}$, which allows to improve constraints on
the coupling by more than a factor 2. Weak lensing gains the most
by a better knowledge of $\sigma_{8}$.

\begin{table}[htbp]
\caption[1-$\sigma$ errors for $\beta^{2}$, for CMB, $P(k)$, WL and
  CMB~+~$P(k)$~+~WL.]{1-$\sigma$ errors for $\beta^{2}$, for CMB, $P(k)$, WL and
  CMB~+~$P(k)$~+~WL. For each line, only the parameter in the left column
  has been fixed to the reference value. The first line corresponds to
  the case in which we have marginalized over all parameters.  Table
  reproduced by permission from~\cite{Amendola:2011ie}, copyright by
  APS.}
  
\label{tab:fixedpar}
\centering
{\small
\begin{tabular}{llllc}
\toprule
Fixed parameter & CMB & $P(k)$ & WL & CMB~+~$P(k)$~+~WL \\
\midrule
(Marginalized on all params) & 0.0094  & 0.0015  & 0.012  & 0.00032 \\
$\alpha$  & 0.0093  & 0.00085  & 0.0098  & 0.00030 \\
$\Omega_{c}$  & 0.0026  & 0.00066  & 0.0093  & 0.00032 \\
$h$  & 0.0044  & 0.0013  & 0.011  & 0.00032 \\
$\Omega_{b}$  & 0.0087  & 0.0014  & 0.012  & 0.00030 \\
$n_{s}$  & 0.0074  & 0.0014  & 0.012  & 0.00028 \\
$\sigma_{8}$  & 0.0094  & 0.00084  & 0.0053  & 0.00030 \\
$\log(A)$  & 0.0090  & 0.0015  & 0.012  & 0.00032 \\
\bottomrule
\end{tabular}}
\end{table}

\subsection{Extra-Euclidean data and priors}
In addition to the baseline Euclid surveys, a possibility may exist for an 
auxiliary Euclid survey, for example focused on Type Ia supernovae. 
Type Ia supernovae used as standardized candles (luminosity distance 
indicators) led to the discovery of cosmic acceleration and they retain 
significant leverage for revealing the nature of dark energy. Their observed 
flux over the months after explosion (the light curve) is calibrated by an 
empirical brightness--light curve width relation into a luminosity distance 
multiplied by a factor involving the unknown absolute brightness and Hubble 
constant. This nuisance factor cancels when supernovae at different redshifts 
are used as a relative distance measure. The relative distance is highly 
sensitive to cosmic acceleration and provides strong complementarity with 
other cosmological probes of dark energy, at the same or different redshifts. 

Another advantageous property of supernovae is their immunity to systematics 
from cosmology theory -- they are purely geometric measures and do not care 
about the matter power spectrum, coupling to dark matter, cosmologically 
modified gravity, dark energy clustering, etc. Their astrophysical 
systematics are independent of other probes, giving important crosschecks. 
The cosmological parameter likelihood function arising from supernovae 
constraints can to a good approximation simply be multiplied with the 
likelihood from other probes. Current supernovae likelihoods are in user 
friendly form from the joint lightcurve analysis (JLA) of the Supernova Legacy 
Survey (SNLS) and Sloan Digital Sky Survey (SDSS) of \cite{2014A&A...568A..22B} 
or the Union2.1 compilation of \cite{Suzuki2012}. 
In the near future the Union3 compilation should merge these sets and all 
other current supernova data, within an improved Bayesian framework. 

The Euclid Supernovae Science Working Group proposed a six month auxiliary 
survey with Euclid, the Dark Energy Supernova InfraRed Experiment (DESIRE) 
\cite{2014A&A...572A..80A}. 
This delivers substantial improvements on dark energy equation of state 
constraints relative to ground-based supernova surveys, with a 50\% higher 
figure of merit, as shown in Table~\ref{tab:cosmo-fom}.

\begin{table}[h]
\caption{Cosmological performance of the simulated surveys. The FoMs assume a 1-D geometrical {\it Planck} prior and flatness.
$z_p$ is the redshift at which the equation of state uncertainty
reaches its minimum $\sigma(w_p)$. The FoM is defined as $[Det(Cov(w_0, w_a))]^{-1/2}= [\sigma(w_a)\sigma(w_p)]^{-1}$ and accounts for a suite of systematic uncertainties (see \cite{2014A&A...572A..80A}).
\label{tab:cosmo-fom}}
\begin{center}
\begin{tabular}{l|rrrr|}
  &  {\boldmath $\sigma(w_a)$} &  {\boldmath $z_p$} &  {\boldmath$\sigma(w_p)$} 
& {\bf FoM} \\ 
\hline
low-z + LSST-DDF  & \multirow{2}{*}{0.22} & \multirow{2}{*}{0.25} & \multirow{2}{*}{0.022} & \multirow{2}{*}{203.2} \\  
        \multicolumn{1}{c|}{+ DESIRE} & & & & \\  
low-z + LSST-DDF &  0.28 &  0.22 &    0.026 &   137.1 \\
LSST-DDF + DESIRE &   0.40 &  0.35 &    0.031 &    81.4 \\
\hline
\end{tabular}
\end{center}
\end{table}

Other dark-energy projects will enable the cross-check of the dark-energy 
constraints from Euclid. These include Planck, BOSS, WiggleZ, HETDEX, DES,
Panstarrs, LSST, BigBOSS and SKA.

Planck will provide exquisite constraints
on cosmological parameters, but not tight constraints on dark energy
by itself, as CMB data are not sensitive to the nature of dark energy
(which has to be probed at $z<2$, where dark energy becomes increasingly
important in the cosmic expansion history and the growth history of 
cosmic large scale structure). Planck data in combination with Euclid
data provide powerful constraints on dark energy and tests of gravity.
In the next Section~\ref{sec:planck-prior}, we will discuss how to
create a Gaussian approximation to the Planck parameter constraints
that can be combined with Euclid forecasts in order to model the
expected sensitivity.

The galaxy redshift surveys BOSS, WiggleZ, HETDEX, and BigBOSS are
complementary to Euclid, since the 
overlap in redshift ranges of different galaxy redshift surveys, both space 
and ground-based, is critical for understanding systematic effects such 
as bias through the use of multiple tracers of cosmic large scale structure. 
Euclid will survey H$\alpha$ emission line galaxies 
at $0.5 < z < 2.0$ over 15,000 square degrees.
The use of multiple tracers of cosmic large scale structure can reduce 
systematic effects and ultimately increase the precision of dark-energy 
measurements from galaxy redshift surveys \citep[see,
  e.g.,][]{Seljak_Hamaus_Desjacques09}.

Currently on-going or recently completed surveys which cover a
sufficiently large volume to measure BAO at several redshifts and thus
have science goals common to Euclid, are the Sloan Digital Sky Survey
III Baryon Oscillations Spectroscopic Survey (BOSS for short) and the
WiggleZ survey.

BOSS%
\epubtkFootnote{\url{http://www.sdss3.org/surveys/boss.php}}
maps the redshifts of 1.5~million Luminous Red Galaxies (LRGs) out to $z\sim0.7$ over 10,000 square degrees, measuring the BAO signal, the large-scale galaxy correlations and extracting information of the growth from redshift space distortions. A simultaneous survey of $2.2 < z < 3.5$ quasars measures the acoustic oscillations in the correlations of the Lyman-$\alpha$ forest. LRGs were chosen for their high bias, their approximately constant number density and, of course, the fact that they are bright. Their spectra and redshift can be measured with relatively short exposures in a 2.4~m ground-based telescope. The data-taking of BOSS will end in 2014. 

The WiggleZ%
\epubtkFootnote{\url{http://wigglez.swin.edu.au/site/index.html}}
survey is now completed, it measured redshifts for almost 240,000 galaxies over 1000 square degrees at $0.2<z<1$. The target are luminous blue star-forming galaxies with spectra dominated by patterns of strong atomic emission lines. This choice is motivated by the fact that these emission lines can be used to measure a galaxy redshift in relatively short exposures of a 4~m class ground-based telescope.

Red quiescent galaxies inhabit dense clusters environments, while blue star-forming galaxies trace better lower density regions such as sheets and filaments. It is believed that on large cosmological scales these details are unimportant and that galaxies are simply tracers of the underlying dark matter: different galaxy type will only have a different `bias factor'. The fact that so far results from BOSS and WiggleZ agree well confirms this assumption.

Between now and the availability of Euclid data other wide-field spectroscopic galaxy redshift surveys will take place. 
Among them, eBOSS will extend BOSS operations focusing on 3100 square degrees using a variety of tracers. Emission line galaxies will be targeted in the redshift window $0.6<z<1$. This will extend to higher redshift and extend the sky coverage of the WiggleZ survey. Quasars in the redshift range $1<z<2.2$ will be used as tracers of the BAO feature instead of galaxies. The BAO LRG measurement will be extended to $z \sim 0.8$, and the quasar number density at $z>2.2$ of BOSS will be tripled, thus improving the BAO Lyman-$\alpha$ forest measure.

HETDEX aims at surveying 1~million Lyman-$\alpha$ emitting galaxies at $1.9 < z < 3.5$ over 420 square degrees. The main science goal is to map the BAO feature over this redshift range. 

Further in the future, we highlight here the proposed BigBOSS survey and SuMIRe survey with HyperSupremeCam on the Subaru telescope. 
The BigBOSS survey will target [OII] emission line galaxies at $0.6 < z < 1.5$ (and LRGs at $z < 0.6$) over 14,000 square degrees.
The SuMIRe wide survey proposes to survey $\sim 2000$ square degrees in the redshift range $0.6<z<1.6$ targeting LRGs and [OII] emission-line galaxies. Both these surveys will likely reach full science operations roughly at the same time as the Euclid launch. 

Wide field photometric surveys are also being carried out and
planned. The on-going Dark Energy Survey (DES)%
\epubtkFootnote{\url{http://www.darkenergysurvey.org}}
will cover 5000 square degrees out to $z\sim1.3$ and is expected to
complete observations in 2017; the Panoramic Survey Telescope \& Rapid
Response System (Pan-STARRS), on-going at the single-mirror stage, The
PanSTARSS survey, which first phase is already on-going, will cover
30,000 square degrees with 5 photometry bands for redshifts up to
$z\sim 1.5$. The second pause of the survey is expected to be competed
by the time Euclid launches. More in the future the Large Synoptic
Survey Telescope (LSST) will cover redshifts $0.3<z<3.6$ over 15,000
square degrees, but is expected to begin operations in 2021, after
Euclid's planned launch date. The galaxy imaging surveys DES,
Panstarrs, and LSST will complement Euclid imaging survey in both the
choice of band passes, and the sky coverage. 

SKA (which is expected to begin operations in 2020 and reach full
operational capability in 2024) will survey neutral atomic hydrogen
(HI) through the radio 21~cm line, over a very wide area of the
sky. It is expected to detect HI emitting galaxies out to $z\sim 1.5$
making it nicely complementary to Euclid. Such galaxy redshift survey
will of course offer the opportunity to measure the galaxy power
spectrum (and therefore the BAO feature) out to $z \sim 1.5$. The well
behaved point spread function of a synthesis array like the SKA should
ensure superb image quality enabling cosmic shear to be accurately
measured and tomographic weak lensing used to constrain cosmology and
in particular dark energy. This weak lensing capability also makes SKA
and Euclid very complementary. For more information see, e.g.,
\cite{Rawlings:2004wk, Blake:2004pb}.

\epubtkImage{contextsp2.png}{%
\begin{figure}
\centerline{\includegraphics[width=0.9\textwidth]{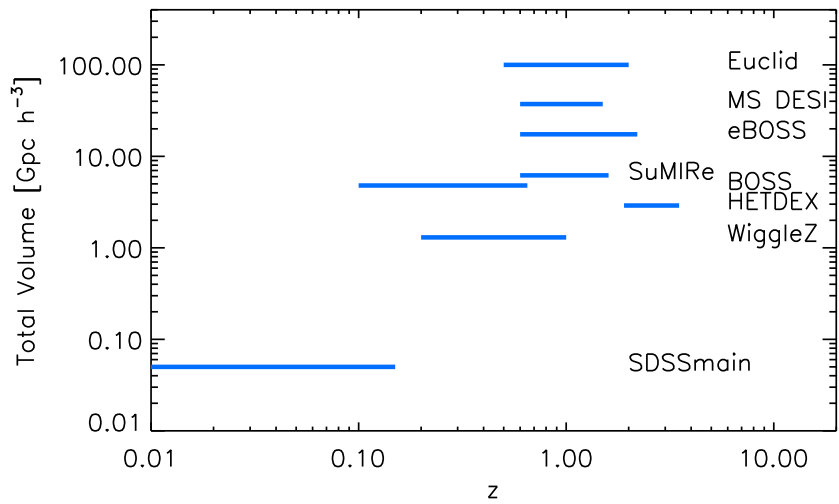}}
\caption{Redshift coverage and volume for the surveys mentioned in the
 text. Spectroscopic surveys only are shown. Recall that while
  future and forthcoming photometric surveys focus on weak
  gravitational lensing, spectroscopic surveys can extract the three
  dimensional galaxy clustering information and therefore measure
  radial and tangential BAO signal, the power spectrum shape and the
  growth of structure via redshift space distortions. The
  three-dimensional clustering information is crucial for BAO. For
  example to obtain the same figure of merit for dark-energy
  properties a photometric survey must cover a volume roughly ten
  times bigger than a spectroscopic one.}
 \label{fig:future_probes_context}
\end{figure}}

Figure~\ref{fig:future_probes_context} puts Euclid into
context. Euclid will survey H$\alpha$ emission line galaxies at $0.5
< z < 2.0$ over 15,000 square degrees. Clearly, Euclid with both
spectroscopic and photometric capabilities and wide field coverage
surpasses all surveys that will be carried out by the time it
is launched. The large volume surveyed is crucial as the number of modes
to sample for example the power spectrum and the BAO feature scales
with the volume. The redshift coverage is also important especially at
$z<2$ where the dark-energy contribution to the density of the
universe is non-negligible (at $z>2$ for most cosmologies the universe
is effectively Einstein--de~Sitter, therefore, high redshifts do not
contribute much to constraints on dark energy). Having a single
instrument, a uniform target selection and calibration is also crucial
to perform precision tests of cosmology without having to build a
`ladder' from different surveys selecting different targets. On the
other hand it is also easy to see the synergy between these
ground-based surveys and Euclid: by mapping different targets (over
the same sky area and ofter the same redshift range) one can gain
better control over issues such as bias. The use of multiple
tracers of cosmic large scale structure can reduce systematic effects
and ultimately increase the precision of dark-energy measurements from
galaxy redshift surveys \citep[see,
  e.g.,][]{Seljak_Hamaus_Desjacques09}.

Moreover, having both spectroscopic and imaging capabilities Euclid is
uniquely poised to explore the clustering with both the three
dimensional distribution of galaxies and weak gravitational lensing.

\subsubsection{The Planck prior}
\label{sec:planck-prior}

Planck will provide highly accurate constraints on many cosmological parameters,
which makes the construction of a Planck Fisher matrix somewhat non-trivial as it
is very sensitive to the detailed assumptions. A relatively robust approach was
used by~\cite{Mukherjee:2008kd} to construct a Gaussian approximation to the
WMAP data by introducing two extra parameters,
\begin{equation}
R \equiv \sqrt{\Omega_m H_0^2} \,r(z_{\mathrm{CMB}}) \,, \qquad
l_a \equiv \pi r(z_{\mathrm{CMB}})/r_s(z_{\mathrm{CMB}}) \,,
\label{eq:planckfish1}
\end{equation}
where $r(z)$ is the comoving distance from the observer to redshift $z$,
and $r_s(z_{\mathrm{CMB}})$ is the comoving size of the sound-horizon at decoupling.

In this scheme, $l_a$ describes the peak location through the angular diameter
distance to decoupling and the size of the sound horizon at that time.
If the geometry changes, either due to non-zero curvature or due to a different
equation of state of dark energy, $l_a$ changes in the same way as the 
peak structure. $R$ encodes similar information, but in addition 
contains the matter density which is connected with the peak height.
In a given class of models (for example, quintessence
dark energy), these parameters are ``observables'' related to 
the shape of the observed CMB spectrum, and constraints on them 
remain the same independent of (the prescription for) the equation of 
state of the dark energy.

As a caveat we note that if some assumptions regarding the evolution
of perturbations are changed, then the corresponding $R$ and $l_a$
constraints and covariance matrix will need to be recalculated under
each such hypothesis, for instance, if massive neutrinos were to be
included, or even if tensors were included in the analysis
\citep{Corasaniti:2007rf}. Further, $R$ as defined in
Eq.~(\ref{eq:planckfish1}) can be badly constrained and is quite
useless if the dark energy clusters as well, e.g., if it has a low
sound speed, as in the model discussed in~\cite{Kunz:2007rk}.

In order to derive a Planck fisher matrix, \cite{Mukherjee:2008kd}
simulated Planck data as described in \cite{Pahud:2006kv} and derived
constraints on our base parameter set $\{R,l_a,\Omega_b h^2,n_s\}$
with a MCMC based likelihood analysis. In addition to $R$ and $l_a$
they used the baryon density $\Omega_bh^2$, and optionally the
spectral index of the scalar perturbations $n_s$, as these are
strongly correlated with $R$ and $l_a$, which means that we will lose
information if we do not include these correlations. As shown
in~\cite{Mukherjee:2008kd}, the resulting Fisher matrix loses some
information relative to the full likelihood when only considering
Planck data, but it is very close to the full analysis as soon as
extra data is used. Since this is the intended application here, it is
perfectly sufficient for our purposes.

The following tables, from~\cite{Mukherjee:2008kd}, give the covariance matrix
for quintessence-like dark energy (high sound speed, no anisotropic stress)
on the base parameters and the Fisher matrix derived from it. Please consult the
appendix of that paper for the precise method used to compute $R$ and $l_a$
as the results are sensitive to small variations.

\begin{table}[htb]
\caption[$R$, $l_a$, $\Omega_bh^2$ and $n_s$ estimated from Planck 
simulated data.]{$R$, $l_a$, $\Omega_bh^2$ and $n_s$ estimated from Planck 
simulated data. Table reproduced by permission
from~\cite{Mukherjee:2008kd}, copyright by APS.}
\centering
{\small
\begin{tabular}{llc}
\toprule
\multicolumn{1}{l}{Parameter} & \multicolumn{1}{c}{mean} & \multicolumn{1}{l}{rms variance} \\
\midrule
\multicolumn{3}{c}{$\Omega_k\neq 0$} \\
\midrule
$R$ & \phantom{00}1.7016 & 0.0055 \\
$l_a$ & 302.108 & 0.098 \\
$\Omega_b h^2$ & \phantom{00}0.02199 & 0.00017 \\
$n_s$ & \phantom{00}0.9602 & 0.0038 \\
\bottomrule
\end{tabular}}
\end{table}

\begin{table}[htb]
\caption[Covariance matrix for $(R, l_a, \Omega_b h^2, n_s)$ from
  Planck.]{Covariance matrix for $(R, l_a, \Omega_b h^2, n_s)$ from
  Planck. Table reproduced by permission from~\cite{Mukherjee:2008kd},
  copyright by APS.}
\centering
{\small
\begin{tabular}{lrrrr}
\toprule
~ & \multicolumn{1}{c}{$R$} & \multicolumn{1}{c}{$l_a$} & \multicolumn{1}{c}{$\Omega_b h^2$} & \multicolumn{1}{c}{$n_s$} \\
\midrule
\multicolumn{5}{c}{$\Omega_k\neq 0$} \\
\midrule
$R$            & 0.303492E--04  & 0.297688E--03  & --0.545532E--06  & --0.175976E--04 \\
$l_a$          & 0.297688E--03  & 0.951881E--02  & --0.759752E--05  & --0.183814E--03 \\
$\Omega_b h^2$ & --0.545532E--06 & --0.759752E-05 &  0.279464E--07  &  0.238882E--06 \\
$n_s$          & --0.175976E--04 & --0.183814E-03 &  0.238882E--06  &  0.147219E--04 \\
\bottomrule
\end{tabular}}
\end{table}

\begin{table}[htb]
\caption[Fisher matrix for ($w_0$, $w_a$, $\Omega_{\mathrm{DE}}$, $\Omega_k$, $\omega_m$, 
$\omega_b$, $n_S$) derived from the covariance matrix for
$(R, l_a, \Omega_b h^2, n_s)$ from Planck.]{Fisher matrix for ($w_0$,
  $w_a$, $\Omega_{\mathrm{DE}}$, $\Omega_k$, $\omega_m$, $\omega_b$,
  $n_S$) derived from the covariance matrix for $(R, l_a, \Omega_b
  h^2, n_s)$ from Planck. Table reproduced by permission
  from~\cite{Mukherjee:2008kd}, copyright by APS.}
\setlength{\tabcolsep}{3pt}
\centering
{\scriptsize
\begin{tabular}{lrrrrrrr}
\toprule
~ & \multicolumn{1}{c}{$w_0$} & \multicolumn{1}{c}{$w_a$} & \multicolumn{1}{c}{$\Omega_{\mathrm{DE}}$} & \multicolumn{1}{c}{$\Omega_k$} & \multicolumn{1}{c}{$\omega_m$} & \multicolumn{1}{c}{$\omega_b$} & \multicolumn{1}{c}{$n_S$} \\
\midrule
$w_0$ &  .172276E+06 &  .490320E+05 &  .674392E+06 & --.208974E+07 &  .325219E+07 & --.790504E+07 & --.549427E+05 \\
$w_a$ &  .490320E+05 & .139551E+05 & .191940E+06 & --.594767E+06 &
.925615E+06 &--.224987E+07 & --.156374E+05\\
$\Omega_{\mathrm{DE}}$ & .674392E+06 & .191940E+06 & .263997E+07 &
--.818048E+07 & .127310E+08 & --.309450E+08 & --.215078E+06\\
$\Omega_k$ & --.208974E+07 & --.594767E+06 & --.818048E+07 &  .253489E+08 & --.394501E+08 &  .958892E+08 &  .666335E+06\\
$\omega_m$ &  .325219E+07 &  .925615E+06 &  .127310E+08 & --.394501E+08 &  .633564E+08 & --.147973E+09 & --.501247E+06\\
$\omega_b$ & --.790504E+07 & --.224987E+07 &--.309450E+08 & .958892E+08 &--.147973E+09  &.405079E+09 & .219009E+07\\
$n_S$ & --.549427E+05 & --.156374E+05 & --.215078E+06  & .666335E+06 & --.501247E+06  & .219009E+07 & .242767E+06\\
\bottomrule
\end{tabular}}
\end{table}

\subsection{Forecasts for model independent observations}\label{s:Clfor}
As discussed in Section \ref{s:model-indep}, it is worth to complement the standard $P(k)$ analysis, see Eq.~(\ref{Fisher}), with the $C_{\ell}(z_1,z_2)$ method, which involves the directly observable redshift and angular separations instead of reconstructed model-dependent comoving distances.
The full relativistic expression of the redshift dependent angular power spectra of galaxy number counts, $C_{\ell}(z_1,z_2)$, which holds for any theory of gravity whose metric can be written as in {Eq.~(\ref{eq:metricmag})} and in which photons and dark matter particles move along geodesics, is given in \cite{Bonvin:2011bg,DiDio:2013bqa}.
In particular, it includes the lensing contribution (see Eq.~(\ref{eq:clstand})) and redshift-space distortions due to peculiar velocities, see~\cite{Kaiser:1987}, as well as other terms depending on the gravitational potentials.

The Fisher matrix is discussed, e.g., in Ref.~\cite{DiDio:2013bqa} and Eq.~(\ref{Fisher}) is replaced by:
\begin{equation}
\label{eq:fisher_Cl_lss}
F_{\alpha \beta} = \sum_{\ell,(ij),(pq)} \frac{\partial C_\ell^{ij} }{\partial p_\alpha} \frac{\partial C_\ell^{pq}}{\partial p_\beta} \text{Cov}^{-1}_{\ell, (ij), (pq)} \,,
\end{equation}
where $C_\ell^{ij}$ is the correlation between redshift bin $z_i$ and $z_j$, and the covariance matrix between different power spectra can be approximated as:
\begin{equation}
\label{eq:cov_Cl_lss}
\text{Cov}_{[\ell,\ell'] [(ij), (pq)]}=\delta_{\ell,\ell'}\frac{C_\ell^{\text{obs},i p} C_\ell^{\text{obs},jq} + C_\ell^{\text{obs},i q} C_\ell^{\text{obs},jp}}{f_\text{sky} \left( 2 \ell + 1 \right) }.
\end{equation}
The observable power spectrum $C_\ell^{\text{obs} ,ij}$ takes into account the fact that we observe a finite number of galaxies instead of a smooth field.
This leads to a shot noise term
\begin{equation}
C_\ell^{\text{obs} ,ij}= C_\ell^{ij} + \frac{\delta_{ij}}{ N(i)}\, ,
\end{equation}
where $ N(i)$ denotes the number of galaxies in the bin around $z_i$.
The power spectra, using $\Delta_{\cal R}^2 =k^3P_{\cal R}/(2\pi^2)$, see Eq.~(\ref{eq:pR}), given by
\begin{equation}
 C_\ell^{ij}  = 4 \pi \int \frac{dk}{k} \Delta_{\cal R}^2(k)  \Delta^i_\ell(k)\Delta^j_\ell(k)~, \label{Cl2}
\end{equation}
are computed in terms of integrals of transfer functions $\Delta_\ell(z,k)$:
\begin{equation}
\Delta_\ell^i(k) = \int dz \frac{dN}{dz} W_i(z) \Delta_\ell(z,k)~, \label{delta}
\end{equation}
which account for the tracer distribution $\frac{dN}{dz}$ and the bin selection function $W_i(z)$.
This can be approximated by a top hat if good redshift estimates are available, or by a Gaussian with standard deviation determined by the photometric redshift errors.
The power spectra are computed, e.g., by the publicly available code {\sc class}gal described in Ref.~\cite{DiDio:2013bqa}.

\begin{figure}[hT]
\centering
 \includegraphics[width=.75\columnwidth]{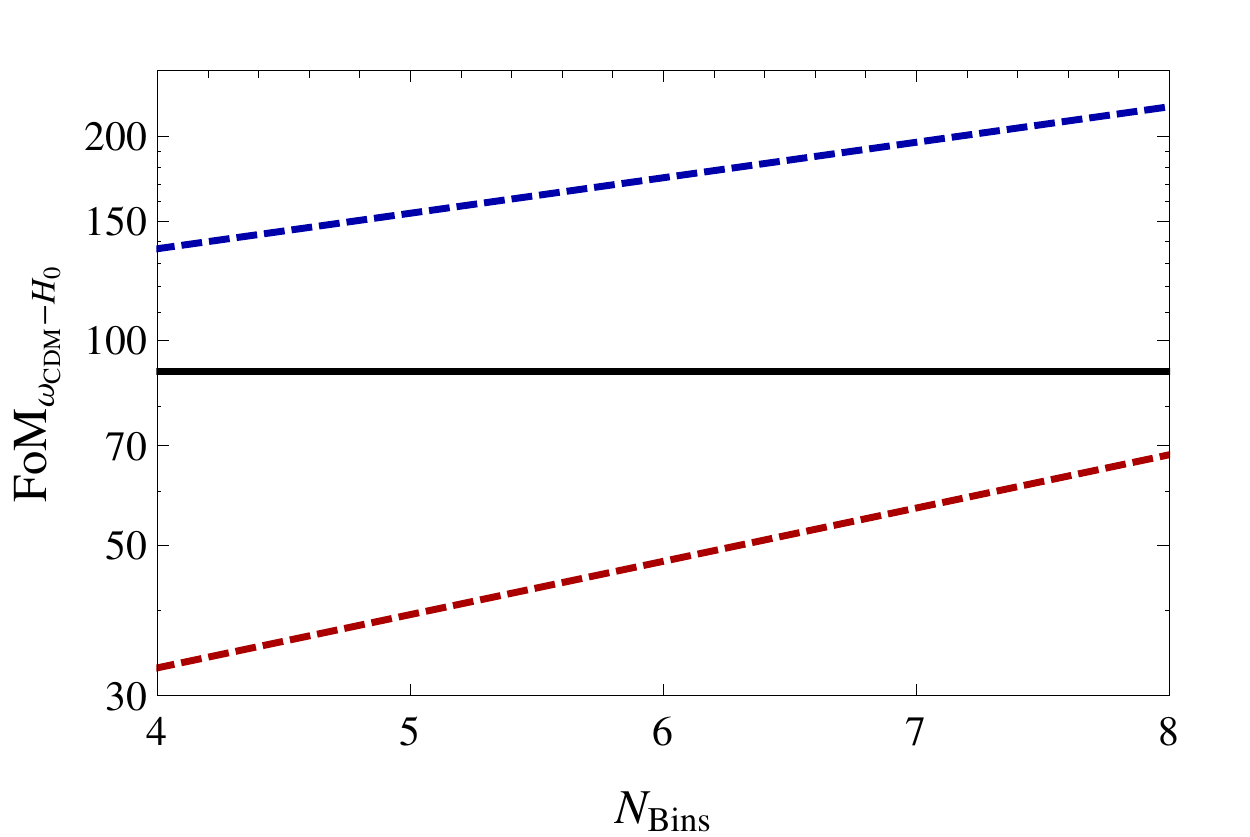} 
 \caption{We show the figure of merit for $\omega_{CDM}=h^2\Omega_{CDM}$ and $H_0$ as a function of the number of bins for the photometric survey of Euclid.
The black line is the $P(k)$ result, the red dashed line is the $C_l(z_1,z_2)$ result for bin auto-correlations only, while the blue line also includes cross-correlations.}
\label{fig:FoM}
\end{figure}

For photometric redshifts, the model independent $C_{\ell}(z_1,z_2)$ method performs significantly better then the standard $P(k)$ analysis~\cite{DiDio:2013sea} as can be seen in Fig.~\ref{fig:FoM}.
 
This is due to the fact that this analysis makes optimal use of the redshift information and does not average over directions.
For spectroscopic redshifts, however, the large number of redshift bins  which would be
needed to fully profit from the redshift information, is severely limited by  shot noise.
In fact when using redshift bins that are significantly thicker than the redshift resolution of
the survey, the $P(k)$ analysis, in principle, has an advantage since it makes use of the full redshift resolution in determining distances of galaxies, while in the $C_{\ell}(z_1,z_2)$ analysis we do not distinguish redshifts of galaxies in the same bin.
However, for spectroscopic surveys we can in principle allow for very slim bins with a thickness significantly smaller than the nonlinearity scale, and the maximal number of useful bins is decided by the shot noise, as well as by numerical limitations related to Markov Chain Monte Carlo data analysis.

The cross correlations from different redshift bins provide an alternative measure of the lensing potential~\cite{Montanari:2015rga} which is complementary to the analysis of shear with completely different systematic errors. This will allow the measurement of $\langle \delta(z_1)\kappa(z_2)\rangle$ for $z_2>z_1$.

\clearpage
\section{Summary and outlook}

This section introduced the main features of the most popular
dark energy/modified gravity models.
Here we summarize the performance of Euclid with respect to these models.
Unless otherwise indicated, we always assume Euclid with no external priors  and all errors
fully marginalized over the standard cosmological parameters. Here RS denotes the redshift survey, WLS the weak lensing one.

\begin{enumerate}

\item Euclid (RS) should be able to measure the main standard
  cosmological parameters to percent or sub-percent level as detailed
  in Table~\ref{tab:cosm_par_errors} (all marginalized errors,
  including constant equation of state and constant growth rate, see
  Table~\ref{tab:w0_w1} and Figure~\ref{fig:w0_w1}).

\item The two CPL parameters $w_0,w_1$ should be measured with errors
  0.06 and 0.26, respectively (fixing the growth rate to fiducial),
  see Table~\ref{tab:w0_w1} and Figure~\ref{fig:w0_w1}.

\item The equation of state $w$ and the growth rate parameter
  $\gamma$, both assumed constant, should be simultaneously
  constrained to within 0.04 and 0.03, respectively.

\item The growth function should be constrained to within 0.01\,--\,0.02
  for each redshift bin from $z=0.7$ to $z=2$ (see
  Table~\ref{tab:sigma_bias_s_bint}).

\item A scale-independent bias function $b(z)$ should be constrained
  to within 0.02 for each redshift bin (see
    Table~\ref{tab:sigma_bias_s_bint}).

\item The growth rate parameters $\gamma_0,\gamma_1$ defined in
  Eq.~\ref{eq:gam_CPL} should be measured to within 0.08, 0.17,
  respectively.

\item Euclid will achieve an accuracy on measurements of the dark
  energy sound speed of $\sigma(c_s^2)/c_s^2=2615$ (WLS) and
  $\sigma(c_s^2)/c_s^2=50.05$ (RS), if $c_s^2=1$, or
  $\sigma(c_s^2)/c_s^2=0.132$ (WLS) and $\sigma(c_s^2)/c_s^2=0.118$
  (RS), if $c_s^2=10^{-6}$.

\item The coupling $\beta^{2}$ between dark energy and dark matter can
  be constrained by Euclid (with Planck) to less than $3\cdot10^{-4}$
  (see Figure~\ref{fig:cmbbaowl-cont} and Table~\ref{tab:combined}).

\item Any departure from GR greater than $\simeq 0.03$ in the growth
  index $\gamma$ will be distinguished by the WLS
  \citep{Heavens-etal:2007}.

\item Euclid WLS can detect deviations between 3\% and 10\% from the
  GR value of the modified-gravity parameter $\Sigma$
  (Eq.~\ref{Sigma-wl-isw}), whilst with the RS there will be a 20\%
  accuracy on both $\Sigma$ and $\mu$ (Eq.~\ref{eq:muQeta}).

\item With the WLS, Euclid should provide an upper limit to the
  present dimensionless scalaron inverse mass $\mu \equiv H_{0}/M_{0}$
  of the $f(R)$ scalar (where the time dependent scalar field mass is
  defined in Eq.~\ref{eq:107}) as $\mu = 0.00 \pm 1.10\times 10^{-3}$
  for $l < 400$ and $\mu = 0.0 \pm 2.10 \times 10^{-4}$ for $l <
  10000$

\item The WLS will be able to rule out the DGP model growth index with
  a Bayes factor $|\ln B|\simeq 50$ \citep{Heavens-etal:2007}, and
  viable phenomenological extensions could be detected at the
  $3\sigma$ level for $1000\lesssim\ell\lesssim4000$
  \citep{Camera:2011mg}.

\item The  photometric survey of Euclid, i.e. the WLS,  is very promising in measuring directly observable angular and redshift dependent power spectra $C_{\ell}(z_1,z_2)$ (and correlation function) as discussed in Ref.~\cite{DiDio:2013sea}. This spectra are truly model independent and especially well suited to  estimate cosmological parameter or test models of modified gravity.

\end{enumerate}

At the same time, there are  several areas of research that we feel
are important for the future of Euclid, both to improve the current
analyses and to maximize its science return. Here we provide a
preliminary, partial list.

\begin{enumerate}

\item The results of the redshift survey and weak lensing surveys
  should be combined in a statistically coherent way

\item The set of possible priors to be combined with Euclid data
  should be better defined

\item The forecasts for the parameters of the modified gravity and
  clustered dark-energy models should be extended to include more
  general cases

\item We should estimate the errors on a general reconstruction of the
  modified gravity functions $\Sigma,\mu$ or of the metric potentials
  $\Psi,\Phi$ as a function of both scale and time.

\item We should use the $C_\ell(z_1,z_2)$--method to constrain modified gravity models.
\end{enumerate}










\newpage
\part{Dark Matter and Neutrinos}
\label{dark-matter}





\section{Introduction}

The identification of dark matter is one of the most important open problems in particle
physics and cosmology. In standard cosmology, dark matter contributes 85\% of all the
matter in the universe, but we do not know what it is made of, as we have never observed
dark matter particles in our laboratories. The foundations of the modern dark matter
paradigm were laid in the 1970s and 1980s, after decades of slow accumulation of evidence. 
Back in the 1930s, it was noticed that the Coma cluster seemed to contain much more mass
than what could be inferred from visible galaxies \citep{Zwicky:1933gu,Zwicky:1937zza}, and a few years later, it became clear
that the Andromeda galaxy M31 rotates anomalously fast at large radii, as if most of its mass 
resides in its outer regions. Several other pieces of evidence provided further support to the dark matter
hypothesis, including the so called timing-argument. In the 1970s, rotation curves were
extended to larger radii and to many other spiral galaxies, proving the presence of large
amounts of mass on scales much larger than the size of galactic disks \cite{Peacock99}.

We are now in the position of determining the total abundance
of dark matter relative to normal, baryonic matter, in the universe with exquisite accuracy; 
we have a much better understanding of how dark matter is distributed in structures ranging from dwarf galaxies
to clusters of galaxies, thanks to gravitational lensing 
observations \citep[see][for a review]{2010RPPh...73h6901M} and theoretically from 
high-resolution numerical simulations made possible by
modern supercomputers (such as, for example, the Millennium or Marenostrum simulations).

Originally, Zwicky thought of dark matter as most likely baryonic -- missing cold gas, or low mass stars. 
Rotation curve observation could be explained by dark matter in  the form of MAssive Compact Halo
Objects (MACHOs, e.g., a halo of black holes or brown dwarfs). 
However, the MACHO and EROS experiments have shown that dark matter cannot be in 
the mass range $0.6\times10^{-7}\,M_{\odot}<M<15\,M_{\odot}$ if it comprises 
massive compact objects \citep{2000ApJ...542..281A,2007AA...469..387T}. 
Gas measurements are now extremely sensitive, ruling out dark matter as
undetected gas (\cite{1997ApJ...479..523B, 2001ApJ...559...29C, 2006AA...445..827R}; but see \cite{1994AA...285...79P}). And the CMB
and Big Bang Nucleosynthesis require the total mass in baryons in the universe
to be significantly less that the total matter density
\citep{2002NuPhS.110...16R,2002PhRvD..65d3510C,2002ApJ...576L.101T}.

This is one of the most spectacular results in cosmology obtained at the end
of the 20th century: dark matter has to be non-baryonic. 
As a result, our expectation of
the nature of dark matter shifted from an astrophysical explanation to particle
physics,
linking the smallest and largest scales that we can probe. 

During the seventies
the possibility of the neutrino to be the dark matter particle with a mass of
tenth of eV was explored, but it was realized that such light
particle would erase the primordial fluctuations on small scales, leading
to a lack of structure formation on galactic scales and below. It was therefore
postulated that the dark matter particle must be cold (low thermal
energy, to allow structures on small scale to
form), collisionless 
(or have a very low interaction cross section, because dark matter is observed
to be pressureless) and  stable 
over a long period of time: such a candidate is referred to as a weakly interacting massive particle (WIMP). 
This is the standard cold dark matter (CDM) picture \citep[see][]{1990ApJ...351...10F,1991Natur.352..769P}.

Particle physicists have proposed several possible dark matter candidates. 
Supersymmetry (SUSY) is an attractive extension of the Standard Model of particle physics. The lightest SUSY particle (the LSP) is stable, uncharged, and weakly interacting, providing a perfect WIMP candidate known as a neutralino. Specific realizations of SUSY each provide slightly different dark matter candidates \citep[for a review see][]{jungman1996}. Another distinct dark matter candidate arising from extensions of the Standard Model is the axion, 
a hypothetical pseudo-Goldstone boson whose existence was postulated to solve the so
called strong $CP$ problem in quantum chromodynamics \citep{Peccei-Quinn:1977}, also arising generically in string theory \citep{witten1984,SvrcekWitten:2006}. They are known to be very well
motivated dark matter candidates \citep[for a review of axions in cosmology see][]{sikivie2008}. Other well-known candidates are sterile neutrinos,
which interact only gravitationally with ordinary matter, apart from a small mixing with
the familiar neutrinos of the Standard Model (which should make them ultimately unstable), and candidates arising from technicolor \citep[see, e.g.,][]{gudnason2006}. A wide array of other possibilities have been discussed in the literature, and they are currently being searched for with a variety of experimental strategies \citep[for a complete review of dark matter in particle physics see][]{pdg}.

There remain some possible discrepancies in the standard cold dark matter model, such as the missing satellites
problem, and the cusp-core controversy (see below for details and references) that have led some
authors to question the CDM model and to propose alternative solutions.
The physical mechanism by which one may reconcile the observations with the standard
theory of structure formation is the suppression of the matter power spectrum at small scales.
This can be achieved with dark matter particles with a strong self-scattering cross section, or
with particles with a non-negligible velocity dispersion at the epoch of structure formation,
also referred to as warm dark matter (WDM) particles.

Another possibility is that the extra gravitational degrees of freedom arising in modified theories of gravity play the role of dark matter.  In particular this happens for the Einstein-Aether, TeVeS and bigravity models. These theories were developed following the idea that the presence of unknown dark components in the universe may be indicating us that it is not the matter component that is exotic but rather that gravity is not described by standard GR.

Finally, we note that only from astrophysical probes can any dark matter candidate found in either direct detection experiments or accelerators, such as the LHC, be confirmed. Any direct dark matter candidate discovery will give Euclid a clear goal to verify the existence of this particle on astrophysical scales. Within this context, Euclid can provide precious information on the nature of dark matter. In this part, we discuss the most relevant results that can be obtained with Euclid, and that can be summarized as follows:
\begin{itemize}
\item 
The discovery of an exponential suppression in the power spectrum at small scales, that
would rule out CDM and favor WDM candidates, or, in absence of it, the determination
of a lower limit on the mass of the WDM particle, $m_{\mathrm{WDM}}$, of 2~keV;
\item 
the determination of an upper limit on the dark matter
self-interaction cross section $\sigma/m\sim 10^{-27}\mathrm{\ cm^2\ GeV^{-1}}$ at 68\% CL, which represents an improvement of three orders of magnitude compared to the best constraint available today, which arises from the analysis of the dynamics of the bullet cluster;
\item 
the measurement of the slope of the dark matter distribution within galaxies and clusters of galaxies with unprecedented accuracy;
\item 
the determination of the properties of the only known -- though certainly subdominant -- non-baryonic dark matter particle: the standard neutrino, for which Euclid can provide information on the absolute mass scale, its normal or inverted hierarchy, as well as its Dirac or Majorana nature;
\item 
the test of unified dark matter (UDM, or quartessence) models, through the detection of characteristic oscillatory features predicted by these theories on the matter power spectrum, detectable through weak lensing or baryonic acoustic oscillations studies;
\item
a probe of the axiverse, i.e., of the legacy of string theory through the presence of ultra-light scalar fields that can affect the growth of structure, introducing features in the matter power spectrum and modifying the growth rate of structures.
\end{itemize}
Finally, Euclid will provide, through gravitational lensing measurement, a map of the dark matter distribution over the entire extragalactic sky, allowing us to study the effect of the dark matter environment on galaxy evolution and structure formation as a function of time. This map will pinpoint our place within the dark universe.

\section{Dark matter halo properties}

Dark matter was first proposed by \citet{Zwicky:1937zza} to explain the anomalously high velocity of galaxies in galaxy clusters. Since then, evidence for dark matter has been accumulating on all scales. The velocities of individual stars in dwarf galaxies suggest that these are the most dark matter dominated systems in the universe \citep[e.g.,][]{1998ARAA..36..435M,2001ApJ...563L.115K,2007ApJ...670..313S,2007MNRAS.380..281M,2007ApJ...667L..53W}. Low surface brightness (LSB) and giant spiral galaxies rotate too fast to be supported by their stars and gas alone, indicating the presence of dark matter \citep{2001ApJ...552L..23D,2005ApJ...621..757S, 2001MNRAS.323..285B, 2002ApJ...573..597K}. Gravitationally lensed giant elliptical galaxies and galaxy clusters require dark matter to explain their observed image distributions \citep[e.g.,][]{1964MNRAS.128..295R,1975ApJ...195...13B,1979Natur.279..381W,1987AA...172L..14S,2006ApJ...648L.109C}. Finally, the temperature fluctuations in the cosmic microwave background (CMB) radiation indicate the need for dark matter in about the same amount as that required in galaxy clusters \citep[e.g.,][]{1992ApJ...396L...1S,1992ApJ...396L..13W,SpergelEtAl2006}. 

While the case for particle dark matter is compelling, until we find direct evidence for such a particle, astrophysics remains a unique dark matter probe. Many varieties of dark matter candidates produce a noticeable change in the growth of structure in the universe \citep{jungman1996,2009EPJC...59..557S}. Warm dark matter (WDM) suppresses the growth of structure in the early universe producing a measurable effect on the small-scale matter power spectrum \citep{2001ApJ...556...93B,2001ApJ...559..516A,2001ApJ...558..482B}. Self-interacting dark matter (SIDM) changes the expected density distribution \textit{within} bound dark matter structures \citep{2001ApJ...561...35D,2000PhRvD..62f3511H}. In both cases, the key information about dark matter is contained on very small scales. In this section, we discuss previous work that has attempted to measure the small scale matter distribution in the universe, and discuss how Euclid will revolutionize the field. We divide efforts into three main areas: measuring the halo mass function on large scales, but at high redshift; measuring the halo mass function on small scales through lens \textit{substructures}; measuring the dark matter density profile within galaxies and galaxy clusters.

\epubtkImage{baryonicmassfunc.png}{%
\begin{figure}[htb]
\centerline{\includegraphics[height=0.6\textwidth]{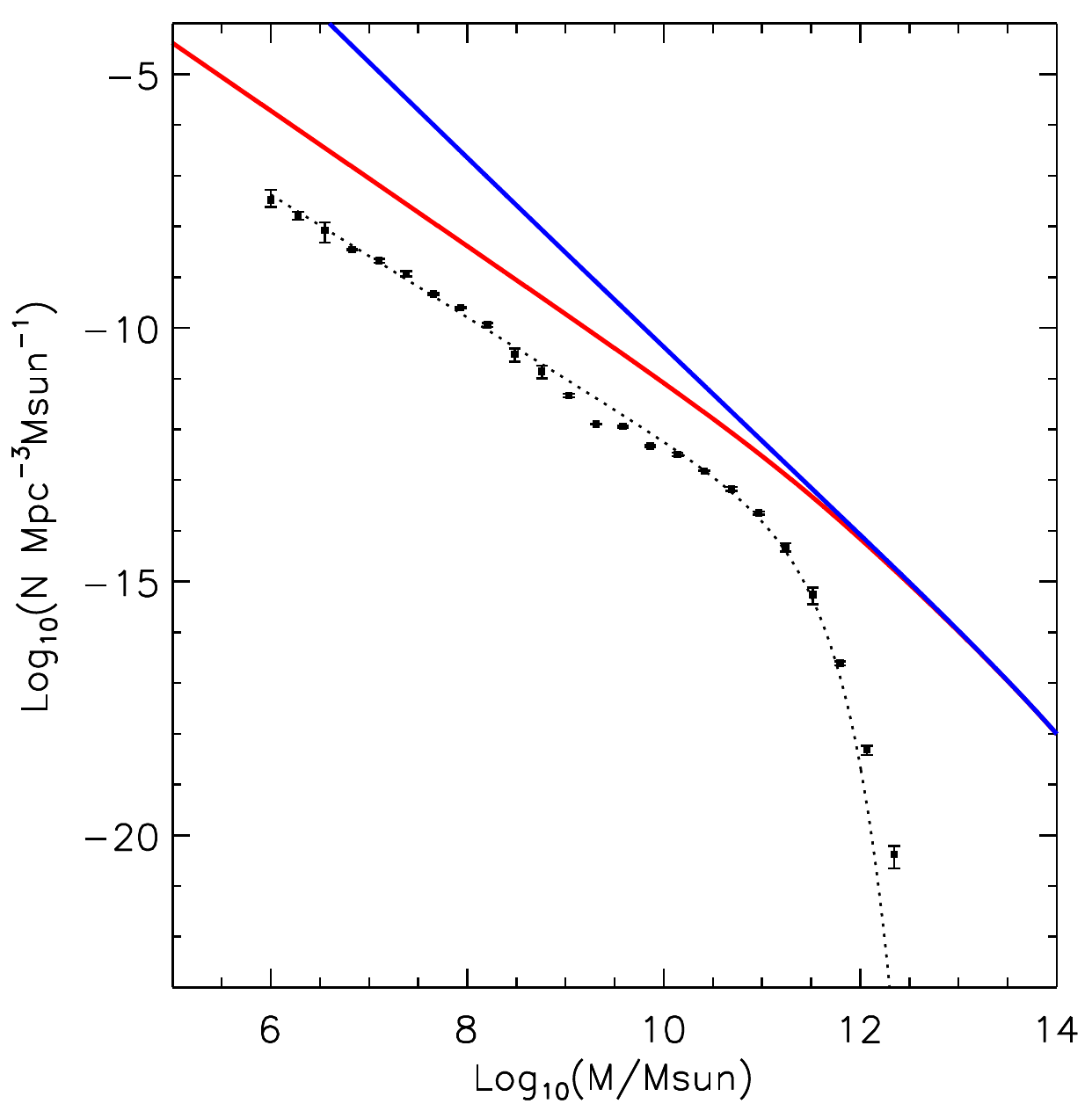}}
\caption{The baryonic mass function of galaxies (data points). The
  dotted line shows a Schechter function fit to the data. The blue
  line shows the predicted mass function of dark matter haloes,
  assuming that dark matter is cold. The red line shows the same
  assuming that dark matter is warm with a (thermal relic) mass of
  $m_{\mathrm{WDM}}=1\mathrm{\ keV}$.}
\label{fig:readtrent}
\end{figure}}

\subsection{The halo mass function as a function of redshift}
\label{sec:matterpower}

Attempts have already been made to probe the small scale power in the
universe through galaxy counts. Figure~\ref{fig:readtrent} shows the
best measurement of the `baryonic mass function' of galaxies to date
\citep{2005RSPTA.363.2693R}. This is the number of galaxies with a
given total mass in baryons normalized to a volume of 1~Mpc. To
achieve this measurement, \citet{2005RSPTA.363.2693R} sewed together
results from a wide range of surveys reaching a baryonic mass of just
$\sim10^6\,M_{\odot}$ -- some of the smallest galaxies observed to
date.

The baryonic mass function already turns up an interesting result. Over-plotted in blue on Figure~\ref{fig:readtrent} is the \textit{dark matter} mass function expected assuming that dark matter is `cold' -- i.e., that it has no preferred scale. Notice that this has a different shape. On large scales, there should be bound dark matter structures with masses as large as $10^{14}\,M_{\odot}$, yet the number of observed galaxies drops off exponentially above a baryonic mass of $\sim 10^{12}\,M_{\odot}$. This discrepancy is well-understood. Such large dark matter haloes have been observed, but they no longer host a single galaxy; rather they are bound collections of galaxies -- galaxy clusters \citep[see e.g.][]{Zwicky:1937zza}. However, there is also a discrepancy at low masses that is not so well understood. There should be far more bound dark matter haloes than observed small galaxies. This is the well-known `missing satellite' problem \citep{1999ApJ...524L..19M,Klypin:1999uc}. 

The missing satellite problem could be telling us that dark matter is
not cold. The red line on Figure~\ref{fig:readtrent} shows the
expected dark matter mass function for WDM with a (thermal relic) mass
of $m_{\mathrm{WDM}}=1\mathrm{\ keV}$. Notice that this gives an
excellent match to the observed slope of the baryonic mass function on
small scales. However, there may be a less exotic solution. It is
likely that star formation becomes inefficient in galaxies on small
scales. A combination of supernovae feedback, reionization and
ram-pressure stripping is sufficient to fully explain the observed
distribution assuming pure CDM
\citep{2004ApJ...609..482K,2006MNRAS.371..885R,2009arXiv0903.4681M}. Such
`baryon feedback' solutions to the missing satellite problem are also
supported by recent measurements of the orbits of the Milky Way's
dwarf galaxies \citep{2010arXiv1001.1731L}.

\subsubsection{Weak and strong lensing measurements of the halo mass
  function}
 
To make further progress on WDM constraints from astrophysics, we must avoid the issue of baryonic physics by probing the halo mass function \textit{directly}. The only tool for achieving this is gravitational lensing. In weak lensing this means stacking data for a very large number of galaxies to obtain an averaged mass function. In strong lensing, this means simply finding enough systems with `good data.' Good data ideally means multiple sources with wide redshift separation \citep{2009ApJ...690..154S}; combining independent data from dynamics with lensing may also prove a promising route \citep[see e.g.][]{2002MNRAS.337L...6T}. 

Euclid will measure the halo mass function down to $\sim 10^{13}\,M_{\odot}$ using weak lensing. It will simultaneously find 1000s of strong lensing systems. However, in both cases, the lowest mass scale is limited by the lensing critical density. This limits us to probing down to a halo mass of $\sim 10^{11}\,M_{\odot}$ which gives poor constraints on the nature of dark matter. However, if such measurements can be made as a \textit{function of redshift}, the constraints improve dramatically. We discuss this in the next Section. 

\subsubsection{The advantage of going to high redshift}
\label{sec:redhigh}

Dark matter constraints from the halo mass function become much stronger if the halo mass function is measured as a function of redshift. This is because warm dark matter \textit{delays} the growth of structure formation as well as suppressing small scale power. This is illustrated in Figure~\ref{fig:fractionbound}, which shows the fraction of mass in bound structures as a function of redshift, normalized to a halo of Milky Way's mass at redshift $z=0$. Marked are different thermal relic WDM particle masses in keV (black solid lines). Notice that the differences between WDM models increase significantly towards higher redshift at a given mass scale. Thus we can obtain strong constraints on the nature of dark matter by moving to higher $z$'s, rather than lower halo mass.

The utility of redshift information was illustrated recently by
observations of the Lyman-$\alpha$ absorption spectra from Quasars
\citep{Viel:2007mv,Seljak:2006qw}. Quasars act as cosmic
`flashlights' shining light from the very distant universe. Some of
this light is absorbed by intervening neutral gas leading to
absorption features in the Quasar spectra. Such features contain rich
information about the matter distribution in the universe at high
redshift. Thus, the Lyman-$\alpha$ forest measurements have been able
to place a lower bound of $m_{\mathrm{WDM}}>4\mathrm{\ keV}$ probing
scales of $\sim1\mathrm{\ Mpc}$. Key to the success of this
measurement is that much of the neutral gas lies in-between galaxies
in filaments. Thus, linear approximations for the growth of structures
in WDM versus CDM remain acceptable, while assuming that the baryons
are a good tracer of the underlying matter field is also a good
approximation. However, improving on these early results means probing
smaller scales where nonlinearities and baryon physics will creep
in. For this reason, tighter bounds must come from techniques that
either probe even higher redshifts, or even smaller scales. Lensing
from Euclid is an excellent candidate since it will achieve both while
measuring the halo mass function directly rather than through the
visible baryons.

\epubtkImage{fractionbound.png}{%
\begin{figure}[htbp]
\centerline{\includegraphics[height=0.6\textwidth]{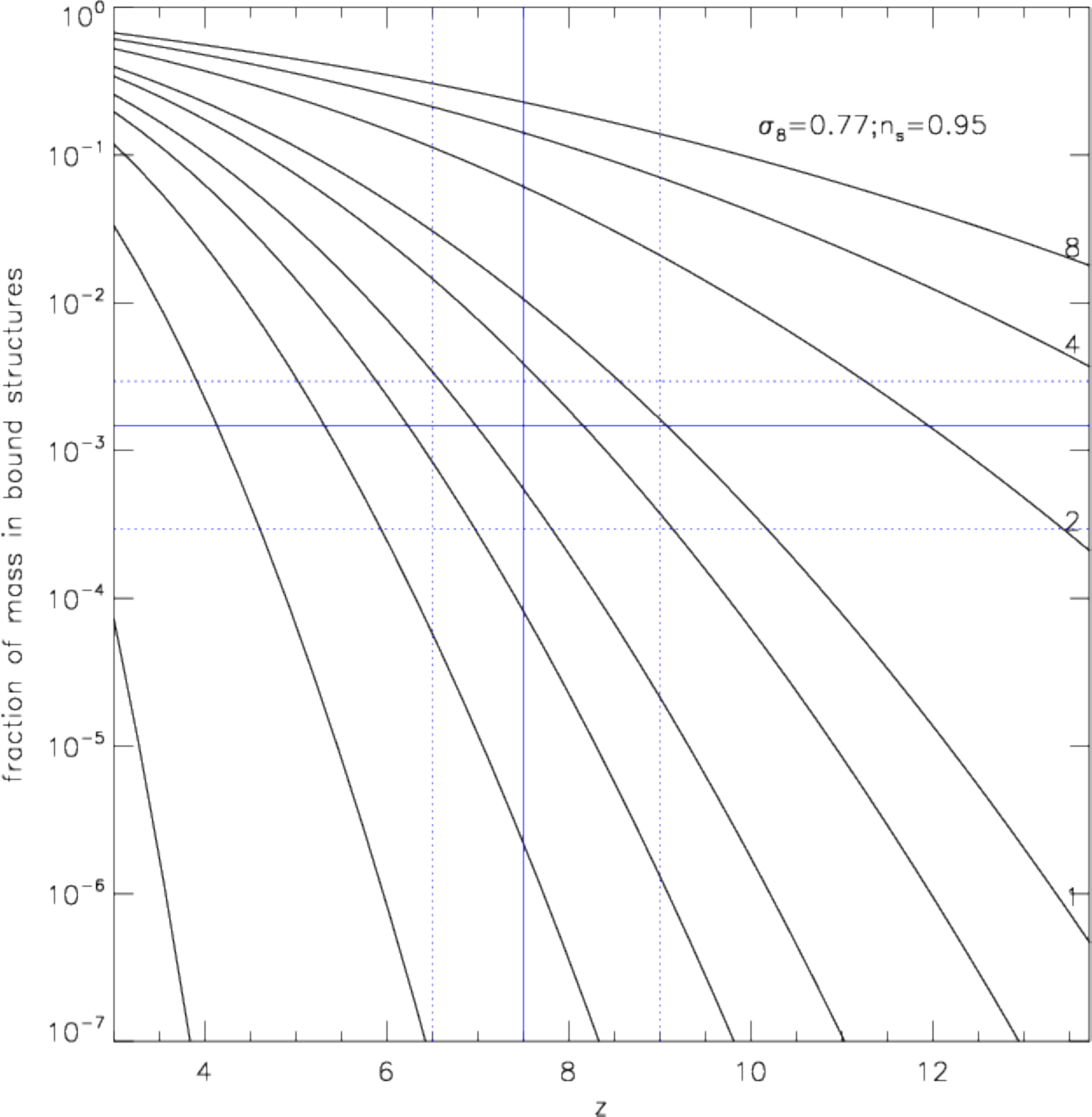}}
\caption{The fraction of mass in bound structures as a function of redshift, normalized to a halo of Milky Way's mass at redshift $z=0$. Marked are different masses of thermal-relic WDM particles in keV (black solid lines). Notice that the differences between different WDM models increases towards higher redshift.}
\label{fig:fractionbound}
\end{figure}}

\subsection{The dark matter density profile}
\label{sec:darkdist}

An alternative approach to constraining dark matter models is to measure the distribution of dark matter \textit{within} galaxies. Figure~\ref{fig:cuspslope} shows the central log-slope of the density distribution for 9 galaxies/groups and 3 lensing clusters as a function of the enclosed lensing mass \citep{2006ApJ...652L...5S,2007arXiv0704.3267R,2009ApJ...690..154S}. Over the visible region of galaxies, the dark matter distribution tends towards a single power law: $\rho \propto r^\alpha$. Marked in red is the prediction from structure-formation simulations of the standard cosmological model, that assume non-relativistic CDM, and that do not include any baryonic matter. Notice that above an enclosed lensing mass of $\sim 10^{12}\,M_{\odot}$, the agreement between theory and observations is very good. This lends support to the idea that dark matter is cold and not strongly self-interacting. However, this result is based on only a handful of galaxy clusters with excellent data. Furthermore, lower mass galaxies and groups can, in principle, give tighter constraints. In these mass ranges, however ($M_{\mathrm{enc}}<10^{12}\,M_{\odot}$), the lensing mass is dominated by the visible stars. Determining the underlying dark matter distribution is then much more difficult. It is likely that the dark matter distribution is also altered from simple predictions by the dynamical interplay between the stars, gas and dark matter during galaxy formation \citep[e.g.,][]{2007arXiv0707.0737D}.

\epubtkImage{alphamass.png}{%
\begin{figure}[htbp]
\centerline{\includegraphics[height=0.6\textwidth]{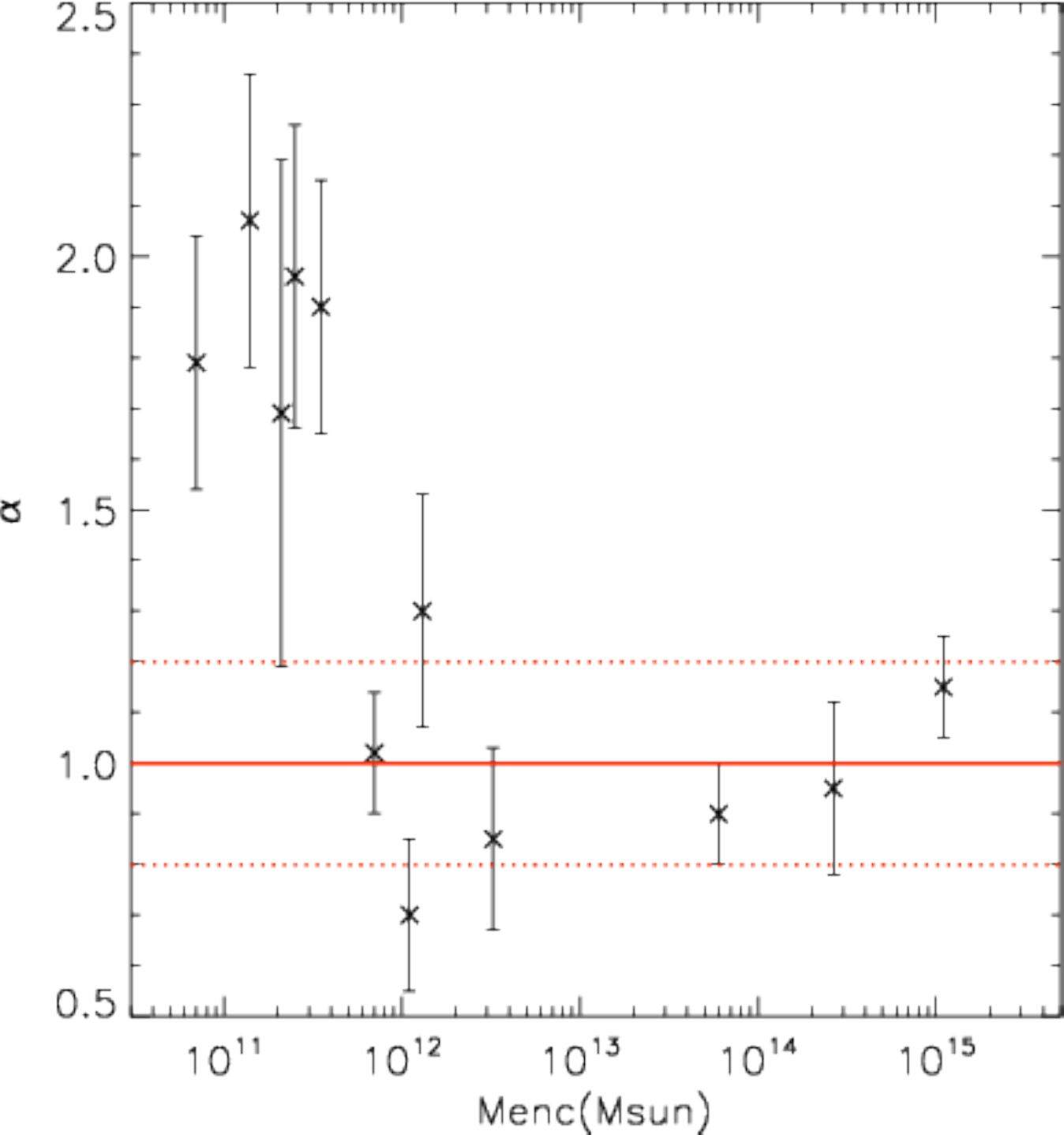}}
\caption{The central log-slope $\alpha$ of the density distribution $\rho \propto r^\alpha$ for 9 galaxies/groups and 3 lensing clusters as a function of the enclosed lensing mass. Marked in red is the prediction from structure formation simulations of the standard cosmological model, that assume non-relativistic CDM, and that do not include any baryonic matter.}
\label{fig:cuspslope}
\end{figure}}


\clearpage
\section{Euclid dark matter studies: wide-field X-ray complementarity}

The predominant extragalactic X-ray sources are AGNs and galaxy clusters. For dark matter studies the latter are the more interesting targets. X-rays from clusters are emitted as thermal bremsstrahlung by the hot intracluster medium (ICM) which contains most of the baryons in the cluster. The thermal pressure of the ICM supports it against gravitational collapse so that measuring the temperature through X-ray observations provides information about the mass of the cluster and its distribution. Hence, X-rays form a complementary probe of the dark matter in clusters to Euclid weak lensing measurements.

The ongoing X-ray missions XMM-Newton and Chandra have good enough
angular resolution to measure the temperature and mass profiles in
$\sim10$ radial bins for clusters at reasonable redshifts, although
this requires long exposures. Many planned X-ray missions aim to
improve the spectral coverage, spectral resolution, and/or collection
area of the present mission, but they are nonetheless mostly suited
for targeted observations of individual objects. Two notable
exceptions are eROSITA%
\epubtkFootnote{\url{http://www.mpe.mpg.de/erosita/}}
\citep{Cappelluti:2010ay} and the Wide Field X-ray
Telescope%
\epubtkFootnote{\url{http://www.wfxt.eu/home/Overview.html}}
\citep[WFXT][proposed] {2009astro2010S..90G, Vikhlinin:2009xb,
  2010MNRAS.407.2339S, 2011MSAIS..17....3R, 2011MSAIS..17...36B,
  2011arXiv1112.0327S} which will both conduct full sky surveys and,
in the case of WFXT, also smaller but deeper surveys of large
fractions of the sky. 

A sample of high-angular resolution X-ray cluster observations can be used to test the prediction from $N$-body simulations of structure formation that dark matter haloes are described by the NFW profile \citep{1996ApJ...462..563N} with a concentration parameter $c$. This describes the steepness of the profile, which is related to the mass of the halo \citep{2007MNRAS.381.1450N}. Weak or strong lensing measurements of the mass profile, such as those that will be provided from Euclid, can supplement the X-ray measurement and have different systematics. Euclid could provide wide field weak lensing data for such a purpose with very good point spread function (PSF) properties, but it is likely that the depth of the Euclid survey will make dedicated deep field observations a better choice for a lensing counterpart to the X-ray observations. However, if the WFXT mission becomes a reality, the sheer number of detected clusters with mass profiles would mean Euclid could play a much more important r\^ole.

X-ray observations of galaxy clusters can constrain cosmology by measuring the geometry of the universe through the baryon fraction $f_{\mathrm{gas}}$ \citep{2008MNRAS.383..879A} or by measuring the growth of structures by determining the high-mass tail of the mass function \citep{2010MNRAS.406.1759M}. The latter method would make the most of the large number of clusters detected in full-sky surveys and there would be several benefits by combining an X-ray and a lensing survey. It is not immediately clear which type of survey would be able to better detect clusters at various redshifts and masses, and the combination of the two probes could improve understanding of the sample completeness. An X-ray survey alone cannot measure cluster masses with the required precision for cosmology. Instead, it requires a calibrated relation between the X-ray temperature and the cluster mass. Such a calibration, derived from a large sample of clusters, could be provided by Euclid. In any case, it is not clear yet whether the large size of a Euclid sample would be more beneficial than deeper observations of fewer clusters.

Finally, X-ray observations can also confirm the nature of possible `bullet-like' merging clusters. In such systems the shock of the collision has displaced the ICM from the dark matter mass, which is identified through gravitational lensing. This offers the opportunity to study dark matter haloes with very few baryons and, e.g., search for signatures of decaying or annihilating dark matter.


\section{Dark matter mapping}

Gravitational lensing offers a unique way to chart dark matter structures in the universe as it is sensitive to all forms of matter. Weak lensing has been used to map the dark matter in galaxy clusters \citep[see for example][]{2006AA...451..395C} with high resolution reconstructions recovered for the most massive strong lensing clusters \citep[see for example][]{2006ApJ...652..937B}. Several lensing studies have also mapped the projected surface mass density over degree scale-fields \citep{2007AA...462..459G,2007AA...462..875S,2009ApJ...702..980K} to identify shear-selected groups and clusters. The minimum mass scale that can be identified is limited only by the intrinsic ellipticity noise in the lensing analysis and projection effects. Using a higher number density of galaxies in the shear measurement reduces this noise, and for this reason the Deep Field Euclid Survey will be truly unique for this area of research, permitting high resolution  reconstructions of dark matter in the field \citep{2007Natur.445..286M,2008MNRAS.385.1431H} and the study of lenses at higher redshift.

There are several non-parametric methods to reconstruct dark matter in 2D which can be broadly split into two categories: convergence techniques \citep{1993ApJ...404..441K} and potential techniques \citep{1996ApJ...464L.115B}. In the former one measures the projected surface mass density (or convergence) $\kappa$ directly by applying a convolution to the measured shear under the assumption that $\kappa\ll1$. Potential techniques perform a $\chi^2$ minimization and are better suited to the cluster regime and can also incorporate strong lensing information \citep{2005AA...437...39B}. In the majority of methods, choices need to be made about smoothing scales to optimize signal-to-noise whilst preserving reconstruction resolution. Using a wavelet method circumvents this choice \citep{2006AA...451.1139S, 2008ApJ...684..794K} but makes the resulting significance of the reconstruction difficult to measure.

In \citep{2013MNRAS.433.3373V} the techniques of weak lensing mass mapping were applied to a wide-field survey for the first time, using the CFHTLenS data set. These mass maps 
were used to generate higher order statistics beyond the two-point correlation function.

\subsection{Charting the universe in 3D}

The lensing distortion depends on the total projected surface mass density along the line of sight with a geometrical weighting that peaks between a given source and observer, while increasing with source distance. This redshift dependence can be used to recover the full 3D gravitational potential of the matter density as described in \citet{2002PhRvD..66f3506H,2003MNRAS.344.1307B} and applied to the COMBO-17 survey in \citet{2004MNRAS.353.1176T} and the COSMOS survey in \citet{2007Natur.445..286M}. This work has been extended in \citet{2009MNRAS.399...48S} to reconstruct the full 3D mass density field and applied to the STAGES survey in  \citet{2011arXiv1109.0932S}.

All 3D mass reconstruction methods require the use of a prior based on the expected mean growth of matter density fluctuations. Without the inclusion of such a prior, \citet{2002PhRvD..66f3506H} have shown that one is unable to reasonably constrain the radial matter distribution, even for densely sampled space-based quality lensing data. Therefore 3D maps cannot be directly used to infer cosmological parameters.

The driving motivation behind the development of 3D reconstruction techniques was to enable an unbiased 3D comparison of mass and light. Dark haloes for example would only be detected in this manner. However the detailed analysis of noise and the radial PSF in the 3D lensing reconstructions presented for the first time in \citet{2011arXiv1109.0932S} show how inherently noisy the process is. Given the limitations of the method to resolve only the most massive structures in 3D the future direction of the application of this method for the Euclid Wide survey should be to reconstruct large scale structures in the 3D density field. Using more heavily spatially smoothed data we can expect higher quality 3D resolution reconstructions as on degree scales the significance of modes in a 3D mass density reconstruction are increased \citep{2009MNRAS.399...48S}. Adding additional information from flexion may also improve mass reconstruction, although using flexion information alone is much less sensitive than shear \citep{2010ApJ...723.1507P}.

\subsection{Mapping Large-Scale Structure Filaments}
\label{sec:mapping-large-scale}

Structure formation theory robustly predicts that matter in the
Universe is concentrated in sheets and filaments and that galaxy
clusters live at the intersection of these filaments. The most
comprehensive analytical framework for describing the emergence of
these structure from anisotropic gravitational collapse is the work
of \citet{1996Natur.380..603B}, which coined the term ``cosmic web''
for them.  It combines the linear evolution of density fluctuations
in the Zeldovich approximation \citep{1970A&A.....5...84Z} with the
statistics of peaks in the primordial density field
\citep{Bardeen:1985tr} using the the peak-patch formalism
\citep{1996ApJS..103....1B,1996ApJS..103...41B,1996ApJS..103...63B}.

Numerically, filaments have been seen since the early days of $N$-body
simulations \citep[e.g.,][]{1983MNRAS.204..891K}. Increasing mass and
spatial resolution of these simulations have refined our understanding
of them and a detailed inventory of the mass distribution over the
different kinds of large-scale structures (galaxy clusters, filaments,
sheets, voids) indicates that a plurality of all mass in the Universe
and thus probably of all galaxies is in filaments
\citep{2010MNRAS.408.2163A}.

Observationally, filaments have been traced by galaxy redshift surveys
from early indications \citep{1978MNRAS.185..357J} of their existence
to conclusive evidence in the CfA redshift survey
\citep{1989Sci...246..897G} to modern day redshift surveys like 2dF,
SDSS, BOSS and VIPERS
\citep{2001MNRAS.328.1039C, 2012ApJS..203...21A,
    2013AJ....145...10D, guzzo2014}. In X-rays the tenuous Warm-Hot
Intergalactic Medium expected to reside in filaments
\citep{2001ApJ...552..473D} has been seen in emission
\citep{2008A&A...482L..29W} and absorption
\citep{2009ApJ...695.1351B,2010ApJ...714.1715F}. Observing the
underlying dark matter skeleton has been much more challenging and
early weak-lensing candidates for direct detections
\citep{1998astro-ph/9809268K,2002ApJ...568..141G} could not be
confirmed by higher quality follow-up observations
\citep{2004A&A...422..407G,2008MNRAS.385.1431H}.

The most significant weak-lensing detection of a large-scale structure
filament yet was presented by \citet{2012Natur.487..202D}, who found a
mass bridge connecting the galaxy clusters Abell~222 and Abell~223 at
$4\sigma$ in a mass reconstruction. This dark matter filament is
spatially coincident with an overdensity of galaxies
\citep{2005A&A...440..453D} and extended soft X-ray emission
\citep{2008A&A...482L..29W}. This study, like the others mentioned
before, makes use of the fact that filaments have a higher density
closer to galaxy clusters and are expected to be particular massive
between close pairs of galaxy clusters \citep{1996Natur.380..603B}.
\citet{2012MNRAS.426.3369J}, reported another weak-lensing filament
candidate at $3\sigma$ significance coming out of a galaxy cluster but
not obviously connecting to another overdensity of galaxies or dark
matter. The works of \citet{2012Natur.487..202D} and
\citet{2012MNRAS.426.3369J} also provide the first direct mass
measurements of filaments. These are in agreement with the prediction
that massive filaments can be as heavy as small galaxy clusters
\citep{2010MNRAS.408.2163A}.

The relative dearth of weak lensing filament observations compared to
galaxy cluster measurements is of course due to their much lower
density contrast. Numerical simulations of weak-lensing observations
accurately predict that filaments will generally be below their
detection threshold \citep{2005A&A...440..453D,2010MNRAS.401.2257M}. A
statistical analysis of large set of ray-tracing simulations indicates
that even with a survey slightly deeper than the Euclid wide-survey,
the vast majority of filaments will not be individually detectable
\citep{2014MNRAS.441..745H}. \citet{2013A&A...559A.112M} propose a
matched filter tuned to the shape of filaments to overcome the
obstacles in filament detections in weak lensing data.

An alternative to lensing by individual filaments is to average (or
``stack'') the lensing signal of many filaments. These filaments could
either be identified in the Euclid spectroscopic survey or one could
use the high probability that neighbouring massive dark matter halos
are often connected by filaments. \citet{2013ApJ...773..115Z}
pioneered this technique of blindly stacking the area between galaxy
cluster pairs to boost the overdensity of filament galaxies with
respect to the field. Their selection of cluster pairs was based on
statistical studies of the abundance and properties of filaments
between cluster pairs
\citep{2004MNRAS.354L..61P,2005MNRAS.359..272C}. This stacking
approach was extended to weak lensing by
\citet{2014arXiv1402.3302C}. They developed a method to measure the
lensing signal of extended structures while at the same time nulling
the contribution of the halo pairs at the endpoints of
filaments. Stacking the lensing signal in the regions between luminous
red galaxies in SDSS, \citet{2014arXiv1402.3302C} were able to put
first constraints on the density profiles of filaments.


\section{Constraints on dark matter interaction cross sections}

We now move towards discussing the particulate aspects of dark matter,
starting with a discussion on the scattering cross-sections of dark
matter. At present, many physical properties of the dark matter
particle remain highly uncertain.  Prospects for studying the
scattering of dark matter with each of the three major constituents of
the universe -- itself, baryons, and dark energy -- are outlined
below.

\subsection{Dark matter--dark matter interactions}

Self-interacting dark matter (SIDM) was first postulated by
\citet{spergel-steinhardt}, in an attempt to explain the apparent
paucity of low-mass haloes within the Local Group.  The
  required cross-section $\sigma/m\sim1\,$cm$^2$/g was initially shown
  to be infeasible (Yoshida et al.\ 2000; Meneghetti et al.\ 2001;
  Gnedin \& Ostriker 2001; Miralda-Escud\'e 2002), but recent
  high-resolution simulations (Peter et al.\ 2012; Rocha et al.\ 2012;
  Vogelsberger et al.\ 2012) have revised the expected impact of
  self-interaction, which now remains consistent with observations of
  cluster halo shapes and profiles.  Indeed, self-interaction within a
  hidden dark sector is a generic consequence of some extensions to
  the Standard Model. For example, atomic, glueballino, and mirror
  dark matter models predict a cross-section
  $\sigma/m\approx0.6\,$cm$^2$/g$=1\,$barn/GeV (similar to nuclear
  cross-sections in the Standard Model).  Note that couplings within
  the dark sector can be many orders of magnitude larger than those
  between dark matter and Standard Model particles, which is of order
  picobarns.  Interactions entirely within the dark sector
are unprobed by direct detection or collider experiments, but leads to
several phyical effects that can potentially be observed by Euclid.

Clusters of galaxies present an interesting environment in which the
dark matter density is sufficiently high for collisions to play a
significant role. If dark matter particles possess even a small
cross-section for elastic scattering, \reviewtext{s}mall-scale
structure can be erased, and cuspy cores can be smoothed. In
particular, collisions between galaxy clusters act as
astronomical-scale particle colliders. Since dark matter and baryonic
matter are subject to different forces, they follow different
trajectories out of the collision. If dark matter's particle
interactions are rare but exchange a lot of momentum (often
corresponding to short-ranged forces), dark matter will tend to be
scattered away and lost. If the interactions are rare but
exchange little momentum (often corresponding to long-ranged
forces),the dark matter will be decelerated by an additional drag
force and become spatially offset \citet{2014MNRAS.437.2865K}.

How do these cosmological constraints relate to the values anticipated
by particle physics? WIMPs are expected to fall in the range of 10~GeV
to a few TeV. The aforementioned values would then correspond to
around $\sigma_p\lesssim10^{-24}\mathrm{\ cm}^2$, at least twenty
order of magnitudes greater than what one might expect to achieve from
neutral current interactions. Therefore in a cosmological context
WIMPs are essentially collisionless, as are axions, since they exhibit
an even smaller cross section. Any cosmological detection of SIDM
would thus point towards the more exotic candidates postulated by
particle physicists, particularly those which are not point particles
but instead comprise of extended objects such as Q-balls. A
measurement of the scattering cross-section would also place an upper
bound on the mass of the dark matter particle, since unitarity of the
scattering matrix forbids extremely large cross sections
\citep{PhysRevLett.86.3467}, i.e.,
\begin{equation}
\sigma_\mathrm{tot}\leq1.76\times10^{-17}\,\mathrm{cm^2}\left(\frac{\mathrm{GeV}}{m_\chi}\right)^2\left(\frac{10\,\mathrm{km\ s^{-1}}}{v_\mathrm{rel}}\right)^2
\, \reviewtext{.}
\end{equation}

\subsubsection{Dark matter evaporation}

As highlighted by \citet{ostriker-cross}, cross-sections large enough
to alleviate the structure formation issues would also allow
significant heat transfer from particles within a large halo to the
cooler sub-haloes. This effect is most prominent close to the centers
of clusters. As the sub-halo evaporates, the galaxy residing within
the halo would be disrupted. Limiting this rate of evaporation to
exceed the Hubble time allows an upper bound to be placed on the
scattering cross-section of approximately
$\sigma_p/m_p\lesssim0.3\mathrm{\ cm^{2}\ g^{-1}}$ (neglecting any
velocity dependence). Note the dependence on particle mass -- a more
massive CDM particle would be associated with a lower number density,
thereby reducing the frequency of collisions.

\subsubsection{Dark matter deceleration}
 Particulate dark matter and baryonic matter may be
  temporarily separated during collisions between galaxy clusters,
  such as 1E~0657-56 \citep{2006ApJ...648L.109C,2006ApJ...652..937B}
  and MACS~J0025.4-1222 \citep{BradacEtAl2008}. These `bullet
  clusters' have provided astrophysical constraints on the interaction
  cross-section of hypothesized dark matter particles
  \citep{randall2008}, and may ultimately prove the most useful
  laboratory in which to test for any velocity dependence of the
  cross-section.  Unfortunately, high-speed collisions between two
  massive progenitors are rare
  \citep{2010MNRAS.406.1134S,2010MNRAS.408.1277S}, and constraints
  from individual systems are limited by uncertainties in their
  collision velocity, impact parameter and angle with respect to the
  plane of the sky.

However, all galaxy clusters grow through almost continual minor merger
accretion. In~\citet{2011MNRAS.413.1709M,2014MNRAS.441..404H}, a statistical
`bulleticity' method has been proposed to exploit every individual infalling
substructure in every cluster. For each piece of infalling substructure, a local
vector from the dark matter peak (identified using weak lensing analysis) and
the baryonic mass peak (from X-rays). 
An average bulleticity signal of zero would imply an equal cross sections for
the dark matter and baryonic matter. By measuring any observed, finite
amplitude of bulleticity, one can empirically measure the ratio between the dark 
matter self-interaction and baryonic self-interaction cross sections.
Since we know the baryonic
cross-section relatively well, we can infer the dark matter-dark
matter cross-section. 

In Figure~\ref{fig:hydro_images} a result from hydrodynamical
simulations of dark and baryonic matter within clusters in shown.
\citet{2011MNRAS.413.1709M,2013MNRAS.433.1517H} have used these
simulations to show that the measurement of a net bulleticity
consistent with the cold dark matter used in the simulations will be
possible with Euclid.

\epubtkImage{fig4.jpg}{%
\begin{figure}
\centerline{\includegraphics[width=0.45\textwidth]{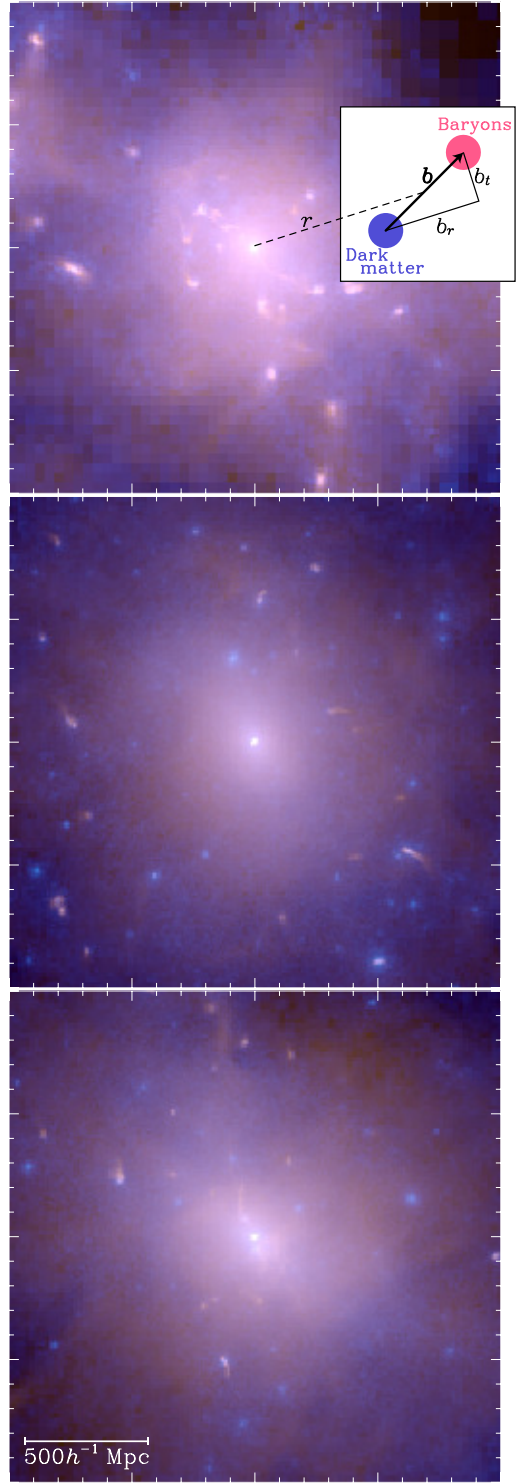}}
\caption{Full hydrodynamical simulations of massive clusters at
  redshift $z=0.6$. Total projected mass is shown in blue, while X-ray
  emission from baryonic gas is in red. The preferential trailing of
  gas due to pressure from the ICM, and its consequent separation from
  the non interacting dark matter, is apparent in much of the
  infalling substructure.}
\label{fig:hydro_images}
\end{figure}}

Finally, a Fisher matrix calculation has shown that, under the
assumption that systematic effects can be controlled, Euclid could use
such a technique to constrain the relative particulate cross-sections
to $6\times10^{-27}\mathrm{\ cm^{2}\ GeV^{-1}}$.

The dark matter-dark matter interaction probed by Euclid using this
technique will be complementary to the interactions constrained by
direct detection and accelerator experiments where the primary
constraints will be on the dark matter-baryon interaction.

\subsubsection{Dark matter halo shapes}

Self-interacting dark matter circularises the centres of dark
  matter halos, removing triaxiality \citep{feng10}(Peter et
  al.\ 2012), and smooths out cuspy cores. These profiles can be
  measeured directly using strong gravitational lensing (e.g.\ Sand et
  al.\ 2008, Newman et al.\ 2012).

\citet{2001MNRAS.325..435M} have performed ray-tracing through
$N$-body simulations, and have discovered that the ability for galaxy
clusters to generate giant arcs from strong lensing is compromized if
the dark matter is subject to just a few collisions per particle. This
constraint translates to an upper bound
{$\sigma_p/m_p\lesssim0.1\mathrm{\ cm^{2}\ g^{-1}}$}. Furthermore,
more recent analyses of SIDM models \citep{mark-bullet,randall2008}
utilize data from the Bullet Cluster to provide another independent
limit on the scattering cross section, though the upper bound remains
unchanged.  \citet{2011MNRAS.413.1709M} have proposed that the
tendency for baryonic and dark matter to become separated within
dynamical systems, as seen in the Bullet Cluster, could be studied in
greater detail if the analysis were to be extended over the full sky
in Euclid.

\subsection{Dark matter--baryonic interactions}

Currently, a number of efforts are underway to directly detect WIMPs
via the recoil of atomic nuclei. The underground experiments such as
CDMS, CRESST, XENON, EDELWEISS and ZEPLIN have pushed observational
limits for the spin-independent WIMP-nucleon cross-section down to the
$\sigma\lesssim10^{-43}\mathrm{cm}^2$ r\'egime.  \epubtkFootnote{It is
  anyway worth noticing the controversial results of DAMA/LIBRA, and
  more recently of CoGeNT.}  A collection of the latest constraints
can be found at \url{http://dmtools.brown.edu}.

Another opportunity to unearth the dark matter particle lies in
accelerators such as the LHC. By 2018 it is possible these experiments
will have yielded mass estimates for dark matter candidates, provided
its mass is lighter than a few hundred GeV. However, the discovery of
more detailed properties of the particle, which are essential to
confirm the link to cosmological dark matter, would have to wait until
the International Linear Collider is constructed.

\subsection{Dark matter--dark energy interactions}

Interactions in the dark sector have provided a popular topic for
exploration, with a view to building models which alleviate the
coincidence and fine-tuning issues associated with dark energy (see
Section~\ref{mg:cde}). The great uncertainty surrounding the physical
nature of dark energy leaves plenty of scope for non-gravitational
physics to play a r\^ole. These models are discussed at length in
other sections of this review (\ref{models-of-modified-gravity} and
\ref{dms:de_dm}). Here, we only mention that \citet{simpscat} have
explored the phenomenology associated with dark matter scattering
elastically with dark energy. The growth rate of large-scale
structures is artificially slowed, allowing a modest constraint of
\begin{equation}
\sigma_p/m_p\lesssim\frac{10}{1+w}\mathrm{\ cm^{2}\ g^{-1}} \,.
\end{equation}

It is clear that such dark sector interactions do not arise in the
simplest models of dark matter and dark energy. However a rigorous
refutation of GR will require not only a robust measure of the growth
of cosmic structures, but confirmation that the anomalous dynamics are
not simply due to physics within the dark sector.



\section{Constraints on warm dark matter}
\label{section_wdm}
 
$N$-body simulations of large-scale structures that assume a
$\Lambda$CDM cosmology appear to over-predict the power on small
scales when compared to observations \citep{Primack:2009jr}: `the
missing-satellite problem'
\citep{Kauffmann:1993gv,Klypin:1999uc,Strigari:2007ma,Bullock:2010uy},
the `cusp-core problem' \citep{Li:2009mp,2005ApJ...621..757S,
  Zavala:2009ms} and sizes of mini-voids
\citep{Tikhonov:2009jq}. These problems may be more or less solved by
several different phenomena \citep[e.g.][]{Diemand:2009bm}, however
one which could explain all of the above is warm dark matter (WDM)
\citep{2001ApJ...556...93B, Colin:2000dn, Boyanovsky:2007ay}. If the
dark matter particle is very light, it can cause a suppression of the
growth of structures on small scales via free-streaming of the dark
matter particles whilst relativistic in the early universe.

\subsection{Warm dark matter particle candidates}

Numerous WDM particle models can be constructed, but there are two
that occur most commonly in literature, because they are most
plausible from particle physics theory as well as from cosmological
observations:

\begin{itemize}
\item Sterile neutrinos may be constructed to extend the standard
  model of particle physics. The standard model active (left-handed)
  neutrinos can then receive the observed small masses through, e.g.,
  a see-saw mechanism. This implies that right-handed sterile
  neutrinos must be rather heavy, but the lightest of them naturally
  has a mass in the keV region, which makes it a suitable WDM
  candidate. The simplest model of sterile neutrinos as WDM candidate
  assumes that these particles were produced at the same time as
  active neutrinos, but they never thermalized and were thus produced
  with a much reduced abundance due to their weak coupling
  \citep[see][and references therein]{Biermann:2007ap}.
\item The gravitino appears as the supersymmetric partner of the
  graviton in supergravity models. If it has a mass in the keV range,
  it will be a suitable WDM candidate. It belongs to a more general
  class of \textit{thermalized} WDM candidates. It is assumed that
  this class of particles achieved a full thermal equilibrium, but at
  an earlier stage, when the number of degrees of freedom was much
  higher and hence their relative temperature with respect to the CMB
  is much reduced. Note that in order for the gravitino to be a good
  dark matter particle in general, it must be very stable, which in
  most models corresponds to it being the LSP
  \citep[e.g.][]{Borgani:1997ds,Cembranos2005}.
\end{itemize}
Other possible WDM candidates exist, for example a non-thermal
neutralino \citep{Hisano:2000dz} or a non-thermal gravitino
\citep{Baltz:2001rq} etc.

\subsection{Dark matter free-streaming}

The modification of the shape of the linear-theory power spectrum of
CDM due to WDM can be calculated by multiplication by a transfer
function \citep{2001ApJ...556...93B}
\begin{equation} \label{eq:WDMtrfnlin}
T(k)\equiv\sqrt{\frac{P_{\mathrm{WDM}}(k)}{P_{\mathrm{CDM}}(k)}}=\left[1+(\alpha k)^{2\mu}\right]^{-5/\mu},
\end{equation}
with suitable parameter $\mu=1.12$ \citep{Viel:2005qj} and with the
scale break parameter, $\alpha$\reviewtext{,} in the case of thermal relic DM
\begin{equation}
\alpha=0.049\left(\frac{m_{\mathrm{WDM}}}{\mathrm{keV}}\right)^{-1.11}\left(\frac{\Omega_{\mathrm{WDM}}}{0.25}
\right)^{0.11}\left(\frac{h}{0.7}\right)^{1.22}\,h^{-1}\,\mathrm{Mpc}.
\end{equation}
This is a fit to the solution of the full Boltzman equations.

There is a one-to-one relation between the mass of the thermalized WDM particle $m_\mathrm{WDM}$ (e.g., gravitino), and the mass of the simplest sterile neutrino $m_\mathrm{\nu s}$, such that the two models have an identical impact on cosmology \citep{Viel:2005qj}
\begin{equation}
m_\mathrm{\nu
  s}=4.43\left(\frac{m_{\mathrm{WDM}}}{\mathrm{keV}}\right)^{4/3}\left(\frac{\omega_{\mathrm{WDM}}}{0.1225}\right)^{-1/3}\mathrm{\ keV} \,,
\end{equation}
where $\omega=\Omega h^2$. The difference comes from the fact that in
the gravitino case the particle is fully thermalized, the number of
effective degrees of freedom being determined by mass and energy
density of dark matter, while in the simplest sterile neutrino case
the number of degrees of freedom is fixed, while abundance is
determined by mass and energy density of dark matter.

\subsection{Current constraints on the WDM particle from large-scale
  structure}

Measurements in the particle-physics energy domain can only reach
masses uninteresting in the WDM context, since direct detectors look
mainly for a WIMP, whose mass should be in the GeV\,--\,TeV
range. However, as described above, cosmological observations are able
to place constraints on light dark matter particles. Observation of
the flux power spectrum of the Lyman-$\alpha$ forest, which can
indirectly measure the fluctuations in the dark matter density on
scales between $\sim100\mathrm{\ kpc}$ and $\sim10\mathrm{\ Mpc}$
gives the limits of $m_{\mathrm{WDM}}>4\mathrm{\ keV}$ or equivalently
$m_{\mathrm{\nu s}}>28\mathrm{\ keV}$ at 95\% confidence level
\citep{Viel:2007mv,Viel:2005qj,Seljak:2006qw}. For the simplest
sterile neutrino model, these lower limits are at odds with the upper
limits derived from X-ray observations, which come from the lack of
observed diffuse X-ray background from sterile neutrino annihilation
and set the limit $m_{\mathrm{\nu s}}<1.8\mathrm{\ keV}$ at the 95\%
confidence limit \citep{Boyarsky:2006jm}. However, these results do
not rule the simplest sterile neutrino models out. There exist
theoretical means of evading small-scale power constraints \citep[see
  e.g.][and references therein]{Boyarsky:2008mt}. The weak lensing
power spectrum from Euclid will be able to constrain the dark matter
particle mass to about $m_{\mathrm{WDM}}>2\mathrm{\ keV}$
\citep{Markovic:2010te}.

\subsection{Nonlinear structure in WDM}

In order to extrapolate the matter power spectrum to later times one
must take into account the nonlinear evolution of the matter density
field.

Several fitting functions have been found to calculate the
  nonlinear power on the small scales of the present-day matter power
  spectrum in the scenario where all dark matter is warm. The most
  basic approach is simply to modify the linear matter power spectrum
  from \autoref{eq:WDMtrfnlin}, which is based on the output of
  Bolzmann codes like \textsc{camb} or \textsc{class}
  \citep{CAMB,CLASS}.  One can then either i) run simulations
  \citep{2005MNRAS.360..282B,Boyanovsky:2007ay,Zavala:2009ms,2007MNRAS.380...93W,2009MNRAS.393..511C,2012MNRAS.421...50V,2012MNRAS.424..684S,2013MNRAS.428.1774B,2013MNRAS.434.3337A,2013arXiv1306.3210S},
  ii) use the halo model
  \citep{2011PhRvD..84f3507S,2012MNRAS.424..684S,2011arXiv1109.6291D}
  or iii) a fit analogous to \autoref{eq:WDMtrfnlin}, where the
  $\Lambda$CDM nonlinear power spectrum is modified by a transfer
  function \citep{2012MNRAS.421...50V} to calculate the present-day
  power on the small scales:
\begin{equation}
T_{\mathrm{nl}}(k) \equiv \sqrt{\frac{P^\mathrm{nl}_{\mathrm{WDM}}(k)}{P^\mathrm{nl}_{\mathrm{CDM}}(k)}} = \left[1+(\alpha\,k)^{\mu l}\right]^{-s/(2\mu)} \ ,
\end{equation}
where
\begin{equation}
\alpha(m_{\mathrm{WDM}},z) = 0.0476 \left(\frac{\mathrm{keV}}{m_{\mathrm{WDM}}}\right)^{1.85}\left(\frac{1+z}{2}\right)^{1.3} \ ,
\end{equation}
and $\mu=3$, $l=0.6$, $s=0.4$ are the fitting parameters. 

Such fits can be used to calculate further constraints on WDM from the
weak lensing power spectrum or galaxy clustering
\citep{Markovic:2010te,2014PASA...31....6M}

It should be noted that in order to use the present day clustering of
structure as a probe for WDM it is crucial to take into account
baryonic physics as well as neutrino effect, which are described in
the following section.  


\section{Neutrino properties}

The first significant evidence for a finite neutrino mass
\citep{SuperK} indicated the incompleteness of the standard model of
particle physics. Subsequent experiments have further strengthened
this evidence and improved the determination of the neutrino mass
splitting required to explain observations of neutrino oscillations.

As a summary of the last decade of neutrino experiments, two
hierarchical neutrino mass splittings and three mixing angles have
been measured. Furthermore, the standard model has three neutrinos:
the motivation for considering deviations from the standard model in
the form of extra sterile neutrinos has disappeared
\citep{mena,miniboone}. Of course, deviations from the standard
effective numbers of neutrino species could still indicate exotic
physics which we will discuss below (Section~\ref{sec:Neff}).

New and future neutrino experiments aim to determine the remaining
parameters of the neutrino mass matrix and the nature of the neutrino
mass. Within three families of neutrinos, and given all neutrino
oscillation data, there are three possible mass spectra: a)
degenerate, with mass splitting smaller than the neutrino masses, and
two non-degenerate cases, b) normal hierarchy (NH), with the larger
mass splitting between the two more massive neutrinos and c) inverted
hierarchy (IH), with the smaller spitting between the two higher mass
neutrinos. Figure~\ref{fig:hierarchy-massnu} \citep{Jimenez:2010ev}
illustrates the currently allowed regions in the plane of total
neutrino mass, $\Sigma$, \label{symbol:Sigma} vs.\ mass of the
lightest neutrino, $m$. Note that a determination of
$\Sigma<0.1\mathrm{\ eV}$ would indicate normal hierarchy and that
there is an expected minimum mass $\Sigma>0.054\mathrm{\ eV}$. The
cosmological constraint is from \citet{Reid-etal:2010}.

\epubtkImage{standardplot.png}{%
\begin{figure}
\centerline{\includegraphics[width=.75\textwidth]{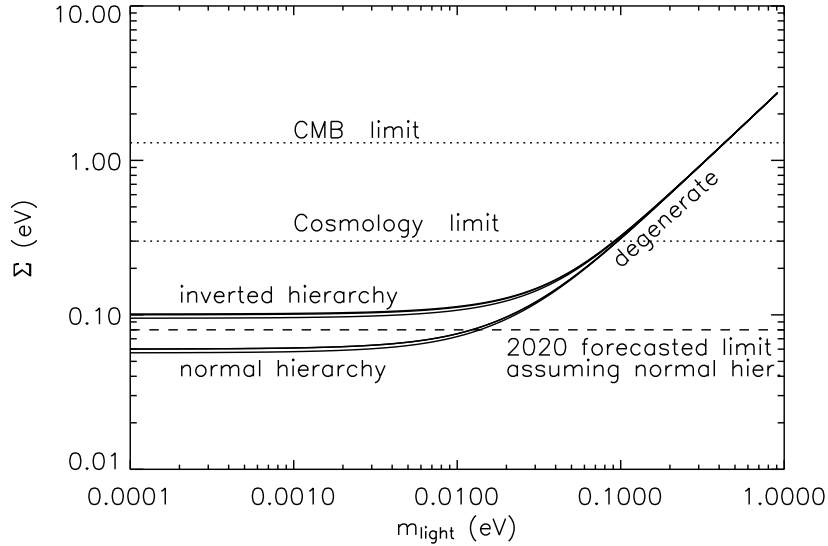}}
\caption{Constraints from neutrino oscillations and from cosmology in
  the $m$-$\Sigma$ plane. Image reproduced by permission
  from~\cite{Jimenez:2010ev}; copyright by IOP and SISSA.}
\label{fig:hierarchy-massnu}
\end{figure}}

Cosmological constraints on neutrino properties are highly
complementary to particle physics experiments for several reasons:

\begin{itemize}
\item \textbf{Relic neutrinos} produced in the early universe are
  hardly detectable by weak interactions, making it impossible with
  foreseeable technology to detect them directly. But new cosmological
  probes such as Euclid offer the opportunity to detect (albeit
  indirectly) relic neutrinos, through the effect of their mass on the
  growth of cosmological perturbations.
\item \textbf{Cosmology remains a key avenue to determine the absolute
  neutrino mass scale}. Particle physics experiments will be able to
  place lower limits on the \textit{effective} neutrino mass, which
  depends on the hierarchy, with no rigorous limit achievable in the
  case of normal hierarchy \citep{murayama}. Contrarily, neutrino free
  streaming suppresses the small-scale clustering of large-scale
  cosmological structures by an amount that depends on neutrino mass.
\item \textbf{``What is the hierarchy (normal, inverted or
  degenerate)?''} Neutrino oscillation data are unable to resolve
  whether the mass spectrum consists in two light states with mass $m$
  and a heavy one with mass $M$ -- normal hierarchy -- or two heavy
  states with mass $M$ and a light one with mass $m$ -- inverted
  hierarchy -- in a model-independent way. Cosmological observations,
  such as the data provided by Euclid, can determine the hierarchy,
  complementarily to data from particle physics experiments.
\item \textbf{``Are neutrinos their own anti-particle?''} If the answer is yes, then neutrinos are Majorana fermions; if not, they are Dirac. If neutrinos and anti-neutrinos are identical, there could have been a process in the early universe that affected the balance between particles and anti-particles, leading to the matter anti-matter asymmetry we need to exist \citep{leptogenesis}. This question can, in principle, be resolved if neutrino-less double-$\beta$ decay is observed \citep[see][and references therein]{murayama}. However, if such experiments \citep[ongoing and planned, e.g.,][]{Cremonesi:2010} lead to a negative result, the implications for the nature of neutrinos depend on the hierarchy. As shown in \citet{Jimenez:2010ev}, in this case cosmology can offer complementary information by  helping determine the hierarchy. 
\end{itemize}

\subsection{Evidence of relic neutrinos}
\label{sec:relictnu}

The hot big bang model predicts a background of relic neutrinos in the
universe with an average number density of
$\sim100\,N_{\nu}\mathrm{\ cm}^{-3}$, where $N_{\nu}$ is the number of
neutrino species. These neutrinos decouple from the CMB at redshift
$z\sim10^{10}$ when the temperature was $T\sim o(\mathrm{MeV})$, but
remain relativistic down to much lower redshifts depending on their
mass. A detection of such a neutrino background would be an important
confirmation of our understanding of the physics of the early
universe.

Massive neutrinos affect cosmological observations in different
ways. Primary CMB data alone can constrain the total neutrino mass
$\Sigma$, if it is above $\sim1\mathrm{\ eV}$ \citep[][finds
  $\Sigma<1.3\mathrm{\ eV}$ at 95\% confidence]{Komatsu:2010fb}
because these neutrinos become non-relativistic before recombination
leaving an imprint in the CMB. Neutrinos with masses
$\Sigma<1\mathrm{\ eV}$ become non-relativistic after recombination
altering matter-radiation equality for fixed $\Omega_mh^2$; this
effect is degenerate with other cosmological parameters from primary
CMB data alone. After neutrinos become non-relativistic, their free
streaming damps the small-scale power and modifies the shape of the
matter power spectrum below the free-streaming length. The
free-streaming length of each neutrino family depends on its mass.

Current cosmological observations do not detect any small-scale power
suppression and break many of the degeneracies of the primary CMB,
yielding constraints of $\Sigma<0.3\mathrm{\ eV}$
\citep{Reid-etal:2010} if we assume the neutrino mass to be a
constant. A detection of such an effect, however, would provide a
detection, although indirect, of the cosmic neutrino background. As
shown in the next section, the fact that  oscillations predict a
minimum total mass $\Sigma\sim0.054\mathrm{\ eV}$ implies that Euclid
has the statistical power to detect the cosmic neutrino background. We
finally remark that the neutrino mass may also very well vary in time
\cite{Wetterich:2007kr}; this might be tested by comparing (and not
combining) measurements from CMB at decoupling with low-$z$
measurements. An inconsistency would point out a direct measurement of
a time varying neutrino mass \cite{Wetterich:2009qf}.

\subsection{Neutrino mass}
\label{sec:numass}

Particle physics experiments are sensitive to neutrino flavours making a determination of the neutrino absolute-mass scales very model dependent. On the other hand, cosmology is not sensitive to neutrino flavour, but is sensitive to the total neutrino mass.

The small-scale power-suppression caused by neutrinos leaves
  imprints on CMB lensing and prior to the experiment forecasts
  indicated that Planck should be able to constrain the sum of
  neutrino masses $\Sigma$, with a $1\sigma$ error of 0.13~eV
  \citep{Knox-Kaplinghat:2003,Lesgourgues-etal:2006,dePutter-etal:2009}. In
  \citep{2013arXiv1303.5076P} Planck reported constraints on the
  $N_{\rm eff}=3.30+/-0.27$ for the effective number of relativistic
  degrees of freedom, and an upper limit of $0.23$ eV for the summed
  neutrino mass. However the Planck cosmological constraints also
  reported a relatively low value of the Hubble parameter with respect
  to previous measurements, that resulted in several papers , for
  example \citep{2014PhRvL.112e1302W}, that investigated the
  possibility that this tension could possibly be resolved by
  introducing an eV-scale (possibly sterile) neutrino.  Combining the
  Planck results with large scale structure measurements or weak
  lensing measurements has resulted in reported claims of even
  stronger constraints on the sum of neutrino masses, for example
  \citep{2014PhRvD..89j3505R} found an upper limit on the sum of
  neutrino masses of $< 0.18$ eV (95\% confidence) by combining with
  WiggleZ data. \citep{2014PhRvL.112e1303B} and
  \citep{2013JCAP...10..044H} combined Planck data with weak lensing
  data from CFHTLenS and found higher values for the sum of neutrino
  masses, as a result of tension in the measured and inferred values
  of $\sigma_8$ between lensing and the CMB where the lensing prefers
  a lower value, however \citep{2014arXiv1401.6842K} find that such a
  lower value of $\sigma_8$ is consistent with Baryon feedback models
  impacting the small-scale distribution of dark matter.
 
Euclid's measurement of the galaxy power spectrum, combined with
Planck (primary CMB only) priors should yield an error on $\Sigma$ of
0.04~eV \citep[for details see][]{Carbone-etal:2010} which is in
qualitative agreement with previous work
\citep[e.g.][]{Saito-etal:2009}), assuming a minimal value for
$\Sigma$ and constant neutrino mass. Euclid's weak lensing should also
yield an error on $\Sigma$ of 0.05~eV
\citep{Kitching-etal:2008}. While these two determinations are not
fully independent (the cosmic variance part of the error is in common
given that the lensing survey and the galaxy survey cover the same
volume of the universe) the size of the error-bars implies more than
$1\sigma$ detection of even the minimum $\Sigma$ allowed by
oscillations. Moreover, the two independent techniques will offer
cross-checks and robustness to systematics. The error on $\Sigma$
depends on the fiducial model assumed, decreasing for fiducial models
with larger $\Sigma$. Euclid will enable us not only to detect the
effect of massive neutrinos on clustering but also to determine the
absolute neutrino mass scale. However, recent numerical
  investigations found severe observational degeneracies between the
  cosmological effects of massive neutrinos and of some modified
  gravity models \citep{Baldi_etal_2014}. This may indicate an
  intrinsic theoretical limit to the effective power of astronomical
  data in discriminating between alternative cosmological scenarios,
  and in constraining the neutrino mass as well. Further
  investigations with higher resolution simulations are needed to
  clarify this issue and to search for possible ways to break these
  cosmic degeneracies \citep[see also][]{LaVacca:2009yp,
    Kristiansen:2009yx, Marulli:2011jk}.

\subsection{Hierarchy and the nature of neutrinos}
\label{sec:hierarchy}

Since cosmology is insensitive to flavour, one might expect that
cosmology may not help in determining the neutrino mass
hierarchy. However, for $\Sigma<0.1\mathrm{\ eV}$, only normal
hierarchy is allowed, thus a mass determination can help disentangle
the hierarchy. There is however another effect: neutrinos of different
masses become non-relativistic at slightly different epochs; the free
streaming length is sightly different for the different species and
thus the detailed shape of the small scale power suppression depends
on the individual neutrino masses and not just on their sum. As
discussed in \citet{Jimenez:2010ev}, in cosmology one can safely
neglect the impact of the solar mass splitting. Thus, two masses
characterize the neutrino mass spectrum: the lightest $m$, and the
heaviest $M$. The mass splitting can be parameterized by
$\Delta=(M-m)/\Sigma$ for normal hierarchy and $\Delta=(m-M)/\Sigma$
for inverted hierarchy. The absolute value of $\Delta$ determines the
mass splitting, whilst the sign of $\Delta$ gives the
hierarchy. Cosmological data are very sensitive to $|\Delta|$; the
direction of the splitting -- i.e., the sign of $\Delta$ -- introduces
a sub-dominant correction to the main effect. Nonetheless,
\citet{Jimenez:2010ev} show that weak gravitational lensing from
Euclid data will be able to determine the hierarchy (i.e., the mass
splitting and its sign) if far enough away from the degenerate
hierarchy (i.e., if $\Sigma<0.13$). 

\epubtkImage{delta_definition-fig1Delta3blowmass.png}{%
\begin{figure}
\centerline{
\includegraphics[width=0.5\textwidth]{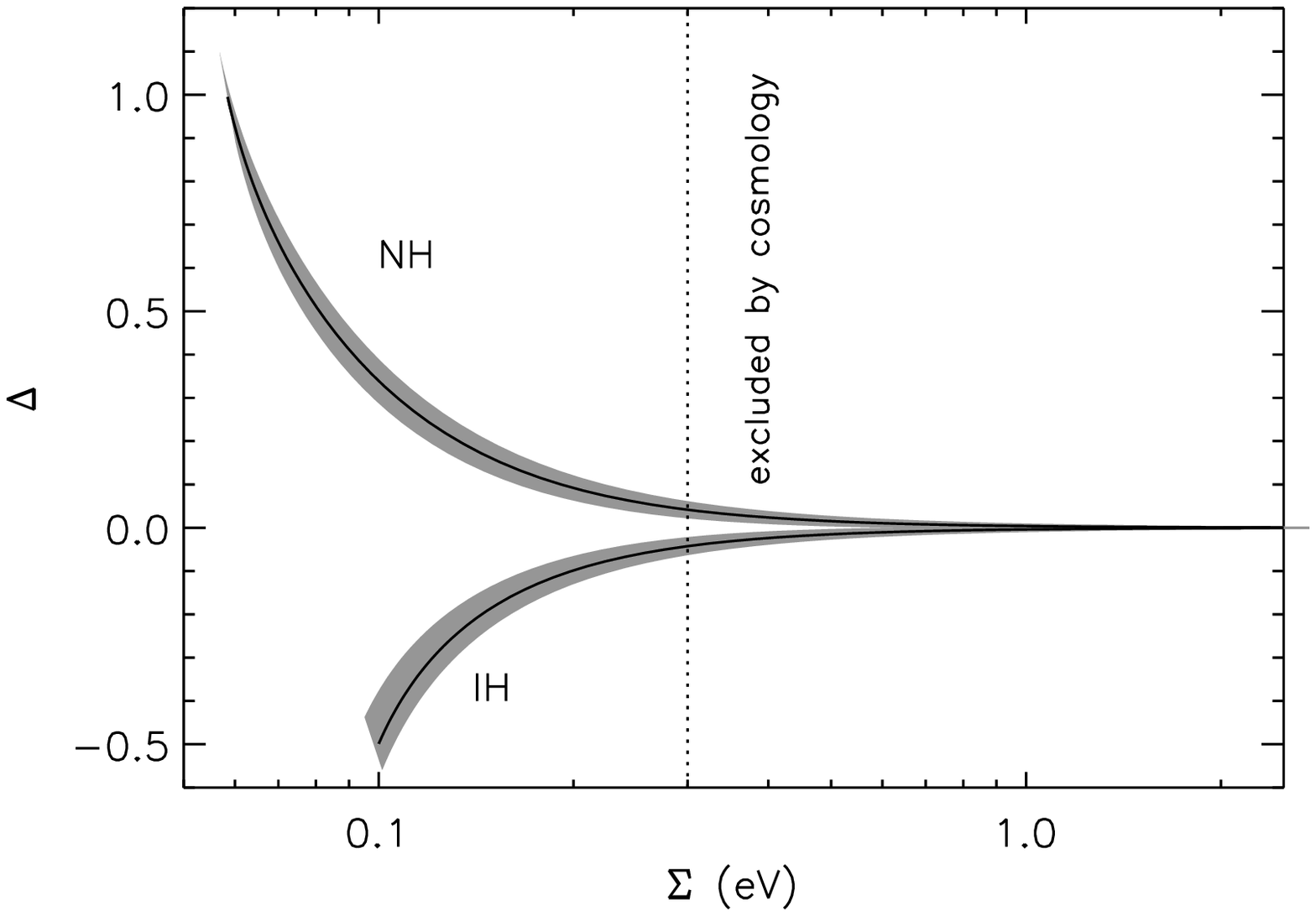}
\includegraphics[width=0.4\textwidth,height=0.45\textwidth]{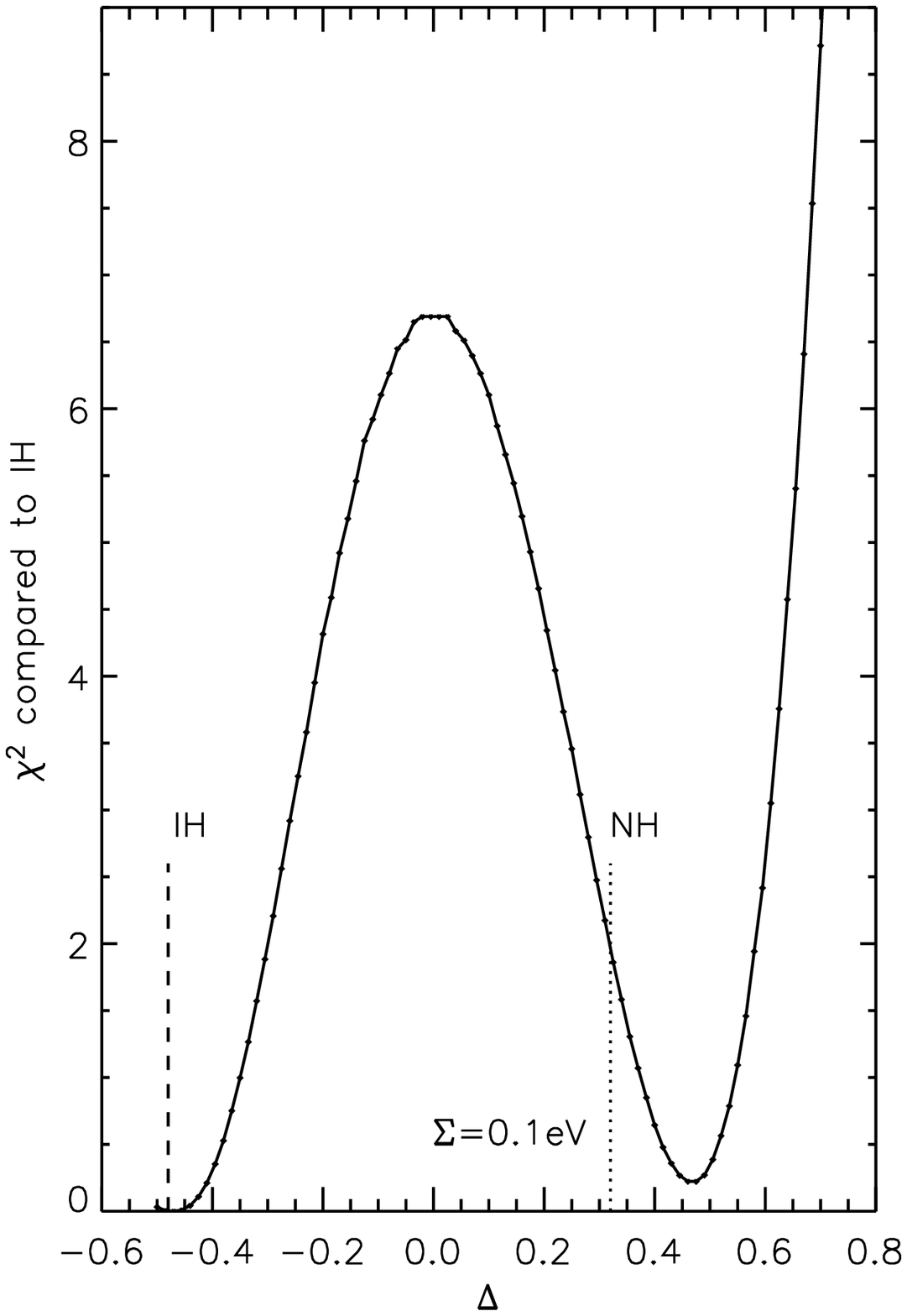}
}
\caption{\emph{Left:} region in the $\Delta$-$\Sigma$ parameter space
  allowed  by oscillations data. \emph{Right:} Weak lensing
  forecasts. The dashed and dotted vertical lines correspond to the
  central value for $\Delta$ given by oscillations data. In this case
  Euclid could discriminate NI from IH with a $\Delta \chi^2=2$. Image
  reproduced by permission from~\cite{Jimenez:2010ev}; copyright by
  IOP and SISSA.}
\label{fig:hierarchy}
\end{figure}}

A detection of neutrino-less double-$\beta$ decay from the next generation experiments would indicate that neutrinos are Majorana particles. A null result of such double-$\beta$ decay experiments would lead to a result pointing to the Dirac nature of the neutrino only for degenerate or inverted mass spectrum. Even in this case, however, there are ways to suppress the double-$\beta$ decay signal, without the neutrinos being Dirac particles. For instance, the pseudo-Dirac scenario, which arises from the same Lagrangian that describes the see-saw mechanism \cite[see e.g.][]{Rodejohann:2012xd}. This information can be obtained from large-scale structure cosmological data, improved data on  the tritium beta decay, or the long-baseline neutrino oscillation experiments. If the small mixing in the neutrino mixing matrix is negligible, cosmology might be the most promising arena to help in this puzzle.

\subsection{Number of neutrino species}
\label{sec:Neff}

Neutrinos decouple early in cosmic history and contribute to a relativistic energy density with an effective number of species $N_{\nu,\mathrm{eff}}=3.046$. Cosmology is sensitive to the physical energy density in relativistic particles in the early universe, which in the standard cosmological model includes only photons and neutrinos: $\omega_{\mathrm{rel}}=\omega_{\gamma}+N_{\nu,\mathrm{eff}}\omega_{\nu}$, where $\omega_{\gamma}$ denotes the energy density in photons and is exquisitely constrained from the CMB, and $\omega_{\nu}$ is the energy density in one neutrino. Deviations from the standard value for $N_{\nu,\mathrm{eff}}$ would signal non-standard neutrino features or additional relativistic species. $N_{\nu,\mathrm{eff}}$ impacts the big bang nucleosynthesis epoch through its effect on the expansion rate; measurements of primordial light element abundances can constrain $N_{\nu,\mathrm{eff}}$ and rely on physics at $T\sim \mathrm{MeV}$ \citep{Bowen:2001in}. In several non-standard models -- e.g., 
decay of dark matter particles, axions, quintessence -- the energy density in relativistic species can change at some later time. The energy density of free-streaming relativistic particles alters the epoch of matter-radiation equality and leaves therefore a signature in the CMB and in the matter-transfer function. However, there is a degeneracy between $N_{\nu,\mathrm{eff}}$ and $\Omega_m h^2$ from CMB data alone (given by the combination of these two parameters that leave matter-radiation equality unchanged) and between $N_{\nu,\mathrm{eff}}$ and $\sigma_8$ and/or $n_s$. Large-scale structure surveys measuring the shape of the power spectrum at large scale can constrain independently the combination $\Omega_m h$ and $n_s$, thus breaking the CMB degeneracy. Furthermore, anisotropies in the neutrino background affect the CMB anisotropy angular power spectrum at a level of $\sim20\%$ through the gravitational feedback of their free streaming damping and anisotropic stress contributions. Detection of this 
effect is now possible by combining CMB and large-scale structure observations. This yields an indication at more than $2\sigma$ level that there exists a neutrino background with characteristics compatible with what is expected under the cosmological standard model \citep{Trotta:2004ty,DeBernardis:2008ys}.

The forecasted errors on $N_{\nu,\mathrm{eff}}$ for Euclid (with a Planck prior) are $\pm0.1$ at $1\sigma$ level \citep{Kitching-etal:2008}, which is a factor $\sim5$ better than current constraints from CMB and LSS and about a factor $\sim2$ better than constraints from light element abundance and nucleosynthesis.

\subsection{Model dependence}
\label{Model_dependence}

A recurring question is how much model dependent will the neutrino
constraints be. It is important to recall that usually
parameter-fitting is done within the context of a $\Lambda$CDM model
and that the neutrino effects are seen indirectly in the
clustering. Considering more general cosmological models, might
degrade neutrino constraints, and vice versa, including neutrinos in
the model might degrade dark-energy constraints. Here below we discuss
the two cases of varying the total neutrino mass $\Sigma$ and the
number of relativistic species $N_{\mathrm{eff}}$,
separately. Possible effects of modified gravity models
  that could further degrade the neutrino mass constraints will not be
  discussed in this section.

\subsection{$\Sigma$ forecasted error bars and degeneracies}
\label{Mnu_cosmology}

In \citet{Carbone-etal:2010} it is shown that, for a general model
which allows for a non-flat universe, and a redshift dependent
dark-energy equation of state, the $1\sigma$ spectroscopic errors on
the neutrino mass $\Sigma$ are in the range 0.036\,--\,0.056~eV,
depending on the fiducial total neutrino mass $\Sigma$, for the
combination Euclid+Planck.

On the other hand, looking at the effect that massive neutrinos have on the dark-energy parameter constraints, it is shown that the total CMB+LSS dark-energy FoM decreases only by $\sim$~15\%\,--\,25\% with respect to the value obtained if neutrinos are supposed to be massless, when the forecasts are computed using the so-called ``$P(k)$-method marginalized over growth-information'' (see Methodology section),  which therefore results to be quite robust in constraining the dark-energy equation of state.

For what concerns the parameter correlations, at the LSS level, the total neutrino mass $\Sigma$ is correlated with all the cosmological parameters affecting the galaxy power spectrum shape and BAO positions. When Planck priors are added to the Euclid constraints, all degeneracies are either resolved or reduced, and the remaining dominant correlations among $\Sigma$ and the other cosmological parameters are $\Sigma$-$\Omega_{\mathrm{de}}$, $\Sigma$-$\Omega_m$, and $\Sigma$-$w_a$, with the $\Sigma$-$\Omega_{\mathrm{de}}$ degeneracy being the largest one.

\subsubsection{Hierarchy dependence}
\label{Hierarchy_dependence}
In addition, the neutrino mass spectroscopic constraints depend also on the neutrino hierarchy. In fact, the $1\sigma$ errors on total neutrino mass for normal hierarchy are $\sim$~17\%\,--\,20\% larger than for the inverted one. It appears that the matter power spectrum is less able to give information on the total neutrino mass when the normal hierarchy is assumed as fiducial neutrino mass spectrum. This is similar to what found in~\citep{Jimenez:2010ev} for the constraints on the neutrino mass hierarchy itself, when a normal hierarchy is assumed as the fiducial one. On the other hand, when CMB information are included, the $\Sigma$-errors decrease by $\sim$~35\% in favor of the normal hierarchy, at a given fiducial value $\Sigma|_{\mathrm{fid}}$. This difference arises from the changes in the free-streaming effect due to the assumed mass hierarchy, and is in agreement with the results of~\citep{0403296}, which confirms that the expected errors on the neutrino masses depend not only on the sum of neutrino 
masses, but also on the order of the mass splitting between the neutrino mass states.

\subsubsection{Growth and incoherent peculiar velocity dependence}
\label{growth+FoG}

$\Sigma$ spectroscopic errors stay mostly unchanged whether
growth-information are included or mar\-gina\-li\-sed over, and
decrease only by 10\%\,--\,20\% when adding $f_g\sigma_8$
measurements. This result is expected, if we consider that, unlike
dark-energy parameters, $\Sigma$ affects the shape of the power
spectrum via a redshift-dependent transfer function $T(k,z)$, which is
sampled on a very large range of scales including the $P(k)$ turnover
scale, therefore this effect dominates over the information extracted
from measurements of $f_g\sigma_8$. This quantity, in turn, generates
new correlations with $\Sigma$ via the $\sigma_8$-term, which actually
is anti-correlated with $M_\nu$ \citep{Marulli2010_inprep}. On the
other hand, if we suppose that early dark-energy is negligible, the
dark-energy parameters $\Omega_{\mathrm{de}}$, $w_0$ and $w_a$ do not
enter the transfer function, and consequently growth information have
relatively more weight when added to constraints from $H(z)$ and
$D_A(z)$ alone.  Therefore, the value of the dark-energy FoM does
increase when growth-information are included, even if it decreases by
a factor $\sim$~50\%\,--\,60\% with respect to cosmologies where
neutrinos are assumed to be massless, due to the correlation among
$\Sigma$ and the dark-energy parameters. As confirmation of this
degeneracy, when growth-information are added and if the dark-energy
parameters $\Omega_{\mathrm{de}}$, $w_0$, $w_a$ are held fixed to
their fiducial values, the errors $\sigma({\Sigma})$ decrease from
0.056~eV to 0.028~eV, for Euclid combined with Planck.

We expect that dark-energy parameter errors are somewhat sensitive also to the effect of incoherent peculiar velocities, the so-called ``Fingers of God'' (FoG). This can be understood in terms of correlation functions in the redshift-space; the stretching effect due to random peculiar velocities contrasts the flattening effect due to large-scale bulk velocities. Consequently, these two competing effects act along opposite directions on the dark-energy parameter constraints (see methodology Section~\ref{statistical}).

On the other hand, the neutrino mass errors are found to be stable again at $\sigma({\Sigma})=0.056$, also when FoG effects are taken into account by marginalising over $\sigma_v(z)$; in fact, they increase only by 10\%\,--\,14\% with respect to the case where FoG are not taken into account.

Finally, in Table~\ref{summary} we summarize the dependence of the
$\Sigma$-errors on the model cosmology, for Euclid combined with
Planck. \epubtkFootnote{In this case we have added the contribution
  from BOSS at redshifts $0.1<z<z_{\min}$, where $z_{\min}=0.5$ is the
  minimum redshift of the Euclid spectroscopic survey.}  We conclude
that, if $\Sigma$ is $>$~0.1~eV, spectroscopy with Euclid will be able
to determine the neutrino mass scale independently of the model
cosmology assumed, provided GR is correct and dark energy
  does not interact with other species
  \citep[][]{Baldi_etal_2014}. If $\Sigma$ is $<$~0.1~eV, the sum of
neutrino masses, and in particular the minimum neutrino mass required
by neutrino oscillations, can be measured in the context of a
$\Lambda$CDM model.

\subsection{$N_{\mathrm{eff}}$ forecasted errors and degeneracies}
\label{Neff_cosmology}

Regarding the $N_{\mathrm{eff}}$ spectroscopic errors, \cite{Carbone-etal:2010} finds $\sigma(N_{\mathrm{eff}})\sim 0.56$ from Euclid, and $\sigma(N_{\mathrm{eff}})\sim 0.086$, for Euclid+Planck. Concerning the effect of $N_{\mathrm{eff}}$ uncertainties on the dark-energy parameter errors, the CMB+LSS dark-energy FoM decreases only by $\sim 5\%$ with respect to the value obtained holding $N_{\mathrm{eff}}$ fixed at its fiducial value, meaning that also in this case the ``$P(k)$-method marginalized over growth--information'' is not too sensitive to assumptions about  model cosmology when constraining the dark-energy equation of state.

About the degeneracies between $N_{\mathrm{eff}}$ and the other
cosmological parameters, it is necessary to say that the number of
relativistic species gives two opposite contributions to the observed
power spectrum $P_{\mathrm{obs}}$ (see methodology
Section~\ref{statistical}), and the total sign of the correlation
depends on the dominant one, for each single cosmological
parameter. In fact, a larger $N_{\mathrm{eff}}$ value suppresses the
transfer function $T(k)$ on scales $k\leq \kmax$. On the other hand, a
larger $N_{\mathrm{eff}}$ value also increases the Alcock--Paczynski
prefactor in $P_{\mathrm{obs}}$. For what concerns the dark-energy
parameters $\Omega_{\mathrm{de}}$, $w_0$, $w_a$, and the dark-matter
density $\Omega_m$, the Alcock--Paczynski prefactor dominates, so that
$N_{\mathrm{eff}}$ is positively correlated to $\Omega_{\mathrm{de}}$
and $w_a$, and anti-correlated to $\Omega_m$ and $w_0$. In contrast,
for the other parameters, the $T(k)$ suppression produces the larger
effect and $N_{\mathrm{eff}}$ results to be anti-correlated to
$\Omega_b$, and positively correlated to $h$ and $n_s$.  The degree of
the correlation is very large in the $n_s$-$N_{\mathrm{eff}}$ case,
being of the order $\sim 0.8$ with and without Planck priors. For the
remaining cosmological parameters, all the correlations are reduced
when CMB information are added, except for the covariance
$N_{\mathrm{eff}}$-$\Omega_{\mathrm{de}}$, as happens also for the
$M_\nu$-correlations. To summarize, after the inclusion of Planck
priors, the remaining dominant degeneracies among $N_{\mathrm{eff}}$
and the other cosmological parameters are $N_{\mathrm{eff}}$-$n_s$,
$N_{\mathrm{eff}}$-$\Omega_{\mathrm{de}}$, and $N_{\mathrm{eff}}$-$h$,
and the forecasted error is $\sigma(N_{\mathrm{eff}})\sim 0.086$, from
Euclid+Planck. Finally, if we fix to their fiducial values the
dark-energy parameters $\Omega_{\mathrm{de}}$, $w_0$ and $w_a$,
$\sigma(N_{\mathrm{eff}})$ decreases from 0.086 to 0.048, for the
combination Euclid+Planck.  However, it has to be
  noticed that if $N_{\mathrm{eff}}$ is allowed to vary, then the
  shape of the matter power spectrum in itself cannot constrain
  $\Omega_m h$. Indeed, in $\Lambda$CDM models, the power spectrum
  constrains $\Omega_m h$ because the turning point $k_\mathrm{eq}$
  corresponds to the comoving Hubble rate at equality. If the
  radiation content is known, then $k_\mathrm{eq}$ depends only on
  $\Omega_m h$. However, if the radiation content is unknown, then
  $k_\mathrm{eq}$ is not linked to a unique value of $\Omega_m h$
  \cite{Abazajian:2012ys}. The fact that one can use a combination of
  CMB (excluding the damping tail) and matter power spectrum data to
  break the $N_{\mathrm{eff}}$--$\Omega_m h^2$ degeneracy is due to a
  decreasing baryon fraction $f_b = \Omega_b h^2/\Omega_m h^2$ when
  $N_{\mathrm{eff}}$ is increased (while keeping $z_\mathrm{eq}$
  fixed) \cite[e.g.][]{Bashinsky:2003tk}.

\begin{table}
\caption{$\sigma(M_\nu)$ and $\sigma(N_{\mathrm{eff}})$ marginalized errors
  from LSS+CMB}
\label{summary}
\setlength{\tabcolsep}{1.5pt}
\centering
{\small
\begin{tabular}{l c c c c c c}
\toprule
~ & ~ & ~ & General cosmology & ~ & ~ & ~ \\
\midrule
fiducial $\to$ &
$\Sigma=0.3\mathrm{\, eV}^a$ &
$\Sigma=0.2\mathrm{\, eV}^a$ &
$\Sigma=0.125\mathrm{\, eV}^b$ &
$\Sigma=0.125\mathrm{\, eV}^c$ &
$\Sigma=0.05\mathrm{\, eV}^b$ &
$N_{\mathrm{eff}}=3.04^d$\\
\midrule
\reviewtext{Euclid}+Planck & 0.0361 & 0.0458 & 0.0322 & 0.0466 & 0.0563 & 0.0862\\
\toprule
~ & ~ & ~ & $\Lambda$CDM cosmology & ~ & ~ & ~ \\
\midrule
\reviewtext{Euclid}+Planck & 0.0176 & 0.0198 & 0.0173 & 0.0218 & 0.0217 & 0.0224\\
\bottomrule
\end{tabular}}
\begin{flushleft}
${}^a$\,\footnotesize{for degenerate spectrum: $m_1\approx m_2\approx
  m_3$};
${}^b$\,\footnotesize{for normal hierarchy: $m_3\neq 0$, $m_1\approx
  m_2\approx 0$}\\
${}^c$\,\footnotesize{for inverted hierarchy: $m_1\approx m_2$,
  $m_3\approx 0$};
${}^d$\,\footnotesize{fiducial cosmology with massless neutrinos}
\end{flushleft}
\end{table}

\subsection{Nonlinear effects of massive cosmological neutrinos on
  bias, P(k) and RSD}

In general, forecasted errors are obtained using techniques, like the
Fisher-matrix approach, that are not particularly well suited to
quantifying systematic effects. These techniques forecast only
statistical errors, which are meaningful as long as they dominate over
systematic errors.  Possible sources of systematic errors of major
concern are the effects of nonlinearities and galaxy bias.

The description of nonlinearities in the matter power spectrum in the
presence of massive neutrinos has been addressed in several different
ways: \citet{2008JCAP...10..035W,Saito-etal:2009,Saito2,Saito3} have
used perturbation theory, \citet{Pietroni} the time-RG flow approach
and \citet{Brandbyge1,Brandbyge2,Brandbyge3,Viel_etal2010} different
schemes of $N$-body simulations.  Another nonlinear scheme that has
been examined in the literature is the halo model.  This has been
applied to massive neutrino cosmologies in
\cite{2005PhRvD..71d3507A,2005JCAP...09..014H,2006JCAP...06..025H}.

On the other hand, galaxy/halo bias is known to be almost
scale-independent only on large, linear scales, but to become
nonlinear and scale-dependent for small scales and/or for very massive
haloes. From the above discussion and references, it is clear that the
effect of massive neutrinos on the galaxy power spectrum in the
nonlinear regime must be explored via $N$-body simulations to
encompass all the relevant effects.

Here below we focus on the behavior of the DM halo mass function (MF),
the DM halo bias, and the redshift-space distortions (RSD), in the
presence of a cosmological background of massive neutrinos. To this
aim, \citet{Brandbyge3} and \citet{Marulli2010_inprep} have analysed a
set of large $N$-body hydrodynamical simulations, developed with an
extended version of the code \textsc{gadget-3} \citep{Viel_etal2010},
which take into account the effect of massive free-streaming neutrinos
on the evolution of cosmic structures.

The pressure produced by massive neutrino free-streaming contrasts the
gravitational collapse which is the basis of cosmic structure
formation, causing a significant suppression in the average number
density of massive structures. This effect can be observed in the high
mass tail of the halo MF in Figure~\ref{fig:MF}, as
compared with the analytic predictions of \citet{sheth2002} (ST),
where the variance in the density fluctuation field, $\sigma(M)$, has
been computed via \textsc{camb} \citep{CAMB}, using the same
cosmological parameters of the simulations. In particular, here the MF
of sub-structures is shown, identified using the \textsc{subfind}
package \citep{Springel_etal2001}, while the normalization of the
matter power spectrum is fixed by the dimensionless amplitude of the
primordial curvature perturbations $\Delta^2_{\cal
  R}(k_0)|_{\mathrm{fid}}=2.3\times 10^{-9}$, evaluated at a pivot
scale $k_0=0.002/\mathrm{Mpc}$ \citep{arXiv:1001.4635}, which has been
chosen to have the same value both in the $\Lambda$CDM$\nu$ and in the
$\Lambda$CDM cosmologies.

In Figure~\ref{fig:MF}, two fiducial neutrino masses have been
considered, $\Sigma=0.3$ and $\Sigma=0.6\mathrm{\ eV}$. From the
comparison of the corresponding MFs, we confirm the theoretical
predictions, i.e., that the higher the neutrino mass is, the larger
the suppression in the comoving number density of DM haloes becomes.
These results have been overall confirmed by recent
  numerical investigations \citep{2014JCAP...03..011V,
    2014JCAP...02..049C, 2013JCAP...12..012C}. Moreover, it was shown
  that an even better agreement with numerical simulations can be
  obtained by using the linear CDM power spectrum, instead of the
  total matter one \citep[see also][]{2012PhRvD..85f3521I}.

Massive neutrinos also strongly affect the spatial
clustering of cosmic structures. A standard statistic generally used
to quantify the degree of clustering of a population of sources is the
two-point auto-correlation function. Although the free-streaming of
massive neutrinos causes a suppression of the matter power spectrum on
scales $k$ larger than the neutrino free-streaming scale, the halo
bias is significantly enhanced. This effect can be physically
explained thinking that, due to neutrino structure suppression, the
same halo bias would correspond, in a $\Lambda$CDM cosmology, to more
massive haloes (than in a $\Lambda$CDM$\nu$ cosmology), which as known
are typically more clustered.

This effect is evident in Figure~\ref{fig:xireal} which shows the
two-point DM halo correlation function measured with the Landy and
Szalay~\citet{Landy-Szalay:1993} estimator, compared to the matter
correlation function. In particular, the clustering difference between
the $\Lambda$CDM and $\Lambda$CDM$\nu$ cosmologies increases at higher
redshifts, as it can be observed from
Figures~\ref{fig:xireal}, \ref{fig:bias} and
  \ref{fig:mean_bias}. Note also the effect of nonlinearities on the
bias, which clearly starts to become scale-dependent for separations
$r<20\mathrm{\ Mpc}/h$ \citep[see
    also][]{2014JCAP...03..011V, 2014JCAP...02..049C,
    2013JCAP...12..012C}.

There are indications from 3D weak lensing in the CFHTLenS
  survey \citep{2014arXiv1401.6842K} that the matter power suppressed
  is suppressed with respect to the $\Lambda$CDM expectation in the
  wavenumber range 1-10 $h$ Mpc$^{-1}$, which may be a hint of either
  massive neutrinos, or feedback from AGN, or both.  Euclid will be
  able to probe this regime with much greater precision, and
  potentially disentangle the two effects.

RSD are also strongly affected by massive
neutrinos. Figure~\ref{fig:xizspace} shows the real and redshift space
correlation functions of DM haloes as a function of neutrino mass. The
effect of massive neutrinos is particularly evident when the
correlation function is measured as a function of the two directions
perpendicular and parallel to the line of sight. The value of the
linear growth rate that can be derived by modelling galaxy clustering
anisotropies can be greatly suppressed with respect to the value
expected in a $\Lambda$CDM cosmology. Indeed, neglecting the relic
massive neutrino background in data analysis might induce a bias in
the inferred growth rate, from which a potentially fake signature of
modified gravity might be inferred. Figure~\ref{fig:beta} demonstrates
this point, showing the best-fit values of $\beta$ and $\sigma_{12}$,
as a function of $\Sigma$ and redshift, where $\beta =
{\frac{f(\Omega_{\mathrm{M}})} {b_\mathrm{eff}}}$, $b_{\mathrm{eff}}$
being the halo effective linear bias factor, $f(\Omega_{\mathrm{M}})$
the linear growth rate and $\sigma_{12}$ the pairwise velocity
dispersion.  \label{symbol:halobias}

\epubtkImage{MF_Sub.png}{%
\begin{figure}[htbp]
\centerline{\includegraphics[width=0.78\textwidth]{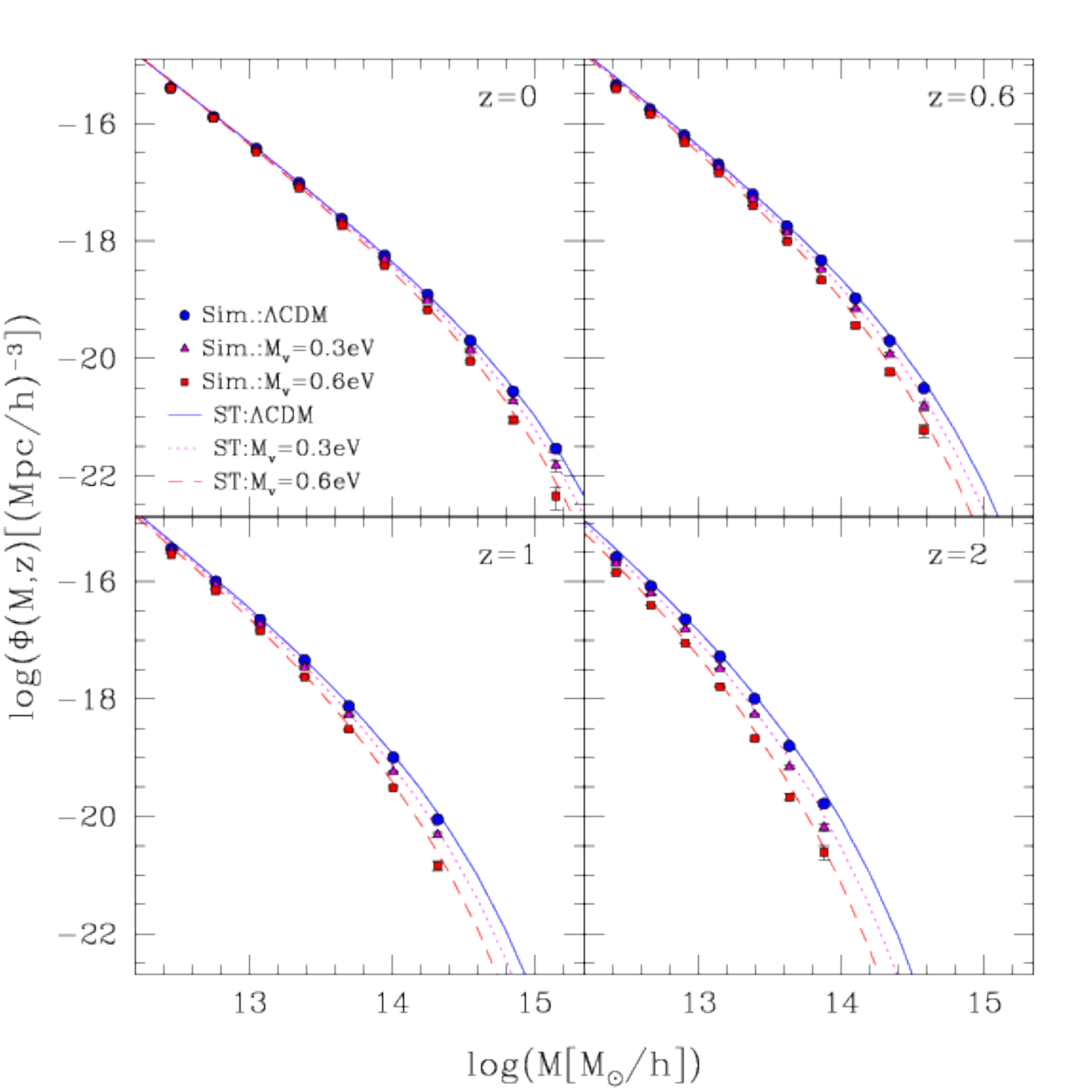}}
\caption{DM halo mass function (MF) as a function of $\Sigma$ and
  redshift. MF of the SUBFIND haloes in the $\Lambda$CDM $N$-body
  simulation (blue circles) and in the two simulations with
  $\Sigma=0.3\mathrm{\ eV}$ (magenta triangles) and
  $\Sigma=0.6\,\mathrm{\ eV}$ (red squares). The blue, magenta and red
  lines show the halo MF predicted by \citet{sheth2002}, where the
  variance in the density fluctuation field, $\sigma(M)$, at the three
  cosmologies, $\Sigma=0,0.3,0.6\mathrm{\ eV}$, has been computed with
  the software \textsc{camb} \citep{CAMB}.}
\label{fig:MF}
\end{figure}}

\epubtkImage{xi_RealSpace_Sub_Mnu06.png}{%
\begin{figure}[htbp]
\centerline{\includegraphics[width=0.78\textwidth]{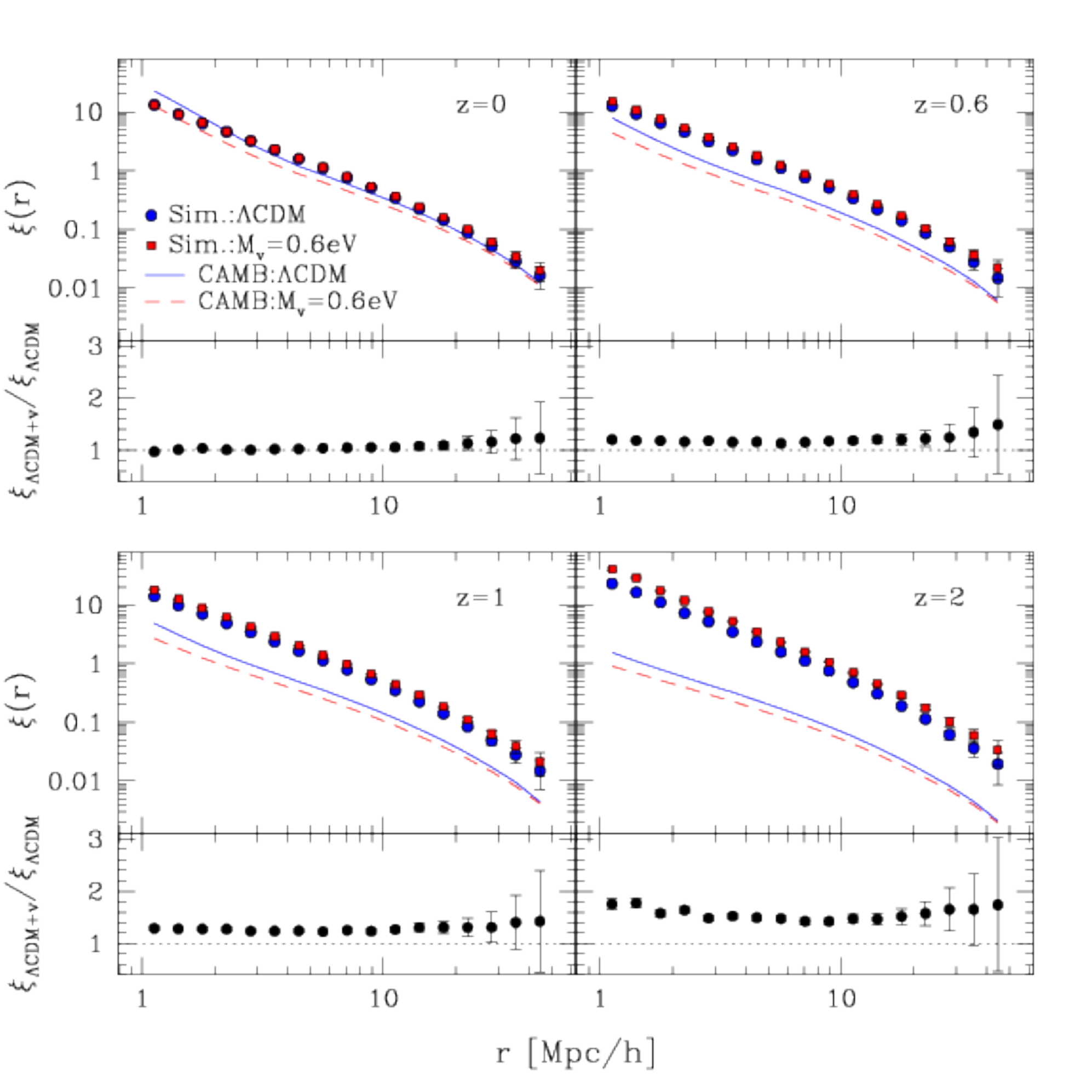}}
\caption{Real space two-point auto-correlation function of the DM
  haloes in the $\Lambda$CDM $N$-body simulation (blue circles) and in
  the simulation with $\Sigma=0.6\mathrm{\ eV}$ (red squares). The
  blue and red lines show the DM correlation function computed using
  the \textsc{camb} matter power spectrum with $\Sigma=0$ and
  $\Sigma=0.6\mathrm{\ eV}$, respectively. The bottom panels show the
  ratio between the halo correlation function extracted from the
  simulations with and without massive neutrinos.}
\label{fig:xireal}
\end{figure}}

\epubtkImage{bias_Sub.png}{%
\begin{figure}[htbp]
\centerline{\includegraphics[width=0.78\textwidth]{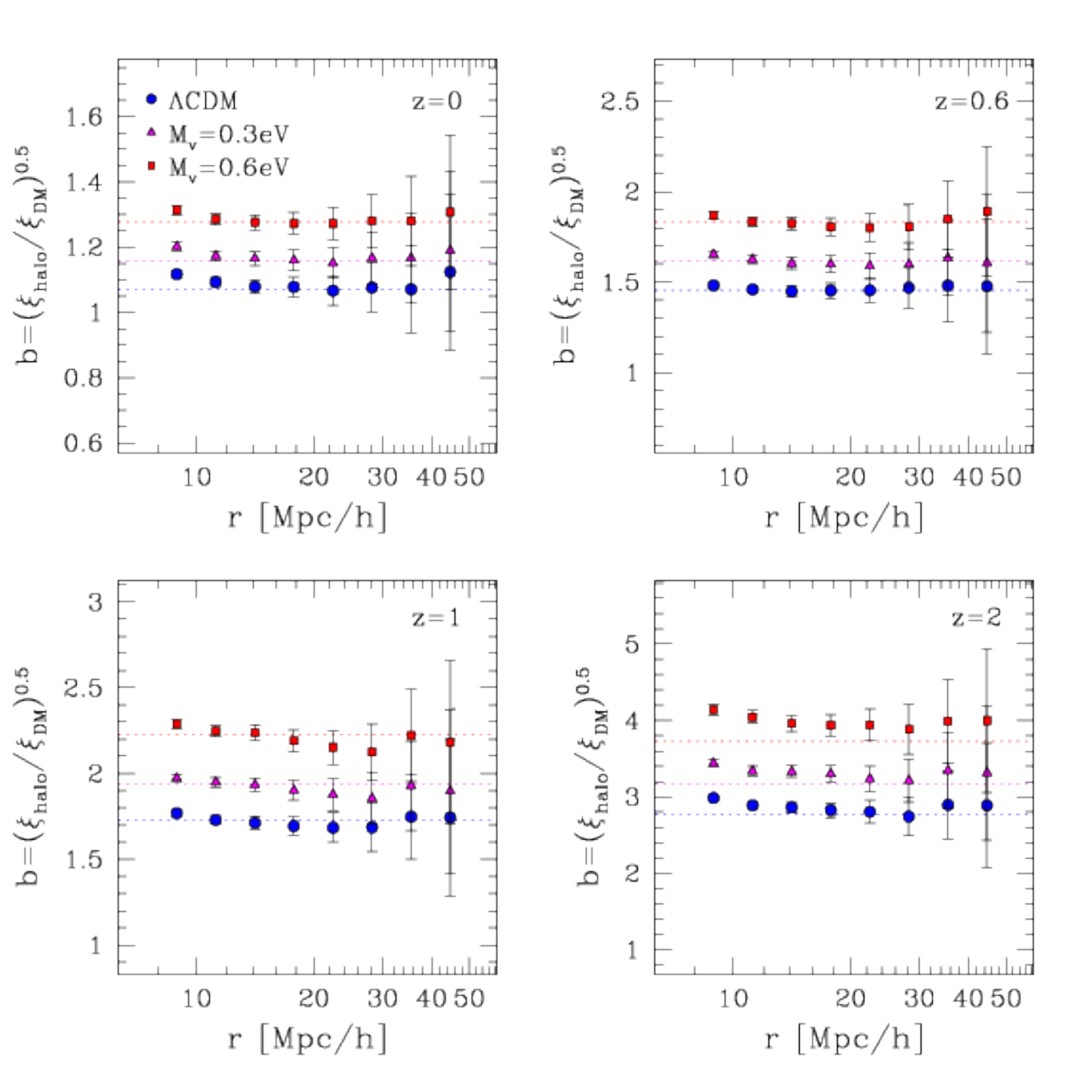}}
\caption{Bias of the DM haloes in the $\Lambda$CDM
    $N$-body simulation (blue circles) and in the two simulations with
    $\Sigma=0.3\mathrm{\ eV}$ (magenta triangles) and
    $\Sigma=0.6\,\mathrm{\ eV}$ (red squares). Dotted lines are the
    theoretical predictions of \citet{sheth_mo_tormen2001}.}
\label{fig:bias}
\end{figure}}

\epubtkImage{bias_Sub_z.png}{%
\begin{figure}[htbp]
\centerline{\includegraphics[width=0.78\textwidth]{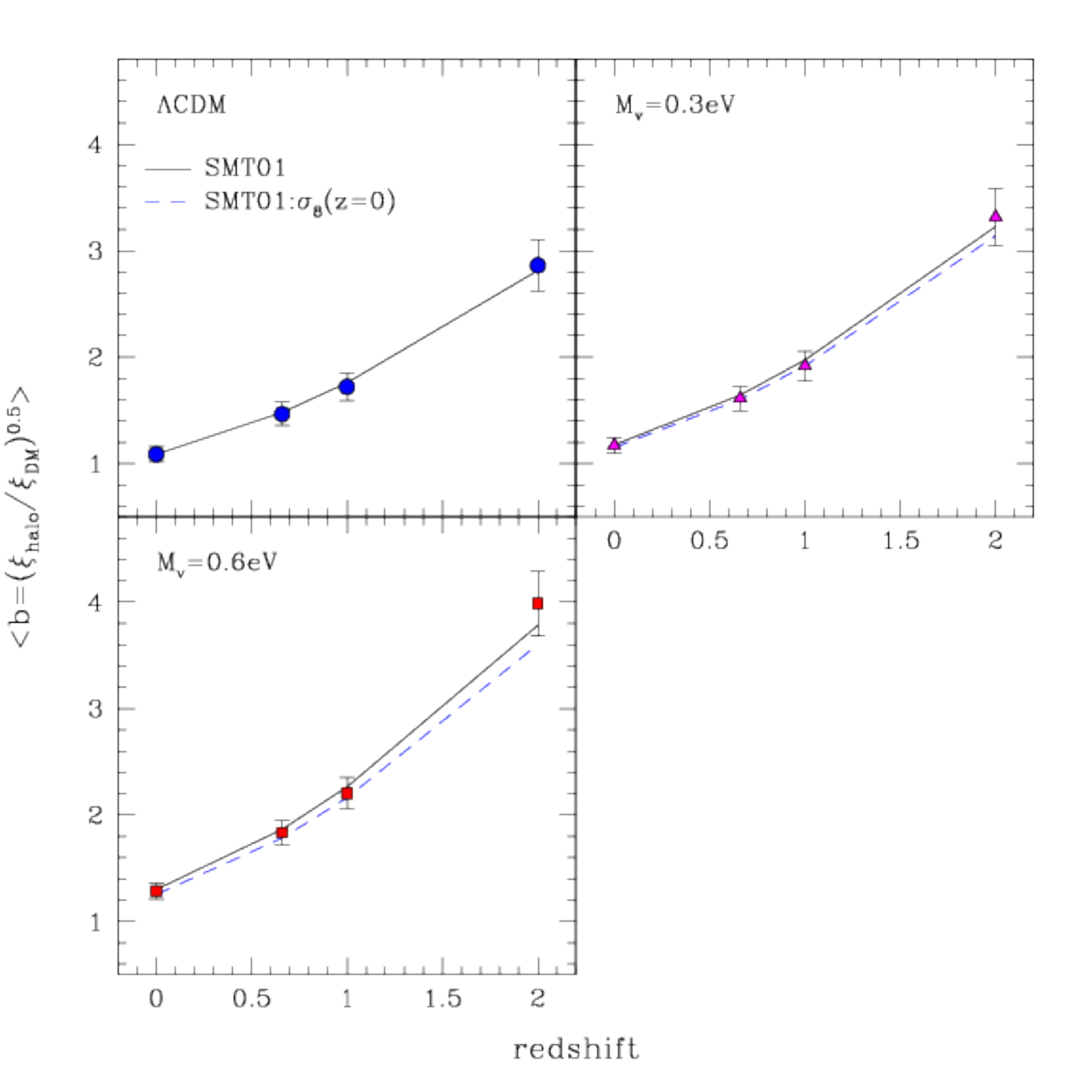}}
\caption{Mean bias (averaged in $10<r\,[\mathrm{Mpc}/h]<50$) as a
  function of redshift compared with the theoretical predictions of
  \citet{sheth2002}. Here the dashed lines represent the theoretical
  expectations for a $\Lambda$CDM cosmology renormalized with the $\sigma_8$
  value of the simulations with a massive neutrino component.}
\label{fig:mean_bias}
\end{figure}}

\epubtkImage{xi_zSpace_Sub_Mnu06.png}{%
\begin{figure}[htbp]
\centerline{\includegraphics[width=0.88\textwidth]{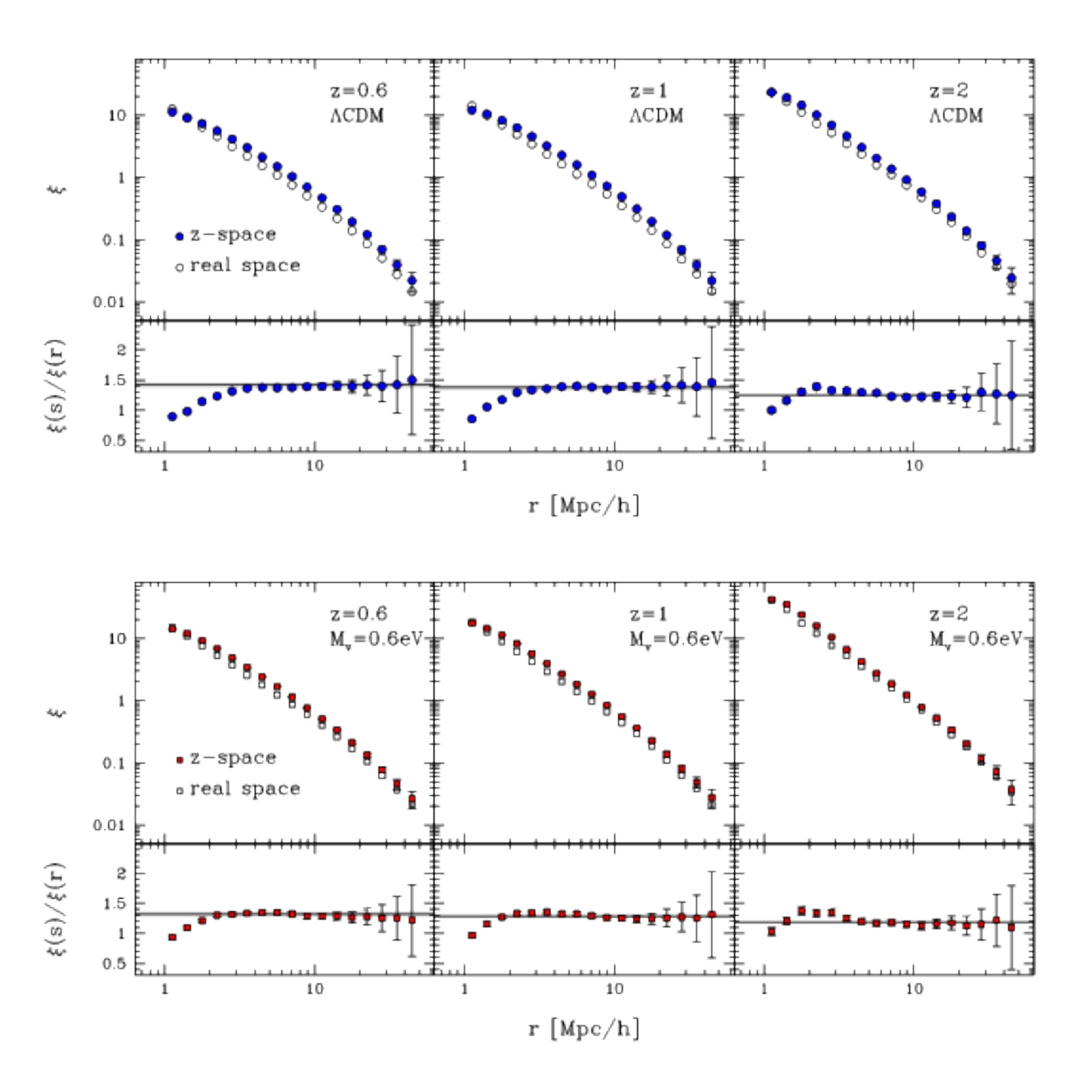}}
\caption{Two-point auto-correlation function in real and redshift
  space of the DM haloes in the $\Lambda$CDM $N$-body simulation (blue
  circles) and in the simulation with $\Sigma=0.6\mathrm{\ eV}$ (red
  squares). The bottom panels show the ratio between them, compared
  with the theoretical expectation.}
\label{fig:xizspace}
\end{figure}}

\epubtkImage{beta_sigma12_Sub.png}{%
\begin{figure}[htbp]
\centerline{\includegraphics[width=0.78\textwidth]{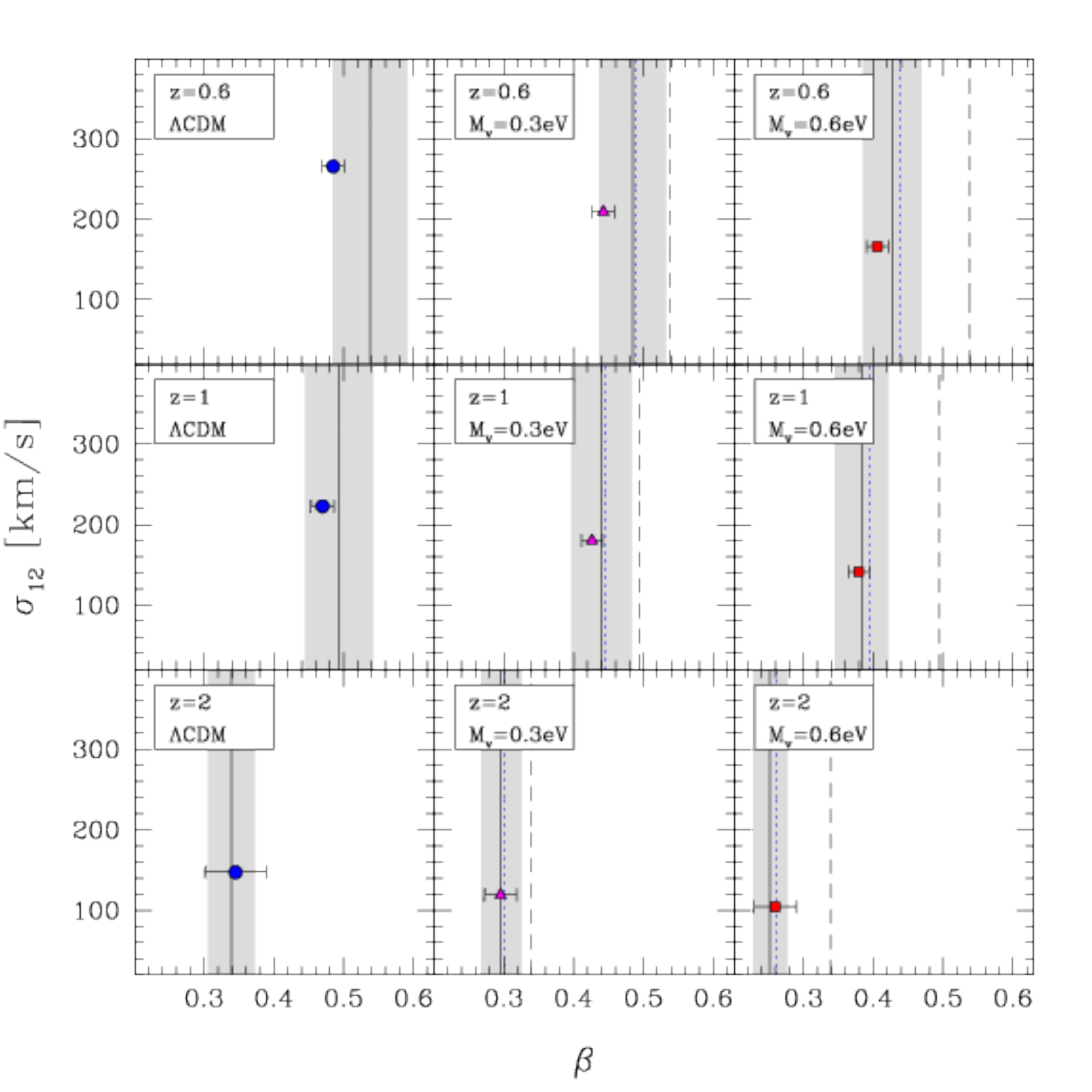}}
\caption{Best-fit values of $\beta$-$\sigma_{12}$, as a function of
  $\Sigma$ and redshift (points), compared with the theoretical
  prediction (grey shaded area). The blue dotted lines
    show the theoretical prediction for a $\Lambda$CDM cosmology
    normalised to the $\sigma_8$ value of the simulation with a
    massive neutrino component.}
\label{fig:beta}
\end{figure}}


\section{Coupling between dark energy and neutrinos}
\label{dms:de_nu}

As we have seen in Section~\ref{mg:cde}, it is interesting to consider the possibility that dark energy, seen as a dynamical scalar field (quintessence), may interact with other components in the universe. In this section we focus on the possibility that a coupling may exist between dark energy and neutrinos.

The idea of such a coupling has been addressed and developed by several authors within MaVaNs theories first \citep{Fardon:2003eh,Peccei:2004sz,Bi:2004ns,Afshordi:2005ym,Weiner:2005ac,Das:2006ht, Takahashi:2006jt,Spitzer:2006hm,Bjaelde:2007ki,Brookfield:2005td,Brookfield:2005bz} and more recently within growing neutrino cosmologies \citep{Amendola2008b,Wetterich:2007kr,Mota:2008nj,Wintergerst:2009fh,Wintergerst:2010ui,Pettorino:2010bv,Brouzakis:2010md}. It has been shown that neutrinos can play a crucial role in cosmology, setting naturally the desired scale for dark energy. Interestingly, a coupling between neutrinos and dark energy may help solving the `why now' problem, explaining why dark energy dominates only in recent epochs. The coupling follows the description illustrated in Section~\ref{mg:cde} for a general interacting dark-energy cosmology, where now $m_\nu=m_\nu(\phi)$. 

Typically, in growing neutrino cosmologies, the function $m_\nu(\phi)$ is such that the neutrino mass grows with time from low, nearly massless values (when neutrinos are non-relativistic) up to present masses in a range in agreement with current observations (see the previous section of this review for latest bounds on neutrino masses). The key feature of growing neutrino models is that the amount of dark energy today is triggered by a cosmological event, corresponding to the transition from relativistic to non-relativistic neutrinos at redshift $z_\mathrm{NR}\sim5-10$. As long as neutrinos are relativistic, the coupling plays no role on the dynamics of the scalar field, which follows attractor solutions of the type described in Section~\ref{mg:cde}. From there on, the evolution of dark energy resembles that of a cosmological constant, plus small oscillations of the coupled dark energy-neutrino fluid. As a consequence, when a coupling between dark energy and neutrinos is active, the amount of dark energy and its equation of state today are strictly connected to the present value of the neutrino mass.

The interaction between neutrinos and dark energy is a nice and concrete example of the significant imprint that dynamical coupled dark energy can leave on observables and in particular on structure formation and on the cosmic microwave background. This is due to the fact that the coupling, playing a role only after neutrinos become non-relativistic, can reach relatively high values as compared to gravitational attraction. Typical values of $\beta$ are order $50-100$ or even more such that even the small fraction of cosmic energy density in neutrinos can have a substantial influence on the time evolution of the quintessence field. During this time the fifth force can be of order $10^2-10^4$ times stronger than gravity. The neutrino contribution to the gravitational potential influences indirectly also dark matter and structure formation, as well as CMB, via the Integrated Sachs--Wolfe effect and the nonlinear Rees--Sciama effect, which is non-negligible at the scales where neutrinos form stable lumps. Furthermore, backreaction effects can substantially modify the growth of large scale neutrino lumps, with effects which are much larger than in the dark matter case. The presence of a fifth force due to an interaction between neutrinos and dark energy can lead to remarkably peculiar differences with respect to a cosmological constant scenario. 

Here, we just recall some of the typical features that can arise when such an interaction is active:
\begin{itemize}
\item existence of very large structures, order $10-500\mathrm{\ Mpc}$ \citep{Afshordi:2005ym,Mota:2008nj,Wintergerst:2009fh,Wintergerst:2010ui,Pettorino:2010bv};
\item enhanced ISW effect, drastically reduced when taking into account nonlinearities \citep{Pettorino:2010bv}: information on the gravitational potential is a good mean to constrain the range of allowed values for the coupling $\beta$;
\item large-scale anisotropies and enhanced peculiar velocities \citep{Watkins:2008hf,Ayaita:2009qz};
\item the influence of the gravitational potential induced by the neutrino inhomogeneities can affect BAO in the dark-matter spectra \citep{Brouzakis:2010md}.
\end{itemize}

Investigation of structure formation at large scales (order
$1-100\mathrm{\ Mpc}$) as well as cross correlation with CMB are
crucial in order to disentangle coupled neutrino-quintessence
cosmologies from a cosmological constant scenario. Detection of a
population of very large-scale structures could pose serious
difficulties to the standard framework and open the way to the
existence of a new cosmological interaction stronger than gravity.



\section{Unified Dark Matter}

The appearance of two unknown components in the standard cosmological model, dark matter and dark energy, has prompted discussion of whether they are two facets of a single underlying dark component. This concept goes under the name of quartessence \citep{Makler2003PhLB}, or unified dark matter (UDM). \emph{A priori} this is attractive, replacing two unknown components with one, and in principle it might explain the `why now?' problem of why the energy densities of the two components are similar (also referred to as the coincidence problem). Many UDM models are characterized by a sound speed, whose value and evolution imprints oscillatory features on the matter power spectrum, which may be detectable through weak lensing or BAO signatures with Euclid.

The field is rich in UDM models \citep[see][for a review and for references
to the literature]{2010AdAst2010E..78B}. The models can grow structure, as
well as providing acceleration of the universe at late times. In many cases,
these models have a non-canonical kinetic term in the Lagrangian, e.g., an
arbitrary function of the square of the time derivative of the field in a
homogeneous and isotropic background. Early models with acceleration driven
by kinetic energy
\citep[$k$-inflation][]{1999PhLB..458..209A,1999PhLB..458..219G,2009PhRvD..80j3508B}
were generalized to more general Lagrangians \citep[$k$-essence;
e.g.,][]{Armendariz-PiconMukhanovSteinhardt2000,2001PhRvD..63j3510A,2004PhRvL..93a1301S}.
For UDM, several models have been investigated, such as the generalized
Chaplygin gas
\citep{2001PhLB..511..265K,2002PhRvD..66d3507B,2002PhLB..535...17B,2006JCAP...01..003Z,2010PhLB..686..211P},
although these may be tightly constrained due to the finite sound speed
\citep[e.g.][]{2003JCAP...07..005A,2003GReGr..35.2063B,2004PhRvD..69l3524S,2004AA...423..421Z}.
Vanishing sound speed models however evade these constraints \citep[e.g.,
the silent Chaplygin gas of][]{2005JCAP...11..009A}. Other models consider a
single fluid with a two-parameter equation of state
\citep[e.g][]{2007PhRvD..76j3519B}), models with canonical Lagrangians but a
complex scalar field \citep{2006PhRvD..74d3516A}, models with a kinetic term
in the energy-momentum tensor
\citep{2010PhRvD..81d3520G,2008PhLB..666..205C}, models based on a DBI
action \citep{2010GReGr..42.1189C}, models which violate the weak
equivalence principle \citep{2007PhRvD..75l3007F} and models with viscosity
\citep{2011AdAst2011E...4D}. Finally, there are some models which try to
unify inflation as well as dark matter and dark energy
\citep{2006PhLB..632..597C,2008arXiv0801.4843N,2008PhRvD..77l1301L,2009arXiv0906.5021L,2009PhRvD..79j3522H}.

A requirement for UDM models to be viable is that they must be able to cluster to allow structure to form. A generic feature of the UDM models is an effective sound speed, which may become significantly non-zero during the evolution of the universe, and the resulting Jeans length may then be large enough to inhibit structure formation. The appearance of this sound speed leads to observable consequences in the CMB as well, and generally speaking the speed needs to be small enough to allow structure formation and for agreement with CMB measurements. In the limit of zero sound speed, the standard cosmological model is recovered in many models. Generally the models require fine-tuning, although some models have a fast transition between a dark matter only behavior and $\Lambda$CDM. Such models \citep{2010JCAP...01..014P} can have acceptable Jeans lengths even if the sound speed is not negligible.

\subsection{Theoretical background}

An action which is applicable for most UDM models, with a single scalar field $\varphi$, is
\begin{equation}
S=\int \mathrm{d}^4x\sqrt{-g}\left[\frac{R}{2}+{\cal L}(\varphi,X)\right],
\end{equation}
where 
\begin{equation}
X\equiv-\frac{1}{2}\nabla_\mu\varphi \nabla^\mu \varphi
\end{equation}
and $\nabla$ indicates covariant differentiation. This leads to an energy density which is $\rho=2X\,\partial p/\partial X-p$, and hence an equation-of-state parameter $w \equiv p/\rho$ (in units of $c=1$) given by
\begin{equation}
w=\frac{p}{2X\,\partial p/\partial X-p} \,,
\end{equation}
and $p={\cal L}$. A full description of the models investigated and Lagrangians considered is beyond the scope of this work; the reader is directed to the review by \citet{2010AdAst2010E..78B} for more details. Lagrangians of the form
\begin{equation}
{\cal L}(\varphi,X) = f(\varphi)g(X)-V(\varphi) \,, 
\label{UDML}
\end{equation}
where $g(X)$ is a Born--Infeld kinetic term, were considered in a Euclid-like context by \citet{Camera:2010wm}, and models of this form can avoid a strong ISW effect which is often a problem for UDM models \citep[see][and references therein]{2008JCAP...10..023B}. This model is parameterized by a late-time sound speed, $c_\infty$, and its influence on the matter power spectrum is illustrated in Figure~\ref{UDMPk}. For zero sound speed $\Lambda$CDM is recovered.

\epubtkImage{P_k-UDM.png}{%
\begin{figure}[htbp]
\centerline{
\includegraphics[width=10cm]{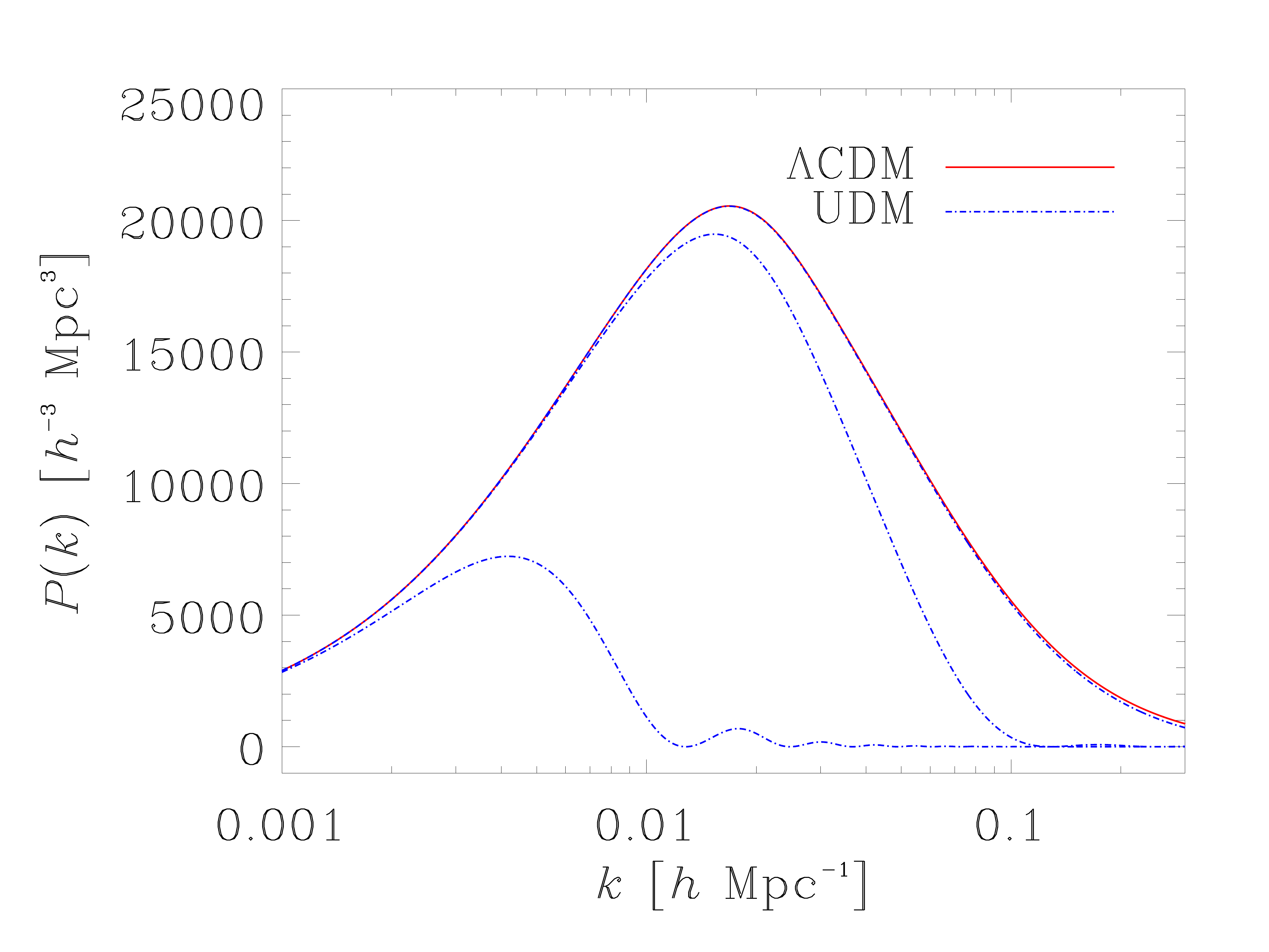}}
\caption{The $z=0$ matter power spectrum arising in UDM models with a
  Lagrangian given by Eq.~(\ref{UDML}). $\Lambda$CDM is solid, and UDM
  models with $c_\infty=10^{-1}, 10^{-2}, 10^{-3}$ are shown from
  bottom to top. Image reproduced by permission from~\cite{Camera:2010wm}.}
\label{UDMPk}
\end{figure}}

\subsection{Euclid observables}

Of interest for Euclid are the weak lensing and BAO signatures of these models, although the supernova Hubble diagram can also be used \citep{2009MNRAS.397.1935T}. The observable effects come from the power spectrum and the evolution of the equation-of-state parameter of the unified fluid, which affects distance measurements. The observational constraints of the generalized Chaplygin gas have been investigated \citep{2010PhRvD..81f3532P}, with the model already constrained to be close to $\Lambda$CDM with SDSS data and the CMB. The effect on BAO measurements for Euclid has been calculated by \citep{Camera:2012sf}, whereas the weak lensing effect has been considered for non-canonical UDM models \citep{2010arXiv1002.4740C}. The change in shape and oscillatory features introduced in the power spectrum allow the sound speed parameter to be constrained very well by Euclid, using 3D weak lensing \citep{2003MNRAS.343.1327H,2007MNRAS.376..771K} with errors $\sim 10^{-5}$ \citep[see also][]{Camera:2009uz,Camera:2012sf}.


\section{Dark energy and dark matter}
\label{dms:de_dm}

In Section~\ref{models-of-modified-gravity}, we have illustrated the
possibility that dark energy, seen as a dynamical scalar field
(quintessence), may interact with other components in the
universe. When starting from an action such as
Eq.~(\ref{mg:cde:action}), the species which interact with
quintessence are characterized by a mass function that changes in time
\citep{Kodama:1985bj,Amendola:1999er,Amendola:2003wa,Pettorino:2008ez}. Here,
we consider the case in which the evolution of cold dark matter (CDM)
particles depends on the evolution of the dark-energy scalar field. In
this case the general framework seen in
Section~\ref{models-of-modified-gravity} is specified by the choice of
the function $m_c=m_c(\phi)$. The coupling is not constrained by tests
of the equivalence principle and solar system constraints, and can
therefore be stronger than the coupling with baryons. Typical values
of $\beta$ presently allowed by observations (within current CMB data)
are within the range $0< \beta < 0.06$ at 95\% CL for a constant
coupling and an exponential potential,
\citep{Bean:2008ac,amendola_etal_2003,Amendola:2003wa,amendola_quercellini_2003},
or possibly more if neutrinos are taken into account or more realistic
time-dependent choices of the coupling are used
\citep{LaVacca:2009yp,Kristiansen:2009yx}. As mentioned in
Section~\ref{mg:cde}, this framework is generally referred to as
`coupled quintessence' (CQ). Various choices of couplings have been
investigated in the literature, including constant $\beta$
\citep{Amendola:1999er,Mangano:2002gg,Amendola:2003wa,Koivisto:2005nr,Guo:2007zk,
  Quartin:2008px,quercellini_etal_2008,Pettorino:2008ez} and varying
couplings \citep{Baldi:2010vv}, with effects on Supernov\ae, CMB and
cross-correlation of the CMB and LSS
\citep{Bean:2008ac,amendola_etal_2003,Amendola:2003wa,amendola_quercellini_2003,
  LaVacca:2009yp,Kristiansen:2009yx,Mainini:2010ng}.

The presence of a coupling (and therefore, of a fifth force acting among dark
matter particles) modifies the expansion of the universe, linear perturbations
and most relevantly, structure formation. Coupled quintessence is a concrete
model in which a non-negligible amount of dark energy is present at early times.
The presence of such an early dark-energy component is accompanied specific
features, as illustrated in Section~\ref{models-of-modified-gravity} for a
general framework:
\begin{enumerate}
\item a fifth force $\mathbf{\nabla} \left[\Phi_\alpha + \beta \phi \right]$ with
an effective $\tilde{G}_{\alpha} = G_{N}[1+2\beta^2(\phi)]$;
\item a velocity-dependent term $\tilde{H}\mathbf{v}_{\alpha} \equiv H \left(1 -
{\beta(\phi)} \frac{\dot{\phi}}{H}\right) \mathbf{v}_{\alpha}$;
\item a time-dependent mass for each particle $\alpha$, evolving according to
Eq.~(\ref{mass_def}).
\end{enumerate} 
All these effects, and in particular the first two, contribute
significantly to structure formation. Note that the second and third
terms are not independent of each other as they are a direct
consequence of momentum conservation. Depending on the function
$m_c(\phi)$, and therefore $\beta(\phi)$, the first two terms can
partially balance: the fifth force increases gravitational attraction
whilst the velocity-dependent term, if the CDM mass decreases with
time, tries to dilute the concentration of the virialized haloes. In
particular, a striking difference between constant and
variable-coupling models concerning the interplay of all these three
effects has been highlighted in \citet{Baldi:2010vv}: whilst for
constant couplings only the latter two effects can alter the virial
equilibrium of an already-collapsed object, for the case of a variable
coupling the time evolution of the effective gravitational constant
can also modify the virial status of a halo, and can either enhance or
counteract the effect of reducing halo concentrations (for decreasing
and increasing couplings, respectively). Nonlinear evolution within
coupled quintessence cosmologies has been addressed using various
methods of investigation, such as spherical collapse
\citep{Mainini:2006zj,Wintergerst:2010ui,Manera:2005ct,Koivisto:2005nr,
  Sutter:2007ky,Abdalla:2007rd,Bertolami:2007tq} and alternative
semi-analytic methods
\citep{Saracco_etal_2010,amendola_quercellini_2004}. $N$-body and
hydro-simulations have also been done
\citep{maccio_etal_2004,Baldi_etal_2010,Baldi:2010vv,Baldi_Pettorino_2010,
    Baldi:2010ks,Li:2010zw,Li:2010eu,Baldi:2010pq,Zhao:2010dz,Marulli:2011jk,Giocoli_etal_2013,Moresco_etal_2014}.

We list here briefly the main observable features typical of this class of
models:
\begin{itemize}
\item enhanced ISW effect
\citep{Amendola:1999er,Amendola:2003wa,Mainini:2010ng}; such effects may be
partially reduced when taking into account nonlinearities, as described in
\citet{Pettorino:2010bv};
\item increase in the number counts of massive clusters at high redshift
\citep{Baldi_Pettorino_2010};
\item scale-dependent bias between baryons and dark matter, which behave
differently if only dark matter is coupled to dark energy
\citep{Baldi_etal_2010,Baldi:2010pq};
\item less steep inner core halo profiles (depending on the interplay between
fifth force and velocity-dependent terms)
\citep{Baldi_etal_2010,Baldi:2010vv,Li:2010zw,Li:2010eu,Baldi:2010pq};
\item lower concentration of the halos
\citep{Baldi_etal_2010,Baldi:2010vv,Li:2010eu};
\item voids are emptier when a coupling is active \citep{Baldi:2010ks}.
\end{itemize}

As discussed in subsection \ref{Nbody_sims}, when a variable coupling $\beta (\phi )$ 
is active the relative balance of the fifth-force and other dynamical effects depends on
 the specific time evolution of the coupling strength. Under such conditions, certain cases
 may also lead to the opposite effect of larger halo inner overdensities and higher concentrations, 
as in the case of a steeply growing coupling function \citep[see][]{Baldi:2010vv}. Alternatively, 
the coupling can be introduced by choosing directly a covariant stress-energy tensor, treating dark 
energy as a fluid in the absence of a starting action \citep{Mangano:2002gg,Valiviita:2008iv,CalderaCabral:2008bx,Schaefer:2008ku,Valiviita:2009nu,
Majerotto:2009np,Gavela:2009cy,CalderaCabral:2009ja,Gavela:2010tm}. For an illustration of nonlinear 
effects in the presence of a coupling see Section~\ref{nonlinear-aspects}.


\section{Ultra-light scalar fields}

Ultra-light scalar fields arise generically in high energy physics, most commonly as axions or other axion-like particles (ALPs). They are the Pseudo-Goldstone bosons (PGBs) of spontaneously broken symmetries. Their mass remains protected to all loop orders by a shift symmetry, which is only weakly broken to give the fields a mass and potential, through non perturbative effects. Commonly these effects are presumed to be caused by instantons, as in the case of the QCD axion, but the potential can also be generated in other ways that give potentials that are useful, for example, in the study of quintessence \citep{panda2010}. Here we will be considering a general scenario, motivated by the suggestions of \citet{axiverse2009} and \citet{hu2000}, where an ultralight scalar field constitutes some fraction of the dark matter, and we make no detailed assumptions about its origin.

\subsection{Phenomenology and Motivation}

If the DM is light then either thermal free-streaming (as for massive neutrinos or WDM) or non-thermal quantum pressure (as is the case for ultra-light scalars) can lead to potentially observable effects in the large scale structure. The mass range of interest for ultra-light scalars is:
\begin{equation}
10^{-33}\mathrm{\ eV} \lesssim m \lesssim 10^{-18}\mathrm{\ eV}\, ,
\label{eqn:alp_mass_range}
\end{equation} 
where the lower end corresponds to the crossover scale to DE, and the upper end is empirically determined as where the ALPs become indistinguishable from pressureless CDM. DM in this mass range has been dubbed ``fuzzy cold dark matter'', or FCDM \citep{hu2000}. 

There may be a small model-dependent thermal population of ALPs, but the majority of the cosmological population will be cold and non-thermally produced. Production of cosmological ALPs proceeds by the vacuum realignment mechanism. When the Peccei--Quinn-like $U(1)$ symmetry is spontaneously broken at the scale $f_a$ \label{symbol:f_a} the ALP acquires a vacuum expectation value, the misalignment angle $\theta_i$, uncorrelated across different causal horizons. However, provided that inflation occurs after symmetry breaking, $f_a>H_I/2pi$, then the field is homogenized over our entire causal volume. This is the scenario we will consider, since large $f_a\gtrsim 10^{16}\mathrm{\ GeV}$ is required for ultra-light scalars to make up any significant fraction of the DM \footnote{In this scenario, isocurvature perturbations in the ALP energy density are produced, and can provide additional constraints.}.

Due to high occupation numbers ALPs can be modelled as a classical scalar field \citep[see][and references therein]{sikivie2009}, $\phi$, rolling in a potential, $V(\phi)$. For simplicity we will take $V(\phi)=m^2\phi^2/2$, and we discuss other possibilities in Section~\ref{sec:alp_models}. The field evolves according to the Klein-Gordon (KG) equation, which for the homogeneous component, $\phi_0(t)$, is:
\begin{equation}
\ddot{\phi_0}+3 H \dot{\phi_0}+m^2\phi_0 = 0 \, ,
\end{equation}
where overdots denote derivatives with respect to physical time. The energy density in axions is given by $\rho_a=\frac{1}{2}(\dot{\phi}^2+m^2\phi^2)$. When $H\gtrsim m$ the axion field is frozen at its initial displacement, the energy density remains constant, $w=-1$, and the density perturbations do not cluster. When $H\lesssim m$ the field begins to oscillate about the potential minimum, and the time averaged energy density on time scales $\gg 1/m$ has equation of state $w=0$. The horizon size when oscillations began imprints a characteristic scale on axion density perturbations. In addition, a WKB approximation leads to a non-zero time averaged sound speed (in e.g. the synchronous or Newtonian gauge) for the density perturbations, even in the $w=0$ phase \citep[e.g.]{hu2000,}:
\begin{equation}
c_s^2\equiv \frac{\delta P}{\delta \rho}=\frac{k^2/4m^2a^2}{1+k^2/4m^2a^2} \, .
\end{equation}
This sound speed leads to a Jeans scale for axion density perturbations, suppressing structure formation. The Jeans scale is given by
\begin{equation}
k_J\approx a^{1/4}\left(\frac{\Omega_mh^2}{0.12}\right)^{1/4}\left(\frac{0.7}{h}\right)\left(\frac{\rho}{\bar{\rho}}\right)^{1/4}\left(\frac{m}{10^{-22}\mathrm{\ eV}}\right)^{1/2}\times 10^2\, h\mathrm{Mpc}^{-1}\, .
\end{equation}
The Jeans scale can also be equated to the de-Broglie wavelength, e.g. in a halo with virial velocity $v$, $r_J \sim \lambda_{\mathrm dB} \sim 1/mv$ \cite{hu2000}.

In the range $m\sim 10^{-22}\mathrm{\ eV}$, due to the existence of the Jeans scale, FCDM may play a role in the resolution of small-scale problems of CDM. Ref.~\cite{Marsh:2013ywa} showed that in this regard it may out-perform WDM and avoid the so-called ``Catch 22'' of WDM whereby one cannot form dwarf galaxies and simultaneously give large core radii to their density profiles. In the mass range $m\sim 10^{-32}\mathrm{\ eV}$ ALPs/FCDM have affects on structure not dissimilar from massive neutrinos \citep{2006PhLB..642..192A,Marsh2012}, however there is no degeneracy when the CMB is used in conjunction with large scale structure, due to the different effects on the expansion rate during the radiation era.

The large phase space density of ultralight scalar fields may cause them to form Bose--Einstein condensates \citep[see][and references therein]{sikivie2009}. This could lead to many interesting, and potentially observable phenomena, such as formation of vortices in the condensate, which may effect halo mass profiles \citep{silverman2002,kain2010}, and black hole super radiance \citep{axiverse2009,arvanitaki2010,rosa2010}. Sikivie has argued \citep{sikivie2010b} that axion dark matter fits the observed caustics in dark matter profiles of galaxies, which cannot be explained by ordinary dust CDM.

Regardless of the specifics of the model,
\citet{tegmark2006} and others have argued that on general
statistical grounds we should expect a scenario where ALPs
make up an order one fraction of the CDM, alongside the standard WIMP
candidate of the lightest supersymmetric particle. However, it must be
noted that there are objections when we consider a population of light
fields in the context of inflation \citep{mack2009a,mack2009b}. The
problem with these objections is that they make some assumptions about
what we mean by ``fine tuning'' of fundamental physical theories,
which is also related to the problem of finding a measure on the
landscape of string theory and inflation models \citep[see,
  e.g.,][]{linde2010}, the so-called ``Goldilocks problem.''

\subsection{Particle Physics and String Theory Models}
\label{sec:alp_models}

Axions and ALPs arise generically in string
theory \citep{SvrcekWitten:2006}. They are similar to the well known
QCD axion
\citep{Peccei-Quinn:1977,Weinberg:1977ma,Wilczek:1977pj,thooft1976a,thooft1976b,dine1981,preskill1983,steinhardt1983,turner1983,abbott1983,dine1983,turner1986,visinelli2009},
and their cosmology has been extensively studied \citep[see, for
  example,][]{banks1996}.  Axions arise from the spontaneous
  breaking of a global $U(1)$ symmetry at some high scale, $f_a$,
  where $f_a$ is known as the axions decay constant. Axion masses are
  protected by a shift symmetry to all orders in perturbation theory,
  but they acquire a potential due to non-perturbative effects which
  preserves a discrete shift symmetry, making the axion field
  periodic. The canonically normalised field is $\phi=f_a\theta$ and
  the potential is:
\begin{equation}
V(\phi)=\Lambda_a^4 U(\theta) \, ,
\end{equation} 
where $U(\theta)$ can be any periodic function of $\theta$, but is commonly taken to be $U(\theta)=(1-\cos \theta)$. The energy scale of non-perturbative physics is $\Lambda_a$. For the QCD axion, QCD instantons provide the potential and we have $\Lambda_a=\Lambda_{\rm QCD}\sim 200\text{ MeV}$. Expanding the potential around the minimum at $\theta=0$, the mass is given by:
\begin{equation}
m=\frac{\Lambda^2}{f_a}\, .
\end{equation}  
Since in the expansion of the potential all self-couplings will be suppressed by the high scale $f_a$, and the specific form of $U$ is model-dependent, this justifies the assumption above to consider only the quadratic mass term as relevant in the cosmological setting. In addition non-periodic potentials have been constructed in string theory via so-called `axion monodromy'.

In a string theory setting, the scale $\Lambda_a$ is unknown, but is expected to scale as $\Lambda_a=\mu e^{-S}$, where $\mu$ is some hard scale possibly related to SUSY breaking, and the action, $S$ scales linearly with the value of some modulus field that fixes the size of a closed cycle in the compact space. The number of axions in string theory models is fixed by the topology of the compact space. Since many known Calabi-Yau manifolds have large Hodge numbers in the range $10$'s to $100$'s or more, the number of axions is expected to be large. The observation of that there are many fields, with masses both protected by a shift symmetry and exponentially sensitive to the details of moduli stabilisation leads naturally to the conclusion that the existence of ultra-light axions is a plausibly general feature of string theory. The resulting scenario has been called the `String Axiverse' \citep{axiverse2009}.

In a field theory model, due to the shift symmetry, additional (to the QCD axion) light axions can also be introduced in a natural way. The non-perturbative effects leading to the potential and a scale $\Lambda_a\neq \Lambda_{\rm QCD}$ require the introduction e.g. of a new strongly coupled sector.

\subsection{Constraints from large scale structure}

Cosmological constraints on ALPs can be obtained across the range of Eq.~\ref{eqn:alp_mass_range}. The combined CMB-large scale structure  likelihood analysis of \citet{2006PhLB..642..192A} has shown that ultra-light fields with mass around $10^{-32}\mbox{\,--\,}10^{-24}\mathrm{\ eV}$ might account for up to 10\% of the dark matter abundance. Outside of this range, the abundance can be larger \footnote{Lyman-alpha forest power spectra can push constraints further at the high mass end, although the results in \citet{2006PhLB..642..192A} were not calibrated with simulations.}. Euclid will be able to place tighter constraints across this entire range, and extend the mass range covered.

Ultra-light fields with $m\lesssim 10^{-29}\mathrm{\ eV}$ are similar in many ways to massive neutrinos \citep{2006PhLB..642..192A}, the major difference being that their non-thermal production breaks the link between the scale of suppression, $k_m$, and the fraction of dark matter, $f_{ax}$, through the dependence of $f_{ax}$ on the initial field value $\phi_i$. Therefore an accurate measurement of the matter power spectrum in the low-$k$ region where massive neutrinos corresponding to the WMAP limits on $\Omega_\nu$ are expected to suppress structure will determine whether the expected relationship between $\Omega_\nu$ and $k_m$ holds. These measurements will limit the abundance of ultra-light fields that begin oscillations in the matter-dominated era.

Another powerful test of the possible abundance of ultralight fields beginning oscillations in the matter era will be an accurate measure of the position of the turn over in the matter power spectrum, since this gives a handle on the species present at equality. Ultra-light fields with masses in the regime such that they begin oscillations in the radiation-dominated era may suppress structure at scales where the BAO are relevant, and thus distort them. Improving the accuracy of the BAO measurement would place severe limits on ultralight fields in this mass regime.

Recently, \citet{Marsh2012} showed that with current and next generation galaxy surveys alone it should be possible to unambiguously detect a fraction of dark matter in axions with $10^{-33}\mathrm{\ eV}\lesssim m\lesssim 10^{-29}\mathrm{\ eV}$ of the order of 1\% of the total. Furthermore, they demonstrated that the tightest constraints on the axion fraction $f_{ax}$ come from weak lensing; when combined with a galaxy redshift survey, constraining $f_{ax}$ to 0.1\% should be possible, see Figure~\ref{fig:axion_constraints}. 
The strength of the weak lensing constraint depends on the photometric redshift measurement, i.e., on tomography. 
Therefore, lensing tomography will allow Euclid -- through the measurement of the growth rate -- to resolve the redshift evolution of the axion suppression of small scale convergence power. Further details can be found in \citet{Marsh2012}.

\epubtkImage{fAxUncert_vs_mAx.png}{%
\begin{figure}
\centerline{\includegraphics[width=0.75\textwidth]{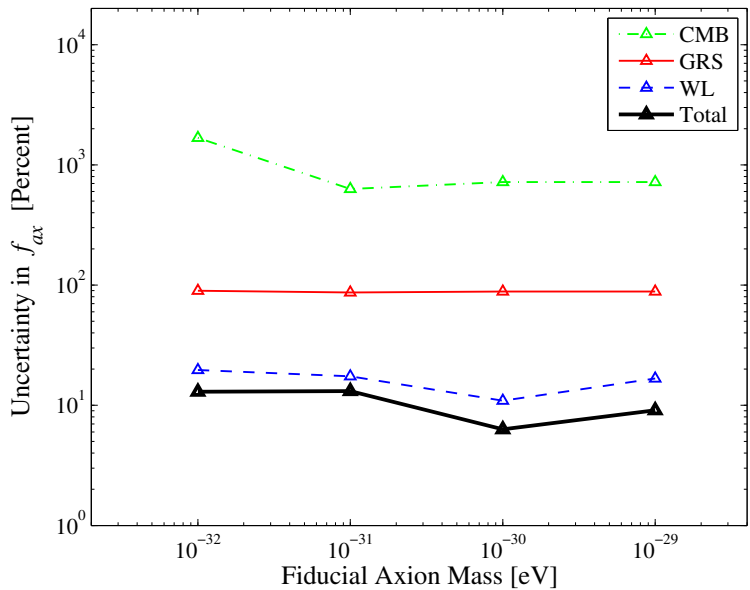}}
\caption{Marginalized uncertainty in $f_{ax}$ for CMB (green), a
  galaxy redshift survey (red), weak lensing (blue) and the total
  (black) evaluated for four different fiducial axion masses, for the
  cosmology $\Lambda$CDM+$f_{ax}$+$\nu$. Image reproduced by
  permission from~\cite{Marsh2012}, copyright by APS.}
\label{fig:axion_constraints}
\end{figure}}

At the heavier end with $m\gtrsim 10^{-24}\mathrm{\ eV}$ axions affect structure formation on scales comparable to WDM with $m_W\gtrsim 0.2\mathrm{\ keV}$ \cite{Marsh:2013ywa}. By matching the scale at which the transfer function is suppressed by factor of 2, one can make a map between axion and WDM masses. This map, adapted from Ref.~\cite{Marsh:2013ywa}, is shown in Fig.~\ref{fig:axion_wdm_map}. Such a map serves as a guide to the constraining power of Euclid on axions based on WDM forecasts and constraints. However it should be noted that the Jeans scale is dynamical for axions, leading to scale dependent growth, and different non-linear behaviour to WDM, so dedicated studies are needed. For example, constraining $m_{\rm WDM}>2\mathrm{\ keV}$ \citep{Markovic:2010te} (and Section~\ref{section_wdm} of this review) can constrain axions to $m\gtrsim 10^{-21}\mathrm{\ eV}$ as the main component of the DM. Reaching this level is significant as it approaches the exclusion limits from black hole super radiance \citep{arvanitaki2010}.
\epubtkImage{axion_wdm_map.png}{%
\begin{figure}
\centerline{\includegraphics[width=0.75\textwidth]{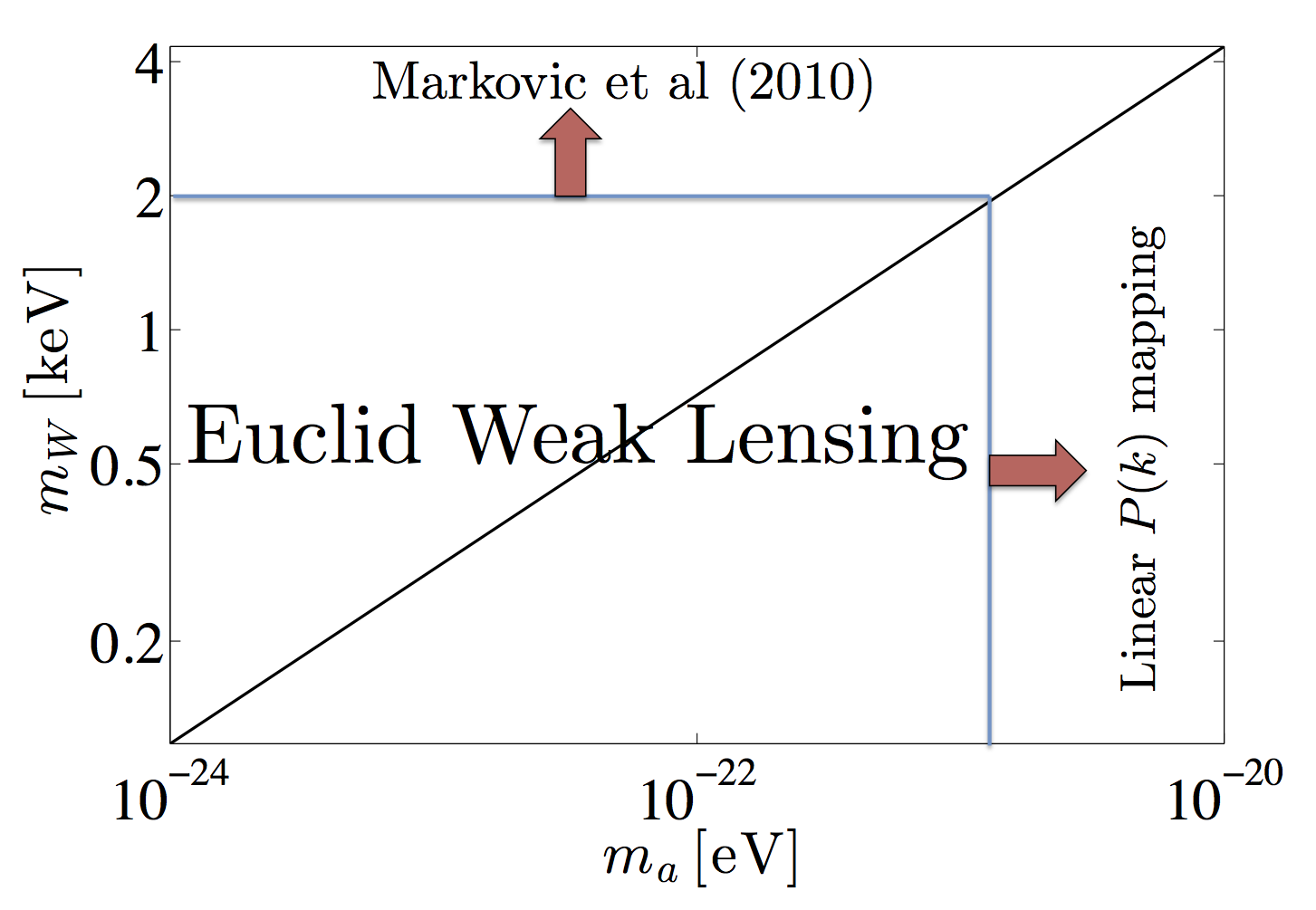}}
\caption{Mapping between axion and WDM mass that suppress power by a factor of two at the same scale. This can be used to approximately the axion mass constraint possible with Euclid on based on WDM forecasts.}
\label{fig:axion_wdm_map}
\end{figure}}

The expected suppression of structure caused by ultralight fields should be properly taken into account in $N$-body simulations. The nonlinear regime of $P(k)$ needs to be explored further both analytically and numerically for cosmologies containing ultra-light fields, especially to constrain those fields which are heavy enough such that $k_m$ occurs around the scale where nonlinearities become significant, i.e., those that begin oscillation deep inside the radiation-dominated regime and are naively degenerate with WDM. For lighter fields the effects in the nonlinear regime should be well-modelled by using the linear $P(k)$ for $N$-body input, and shifting the other variables such as $\Omega_c$ accordingly.

In conclusion, with proper numerical modelling, it should be possible for Euclid to put powerful, percent-level constraints on an ultra-light scalar field contribution to the DM over 12 to 13 orders of magnitude in mass in the range $10^{-33}\mathrm{\ eV}\lesssim m\lesssim 10^{-20}\mathrm{\ eV}$.


\section{Dark-matter surrogates in theories of modified gravity}

\subsection{Extra fields in modified gravity}

The idea that the dark universe may be a signal of modified gravity has led to the development of a plethora of theories. From polynomials in curvature invariants, preferred reference frames, UV and IR modifications and extra dimensions, all lead to significant modifications to the gravitational sector. A universal feature that seems to emerge in such theories is the existence of fields that may serve as a proxy to dark matter. This should not be unexpected. On a case by case basis, one can see that modifications to gravity generically lead to extra degrees of freedom. 

For example, polynomials in curvature invariants lead to higher-derivative theories which inevitably imply extra (often unstable) solutions that can play the role of dark matter. This can be made patently obvious when mapping such theories onto the Einstein frame with an addition scalar field (Scalar-Tensor theories). Einstein-Aether theories \citep{ZlosnikFerreiraStarkman2008} explicitly introduce an extra time-like vector field. The time-like constraint locks the background, leading to modifications to the background expansion; perturbations in the vector field can, under certain conditions, lead to growth of structure, mimicking the effect of pressureless dark matter. The vector field plays the same role in TeVeS \citep{Bekenstein2004a}, where two extra fields are introduced to modify the gravitational dynamics. And the same effects come into play in bigravity models \citep{Banadosetal2009} where two metrics are explicitly introduced -- the scalar modes of the second metric can play the role of dark matter.

In what follows we briefly focus on three of the above cases where
extra gravitational degrees of freedom play the role of dark matter:
Einstein-Aether models, TeVeS models and bigravity models. We will
look at the Einstein-Aether model more carefully and then briefly
discuss the other two cases.

\subsection{Vector dark matter in Einstein-Aether models}

As we have seen in a previous section, Einstein-Aether models introduce a time-like vector field $A^{a}$ into gravitational dynamics. The four vector $A^{a}$ can be expanded as $A^{\mu}=(1+\epsilon X,\epsilon\partial^{j}Z)= (1+\epsilon X,\frac{\epsilon}{a^{2}}\partial_{j}Z)$ \citep{ZlosnikFerreiraStarkman2008}. In Fourier space we have $A^{\mu}=(1-\epsilon \Psi,i\frac{\epsilon}{a} k_{j}V)$, where, for computational convenience, we have defined $V\equiv Z / a$ and have used the fact that the constraint fixes $X=-\Psi$.

The evolution equation for the perturbation in the vector field becomes (where primes denote derivatives with respect to conformal time)
\begin{align}
\label{vecp}
0 &= c_{1}[V''+k^{2}V+2{\cal H}V'+2{\cal H}^{2}V+\Psi'+\Phi '+2{\cal H}\Psi] \nonumber \\ &\quad +c_2[k^{2}V+6{\cal H}^{2}V-3\frac{a''}{a}V+3\Phi'+3{\cal H}\Psi] \nonumber \\ &\quad
 +c_3[k^{2}V+2{\cal H}^{2}V-\frac{a''}{a}V+\Phi'+{\cal H}\Psi] \nonumber \\ &\quad +\frac{F_{KK}}{F_{K}}[-K^{\epsilon}\alpha{\cal H}-K^{0'}(-c_{1}(V'+\Psi)+3c_{2}{\cal H} V+c_{3}{\cal H} V)].
\end{align}
The perturbation in the vector field is sourced by the two gravitational potentials $\Phi$ and $\Psi$ and will in turn source them through Einstein's equations. The Poisson equation takes the form
\begin{align}
\label{eqn:Poisson}
k^{2}\Phi&=-\frac{1}{2}F_{K}c_{1}k^{2}[V'+\Psi+(3+2\tilde{c}_{3}){\cal H}V] \nonumber \\ &\quad -4\pi Ga^{2}\sum_{a}(\bar{\rho}_{a}\delta_{a}+3(\bar{\rho}_{a}+\bar{P}_{a}){\cal H}\frac{\theta_{a}}{k^{2}}).
\end{align}

To understand why the vector field can play the role of dark matter it is instructive to study the effect of the vector field during matter domination. It should give us a sense of how in the generalized Einstein-Aether case, the growth of structure is affected. Let us consider the simplest case in which the the dominant remaining contribution to the energy density is baryonic, treated as a pressureless perfect fluid with energy-momentum tensor $\mathbf{T}$ and let us introduce the variable $V'\equiv E$. For ease of illustration we will initially consider only the case where $V$ is described by a growing monomial, i.e. $V=V_{0}(\eta/\eta_0)^p$. During the matter era we have 
\begin{align}
a^{2}\delta T^{0}_{\phantom{0}0} &\simeq -l_{E}\xi(k)k^{2}\eta^{5+p-6n} \label{too} \\ 
k^{2}(\Psi-\Phi) &\simeq -l_{S}\xi(k)k^{2}\eta^{5+p-6n} 
\end{align}
with $l_{E} \equiv -(c_{1}(2+p)n+2\alpha(1-2n)n)$, $l_{S} \equiv -(c_{1}+c_{3})n(6n-p-10)$, and
\begin{equation}
\xi(k) \sim \gamma
V_{0}(k)\eta_0^{-p}k_\mathrm{hub}^{6-6n}\left[3\alpha\Omega_{m}\left(\frac{H_{0}}{M}\right)^{2}\right]^{n-1} \,,
\label{xixi}
\end{equation}
where $k_\mathrm{hub}\equiv1/\eta_\mathrm{today}$. Hence, the vector field affects our evolution equations for the matter and metric perturbations only through its contribution to the energy density and its anisotropic stress. On large scales, $k \eta \ll1$, and assuming adiabatic initial conditions for the fields $\delta$, $\Phi$ and $\theta$, this leads to 
\begin{equation}
\delta = C_{1}(k)+\frac{6l_{S}\xi(k)}{(10+p-6n)}\eta^{5+p-6n} \,,
\end{equation}
where $C_{1}$ is a constant of integration and we have omitted the decaying mode. Therefore, even before horizon crossing, the anisotropic stress term due to the vector field can influence the time evolution of the baryon density contrast.

On small scales ($k\eta \gg1$), we find
\begin{equation}
\delta(k,\eta) = C_{2}(k)\eta^{2}
+\frac{(\frac{1}{2}l_{E}+l_{S})}{(5+p-6n)(10+p-6n)}\xi(k)(k\eta)^{2}\eta^{5+p-6n} \,,
\end{equation}
where $C_{2}(k)$ is a constant of integration. Hence, for sub-horizon modes, the influence of the vector field on the evolution of $\delta$ is a combination of the effect of the energy density and anisotropic stress contributions though both, in this limit, result in the same contributions to the scale dependence and time evolution of the density contrast. The net effect is that, for particular choices of parameters in the action, the perturbations in the vector field can enhance the growth of the baryon density contrast, very much along the lines of dark matter in the dark matter dominated scenario.

\subsection{Scalar and tensors in TeVeS}

We have already come across the effect of the extra fields of TeVeS. Recall that, in TeVeS, as well as a metric (tensor) field, there is a time-like vector field and a scalar field both of which map the two frames on to each other. While at the background level the extra fields contribute to modifying the overall dynamics, they do not contribute significantly to the overall energy density. This is not so at the perturbative level. The field equations for the scalar modes of all three fields can be found in the conformal Newtonian gauge in \citet{SkordisEtAl2006}. While the perturbations in the scalar field will have a negligible effect, the space-like perturbation in the vector field has an intriguing property: it leads to growth. \citet{DodelsonLiguori2006} have shown that the growing vector field  feeds into the Einstein equations and gives rise to a growing mode in the gravitational potentials and in the baryon density. Thus, baryons will be aided by the vector field leading to an effect akin to that of pressureless dark matter. The effect is very much akin to that of the vector field in Einstein-Aether models -- in fact it is possible to map TeVeS models onto a specific subclass of Einstein-Aether models. Hence the discussion above for Einstein-Aether scenarios can be used in the case of TeVeS.

\subsection{Tensor dark matter in models of bigravity}

In bigravity theories \citep{Banadosetal2009}, one considers two metrics: a dynamical metric $g_{\mu\nu}$ and a background metric, ${\tilde g}_{\alpha\beta}$. As in TeVeS, the dynamical metric is used to construct the energy-momentum tensor of the non-gravitational fields and is what is used to define the geodesic equations of test particles. The equations that define its evolution are usually not the Einstein field equations but may be defined in terms of the background metric.

Often one has that ${\tilde g}_{\alpha\beta}$ is dynamical, with a corresponding term in the gravitational action. It then becomes necessary to link ${\tilde g}_{\alpha\beta}$ to $g_{\mu\nu}$ with the background metric determining the field equations of the dynamical metric through a set of interlinked field equations. In bigravity models both metrics are used to build the Einstein--Hilbert action even though only one of them couples to the matter content. A complete action is of the form
\begin{equation}
S=\frac{1}{16 \pi G}\int \mathrm{d}^4 x\left [\sqrt{-g}(R-2\Lambda)+\sqrt{-{\tilde g}}
({\tilde R}-2{\tilde \Lambda})-\sqrt{-{\tilde g}}\frac{1}{\ell^2}({\tilde
    g}^{-1})^{\alpha\beta}g_{\alpha\beta}\right ] ,
\end{equation}
where $\Lambda$ and ${\tilde \Lambda}$ are two cosmological constant terms and $\ell^2$ defines the strength of the linking term between the two actions. The cosmological evolution of perturbations in these theories has been worked out in some detail. It turns out that perturbations in the auxiliary field can be rewritten in the form of a generalized dark matter fluid \citep{Hu1998} with fluid density, momentum, pressure and shear that obey evolution equations which are tied to the background evolution. As a result, it is possible to work out cosmological observables such as perturbations in the CMB and large scale structure. If we restrict ourselves to a regime in which ${\tilde \rho}$ simply behaves as dark matter, then the best-fit bimetric model will be entirely indistinguishable from the standard CDM scenario.



 
\section{Outlook}

Dark matter dominates the matter content of the universe, and only through astrophysical and cosmological observations can the nature of dark matter on large scales be determined. In this review, we have discussed a number of observational techniques available to Euclid: dark matter mapping, complementarity with other astronomical observations (e.g., X-ray and CMB experiments); cluster and galaxy scale dark matter halo mapping; and power spectrum analyses. The techniques described will allow Euclid to constrain a variety of dark matter candidates and their microphysical properties. We have discussed Warm Dark Matter scenarios, axion-like dark matter, scalar field dark matter models (as well as the possible interactions between dark energy and scattering with ordinary matter) and massive neutrinos (the only known component of dark matter). 

Here, we briefly list the main dark matter constraints so far forecasted for Euclid:
\begin{itemize}
\item The weak lensing power spectrum from Euclid will be able to
  constrain warm dark matter particle mass to about
  $m_\mathrm{WDM}>2\mathrm{\ keV}$ \citep{Markovic:2010te};
\item The galaxy power spectrum, with priors from Planck (primary CMB
  only), will yield an error on the sum of neutrino masses $\Sigma$ of
 0.03-0.06~eV, depending on the fiducial value of
    $\Sigma$, and assuming GR and non-interacting dark energy (see
  Table~\ref{summary}; \cite{Carbone-etal:2010});
\item Euclid's weak lensing should also yield an error on $\Sigma$ of 0.05~eV \citep{Kitching-etal:2008};
\item \citet{Jimenez:2010ev} have shown that weak gravitational lensing from Euclid data will be able to determine neutrino hierarchy (if $\Sigma<0.13$);
\item The forecasted errors on the effective number of neutrino species $N_{\nu,\mathrm{eff}}$ for Euclid (with a Planck prior) are $\pm0.1$ \citep[for weak lensing][]{Kitching-etal:2008} and $\pm0.086$ \citep[for galaxy clustering][]{Carbone-etal:2010};
\item The sound speed of unified dark energy-dark matter can be constrained with errors $\sim10^{-5}$ by using 3D weak lensing \citep{2010arXiv1002.4740C};
\item Recently, \citet{Marsh2012} showed that with current and next generation galaxy surveys alone it should be possible to unambiguously detect a fraction of dark matter in axions of the order of 1\% of the total;
\end{itemize}

We envisage a number of future scenarios, all of which give Euclid an imperative to confirm or identify the nature of dark matter. In the event that a dark matter candidate is discovered in direct detection experiments or an accelerator (e.g.\ LHC) a primary goal for Euclid will be to confirm, or refute, the existence of this particle on large scales. In the event that no discovery is made directly, then astronomical observations will remain our only way to determine the nature of dark matter.


\newpage
\part{Initial Conditions}
\label{ini-cond}
 
\section{Introduction}

The exact origin of the primordial perturbations that seeded the formation of the large-scale 
structure in the universe is still unknown. 
Our current understanding of the initial conditions is based on inflation, a phase of accelerated 
expansion preceding the standard evolution of the universe \cite{Guth:1981,Starobinsky:1979,Starobinsky:1982,Sato:1981}. 
In particular, inflation explains why the universe is so precisely flat, homogeneous and isotropic.
During this phase, scales much smaller than the Hubble radius are inflated to super-horizon sizes, 
so that regions appearing today as causally disconnected were in fact very close in the past. 
This mechanism is also at the origin of the cosmic large-scale structure. Vacuum quantum 
fluctuations of any light field present during inflation are amplified by the accelerated 
expansion and \emph{freeze-out} on super-Hubble scales acquiring a quasi-scale invariant 
spectrum \cite{Mukhanov-Chibisov1981,Hawking:1982,Starobinsky:1982,Guth-Pi:1982,Bardeen-Steinhardt-Turner:1983}. 

From the early development of inflation, the simplest proposal based on a weakly-coupled single scalar field slow-rolling 
along its potential \cite{Linde:1982,Albrecht-Steinhardt:1982} has gained strength and many models have been 
built based on this picture (see for instance~\cite{Linde:2008} for a review). 
The simplest inflationary models based on a weakly-coupled single scalar field 
are compatible with Planck 2015 cosmological results. The geometry is compatible with being flat 
($\Omega_K = -0.005^{+0.016}_{-0.017}$ at 95\% CL by combining  Planck 2015 temperature and
lensing information \citep{planck2013-p11,planck2013-p12}). The most recent Planck 2015 determination of the scalar spectral index is  
$n_\mathrm{s}=0.968 \pm 0.06$ at 68$\%$ CL, disfavouring exact scale invariance at 6 $\sigma$. There is no statistical evidence of a running of the spectral index, i.e. $\mathrm{d} n_{\mathrm s}/\mathrm{d}\ln k = -0.003 \pm 0.007$ at 68$\%$CL.
Although many inflationary 
potentials are now excluded by Planck 2015 data \citep{planck2013-p17}, 
the simplest inflationary predictions 
have been extremely successful in passing many observational tests: it predicts perfectly adiabatic 
and almost Gaussian fluctuations with a quasi scale-invariant spectrum and a small amount of gravitational waves. 

While current data have ruled out some classes of inflationary models, the next qualitative step 
forward is investigating the physics responsible for inflation: we still lack a complete understanding 
of the high energy physics describing it.   In fact, most likely the physics of inflation is far out of 
reach of terrestrial experiments, many orders of magnitude  
larger than the center-of-mass energy at the Large Hadron Collider (LHC).
Thus, cosmological tests of inflation offer a unique  opportunity to  learn about ultra-high energy physics.
We can do this by targeting observations which directly probe the dynamics of inflation. One route is  to accurately 
measure the shape of the primordial power spectrum of scalar perturbations produced during the phase of accelerated
expansion, which is directly related to the shape of the inflaton potential, and to constrain the amplitude of the 
corresponding stochastic gravitational-wave background, which is related instead to the energy-scale of inflation.
Planck, in combination with the BICEP2/Keck Array data, has constrained the tensor-to-scalar ratio to $r_{0.05} < 0.08$ at 95$\%$CL \cite{2015PhRvL.114j1301B}. With the most recent addition of the Keck Array 95 GHz channel, the constraint has been further tightened to $r_{0.05} < 0.07$ at 95$\%$CL \cite{2016PhRvL.116c1302B}.

A complementary approach is offered by constraining -- or exploring -- how much the distribution of primordial density 
perturbations departs from Gaussian statistics and purely adiabatic fluctuations. Non-Gaussianity is a very sensitive probe of self-couplings
and interactions between the fields generating the primordial perturbations, whereas the presence of isocurvature modes can
teach us about the number of fields present during inflation and their role in reheating and generating the matter in the universe.
At present, Planck establish that the statistics of 
primordial perturbations is compatible with a Gaussian one \citep{2015arXiv150202114P, 2015arXiv150201592P} and that the preference of 
isocurvature initial conditions is not statistically significant. 
Nevertheless, future large-scale structure 
surveys like Euclid can probe these features with an unprecedented accuracy, thus providing a way to test aspects of 
inflationary physics that are not easily accessible otherwise.

Non-minimal scenarios or proposals even radically different
from single-field inflation are still compatible with the data.
In order to
learn something about the physics of the early universe we need to rule out
or confirm the conventional slow-roll scenario and possibly discriminate
between non-conventional models. Non-Gaussianities and isocurvature
perturbations currently represent the best tools that we have to accomplish
this task. Any deviation from the conventional Gaussian and adiabatic
initial perturbations would represent important breakthroughs in our
understanding of the early universe. In this section we are going to review
what we can learn by constraining the initial conditions with a large-scale
structure survey like Euclid.

\section{Constraining inflation}

The spectrum of cosmological perturbations represents an important source of information on the early universe. During inflation scalar (compressional) and tensor (purely gravitational) fluctuations are produced. The shape and the amplitude of the power spectrum of scalar fluctuations can be related to the dynamics of the inflationary phase, providing a window on the inflaton potential. Inflation generically predicts a deviation from a purely scale-invariant spectrum. Together with future CMB experiments such as Planck, Euclid will improve our constraints on the scalar spectral index and its running, helping to pin down the model of inflation.

\subsection{Primordial perturbations from inflation}

It is convenient to describe primordial perturbations using the
curvature perturbation on uniform density hypersurfaces
$\zeta$ introduced in~\cite{Bardeen-Steinhardt-Turner:1983}. An
important property of this quantity is that for adiabatic
perturbations -- i.e., in absence of isocurvature perturbations,
discussed in Section~\ref{sec:isocurvature} --  it remains constant on
super-Hubble scales, allowing us to connect the early inflationary
phase to the late-time universe observations, regardless of the
details of reheating. In a gauge where the energy density of the
inflaton vanishes, we can define $\zeta$ from the spatial part of the
metric (assuming a flat FRW universe), as
\citep{Salopek-Bond:1990,Maldacena:2003}
\begin{equation}\label{eq:zeta}
g_{ij} = a^2(t) \exp \left( 2 \zeta \right) \delta_{ij} \,.
\end{equation}
This definition, where $\zeta$ enters the metric in the exponential form, has the advantage that it is valid also beyond linear order and can be consistently used when discussing non-Gaussian fluctuations, such as in Section~\ref{sec:NG-TH}. 

The power spectrum of primordial perturbations is given by 
\begin{equation}
\langle \zeta_{\mathbf{k}} \zeta_{\mathbf{k}'} \rangle = (2 \pi)^3 \delta(\mathbf{k}+\mathbf{k}') P_{\zeta}(k) \,,
\end{equation}
where $\langle \ldots \rangle$ denotes the average over an ensemble of
realizations. It is useful to define a dimensionless spectrum as ${\cal P}_s(k) \equiv \frac{k^3}{2\pi^2} P_\zeta (k)\;$, where the index $s$ stands for scalar, to distinguish it from the spectrum of tensor perturbations, defined below.
The deviation from scale-invariance of the scalar spectrum is
characterized by the spectral index $n_s$, defined by (see, for
instance, \cite{Liddle-Lyth:2000})
\begin{equation}
\label{eq:index_ns}
n_s \equiv 1 + \frac{d  \ln {\cal P}_s}{d \ln k} \,,
\end{equation}
where $n_s=1$ denotes a purely scale-invariant spectrum. We also define the running of the spectral index $\alpha_s$ as
\begin{equation}
\label{eq:running}
\alpha_s \equiv \frac{d n_s}{d \ln k} \,.
\end{equation}
These quantities are taken at a particular pivot scale. For our
analysis we chose it to be $k_* \equiv 0.05\mathrm{\ Mpc}^{-1}$. Thus,
with these definitions the power spectrum can be written as
\begin{equation}
P_\zeta(k) = \frac{2 \pi^2}{k^3} A_s(k_*) (k/k_*)^{n_s(k_*) -1 + \frac12 \alpha_s(k_*) \ln (k/k_*) } \,,
\label{eq:shape}
\end{equation}
where $A_s$ is the normalization parameterising the amplitude of the fluctuations. 

During inflation tensor modes are also generated. They are described by the gauge invariant metric perturbation $h_{ij}$, defined from the spatial part of the metric as
\begin{equation}
g_{ij} = a^2(t) \left( \delta_{ij} + h_{ij} \right) \,, \quad h^j_{i,j} = 0 = h_{i}^i \,.
\end{equation}
Each mode has $2$ polarizations, $h_+$ and $h_\times$, each with power spectrum given by 
\begin{equation}
\langle h_{\mathbf{k}} h_{\mathbf{k}'} \rangle = (2 \pi)^3 \delta(\mathbf{k}+\mathbf{k}') P_h(k) \,.
\end{equation}
Defining the dimensionless power spectrum of tensor fluctuations as ${\cal P}_t(k) \equiv 2 \frac{k^3}{2\pi^2} P_h (k)\;$, where the factor of 2 comes from the two polarizations, it is convenient to define the ratio of tensor to scalar fluctuations as 
\begin{equation}
\label{eq:r_ts}
r\equiv {{\cal P}_t (k_*)}/{{\cal P}_s (k_*)} \,.
\end{equation}

The form of the power spectrum given in Eq.~\eqref{eq:shape} approximates very well power spectra of perturbations generated by slow-roll models. In particular, the spectrum of scalar fluctuations  is given in terms of the Hubble rate $H$ and the first slow-roll parameter $\epsilon \equiv - \dot H/H^2$, both evaluated at the time when the comoving scale $k$ crosses the Hubble radius during inflation,
\begin{equation}
{\cal P}_s(k) = \left.\frac{1}{8 \pi^2 \epsilon}\frac{H^2}{M_{\mathrm{Pl}}^2} \right|_{k=aH} \,.
\end{equation}
During slow-roll, $\epsilon$ is related to the first derivative of the inflaton potential $V(\phi)$, $\epsilon \approx \frac{M_{\mathrm{Pl}}^2}{2} \left( \frac{V'}{V} \right)^2$, where the prime denotes differentiation with respect to $\phi$. As $H$ and $\epsilon$ vary slowly during inflation, this spectrum is almost scale-invariant. Indeed, the scalar spectral index $n_s$ in Eq.~\eqref{eq:index_ns} reads
\begin{equation}
n_s=1 - 6 \epsilon + 2 \eta_V  \,,
\end{equation}
where the second slow-roll parameter $\eta_V \equiv M_{\mathrm{Pl}}^2 \frac{V''}{V} $ must be small for inflation to yield a sufficient number of $e$-foldings. The running of the spectral index defined in Eq.~\eqref{eq:running} is even smaller, being second-order in the slow-roll parameters. It is given by $\alpha_s = 16\epsilon \eta_V - 24 \epsilon^2- 2\xi_V$ where we have introduced the third slow-roll parameter $\xi_V \equiv M_{\mathrm{Pl}}^4 \frac{V' V''' }{V^2} $.

The spectrum of tensor fluctuations is given by
\begin{equation}
{\cal P}_t(k) = \left.\frac{2}{\pi^2 }\frac{H^2}{M_{\mathrm{Pl}}^2} \right|_{k=aH} \,,
\end{equation}
which shows that the ratio of tensor to scalar fluctuations in Eq.~\eqref{eq:r_ts} is simply related to the first slow-roll parameter by $r =  16 \epsilon$.

As a fiducial model, in the next section we will consider chaotic inflation~\citep{Linde:1983}, based on the quadratic inflaton potential $V = \frac12 m^2 \phi^2$. In this case, the first two slow-roll parameters are both given in terms of the value of the inflaton field at Hubble crossing $\phi$ or, equivalently, in terms of number of $e$-folds from Hubble crossing to the end of inflation $N$, as $\epsilon=\eta_V = 2 M_{\mathrm{Pl}}^2 /\phi^2 = 1/2 N$, while $\xi_V =0$. This implies \label{symbol:efolds}
\begin{equation}
n_s = 1 - 2/N_* \,, \quad r = 8/N_* \,, \quad \alpha_s = -2 /N_*^2 \,,
\end{equation}
where the star denotes Hubble crossing of the pivot scale $k_*$.
Choosing $N_*=62.5$, this yields $n_s= 0.968$, $r=0.128$ and $\alpha_s =0$ as our fiducial model.

\subsection{Forecast constraints on the power spectrum}

We will now study how much Euclid will help in improving the already very
tight constraints on the power spectrum given by the Planck satellite. Let
us start discussing the forecast for Planck. We assume 2.5~years (5 sky
surveys) of multiple CMB channel data, with instrument characteristics for
the different channels listed in Table~\ref{tbl:PlanckSpec}. We take the
detector sensitivities and the values of the full width half
maximum from the Planck ``Blue Book''
\citep{PlanckBlueBook}. In this analysis we use three channels for
Planck mock data and we assume that the other channels are used for
foreground removal and thus do not provide cosmological information.

\begin{table}[htbp]
\caption{Instrument specifics for the Planck satellite with 30 months
  of integration.}
\label{tbl:PlanckSpec}
\centering
\begin{tabular}{llll}
\toprule
Channel Frequency (GHz) & 70 & 100 & 143 \\
\midrule
Resolution (arcmin) & 14 & 10 & 7.1 \\
Sensitivity - intensity ($\mu K$) &  8.8 & 4.7 & 4.1 \\
Sensitivity - polarization ($\mu K$) & 12.5 & 7.5  & 7.8  \\
\bottomrule
\end{tabular}
\end{table}

For a nearly full-sky CMB experiment (we use $f_{\mathrm{sky}}=0.75$), the
likelihood ${\cal L}$ can be approximated
by~\citep{Verde-Peiris-Jimenez:2006}
\begin{equation}
\begin{split}
-2 \ln{{\cal L}} =& \sum_{\ell=\ell_{\min}}^{\ell_{\max}} (2\ell+1)f_{\mathrm{sky}}\, \left[ -3 + \frac{\hat{C}_\ell^{BB}}{C_\ell^{BB}} + \ln\left(\frac{C_\ell^{BB}}{\hat{C}_\ell^{BB}}\right) \right. \\
& \left.  + \frac{\hat{C}_\ell^{TT}C_\ell^{EE} + \hat{C}_\ell^{EE}C_\ell^{TT} - 2\hat{C}_\ell^{TE}C_\ell^{TE}}{C_\ell^{TT}C_\ell^{EE}-(C_\ell^{TE})^2}  +\ln{\left(\frac{C_\ell^{TT}C_\ell^{EE}-(C_\ell^{TE})^2}{\hat{C}_\ell^{TT}\hat{C}_\ell^{EE}-(\hat{C}_\ell^{TE})^2}\right)}\right] , 
\end{split}
\end{equation}
where we assume $l_{\min}=3$ and $l_{\max}=2500$. \label{symbol:Cell}
Here, $C_\ell$ is the sum of the model-dependent theoretical power
spectrum $C_\ell^{\mathrm{theory}}$ and of the noise spectrum
$N_\ell$, which we assume perfectly known. The mock data
$\hat{C}_\ell$ is $C_\ell$ for the fiducial model, with
$C_\ell^{\mathrm{theory}}$ calculated using the publicly available
code \textsc{camb} \citep{CAMB} and $N_\ell$ calculated
assuming a Gaussian beam.
We use the model described in~\cite{Verde-Peiris-Jimenez:2006,Baumann-etal:2009} to propagate the effect of polarization foreground residuals into the estimated uncertainties on the cosmological parameters. For simplicity, in our simulation we consider only the dominating components in the frequency bands that we are using, i.e., the synchrotron and dust signals. The fraction of the residual power spectra are all assumed to be 5\%.

Let us turn now to the Euclid forecast based on the spectroscopic redshift survey. We will model the galaxy power spectrum in redshift space as (\cite{Kaiser:1987, Peacock:1992, Peacock-Dodds:1994}; see also discussion in Section~\ref{dark-energy-and-redshift-surveys})
\begin{equation}
P_g(k,z,\mu) = \left(b+ f_g \mu^2\right)^2 G^2(z)P_{\mathrm{matter}}(k;z=0) e^{-k^2\mu^2\sigma_r^2}, \label{eq:Pgini}
\end{equation}
where $\mu$ is the cosine of the angle between the wavenumber $\mathbf{k}$ 
and the line of sight, $G(z)$ is the linear growth factor defined in Eq.~\eqref{def_gf},
 $f_g \equiv {d\ln G}/{d\ln a}$ is the linear growth rate (see Eq.~\eqref{def-growth-rate}) 
and $P_{\mathrm{matter}}(k;z=0)$ is the matter power spectrum at redshift $0$. The term $f_g \mu^2$  
comes for the redshift distortions due to the large-scale peculiar velocity field \citep{Kaiser:1987}, 
which is correlated with the matter density field. The factor $e^{-k^2\mu^2\sigma_r^2}$ accounts for 
the radial smearing due to the redshift distortions that are uncorrelated with the large-scale structure. 
We consider two contributions. The first is due to the redshift uncertainty of the spectroscopic galaxy 
samples. Assuming a typical redshift uncertainty $\sigma_z=0.001(1+z)$, this turns into a contribution 
to $\sigma_r$ given by $\partial r /\partial z \, \sigma_z = H^{-1} \, \sigma_z
$, where $r(z)=\int_0^z cdz'/H(z')$ is the comoving distance of a flat FRW
universe and $H$ is the Hubble parameter as a function of the redshift. The
second contribution comes from the Doppler shift due to the
virialized motion of
galaxies within clusters, which typically have a pairwise velocity dispersion
$v_p$ of the order of few hundred kilometers per second. This term can be
parameterized as $\frac{v_p}{\sqrt{2}}H^{-1} (1+z)$ \citep{Peacock-Dodds:1994}.
Taking the geometric mean of the two contributions, we obtain 
\begin{equation}
\sigma_r^2 = \frac{(1+z)^2}{H^2}  \left(10^{-6} + {v_p^2}/{2} \right) \,, \label{eq:sigma2}
\end{equation}
where the two velocities in the parenthesis contribute roughly the same.
Practically neither the redshift measurement nor the virialized
virialized motion of 
galaxies can be precisely quantified. In particular, the radial smearing due 
to peculiar velocity is not necessarily close to Gaussian. Thus, Eq.~(\ref{eq:Pgini}) 
should not be used for wavenumbers  $k>\frac{H(z)}{v_p(1+z)}$, where the radial smearing effect is important. 

On large scales the matter density field has, to a very good approximation, Gaussian statistics and uncorrelated Fourier modes. Under the assumption that the positions of observed galaxies are generated by a random Poissonian point process, the band-power uncertainty is given by (\cite{Tegmark-etal:1998}; see also Eq.~\eqref{eqn:pkerror} in Section~\ref{dark-energy-and-redshift-surveys})
\begin{equation}
\Delta  P_g = \left[ \frac{2 (2\pi)^3}{(2\pi k^2 \,dk \,d\mu) (4\pi
    r^2f_{\mathrm{sky}} \,
    dr)}\right]^{1/2}\left(P_g+\frac{1}{\bar{n}}\right). 
\label{eq:dPg}
\end{equation}
Here $f_{\mathrm{sky}}$ is the observed fraction of sky, $r$ the comoving distance defined above, and $\bar{n}$ is the expected number density of galaxies that can be used.

Finally, we ignore the band-band correlations and write the likelihood as 
\begin{equation}
-2 \ln {\cal L} =  \sum_{k,\mu,z\ \mathrm{bins}} \left(\frac{
  P_g^{\mathrm{model}} - P_g^{\mathrm{fiducial}} }{\Delta
  P_g^{\mathrm{fiducial}} }\right)^2 .
\end{equation}

To produce the mock data we use a fiducial $\Lambda$CDM model with
$\Omega_ch^2=0.1128$, $\Omega_bh^2 = 0.022$, $h=0.72$, $\sigma_8 =
0.8$ and $\tau=0.09$, where $\tau$ is the reionization optical depth.
As mentioned above, we take the fiducial value for the spectral index,
running and tensor to scalar ratio, defined at the pivot scale
$k_*=0.05\,{\mathrm{Mpc}}^{-1}$, as given by chaotic inflation with
quadratic potential, i.e., $n_s = 0.968$, $\alpha_s=0$ and
$r=0.128$. We have checked that for Planck data $r$ is almost
orthogonal to $n_s$ and $\alpha_s$. Therefore our result is not
sensitive to the fiducial value of $r$.

The fiducial Euclid spectroscopically selected galaxies are split into 14
redshift bins. The redshift ranges and expected numbers of observed galaxies per
unit volume $\bar{n}_{\mathrm{obs}}$ are taken from~\cite{euclidredbook} and shown in the third column 
of Table~\ref{tab:n_z} in Section~\ref{sec:baofm_survey} ($n_2(z)$). The number density of
galaxies that can be used is $\bar{n}=\varepsilon\bar{n}_{\mathrm{obs}}$, where
$\varepsilon$ is the fraction of galaxies with measured redshift. 
The boundaries of the wavenumber range used in the analysis, labeled
$k_{\min}$ and $k_{\max}$, vary  in the ranges
$(0.00435\mbox{\,--\,}0.00334)h\mathrm{\ Mpc}^{-1}$ and
$(0.16004\mbox{\,--\,}0.23644)h\mathrm{\ Mpc}^{-1}$ respectively, for
$0.7\leq z\leq 2$. The IR cutoff $k_{\min}$ is chosen such that
$k_{\min}r = 2\pi$, where $r$ is the comoving distance of the redshift
slice. The UV cutoff is the smallest between $\frac{H}{v_p(1+z)}$ and
$\frac{\pi}{2R}$. Here $R$ is chosen such that the r.m.s.\ linear
density fluctuation of the matter field in a sphere with radius $R$ is
0.5. In each redshift bin we use 30 $k$-bins uniformly in $\ln k$
and 20 uniform $\mu$-bins.

For the fiducial value of the bias, in each of the 14 redshift bins of
width $\Delta z=0.1$ in the range (0.7\,--\,2), we use those derived from
\cite{orsi10},
i.e.\ (1.083, 1.125, 1.104, 1.126, 1.208, 1.243, 1.282, 1.292, 1.363, 1.497, 1.486, 1.491, 1.573, 1.568), 
and we assume that $v_p$ is redshift dependent
choosing $v_p=400\mathrm{\ km/s}$ as the fiducial value in each redshift bin. Then
we marginalize over $b$ and $v_p$ in the 14 redshift bins, for a total of 28
nuisance parameters. 

\begin{table}[htbp]
\caption{Cosmological parameters}
\label{tbl:cosmoparams}
\renewcommand{\arraystretch}{1.2}
\centering
\begin{tabular}{llll}
\toprule
parameter & Planck  2015 constraint & EUCLID fiducial cosmology & Planck + Euclid constraint\\
\midrule
$\Omega_bh^2$ & $0.02238 \pm 0.00027$ & $0.022$ & $0.02227^{+0.00008}_{-0.00008}$ \\
$\Omega_ch^2$ & $0.1180\pm 0.0021$ & $0.12$ & $0.1116^{+0.0002}_{-0.0002}$ \\
$100 \theta$ & $1.04111 \pm 0.00047$ & $1.041$ & $1.0392^{+0.0002}_{-0.0002}$ \\
$\tau_{\mathrm{re}}$ & $0.071\pm 0.018$ & $0.09$ & $0.085^{+0.003}_{-0.003}$ \\
$n_s$ & $0.9690\pm 0.0063$ & $0.96$ & $0.966^{+0.002}_{-0.002}$ \\
$\alpha_s$ & $-0.008^{+0.009}_{-0.008}$ & $0$ & $-0.000^{+0.003}_{-0.003}$ \\
$\ln(10^{10}A_s)$ & $3.073\pm 0.033$ & $3.098$ & $3.077^{+0.006}_{-0.006}$ \\
$r_{0.05}$ & $< 0.16 (95\% {\mathrm CL})$ & $0$ & $0.127^{+0.019}_{-0.018}$ \\
$\Omega_m$ & $0.304\pm 0.013$ & $0.32$ & $0.271^{+0.001}_{-0.001}$ \\
$\sigma_8$ & $0.816 \pm 0.010$ & $0.83$ & $0.808^{+0.003}_{-0.003}$ \\
$h$ & $0.682 \pm 0.010$  & $0.67$ & $0.703^{+0.001}_{-0.001}$ \\
\bottomrule
\end{tabular}
\end{table}
In these two cases, we consider the forecast constraints on eight
cosmological parameters, i.e., $\Omega_bh^2$, $\Omega_ch^2$, $\theta$, $\tau$, $\ln A_s$, $n_s$, $\alpha_s$, and $r$. Here $\theta$ is the angle subtended by the sound horizon on the last scattering surface, rescaled by a factor 100. We use the publicly available code \textsc{CosmoMC} \citep{Lewis-Bridle:2002} to perform Markov Chain Monte Carlo calculation.  The nuisance parameters are marginalized over in the final result. The marginalized 68.3\% confidence level (CL) constraints on cosmological parameters for Planck only (Planck TT + lowP + lensing, for LCDM with r and running of the spectral index \cite{2015arXiv150201589P}), and Planck and Euclid forecast are  listed in the second and third columns of Table~\ref{tbl:cosmoparams}, respectively. 

\epubtkImage{nsnrun_redbook-sigma8omm_redbook.png}{%
\begin{figure}
\centerline{
\includegraphics[width=0.5\textwidth]{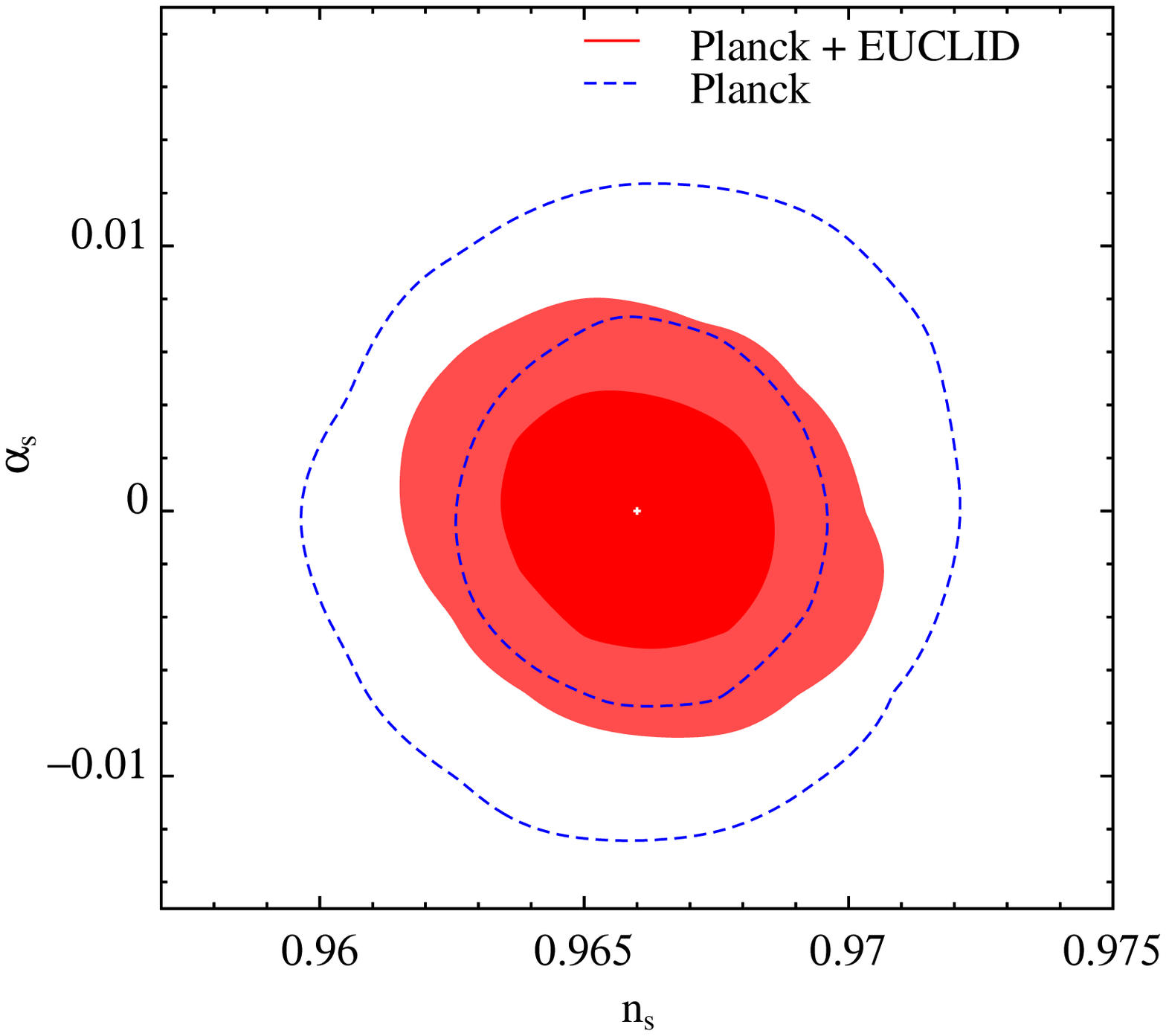}
\includegraphics[width=0.5\textwidth]{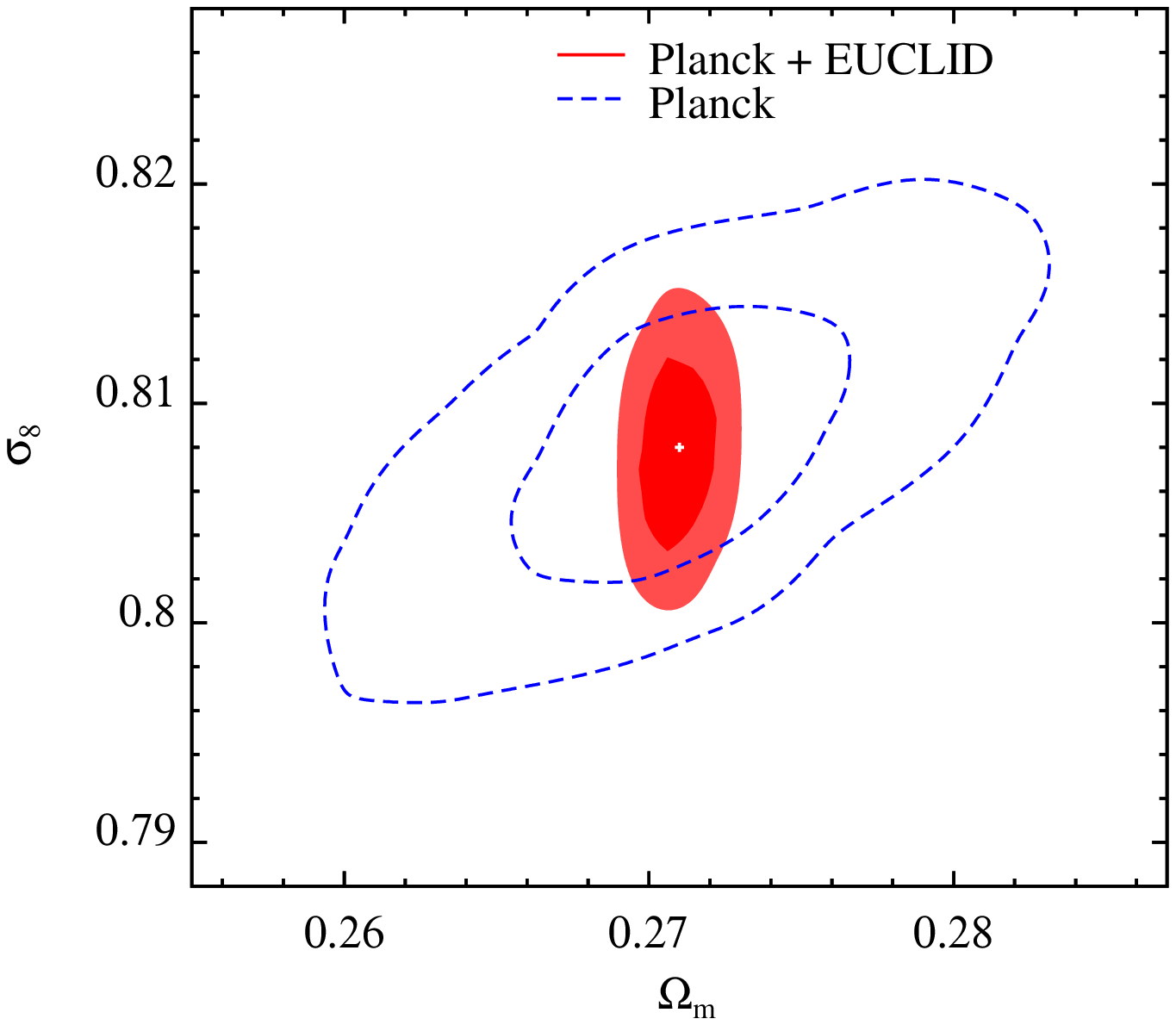}
}
\caption{The marginalized likelihood contours (68.3\% and 95.4\% CL)
  for Planck forecast only (blue dashed lines) and Planck plus Euclid
  pessimistic (red filled contours). The white points correspond to
  the fiducial model.}
 \label{fig:2D_Planck_EUCLID}
\end{figure}}

Euclid can improve the `figure of merit' on the $n_s$-$\alpha_s$ plane by a factor of 2.2, as shown in the left panel of Figure~\ref{fig:2D_Planck_EUCLID}. Because the bias is unknown, the LSS data do not directly measure $A_s$ or $\sigma_8$. However, Euclid can measure $\Omega_m$ to a much better accuracy, which can break the degeneracy between $\Omega_m$ and $\sigma_8$ that one typically finds using CMB data alone. This is shown in the right panel of Figure~\ref{fig:2D_Planck_EUCLID}.

A more extensive and in depth analysis of what constraints on inflationary models a survey like Euclid can provide is presented in~\cite{Huang:2012mr}.  In particular they find that  for models  where the primordial power spectrum  is not featureless (i.e., close to  a power law  with small running) a survey like Euclid will be crucial to detect and measure features. Indeed, what we measure with the CMB is the angular power spectrum of the anisotropies in the 2-D multipole space, which is a projection of the power spectrum in the 3-D momentum space. Features at large $\ell$'s and for small width in momentum space get smoothed during this projection but  this does not happen for large-scale structure surveys.  The main limitation on the width of features measured using  large-scale structure comes from the size of the volume of the survey:  the smallest detectable feature being of the order of the inverse cubic root of this volume and the error  being determined by number of modes contained in this volume. 
Euclid, with the large volume surveyed and the  sheer number of modes that are sampled and cosmic variance dominated   offers a unique  opportunity to  probe inflationary models where the potential is not featureless. In addition the increased statistical power would  enable  us to perform a Bayesian  model selection on the space of inflationary models (e.g.,~\cite{Easther:2011yq, Norena:2012rs} and references therein).

\section{Probing the early universe with non-Gaussianities}
\label{sec:NG-TH}

The workhorse for primordial non-Gaussianity  has been so far the ``local model''
\citep{Salopek-Bond:1990, Gangui-etal:1994, Verde-etal:2000, Komatsu-Spergel:2001,Bartolo-etal:2004}:
\begin{equation}
\Phi=\phi+f_{\mathrm{NL}}\left( \phi^2-\langle \phi^2\rangle \right) .
\label{eq:fnl}
\end{equation}
Here $\phi$ is a Gaussian random field while $\Phi$ denotes 
Bardeen's gauge-invariant potential, which, 
on sub-Hubble scales reduces to the usual Newtonian peculiar 
gravitational potential, up to a minus sign. On large scales it is related to the conserved variable $\zeta$ by
\begin{equation}
\zeta = \frac{5 + 3 w}{3 + 3w} \Phi \,,
\end{equation}
where $w$ is the equation of state of the dominant component in the universe.
The amount of primordial non-Gaussianity is quantified by the nonlinearity parameter
$f_{\mathrm{NL}}$. Note that, since $\Phi\simeq \phi\simeq 10^{-5}$, $f_{\mathrm{NL}}\sim 100$ corresponds
to relative non-Gaussian corrections of order $10^{-3}$.
While $\zeta$ is constant on large scales, $\Phi$ is not. For this reason, 
in the literature there are two conventions for 
Eq.~(\ref{eq:fnl}): the large-scale structure (LSS) and the cosmic microwave background (CMB) 
one. In the LSS convention, $\Phi$ is linearly extrapolated at $z=0$; 
in the CMB convention $\Phi$ is instead primordial:  thus $f^{\mathrm{LSS}}_{\mathrm{NL}}=g(z=\infty)/g(0) f_{\mathrm{NL}}^{\mathrm{CMB}}\sim 1.3 \,f^{\mathrm{CMB}}_{\mathrm{NL}}$, 
where $g(z)$ denotes the linear growth suppression factor relative to an 
Einstein--de~Sitter universe. In the past few years it has become customary to always report 
$ f_{\mathrm{NL}}^{\mathrm{CMB}}$ values even though, for simplicity as it will be clear below, 
one carries out the calculations with $f_{\mathrm{NL}}^{\mathrm{LSS}}$.

In this section we review the theoretical motivations and implications for looking into primordial non-Gaussianity; the readers less theoretically oriented can go directly to Section~\ref{sec:NG-LSS}.

\subsection{Local non-Gaussianity}

The non-Gaussianities generated in the conventional scenario of inflation (single-field with standard kinetic term, in slow-roll, initially in the Bunch--Davies vacuum) are predicted to be extremely small. Earlier calculations showed that $f_{\mathrm{NL}} $ would be of the order of the slow-roll parameters \cite{Salopek-Bond:1990,Falk-etal:1993,Gangui-etal:1994}. More recently, with an exact calculation~\cite{Maldacena:2003} confirmed this result and showed that the dominant contribution to non-Gaussianity comes from gravitational interaction and it is thus independent of the inflaton potential. More precisely, in the squeezed limit, i.e.~when one of the modes is much smaller than the other two, the bispectrum of the primordial perturbation $\zeta$ is given by
\begin{equation}
B_\zeta(k_1 \ll k_2,k_3) = 4 f_{\mathrm{NL}}^{\mathrm{local}} P_\zeta(k_2) P_\zeta(k_3) \,, \label{local_sl}
\end{equation}
where $f_{\mathrm{NL}}^{\mathrm{local}}$ is proportional to the tilt of scalar fluctuations, $f_{\mathrm{NL}}^{\mathrm{local}} = -(5/12) (n_s-1)$, a value much too small to be observable. Thus, any deviation from this prediction would rule out a large class of models based on the simplest scenario. 

Furthermore, \cite{Creminelli-Zaldarriaga:2004} showed that irrespective of slow-roll and of the particular inflaton Lagrangian or dynamics, in single-field inflation, or more generally when only adiabatic fluctuations are present, there exists a consistency relation involving the 3-point function of scalar perturbations in the squeezed limit (see also~\cite{Seery-Lidsey:2005,Chen-etal:2007,Cheung-etal:2008a}). In this limit, when the short wavelength modes are inside the Hubble radius during inflation, the long mode is far out of the horizon and its only effect on the short modes is to rescale the unperturbed history of the universe. This implies that the 3-point function is simply proportional to the 2-point function of the long wavelength modes times the 2-point function of the short wavelength mode times its deviation from scale invariance. In terms of local non-Gaussianity this translates into the same $f_{\mathrm{NL}}^{\mathrm{local}}$ found in~\cite{Maldacena:2003}. Thus, a convincing detection of 
local non-Gaussianity would rule out all classes of inflationary single-field models.

To overcome the consistency relation and produce large local non-Gaussianity
one can go beyond the single-field case and consider scenarios where a
second field plays a role in generating perturbations. In this case, because
of non-adiabatic fluctuations, scalar perturbations can evolve outside the
horizon invalidating the argument of the consistency relation and possibly
generating a large $f_{\mathrm{NL}}^{\mathrm{local}}$ as in
\cite{Linde-Mukhanov:1997}. The curvaton scenario is one of such mechanisms.
The curvaton is a light scalar field that acquires scale-invariant
fluctuations during inflation and decays after inflation but well before
nucleosynthesis \cite{Mollerach:1990,Moroi-Takahashi:2001,Lyth-Wands:2002,Enqvist-Sloth:2002}. 
During the decay it dominates the universe affecting its expansion history thus imprints its 
perturbations on super-horizon scales. The way the expansion history depends on the value of 
the curvaton field at the end of the decay can be highly nonlinear, leading to large non-Gaussianity. 
Indeed, the nonlinear parameter $f_{\mathrm{NL}}^{\mathrm{local}}$ is inversely proportional to the 
curvaton abundance before the decay  \cite{Lyth-Ungarelli-Wands:2003}. 

Models exists where both curvaton and inflaton fluctuations contribute 
to cosmological perturbations \cite{Langlois-Vernizzi:2004}. 
Interestingly, curvaton fluctuations could be negligible in the 2-point 
function but detectable through their non-Gaussian signature in the 3-point function, 
as studied in~\cite{Boubekeur-Lyth:2006}. We shall come back on this point when discussing isocurvature perturbations. 
Other models generating local non-Gaussianities are the so called modulated reheating models, in which one light field 
modulates the decay of the inflaton field \cite{Dvali-Gruzinov-Zaldarriaga:2004,Kofman:2003}. Indeed, non-Gaussianity 
could be a powerful window into the physics of reheating and preheating, the phase of transition from inflation to the 
standard radiation dominated era (see e.g.,~\cite{Bond-etal:2009,Chambers-Nurmi-Rajantie:2010}).

In the examples above only one field is responsible for the dynamics of inflation, while the others are spectators. When the inflationary dynamics is dominated by several fields along the $\sim 60$ e-foldings of expansion from Hubble crossing to the end of inflation we are truly in the multi-field case. For instance, a well-studied model is double inflation with two massive non-interacting scalar fields \cite{Polarski-Starobinky:1992}. In this case, the overall expansion of the universe is affected by each of the field while it is in slow-roll; thus, the final non-Gaussianity is slow-roll suppressed, as in single field inflation \cite{Rigopoulos-Shellard-vanTent:2006,Alabidi-Lyth:2006,Vernizzi-Wands:2006}. 

Because the slow-roll conditions are enforced on the fields while they dominate the inflationary dynamics, it seems difficult to produce large non-Gaussianity in multi-field inflation; however, by tuning the initial conditions it is possible to construct models leading to an observable signal (see~\cite{Byrnes-Choi-Hall:2008,Tanaka-Suyama-Yokoyama:2010}). 
Non-Gaussianity can be also generated at the end of inflation, where large-scale perturbations may have a nonlinear dependence on the non-adiabatic modes, especially if there is an abrupt change in the equation of state (see e.g.,~\cite{Bernardeau-Uzan:2002,Lyth:2005}). Hybrid models \cite{Linde:1994}, where inflation is ended by a tachyonic instability triggered by a waterfall field decaying in the true vacuum, are natural realizations of this mechanism \cite{Enqvist-Vaihkonen:2004,Barnaby-Cline:2006}.

\subsection{Shapes: what do they tell us?}
\label{nongaussianshapes}

As explained above, local non-Gaussianity is expected for models where nonlinearities develop outside the Hubble radius. However, this is not the only type of non-Gaussianity. Single-field models with derivative interactions
yield a negligible 3-point function in the squeezed limit, yet leading to possibly observable non-Gaussianities. Indeed, as the interactions contain time derivatives and gradients, they vanish outside the horizon and are unable to produce a signal in the squeezed limit. Correlations will be larger for other configurations, for instance between modes of comparable wavelength. In order to study the observational signatures of these models we need to go beyond the local case and study the \emph{shape} of non-Gaussianity \cite{Babich-Creminelli-Zaldarriaga:2004}.

Because of translational and rotational invariance, the 3-point function is characterized by a function of the modulus of the three wave-vectors, also called the bispectrum $B_{\zeta}(k_1,k_2,k_3)$, defined as
\begin{equation}\label{eq:bispectrumdef}
\langle \zeta_{\mathbf{k_1}} \zeta_{\mathbf{k_2}} \zeta_{\mathbf{k_3}} \rangle = (2 \pi)^3 \delta(\mathbf{k_1}+\mathbf{k_2}+\mathbf{k_3}) B_{\zeta}(k_1,k_2,k_3) \,.
\end{equation}
Relaxing the assumption of a local $f_{\mathrm{NL}}$, this function is a rich object which can contain a wealth of information, depending on the size and shape of the triangle formed by $k_1$, $k_2$ and $k_3$. 
Indeed, the dependence of the bispectrum on configuration in momentum space is related to the particular inflationary model generating it. Namely, each third-order operator present in the field action gives rise to a particular shape of the bispectrum.

An example of models containing large derivative interactions has been proposed by~\cite{Silverstein-Tong:2004,Alishahiha-Silverstein-Tong:2004}. Based on the Dirac--Born--Infeld Lagrangian,
${\cal L} = f(\phi)^{-1} \sqrt{1- f(\phi) X} +V(\phi)$, with $X= -g^{\mu \nu} \partial_\mu \phi \partial_\nu \phi$, it is called DBI inflation. 
This Lagrangian is string theory-motivated and $\phi$ describes the low-energy radial dynamics of a D3-brane in a warped throat: $f(\phi)^{-1}$ is the warped brane tension and $V(\phi)$ the interaction field potential. In this model the non-Gaussianity is dominated by derivative interactions of the field perturbations so that we do not need to take into account mixing with gravity. 
An estimate of the non-Gaussianity is given by the ratio between the third-order and the second order Lagrangians, respectively ${\cal L}_3$ and ${\cal L}_2$, divided by the amplitude of scalar fluctuations. This gives $f_{\mathrm{NL}} \sim  ({{\cal L}_3}/{{\cal L}_2}) \Phi^{-1} \sim -1/{c_s^2}$, where $c_s^2 = [1+ 2 X (\partial^2 {\cal L} / \partial X^2) /(\partial {\cal L} / \partial X)]^{-1} <1$ is the speed of sound of linear fluctuations and we have assumed that this is small, as it is the case for DBI inflation.
Thus, the non-Gaussianity can be quite large if $c_s \ll 1$. 

However, this signal vanishes in the squeezed limit due to the derivative interactions. More precisely, the particular momentum configuration of the bispectrum is very well described by
\begin{eqnarray}
B_{\zeta}(k_1,k_2,k_3) =  6 f_{\mathrm{NL}}^{\mathrm{equil}}  \bigg(
\frac{P_\zeta(k_1) P_\zeta(k_2)}{2 }  + \frac{\left[ P_\zeta(k_1)
    P_\zeta(k_2) P_\zeta(k_3) \right]^{\frac{2}{3}}}{3} \nonumber\\ 
- P_\zeta(k_1)^{\frac{1}{3}} P_\zeta(k_2)^{\frac{2}{3}} P_\zeta(k_3) + 5 \ {\mathrm{perms.}} \bigg) \,,
\end{eqnarray}
where, up to numerical factors of order unity, $f_{\mathrm{NL}}^{\mathrm{equil}}
\simeq -1/ c_s^2 $. The function of momenta inside the parenthesis is the
\emph{equilateral} shape \cite{Creminelli-etal:2006}, a \emph{template} used to approximate a large class of inflationary models. It is
defined in such a way as to be factorisable, maximized for equilateral configurations and vanishing  in the squeezed limit faster than the local shape, see Eq.~(\ref{local_sl}).

To compare  two shapes $F_1$ and $F_2$, it is useful to define a 3-dimensional scalar product between them as \cite{Babich-Creminelli-Zaldarriaga:2004}
\begin{equation}
F_1 \cdot F_2 = \sum F_1 (k_1,k_2,k_3) F_2 (k_1,k_2,k_3) / (P_\zeta(k_1)P_\zeta(k_2)P_\zeta(k_3)) \,,
\end{equation}
where the sum is over all configurations forming a triangle. Then, $\cos \theta = F_1 \cdot F_2/ \sqrt{(F_1 \cdot F_1)(F_2 \cdot F_2)}$ defines a quantitative measure of how much two shapes ``overlap'' and their signal is correlated. The cosine is small between the local and equilateral shapes. Two shapes with almost vanishing cosine are said to be orthogonal and any estimator developed to be sensitive to a particular shape will be completely blind to its orthogonal one. Note that the observable signal could actually be a combination of different shapes. For instance, multi-field models base on the DBI action \cite{Langlois-etal:2008} can generate a linear combination of local and equilateral non-Gaussianities \cite{Renaux-Petel:2009}.

The interplay between theory and observations, reflected in the relation
between derivative interactions and the shape of non-Gaussianity, has
motivated the study of inflation according to a new approach, the \emph{effective field theory of inflation} (\cite{Cheung-etal:2008}; see also
\cite{Weinberg:2008}). Inflationary models can be viewed as effective field
theories in presence of symmetries. Once symmetries are defined, the
Lagrangian will contain each possible operator respecting such symmetries.
As each operator leads to a particular non-Gaussian signal, constraining
non-Gaussianity directly constrains the coefficients in front of these
operators, similarly to what is done in high-energy physics with particle accelerators. For instance, the operator ${\cal L}_3$ discussed in the context of DBI inflation leads to non-Gaussianity  controlled by the speed of sound of linear perturbations. This operator can be quite generic in single field models. Current constraints on  non-Gaussianity allow to constrain the speed of sound of the inflaton field during inflation to be $c_s \ge 0.01$ \cite{Cheung-etal:2008,Senatore-Smith-Zaldarriaga:2010}.
Another well-studied example is ghost inflation
\cite{Arkani-Hamed-etal:2004b}, based on the ghost condensation, a model
proposed by~\cite{Arkani-Hamed-etal:2004} to modify gravity in the infrared. This model is motivated by shift symmetry and exploits the fact that in the limit where this symmetry is exact, higher-derivative operators play an important role in the dynamics, generating large non-Gaussianity with approximately equilateral shape. 

Following this approach has allowed to construct operators or
combination of operators leading to new shapes, orthogonal to the
equilateral one. An example of such a shape is the \emph{orthogonal}
shape proposed in~\cite{Senatore-Smith-Zaldarriaga:2010}. This shape
is generated by a particular combination of two operators already
present in DBI inflation. It is peaked both on equilateral-triangle
configurations and on flattened-triangle configurations (where the two
lowest-$k$ sides are equal exactly to half of the highest-$k$ side) --
the sign in this two limits being opposite. The orthogonal and
equilateral are not an exhaustive list. For instance,
\cite{Creminelli-etal:2010} have shown that the presence in the
inflationary theory of an approximate Galilean symmetry (proposed
by~\cite{Nicolis:2008in} in the context of modified
gravity) generates third-order operators with two derivatives on each
field. A particular combination of these operators produces a shape
that is approximately orthogonal to the three shapes discussed above.

Non-Gaussianity is also sensitive to deviations from the initial
adiabatic Bunch--Davies vacuum of inflaton fluctuations. Indeed,
considering excited states over it, as done in
\cite{Chen-etal:2007,Holman-Tolley:2008,Meerburg-vanderSchaar-Corasaniti:2009},
leads to a shape which is maximized in the collinear limit,
corresponding to enfolded or squashed triangles in momentum space,
although one can show that this shape can be written as a combination
of the equilateral and orthogonal ones
\cite{Senatore-Smith-Zaldarriaga:2010}.

At present, Planck 2015 data are compatible with Gaussian statistics and 
set tight constraints on primordial non-Gaussian shapes. The constraints at $68 \%$ CL are: 
$f_\textrm{NL}^\textrm{local}=2.5 \pm 5.7$ (after subtraction of the ISW-lensing contribution), $f_\textrm{NL}^\textrm{equil}= - 16 \pm 70$, 
and $f_\textrm{NL}^\textrm{ortho}= - 34 \pm 33$ \citep{2015arXiv150201592P}.

\subsection{Beyond shapes: scale dependence and  the squeezed limit}

There is a way out to generate large non-Gaussianity in single-field inflation. Indeed, one can temporarily break scale-invariance, for instance by introducing features in the potential as in~\cite{Chen-Easther-Lim:2007}. This can lead to large non-Gaussianity typically associated with scale-dependence. These signatures could even teach us something about string theory. Indeed, in axion monodromy, a model recently proposed by~\cite{Silverstein-Westphal:2008} based on a particular string compactification mechanism,  the inflaton potential is approximately linear, but periodically modulated. These modulations lead to tiny oscillations in the power spectrum of cosmological fluctuations and to large non-Gaussianity (see for instance~\cite{Flauger:2010J}).

This is not the only example of scale dependence.
While in general the amplitude of the non-Gaussianity signal  is considered constant, there are  several models, beside the above example,  which predict a scale-dependence. For example models like the Dirac--Born--Infeld (DBI) inflation, e.g.,~\cite{Alishahiha-Silverstein-Tong:2004,Chen:2005a,Chen2005b,Bean-etal:2008} can be characterized by a primordial bispectrum whose amplitude varies significantly over the range of scales accessible by cosmological probes.  

In view of measurements from observations it is also worth considering
the so-called squeezed limit of non-Gaussianity that is the limit in
which one of the momenta is much smaller than the other
two. Observationally this is because some probes (like, for example, the halo bias Section~\ref{sec:halobias}, accessible by large-scale structure surveys like Euclid) are sensitive to this limit. Most importantly, from the theoretical point of view, there are consistency relations valid in this limit that identify different classes of inflation, e.g.,~\cite{Creminelli:2012ed,Creminelli:2011rh} and references therein.

Similar relationships can also be derived for multi-field models (e.g. Byrnes et al. 2012, Elliston et al. 2012). Particularly interesting is the so-called Suyama-Yamaguchi inequality (e.g. Suyama \& Yamaguchi 2008, Smith, LoVerde \& Zaldarriaga 2011, Assassi, Baumann \& Green 2012, Kehegias \& Riotto 2012, Beltran Almeida, Rodriguez \& Valenzuela-Toledo 2013) which connects the amplitudes of the  bi- and tri-spectra in models where non-Gaussianity is seeded by light fields different from the inflaton. Observational tests of the inequality could then provide information regarding the multi-field nature of the inflationary process. Forecasts for the Euclid weak-lensing survey have been presented by Grassi et al. (2013).

The scale dependence of non-gaussianity, the shapes of non-gaussianity and the  behavior of the squeezed limit are all  promising avenues, where the combination of CMB data and large-scale structure surveys such as Euclid can provide powerful constraints as illustrated, e.g., in~\cite{Sefusattietal:2009, Norena-etal:2012, Sefusatti-etal:2012}.

\subsection{Beyond inflation}

As explained above, the search of non-Gaussianity could represent a unique
way to rule out the simplest of the inflationary models and distinguish
between different scenarios of inflation. Interestingly, it could also open
up a window on new scenarios, alternative to inflation. There have been
numerous attempts to construct  models alternative to inflation able to
explain the initial conditions of our universe. In order to solve the
cosmological problems and generate large-scale primordial fluctuations, most
of them require a phase during which observable scales today have exited the
Hubble size. This can happen in bouncing cosmologies, in which the present
era of expansion is preceded by a contracting phase. 
Examples are the pre-big bang \cite{Gasperini-Veneziano:1993} and the ekpyrotic scenario \cite{Khoury-etal:2001}. 

In the latter, the 4-d effective dynamics corresponds to a cosmology driven by a scalar field 
with a steep exponential potential $V(\phi) = \exp(-c \phi)$, with $c \gg 1$.
Leaving aside the problem of the realization of the bounce, it has been
shown that the adiabatic mode in this model generically leads to a steep blue spectrum 
for the curvature perturbations \cite{Lyth:2002,Brandenberger:2001bs,Creminelli-Zaldarriaga:2005}. 
Thus, at least a second field is required to generate an almost scale-invariant 
spectrum of perturbations \cite{Finelli:2002,Creminelli-Senatore:2007,Buchbinder-Khoury-Ovrut:2007,Koyama-Wands:2007,Fertig:2013kwa}. 
If two fields are present, both with exponential potentials and steepness coefficients $c_1$ and $c_2$, the non-adiabatic 
component has negative mass and acquires a quasi invariant spectrum of fluctuations with tilt 
$n_s-1 = 4(c_1^{-2} + c_2^{-2})$, with $c_1,c_2 \gg 1$. Then one needs to convert the non-adiabatic 
fluctuation into curvature perturbation, similarly to what the curvaton
mechanism does. If the two fields $\phi$ and 
$\chi$ are coupled through the 
kinetic term $\exp(-b \phi) g^{\mu \nu} \partial_\mu \partial_\nu \chi$ \citep{DiMarco:2002eb} then nearly 
scale invariant perturbations are generated when $b \sim c$ and $\chi$ has no potential \citep{Fertig:2013kwa}.

As the Hubble rate increases during the collapse, one expects nonlinearities in the fields to become more and more important, 
leading to non-Gaussianity in the produced perturbations. As nonlinearities grow larger on super-Hubble scales, one expects 
the signal to be of local type. The particular amplitude of the non-Gaussianity in the observable curvature perturbations 
depends on the conversion mechanism from the non-adiabatic mode to the observable perturbations. The tachyonic instability 
itself can lead to a phase transition to an ekpyrotic phase dominated by just one field $\phi_1$. 
If the two fields have both an exponential potential, ~\cite{Koyama-etal:2007} have found that 
$f_{\mathrm{NL}}^{\mathrm{local}} = -(5/12)c_1^2$. 
Current constraints on $f_{\mathrm{NL}}^{\mathrm{local}}$ (Planck 2015 data imposes 
$f_{\mathrm{NL}}^{\mathrm{local}} = 2.5 \pm 5.7$ 
at 68\% CL after subtraction of the ISW-lensing contribution
\cite{planck2013-p09a}) gives a value at odd with 
Planck 2015 measurements of the spectral index.
In fact in this model, 
even  for $f_{\mathrm{NL}}=-10$, $c_2\simeq 5$ which implies a too large value of the scalar spectral index  
$n_s-1 > 0.17$) which is  excluded by Planck measurements.
Thus, one needs to modify the potential to accommodate a red spectrum or consider alternative conversion mechanisms to change the 
value of the generated non-Gaussianity \cite{Buchbinder-Khoury-Ovrut:2008,Lehners-Steinhardt:2008}. 
If the kinetic term of the second field is coupled to the first one as in \citep{Fertig:2013kwa,Ijjas:2014fja}, then the generated 
non-Gaussianity is compatible with Planck 2015 constraints on $f_{\mathrm{NL}}$.

\newpage
\section{Primordial Non-Gaussianity and Large-Scale Structure}
\label{sec:NG-LSS}

As we have seen, even the simplest inflationary models predict deviations from Gaussian initial conditions. 
Confirming or ruling out the simplest inflationary model 
is an important goal and in this section we will show how Euclid can help achieving this.
Moreover, Euclid data (alone or in combination with CMB experiments like Planck) can be used to 
explore the primordial bispectrum and thus explore the interaction of the fields during inflation.
  
\subsection{Constraining primordial non-Gaussianity and gravity from 3-point statistics}

Contrary to CMB research which mainly probes the high-redshift universe, 
current studies of the LSS focus on data at much lower redshifts and are 
more heavily influenced by cosmic evolution.
Even if the initial conditions were Gaussian,
nonlinear evolution due to gravitational instability generates a non-zero bispectrum
for the matter distribution.  
The  first non-vanishing term in perturbation theory (e.g., 
\cite{Catelan-etal:1995}) gives
\begin{equation}
B(\vk_1,\vk_2,\vk_3)=2(P(k_1) P(k_2)J(\vk_1,\vk_2)+  {\mathrm{cyclic\,\,
permutations}})
\end{equation}
where   $J(\vk_1,\vk_2)$ is the gravitational instability ``kernel" which depends very weakly on 
cosmology and for an Einstein-de-Sitter universe assumes the form:
\begin{equation}
J(\vk_1,\vk_2)=\frac{5}{7}+\frac{\vk_1\cdot \vk_2}{2k_1k_2}\left(\frac{k_1}{k_2}+\frac{k_2}{k_1}\right)+\frac{2}{7}\left(\frac{\vk_1\cdot \vk_2}{k_1 k_2}\right)^2\,.
\end{equation}
This  kernel represents  the ``signature'' of gravity as we know it on the large-scale structure
of the universe. Either a modification of the gravitational law or the introduction of 
a coupling between dark matter and another component (say dark energy) 
would alter the bispectrum shape from the standard form.
The volume covered by Euclid will enable us to exploit this.

It was recognized a decade ago \citep{Verde-etal:2000} that the contribution to the matter
bispectrum generated by gravitational instability
is large compared to the fossil signal due to primordial non-Gaussianity
and that the primordial signal ``redshifts away''  compared to the gravitational signal. 
In fact, primordial non-Gaussianity of the local type
would affect the late-time dark matter density bispectrum with a 
contribution of the form
\begin{equation}
B^{\fnl\, {\mathrm{local}}}(\vk_1,\vk_2,\vk_3,z)=
2(\fnl P(k_1)P(k_2)\frac{{\cal F}(\vk_1,\vk_2)}{D(z)/D(z=0)}+  {\mathrm{cyclic\,\,
permutations}}).\end{equation}
where $D(z)$ is the linear growth function which in an Einstein--de~Sitter universe goes like $(1+z)^{-1}$ and 
\begin{equation}
{\cal F}= \frac{{\cal M}(k_3)}{{\cal M}(k_1){\cal M}(k_2)}\,; \,\,\,{\cal M}(k)=\frac{2}{3}\frac{k^2 T(k)}{H_0^2\Omega_{m,0}}\,,
\end{equation}
$T(k)$ denoting the matter transfer function, \label{symbol:Tk} 
$H_0$ the Hubble constant and $\Omega_{m,0}$ the matter density parameter.
Clearly the contributions due to primordial non-Gaussianity and gravitational instability 
have different scale and redshift dependence and the two 
kernel shapes in configuration space are different, thus,  making the two components, at least 
in principle and for high signal-to-noise, separable.
This is particularly promising for high-redshift probes of the matter distribution
like the 21-cm background which should simultaneously provide competing measures of
$f_{\mathrm{NL}}$ and a test of the gravitational law \citep{Pillepichetal2007}.
Regrettably,
these studies require using a long-wavelength radio telescope 
above the atmosphere (e.g., on the Moon) and will certainly come well after Euclid.

Galaxy surveys do not observe the dark matter distribution directly. However,
dark matter halos are believed to host galaxy formation, and different galaxy types at different 
redshifts are expected to populate halos in disparate ways 
\citep{Magliocchetti-Porciani:2003,Zehavietal2005}.
A simple (and approximate) way to account for galaxy biasing is to assume that
the overdensity in galaxy counts can be written as a truncated power expansion
in terms of the mass overdensity (smoothed on some scale):
$\delta_g(x)=b_1\delta_{\mathrm{DM}}(x)+b_2(\delta_{\mathrm{DM}}^2-\langle \delta_{\mathrm{DM}}^2 \rangle)$.
The linear and quadratic bias coefficient $b_1$ and $b_2$ are assumed to be scale-independent 
(although this assumption must break down at some point) but they can vary with redshift
and galaxy type. Obviously, a quadratic bias will introduce non-Gaussianity even on an initially 
Gaussian field. In summary, for local non-Gaussianity and scale-independent quadratic bias we 
have \citep{Verde-etal:2000}:
\begin{equation}
B(\vk_1,\vk_2,\vk_3,z)=2 P(k_1)P(k_2) b_1(z)^3\times  \left[ \fnl \frac{{\cal F}(\vk_1,\vk_2)}{D(z)} + J(\vk_1,\vk_2) +\frac{b_2(z)}{2 b_1(z)}\right]+ \mathrm{cyc.}\\ 
\end{equation}
Before the above expression can be compared against observations, it needs to be further 
complicated to account for redshift-space distortions and shot noise. 
Realistic surveys use galaxy redshifts as a proxy for distance, but gravitationally-induced 
peculiar velocities distort the redshift-space galaxy distribution. 
At the same time, the discrete nature of galaxies gives rise to corrections 
that should be added to the bispectrum computed in the continuous limit.
We will not discuss these details here as including redshift space distortions and shot noise 
will not change the gist of the message.

From the observational point of view, it is important to note that photometric surveys
are not well suited for extracting a primordial signal out of the galaxy bispectrum.
Although in general they can cover larger volumes than spectroscopic surveys, 
the projection effects due to the photo-z smearing along the line-of-sight is 
expected to suppress significantly the sensitivity of the measured bispectrum to the shape of 
the primordial one (see e.g.,~\cite{Verde-Heavens-Matarrese:2000}).
\cite{Sefusatti-Komatsu:2007}
have shown that, if the evolution of the bias parameters
is known a priori, spectroscopic surveys like Euclid would be able to give 
constraints on the $f_{\mathrm{NL}}$ parameter that are competitive with CMB studies.
While the gravitationally-induced non-Gaussian signal in the bispectrum has been detected  
to high statistical significance (see, e.g., \cite{Verde02, Kulkarni-etal:2007} and references 
therein),  the identification of nonlinear biasing (i.e., $b_2\neq 0$) is still controversial, 
and there has been so far no detection of any extra (primordial) bispectrum contributions.

Of course, one could also consider higher-order correlations. One of the advantages of considering,
 e.g., the trispectrum is that, contrary to the bispectrum, it has very weak nonlinear growth 
\cite{Verde-Heavens:2001}, but it has the disadvantage that the signal is de-localized: the number
of possible configurations grows fast with the dimensionality $n$ of the $n$-point function!  

Finally, it has been proposed to 
measure the level of primordial non-Gaussianity using 
Minkowski functionals applied either to the galaxy distribution or the weak
lensing maps (see, e.g., \cite{Hikageetal2008, Munshietal2011} and references therein).
The potentiality of this approach compared to more traditional methods
needs to be further explored in the near future. 
 
\subsection{Non-Gaussian halo bias}
\label{sec:halobias}

The discussion above neglects an important fact which went unnoticed until year 2008:
the presence of small non-Gaussianity can have a large effect on the clustering of dark matter 
halos \citep{Dalal-etal:2008,Matarrese-Verde:2008}. The argument goes as follows.
The clustering of the peaks in a Gaussian random field is completely specified by the field power 
spectrum. Thus, assuming that halos form out of linear density peaks, 
for Gaussian initial conditions the clustering of the dark matter halos is 
completely specified by the linear matter power spectrum.
On the other hand, for a non-Gaussian field, the clustering of the peaks depends on all 
higher-order correlations, not just on the power spectrum. Therefore, for non-Gaussian initial 
conditions,  the clustering of dark matter halos  depends on the 
linear bispectrum (and higher-order moments).

One can also understand the effect in the peak-background-split framework:
overdense patches of the (linear) universe collapse to form dark matter halos if their 
overdensity lies above a critical collapse threshold.  
Short-wavelength modes define the overdense patches while the long-wavelength modes 
determine the spatial distribution of the collapsing ones by modulating their height above and 
below the critical threshold.  
In the  Gaussian case, long- and short-wavelength modes are uncorrelated, 
yielding  the well  known linear, scale-independent peak bias. In the non-Gaussian case, 
however, long and short wavelength modes are coupled, yielding a different spatial pattern of 
regions that cross the collapse threshold.

In particular, for primordial non-Gaussianity of the local type, 
the net effect is that the halo distribution on very large scales relates to the underlying
dark matter in a strongly scale-dependent fashion.
For $k\lesssim 0.02\,h\mathrm{\ Mpc}^{-1}$,
the effective linear bias parameter scales as $k^{-2}$. 
\citep{Dalal-etal:2008,Matarrese-Verde:2008,Giannantonio-Porciani:2010}.
This is because the halo overdensity depends not only
on the underlying matter density but also on the value of the auxiliary Gaussian potential $\phi$
\cite{Giannantonio-Porciani:2010}. 

\epubtkImage{Ngpk.png}{%
\begin{figure}[htbp]
\centerline{\includegraphics[width=0.75\textwidth]{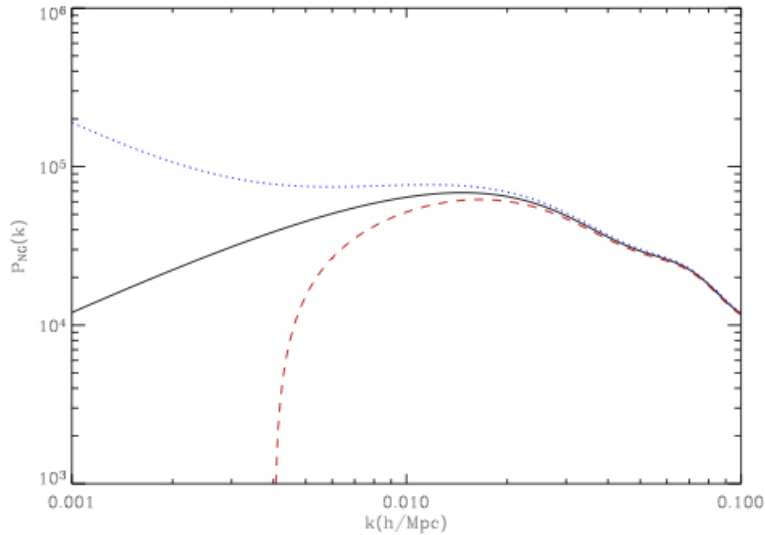}}
\caption{For illustration purposes this is the effect of a local
  $f_{\mathrm{NL}}$ of $\pm 50$ on the $z=0$ power spectrum of halos with
  mass above $10^{13}\,M_{\odot}$.} 
\label{fig:Pkng}
\end{figure}}

The presence of this effect is extremely important for observational studies 
as it allows to detect primordial non-Gaussianity from 2-point
statistics of the galaxy distribution like the power spectrum.
Combining current LSS data gives constraints on $f_{\mathrm{NL}}$ which are comparable to the \newtext{pre-Planck} CMB ones
 \citep{Slosar-etal:2008,Xia-etal:2010,Giannantonio-etal:2013}. Similarly, planned galaxy surveys are expected
to progressively improve upon existing limits
\citep{Carbone-Verde-Matarrese:2008,Carbone-Mena-Verde:2010,Giannantonioetal2010}. 
For example, Euclid could reach an error on $\fnl$ of $\sim 5$ (see below for further details)
which is comparable with the BPol forecast errors.  

The scale dependence of the halo bias changes considering different
shapes of primordial non-Gaussianity \citep{Schmidt-Kamion:2010,Wagneretal2010}. 
For instance, orthogonal and folded models produce an effective bias that scales as $k^{-1}$
while the scale dependence becomes extremely weak for equilateral models.\footnote{A clarification is in order here regarding the usage of the term ``orthogonal". Senatore et al. (2010) introduced two different bispectrum shapes with the same denomination: their equation (53) generates the $k^{-1}$ dependence in the linear bias parameter while their equation (113), which provides the correct large-scale behavior, causes a much weaker effect. Albeit unrelated to the orthogonal template, quasi-single-field models (Chen et al. 2007) also give a $k^{-1}$ scaling of the bias.}
Therefore, measurements of the galaxy power spectrum on the largest possible scales
have the possibility to constrain the shape and the amplitude of primordial non-Gaussianity
and thus shed new light on the dynamics of inflation.

On scales comparable with the Hubble radius, matter and halo clustering
are affected by general-relativity effects: 
the Poisson equation gets a quadratic correction that acts effectively as a non-zero local
$\fnl$ \citep{Bartoloetal2005,Pillepichetal2007}.
This contribution is peculiar to the inflationary initial conditions because it 
requires perturbations on super-horizon scales and it is mimicked in the halo bias by a 
local $\fnl=-1.6$ \citep{Verde-Matarrese:2009}. This is at the level of detectability by 
a survey like Euclid.
 
\subsection{Number counts of nonlinear structures}
\label{sec:massfn}

Even a small deviation from Gaussianity in the initial conditions can have a strong impact on those 
statistics which probe the tails of the linear density distribution. 
This is the case for the abundance of the most extreme  
nonlinear objects existing at a given cosmic epoch, massive dark matter halos and voids,
as they correspond to the highest and lowest density peaks (the rarest events) in the
underlying linear density field.

Thus small values of $f_{\mathrm{NL}}$ are potentially detectable by measuring 
the abundance of massive dark matter halos as traced by galaxies and galaxy clusters
at $z \gtrsim 1$ 
\citep{Matarrese-Verde-Jimenez:2000}. This approach has recently received  
renewed attention (e.g., \cite{Loverde-etal:2008,Grossi-etal:2009,Pillepich-Porciani-Hahn:2010, 
Maggiore-Riotto:2010,Damico-etal:2010,Verdereview:2010,Pillepich-Porciani-Reiprich2011} and references therein)
and might represent a promising tool for Euclid science.
In Euclid, galaxy clusters at high redshift can be identified either by lensing studies or by
building group catalogs based on the spectroscopic and photometric galaxy data. The main
challenge here is to determine the corresponding halo mass with sufficient accuracy to allow
comparison with the theoretical models. 

While galaxy clusters form at the highest overdensities of the primordial density field and 
probe the high-density tail of the PDF, voids form in the low-density regions and thus probe the 
low-density tail of the PDF. Most of the volume of the evolved universe is underdense, so 
it seems interesting to pay attention to the distribution of underdense regions.  
For the derivation of the non-Gaussian void probability function one proceeds in parallel to the 
treatment for halos with the only subtlety that the critical threshold is not negative and that 
its numerical value depends on the precise definition of a void (and may depend on the 
observables used to find voids), e.g., \cite{Kamionkowski-Verde-Jimenez:2009}.
Note that while a positive skewness ($\fnl>0$) boosts the number of halos at the high mass end 
(and slightly suppress the number of low-mass halos), it is a negative skewness that will 
increase the voids size distribution at  the largest voids end (and slightly decrease it for 
small void sizes).  In addition voids may probe slightly larger scales than halos, making the 
two approaches highly complementary.

Even though a number of observational techniques to detect voids in galaxy surveys 
have been proposed
(see, e.g., \citet{Colbergetal2008} and references therein), the challenge here is to match
the theoretical predictions to a particular void-identification criterion based on a specific
galaxy sample. We envision that mock galaxy catalogs based on numerical simulations will
be employed to calibrate these studies for Euclid.

\subsection{Forecasts for Euclid}

A number of authors have used the Fisher-matrix formalism to
explore the potentiality of Euclid in determining the level and
the shape of primordial non-Gaussianity \citep{Carbone-Verde-Matarrese:2008,Carbone-Mena-Verde:2010,Giannantonioetal2010}. 
In what follows, unless specifically mentioned, we will focus on the local 
type of non-Gaussianity which has been more widely studied so far. 

The most promising avenue is exploiting the scale-dependent bias on very large scales in
studies of galaxy clustering at the two-point level.
Early Fisher forecasts for the Euclid redshift survey found that,
for a fiducial model with $f_{\mathrm{NL}}=0$, this gives a marginalized $1\sigma$ error
 on the nonlinearity parameter of $\Delta f_{\mathrm{NL}}\simeq 2$ \citep{Carbone-Verde-Matarrese:2008,Carbone-Mena-Verde:2010}.
Forecasts based on the most recent specifics for the Euclid surveys 
(see Table~\ref{tbl:EucludSpecfnl})
are presented
in~\cite{Giannantonioetal2010} and summarized in Table~\ref{tbl:EUCLIDFNL} 
below.
Updated values of the galaxy number counts and of the efficiency in measuring 
spectroscopic redshifts correspond to a marginalized $1\sigma$ error
of $f_{\mathrm{NL}} \simeq 4\mbox{\,--\,}5$ 
(depending a little on the detailed assumptions of the Fisher
matrix calculation), 
with a slightly better result obtained using the 
Euclid spectroscopic sample rather than the
photometric one
(complemented with multi-band ground-based photometry), 
at least for a fiducial value of $f_{\mathrm{NL}}=0$ \citep{Giannantonioetal2010}.
The forecast errors further improve by nearly a few per cent 
using Planck priors on the cosmological
parameters determined with the power spectrum of CMB temperature anisotropies.  

\begin{table}[htbp]
\caption[Specifications of the surveys used in the Euclid forecasts given
in Table~\ref{tbl:EUCLIDFNL}.]{Specifications of the surveys used in
  the Euclid forecasts given in Table~\ref{tbl:EUCLIDFNL}. The
  redshift distributions of the different galaxy samples are as in
  Section~\ref{sec:baofm_survey} (see also~\cite{Giannantonioetal2010}).}
\label{tbl:EucludSpecfnl}
\centering
{\small
\begin{tabular}{lcc}
\toprule
~ & Photometric survey  & Spectroscopic survey \\
\midrule
Surveyed area (deg$^2$)& 15,000 & 15,000 \\
Galaxy density (arcmin$^{-2}$)& 30 & 0.56 \\
Median redshift & 0.8 & 1.0 \\
Number of redshift bins & 12 & 12 \\
Redshift uncertainty $\sigma_z/(1+z)$ & 0.05 & 0.001 \\
Intrinsic ellipticity noise $\gamma$ & - & 0.247 \\
Gaussian linear bias param. & $\sqrt{1+z}$ & $\sqrt{1+z}$ \\
\bottomrule
\end{tabular}}
\end{table}

The amplitude and shape of the matter power spectrum in the mildly 
nonlinear regime depend (at a level of a few per cent)
on the level of primordial non-Gaussianity \citep{Taruyaetal2008,Pillepich-Porciani-Hahn:2010,Giannantonio-Porciani:2010}. Measuring this signal with the Euclid weak-lensing survey 
gives $\Delta f_{\mathrm{NL}}\simeq 70$ (30 with Planck priors) \citep{Giannantonioetal2010}. 
On the other hand, counting nonlinear structures in terms of 
peaks in the weak-lensing maps (convergence or shear)
should give limits in the same ballpark 
(\cite{Marianetal2010} find $\Delta f_{\mathrm{NL}}=13$
assuming perfect knowledge of all the cosmological parameters).

Finally, by combining lensing and angular power spectra 
(and accounting for all possible cross-correlations)
one should achieve $\Delta f_{\mathrm{NL}}\simeq 5$
(4.5 with Planck priors) \citep{Giannantonioetal2010}. 
This matches what is expected from both the Planck mission and 
the proposed BPol satellite.

Note that the forecast errors on $f_{\mathrm{NL}}$ are somewhat sensitive to 
the assumed fiducial values of the galaxy bias. 
In our study we have adopted the approximation 
$b(z)=\sqrt{1+z}$ \citep{rassat08}. On the other hand, using 
semi-analytic models of galaxy formation, \citet{orsi10} found
bias values which are nearly 10\,--\,15\% lower at all redshifts.
Adopting this slightly different bias, 
the constraint on $f_{\mathrm{NL}}$ already degrades by 50\% with respect to our 
fiducial case. 

Euclid data can also be used to constrain the scale dependence of the 
nonlinearity parameter (see Table~\ref{tbl:Euclidfnlrun}).
To this purpose, we consider a local model of primordial non-Gaussianity where
\begin{equation}
f_{\mathrm{NL}}=f_{\mathrm{NL}}^{\mathrm{(piv)}}\cdot \left( \frac{k}{k_{\mathrm{piv}}}\right)^{n_{f_{\mathrm{NL}}}} \,,
\end{equation}
 with fiducial values $k_{\mathrm{piv}}=0.02\, h\mathrm{\ Mpc}^{-1}$, 
$f_{\mathrm{NL}}^{\mathrm{(piv)}}=30$, and $n_{f_{\mathrm{NL}}}=0$.
In this case, the combination of lensing and clustering data gives 
$\Delta \alpha_{\mathrm{s,m}}=0.18$ (0.14 with Planck priors) 
and $\Delta f_{\mathrm{NL}}^{\mathrm{(piv)}}\simeq 9$ (7 with Planck priors)
\citep{Giannantonioetal2010}. These constraints are similar to what is
expected from future studies of the CMB bispectrum with Planck 
\citep{Sefusattietal:2009}.

\begin{table}[htb]
\caption[Forecast $1\sigma$ errors for the nonlinearity parameter 
$f_{\mathrm{NL}}$ based on two-point statistics (power spectra)
of the Euclid redshift and weak-lensing surveys.]{Forecast $1\sigma$
  errors for the nonlinearity parameter $f_{\mathrm{NL}}$ based on
  two-point statistics (power spectra) of the Euclid redshift and
  weak-lensing surveys. Results are obtained using the Fisher-matrix
  formalism and marginalizing over eight cosmological parameters
  ($\Omega_\Lambda$, $\Omega_m$, $\Omega_b$, $h$, $n_s$, $\sigma_8$,
  $w_0$, $w_a$) plus a large number of nuisance parameters to account
  for galaxy biasing, nonlinear redshift-space distortions and shot
  noise (see~\cite{Giannantonioetal2010} for details). Results within
  parentheses include the forecast priors for the cosmological
  parameters from the power spectrum of CMB temperature anisotropies
  measured with the Planck satellite (note that no prior is assumed on
  $f_{\mathrm{NL}}$).
The label ``Galaxy clustering'' refers to the anisotropic power
spectrum $P(k_\parallel,k_\perp)$ for spectroscopic data and to 
the angular power spectrum $C_\ell$ for photometric data.
The combined analysis of clustering and lensing data is based on
angular power spectra and includes all possible cross-correlations between
different redshift bins and probes.
nonlinear power spectra are computed using the halo model.
This introduces possible inaccuracies in the forecasts for weak lensing
data in the equilateral and orthogonal shapes (see main text for details).}
\label{tbl:EUCLIDFNL}
\centering
{\small
\begin{tabular}{ccccc}
\toprule
Bispectrum shape  & local & orthogonal & equilateral \\
Fiducial $f_{\mathrm{NL}}$ & 0 & 0 & 0 \\
\midrule
Galaxy clustering (spectr. $z$) & 4.1 (4.0) & 54 (11) & 220 (35) \\
Galaxy clustering (photom. $z$) & 5.8 (5.5) & 38 (9.6)& 140 (37) \\
Weak lensing                    & 73 (27)   & 9.6 (3.5) & 34 (13) \\
Combined                        & 4.7 (4.5) & 4.0 (2.2) & 16 (7.5)\\
\bottomrule
\end{tabular}}
\end{table}

\begin{table}[htb]
\caption[Forecast $1\sigma$ errors for a scale-dependent local model
of primordial non-Gaussianity.]{Forecast $1\sigma$ errors for a
  scale-dependent local model of primordial non-Gaussianity 
\citep{Giannantonioetal2010}. 
Details of the forecasts are as in the previous Table~\ref{tbl:EUCLIDFNL}.}
\label{tbl:Euclidfnlrun}
\centering
{\small
\begin{tabular}{ccc}
\toprule
& $\Delta f_{\mathrm{NL}}^{\mathrm{(piv)}}$ & $\Delta n_{f_{\mathrm{NL}}}$ \\
\midrule
Galaxy clustering (spectr. $z$) & 9.3 (7.2) & 0.28 (0.21) \\
Galaxy clustering (photom. $z$) & 25 (18)   & 0.38 (0.26) \\
Weak lensing                    & 134 (82)  & 0.66 (0.59)   \\
Combined                        & 8.9 (7.4) & 0.18 (0.14) \\
\bottomrule
\end{tabular}}
\end{table}

In the end, we briefly comment on how well Euclid data could constrain
the amplitude of alternative forms of primordial non-Gaussianity than
the local one. In particular, we consider the equilateral and orthogonal shapes
introduced in Section~\ref{nongaussianshapes}.
Table~\ref{tbl:EUCLIDFNL} summarizes the resulting constraints on 
the amplitude of the primordial bispectrum, $f_{\mathrm{NL}}$. 
The forecast errors from galaxy clustering grow larger and larger
when one moves from the local to the orthogonal and finally to the equilateral
model.
This reflects the fact that the scale-dependent part of the galaxy bias
for $k\to 0$
approximately scales as $k^{-2}$, $k^{-1}$, and $k^0$ 
for the local, orthogonal, and equilateral shapes, respectively  
\citep{Schmidt-Kamion:2010,Wagneretal2010,Scoccimarroetal2011,DJS2011,DJSb2011}. 
On the other hand, 
the lensing constraints (that, in this case, come from the 
very nonlinear scales) 
appear to get much stronger for the non-local shapes.
A note of caution is in order here.
In~\cite{Giannantonioetal2010}, the nonlinear matter power spectrum
is computed using a halo model which has been tested
against $N$-body simulations only for non-Gaussianity of the local
type.%
\epubtkFootnote{Very few $N$-body simulations of the non-local models
  are currently available and none of them has very high spatial
  resolution.}
In consequence, the weak-lensing forecasts might be less reliable than
in the local case (see the detailed discussion
in~\citet{Giannantonioetal2010}). This does not apply for the
forecasts based on galaxy clustering which are always robust as they
are based on the scale dependence of the galaxy bias on very large
scales.

\subsection{Complementarity} 

The CMB bispectrum is very sensitive to the shape of non-Gaussianity; 
halo bias and mass function, the most promising approaches to constrain $\fnl$ with a survey 
like Euclid, are much less sensitive.
However, it is the complementarity between CMB and LSS that matters. 
One could  envision different scenarios. 
If non-Gaussianity is local with negative $\fnl$ and CMB 
obtains a detection, then the halo bias approach should 
also give a high-significance detection (GR correction and 
primordial contributions add up), while if it is local but 
with positive $\fnl$, the halo-bias approach could give a 
lower statistical significance as the GR correction contribution has the opposite sign. 
If CMB detects $\fnl$ at the level of  10 and a form 
that is close to local, but halo bias does not detect it, 
then the CMB bispectrum is given by secondary effects  (e.g.,~\cite{Mangilli-Verde:2009}). 
If CMB detects non-Gaussianity that is not of the local 
type, then halo bias can help discriminate between equilateral and enfolded shapes: if halo bias sees a signal, it  indicates the enfolded type, and if halo bias does not see  a signal, it indicates the equilateral type. Thus even a  non-detection of the halo-bias effect, in combination with 
CMB constraints, can have an important discriminative  power.

\section{Isocurvature modes}
\label{sec:isocurvature}

At some time well after inflation but deep into the radiation era the universe
is filled with several components. For instance, in the standard picture right
before recombination there are four components: baryons, cold dark matter,
photons and neutrinos. One can study the distribution of \emph{super-Hubble}
fluctuations between different species, which represent the initial conditions
for the subsequent evolution. So far we have investigated mostly the adiabatic initial conditions;
in this section we explore more generally the possibility of isocurvature initial conditions.
Although CMB data are the most sensitive to constrain isocurvature perturbations,
we discuss here the impact on Euclid results.

\subsection{The origin of isocurvature perturbations}

Let us denote by $\rho_\alpha$ the energy density of the component $\alpha$. Perturbations are purely adiabatic when for each component $\alpha$ the quantity $\zeta_\alpha \equiv - 3 H \delta \rho_\alpha/\dot \rho_\alpha$ is the same \cite{Weinberg:2003,Malik-Wands-Ungarelli:2003}. Let us consider for instance cold dark matter and photons. When fluctuations are adiabatic it follows that $\zeta_{\mathrm{cdm}} = \zeta_\gamma$. Using the energy conservation equation, $\dot \rho_\alpha = - 3 H (\rho_\alpha + p_\alpha)$ with $p_{\mathrm{cdm}}=0$ and $p_\gamma = \rho_\gamma/3$, one finds that the density contrasts of these species are related by
\begin{equation}
\frac{\delta \rho_{\mathrm{cdm}}}{\rho_{\mathrm{cdm}}} = \frac34 \frac{\delta \rho_\gamma}{\rho_\gamma} \,.
\end{equation}
Using that $n_{\mathrm{cdm}} \propto \rho_{\mathrm{cdm}}$ and $n_\gamma \propto
\rho_\gamma^{3/4}$, this also implies that particle number ratios
between these species is fixed, i.e., $\delta(n_{\mathrm{cdm}}/n_\gamma) = 0$. 

When isocurvature perturbations are present, the condition described
above is not satisfied.%
\epubtkFootnote{Strictly speaking, isocurvature perturbations are
  defined by the condition that their total energy density in the
  total comoving gauge vanishes, i.e., $\sum_\alpha \delta \rho^{({\mathrm{com.}})}_\alpha=0$. Using the relativistic Poisson equation, one can
  verify that this implies that they do not contribute to the
  ``curvature'' potential.}
In this case one can define a non-adiabatic or entropic perturbation
between two components $\alpha$ and $\beta$ as ${\cal S}_{\alpha,
  \beta} \equiv \zeta_\alpha -\zeta_\beta$, so that, for the example
above one has
\begin{equation}
{\cal S}_{{\mathrm{cdm}},r } = \frac{\delta \rho_{\mathrm{cdm}}}{ \rho_{\mathrm{cdm}}} - \frac34 \frac{\delta \rho_\gamma}{ \rho_\gamma} = \frac{\delta (n_{\mathrm{cdm}}/n_\gamma)}{n_{\mathrm{cdm}}-n_\gamma} \,.
\end{equation}

A sufficient condition for having purely adiabatic perturbations is that all
the components in the universe were created by a
single degree of freedom, such as during reheating after single field
inflation.%
\epubtkFootnote{In this case in the flat gauge one finds, for each
  species $\alpha$, $\zeta_\alpha = \zeta$, where $\zeta$ is the
  Bardeen curvature perturbation conserved on super-Hubble scales.}
Even if inflation has been driven by several fields, thermal equilibrium may erase isocurvature perturbations if it is established before any non-zero conserving quantum number was created (see~\cite{Weinberg:2004}). Thus, a detection of non-adiabatic fluctuations would imply that several scalar fields where present during inflation \emph{and} that either some of the species were not in thermal equilibrium afterwards or that some non-zero conserving quantum number was created before thermal equilibrium. 

The presence of many fields is not unexpected. Indeed, in all the
extension of the Standard Model scalar fields are rather
ubiquitous. In particular, in String Theory dimensionless couplings
are functions of moduli, i.e., scalar fields describing the compactification. 
Another reason to consider the relevant role of a second field other than the 
inflaton is that this can allow to circumvent the necessity of slow-roll (see, e.g.,~\cite{Dvali-Kachru:2003}) 
enlarging the possibility of inflationary models. 

Departure from thermal equilibrium is one of the necessary conditions for the generation of baryon asymmetry and 
thus of the matter in the universe. Interestingly, the oscillations and decay of a scalar field requires departure 
from thermal equilibrium. Thus, baryon asymmetry can be generated by this process; examples are the decay of a 
right-handed sneutrino \cite{Hamaguchi-Murayama-Yanagida:2002} or the~\cite{Affleck-Dine:1985} scenario. 
If the source of the baryon-number asymmetry in the universe is the condensation of a scalar field after 
inflation, one expects generation of baryon isocurvature perturbations \cite{Moroi-Takahashi:2001}. 
This scalar field can also totally or partially generate adiabatic density perturbations through the curvaton mechanism.

In summary, given our ignorance about inflation, reheating, and the generation of matter in the universe, 
a discovery of the presence of isocurvature initial conditions would have radical implications on both the 
inflationary process and on the mechanisms of generation of matter in the universe. 

Let us concentrate on the non-adiabatic perturbation between cold dark matter (or baryons, which are also non-relativistic) and radiation ${\cal S} = {\cal S}_{{\mathrm{cdm}}, \gamma}$. Constraints on the amplitude of the non-adiabatic component are given in terms of the parameter $\alpha$, defined at a given scale $k_0$, by $P_{\cal S}-P_\zeta \equiv \alpha-(1-\alpha)$, see e.g.,~\cite{Beltran-etal:2004,Bean-Dunkley-Pierpaoli:2006,KomatsuWMAP7}. As discussed in~\cite{Langlois:1999}, adiabatic and entropy perturbations may be correlated. To measure the amplitude of the correlation one defines a cross-correlation coefficient, $\beta \equiv  - P_{{\cal S}, \zeta} / \sqrt{P_{\cal S} P_\zeta}$. Here $P_{{\cal S}, \zeta}$ is the cross-correlation power-spectrum between ${\cal S}$ and $\zeta$ and for the definition of $\beta$ we have adopted the sign convention of~\cite{KomatsuWMAP7}.
Observables, such as for instance the CMB anisotropies, depend on linear combinations of $\zeta$ and ${\cal S}$. Thus, constraints on $\alpha$ will considerably depend on the cross-correlation coefficient $\beta$ (see for instance discussion in~\cite{Gordon-Lewis:2003}).

If part of the cold dark matter is created out of equilibrium from a field other than the inflaton, 
totally uncorrelated isocurvature perturbations, with $\beta=0$, are produced, as discussed for 
instance in~\cite{Efstathiou-Bond:1986,Linde-Mukhanov:1997}. The axion is a well-known example of such a field. 
The axion is the Nambu--Goldstone boson associated with the
\cite{Peccei-Quinn:1977} mechanism to solve the strong-CP problem in QCD. As
it acquires a mass through QCD non-perturbative effects, when the Hubble
rate drops below its mass the axion starts oscillating coherently, behaving
as cold dark matter \cite{preskill1983,abbott1983,dine1983}.
During inflation, the axion is practically massless and acquires fluctuations 
which are totally uncorrelated from photons, produced by the inflaton decay 
\cite{Seckel-Turner:1985,Linde:1985,Linde:1991,Turner-Wilczek:1991}. 
Planck data 2015 set a tight constrain on CDM uncorrelated isocurvature fluctuations:
$\alpha_{\beta=0} < 0.037$ at 95\% CL \citep{2015arXiv150202114P}.

Totally uncorrelated isocurvature perturbations can also be produced in the curvaton mechanism, if the dark matter or 
baryons are created from inflation, before the curvaton decay, and remain decoupled from the product of curvaton 
reheating \cite{Langlois-Vernizzi-Wands:2008}. This scenario is ruled out if the curvaton is entirely responsible 
for the curvature perturbations. 
However, in models when the final curvature perturbation is a mix of the inflaton 
and curvaton perturbations \cite{Langlois-Vernizzi:2004}, such an entropy contribution is still allowed.

When dark matter or baryons are produced solely from the curvaton decay,
such as discussed by~\cite{Lyth-Ungarelli-Wands:2003}, the isocurvature
perturbations are totally anti-correlated, with $\beta=-1$ \footnote{Note that this 
convention is the opposite of the one used in \cite{planck2013-p17}, where ``fully correlated'' 
means that the cross-correlation term contributes constructively
to the Sachs-Wolfe component of the total temperature spectrum.}. 
 The Planck 2015 constrains the fraction of this isocurvature CDM 
modes to $\alpha_{\beta=0} < 0.0018$ at 95\% CL \citep{2015arXiv150202114P},
which implies that the curvaton has decayed in CDM when it contributed to most
of the energy content of the Universe.

For instance, some fraction of the curvaton decays 
to produce CDM particles or the out-of-equilibrium curvaton decay generates the primordial baryon asymmetry 
\cite{Hamaguchi-Murayama-Yanagida:2002,Affleck-Dine:1985}.  
 
If present, isocurvature fields are not constrained by the slow-roll conditions imposed 
on the inflaton field to drive inflation. Thus, they can be highly 
non-Gaussian~\cite{Linde-Mukhanov:1997,Bernardeau-Uzan:2002}. 
Even though negligible in the two-point function, their presence could be detected in the three-point 
function of the primordial curvature and isocurvature perturbations and their cross-correlations, 
as studied in~\cite{Kawasaki-etal:2008,Langlois-Vernizzi-Wands:2008}.

\subsection{Constraining isocurvature perturbations}

Even if pure isocurvature models have been ruled out, current observations allow for mixed adiabatic and isocurvature contributions (e.g.,~\cite{CrottyEtAl2003,Trotta:2006ww,KomatsuWMAP7, Valiviita-Giannantonio:2009}). 
As shown in~\cite{Trotta-Riazuelo-Durrer:2001,2002PhRvL..88u1302A,Valiviita-Giannantonio:2009,Langlois-Riazuelo:2000, Bucher-Moodly-Turok:2001,Sollom-Challinor-Hobson:2009}, the initial conditions issue is a very delicate problem: in fact, for current cosmological data, relaxing the assumption of adiabaticity reduces our ability to do precision cosmology since it compromises the accuracy of parameter constraints.   
 Generally, allowing for isocurvature modes introduces new degeneracies in
 the parameter space which weaken  constraints considerably.
  
The cosmic microwave background radiation (CMB), being our window on
the early universe, is the preferred data set to learn about initial
conditions. Up to now, however, the CMB temperature power spectrum
alone, which is the CMB observable better constrained so far, has not
been able to  break the degeneracy between the nature of initial
perturbations (i.e., the amount  and properties of an  isocurvature
component) and cosmological parameters,
e.g.,~\cite{Kurki-etal:2005,Trotta-Riazuelo-Durrer:2001}. Even if the
precision measurement of the CMB first acoustic peak at $\ell \simeq
220$  ruled out the possibility of a dominant isocurvature mode,
allowing for isocurvature perturbations together with the adiabatic
ones introduce additional degeneracies in the interpretation of the
CMB data that current experiments could not break. Adding external
data sets somewhat alleviates the issue for some degeneracy
directions,
e.g.,~\cite{Trotta:2002iz,Beltran-etal:2004,Dunkley-etal:2005}. As
shown in~\cite{Bucher-Moodly-Turok:2001}, the precision polarization
measurement of the next CMB experiments like Planck will be crucial to
lift such degeneracies, i.e., to distinguish the effect of the
isocurvature modes from those due to the variations of the
cosmological parameters.
  
It is important to keep in mind that analyzing the CMB data with the
prior assumption of purely
adiabatic initial conditions when the real universe contains even a
small isocurvature contribution,
could lead to an incorrect determination of the cosmological
parameters and on the inferred value
of the sound horizon at radiation drag. The sound horizon at radiation
drag is the standard ruler
that is used to extract information about the expansion history of the
universe from measurements
of the baryon acoustic oscillations.
Even for a CMB experiment like Planck, a small but non-zero
isocurvature contribution, still allowed by Planck data - see \cite{planck2013-p17} - , if ignored,
can introduce a systematic error in the interpretation of the BAO signal that is comparable if not
larger than the statistical errors.
In fact, \cite{Mangilli-Verde-Beltran:2010} shows that even a tiny amount of
isocurvature perturbation, if not accounted for, could  affect
standard rulers calibration from  CMB observations such as those
provided by the Planck mission, 
affect BAO interpretation, and introduce biases in the recovered dark energy
properties that are larger than forecast statistical errors from future surveys.
In addition it will introduce a mismatch of the expansion history as
inferred from CMB and as measured by BAO surveys. The mismatch between
CMB predicted and the measured expansion histories has been proposed
as a signature for deviations from a DM cosmology in the form of
deviations from Einstein's gravity (e.g.,~\cite{Acquaviva-Verde:2007, Ishak-etal:2006}), couplings in 
the dark sector (e.g.,~\cite{Lopez-etal:2010}) or time-evolving dark energy.

For the above reasons, extending on the work of~\cite{Mangilli-Verde-Beltran:2010}, \cite{Carbone_etal2011} adopted a general fiducial cosmology
which includes a varying dark energy equation of state parameter and
curvature. In addition to BAO measurements, in this case the information from the shape of the galaxy power spectrum
are included and a joint analysis of a Planck-like CMB probe and a
Euclid-type survey is considered.
This allows one to break the degeneracies that affect the
CMB and BAO combination. As a result, most of the cosmological
parameter systematic biases arising from an incorrect assumption on
the isocurvature fraction parameter $f_{\mathrm{iso}}$, become negligible with
respect to the statistical errors.   The combination of CMB and LSS gives a statistical
error $\sigma(f_{\mathrm{iso}}) \sim 0.008$,
even when curvature and a varying dark energy  equation of state are
included, which is smaller than the error obtained from CMB  alone when flatness and cosmological constant are
assumed. These results confirm the synergy and complementarity
between  CMB and LSS, and the great potential of   future  and planned galaxy surveys.

\section{Summary and outlook}

We have  summarized  aspects of the initial conditions  for the growth of
cosmological perturbations that Euclid will enable us to probe. In particular we
have considered the shape of the primordial power spectrum and its connection to
inflationary models, primordial non-Gaussianity and isocurvature perturbations.

A survey like Euclid will greatly improve our knowledge of the initial
conditions for the growth of perturbations and  will help shed light on the
mechanism  for the generation of primordial perturbations.
The addition of Euclid data will improve the Planck satellite's cosmic microwave
background constraints on parameters describing  the shape of the primordial
power spectrum by a factor of 2\,--\,3.  

Primordial non-Gaussianity can be tested by Euclid in three different and
complementary ways: via the galaxy bispectrum, number counts of nonlinear
structures and the non-Gaussian halo bias.  These approaches are also highly 
competitive with and complementary to CMB constraints. In combination with
Planck, Euclid will not only test a possible scale-dependence of non-Gaussianity
but also its shape. The shape of non-Gaussianity  is the  key to constrain and
classify possible deviations for the simplest single-field slow roll inflation. 

Isocurvature modes affect the  interpretation of large-scale structure
clustering in two ways.
The power spectrum shape is modified on small scales due to the extra
perturbations although this effect however can be mimicked by  scale-dependent
bias. More importantly  isocurvature modes can lead to an incorrect inferred
value for the sound horizon at radiation drag from CMB data. This then predicts
an incorrect   location of the baryon acoustic feature. It is through this
effect that  Euclid BAO measurements improve constraints on isocurvature modes.

\newpage
\part{Testing the Basic Cosmological Hypotheses}
\label{testing}
 \section{Introduction}

The standard cosmological analyses implicitly make several
assumptions, none of which are seriously challenged by current data.
Nevertheless, Euclid offers the possibility of testing some of these
basic hypotheses. Examples of the standard assumptions are that
photon number is conserved, that the Copernican principle holds
(i.e., we are not at a special place in the universe) and that the
universe is homogeneous and isotropic, at least on large enough
scales. These are the pillars on which standard cosmology is built,
so it is important to take the opportunity offered by Euclid
observations to test these basic hypotheses.

\section{Photon Number Conservation, Transparency and the Etherington Relation} 
\label{transparency-and-Etherington-relation}

There are several examples of non-standard---but nevertheless theoretically 
well-motivated---physical processes in which photon number is not conserved. 
A non-exhaustive list includes a non perfectly transparent Universe, decaying 
vacuum cosmologies/photon injection mechanisms, models in which the fine-structure 
constant varies, physically motivated scenarios where photons mix with other particles 
(such as axions), modified gravity scenarios, and so on. There are two basic ways 
to observationally constrain these scenarios: the temperature-redshift relation and 
the distance duality (or Etherington) relation.

The Etherington relation \citep{Etherington:1933} implies that, in a cosmology based on a metric theory 
of gravity, distance measures are unique: the luminosity distance is $(1 + z)^2$ times the 
angular diameter distance. This is valid in any cosmological background where photons 
travel on null geodesics and where, crucially, photon number is conserved. 

On the other hand, if the expansion of the Universe is adiabatic and the CMB spectrum 
was a black-body at the time it originated, this shape will be preserved with its 
temperature evolving as $T(z) = T_0 (1 + z)$. This is a robust prediction of standard 
cosmology, but it is violated in many non-standard models, including scenarios involving 
photon mixing/violation of photon number conservation. 

An optical/IR survey like Euclid will only be able to test the distance duality relation. 
However, we argue here, complementary information from the CMB temperature-redshift relation, 
which will be available by the year 2020, can greatly enhance the interpretation of Euclid results.

Specific scenarios in which the Etherington relation would be violated include deviations 
from a metric theory of gravity, photons not travelling along unique null geodesics, 
variations of fundamental constants, etc. Following~\cite{Avgoustidis-etal:2009,Avgoustidis-etal:2010} 
generic deviations from the Etherington relation can be parametrized as:
\begin{equation}\label{symbol:tau}
D_L(z) = D_A(z)(1 + z)^{2+\epsilon}\,,
\end{equation}
with current constraints being at the few percent level~\cite{Avgoustidis-etal:2010}.
Similarly, deviations from the standard evolution of the CMB temperature with redshift 
can be parametrized phenomenologically by~\cite{Avgoustidis:2011aa}
\begin{equation}\label{symbol:Tzbeta}
T(z) = T_0 (1 + z)^{1-\beta}\,.
\end{equation}
At low redshifts (say $z < 1$), the CMB temperature can be measured via the Sunyaev-Zeldovich 
effect towards galaxy clusters~\cite{Luzzi:2009ae}, while at higher redshifts ($z > 1$) it can 
be obtained from high-resolution absorption spectroscopy of atomic, ionic or molecular levels 
excited by the photon absorption of the CMB radiation~\cite{Noterdaeme:2010tm}. 
Up to redshifts $z\sim 3$, current direct constraints on the parameter $\beta$ are at the few percent level \cite{Noterdaeme:2010tm}, while combining all currently available direct and indirect $T(z)$ measurements yields a sub-percent constraint on $\beta$ \cite{Avgoustidis:2015xhk}.

Therefore, in models where photon number is not conserved, both the temperature-redshift relation 
and the Etherington relation will be violated, providing two independent observables. Crucially, 
in a broad range of models the two violations are not independent: for example assuming that 
deviations from the standard behaviour only depend on redshift (but not on frequency) one can 
show~\cite{Avgoustidis:2011aa} that if
\begin{equation}
T(z) = T_0 (1 + z) f(z)
\end{equation}
then
\begin{equation}
D_L(z) = D_A(z)(1 + z)^2 f(z)^{3/2}\,.
\end{equation}
In particular, for the two simple parametrizations above, $\epsilon = - 3\beta/2$. 

Therefore, distance duality data can be used to constrain $\beta$ (i.e. deviations from the 
standard CMB temperature evolution), leading to a 40\% improvement in existing constraints, 
as discussed in~\cite{Avgoustidis:2011aa}.

\subsection{Transparency}

A change in the photon flux during propagation towards the Earth will affect
the supernovae (SNe) luminosity distance measures $D_L(z)$
\label{symbol:luminosity}but not the determinations of the angular diameter
distance. BAO will not be affected so $D_A(z)$ and $H(z)$ measurements
from BAO could be combined with supernovae measurements of $D_L(z)$ to
constrain deviations from photon number conservation. Photon conservation
can be violated by simple astrophysical effects or by exotic physics.
Amongst the former we find, for instance, attenuation due to interstellar
dust, gas and/or plasmas. Most known sources of attenuation are expected to
be clustered and can be typically constrained down to the 0.1\% level
\citep{Menard-etal:2008, More-etal:2009}.

Unclustered sources of attenuation
are however much more difficult to constrain. For example, 
grey dust \citep{Aguirre:1999} has been invoked to explain the
observed dimming of Type~Ia supernovae without resorting to cosmic
acceleration.
While the latter scenario has been ruled out by observations~\cite{Aguirre:1999uu,Bassett:2003vu}, it has been shown~\cite{Corasaniti:2006cv} that the effect of grey dust could cause an extinction as large as 0.08 mag at $z = 1.7$, thus potentially affecting dark energy parameter inference from future supernova surveys.
More exotic sources of photon conservation violation involve a
coupling of photons to particles beyond the standard model of particle
physics. Such couplings would mean that, while passing through the
intergalactic medium, a photon could disappear
 --or even (re)appear-- while interacting 
with such exotic particles, modifying the apparent luminosity of sources. 
Recently, \cite{Avgoustidis-etal:2010} considered the mixing 
of photons with scalars, known as axion-like particles, chameleons, and the 
possibility of mini-charged particles which have a tiny, and unquantized 
electric charge. In particular, the implications of these particles on the 
SN luminosity have been described in a number of 
publications~\citep{Csaki-etal:2002, Mortsell-etal:2002,Burrage:2008, Ahlers:2009} 
and a detailed discussion of the proposed approach can be found 
in~\cite{BassettKunz1:2004,BassettKunz2:2004,Avgoustidis-etal:2009,Avgoustidis-etal:2010}.

Any systematic violations in photon conservation can then be interpreted as
an opacity effect in the observed luminosity distance, parametrized through
a generic opacity parameter, $\tau(z)$, as: 
\begin{equation}
D_{L,\mathrm{obs}}^2=D^2_{L,\mathrm{true}}\exp[\tau(z)]\,. 
\end{equation}
Note that a negative $\tau(z)$
allows for apparent brightening of light sources, as would be the case, for
example, if exotic particles were also emitted from the source and converted
into photons along the line of sight~\citep{Burrage:2008}. 
For specific models of exotic matter-photon coupling, such as axion-like 
particles (ALPs), chameleons, and mini-charged particles (MCPs), the appropriate 
parametrization of $\tau(z)$ can be used~\cite{Avgoustidis-etal:2010}.
In order to discuss generic forecasts for Euclid, 
we return to the parametrization of~\cite{Avgoustidis-etal:2009}, 
equation (\ref{symbol:tau}) above.

Forecast Euclid constraints are shown in Figure~\ref{fig:transparency}, taken 
from~\cite{Avgoustidis-etal:2010},
and adapted to also show the corresponding constraint on 
the temperature-redshift violation discussed in~\cite{Avgoustidis:2011aa}. 
The lower x-axis displays the parameter $\epsilon$ of equation (\ref{symbol:tau}), 
while the upper x-axis shows parameter $\beta$ of equation (\ref{symbol:Tzbeta}), 
assuming the linear relation $\epsilon = - 3\beta/2$. Current constraints are shown in 
blue, while the orange contours represent the expected improvement from Euclid, assuming 
it is accompanied by a supernova sample with the characteristic of a Dark Energy Task 
Force stage IV survey.

\epubtkImage{tau_omm_constr_SNBAOforec.png}{%
\begin{figure}[htbp]
\centerline{\includegraphics[width=.75\textwidth]{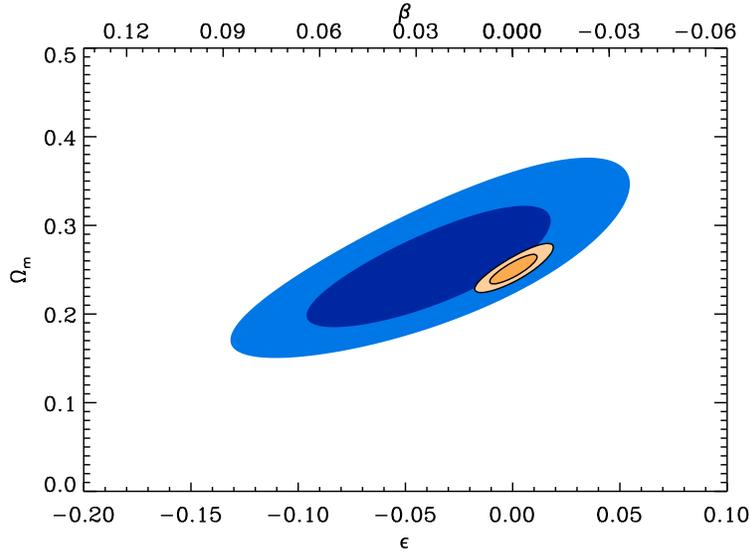}}
\caption{Constraints on possible violation of the Etherington
relation in the form of deviations from a perfectly transparent universe
($\epsilon=0$). The corresponding constraint on the parameter 
$\beta$, quantifying violations from the standard temperature-redshift relation, 
can be read in the upper x-axis. Blue regions represent current constraints 
while orange are forecast Euclid constraints assuming it is accompanied by a 
Dark Energy Task Force stage IV supernovae sample. Figure from \cite{Avgoustidis-etal:2010}.}
\label{fig:transparency}
\end{figure}}

If considering Euclid data in combination with SN brightness measurements, allowing 
for the coupling of scalar fields to photons will drastically weaken constraints on 
cosmological parameters. A remedy is to independently constrain photon number 
non-conservation arising from the possible coupling of photons to a putative dark 
energy/modified gravity scalar field. As shown in~\cite{Avgoustidis:2013bqa} and 
illustrated in Figure~\ref{fig:DEphotoncoupl} below, improved direct measurements 
of the CMB temperature at different redshifts (such as those expected in the coming 
years from ALMA and ESPRESSO, and ultimately ELT-HIRES) can be used in combination 
with distance measures and SN data to break parameter degeneracies and significantly 
improve constraints on physical processes in the early universe. Note in particular 
that the combined BAO+SN constraint is dominated by the BAO data if the coupling 
between photons and dark energy is included at the currently allowed level, but 
gets dominated by SN (and the combined constraint improves dramatically) if 
constraints on this coupling improve by more than one order of magnitude. 

In these forecasts we are assuming that the SN data will come from a SNAP-like or 
other DETF IV dataset. However, the need for this external dataset may be alleviated 
if Euclid carries out its own SN survey; this scenario is also discussed 
in~\cite{Avgoustidis:2013bqa}.

\epubtkImage{DEphotoncoupl.png}{%
\begin{figure}[htbp]
\centerline{\includegraphics[width=.5\textwidth]{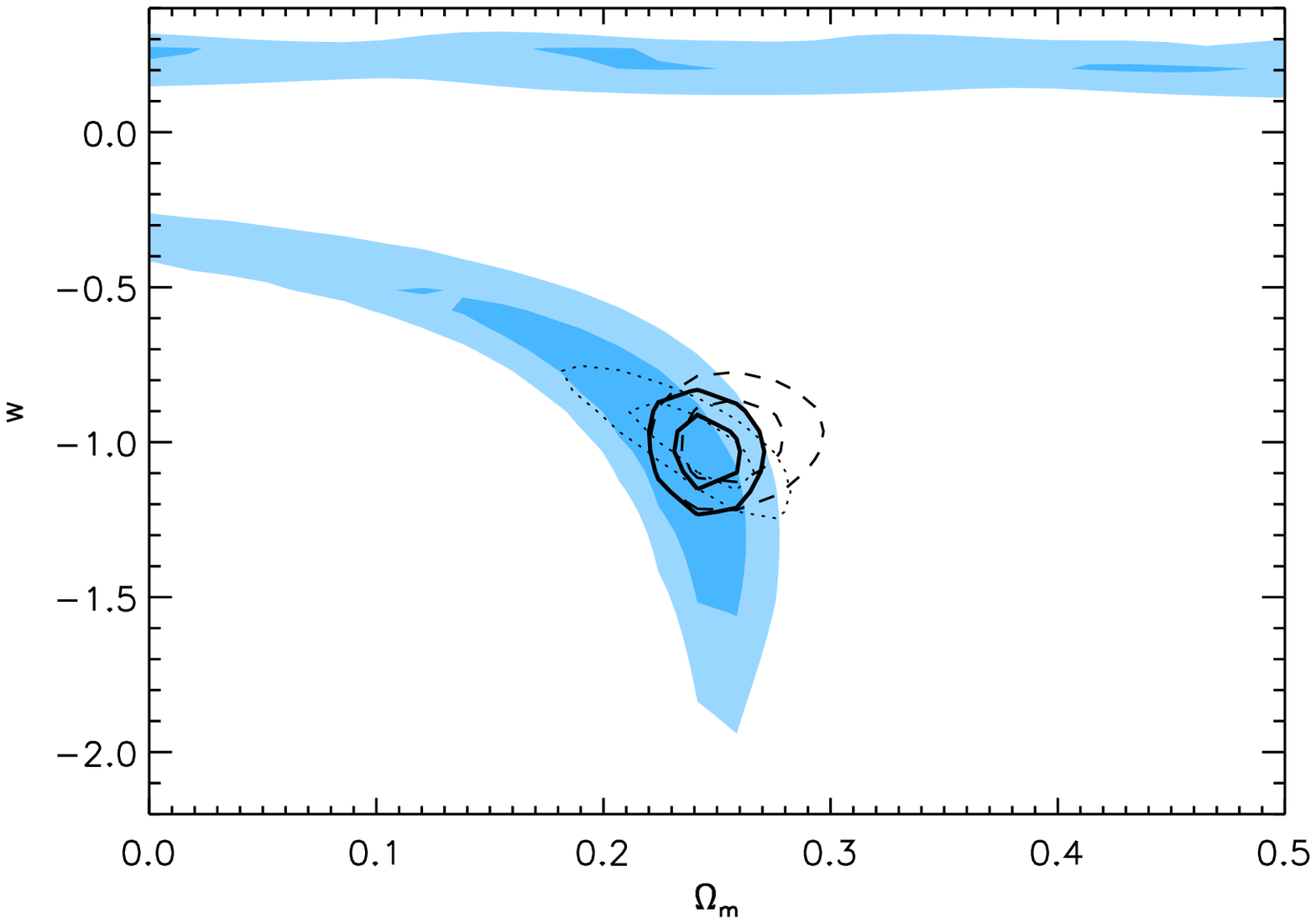}
            \includegraphics[width=.5\textwidth]{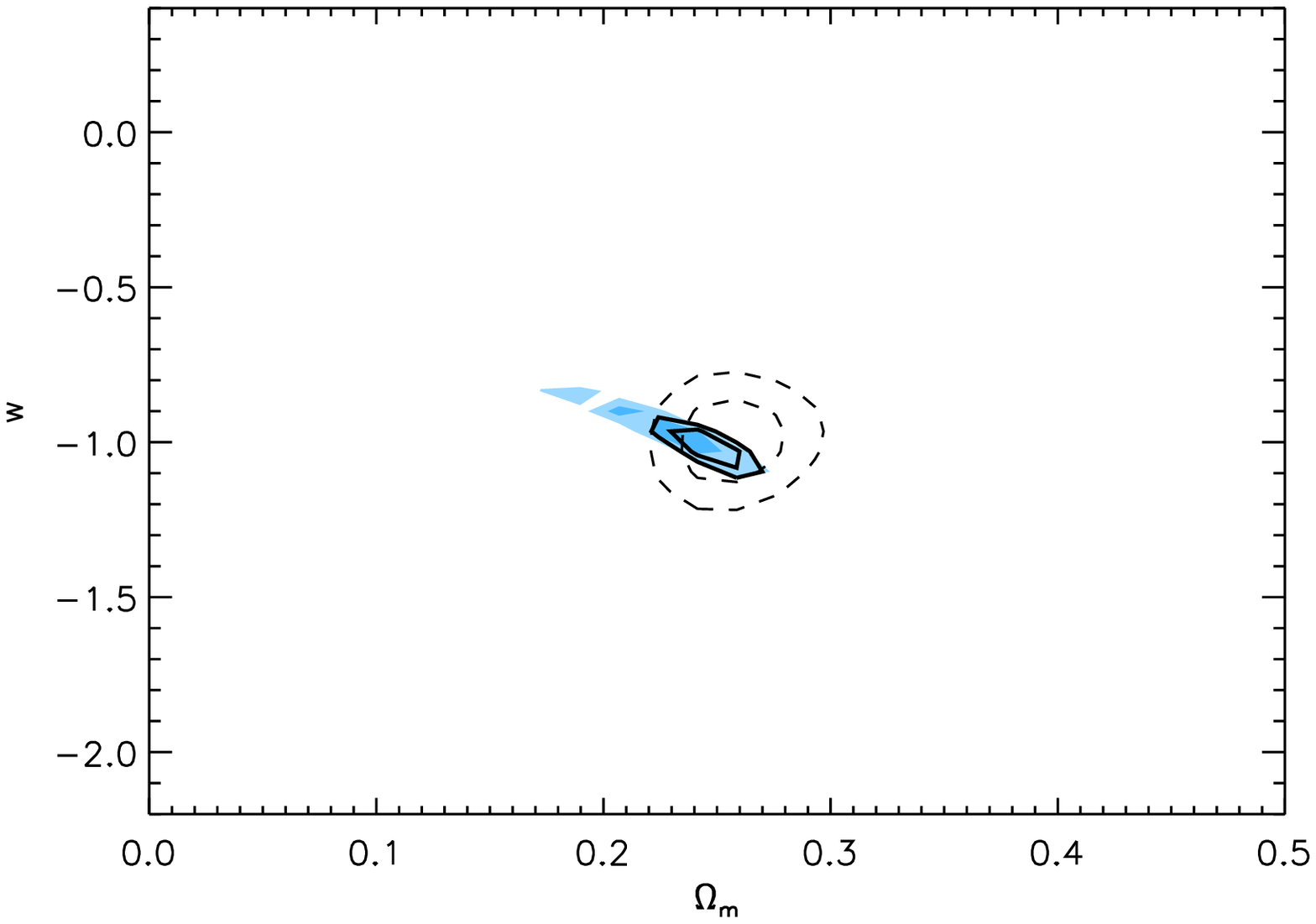}}
\caption{ Forecasted 68\% and 95\% likelihood contours for SN (filled blue), 
H(z) (dashed line transparent) and combined SN+H(z) (solid line transparent), assuming 
Euclid BAO is accompanied by a SNAP-like (or Dark Energy Task Force stage IV) supernova 
sample. We show constraints on the $\Omega_m-w$ plane, having marginalised over all other 
parameters in the context of flat wCDM models. On the left, we have allowed a coupling 
between photons and a putative dark energy scalar at the level allowed by current data, 
while on the right we have set this coupling to zero. The dotted contours on the left 
show SN contours assuming constraints on this coupling can be improved by an order of 
magnitude. Note how the joint contours become  dominated by the SN data if this coupling 
is strongly constrained. }
\label{fig:DEphotoncoupl}
\end{figure}}

\subsection{Axion-like particles}

Axion-like particles (ALP) can arise from field theoretic extensions of the standard model as Goldstone bosons when a global shift symmetry, present in the high energy sector, is spontaneously broken. Interestingly, these fields also arise naturally in string theory (for a review see~\cite{SvrcekWitten:2006}). Chameleon scalar fields are another very interesting type of ALPs \citep{Brax-etal:2010}. They were originally invoked to explain the current accelerated expansion of the universe with a quintessence field which can couple to matter without giving rise to large fifth forces or unacceptable violations of the weak equivalence principle. A chameleon model with only matter couplings will induce a coupling to photons.

The presence of ALPs will have an impact on observations of SNe if their
observed light passes through (intergalactic) magnetic fields. The net
effect depends on the ratio of the transition probability to the length
travelled through a magnetic field, and a parameter
$A$ describing the degree of thermalization of the initial flux ($A=1$ means
thermalized flux where the photon to ALP transition is compensated by the
inverse ALP to photon, making the photon number constant). For the simplest
ALP model $A=2/3$, the present and forecast constraints are shown in Figure~\ref{fig:axion2} taken from~\cite{Avgoustidis-etal:2010}. 

\epubtkImage{POmm_constrs_strongandweak_A23_SNBAOforec.png}{%
\begin{figure}[htbp]
  \centerline{\includegraphics[width=.75\textwidth]{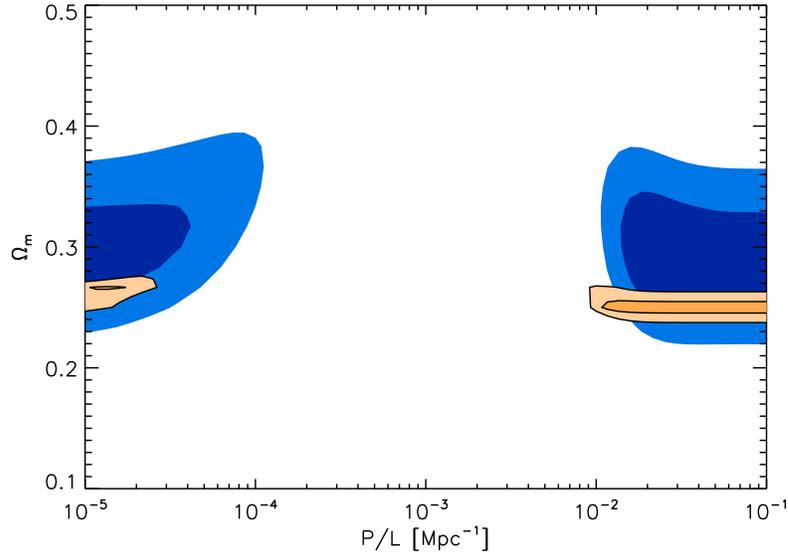}}
  \caption{Constraints on the simplest Axion-like particles models. 
Blue regions represent current constraints while orange are 
forecast
Euclid constraints assuming it is accompanied by a Dark Energy Task Force stage IV supernovae sample.
Here $P/L$ is the conversion probability per unit length
and is the relevant parameter for $\tau(z)$ (see~\cite{Avgoustidis-etal:2010}).}
  \label{fig:axion2}
\end{figure}}

\subsection{Mini-charged particles}

New particles with a small unquantized charge have been investigated in several 
extensions of the standard model \citep{Holdom:1986, BatellGherghetta:2006}. In particular, they arise naturally in 
extensions of the standard model which contain at least one additional U(1) hidden 
sector gauge group \citep{Holdom:1986, Bruemmer-etal:2009}. The gauge boson of this additional U(1) is known as 
a hidden photon, and hidden sector particles, charged under the hidden U(1), get an 
induced electric charge proportional to the small mixing angle between the kinetic terms 
of the two photons. In string theory, such hidden U(1)s and the required kinetic mixing 
are a generic feature \citep{Abel-etal:2008a, Abel-etal:2008b, Dienes-etal:1997, AbelSchofield:2004, Goodsell-etal:2009}. Hidden photons are not necessary however to 
explain mini-charged particles, and explicit brane-world scenarios have been constructed 
\citep{BatellGherghetta:2006} where MCPs arise without the need for hidden photons. 

More interestingly, \citet{Ahlers:2009,Gies-etal:2006,Ahlers-etal:2008} pointed
out that photons propagating in a background magnetic field can actually
pair-produce 
MCPs without the need for a second photon in the initial state.
The opacity in this case is parametrized by $\kappa y(z)$ where $y$ is the 
comoving distance to the source and $\kappa$ encloses information on the MCP electric
charge and the intervening magnetic field strength. 
Figure~\ref{fig:mcp} shows current and forecast Euclid's constraints, taken from
\cite{Avgoustidis-etal:2010} assuming Euclid is accompanied by a supernova
sample with the characteristic of a Dark Energy Task Force stage IV survey. 

\epubtkImage{Omm_vs_kappa_A0_SNBAOforec.png}{%
\begin{figure}[htbp]
  \centerline{\includegraphics[width=.75\textwidth]{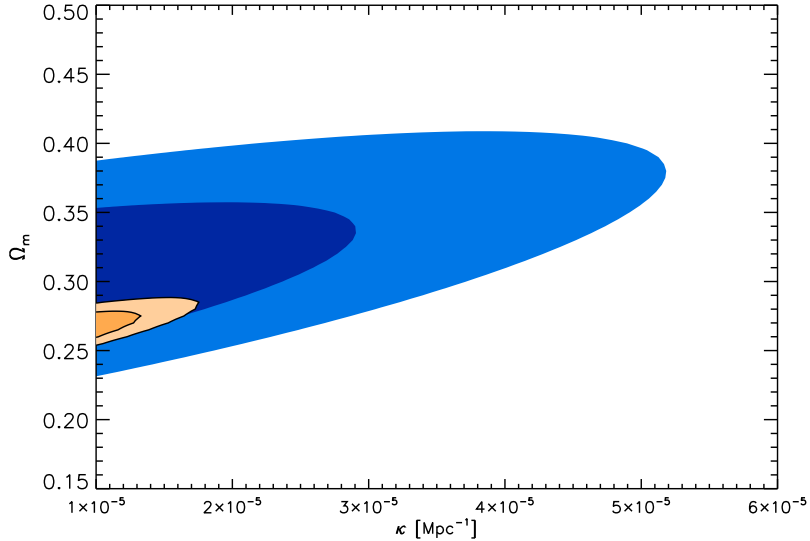}}
  \caption{Constraints on MCP models. Blue regions represent current  constraints while orange are forecast Euclid 
constraints assuming it is accompanied by a Dark Energy Task Force stage IV supernovae sample. Figure from \cite{Avgoustidis-etal:2010}.}
  \label{fig:mcp}
\end{figure}}

\section{Beyond homogeneity and isotropy}

The crucial ingredient that kickstarted dark energy research was the interpretation in 1998 of standard 
candle observations in terms of cosmic acceleration required to explain the data in the context of the 
FLRW metric. What we observe is however merely that distant sources ($z>0.3$) are dimmer than we 
would predict in a matter-only universe calibrated through ``nearby'' sources. That is, we observe a different 
evolution of luminosity rather than directly an increase in the expansion rate. Can this be caused by a 
strong inhomogeneity rather than by an accelerating universe?

In addition, cosmic acceleration seems to be a recent phenomenon at least for standard dark-energy 
models, which gives rise to the coincidence problem. The epoch in which dark energy begins to play a 
role is close to the epoch in which most of the cosmic structures formed out of the slow linear gravitational 
growth. We are led to ask again: can the acceleration be caused by strong inhomogeneities rather than 
by a dark energy component?

Finally, one must notice that in all the standard treatment of dark energy one always assumes a perfectly 
isotropic expansion. Could it be that some of the properties of acceleration depend critically on this assumption?

In order to investigate these issues, in this section we explore radical deviations from homogeneity and 
isotropy and see how Euclid can test them.

\subsection{Anisotropic models}
\label{sec:anisotropicmodels}

In recent times, there has been a resurgent interest towards anisotropic
cosmologies, classified in terms of Bianchi solutions to general relativity.
This has been mainly motivated by hints of anomalies in the cosmic
microwave background (CMB) distribution observed on the full sky by
the WMAP satellite
\citep{de-Oliveira:2004,2004ApJ...609...22V,2005MNRAS.356...29C,
2004ApJ...605...14E}
and Planck satellite
\cite{Ade:2013nlj,Ade:2013ydc,Ade:2013xla,Ade:2013vbw}.

While the CMB is very well described as a highly isotropic (in a statistical sense) Gaussian random 
field, and the anomalies are \textit{a posteriori} statistics and therefore their statistical significance
should be corrected at least for the so-called \textit{look elsewhere effect} (see, e.g., 
\cite{Pontzen:2010yw, Bennet:2011} and references therein), recent analyses have shown that local 
deviations from Gaussianity in some directions (the so called cold spots, 
see~\cite{2005MNRAS.356...29C,Ade:2013nlj}) 
cannot be excluded at high confidence levels. Furthermore, the CMB angular power spectrum 
extracted from the WMAP and Planck maps has shown in the past a quadrupole power lower than 
expected from the best-fit cosmological model \citep{2004MNRAS.348..885E,Ade:2013zuv}. 
Several explanations for this anomaly have been proposed (see, e.g.,
\cite{2003PhLB..574..141T,2003JCAP...09..010C,2003PhRvD..67j3509D,2007PhRvD..76f3007C,2007PhRvD..76h3010G})
including the fact that the universe is expanding with different velocities along different directions. 
While deviations from homogeneity and isotropy are constrained to be very small from cosmological 
observations, these usually assume the non-existence of anisotropic sources in the late universe. 
Conversely, as suggested 
in~\cite{2008JCAP...06..018K,2008ApJ...679....1K,2006PhRvD..74d1301B,2006PhRvD..73f3502C,2008arXiv0812.0376C},
dark energy with anisotropic pressure acts as a late-time source of anisotropy. Even if one considers 
no anisotropic pressure fields, small departures from isotropy cannot be excluded, and it is interesting
to devise possible strategies to detect them.

The effect of assuming an anisotropic cosmological model on the CMB pattern has been studied by
\cite{1973MNRAS.162..307C,1985MNRAS.213..917B,1995AA...300..346M,
1996AA...309L...7M,1996PhRvL..77.2883B,1997PhRvD..55.1901K}.
The Bianchi solutions describing the anisotropic line element were treated as small perturbations to a 
Friedmann--Robertson--Walker (FRW) background. Such early studies did not consider the possible 
presence of a non-null cosmological constant or dark energy and were upgraded \reviewtext{updated} recently 
by~\cite{2006MNRAS.369.1858M,2006ApJ...644..701J}.

One difficulty with the anisotropic models that have been shown to fit the large-scale CMB pattern, is 
that they have to be produced according to very unrealistic choices of the cosmological parameters. 
For example, the Bianchi VIIh template used in~\cite{2006ApJ...644..701J} requires an open universe, 
a hypothesis which is excluded by most cosmological observations. 

Bianchi models are described by homogeneous and anisotropic metrics. If anisotropy is slight, the 
dynamics of any Bianchi model can be decomposed into an isotropic FRW background linearly 
perturbed to break isotropy; on the other side, homogeneity is maintained with respect to three Killing 
vector fields. 
Anisotropy can also be studied by means of inhomogeneous models such as the Szekeres metric. In particular, the quasi-spherical Szekeres models feature a mass-density distribution which can be interpreted as a superposition of a mass monopole and a mass dipole (see~\cite{Bolejko:2011jc} and references therein), thus generalizing the case of the isotropic LTB metric (see Section~\ref{LTBmetric} below).

The geometry of Bianchi models is set up by the structure constants
$C^k_{ij}$, defined by the commutators of (these) three Killing fields
$\vec{\xi}_i$: 
\begin{equation}
\label{eq:killing}
\left[ \vec{\xi}_i, \vec{\xi}_j \right] = C^k_{ij} \vec{\xi}_k.
\end{equation}
The structure constants are subject to the antisymmetry relation $C^k_{ij} = - C^k_{ji}$ and the 
Jacobi identities $C^a_{[bc} C^d_{e]a}=0$. As a consequence, their attainable values are restricted 
to only four of the initial 27 necessary to describe a given space. In~\cite{1969CMaPh..12..108E} 
these four values are dubbed as $n_1, n_2, n_3$ and $a_1$ \reviewtext{$n_1, n_2, n_3$ and $a$}. The categorization of Bianchi models 
into different types relies on classifying the inequivalent sets of these four constants. In
Table~\ref{tab:bianchi} the subclass of interest containing the FRW limit is shown. Bianchi types 
VIIh and IX contain the open and closed FRW model, respectively. Type VII$_{0}$ contains the flat 
FRW; types I and V are just particular subcases of the VII$_{0}$ and VIIh. In type I no vertical motions 
are allowed and the only extension with respect to the FRW case is that there are three different scale 
factors. The metric in general can be written as
\begin{equation}
\label{eq:bianmetr}
g_{\mu\nu}=-n_{\mu}n_{\nu}+g_{ab}\xi_{\mu}^{a}\xi_{\nu}^{b},
\end{equation}
where $g_{ab}$ is a $3\times3$ metric depending on $t$. It can be decomposed as $g_{ab}=e^{2\alpha}
[e^{2\beta}]_{ab}$, where the first term represents the volumetric expansion and the second term includes 
the anisotropy.

\begin{table}
\caption{Bianchi models containing FRW limit and their structure
constants.}
\label{tab:bianchi}
\centering
\begin{tabular}{c|ccccl}
Type & $a$ & $n_1$ & $n_2$ & $n_3$ \\
\hline
I & 0 & 0 & 0 & 0 \\
V & $1$ & 0 & 0 & 0 \\
VII$_{0}$ & 0 & 0 & 1 & 1 \\
VII$_h$ & $\sqrt{h}$ & 0 & 1 & 1 \\
IX & 0 & 1 & 1 & 1 \\
\end{tabular}
\end{table}

\subsubsection{Late-time anisotropy}
\label{sec:ltanisotr}

While deviations from homogeneity and isotropy are constrained to be very small from cosmological 
observations, these usually assume the non-existence of anisotropic sources in the late universe. 
The CMB provides very tight constraints on Bianchi models at the time of
recombination~\citep{1996PhRvL..77.2883B,1997PhRvD..55.1901K,1995AA...300..346M}
of order of the quadrupole value, i.e., $\sim10^{-5}$. Usually, in standard cosmologies with a 
cosmological constant the anisotropy parameters scale as the inverse of the comoving volume. 
This implies an isotropization of the expansion from the recombination up to the present, leading
to the typically derived constraints on the shear today, namely $\sim10^{-9}\div10^{-10}$. However, 
this is only true if the anisotropic expansion is not generated by any anisotropic source arising after 
decoupling, e.g., vector fields representing anisotropic dark energy~\citep{2008ApJ...679....1K}.

An additional problem is that an inflationary phase -- required to explain a number of \reviewtext{features} of the \reviewtext{standard}
cosmological model -- isotropizes the universe very efficiently, leaving a residual anisotropy that is 
negligible for any practical application. These difficulties vanish if an anisotropic expansion takes 
place only well after the decoupling between matter and radiation, for example at the time of dark 
energy domination~\citep{2008JCAP...06..018K,2008ApJ...679....1K,2006PhRvD..74d1301B,
2006PhRvD..73f3502C,2008arXiv0812.0376C}. In these references it is proposed that dark energy 
with anisotropic pressure acts as a late-time source of anisotropy. 

It would be great if one could measure deviations from isotropic expansion. For example, the effect 
of cosmic parallax~\citep{2008arXiv0809.3675Q} has been recently proposed as a tool to assess the 
presence of an anisotropic expansion of the universe. It is essentially the change in angular
separation in the sky between far-off sources, due to an anisotropic expansion. 

A common parameterization of an anisotropically distributed dark energy component is studied in a 
class of Bianchi~I type, where the line element is
\begin{equation}
\mathrm{d}s^{2} = -\mathrm{d}t^{2}+a^{2}(t) \, \mathrm{d}x^{2}+b^{2}(t) \, \mathrm{d}y^{2}+c^{2}(t) \, \mathrm{d}z^{2} \,.
\label{metric}
\end{equation}
The expansion rates in the three Cartesian directions $x$, $y$ and $z$ are defined as $H_{X}=\dot{a}/a$, 
$H_{Y}=\dot{b}/b$ and $H_{Z}=\dot{c}/c$, where the dot denotes the derivative with respect to coordinate time. 
In these models they differ from each other, but in the limit of $H_{X}=H_{Y}=H_{Z}$ the flat FRW isotropic 
expansion is recovered. Among the Bianchi classification models the type I exhibits flat geometry and no 
overall vorticity; conversely, shear components $\Sigma_{X,Y,Z}=H_{X,Y,Z}/H-1$ are naturally generated, 
where $H$ is the expansion rate of the average scale factor, related to the volume expansion as 
$H=\dot{A}/A$ with $A=(abc)^{1/3}$.

The anisotropic expansion is caused by the anisotropically stressed dark energy fluid whenever its energy 
density contributes to the global energy budget. If the major contributions to the overall budget come from 
matter and dark energy, as after recombination, their energy-momentum tensor can be parametrized as:
\begin{eqnarray}
T_{(m)\nu}^{\mu} & = & \mbox{diag}(-1,w_{m},w_{m},w_{m})\rho_{m}\\
T_{(\mathrm{DE})\nu}^{\mu} & = &
\mbox{diag}(-1,w,w+3\delta,w+3\gamma)\rho_{\mathrm{DE}} \,,
\label{eq:Tmunu-de}
\end{eqnarray}
respectively, where $w_{m}$ and $w$ are the equation of state parameters of matter and dark energy and 
the skewness parameters $\delta$ and $\gamma$ can be interpreted as the difference of pressure along the
$x$ and $y$ and $z$ axis. Note that the energy-momentum tensor~\eqref{eq:Tmunu-de} is the most general 
one compatible with the metric~\eqref{metric}~\citep{2008ApJ...679....1K}. Two quantities are introduced to 
define the degree of anisotropic expansion:
\begin{equation}
\begin{aligned}R &
\,\equiv\,(\dot{a}/a-\dot{b}/b)/H\;=\;\Sigma_{x}-\Sigma_{y} \,,\\
S & \,\equiv\,(\dot{a}/a-\dot{c}/c)/H\;=\;2\Sigma_{x}+\Sigma_{y} \,.
\end{aligned}
\label{dom}
\end{equation}

Considering the generalized Friedmann equation, the continuity equations for matter and dark energy and 
no coupling between the two fluids, the derived autonomous system reads
\citep{2008JCAP...06..018K,2008ApJ...679....1K}:
\begin{equation}
\begin{aligned}
U'= & U(U-1)[\gamma(3+R-2S)\;+\,\delta(3-2R+S)\,+\,3(w-w_{m})]\\
S'= &
\frac{1}{6}(9-R^{2}+RS-S^{2})\big\{S[U(\delta+\gamma+w-w_{m})+w_{m}-1]-6\,
\gamma\, U\big\}\\
R'= &
\frac{1}{6}(9-R^{2}+RS-S^{2})\big\{R[U(\delta+\gamma+w-w_{m})+w_{m}-1]-6\,
\delta\, U\big\},
\end{aligned}
\label{sys}
\end{equation}
where $U\equiv\rho_{\mathrm{DE}}/(\rho_{\mathrm{DE}}+\rho_{m})$ and the derivatives are taken with 
respect to $\log(A)/3$.  System~(\ref{sys}) exhibits many different fixed points, defined as the solutions of the 
system $S'=R'=U'=0$. Beside the Einstein--de~Sitter case ($R_{*}=S_{*}=U_{*}=0$), the most physically 
interesting for our purposes are the dark energy dominated solution
\begin{equation}
R_{*} = \frac{6\delta}{\delta+\gamma+w-1} \,, \quad
S_{*} = \frac{6\gamma}{\delta+\gamma+w-1} \,, \quad
U_{*} = 1 \,,
\label{eq:de-domination}
\end{equation}
and the scaling solution
\begin{equation}
\begin{aligned} &
R_{*} = \frac{3\delta(\delta+\gamma+w)}{2(\delta^{2}-\delta\gamma+\gamma^{2})}
\,,\quad
S_{*} = \frac{3\gamma(\delta+\gamma+w)}{2(\delta^{2}-\delta\gamma+\gamma^{2})}
\,,\quad
U_{*} = \frac{w+\gamma+\delta}{w^{2}-3(\gamma-\delta)^{2}+2w(\gamma+\delta)}
\,,
\end{aligned}
\label{scal}
\end{equation}
in which $\rho_{\mathrm{DE}}/\rho_{m}={\mathrm{const.}}$, i.e., the fractional dark energy contribution to 
the total energy density is constant.

Anisotropic distribution of sources in Euclid survey might constrain the anisotropy at present, when the
dark energy density is of order 74\%, hence not yet in the final dark energy dominant attractor 
phase~(\ref{eq:de-domination}).

\subsubsection{Early-time anisotropy}
\label{sec:etanisotr}

An alternative (and, arguably, more speculative) possibility is that anisotropy may be generated by the presence of
anisotropic fields at inflation. Such fields could be spinors, vectors or higher order
forms which modify the properties of fluctuations in a direction-dependent way, 
either directly through perturbation dynamics or by causing the background to
inflate slightly anisotropically. The most common alternative is vector
fields (which will be further discussed in Section~\ref{VSect}).

Whereas a canonical scalar field easily inflates the universe if suitable
initial conditions are chosen, it turns out that it is much less straightforward to
construct vector field alternatives. In particular, one must maintain a
sufficient level of isotropy of the universe, achieve slow roll and keep
perturbations stable. Approaches to deal with the anisotropy have
been based on a
``triad'' of three identical vectors aligned with the three axis
\citep{ArmendarizPicon:2004pm}, a large number of randomly oriented fields
averaging to isotropy \citep{Golovnev:2008cf}, time-like \citep{Koivisto:2008xf}
or sub-dominant \citep{Dimopoulos:2008} fields. There are many variations of
inflationary scenarios involving vector fields, and in several cases the
predictions of the primordial spectra of perturbations have been worked out in
detail, see e.g., \cite{Watanabe:2010fh}. The generic prediction is that the
primordial perturbation spectra become statistically anisotropic, see
e.g., \cite{Ackerman:2007}.

Anisotropy could be also regarded simply as a trace of the initial conditions
set before inflation. One then assumes that inflation has lasted just about the
60 e-folds so that the largest observable scales were not yet smoothed out, or
isotropized, by the early inflationary expansion \citep{Pitrou:2008gk}. Such a
scenario can also be linked to various speculative ideas of pre-inflationary
physics such as gravitational tunnelling into an anisotropic universe, see e.g.,
\cite{Adamek:2010sg}. 

Also in this case the interest in such possibilities has been stimulated by several anomalies observed in the temperature 
WMAP maps, see~\cite{Copi:2010} for a recent review 
(some of them were also present in the COBE maps). Their statistical evidence is quite robust w.r.t.\ the increase of the 
signal-to-noise ratio over the years of 
the WMAP mission and to independent tests by the international scientific community, although the a posteriori choice of statistics could make their 
interpretation difficult, see~\cite{Bennet:2011}. Apart from those already mentioned in Section~\ref{sec:anisotropicmodels}, 
these anomalies include
an alignment between the harmonic quadrupole and octupole modes in the temperature anisotropies \cite{de-Oliveira:2004}, 
an asymmetric distribution of CMB power between two hemispheres, or dipole asymmetry~\cite{Eriksen:2004}, the lack of power 
of the temperature two-point correlation function 
on large angular scales ($> 60^{\circ}$), asymmetries in the even vs.\ odd multipoles of the CMB power spectra (parity symmetry breaking), both at large 
\cite{Kim:2010a, Gruppuso:2010} and intermediate angular scales \cite{Bennet:2011}. 
Some of the anomalies could be connected among each other, e.g., the CMB parity breaking has been recently linked to the lack of large-scale power
\cite{Maris:2010, Copi:2007, Kim:2010b}.

\subsection{Inhomogeneous models}
\label{LTBmetric}

Nonlinear inhomogeneous models are traditionally studied either with higher-order perturbation theory or 
with $N$-body LSS simulation codes. Both approaches have their limits. A perturbation expansion obviously 
breaks down when the perturbations are deeply in the nonlinear regime. $N$-body codes, on the other hand, 
are intrinsically Newtonian and, at the moment, are unable to take into account full relativistic effects, although
there are recent attempts to address these issues, see~\cite{Adamek:2013wja}.

Nevertheless, $N$-body codes can still account for the general relativistic behavior of gravitational collapse in the specific
case of inhomogeneous spherically symmetric models, as shown recently in~\cite{AGBTV2010}, where the growth of a 
void follows the full nonlinear GR solution down to large density contrasts (of order one). A possibility to make 
progress is, therefore, to proceed with the most extreme simplification: spherical symmetry. By assuming that the inhomogeneity 
is radial the general relativistic equations can be solved exactly and the suite of available $N$-body techniques can be applied, and one can make definite observable predictions both at the background and perturbation level.

It is important to stress that the fact that the model is spherically symmetric does not necessarily mean that the observer is at the center of a large void or super structure, in gross violation of the Copernican principle. 
The universe is indeed inhomogeneous and a spherically symmetric description of the density field is legitimate if the observer averages over angles. Clearly, which picture is valid is determined by the parameters used to model the inhomogeneity. If the inhomogeneity is standard -- i.e.~stemming from standard perturbations -- then the latter Copernican view is applicable. However, if the inhomogeneity is unconstrained, then the non-Copernican view is the only viable.

Historically, large-void models have been investigated as an alternative to the dark-energy scenario.
Indeed, a void creates an apparent acceleration field that can in principle match any supernova 
observations (see the review article~\cite{Marra:2011ct} and references therein).
This effect is easy to understand: standard candles are confined to the light cone and hence temporal changes can be replaced by spatial changes along photon geodesics. In this case, ``faster expansion now than before" is simply replaced by ``faster expansion here than there".  This is why a void model can mimic the effect of dark energy if it extends to the point in spacetime where the dark energy becomes subdominant: a typical scenario that can mimic the late-time acceleration of the concordance $\Lambda$CDM model consists of a deep void extending for 1-3~Gpc, corresponding to a transition redshift 
$z_{e}$ (i.e., the void edge) of about 0.3\,--\,1.
Of course, these models strongly violate the Copernican principle (CP) as the observer needs to be confined within few tens of Mpc from the center of a Gpc-scale void in order not to predict a dipole anisotropy that it is not observed in the CMB~\cite{Marra:2011ct}.

The consistent way to realize such a spherical inhomogeneity has been studied since the 1930s in the relativistic 
literature: the Lema\^{\i}tre--Tolman--Bondi (LTB)\index{Lemaitre-Tolman-Bondi (LTB)
metric@Lema\^{i}tre-Tolman-Bondi (LTB) metric} metric. This is the generalization of an FLRW metric in which the 
expansion factor along the radial coordinate $r$ is different relative to the surface line element 
$\d\Omega^{2}=\d\theta^{2}+\sin^{2}\theta\,\d\phi^{2}$. If we assume the inhomogeneous metric (this subsection 
follows closely the treatment in~\cite{Enqvist:2006cg,Amendola2010}, see~\citep[][Appendix B]{Marra:2011zp} for an alternative approach)
\begin{equation}
\mathrm{d}s^{2}=-\mathrm{d}t^{2}+X^{2}(t,r)\,\mathrm{d}r^{2}
+R^{2}(t,r)\,\mathrm{d}\Omega^{2}\,,
\label{eq:inh-met}
\end{equation}
and solve the $(0,1)$ Einstein equation for a fluid at rest 
 we find that the LTB metric is given by 
\begin{equation}
\mathrm{d}s^{2}=-\mathrm{d}t^{2}+\frac{\left[R'(t,r)\right]^{2}}
{1+\beta(r)}\mathrm{d}r^{2}+R^{2}(t,r)\mathrm{d}\Omega^{2}\,,
\label{eq:LTB}
\end{equation}
where $R(t,r),\beta(r)$ are arbitrary functions. Here primes and dots refer to partial space and time derivatives, 
respectively. The function $\beta(r)$ can be thought of as a position-dependent spatial curvature. If $R$ is
factorized so that $R(t,r)=a(t)f(r)$ and $\beta(r)=-Kf^{2}(r)$, then we recover the FLRW metric (up to a redefinition 
of $r$: from now on when we seek the FLRW limit we put $R=a(t)r$ and $\beta=-Kr^{2}$). Otherwise, we have a 
metric representing a spherical inhomogeneity centered on the origin. An observer located at the origin will observe 
an isotropic universe. We can always redefine $r$ at the present time to be $R_{0}\equiv R(t_{0},r)=r$, so that the 
metric is very similar to an FLRW today.

Considering the infinitesimal radial proper length $D_{||}=R'\d r/\sqrt{1+\beta}$, we can define the \emph{radial} 
\emph{Hubble function} \index{Hubble parameter!radial} as
\begin{equation}
H_{||}\equiv\dot{D}_{||}/D_{||}=\dot{R}'/R'\,,
\end{equation}
and similarly the \emph{transverse Hubble function}:\index{Hubble parameter!transverse}
\begin{equation}
H_{\perp}=\dot{R}/R\,.
\end{equation}
Of course the two definitions coincide for the FLRW metric. The non-vanishing components of the Ricci tensor for 
the LTB metric\index{Ricci tensor!for LTB metric} are 
\begin{eqnarray}
& & R_{0}^{0}=\frac{2\ddot{R}}{R}+\frac{\ddot{R}'}{R'}\,,\\
& & R_{1}^{1}= \frac{2\dot{R}\dot{R'}+R\ddot{R}'-\beta'}{RR'}\,,\\
& & R_{2}^{2}=R_{3}^{3}= \frac{\dot{R}^{2}-\beta}{R^{2}}+
\frac{\dot{R}\dot{R}'+R'\ddot{R}-\beta'/2}{RR'}\,.
\end{eqnarray}

In terms of the two Hubble functions, we find that the Friedmann equations are given by
\begin{eqnarray}
H_{\perp}^{2}+2H_{||}H_{\perp}-\frac{\beta}{R^{2}}-\frac{\beta'}{RR'} & = & 8\pi
G(\rho_{m}+\rho_{\Lambda})\,,\label{eq:ltb-fri1}\\
6\frac{\ddot{R}}{R}+2H_{\perp}^{2}-2\frac{\beta}{R^{2}}-2H_{||}H_{\perp}+\frac{
\beta'}{RR'} & = & -8\pi G (\rho_{m} - 2 \rho_{\Lambda})\,,\label{eq:ltb-fri2}
\end{eqnarray}
where $\rho_{m}(t,r)$ is the pressureless matter density and  $\rho_{\Lambda}= \Lambda/8\pi G$ is the energy density associated with a cosmological constant.
Adding Eqs.~(\ref{eq:ltb-fri1}) and (\ref{eq:ltb-fri2}), it follows that $2R\ddot{R}+\dot{R}^{2}=\beta +\Lambda R^2$. Integrating this 
equation, we obtain a Friedmann-like equation
\begin{equation}
H_{\perp}^{2}=\frac{\alpha(r)}{R^{3}} + \frac{8\pi G}{3}\rho_{\Lambda}  +\frac{\beta(r)}{R^{2}}\,,
\label{eq:hperpeq}
\end{equation}
where $\alpha(r)$ is a free function that we can use along with $\beta(r)$ to describe the inhomogeneity.
Using Eq.~(\ref{eq:hperpeq}) we can define the effective density parameters $\Omega_{i}^{(0)}(r)=\Omega_{i}(r,t_{0})$ today: 
\begin{eqnarray} \label{eq:ltb-curv}
\Omega_{m}^{(0)}(r)&\equiv&\frac{\alpha(r)}{R_{0}^{3}H_{\perp0}^{2}}\,, \\
\Omega_{\Lambda}^{(0)}(r)&\equiv&   \frac{8\pi G}{3 H_{\perp0}^{2}}\rho_{\Lambda}  \,, \\
\Omega_{K}^{(0)}(r)&=&1-\Omega_{m}^{(0)}(r)-\Omega_{\Lambda}^{(0)}(r)=\frac{\beta(r)}{R_{0}^{2}H_{\perp0}^{2}} \,.
\end{eqnarray}
where $R_{0}\equiv R(r,t_{0})=r$ and $H_{\perp0}\equiv H_{\perp}(r,t_{0})$
(the superscript $(0)$ denotes the present value).
Hence, we see that the initial condition at some time $t_{0}$ (which here we take as the present time) must 
specify two free functions of $r$, for instance $\alpha(r),\beta(r)$ or $\Omega_{m}^{(0)}(r),H_{\perp0}^{}(r)$. 
The latter choice shows that the inhomogeneity can be in the matter distribution or in the expansion rate or 
in both. This freedom can be used to fit simultaneously for any expansion rate (and therefore luminosity and 
angular diameter distances and for any source number density \citep{musta97}.

If one imposes the additional constraint that the age of the universe is the same for every observer, then only 
one free function is left \citep{garcia-haug08}. The same occurs if one chooses $\Omega_{m}^{(0)}(r)=
{\mathrm{constant}}$ (notice that this is different from $\rho_{m}^{(0)}(r)={\mathrm{constant}}$, which is 
another possible choice), i.e., if the matter density fraction is assumed homogeneous today (and only today) 
\citep{enq07}. The choice of a homogeneous universe age guarantees against the existence of diverging 
inhomogeneities in the past.
In particular, a simultaneous big bang excludes decaying modes which would be strongly in contradiction with the inflationary paradigm \cite{Zibin:2008vj}.

In the case of zero cosmological constant, Eq.~(\ref{eq:hperpeq}) is a classical cycloid equation\index{cycloid equation} whose solution for $\beta>0$ 
is given parametrically by
\begin{align}
R(r,\eta)=\,\frac{\alpha(r)}{2\beta(r)} &
(\cosh\eta-1)=\frac{R_{0}\Omega_{m}^{(0)}(r)}{2[1-\Omega_{m}^{(0)}(r)]}
(\cosh\eta-1)\,,\label{eq:sol-R}\\
t(r,\eta)-t_{B}(r)=\, &
\frac{\alpha(r)}{2\beta^{3/2}(r)}(\sinh\eta-\eta)=\frac{\Omega_{m}^{(0)}(r)}{2[
1-\Omega_{m}^{(0)}(r)]^{3/2}H_{\perp0}}(\sinh\eta-\eta)\,,\label{eq:sol-beta-t}
\end{align}
where $t_{B}(r)=t(r,\eta=0)$ is the inhomogeneous ``big-bang'' time, i.e., the time for which $\eta=0$ and $R=0$ 
for a point at comoving distance $r$. This can be put to zero in all generality by a redefinition of time. The 
``time'' variable $\eta$ is defined by the relation 
\begin{equation}
\eta=\int^{t}_0\frac{\beta(r)^{1/2}}{R(\tilde{t},r)}\d \tilde{t}\,.
\end{equation}
Notice that the ``time'' $\eta$ that corresponds to a given $t$ depends on $r$; so $R(r,t)$ is found by solving 
numerically $\eta(t,r)$ from Eq.~(\ref{eq:sol-beta-t}) and then substituting $R[r,\eta(r,t)]$. The present epoch 
$\eta_{0}(r)$ is defined by the condition $R=R_{0}$. One can derive the age of the universe  
$t_{\mathrm{age}}(r)=t(r,\eta_{0})-t_{B}(r)$in terms of $\Omega_{m}^{(0)},H_{\perp0}$. For $\beta<0$ the 
$\eta$ functions in Eqs.~(\ref{eq:sol-R}\,--\,\ref{eq:sol-beta-t}) become $(1-\cos\eta)$ and $(\eta-\sin\eta)$ for 
$R$ and $t$, respectively, while for $\beta=0$ they are $\eta^{2}/2$ and $\eta^{3}/6$.
In the case of a nonzero cosmological constant, the differential equation of \eqref{eq:hperpeq} can be integrated numerically or approached semi-analytically with the method presented in~\cite{Valkenburg:2011tm}.

As anticipated, if one wants to have faster expansion inside some radius in order to mimic cosmic acceleration, 
one needs to impose to the solution the structure of a void. An example of the choice of
$\Omega_{m}^{(0)}(r)$ and $h^{(0)}(r)\equiv
H_{\perp0}/(100\mathrm{\ km\ s}^{-1}\mathrm{\ Mpc}^{-1}$) is~\cite{alnes-amarz,garcia-haug08}
\begin{eqnarray}
\Omega_{m}^{(0)}(r) & = & \Omega_{\mathrm{out}}+(\Omega_{\mathrm{in}}-\Omega_{\mathrm{out}})f(r,r_{0},\Delta)\,,\\
h^{(0)}(r) & = &
h_{\mathrm{out}}+(h_{\mathrm{in}}-h_{\mathrm{out}})f(r,r_{0},\Delta)\,,
\end{eqnarray}
with
\begin{equation}
f(r,r_{0},\Delta)=\frac{1-\tanh[(r-r_{0})/2\Delta]}{1+\tanh(r_{0}/2\Delta)}\,,
\end{equation}
representing the transition function of a shell of radius $r_{0}$ and thickness $\Delta$. The six constants 
$\Omega_{\mathrm{in}},\Omega_{\mathrm{out}},h_{\mathrm{in}},h_{\mathrm{out}},r_{0},\Delta$
completely fix the model. If $h_{\mathrm{in}}>h_{\mathrm{out}}$ one can mimic the accelerated expansion.

In order to compare the LTB model to observations we need to generalize two familiar concepts: redshift and 
luminosity distance. The redshift can be calculated through the equation
\begin{equation}
\frac{\mathrm{d}z}{\mathrm{d}r}=(1+z)\frac{\dot{R}'}{\sqrt{1+\beta}}\,,
\label{eq:ltb-red}
\end{equation}
where we must impose $z(r=0)=0$ and $R(t,r)$ must be calculated on the radial geodesic $t_{p}(r)$:
\begin{equation}
\frac{\mathrm{d}t_p}{\mathrm{d}r}=-\frac{R'(t_p(r),r)}{\sqrt{1+\beta(r)}}\,.
\label{eq:geoltb}
\end{equation}
Every LTB function, e.g.~$H_{\perp}(t,r),R(t,r)$ etc., can be converted into a line-of-sight function of redshift by evaluating 
the arguments $r_{p}(z),t_{p}(z)$ along the past light cone.

The proper area of an infinitesimal surface at $r,t={\mathrm{constant}}$ is given by $A=R^{2}(r,t)\sin\theta\,
\d\theta\,\d\phi$. The angular diameter distance is the square root of $A/(\sin\theta\,\d\theta\,\d\phi)$
so that $d_{A}(z)=R(t_{p}(z),r_{p}(z))$. Since the Etherington duality relation\index{duality relation|see{Etherington
relation}} $d_{L}=(1+z)^{2}d_{A}$ remains valid in inhomogeneous models, we have \citep{sachs66}
\begin{equation}
d_{L}(z)=(1+z)^{2}R(t_{p}(z),r_{p}(z))\,.
\label{eq:dlvoid}
\end{equation}
This clearly reduces to $d_{L}=(1+z)r(z)$ in the FLRW background. Armed with these observational tools, 
we can compare any LTB model to the observations.

Predictions of the $\Lambda$LTB model can be obtained using the packages {\sc ColLambda}\footnote{\url{http://www.wesselvalkenburg.nl/software/collambda/}} \cite{Valkenburg:2011tm} and {\sc VoidDistancesII}{\footnote{\url{http://www.wesselvalkenburg.nl/software/voiddistances/}}} \cite{Marra:2012pj}. The latter acts as a wrapper around {\sc camb} \cite{Lewis:1999bs}, necessitating no changes to {\sc camb}'s source code and minimal changes to {\sc CosmoMC}'s source code.

\subsubsection{Void models as alternative to dark energy}

As said earlier, the LTB model can potentially fit distance-redshift relations such as the SN~Ia Hubble diagram, the CMB 
power spectrum and BAO scale~\cite{Moss:2010jx,Biswas:2010xm}. However, LTB models as alternative to dark energy 
(without a cosmological constant) have been ruled out by basically two effects that involve the velocity field caused by the 
inhomogeneity: Compton y-distortion and kSZ effect.

Off-center observers see a large dipole in the CMB in an inhomogeneous universe with simultaneous big bang, which affects the observed CMB  through the kinetic Sunyaev-Zel'dovich (kSZ) effect~\cite{garcia-haug08b}: hot electrons inside an overdensity distort the CMB spectrum through inverse Compton scattering, in which low energy CMB photons receive energy boosts during collisions with the high-energy electrons.
By considering the `linear kSZ effect' \cite{Zhang:2010fa,Moss:2011ze}, in which the effect due to all free electrons in the reionized universe is taken into account, it has been shown that void models able to mimic acceleration predict a kSZ signal by far larger than the upper limits from SPT~\cite{Shirokoff:2010cs} and ACT~\cite{Das:2010ga} at $\ell=3000$. Consequently, void models with homogeneous big bang are ruled out.

On the other hand, if the big bang is not homogeneous (hypothesis which is, as said earlier, strongly in contradiction with the inflationary paradigm), it may be possible to reduce the kSZ signal so as to satisfy the experimental limits. However, in this case, the CMB undergoes large distortions because CMB photons are scattered from inside our past light-cone into our line-of-sight by off-centre reionized structures which act as mirrors.
The spectrum observed by the central observer is then a mixture of blackbody spectra with different temperatures, producing a distorted blackbody spectrum. It has been shown~\cite{Caldwell:2007yu, Zibin:2011ma} that such models generate a Compton y-distortion which exceeds the limits placed by the COBE satellite~\cite{Fixsen:1996nj}.

\subsubsection{Late-time inhomogeneity}
\label{latetime-inho}

\begin{figure}
\begin{center}
\includegraphics[width=9 cm]{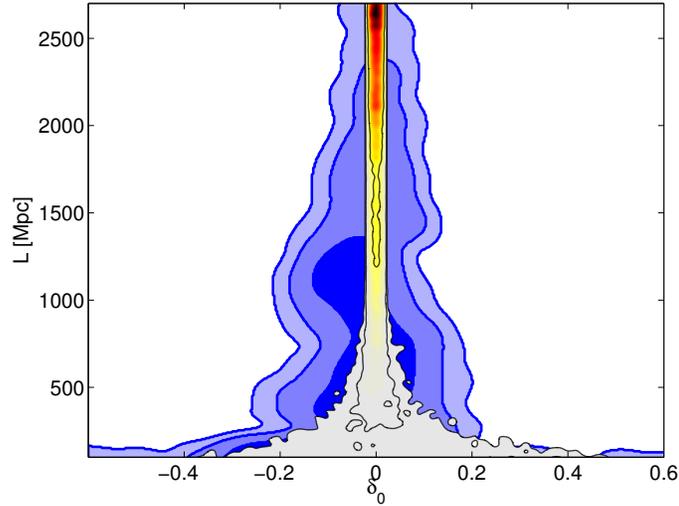}
\caption{Marginalized constraints 
on the amount of radial inhomogeneity $\delta_0$ at a given scale $L$ from local Hubble parameter measurements, supernova Ia data, CMB anisotropies, BAO observations, Compton y-distortion and kSZ constraints and age data (as lower bounds only)
at 68\%, 95\% and 99\% confidence level (blue contours), compared to the expected level of inhomogeneity within a $\Lambda$CDM model that satisfies the Copernican principle (red-to-gray contours).
See \cite{Valkenburg:2012td} for more details.}
\label{fig:awesome}
\end{center}
\end{figure}

The $\Lambda$CDM model is a limiting case of the $\Lambda$LTB metric. This makes the latter a useful tool to constrain departure from the homogeneous FLRW paradigm. This program is similar in spirit to one aimed at constraining departures from isotropy using the Bianchi models of Section~\ref{sec:anisotropicmodels}.
In \cite{Valkenburg:2012td} available observations were used to constrain the amount of radial inhomogeneity $\delta_0$ at a given scale $L$. As stressed earlier, in this context, spherical symmetry is interpreted as the result of averaging the density field around us with a spherically symmetric filter. It is not assumed that the universe is spherically symmetric, rather that the observer averages over all angles, which is equivalent to expanding the full matter field in spherical harmonics, and throwing away all other information than the radially dependent monopole.
The result of this analysis is shown in Fig.~\ref{fig:awesome}, blue contours. The red-to-gray contours shows the expected level of inhomogeneity within a $\Lambda$CDM model that satisfies the Copernican principle.
Forthcoming data from Euclid should be able to help improving these constraints.

One can improve upon these constraints by including observables sensitive to the interior of our past light cone 
such as the fossil record of a large number of galaxies~\citep{Heavens:2011}. One can use the average star 
formation rate at a fixed lookback time as a diagnostic test for homogeneity, which is valid also outside the LTB 
framework. The lookback time has two elements to it -- the lookback time of the emission of the light, plus the time 
along the past world line. The last of these can be probed using the integrated stellar spectra of the galaxies, 
using a code such as \textsc{vespa}~\citep{Tojeiro:2007}, and this is evidently dependent only on atomic 
and nuclear physics, independent of homogeneity. The lookback time can also be computed, surprisingly 
simply, without assuming homogeneity from \citep{Heavens:2011} 
\begin{equation}
\Delta t = \int_0^z \frac{dz'}{(1+z')H_r(z')} \,,
\end{equation}
where $H_r$ is the radial Hubble constant. In principle, this can be obtained from radial BAOs, assuming 
early-time homogeneity so that the physical BAO scale is fixed. The spectroscopic part of Euclid could 
estimate both the star formation histories from stacked spectra, and the radial expansion rate.

Inhomogeneous models such as LTB are very useful in order to test various consistency relations which are valid 
only within the homogenous FLRW paradigm, see Section~\ref{glocur} and \cite{Clarkson:2012bg} for a review.
For example, while in FLRW the function $H(z)$ fixes the comoving distance $\chi(z)$ up to a 
constant curvature (and consequently also the luminosity and angular diameter distances), in the LTB model
the relation between $\chi(z)$ and $H_{\perp}(z)$ or $H_{\Vert}(z)$ can be arbitrary. That is, one can choose the 
two spatial free functions to be for instance $H_{\perp}(r,0)$ and $R(r,0)$, from which the line-of-sight values 
$H_{\perp}(z)$ and $\chi(z)$ would also be arbitrarily fixed. This shows that the ``consistency'' FLRW relation
between $\chi(z)$ and $H(z)$ is violated in the LTB model, and in general in any strongly inhomogeneous universe.
Further below we discuss how this consistency test can be exploited by Euclid to test for large-scale inhomogeneities. 
Recently, there has been an implementation of LTB models in large-scale structure $N$-body 
simulations~\citep{AGBTV2010}, where inhomogeneities grow in the presence of a large-scale void and seem to 
follow the predictions of linear perturbation theory. Since the LSS simulations capture the full relativistic behavior 
of matter as it falls into potential wells in the presence of a large void, one can use such simulations to test the
non-linear dynamics of LTB models. In particular, one can follow the clustering of galaxies, the shear induced by 
the inhomogeneous background and the gravitational lensing induced by the whole web of cosmic structures.

An interesting class of tests on large-scale inhomogeneities involve probes of the growth of structure. However, 
progress in making theoretical predictions has been hampered by the increased complexity of cosmological 
perturbation theory in the LTB spacetime, where scalar and tensor perturbations couple, see for 
example~\cite{Zibin:2008vj,Clarkson:2009sc,Nishikawa:2012we,February:2013qza} for recent progress. 
Nevertheless, a number of promising tests of large-scale inhomogeneity using 
the growth of structure have been proposed. Ref.~\cite{Alonso:2012ds} used $N$-body simulations to modify the 
Press--Schechter halo mass function, introducing a sensitive dependence on the background shear. The shear 
vanishes in spatially-homogeneous models, and so a direct measurement of this quantity would put stringent 
constraints on the level of background inhomogeneity, independent of cosmological model assumptions.

Purely geometric tests involving large-scale structure have been proposed, which neatly side-step the perturbation 
theory issue. The Baryon Acoustic Oscillations (BAO) measure a preferred length scale, $d(z)$, which is a 
combination of the acoustic length scale, $l_s$, set at matter-radiation decoupling, and projection effects due to the 
geometry of the universe, characterized by the volume distance, $D_V(z)$. In general, the volume distance in an 
LTB model will differ significantly from that in the standard model, even if the two predict the same SN~Ia Hubble 
diagram and CMB power spectrum. Assuming that the LTB model is almost homogeneous at the decoupling 
epoch, $l_s$ may be inferred from CMB observations, allowing the purely geometric volume distance to be 
reconstructed from BAO measurements. It has been shown by~\cite{Zumalacarregui:2012pq} that, based on these 
considerations, recent BAO measurements effectively rule out giant void models, independent of other observational 
constraints.

Another important consequence of late-time inhomogeneity is the impact of cosmic variance on cosmological parameters.
Indeed, we can only observe the Universe from our own position which is fixed, in terms of cosmological scales.
If we could move around in the Universe, we would measure the variation of local parameters, a variation caused 
by observing from locations with different values of the gravitational potential. However, as we cannot probe this 
unavoidable variation, there is a cosmic variance on the parameters inferred from observations which is systematic 
in nature. Particularly important for the dark-energy quest are the cosmic-variance errors on the dark-energy equation 
of state~\cite{Valkenburg:2013qwa} and the Hubble parameter~\cite{Marra:2013rba}. Forthcoming probes such as Euclid 
will have to deal with this issue as the presence of unaccounted for inhomogeneity (either within the FLRW paradigm or not) 
could affect forecasts (see Section \ref{euclid-forecasts}) and cosmological inference.

Finally, the ``cold spot'' in the CMB sky could be attributed to a very large underdensity~\citep{coldspot,coldspot-notari} 
allowing to directly test the assumption of local homogeneity. The recent discovery~\cite{Finelli:2014yha} of a large 
underdensity ($\delta_0\simeq-0.2,\ r_0\simeq200$ Mpc) in WISE-2MASS, at a median redshift of $z\simeq0.2$, with 
an LTB profile that matches both the local galaxy density and the CMB cold spot anisotropy, opens the possibility to 
explore the local inhomogeneous Universe by cross-correlating the CMB secondary anisotropies and the local voids 
(and other structures) on large scales. We are thus entering the realm of the inhomogeneous universe beyond FRW, 
where a full understanding of local structures is necessary for a comprehensive picture of the universe.

\subsubsection{Measuring the transition to homogeneity at different redshifts}
\label{scalehomogeneity}

Large-scale homogeneity is usually assumed without proof when analyzing certain cosmological probes (Durrer
2011). This is often a reasonable approach, since it would not be possible to obtain many observational constraints without doing so. However, in order to be able to rely on these constraints, we must verify the validity of the Cosmological 
Principle independently in an unbiased way. Along these lines, different groups have argued that the Universe might
in fact not reach a homogeneous regime on large scales, and that instead it behaves like a fractal \citep{1992PhR...213..311C,1997cdc..conf...24P,1997EL.....39..103M,
1998PhR...293...61S,1999ApJ...514L...5J,2009A&A...508...17S,
2011EL.....9659001S},
while other groups claim the opposite result: the predictions of the standard $\Lambda$CDM model \citep{1994ApJ...437..550M,1997NewA....2..517G,
1997ApL&C..36...59S,1998MNRAS.298.1212M,1998A&A...334..404S,
1999ApJ...514L...1A,2000MNRAS.318L..51P,2001A&A...370..358K,
2005MNRAS.364..601Y,2009MNRAS.399L.128S,
Scrimgeour:2012wt,2013MNRAS.434..398N}.

The disparity between these two results seems to stem from the differences in the analysis methods. On the one
hand it is desirable to use methods that are, as far as possible, free of assumptions, especially regarding the property
you want to measure. However, in the case of the validity of the Cosmological Principle, this is not an easy task,
since homogeneity must sometimes be assumed in order to cope with certain observational effects. These issues will be
further explained in section 2. At the end of the day, we must ensure that the method used is able to distinguish 
homogeneous from non-homogeneous models to a reasonable level of precision. A robust and popular method to study the
transition to homogeneity in the matter density field at late times is to analyze the fractality of the galaxy distribution
in a redshift survey. Furthermore, fractal dimensions can be used to quantify clustering, since they depend on the scaling
of the different moments of galaxy counts in spheres, which in turn are related to the n-point correlation functions.
As has been said, the homogeneous regime is reached, within the standard $\Lambda$CDM model, at very large scales, and
therefore a large survey volume is necessary in order to safely
claim a detection of this transition. In this sense, photometric galaxy redshift surveys such as DES (The Dark Energy
Survey Collaboration 2005) provide a unique opportunity for this study, since they are able to observe large numbers of
objects distributed across wide areas and to further redshifts than their spectroscopic counterparts. The main caveat of
these surveys is that, due to the limited precision in the redshift determination, much of the radial information is
lost, and we are only able to study angular clustering in different thick redshift slices. Hence, in order to study the
fractality of the galaxy distribution with a photometric survey, the methods and estimators used in previous analyses
must be adapted to draw results from angular information alone. One advantage of this approach is that, since angular 
positions are pure observables (unlike three-dimensional distances, which can only be calculated assuming a fiducial 
cosmology), the results obtained are completely model independent. 

In Ref.~\cite{Alonso:2013boa} we \reviewtext{the authors} proposed an observable, the angular homogeneity index H2, which could be used by 
photometric surveys in the near-future to study the fractal structure of the galaxy distribution.

\subsubsection{Reconstructing the global curvature at different redshifts}
\label{glocur}

Clarkson et al.~\cite{Clarkson-Bassett-Lu:2008} presented an observational test for the Copernican principle which 
relies on the consistency relation between expansion rate and angular diameter distance. This test is valid also outside the LTB framework discussed earlier. Here we discuss the 
implications for Euclid.

Let us recall that the angular diameter distance in a FLRW model can be written as (see eq.(\ref{eq:ang}) and the surrounding discussion):
\begin{equation}
D_{A}(z)=\frac{1}{1+z}\frac{1}{H_{0}\sqrt{-\Omega^{(0)}_{K}}}\sin\left(\sqrt{-\Omega^{(0)}_{K}}\int_{0}^{z}{dz'\frac{H_{0}}{H(z')}}\right)\,.
\label{eq:distanza-angolare-curv}
\end{equation}
where $\Omega^{(0)}_{K}$ is the curvature parameter \textit{today}.
We can invert the last equation to obtain an expression for the curvature parameter that depends on the Hubble 
parameter $H$ and comoving angular diameter distance  $D\left(z\right)=\left(1+z\right)D_{A}\left(z\right)$ only, 
see~\cite{Clarkson-Bassett-Lu:2008}:
\begin{equation}
\Omega^{(0)}_{K}=\frac{\left[H\left(z\right)D'\left(z\right)\right]^{2}-1}{\left[H_{0}D\left(z\right)\right]^{2}} \,,
\label{eq:omegak}
\end{equation}
where here the prime refers to the derivative with respect the redshift. Then Eq.~(\ref{eq:omegak}) tells us how the 
curvature parameter can be measured from the distance and the Hubble rate observations, in a model-independent way.
The idea is then to measure the curvature parameter $\Omega^{(0)}_{K}$ at different redshifts. Let us consider again 
Eq.~(\ref{eq:omegak}); if we are in a FLRW universe then $\Omega^{(0)}_K$ should be independent of redshift, i.e., 
its derivative with respect to $z$ should be zero 
\begin{equation}
\mathcal{C}(z) = \frac{{\mathrm{d}}\Omega^{(0)}_{K}}{{\mathrm{d}}z}=0\,.
\label{eq:domegak}
\end{equation}
If it happens that $\mathcal{C}(z)\neq 0$ even at a single redshift then this means the large-scale universe is not 
homogeneous.

A possible test to measure $\Omega^{(0)}_K$ at various redshifts is provided by baryon acoustic oscillations. 
Observing the features of BAO in the galaxy power spectrum in both angular (orthogonal to the line of sight $L_{\perp}$) 
and radial direction (along the line of sight $L_{\parallel}$) allows us to measure with a great accuracy both $D_{A}(z)$ and $H(z)$, respectively. 
If the geometry is not FLRW, then the standard BAO will be deformed in three different ways:
\begin{enumerate}
\item The sound horizon scale, which is the characteristic ruler, will be different 
in the $\perp$ and $\parallel$ directions and it will be also different from that for
 the FLRW universe.
\item Even if the sound horizon were isotropic at decoupling, the subsequent 
expansion in the $\perp$ and $\parallel$ directions will be different just 
because they will be governed by two distinct Hubble parameters: $H_{\perp}$ 
and $H_{\parallel}$.
\item The redshift distortion parameter will be different because 
it will depend on the background expansion.
\end{enumerate} 

Also the growth factor will be modified, perhaps in a scale dependent way. If the true underlying model is radically 
inhomogeneous, but we assume a FLRW in interpreting the observations, the derived cosmological parameters will
be biased (or unphysical) and the parameters derived from BAO data will be different from those measured by SN~Ia 
and/or lensing. As argued also in different contexts, a mismatch on the value of one of more parameters may indicate 
that we are assuming a wrong model.

We show here the sensitivity that can be reached with an experiment like Euclid for the curvature parameter 
$\Omega^{(0)}_{K}$. We choose a redshift survey with a depth of $z=1.6$ and consider different redshift bins.
In Figure~\ref{fig:omk} we show the first $1\sigma$ absolute errors on the curvature parameter for different redshift 
bins that can be obtained measuring the Hubble parameter and the angular diameter distance. In obtaining these 
errors we used Fisher-based forecasts for the radial and angular BAO signal following~\cite{seo03,eisenstein07}, 
as discussed in Section~\ref{dark-energy-and-redshift-surveys}.
The sensitivity that can be reached with an experiment like Euclid is extremely high; we can measure the curvature 
parameter better than 0.02 at redshift of $z\simeq 1$. This will allow us to discriminate between FLRW 
and averaged cosmology as for example illustrated in Figure~\ref{fig:newfigomk}.

\epubtkImage{errorbars-optim.png}{%
\begin{figure}
\centerline{\includegraphics[width=0.48 \columnwidth]{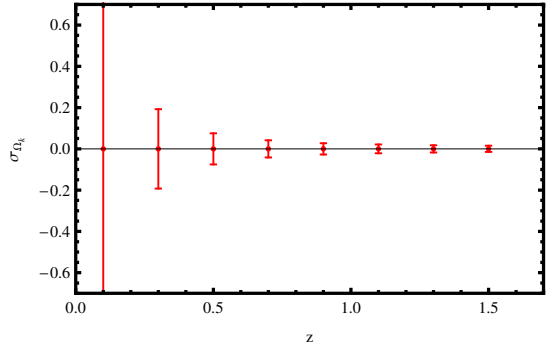}}
\caption{Relative errors on $\Omega_{K}$ for our benchmark survey for different 
redshifts.}
\label{fig:omk}
\end{figure}}

\epubtkImage{omegak-tot-omegak-tot-zoom.png}{%
\begin{figure}
\centerline{
\includegraphics[width=0.48 \columnwidth]{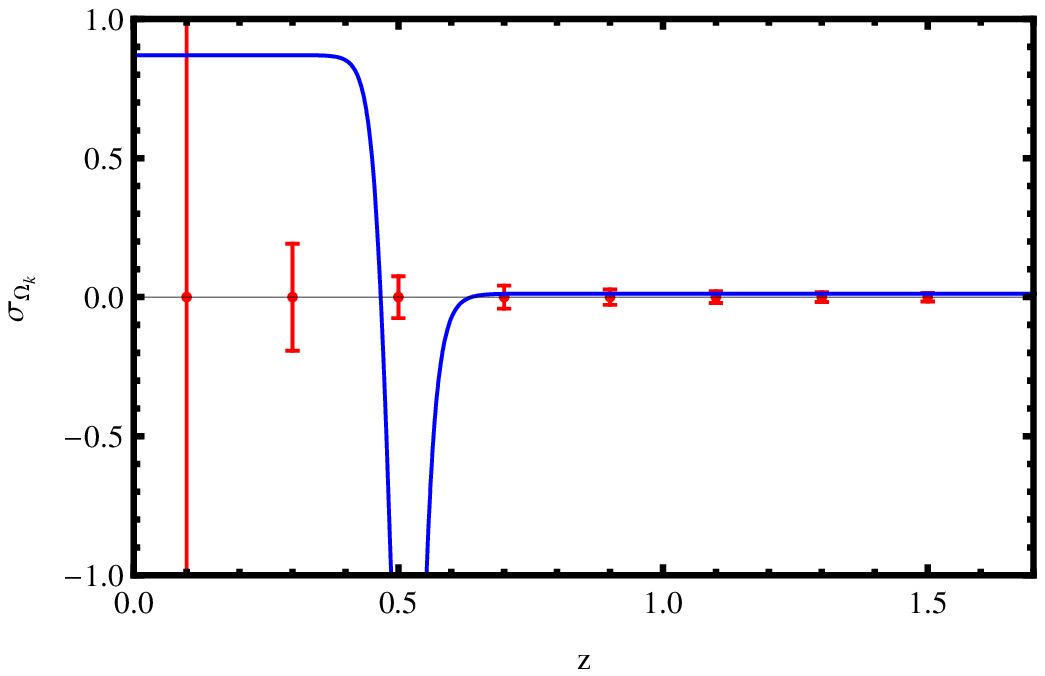}
\includegraphics[width=0.48 \columnwidth]{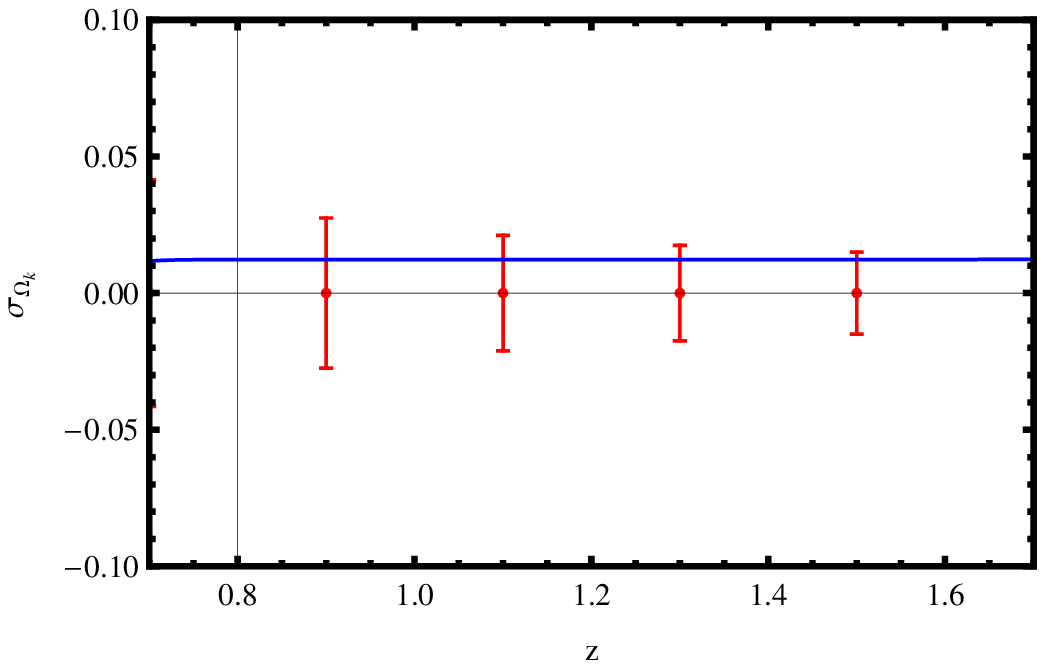}
}
\caption{\emph{Left:} same as Figure~\ref{fig:omk} but now with
  superimposed the prediction for the Lema\^{\i}tre--Tolman--Bondi model
  considered by~\cite{garcia-haug08}. \emph{Right:}
  zoom in the high-redshift range.}
\label{fig:newfigomk}
\end{figure}}

An alternative to measuring the global curvature is to measure the shear of the background geometry. If there is a large 
inhomogeneous void then a congruence of geodesics will not only expand but also suffer shear~\citep{garcia-haug08c}. 
The amount of shear will depend on the width and magnitude of the transition between the interior of the void and the 
asymptotic Einstein--de~Sitter universe. Normalizing the shear w.r.t.\ the overall expansion, one finds~\citep{garcia-haug08c}
\begin{equation}\label{eq:normshear}
\varepsilon = \frac{H_\perp(z)-H_{||}(z)}{2H_\perp+H_{||}} \simeq
\frac{1 - H_{||}(z)\,\partial_z\Big[(1+z)\,D_A(z)\Big]}{3H_{||}(z)D_A(z) + 2\Big(1 - H_{||}(z)\,\partial_z\Big[(1+z)\,D_A(z)\Big]\Big)}\,.
\end{equation}
Clearly, in homogeneous FRW universes the shear vanishes identically since $H_\perp = H_{||} = H$. Also note that the 
function $H_{||}(z)D_A(z)$ is nothing but the Alcock--Paczynski factor, which is normally used as a geometric test for the 
existence of vacuum energy in $\Lambda$CDM models.

\subsection{Backreaction of inhomogeneities on overall expansion}

Besides the observational effects of inhomogeneities on the propagation of photons, it has been argued~\cite{Rasanen:2003fy, Kolb:2004am,Ellis:2005uz} that large-scale structures could -- due to the nonlinear nature of gravity -- affect the way the homogeneous and isotropic background metric itself evolves. In other words, there could be a backreaction of inhomogeneities on the background which would evolve according to modified Friedmann equations.
The motivation -- as said earlier -- stems from the coincidence that large-scale structures became nonlinear recently, exactly when a primary dark energy is supposed to start dominating the energy content of the universe and cause acceleration.

In general, we would like to compute directly the impact of the inhomogeneities, without requiring an exact and 
highly symmetric solution of Einstein's equations like FLRW or even LTB. Unfortunately there is no easy way to 
approach this problem. One ansatz tries to construct average quantities that follow equations similar to those of 
the traditional FLRW model, see~\cite{Buchert:1999er,Buchert:2007ik}.
In this framework, it is possible to obtain a set of equations, often called the Buchert equations,
that look surprisingly similar to the Friedmann equations for the averaged scale factor $a_\CD$, with extra contributions:
\begin{eqnarray}
3\left( \frac{{\dot a}_\CD}{a_\CD}\right)^2 - 8\pi G \average{\varrho}-\Lambda &=& - \frac{\average{\CR}+{\CQ}_\CD }{2} \;, \\
3\frac{{\ddot a}_\CD}{a_\CD} + 4\pi G \average{\varrho} -\Lambda &=& {\CQ}_\CD \,.
\end{eqnarray}
Here $\CR$ is the 3-Ricci scalar of the spatial hypersurfaces and the kinematical backreaction
$\CQ$ is given by
\begin{equation}
\label{Q} 
{\cal Q}_\CD =
\frac{2}{3}\average{\left(\theta - \average{\theta}\right)^2 } - 2\average{\sigma^2} \,,
\end{equation}
i.e., it is a measure of the variance \reviewtext{of the} expansion rate $\theta$ and of the shear $\sigma_{ij}$. We see that this 
quantity, if positive, can induce an accelerated growth of $a_\CD$, which suggests that observers would 
conclude that the universe is undergoing accelerated expansion.

Unfortunately, the Buchert equations are not closed and further input is needed. There has been extensive work trying to determinate if the backreaction effect is negligible or not.
For example, \cite{Clifton:2013jpa,Yoo:2014boa, Bruneton:2012cg} considered approaches which model the universe as a lattice of Schwarzschild patches.
However, after more than 10 years of research it is not yet clear if backreaction can account for dark energy or not, see e.g.~\cite{Kolb:2011zz,Rasanen:2011ki,Green:2013yua} and references therein.

An alternative route is to avoid the fact that the Buchert equations are not closed by imposing by 
hand an effective, average geometry with the help of a template metric that only holds on average. The probably 
simplest first choice is to impose on each spatial hypersurface a spatial metric with constant curvature, by imagining 
that the inhomogeneities have been smoothed out. But in general the degrees of freedom of this metric (scale factor
and spatial curvature) will not evolve as in the FLRW case, since the evolution is given by the full, inhomogeneous 
universe, and we would not expect that the smoothing of the inhomogeneous universe follows exactly the evolution 
that we would get for a smooth (homogeneous) universe. For example, the average curvature could grow over time, 
due to the collapse of overdense structure and the growth (in volume) of the voids. Thus, unlike in the FRLW case,
the average curvature in the template metric should be allowed to evolve. This is the case that was studied 
in~\cite{Larena:2008be}.

However, imposing a template metric is not enough. Firstly, although there is an integrability condition linking the evolution of 
$\average{\CR}$ and $\CQ_\CD$ and in addition a consistency requirement that the effective curvature $\kappa(t)$ 
in the metric is related to $\average{\CR}$, we still need to impose an overall evolution by hand as it was not yet 
possible to compute this from first principles. \cite{Larena:2008be} assumed a scaling solution $\average{\CR}\propto a_\CD^n$, 
with $n$ a free exponent. In a dark energy context, this scaling exponent $n$ corresponds to an effective dark energy 
with $w_\CD = -(n+3)/3$, but in the backreaction case with the template metric the geometry is different from the usual 
dark energy case. A perturbative analysis \citep{Li:2007ci} found $n=-1$, but of course this is only an indication of the 
possible behavior as the situation is essentially non-perturbative.
The second choice concerns the computation of observables. \cite{Larena:2008be} studied distances to supernovae 
and the CMB peak position, effectively another distance. The assumption taken was that distances could be computed 
within the averaged geometry as if this was the true geometry, by integrating the equation of radial null geodesics. In 
other words, the effective metric was taken to be the one that describes distances correctly. The resulting constraints 
are shown in Figure~\ref{fig:larena1}. We see that the leading perturbative mode ($n=1$) is marginally consistent with 
the constraints. These contours should be regarded as an indication of what kind of backreaction is needed if it is to 
explain the observed distance data.

\epubtkImage{larena1.png}{%
\begin{figure}[htbp]
\centerline{\includegraphics[width=8cm]{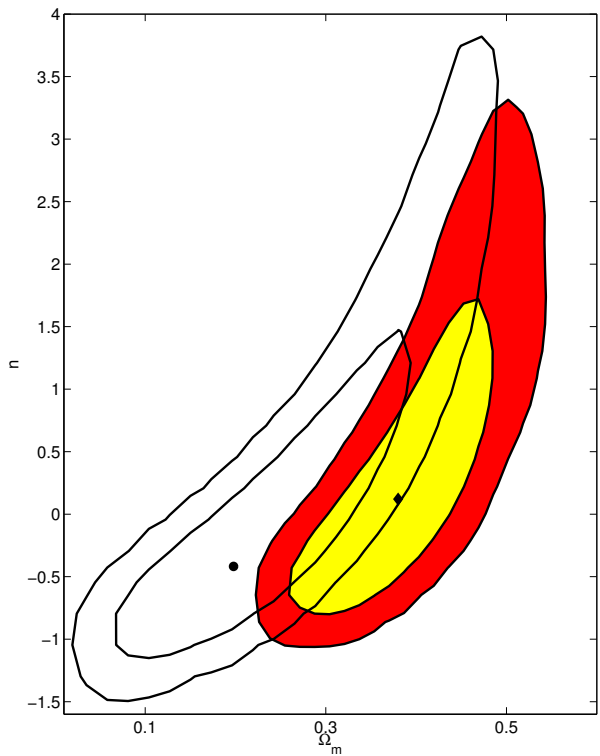}}
\caption{Supernova and CMB constraints in the $(\Omega^{\CD_0}_{m},n)$ plane for the averaged effective model 
with zero Friedmannian curvature (filled ellipses) and for a standard flat FLRW model with a quintessence field with 
constant equation of state $w=-(n+3)/3$ (black ellipses). The disk and diamond represent the absolute best-fit
models respectively for the standard FLRW model and the averaged effective model.}
\label{fig:larena1}
\end{figure}}

One interesting point, and maybe the main point in light of the discussion of Section~\ref{glocur}, is that the averaged 
curvature needs to become necessarily large at late times due to the link between it and the backreaction term $\CQ$, 
in order to explain the data. Just as in the case of a large deviation from homogeneity, this effective curvature makes the backreaction scenario testable to some degree with future large surveys like Euclid.

\newpage
\section{Speculative avenues: non-standard models of primordial fluctuations}

In this section we explore other non-conventional scenarios that challenge our understanding of the universe. Here we
present models that include mechanisms for primordial anisotropy in the fluctuation spectrum, due to
spacetime non-commutativity, to inflationary vector fields or to
super-horizon fluctuations. Since inflation can occur at high energies for which we lack robust direct 
experimental probes, it is reasonable to pay attention on possible deviations from some standard properties of low 
energy physics. We review these here and point out possible observables for the Euclid project.

\subsection{Probing the quantum origin of primordial fluctuations}

Conventionally, the 2-point correlation function of a random variable
$X(\vec k,t)$ is regarded as a classical object, related to the power
spectrum $P_{X}$ via the relation
\ba
\langle X(\vec k,t)X(\vec k',t)\rangle=(2\pi)^{3}\delta(\vec k-\vec k')P_{X}(k)\ ,
\ea
where $k=|\vec k|$.

When we look at $X(\vec k,t)$ in terms of a \emph{quantum field} in momentum
space, we need to reinterpret the average $\langle\ldots \rangle$ as the
expectation value of the 2-point function over a determined quantum state.
This raises several issues that are usually ignored in a classical analysis.
For instance, the value of the expectation value depends in the algebra of
the annihilation and creation operators that compose the field operator.
Any non-trivial
algebra such as a non-commutative one, leads to non-trivial power spectra. Also, the quantum expectation value depends on the state of the field, and different choices can lead to radically different results.

Suppose that $\varphi(\vec x,t)$ represents a perturbation propagating
on an inflationary background. Upon quantization, we have
\ba
\hat\varphi(\vec x,t)=(2\pi)^{-3/2}\int \mathrm{d}^{3}k\left[\varphi_{k}(t)\hat a_{\vec k}\,e^{i\vec k\cdot t}+\varphi_{k}^{*}(t)\hat a_{\vec k}^{\dagger}\,e^{-i\vec k\cdot t}\right]\ ,
\ea
where $\hat a_{\vec k}$ is the usual annihilation operator. When calculated
in the limit $\vec k\rightarrow \vec k'$, the expectation value of the
two-point function in coordinate space diverges, 
signalling the breakdown of the theory at short distances. From
the quantum field theory perspective, this means that the expectation value
needs to be regularized in the ultraviolet (UV). It has been argued that
this has in specific scenarios sizeable effects on the
observable spectrum -- see e.g., \cite{renorm1}, see however e.g., \cite{DMR} for a contrary viewpoint.

In addition to UV divergences, there are infrared (IR) ones in long-range
correlations. Usually, one tames these by putting the universe in
a box and cutting off super-horizon correlations. However,
several authors have recently proposed more sensible IR regulating
techniques, see e.g., \cite{sloth,Koivisto:2010pj}. Very natural 
ways to obtain IR finite results are to take into account the presence of tiny spatial curvature or a pre-inflationary phase which alters the initial conditions~\cite{proko,Koivisto:2010pj}. In principle these regularizations will leave an imprint in the large-scale structure data, in the case that regularization scale is not too far beyond the present horizon scale. If this pre-inflationary phase is characterized by modified field theory, such as modified dispersion relations or lower dimensional effective gravity, the scalar and tensor power spectra show a modification whose magnitude is model-dependent, see e.g., \cite{maxnew}.

The two-point function of a scalar field is constructed from basic quantum
field theory, according to a set of rules determined in the context of
relativistic quantum mechanics. In particular, the usual commutation rules
between position and momentum are promoted to commutation rules between the
field and its canonical conjugate. A modification of the fundamental quantum
mechanical commutation rules can easily be generalized to field theory. The
most popular case is represented by non-commutative geometry, which implies that coordinate operators do not commute, i.e.,
\ba
 \label{canonical_nc}
[\hat x^{\mu},\hat x^{\nu}]=i\theta^{\mu\nu} \,,
\ea
where $\theta^{\mu\nu}$ is an anti-symmetric matrix, usually taken to be
constant, see e.g., \cite{ncfund1,ncfund2}. There are many fundamental
theories that phenomenologically reduce to an ordinary field theory over a
non-commutative manifold, from string theory to quantum gravity. It is
therefore important to consider the possibility that non-commutative effects took place during the inflationary era and try to extract some prediction.

One can construct models where the inflationary expansion of the universe is
driven by non-commutative effects, as in~\cite{mag,ncmax}. In this kind of
models, there is no need for an inflaton field and non-commutativity modifies the equation of state in the radiation-dominated universe in a way that it generates a quasi-exponential expansion. The initial conditions are thermal and not determined by a quantum vacuum. For the model proposed in~\cite{mag}, the predictions for the power spectra have been worked out in~\cite{koh}. Here, Brandenberger and Koh find that the spectrum of fluctuations is nearly scale invariant, and shows a small red tilt, the magnitude of which is different from what is obtained in a usual inflationary model with the same expansion rate.

On the other hand, non-commutativity could introduce corrections to standard
inflation. Such a, perhaps less radical approach, consists in assuming the usual
inflaton-driven background, where scalar and tensor perturbations propagate
with a Bunch and Davies vacuum as initial condition, but are subjected to
non-commutativity at short distance. It turns out that the power spectrum is modified according to (see e.g., \cite{Koivi1}, and references therein)
\ba
\label{nc_spectrum}
P=P_{0}\,e^{H\vec\theta\cdot \vec k} \,,
\ea
where $H$ is the Hubble parameter, $P_{0}$ is the usual commutative
spectrum, and $\vec \theta$ is the vector formed by the $\theta^{0i}$
components of $\theta^{\mu\nu}$. This prediction can be obtained by using a
deformation of statistics in non-commutative spacetime on the usual
inflationary computation. It can be also derived in an alternative way
beginning from an effective deformation of the Heisenberg algebra of the
inflaton field. The most important aspect of the result is that the spectrum
becomes direction-dependent. The perturbations thus distinguish a preferred
direction given by the vector $\vec\theta$ that specifies the non-commutativity between space and time.

Furthermore, it is interesting that the violation of isotropy can also
violate parity. This could provide what seems a quite unique property of
possible signatures in the CMB and large-scale structure. However, there is
also an ambiguity with the predictions of the simplest models, which is
related to interpretations of non-commuting quantum observables at the
classical limit. This is evident from the fact that one has to consider an
effectively imaginary $\vec{\theta}$ in the above formula
(\ref{nc_spectrum}). Reality of physical observables requires the odd parity
part of the spectrum (\ref{nc_spectrum}) to be imaginary. The appearance of
this imaginary parameter $\vec \theta$ into the theory may signal the
unitary violation that has been reported in theories of time-space
non-commutativity. It is known that the Seiberg--Witten map to string theory
applies only for space-space non-commutativity~\cite{Seiberg:1999vs}.
Nevertheless, the phenomenological consequence that the primordial
fluctuations can distinguish handedness, seems in principle a physically
perfectly plausible -- though speculative -- possibility, and what ultimately
renders it very interesting is that we can test by cosmological
observations. Thus, while lacking the completely consistent and unique
non-commutative field theory, we can parametrize the ambiguity by a phenomenological parameter whose correct value is left to be determined observationally. The parameter $\alpha \in [0,1]$ can be introduced~\cite{Koivi1} to quantify the relative amplitude of odd and even contributions in such a way that $P = \alpha P^+ + i(1-\alpha)P^-$, where $P^\pm = (P(\vec k)\pm P(-\vec k))/2$.

The implications of the anisotropic power spectra, such as
(\ref{nc_spectrum}), for the large-scale structure measurements, is discussed
below in Section~\ref{anisotropicconstraints_cmb_pk}. Here we proceed to
analyse some consequences of the non-commutativity
relation (\ref{canonical_nc}) to the higher order correlations of
cosmological perturbations. We find that they can violate both isotropy and
parity symmetry of the FRW background. In particular, the latter effect
persists also in the case $\alpha=1$. This case
corresponds to the prescription in~\cite{Akofor:2007fv} and in the
remainder of this subsection we restrict to this case for simplicity. Thus,
even when we choose this special prescription where the power spectrum is
even, higher order correlations will violate parity. This realizes the
possibility of an odd bispectrum that was recently contemplated
upon in~\cite{Kamionkowski:2010rb}.

More precisely, the functions $B$ defined in
Eq.~(\ref{eq:bispectrumdef}) for the three-point function of the
curvature perturbation can be shown to have the form
\ba
\label{bispectrum}
B_\Phi(\vec{k}_1,\vec{k}_2,\vec{k}_3) & = &
2\cos{\left( \vec{k}_1\wedge\vec{k}_2 \right)}\Big(\cosh(2H\vec{\theta}\cdot\vec{k}_3)P_0(\vec{k}_1)P_0(\vec{k}_2)f_s(\vec{k}_3) + 2\,{\mathrm{perm.}}\Big) \nonumber \\
& - & 2i\sin{\left( \vec{k}_1\wedge\vec{k}_2 \right)}\Big(\sinh(2H\vec{\theta}\cdot\vec{k}_3)P_0(\vec{k}_1)P_0(\vec{k}_2)f_s(\vec{k}_3) + 2\,{\mathrm{perm.}}\Big)\,,
\ea
where the function $f_s(k)$ is
\be
f_s(k)=\frac{N''}{2N'^2}\left(1+{n_{f_{{\mathrm{NL}},0}}}\,\ln\frac{k}{k_p}\right)\,,
\ee
$k_p$ being a pivot scale and primes denoting derivatives with respect to the inflaton field. The quantity ${n_{f_{{\mathrm{NL}},0}}}$ is the scale dependence in the commutative case explicitly given by
\be
\label{nf}
{n_{f_{{\mathrm{NL}},0}}}
  =\frac{N'}{N''}\left(
  -3\eta+
  \frac{V'''}{3H^2}\right)\,.
\ee
The spatial components of the non-commutativity
matrix $\theta_{ij}$ enter the bispectrum through the phase $\vec{k}_1\wedge\vec{k}_2= k_1^i k_2^j\,\theta_{ij}$.
They do not appear in the results for the spectrum and therefore affect only
the non-Gaussian statistics of primordial perturbations.

We now focus on this part in the following only and set all components of $\vec{\theta}$ equal to
zero. This gives
\ba
  \label{fnl_theta_ij}
  f_{{\mathrm{NL}},\theta}
  &=&\frac{5}{3}\cos{\left( \vec{k}_1\wedge\vec{k}_2
  \right)}\frac{P_{0}(k_1)P_{0}(k_2)f_s(k_3)+2\,{\mathrm{perm.}}}{P_{0}(k_1)P_{0}(k_2)+2\,{\mathrm{perm.}}}\ ,
\ea
where the only contribution from the non-commutativity is the pre-factor
involving the wedge product. This affects the scale dependence of
${n_{f_{{\mathrm{NL}},\theta}}}$ and can hence be constrained observationally. For example,
computing the scale-dependence for shape preserving variations of
the momentum space triangle, $\vec{k}_i\rightarrow \lambda \vec{k}_i$,
defined as
\be
{n_{f_{{\mathrm{NL}},\theta}}} = \frac{\partial \ln |f_{{\mathrm{NL}},\theta}(\lambda\vec{k}_1,\lambda\vec{k}_2,\lambda\vec{k}_3)|}{\partial \ln \lambda}\Big|_{\lambda=1}\,,
\ee
we find, in the present case
\be
\label{nfnl_result}
{n_{f_{{\mathrm{NL}},\theta}}} =
 -2k_1^ik_2^j\theta_{ij}\tan(k_1^ik_2^j\theta_{ij})+{n_{f_{{\mathrm{NL}},0}}}\,,
\ee
where ${n_{f_{{\mathrm{NL}},0}}}$ given by (\ref{nf}) is the result in the commuting
case. The part dependent on $\theta_{ij}$ arises purely from
non-commutative features.
The Euclid data can be used to constrain the scale dependence of the nonlinearity parameter $f_{{\mathrm{NL}},\theta}$, and the scale dependence could therefore place interesting bounds on $\theta_{ij}$.
We note however that the amplitude of the nonlinearity is not enhanced by
the purely spatial non-commutativity, but is given by the underlying
inflationary model. The amplitude on the other hand is exponentially
enhanced by the possible timespace non-commutativity.

Moreover, it is worth noting
that the result (\ref{nfnl_result}) depends on the wave vectors
$\vec{k}_1$ and $\vec{k}_2$ and hence on the shape of the momentum space
triangle. This is in contrast with the commutative case, where the
scale dependence is given by the same result (\ref{nf}) for all shape
preserving variations, $\vec{k}_i\rightarrow \lambda \vec{k}_i$, regardless of
triangle shape. This allows, in principle, to distinguish between
the contributions arising from the non-commutative properties of the
theory and from the standard classical inflationary physics or gravitational clustering.

To recapitulate, parity violations in the statistics of large-scale
structures would be a smoking gun signature of timespace non-commutativity
at work during inflation. Moreover, purely spatial non-commutativity
predicts peculiar features in the higher order correlations of the
perturbations, and in particular these can be most efficiently detected by
combining information of the scale- and shape-dependence of non-Gaussianity.
As discussed earlier in this document, this information
is extractable from the Euclid data.

\subsection{Vector field models and modulated perturbations}
\label{VSect}

Various inflationary models populated by vector fields can be
described with a Lagrangian of the following form 
\begin{equation}
L_{\mathrm{vector}}=-\frac{1}{4} f(\varphi)
F_{\mu\nu}F^{\mu\nu}+\frac{1}{2}m^2 B_{\mu}B^{\mu} \,,
\end{equation}
where $F_{\mu\nu}\equiv \partial_{\mu}B_{\nu}-\partial_{\nu}B_{\mu}$, and $f(\varphi)$ is a suitable function of the inflaton field. 
A Lagrangian containing just the standard kinetic term $F_{\mu\nu}F^{\mu\nu}$ would be conformally invariant 
thus preventing fluctuations of the vector field 
$B_{\mu}$ to be excited on super-horizon scales. Contrary to the case of a light scalar 
field, large-scale primordial perturbations of the vector field can be generated during inflation if the vector field 
is sufficiently massive (with $m^2\approx -2H^2$). 
This Lagrangian includes the case of a massive (curvaton) vector field (when $f\equiv 1$) studied by~\cite{Dimopoulos:2006,Dimopoulos:2008} and 
where the mass of the vector field is acquired via a non-minimal coupling to gravity to break conformal invariance. For some of these models 
there are actually some 
instability issues about the evolution of the primordial longitudinal perturbation modes of the vector field~\cite{Himmetoglou:2009a, Himmetoglou:2009b}. 
The models with varying kinetic function (when $f(\varphi)$ is switched on) allows to overcome these difficulties, 
since in this case the longitudinal mode is gauged away. They have been
studied in various contexts 
(e.g., \cite{Soda:2008, Dimopoulos:2010a}). 
The Ackerman--Carroll--Wise models, \cite{Ackerman:2007}, employ a different Lagrangian of the form 
$L_{\mathrm{vector}}=-\frac{1}{4} F_{\mu\nu}F^{\mu\nu}+\lambda(B^\mu B_\mu-m^2)$, so that the norm of the vector field is fixed by the Lagrangian 
multiplier $\lambda$. 
In these models (where inflation is driven by an inflaton field) the main effect of the vector field is a slightly anisotropic 
background evolution described by a metric, with $c(t)=b(t)$) with a backreaction on the inflaton 
field fluctuations, rather than the vector field perturbations themselves. Another possibility that has been 
explored is based on a non-Abelian gauge $SU(2)$ vector multiplet \cite{Bartolo:2009a, Bartolo:2009b}, 
providing a realistic model of gauge interactions neglected so far. 

A general prediction from all these scenarios is that the power spectrum of primordial perturbations can be written as 
\begin{equation}
\label{Panis}
P(\mathbf{k})=P(k)\left[1+g(k) (\hat{\mathbf{k}} \cdot \hat{\mathbf{n}})^2 \right]\, ,
\end{equation}
where $g(k)$ is the amplitude of the rotational invariance breaking (statistical isotropy breaking) induced by a 
preferred direction $\mathbf{n}$. Thus, the power spectrum 
is not just a function of $k$ but it depends on the wave vector $\mathbf{k}$. 
Usually the preferred direction is related to the vector fields $n^i \propto B^i$ 
while the amplitude is related to the contribution of the vector field perturbations 
to the total curvature perturbation $g \sim P_{\zeta_B}/P_\zeta$. 

However, beyond the various concrete realizations, the expression~(\ref{Panis}), 
first introduced in~\cite{Ackerman:2007}, provides a robust and useful way to study observable 
consequences of a preferred direction during inflation and also a practical
template for comparison with observations (see below). 
Usually the amplitude $g(k)$ is set to a 
constant $g_*$. A generalization of the above 
parametrization is $P(\mathbf{k})=P(k)\left[1+\sum_{LM} g_{\mathrm{LM}}(k) Y_{LM}(\hat{\mathbf{k}}) \right]$, 
where $Y_{\mathrm{LM}}(\hat{\mathbf{k}})$ are spherical harmonics with only even multipoles $L \ge 2$ \cite{Pullen:2007}.
Throughout, we use upper-case indices $LM$ for power anisotropies, and
lower-case indices $lm$ for temperature/polarization anisotropies. 
Interestingly enough, inflationary models with vector fields can also generate 
higher-order correlators, such as bispetrum and trispectrum, which display
anisotropic features as well 
(e.g., \cite{Soda:2008, Kar:2009, Bartolo:2009a, Bartolo:2009b}). 

The alignment of low CMB multipoles and the hemispherical power asymmetry observed in the CMB anisotropies can find an explanation in some models 
where the primordial 
gravitational perturbation is the result of fluctuations within our Hubble volume, modulated by super-horizon fluctuations. The primordial 
gravitational perturbation can thus be thought of as a product of two fields
$\Phi_1(\mathbf{x})$ and $\Phi_2(\mathbf{x})$ (\cite{Dvorkin:2008}, and references
therein) 
\begin{equation}
\Phi(\mathbf{x})=\Phi_1(\mathbf{x})\left[1+\Phi_2(\mathbf{x}) \right]\, , 
\label{modulated}
\end{equation}
where $\Phi_2(\mathbf{x})$ has only super-horizon fluctuations, so that within a given Hubble volume it takes a fixed value, while 
$\Phi_1(\mathbf{x})$ has sub-horizon stochastic fluctuations within that volume. The result is that an observer within our Hubble volume 
would see broken statistical 
homogeneity from the modulation on large scales of $\Phi_1(\mathbf{x})$, 
and also broken statistical isotropy from the gradient of the modulating field $\Phi_2(\mathbf{x})$. 
The dipole modulation $\delta T (\hat{\mathbf{p}})/T = S(\hat{\mathbf{p}})\left[1+A(\hat{\mathbf{p}} \cdot \hat{\mathbf{n}}) \right]$
used for CMB by, e.g., \cite{Eriksen:2007pc} and \cite{Hanson:2009} (or for LSS \cite{Hirata:2009ar}) to explain the hemispherical asymmetry
falls within the parametrization of Eq.~(\ref{modulated}).
A scenario with a dipole modulation has been realized in some concrete and detailed models, 
such as those involving adiabatic and isocurvature modulating perturbations from 
a curvaton field \cite{Erickcek:2008, Erickcek:2009}.

\subsection{Current and future constraints from CMB and LSS on an anisotropic power spectrum} \label{anisotropicconstraints_cmb_pk}

Groeneboom and Eriksen~\cite{Groeneboom:2009}, using WMAP5 year data (up to multipoles $\ell =400$), claimed a detection of 
a quadrupolar power spectrum of the form of Eq.~(\ref{Panis}) at more than $3 \sigma$ ($g_*=0.15 \pm 0.039$) with preferred direction 
$(l,b)=(110^{\circ}, 10^{\circ})$. Subsequently this result has been put under further check. \cite{Hanson:2009} confirmed 
this effect at high statistical 
significance, pointing out however that beam asymmetries could be a strong contaminant (see also~\cite{Hanson:2010}). 
The importance of this systematic effect is 
somewhat debated: \cite{Groeneboom:2010}, including polarization and beam asymmetries analysis excluded that the latter 
can be responsible for the observed effect. 
Their claim is a $9 \sigma$ detection with $g_*=0.29 \pm 0.031$. However, the preferred direction shifted much closer to 
the ecliptic poles, which is probably 
an indication that some unknown systematic is involved and must be corrected in order to obtain true constraints on any primordial modulation. 
Foregrounds and noise are disfavored as possible systematic effects~\citep{Bennet:2011,Groeneboom:2009}. Thus the cause of this kind of 
asymmetry is not definitely known. Planck should be able to detect a power quadrupole as 
small as 2\% (at $3 \sigma$)~\citep{Pullen:2007,Groeneboom:2009,Groeneboom:2010}. It is of course desirable to test this (and other anisotropic effects) with other techniques. 

What about large-scale structure surveys? Up to now there are just
a few analyses testing anisotropies in large-scale structure surveys, but all of them have been crucial, indicating that large-scale structure surveys such as Euclid offer a promising avenue to constrain these features.

Hirata~\cite{Hirata:2009ar} used high-redshift quasars from the Sloan Digital Sky Survey to rule out the simplest version of dipole modulation of the 
primordial power spectrum. In comparison the Planck mission using the CMB hemispherical asymmetry would only marginally distinguish it from the standard case \cite{Eriksen:2007pc}. 
The constraints obtained by high-redshift quasars require an amplitude for the dipole modulation 6 times smaller than the one required by CMB.
This would disfavor the simple curvaton spatial 
gradient scenario \citep{Erickcek:2008} proposed to generate this dipole modulation. 
Only a curvaton scenario with a non-negligible fraction of isocurvature perturbations at late 
times could avoid this constraint from current high-redshift quasars \citep{Erickcek:2009}.

Pullen and Hirata~\cite{Pullen:2010} considered a sample of photometric luminous red galaxies from the SDSS survey to assess the quadrupole anisotropy 
in the primordial power spectrum
of the type described by Eq.~(\ref{Panis}). The sample is divided into eight redshift slices 
(from $z=0.2$ up to $z=0.6$), 
and within each slice the galaxy angular power spectrum is analysed. They also accounted for 
possible systematic effects (such as a modulation of the signal and noise due to a slow variation 
of the photometric calibration errors across the survey) and redshift-space
distortion effects.
In this case \citep{Pullen:2010}
\begin{eqnarray}
\label{Cg}
C_g(\mathbf{n},\mathbf{n'}) = \langle \delta_g(\mathbf{n}) \delta_g(\mathbf{n'}) \rangle &=& \sum_l \frac{2l+1}{4\pi} C_{g,l}P_l(\mathbf{n} \cdot \mathbf{n'})
\nonumber \\ &&
+\sum_{LM}\sum_{lml'm'}D_{g,ll'}^{LM}X_{lml'm'}^{LM}R_{lm}(\mathbf{n})R_{l'm'}(\mathbf{n'})\, .
\end{eqnarray}
Here, the set of $C_{g,l}$s are given by the usual galaxy angular power spectrum for the case of statistical isotropy. 
Statistical anisotropy produces the second term 
\begin{eqnarray}
\label{Dg}
D_{g,ll'}^{LM}={\mathrm{i}}^{l-l'}\frac{2}{\pi}\int_0^\infty {\mathrm{d}}k\,k^2 P_g(k)g_{LM}W_l(k)W_{l'}(k)\,,
\end{eqnarray}
where $X_{lml'm'}^{LM}$ are geometric coefficients related to Wigner $3-j$ symbols, $R$ denotes the real spherical harmonics (see Eqs.~(3) and (13) of~\cite{Pullen:2007} for more details), $P_g(k)=b^2_g P(k)$ is the isotropic galaxy power spectrum and 
$W_l(k) = \int \mathrm{d}\chi f(\chi)j_l(k\chi)$ 
is the window function ($\chi$ is the comoving
distance, and $f(\chi)$ is the selection function, i.e., the normalized redshift distribution for 
a redshift slice). 
 
Assuming the same preferred direction singled out by \cite{Groeneboom:2009}, they derive a constraint on the anisotropy amplitude 
$g_*=0.006 \pm 0.036$ ($1 \sigma$), 
thus finding no evidence for anisotropy. 
Marginalizing over $\mathbf{n}$ with a uniform prior they find $-0.41 < g_* < 0.38$ at 95\% C.L. These results could confirm that the signal 
seen in CMB data is of systematic nature. However, it must be stressed that CMB and LSS analyses probe different scales, 
and in general the amplitude of the anisotropy is 
scale dependent $g=g(k)$, as in the model proposed in \cite{Erickcek:2009}. 
An estimate for what an experiment like Euclid can achieve is to consider how the uncertainty in $g_*$ 
scale in terms of number of modes measured 
and the number of redshift slices. Following the arguments of~\cite{Pullen:2010}, the uncertainty will scale roughly as $\ell_{\max}^{-1} N_z^{-1/2}$, 
where $\ell_{\max}$ is the maximum multipole at which the galaxy angular
power spectrum is probed, and $N_z$ is the number of redshift slices. 
Considering that the redshift survey of Euclid will cover 
redshifts $0.4 < z <2$, there is an increase by a factor of 3 in distance of the survey and hence a factor 3 increase in $l_{\max}$ 
($l_{\max} \sim k_{\max} \chi(z)$, see the expression for the selection function after Eq.~(\ref{Dg})). 
Taking $k_{\max}= 0.2h\, {\mathrm{Mpc}}^{-1}$ the effective number of redshift slices 
is also increased of a factor of $\sim 3$ ($N_z \sim k_{\max} \
\Delta \chi /\pi$, with $\Delta \chi$ the radial width of the survey). Therefore, one could expect that for a mission like 
Euclid one can achieve an uncertainty (at $1\sigma$)
$\sigma_{g_*} \sim 10^{-3}\mbox{\,--\,}10^{-2}$ or $\sigma_{g_*} \sim 10^{-2}$, for a fixed anisotropy axis or marginalizing over $\mathbf{n}$, respectively. 
This will be competitive with Planck measurements and highly complementary to it \citep{Paci:2010wp,Gruppuso:2010}.
Notice that these constraints apply to an analysis of the galaxy angular 
power spectrum. An analysis of the 3-dimensional power spectrum $P(\mathbf{k})$ could improve the sensitivity further. 
In this case the uncertainty would scale as $\Delta g_* \sim N^{-1/2}_{\mathrm{modes}}$, where $N_{\mathrm{modes}}$ is the number of independent Fourier modes.

\newpage
\part{Statistical Methods for Performance Forecasts}
\label{statistical}

\section{Introduction}

As cosmology becomes increasingly dominated by results emerging from
large-scale observational programmes, it is imperative to be able to
justify that resources are being deployed as effectively as possible.
In recent years it has become standard to quantify the expected
outcome of cosmological surveys to enable comparison, a procedure
exemplified by the Figure of Merit (FoM) introduced
by~\cite{Huterer:2000mj} and later used in the influential Dark Energy
Task Force (DETF) report about dark-energy surveys~\citep{Albrecht2006,Albrecht:2009ct}. 

The idea is to be able to capture in one single number the scientific return of a future mission, in order to be able to rank competing proposals and to forecast their ability to answer relevant scientific questions, such as: is dark energy a cosmological constant or does it evolve with time? Is it an expression of modified gravity? How well can a time-evolution of dark energy be constrained? 

Encapsulating the entire value of a proposed cosmological survey in one single number is of course highly reductive, and the ensuing conclusions should therefore be taken with a large grain of salt. Having said that, work in recent years has focused on attempts to devise Figures of Merit (FoMs) that represent in an increasingly realistic way future missions. It is perhaps obvious that, to a certain extent, the assessment of a future probe will depend on the scientific question one is most interested in: parameter constraints, model selection, robustness to systematics are but a few examples of the different levels on which a proposed mission can be evaluated and optimized.

This gives an overview of some of the approaches recently adopted in the field, and used elsewhere in this document to produce forecasts for Euclid. Useful references and background material to some of the concepts discussed below are: \cite{Trotta:2008qt,BMIC} for an introduction to Bayesian methods in cosmology,  \cite{Sivia:1996, MacKay:2003} for introductions to the Bayesian approach in data analysis, \cite{MCMC:1996} for an introduction to Markov Chain Monte Carlo (MCMC) methods.

\section{Predicting the science return of a future experiment}

We consider a toy Gaussian linear model in order to illustrate the different
approaches to performance forecast. We notice that, although motivated by
computational simplicity and the ability to obtain analytical results, a
Gaussian model is actually a fairly close representation of many cases of
interest. In Figure~\ref{fig:Gauss_1} we illustrate this point by plotting
the parameter constraints expected from a Euclid-like survey and the
corresponding Gaussian approximation in the Fisher-matrix approach to the
likelihood (described below). In these cases, it seem clear that the
Gaussian model captures fairly well the full probability distribution.
Another example shown in Figure~\ref{fig:Gauss_2} are cosmological
constraints from  WMAP and SDSS data, where a Gaussian approximation to the
likelihood (\reviewtext{the} so-called Laplace approximation) is seen to give an excellent
description of the full distribution obtained numerically via MCMC. 

\epubtkImage{multi_like.png}{%
\begin{figure}[htbp]
\centerline{\includegraphics[width=0.9\textwidth]{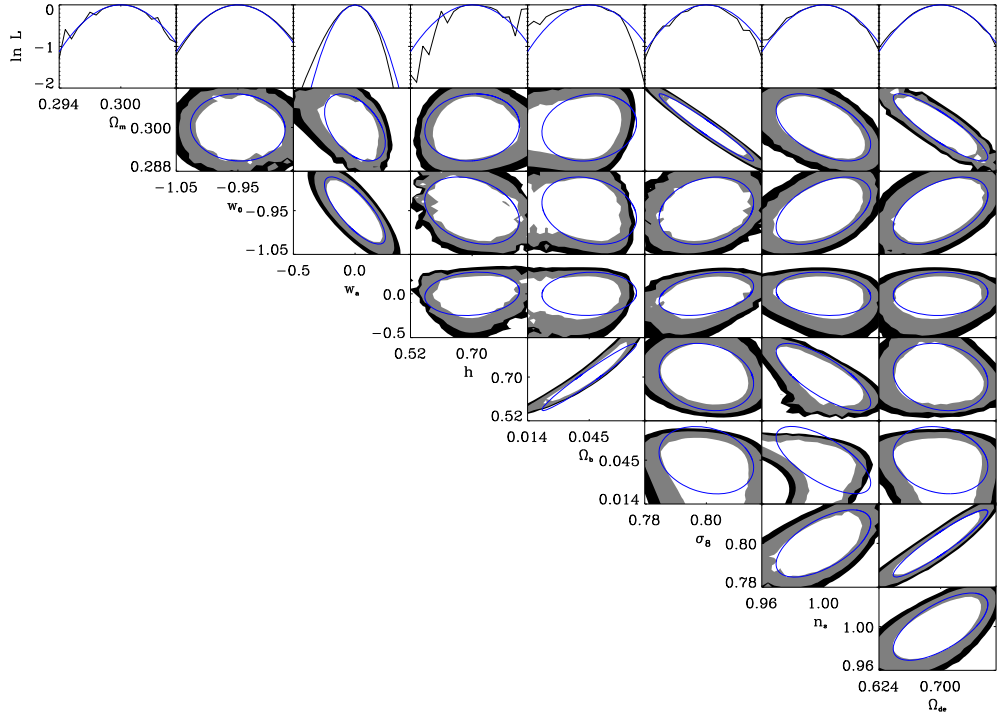}}
 \caption{Projected cosmological 8-parameter space for a 15,000 square
   degrees, median redshift of $z=0.8$, 10 bin tomographic cosmic
   shear survey. Specifications are based on Euclid Yellow book
   \cite{euclidyellowbook} as this figure is representative of a
   method, rather than on forecast analysis; the discussion is still
   valid with more updated \cite{euclidredbook} Euclid specifications.
   The upper panel shows the 1D parameter constraints using analytic
   marginalization (black) and the Gaussian approximation (Fisher
   matrix, blue, dark grey). The other panels show the 2D parameter
   constraints. Grey contours are 1- 2- and 3-$\sigma$ levels using
   analytic marginalization over the extra parameters, solid blue
   ellipses are the 1-$\sigma$ contours using the Fisher-matrix
   approximation to the projected likelihood surface, solid red
   ellipses are the 1-$\sigma$ fully
   marginalized. Image reproduced by permission
   from~\cite{Taylor-Kitching:2010}.}
 \label{fig:Gauss_1}
\end{figure}}

\epubtkImage{St_K14_CMB_sdss_r2_tri.png}{%
\begin{figure}[htbp]
\centerline{\includegraphics[width=0.8\textwidth]{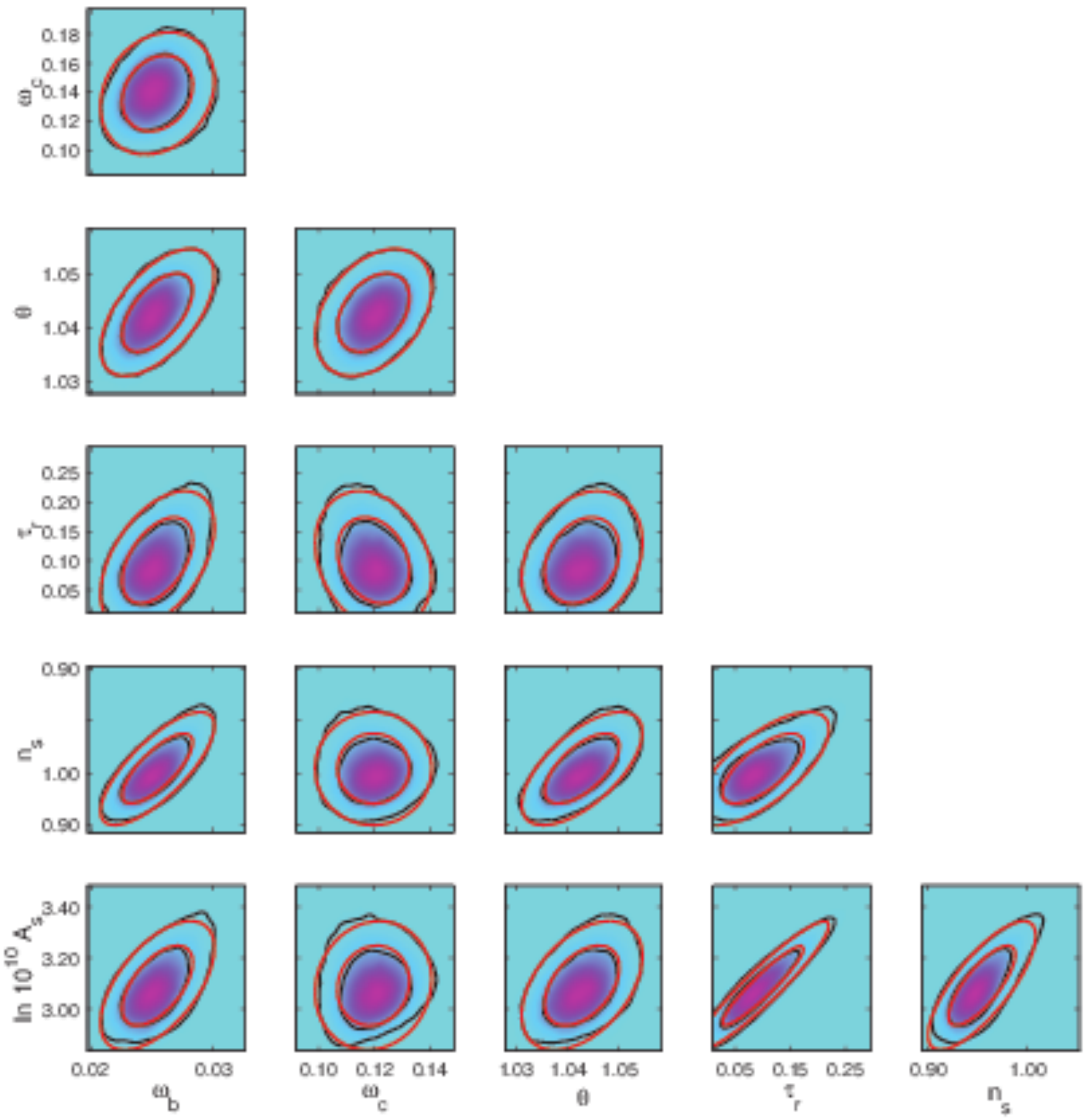}}
\caption{Gaussian approximation (Laplace approximation) to a
  6-dimensional posterior distribution for cosmological parameters,
  from WMAP1 and SDSS data. For all couples of parameters, panels show
  contours enclosing 68\% and 95\% of joint probability from
  $2\cdot10^5$ MC samples (black contours), along with the Laplace
  approximation (red ellipses). It is clear that the Laplace
  approximation captures the bulk of the posterior volume in parameter
  space in this case where there is little non-Gaussianity in the
  posterior PDF. Image reproduced from 2005 preprint of~\cite{Trotta:2005ar}.}
\label{fig:Gauss_2}
\end{figure}}

\subsection{The Gaussian linear model}

Suppose we have $N$ cosmological probes, whose likelihood function is
assumed to be a multi-dimensional Gaussian, given by: 
\begin{equation} \label{eq:likelihood_i}
\mathcal{L}_i (\params) \equiv p(D_i |\bf{\Theta})=
\mathcal{L}_{0}^{i}\exp\left(-\frac{1}{2}(\bf{\mu}_{i}-\params)^{t}C^{-1}_{i}(\bf{\mu}_{i}-\bf{\Theta})\right).
\end{equation}
where $\bf\Theta$ are the parameters one is interested in constraining, $D_i$
are the available data from probe $i$ and $\bf\mu_i$ is the location of the
maximum likelihood value in parameter space. The matrix is the covariance matrix of the parameters.

The posterior distribution for the parameters from each probe, $p(\params|D_i)$, is obtained by Bayes' theorem as 
\begin{equation} \label{eq:bayes}
p(\params|D_i)=\frac{p(\bf{\Theta})p(D_i|\bf{\Theta})}{p(D_i)},
\end{equation}
where and $p(\params)$ is the prior and $p(D_i)$ is a normalizing constant (the Bayesian evidence). 
If we assume a Gaussian prior centered around
the maximum likelihood value with inverse covariance
matrix $\Sigma$, the posterior from each probe is also a Gaussian, with
inverse covariance matrix \label{symbol:invcov}
\begin{equation}
F_i = C^{-1}_i + \Sigma 
\end{equation}
and posterior mean 
\begin{equation}
\overline{\mu}_{i} = F_{i}^{-1}(C^{-1}_{i}\mu_{i}).
\end{equation} 
Tighter constraints on the parameters can be usually obtained
by combining all available probes together (provided there are no
systematics, see below). If we combine all probes together 
 (assuming they are independent), we obtain a
Gaussian posterior with inverse covariance matrix
\begin{equation} \label{eq:post_Fisher}
F = \sum_{i=1}^N C^{-1}_i  + \Sigma
\end{equation}
and mean  
\begin{equation}
\overline{\mu} =F^{-1}\sum_{i=1}^N C^{-1}_{i}\mu_{i}.  \label{eq:post_mean}
\end{equation}
Notice that the precision of the posterior (i.e., the inverse covariance matrix) does not
 depend on the degree of overlap of the likelihoods from the individual
 probes. This is a property of the Gaussian linear model. 
 
For future reference, it is also useful to write down the general expression
for the Bayesian evidence. For a normal prior
$p(\params) \propto {\mathcal{N}}(\theta_{\pi},\Sigma)$ and a
likelihood 
\begin{equation}
  \mathcal{L}(\params)=\mathcal{L}_{0}
  \exp\left(-\frac{1}{2}(\theta_{0}-\params)^{t}L(\theta_{0}-\bf{\Theta})\right),\end{equation}
the evidence for data $d$ is given by
\begin{equation}
p(d) \equiv\int{\mathrm{d}}\params p(d|\bf{\Theta})p(\bf{\Theta})=
\mathcal{L}_{0}\frac{|\Sigma|^{1/2}}{|F|^{1/2}} 
 \exp\left[-\frac{1}{2}\left(\theta_{0}^{t}L\theta_{0}+
    \theta_{\pi}^{t}\Sigma\theta_{\pi}-\overline{\theta}^{t}F\overline{\theta}\right)\right],
\label{eq:evidence}
\end{equation}
where $F$ is given by Eq.~\eqref{eq:post_Fisher} with $N=1$ and
$\overline{\theta} = F^{-1}L\theta_0$.

\subsection{Fisher-matrix error forecast}
\label{Fisher_matrix}

A general likelihood function for a future experiment (subscript $i$) can be Taylor-expanded around its maximum-likelihood value, $\mu_i$. By definition, at the maximum the first derivatives vanish, and the shape of the log-likelihood in parameter space is approximated by the Hessian matrix
$H_i$,
 \begin{equation}
 \label{eq:like_second_order_expansion}
 \lnlike_i(\params) \approx \lnlike_i(\mu_i)
 + \frac{1}{2}(\params-\mu_i)^t H_i (\params-\mu_i),
 \end{equation}
where $H_i$ is given by
 \begin{equation}
 \left(H_i\right)_{\alpha \beta} \equiv
\frac{\partial^2 \lnlike_i}{\partial \params_\alpha \partial \params_\beta} {{\Big\arrowvert}_{\mu_i}} ,
 \end{equation}
and the derivatives are evaluated at the maximum-likelihood
point. By taking the expectation of equation
\eqref{eq:like_second_order_expansion} with respect to many data
realizations, we can replace the maximum-likelihood value $\mu_i$ with
the true value, $\fid$, as the maximum-likelihood estimate is
unbiased (in the absence of systematics), i.e., $\langle \mu_i \rangle = \fid$. We then define
the Fisher information matrix as the expectation value of the
Hessian,
 \begin{equation} \label{eq:Fisher_Matrix}
 F_i \equiv
\langle H_i \rangle. 
 \end{equation}
The inverse of the Fisher matrix, $F^{-1}$, is an estimate of the
covariance matrix for the parameters, and it describes how fast
the log-likelihood falls (on average) around the maximum
likelihood value, and we recover the Gaussian expression for the likelihood, Eq.~\eqref{eq:likelihood_i}, with the maximum likelihood value replaced by the true value of the parameters and the inverse covariance matrix given by the Fisher matrix, $L_i = F_i^{-1}$~\citep{KendallStuart}.
In general, the derivatives depend on where in parameter space we
take them (except for the simple case of linear models), hence it
is clear that $F_i$ is a function of the fiducial parameters.

Once we have the Fisher matrix, we can give estimates for the
accuracy on the parameters from a future measurement, by computing the posterior as in Eq.~\eqref{eq:bayes}. If we are only interested in a subset of the parameters, then
we can marginalize easily over the others: computing the Gaussian
integral over the unwanted parameters is the same as inverting the
Fisher matrix, dropping the rows and columns corresponding to
those parameters (keeping only the rows and columns containing the
parameters of interest) and inverting the smaller matrix back. The
result is the marginalized Fisher matrix $\mathcal{F}_i$. For example,
the $1\sigma$ error for parameter $\alpha$ from experiment $i$, marginalized over all other
parameters, is simply given by $\sigma_\alpha =
\sqrt{(F_i^{-1})_{\alpha \alpha}}$. 

It remains to compute the Fisher matrix for the future experiment.
This can be done analytically for the case where the likelihood
function is approximately Gaussian in the data, which is a good
approximation for many applications of interest. We can write for
the log-likelihood (in the following, we drop the subscript $i$ denoting the experiment under consideration for simplicity of notation)
 \begin{equation}
 -2 \lnlike = \ln |C| + (D-\mu)^t C^{-1} (D-\mu),
 \end{equation}
where $D$ are the (simulated) data that would be observed by the
experiment and in general both the mean $\mu$ and covariance
matrix $C$ may depend on the parameters $\params$ we are trying to
estimate. The expectation value of the data corresponds to the
true mean, $\langle D \rangle = \mu$, and similarly the
expectation value of the data matrix $\Delta \equiv
(D-\mu)^t(D-\mu)$ is equal to the true covariance, $\langle \Delta
\rangle = C$. Then it can be shown (see e.g.~\cite{Tegmark:1996bz})
that the Fisher matrix is given by
\begin{equation}\label{eq:Falphabeta}
    F_{\alpha \beta} = \frac{1}{2} {\mathrm{tr}}\left[ A_{\alpha} A_{\beta} +
    C^{-1} \langle \Delta_{,\alpha\beta} \rangle \right],
\end{equation}
where $A_\alpha \equiv C^{-1} C_{,\alpha}$ and the comma denotes a
derivative with respect to the parameters, for example  $C_{,\alpha}
\equiv
\partial C/\partial \params_\alpha$. The fact that this
expression depends only on \emph{expectation values} and not on the
particular data realization means that the Fisher matrix can be
computed from knowledge of the noise properties of the experiment
without having to go through the step of actually generating any
simulated data. The specific form of the Fisher matrix then
becomes a function of the type of observable being considered and
of the experimental parameters.

Explicit expressions for the Fisher matrix for cosmological
observables can be found in~\cite{Tegmark:1996bz} for cosmic
microwave background data, in~\cite{Tegmark:1997rp} for the matter power
spectrum from galaxy redshift surveys (applied to baryonic
acoustic oscillations in~\cite{seo03} and in~\cite{Hu:2003pt} for
weak lensing. These approaches have been discussed in
Section~\ref{observational-properties-of-modified-gravity}. A useful
summary of Fisher matrix technology is given in the Dark Energy Task
Force report~\citep{Albrecht2006} and in \cite{Verde:2009tu}. A useful
numerical package which includes several of the above calculations is
the publicly available Matlab code%
\epubtkFootnote{Available from \url{http://www.cosmology.org.za}}
\textsc{Fisher4Cast}~\citep{Bassett:2009uv,Bassett:2009tw}. Attempts
to include systematic errors modelling in this framework can be found
in \cite{Kitching-Taylor:2010,Taylor-Kitching:2010,
  Kitching-etal:2009,MNR:MNR13419}. 
  
For Gaussian data, the Fisher matrix can be obtained also by expanding the
mean $\mu$ and the correlation matrix $C$ to first order in a Taylor
series around the posterior peak in parameter space. The credible
regions obtained from the Fisher matrix approach are a good approximation
to the exact results only when this linear expansion is acceptable.
A way to test for this assumption and to correct it when it fails
is obtained by expanding the mean and correlation to higher orders.
If the parameters are only in the mean, one can show that the posterior
remains normalizable at all orders; additional conditions are needed
in the more general case. Expanding to second order in the derivatives
one obtains \cite{2014arXiv1401.6892S}

\begin{equation}
\begin{aligned}\mathrm{\mathcal{L}}(\params)=N\exp\bigg[ & \left.-\frac{1}{2}\boldsymbol{\mu}_{,\alpha}C^{-1}\boldsymbol{\mu}_{,\beta}\Delta\theta_{\alpha}\Delta\theta_{\beta}-\frac{1}{2}\boldsymbol{\mu}_{,\alpha\beta}C^{-1}\boldsymbol{\mu}_{,\gamma}\Delta\theta_{\alpha}\Delta\theta_{\beta}\Delta\theta_{\gamma}\right.\\
& -\frac{1}{8}\boldsymbol{\mu}_{,\delta\gamma}C^{-1}\boldsymbol{\mu}_{,\beta\alpha}\Delta\theta_{\alpha}\Delta\theta_{\beta}\Delta\theta_{\gamma}\Delta\theta_{\delta}\,\bigg]\,,
\end{aligned}
\label{eq:exp-1}
\end{equation}
where $\Delta\theta_{\alpha}=\params_{\alpha}-\theta_{\alpha}$. The
first term is the standard Fisher term (when the parameters are only
in the mean), the other terms express the deviations from Gaussianity
in parameter space. As long as the deviation from Gaussianity is not
too large, the additional terms describe the common ``banana'' shaped
posteriors often encountered in the data analysis. Clearly, if these
terms dominate over the Fisher term then a full exploration of the
posterior, for instance through Monte Carlo methods, is advisable. 

\subsection{Figure of merits}

It has become customary to describe the statistical power of a future dark
energy probe by the inverse area enclosed by the 68\% covariance ellipse marginalized down to the dark-energy parameter space. This measure of
statistical performance for probe $i$ (widely known as the DETF
FoM~\cite{Albrecht2006,Huterer:2000mj}) is usually defined (up to multiplicative
constants) as 
\begin{equation}
\text{FoM} = |F_i|^{1/2} \,,  
\end{equation}
where the Fisher matrix $F_i$ is given in Eq.~\eqref{eq:Fisher_Matrix}. 
\cite{Albrecht2006} suggested to use the inverse area of the 95\% error ellipse of $w_0-w_a$ (where $w_0$ and $w_a$ are defined in \cite{Linder:2002et}, \cite{chevallier01}). This definition was inspired by \cite{Huterer:2000mj}. In \cite{Albrecht:2009ct} it is suggested to model $w(a)$ as piecewise constant values of $w(a)$ defined in many small redshift bins ($\Delta a = 0.025$). The suggestion is then to apply a principal component approach \cite{Huterer:2002hy} in order to understand the redshifts at which each experiment has the power to constrain $w$.

A closely related but more statistically motivated measure of the information gain is the Kullback--Leibler divergence (KL) between the
posterior and the prior, representing the information gain
obtained when upgrading the prior to the posterior via Bayes' theorem:
\begin{equation} 
D_{\mathrm{KL}} \equiv\int
  p(\params|D)\ln\frac{p(\bf{\Theta}|D)}{p(\bf{\Theta})}d\bf{\Theta}.
\label{eq:def_KL}
\end{equation}
The KL divergence measures the relative entropy between the two distributions:
it is a dimensionless quantity which expresses the information gain obtained via
the likelihood. 
For the Gaussian likelihood and prior introduced above, the information gain
(w.r.t.\ the prior $\Sigma$) from the combination of all probes is given
by~\cite{BMICRTetal}
\begin{equation} \label{eq:DKL}
  D_{\mathrm{KL}}=\frac{1}{2}\left(\ln|F|-\ln|\Sigma|-{{\mathrm{tr}}[1-\Sigma
      F^{-1}]}\right).
\end{equation} 

A discussion of other, alternative FoMs (D-optimality, A-optimality) can be
found in~\cite{Bassett05}. In~\cite{Wang:2008zh} a different FoM for dark energy
is suggested. For a set of DE parameters $\params$, the FoM is defined
as $\mathrm{FoM} = 1/\sqrt{\det C(\params)}$, where $C(\params)$
is the covariance matrix of $\params$. This definition is more
flexible since one can use it for any DE parametrization~\citep{Wang2010}. 

Given that Euclid can constrain both the expansion history and the growth of
structure, it is also useful to introduce a new FoM for the growth of
perturbations. Similarly to the DETF FoM, one can define this new FoM as the
inverse area of the 95\% error ellipse of $\Omega_m-\gamma$, where $\gamma$ is
the growth index, defined starting from the growth rate $f_G(z) \equiv \frac{d\ln
G(z)}{d\ln a} = \Omega_m^\gamma$, or as $1/\sqrt{\det C(w_0,w_a,\gamma)}$ or similar variants
\citep{majerotto11,diporto10}. Instead of $\gamma$, other
parameters describing the growth can also be employed.

A FoM targeted at evaluating the robustness of a future probe to potential systematic 
errors has been introduced in~\cite{MarchRobustness}. The robustness of a
future probe is defined via the degree of overlap between the posterior
distribution from that probe and the posterior from other, existing probes.
The fundamental notion is that maximising statistical power (e.g., by
designing a future probe to deliver orthogonal constraints w.r.t.\ current probes) will in general reduce its robustness (by increasing the probability of an incompatible results, for example because of systematic bias). Thus in evaluating the strength of a probe, both its statistical power and its resilience to plausible systematics ought to be considered.

\subsection{The Bayesian approach} 

When considering the capabilities of future experiments, it
is common stance to predict their performance in terms of
constraints on relevant parameters, assuming a fiducial point in
parameter space as the true model (often, the current best-fit
model), as explained above. While this is a useful indicator for parameter inference
tasks, many 
questions in cosmology fall rather in the model
comparison category. Dark energy is a case in point, where
the science driver for many future probes (including Euclid)
is to detect possible departures from a cosmological constant,
hence to gather evidence in favor of an evolving dark-energy
model. Therefore, it is preferable to assess the capabilities of
future experiments by their ability to answer model selection
questions.

The procedure is as follows (see~\cite{Mukherjee:2005tr} for
details and the application to dark-energy scenarios). At every
point in parameter space, mock data from the future observation
are generated and the Bayes factor between the competing models is
computed, for example between an evolving dark energy and a
cosmological constant. Then one delimits in parameter space the
region where the future data would \emph{not} be able to deliver a
clear model comparison verdict, for example $\vert \ln B_{01} \vert <
5$ (evidence falling short of the ``strong'' threshold). Here, $B_{01}$ is the Bayes factor, which
is formed from the ratio of the Bayesian evidences of the two models
being considered:
\begin{equation}
B_{01} = \frac{p(\data | \mdl_0)}{p(\data | \mdl_1)},
\label{eq:bayesfactor}
\end{equation}
where the Bayesian evidence is the average of the likelihood under the prior in each model (denoted by a subscript $m$):
\begin{equation}
p(\data | \mdl_m) = \int \dr\params_m p(\data | \params_m, \mdl_m)p(\params_m | \mdl_m).
\end{equation}
 The Bayes factor updates the prior probability ratio
of the models to the posterior one, indicating the extent to which the
data have modified one's original view on the relative probabilities
of the two models. The
experiment with the smallest ``model-confusion'' volume in
parameter space is to be preferred, since it achieves the highest
discriminative power between models. An application of a related
technique to the spectral index from the Planck satellite is
presented in~\cite{Pahud:2007gi,Pahud:2006kv}.

Alternatively, we can investigate the full probability
distribution for the Bayes factor from a future observation. This
allows to make probabilistic statements regarding the outcome of a
future model comparison, and in particular to quantify the
probability that a new observation will be able to achieve a
certain level of evidence for one of the models, given current
knowledge. This technique is based on the \emph{predictive
distribution} for a future observation, which gives the expected
posterior for an observation with a certain set of experimental
capabilities (further details are given in~\cite{Trotta:2007hy}). This method
is called PPOD, for \emph{predictive posterior odds distribution}
and can be useful in the context of experiment design and
optimization

Hybrid approaches have also been attempted, i.e., to defined model-selection
oriented FoMs while working in the Fisher-matrix
framework, such as the Bayes
factor~\cite{Heavens-etal:2007, Amara-Kitching:2010}.

The most general approach to performance forecasting involves the use of a
suitably defined utility function, and it has recently been presented in
\cite{Trotta:2010ug}. Consider the different levels of uncertainty that are
relevant when
predicting the probability of a certain model selection outcome from a
future probe, which can be summarized as follows:
\begin{itemize}
\item \textbf{Level 1:} current uncertainty about the correct model
  (e.g., is it a cosmological constant or a dark-energy model?). 
\item \textbf{Level 2:} present-day uncertainty in the value of the
  cosmological parameters for a given model (e.g., present error on
  the dark-energy equation of state parameters assuming an evolving
  dark-energy model). 
\item \textbf{Level 3:} realization noise, which will be present in
  future data even when assuming a model and a fiducial choice
  for its parameters.    
\end{itemize}
The commonly-used Fisher matrix forecast ignores the uncertainty arising from Levels~1 and 2, as it assumes a fiducial model (Level~1) and fiducial
parameter values (Level~2). It averages over realization noise
(Level~3) in the limit of an infinite number of realizations. Clearly,
the Fisher matrix procedure provides a very limited assessment of what
we can expect for the scientific return of a future probe, as it
ignores the uncertainty associated with the choice of model and parameter values. 

The Bayesian framework allows improvement on the usual Fisher matrix error
forecast thanks to a general 
procedure which fully accounts for all three levels of uncertainty given above. 
Following~\cite{Loredo:2003nm}, we think of
future data $\Df$ as \emph{outcomes}, which arise as consequence of our
choice of experimental parameters $e$ (\emph{actions}). For each action
and each outcome, we define a utility function $\Uf(\Df, e)$. Formally, the
utility only depends on the future data realization $\Df$. However, as will
become clear below, the data $\Df$ are realized from a fiducial model and model
parameter values. Therefore, the utility function implicitly depends on the
assumed model and parameters from which the data $\Df$ are generated. The best
action is the one that maximizes the expected utility, i.e., the
utility averaged over possible outcomes:
\begin{equation} \label{def:EU}
\EU (e) \equiv \int \dr \Df p(\Df | e, \data) \Uf(\Df, e).
\end{equation}
Here, $p(\Df | e, \data) $ is the predictive distribution
for the future data, conditional on the experimental setup ($e$) and
on current data ($\data$). For a single fixed model the
predictive distribution is given by 
\begin{align} \label{eq:predictive_general}
p(\Df | e, \data) &  = \int \dr\params \, p(\Df, \params | e, \data) \nonumber\\
 			 & =  \int \dr\params \, p(\Df | \params ,  e,
 			 \data) p(\params|e,\data) \nonumber\\ 
			 & = \int \dr\params \, p(\Df | \params ,  e)
 			 p(\params|\data) \,, 
\end{align}
where the last line follows because $p(\Df | \params ,  e,  \data) =
p(\Df | \params ,  e)$ (conditioning on current data is irrelevant
once the parameters are given) and  $p(\params|e,\data)  = p(\params|
\data)$ (conditioning on future experimental parameters is irrelevant
for the present-day posterior). So we 
can predict the probability distribution for future data $\Df$ 
by averaging the likelihood function for the future measurement
(Level~3 uncertainty) over the current posterior on the parameters
(Level~2 uncertainty). The expected utility then becomes
\begin{equation} \label{eq:exp_utility_1}
\EU (e) = \int \dr \params p(\params| o, \data) \int \dr \Df p(\Df |
\params, e) \Uf(\Df, e) \,. 
\end{equation}

So far, we have tacitly assumed that only one model was being considered for the
data. In practice, there will be several models that one is interested in
testing (Level 1 uncertainty), and typically there is uncertainty over which one
is best. This is in fact one of the main motivations for designing a new dark
energy probe. If $M$ models $\{ \mdl_1, \dots, \mdl_M \}$ are being considered,
each one with parameter vector $\params_m$ ($m=1,\dots, M$), the current
posterior can be further extended in terms of model
averaging (Level 1), weighting each model by its current model
posterior probability, $p(\mdl_m | \data)$, obtaining from
Eq.~\eqref{eq:exp_utility_1} the model-averaged expected utility
\begin{align} \label{eq:exp_utility_2}
\EU (e) & = \sum_{m=1}^M p(\mdl_m| \data) \int \dr \params_m p(\params_m|
\data ,\mdl_m) \nonumber\\ 
        & \times \int \dr \Df p(\Df | \params_m,
e,\mdl_m) \Uf(\Df, e,\mdl_m) \,.
\end{align}
This expected utility is the most general definition of a FoM for a future
experiment characterized by experimental parameters $e$. The usual Fisher matrix
forecast is recovered as a special case of Eq.~\eqref{eq:exp_utility_2}, as are
other \emph{ad hoc} FoMs that have been defined in the literature. Therefore
Eq.~\eqref{eq:exp_utility_2} gives us a formalism to define in all generality
the scientific return of a future experiment. This result clearly accounts for
all three levels of uncertainty in making our predictions: the utility function
$\Uf(\Df, e,\mdl_m)$ (to be specified below) depends on the future data
realization, $\Df$, (Level~3), which in turn is a function of the fiducial
parameters value, $\params_m$, (Level~2), and is averaged over present-day model
probabilities (Level~1).

This approach is used in~\cite{Trotta:2010ug} to define two model-selection
oriented Figures of Merit: the decisiveness $\mathcal{D}$, which quantifies the
probability that a probe will deliver a decisive result in favor or against the
cosmological constant, and the expected strength of evidence, $\mathcal{E}$, that returns a measure of the expected power of a probe for model selection.

\section{Survey design and optimization}

Although the topic of survey design is still in its infancy, the basic idea is to carry
out an optimization of survey parameters (such as for example choice of targets,
depth of field, number of spectroscopic fibers, etc.) in order to identify the
configuration that is more likely to return a high FoM for the scientific
question being considered. Example of this approach applied to dark-energy parameters can be found
in~\cite{Bassett05,Parkinson:2007cv,Parkinson:2009zi,Bassett:2004np}, while \cite{Loredo:2003nm} discussed a more general methodology.  In \cite{Bassett05} a method is defined to
optimize future surveys, in the framework of Bayesian statistics and without
necessarily assuming a dark-energy model. In \cite{Bassett:2004np},
\cite{Parkinson:2007cv} and \cite{Parkinson:2009zi} this method is used to
produce forecasts for future weak lensing and galaxy redshift surveys.

The optimization process is carried out subject to constraints, such as for
example design parameter ranges and/or cost constraints. This is generally a
numerically complex and computationally expensive procedure. It typically
requires to explore the design parameters space (e.g., via MCMC), generating at
each point a set of pseudo-data that are analysed as real data would, in order
to compute their FoM. Then the search algorithm moves on to maximize the FoM. 

In order to carry out the optimization procedure, it might be useful to
adopt a principal component analysis (PCA) to determine a suitable
parametrization of $w(z)$~\citep{Huterer:2002hy,Simpson:2006bd}. The redshift range of the survey can be split into $N$ bins, with the equation of state taking on a value $w_i$ in the $i$-th bin:
\begin{equation}
w(z) = \sum_{i = 1}^N w_i b_i(z) \,.
\end{equation}
where the basis functions $b_i$ are top-hats of value 1 inside the bin, and 0 elsewhere. If $F$ is the Fisher matrix for the $N$ parameters $w_i$, one can diagonalize it by writing $F = W^T \Lambda W$, where $\Lambda$ is a diagonal matrix, and the rows of $W$ are the eigenvectors $e_i(z)$ or the so-called principal components. These define a new basis (in which the new coefficients $\alpha_i$ are uncorrelated) so the equation of state can be written as
\begin{equation}
w(z) = \sum_{i = 1}^N \alpha_i e_i(z) \,.
\end{equation}
The diagonal elements of $\Lambda$ are the eigenvalues $\lambda_i$ and define the variance of the new parameters, $\sigma^2(\alpha_i)  = 1/\lambda_i$.

One can now reconstruct $w(z)$ by keeping only a certain number of the most accurately determined modes, i.e., the ones with largest eigenvalues. The optimal number of modes to retain can be estimated by minimizing the risk, defined as the sum of the bias squared (how much the reconstructed equation of state departs from the true one by neglecting the more noisy modes) plus the variance of the estimate~\citep{Huterer:2002hy}.

\section{Future activities and open challenges}

As outlined in the previous sections, several approaches are available to capture the expected scientific performance of Euclid. As part of future theoretical activities, it will be necessary to build on the above concepts in 
order to obtain a realistic assessment of the science return of Euclid. Operationally, this means that the following tasks will need to be carried out:
\begin{itemize}
\item Estimation of likelihood contours around the maximum likelihood peak beyond the Fisher matrix approach. We envisage here a programme where  
simulated mock data will be generated and then used to blindly reconstruct the likelihood surface to sufficient accuracy. 

\item Estimation of Bayesian posterior distributions and assessment of impact of various priors. Bayesian inference is a mature field in cosmology and we now have at our disposal a number of efficient and reliable numerical algorithms based on Markov Chain Monte Carlo or nested sampling methods.

\item Comparison of Bayesian inferences with frequentist inferences based on profile likelihoods. Discrepancies might occur in the presence of large ``volume and sampling effects'' arising from insufficiently constraining data sets and highly multi-modal likelihoods~\citep{Trotta:2008bp}. Based on our experience so far, this is unlikely to be a problem for most of the statistical quantities of interest here but we recommend to check this explicitly for the more complicated distributions.
  
\item Investigation of the coverage properties of Bayesian credible and frequentist confidence intervals. Coverage of intervals is a fundamental property in particle physics, but rarely discussed in the cosmological setting. We recommend a careful investigation of coverage from realistically simulated data sets (as done recently in \cite{March:2011xa}). Fast neural networks techniques might be required to speed up the inference step by several orders of magnitude in order to make this kind of studies computationally feasible~\citep{shaw07,Bridges:2010de}.

\item Computation of the Bayesian evidence to carry out Bayesian model
selection~\citep{Trotta:2008qt,Mukherjee:2005tr}. Algorithms based on nested
sampling, and in particular, \textsc{MultiNest}~\citep{feroz08}, seem to be ideally
suited to this task, but other approaches are available, as well, such as
population Monte Carlo~\citep{2011arXiv1101.0950K} and semi-analytical
ones~\citep{Trotta:2005ar,Heavens-etal:2007}. A robust Bayesian model
selection will require a careful assessment of the impact of priors.
Furthermore, the outcome of Bayesian model selection is dependent on the
chosen parametrization if different nonlinearly related reparametrizations can
equally plausibly be chosen from physical consideration (relevant examples
include parametrizations of the isocurvature fraction~\citep{Beltran:2005xd}, the tensor-to-scalar ratio~\citep{Parkinson:2006ku} and the inflaton potential~\citep{Martin:2010hh}). It will be important to cross check results with frequentist hypothesis testing, as well. The notion of Bayesian doubt, introduced in~\cite{March:2010ex}, can also be used to extend the power of Bayesian model selection to the space of unknown models in order to test our paradigm of a $\Lambda$CDM cosmological model.

\item Bayesian model averaging~\citep{Liddle:2006kn,Parkinson:2010zr} can also be used to obtain final inferences parameters which take into account the residual model uncertainty. Due to the concentration of probability mass onto simpler models (as a consequence of Occam's razor), Bayesian model averaging can lead to tighter parameter constraints than non-averaged procedures, for example on the curvature parameter~\citep{Vardanyan:2011in}.

\item In cosmology and astronomy it is intrinsically difficult to deal with systematic errors that can bias an experiment. Particularly important can be the issue of dataset contamination. For instance, spurious transients such as core-collapse supernovae could end up in SNIa catalogs with the effect of biasing the best fit (i.e.~affecting the accuracy of an experiment) rather than increasing the errors (i.e.~affecting the precision)~\citep[see][Figure 3]{Amendola:2012wc}.
The way out is to look for deviating subpopulations in data in order to purge the dataset of possible contaminations.
A possible tool for this task is the \emph{internal robustness} introduced in \cite{Amendola:2012wc}, which uses bayesian evidences to assess if the dataset at hand is better described by a single cosmological model or rather by two models, a cosmological one for unbiased data and a systematical one for contaminated data.
It will be necessary to develop fast algorithms in order to scan the vast datasets that will be generated by Euclid in order to find partitions that minimize the value of internal robustness.

\end{itemize}

\newpage

\ifpdf
\phantomsection\addcontentsline{toc}{section}{Acknowledgments}\else
\fi
\section*{Acknowledgments}

It is a pleasure to thank the Euclid theory science working group, all
science working group members, the Euclid Consortium Board and the whole
Euclid consortium for fruitful discussions. We also kindly thank Rene
Laureijs, Ana Heras, Philippe Gondoin, Ludovic Duvet and Marc Sauvage for
their continuous work and support, and Thomas Buchert
for comments on the draft.

Stefano Camera is funded by FCT-Portugal under Post-Doctoral Grant No.
SFRH/BPD/80274/2011.

Carlos J. Martins is supported by an FCT Research Professorship, contract
reference IF/00064/2012, funded by FCT/MCTES (Portugal) and POPH/FSE (EC).

Valeria Pettorino acknowledges the Transregio TRR33 grant on `The Dark Universe'.

Claudia de Rham is supported by a Department of Energy grant DE-SC0009946.

Valerio Marra is supported by the research agency CNPq.


\def\ref@jnl#1{{#1}}

\def\aj{\ref@jnl{AJ}}                   
\def\actaa{\ref@jnl{Acta Astron.}}      
\def\araa{\ref@jnl{ARA\&A}}             
\def\apj{\ref@jnl{ApJ}}                 
\def\apjl{\ref@jnl{ApJ}}                
\def\apjs{\ref@jnl{ApJS}}               
\def\ao{\ref@jnl{Appl.~Opt.}}           
\def\apss{\ref@jnl{Ap\&SS}}             
\def\aap{\ref@jnl{A\&A}}                
\def\aapr{\ref@jnl{A\&A~Rev.}}          
\def\aaps{\ref@jnl{A\&AS}}              
\def\azh{\ref@jnl{AZh}}                 
\def\baas{\ref@jnl{BAAS}}               
\def\bac{\ref@jnl{Bull. astr. Inst. Czechosl.}}
\def\caa{\ref@jnl{Chinese Astron. Astrophys.}}
\def\cjaa{\ref@jnl{Chinese J. Astron. Astrophys.}}
\def\icarus{\ref@jnl{Icarus}}           
\def\jcap{\ref@jnl{J. Cosmology Astropart. Phys.}}
\def\jrasc{\ref@jnl{JRASC}}             
\def\memras{\ref@jnl{MmRAS}}            
\def\mnras{\ref@jnl{MNRAS}}             
\def\na{\ref@jnl{New A}}                
\def\nar{\ref@jnl{New A Rev.}}          
\def\pra{\ref@jnl{Phys.~Rev.~A}}        
\def\prb{\ref@jnl{Phys.~Rev.~B}}        
\def\prc{\ref@jnl{Phys.~Rev.~C}}        
\def\prd{\ref@jnl{Phys.~Rev.~D}}        
\def\pre{\ref@jnl{Phys.~Rev.~E}}        
\def\prl{\ref@jnl{Phys.~Rev.~Lett.}}    
\def\pasa{\ref@jnl{PASA}}               
\def\pasp{\ref@jnl{PASP}}               
\def\pasj{\ref@jnl{PASJ}}               
\def\rmxaa{\ref@jnl{Rev. Mexicana Astron. Astrofis.}}%
\def\qjras{\ref@jnl{QJRAS}}             
\def\skytel{\ref@jnl{S\&T}}             
\def\solphys{\ref@jnl{Sol.~Phys.}}      
\def\sovast{\ref@jnl{Soviet~Ast.}}      
\def\ssr{\ref@jnl{Space~Sci.~Rev.}}     
\def\zap{\ref@jnl{ZAp}}                 
\def\nat{\ref@jnl{Nature}}              
\def\iaucirc{\ref@jnl{IAU~Circ.}}       
\def\aplett{\ref@jnl{Astrophys.~Lett.}} 
\def\apspr{\ref@jnl{Astrophys.~Space~Phys.~Res.}}
\def\bain{\ref@jnl{Bull.~Astron.~Inst.~Netherlands}} 
\def\fcp{\ref@jnl{Fund.~Cosmic~Phys.}}  
\def\gca{\ref@jnl{Geochim.~Cosmochim.~Acta}}   
\def\grl{\ref@jnl{Geophys.~Res.~Lett.}} 
\def\jcp{\ref@jnl{J.~Chem.~Phys.}}      
\def\jgr{\ref@jnl{J.~Geophys.~Res.}}    
\def\jqsrt{\ref@jnl{J.~Quant.~Spec.~Radiat.~Transf.}}
\def\memsai{\ref@jnl{Mem.~Soc.~Astron.~Italiana}}
\def\nphysa{\ref@jnl{Nucl.~Phys.~A}}   
\def\physrep{\ref@jnl{Phys.~Rep.}}   
\def\physscr{\ref@jnl{Phys.~Scr}}   
\def\planss{\ref@jnl{Planet.~Space~Sci.}}   
\def\procspie{\ref@jnl{Proc.~SPIE}}   

\let\astap=\aap
\let\apjlett=\apjl
\let\apjsupp=\apjs
\let\applopt=\ao

\newpage
\bibliography{refs}


\end{document}